\documentclass[modern]{aastex62}

\usepackage{subfigure}
\usepackage{comment}
\usepackage{threeparttable}
\usepackage{longtable}
\usepackage{afterpage,booktabs}
\usepackage{xcolor}

\received{ddmmyyyy}
\revised{ddmmyyyy}
\accepted{ddmmyyyy}
\submitjournal{ApJS}

\shorttitle{Bayesian Time-Resolved Spectra of GRB Pulses}
\shortauthors{Yu et al.}
\begin{document}

\title{Bayesian Time-Resolved Spectroscopy of GRB Pulses}

\correspondingauthor{Hoi-Fung Yu}
\email{davidyu@hku.hk}

\author[0000-0001-5643-7445]{Hoi-Fung Yu}
\affil{Department of Physics, KTH Royal Institute of Technology, AlbaNova, 10691 Stockholm, Sweden}
\affil{Oskar Klein Centre for Cosmoparticle Physics, AlbaNova, 10691 Stockholm, Sweden}
\affil{Faculty of Science, The University of Hong Kong, Pokfulam, Hong Kong}

\author{H\"usne Dereli-B\'egu\'e}
\affil{Department of Physics, KTH Royal Institute of Technology, AlbaNova, 10691 Stockholm, Sweden}
\affil{Oskar Klein Centre for Cosmoparticle Physics, AlbaNova, 10691 Stockholm, Sweden}
\affil{Max Planck Institute for Extraterrestrial Physics, Giessenbachstrasse 1, 85748 Garching, Germany}

\author[0000-0002-9769-8016]{Felix Ryde}
\affil{Department of Physics, KTH Royal Institute of Technology, AlbaNova, 10691 Stockholm, Sweden}
\affil{Oskar Klein Centre for Cosmoparticle Physics, AlbaNova, 10691 Stockholm, Sweden}

\begin{abstract}

We performed time-resolved spectroscopy on a sample of 38 single pulses from 37 gamma-ray bursts detected by the \textit{Fermi}/Gamma-ray Burst Monitor during its first 9 years of mission. For the first time a fully Bayesian approach is applied. A total of 577 spectra are obtained and their properties studied using two empirical photon models, namely the cutoff power law and Band model. We present the obtained parameter distributions, spectral evolution properties, and parameter relations. We also provide the result files containing this information for usage in further studies.
It is found that the cutoff power law model is the preferred model, based on the deviance information criterion and the fact that it 
consistently provides constrained posterior density maps. In contrast to previous works, the high-energy power-law index of the Band model, $\beta$, has in general a lower value for the single pulses in this work. In particular, we investigate the individual spectrum in each pulse, that has the largest value of the low-energy spectral indexes, $\alpha$. For these 38 spectra, we find that 60 \% of the $\alpha$ values are larger than $-2/3$, and thus incompatible with synchrotron emission.
Finally, we find that the parameter relations show a variety of behaviours. 
Most noteworthy is the fact that the relation between $\alpha$ and the energy flux is similar for most of the pulses, independent of any evolution of the other parameters. 

\end{abstract}


\section{Introduction}\label{sect:intro}

The study of spectral shapes of the photon flux observed from astrophysical objects is a powerful tool to investigate the underlying physical processes. However, even after half a century of observations, the intrinsic spectral shape of the prompt emission of gamma-ray bursts (GRBs) remains unknown. Although during the past few decades many attempts have been made to fit the spectra with empirical, semi-physical, and physical photon models, we still have not found a comprehensive explanation of the emission mechanism of the prompt emission phase in GRBs. This is partly due to the large diversity in spectral shapes that is observed, which prevents a single and simple  explanation to be found, and partly due to  the inherent difficulties of performing gamma-ray spectroscopy.

Conventionally, mathematical functions (aka.~models) are fit to the observed photon counts. These are usually empirical models with the least possible number of parameters. Physical meaning of the parameters can be interpreted by comparing the values resulted from the fit to the predicted values from the theories. Among the frequently used models are the simple power law, cutoff power law, Band function \citep[e.g.,][]{Band1993}, smoothly broken power law, and the Planck function \citep[aka.~the blackbody spectrum, e.g.,][]{Ghirlanda2003,Ryde2004}. Power laws are usually attributed to non-thermal processes, the Planck function indicates a thermal origin, and the Band function and broken power laws can be either thermal or non-thermal depending on the values of their parameters (i.e., the values of their spectral slopes).

Composite models have also been used to fit GRB spectra. For instance, \citet{Gonzalez2003} first found that including a broader energy range beyond a few MeV, one burst observed by the Compton Gamma-Ray Observatory ({\it CGRO}) could be fitted by a power law, in addition to the Band function which dominates the emission at low energy. Moreover, \citet{Ryde2005} fitted a Planck function plus a power law to {\it CGRO}/BATSE GRBs and found that the Planck component dominates. Similarly, using {\it Fermi} data, \citet{Abdo2009} fitted a Band function plus power law to GRB090902B, while \citet{Ryde2010} fitted a multi-color blackbody instead of the Band component to the same burst. GRB090902B is the most prominent example with the thermal Band or multi-color blackbody dominating over a non-thermal power law. The {\it Fermi}/GBM later confirmed the existence of an additional higher energy power-law component in a number of bursts \citep[e.g.,][]{Ackermann2010,Guiriec2010}. Furthermore, it was also shown that if a blackbody component is added to the non-thermal Band function
the fit quality improves significantly in many cases \citep[e.g.,][]{Guiriec2011,Axelsson2012, Guiriec2013,Burgess2014b,Nappo2017}. Later, \citet{Guiriec2015a} introduced a three-component model, which could be fitted to many bursts. Moreover, \citet{Vianello2018b} reported detection of a high-energy break in two long GRBs \citep[see also,][]{Barat}, and \citet{Oganesyan2018} reported an additional low-energy break in the spectrum of several GRBs.

In contrast to time-integrated spectroscopy \citep[e.g.,][]{Goldstein2012,Gruber2014}, i.e., the whole period of emission (or pulsation in the light curve) is treated as a single time bin, spectroscopy can also be done in a time-resolved manner \citep[e.g.,][]{Yu2016}, i.e., the light curve of the pulsation period is grouped into multiple time bins and spectral analysis is performed in each time bin individually. Indubitably, a burst often displays a varying behavior in its time-resolved emission. As an example of this, \citet{Guiriec2015b} found a pure blackbody at the beginning time of GRB131014A, followed by mixed thermal and non-thermal components in latter time, a property similarly demonstrated in other bursts as well \citep[e.g.,][]{Ghirlanda2003, Ryde2011, Zhang2018}. However, there is no single empirical model found to be preferred for every GRB spectrum.

Gamma-ray burst spectra were early noted to evolve significantly within each pulse \citep{Golenetskii1983,Norris1986}. Therefore, time-integrated spectra (as they are usually called) are actually averaged spectra, hence only time-resolved spectra should be used to directly infer Physics. Alternatively, though, indirect methods can still be used as shown in \citet{Ryde1999}. Several time-resolved spectral catalogues of GRBs exist in the literature \citep[e.g.,][]{Kaneko2006,Yu2016}, but they all make use of the frequentist approach. Similarly, spectra from overlapping pulses are likely to show averaged behaviors, so that separated pulses must be used in order to obtain the cleanest possible spectral results that are suitable to be used to draw physical conclusions. On the other hand, the temporal binning also affects the results of the spectral analysis. If the time bins are too coarse, there is spectral evolution within the bins; if the time bins are too fine, the signal-to-noise (or statistical significance) decreases. Therefore, the time bins must be defined in such a way that they capture the intrinsic variability of the light curve (i.e., they can be treated as ``instantaneous'') while maximising the signal in each bin. The Bayesian block method \citep{Scargle2013, Burgess2014c} that identifies statistically  significant intensity changes in the light curve has shown to be an adequate method for this task. This method results in timebins that only has a small, observed, intensity variation across their duration.

In the current study, we employ Bayesian inference which accounts for relevant prior information. During this process the background is incorporated into the model as nuisance parameter which can be marginalised out. The resulting posterior probability distributions of parameters are obtained by the technique of Markov chain Monte Carlo (MCMC). All parameter uncertainties are characterised by the highest posterior density credible intervals.

In this paper, we present the first systematic study of the time-resolved spectra of individual GRB pulses using full Bayesian analysis method. Our sample is observed by the {\it Fermi}/Gamma-ray Burst Monitor (GBM) during its first 9 years of mission and consists of 38 pulses from 37 bursts. The analysis methods and results are presented in Sect.~\ref{sec:meth} and \ref{sect:results} respectively. We summarise and conclude our findings in Sect.~\ref{sec:summary}. Unless otherwise stated, all error bars are given at the 68\% (1-$\sigma$) Bayesian credible level.

\section{Methods and Results}\label{sec:meth}

\subsection{Burst, Detector, and Pulse Selection}

The {\it Fermi}/GBM has triggered on 2,050 GRBs from July 2008 until March 2017. The GBM consists of 14 detectors, of which 12 are sodium iodide (NaI, named from n0 to n9, na and nb) detectors which cover roughly 8~keV to 1~MeV, and two are non-directional bismuth germanium oxide (BGO, named b0 and b1) detectors which cover roughly 200~keV to 40~MeV \citep{Bissaldi2009,Meegan2009}. This arrangement makes the {\it Fermi}/GBM a powerful all-sky ($\gtrsim8$~sr that is not occulted by the Earth) surveying monitor with a wide energy range over 3 orders of magnitudes. Preliminary GRB data is uploaded to the NASA/HEASARC database minutes after the trigger, including the trigger file and quick-look light curves. Detailed data files with the highest temporal (CTIME and TTE files) and spectral (CSPEC and TTE files) resolutions are downloaded from the spacecraft within hours. This makes the online {\it Fermi}/GBM GRB {  database} a near real-time and most up-to-date GRB data repository.\footnote{The data can be obtained by either visiting \url{https://heasarc.gsfc.nasa.gov/W3Browse/fermi/fermigbrst.html} or using the built-in command of {\tt 3ML} \citep{Vianello2015}.} For the spectral analysis described in Sect.~\ref{subsec:source}, we used the standard \textit{Fermi}/GBM analysis energy ranges: 8~keV to 30~keV and 40~keV to $\sim$850~keV for the NaIs (avoiding the K-edge at 33.17~keV)\footnote{\url{https://fermi.gsfc.nasa.gov/ssc/data/analysis/GBM_caveats.html}}, and $\sim$250~keV to 40~MeV for the BGOs.

The purpose of our study is to follow the spectral evolution during individual emission episodes in the jet environment of the GRB. Therefore, we searched specifically for structures  in the light curve that can be characterised as connected emission activities. We visually inspected {  all} the 256~ms, 512~ms, and 1,024~ms TTE (Time-Tagged Events) light curves and searched for such structures from all of the 2,050 GRBs.  We used those NaI detectors with viewing angles of less than 60 degrees in order to maximise effective area \citep[see][]{Goldstein2012}. In many of the selected cases, the emission episodes consist of individual pulses that are clearly separated by intervals of background level, which is identified as flat or monotonic inter-pulse signal. However, since the shape of any connected emission activity from the jet is not, a priori, known \citep[see, e.g., ][]{Lazzati2013}, we  want to avoid to be too restrictive in our selection. Therefore, we also include in our sample  emission episodes with additional features that can be interpreted as subpulses (that are more prominent than statistical fluctuation). However, these features should clearly be subdominant and  be temporally connected with the main change in intensity. 
{  The variety of connected emission activities that were selected, for this step in defining the sample, are illustrated
by the light curves shown in Figures~\ref{fig:evolution_group1} to \ref{fig:evolution_group10} in the appendix.}
We note that another selection criterion {  to identify emission episodes} could have been chosen, for instance,  requiring a certain shape of the pulses { , described by analytical functions} \citep{Norris1996, Hakkila}. However, any such criteria {  are unnecessarily restrictive, since they} assumes a {  particular, analytical} shape of the pulse, which we want to avoid.
Finally, sometimes solar flares could also contribute to the low-energy channels which cause a broad pulse, however, these are easily identified by their emission characteristics. Such background events are identified and excluded in our study.

We selected 290 long bursts that were identified with at least one of these emission episode, clearly separated by the non-emission background intervals. {  The next step in the selection process is to apply the method of Bayesian Blocks in order to identify spectra for which time-resolved spectroscopy can be performed.} {  This step is the most restrictive and important and  is therefore discussed in Sect.~\ref{subsec:source}.} The sample of {  290 bursts is thus further reduced to 37 bursts}, resulting in a sample of 577 spectra, {  which defines the final sample}.

For the spectroscopy, we follow the procedure of the {\it Fermi}/GBM GRB time-integrated \citep{Goldstein2012,Gruber2014} and time-resolved spectral catalogues \citep{Yu2016} to select at most three NaIs and one BGO for the spectral analysis. The respective TTE and spectral response files are used for the sets of detectors selected. We followed the standard \textit{Fermi}/GBM catalog analysis method to use the suitable response files. All the response files used in this study are automatically generated by the \textit{Fermi}/GBM repository\footnote{\url{https://heasarc.gsfc.nasa.gov/W3Browse/fermi/fermigbrst.html}} according to the location obtained by the Burst Advocate (BA) from the GBM Team within weeks of the detection of the burst. The choice of the burst location and systematics that might affect the spectral analysis results are discussed in \citet{Connaughton2015}.

\subsection{Background Fitting}

One of the NaI detectors {  that recorded the largest value photon counts per second} is used to define the background intervals pre- and post-emission (i.e., the before and after the pulse). These intervals are then applied to all detectors. {  As a standard procedure in GRB background fitting of GBM data,} we fit a polynomial, of order 0 to 4, to each energy channel (128 channels for TTE) of each of the detectors. The optimal order of the polynomial is determined by a likelihood ratio test independently for each energy channel. The polynomial is interpolated into the source and integrated over the source interval to obtain the background photon count flux. The error of the flux in each channel is also computed.

For some bursts, selection of two background intervals were not possible. For two cases in our sample (GRB110817 and GRB130305; Table \ref{tab:pulses}), only one background interval was selected. This happens when the burst occurs right before the entrance of the South Atlantic Anomaly (SAA) region where the detectors must be shut down to avoid damage, or right after the exit of it. For one case (GRB081009, see Table \ref{tab:pulses}) there are several pulses separated by intervals that are background dominated. In order to better constraint the background polynomial shape, in this case, three background intervals were selected, instead. 

\subsection{Light Curve Binning \& Spectral Fitting}\label{subsec:source}

The spectral analysis is done using the Bayesian spectral analysis package {\tt 3ML}\footnote{\url{https://threeml.readthedocs.io/en/latest/notebooks/Bayesian_tutorial.html}} \citep{Vianello2015}. As a first step in time-resolved spectroscopy the light-curves have to be rebinned into adequate intervals. Different methods can be used, for instance, binning by constant time interval ${dt}$, binning by statistical significance, and binning by Bayesian Blocks \citep{Scargle2013}. 
\citet{Burgess2014c} argued that in order to obtain finest time bins (therefore the highest number of time bins) while minimising the effect of mixed spectra caused by intrinsic spectral evolution (photons coming from distinct emission regions in the ejecta could arrive the detector at the same time), the Bayesian Block method should be used. Therefore, for each burst, we rebinned the TTE light curve of the brightest NaI detector into Bayesian Blocks with a correct detection rate for single change point of $p_0=0.01$ \citep[see Eqn.~(11) of][]{Scargle2013}. The Bayesian Block binning is then transferred and applied to all other detectors. We note, though, that the Bayesian Block method assumes that the variability in the light curve is the same over the whole energy range. However, the variability of the light curve might be dominated by the variability in the lowest energy photons, since GRB spectra are, in general, soft \citep[e.g.,][see also Sect.~\ref{sect:results}]{Kaneko2006,Goldstein2012,Gruber2014,Yu2016}. Therefore, there is a possibility that spectral changes in the high energy channels could be missed due to lower signal strength \citep{Guiriec2015a}. Moreover, we note that there is an implicit assumption that spectral variations is small when the variation in the light curve is small. This assumption is based on early studies, e.g. \citet{Golenetskii1983}.

Since our aim is to study the time-resolved spectra of individual pulses, we need at least a few time bins in order to study the spectral evolution within the pulse. \citet{Yu2016} used a similar criterion that required the bursts to have at least five time bins with signal-to-noise ratio $\geq 30$ \citep[see][for detailed derivation and discussion]{Vianello2018}. The statistical significance, $S$, adopted in the current work is a test statistic that incorporates the information of signal-to-noise ratio and suitable for Poisson sources with Gaussian backgrounds \citep[see][for the definition of $S$]{Vianello2018}. We found that the spectral parameters are typically well constrained for bins with statistical significance $S \geq 20$. Therefore, among the (initially selected) 290 bursts, we further require pulses to have at least five Bayesian Block time bins with $S \geq 20$ in order to study their time-resolved spectral evolution. This results in a sample of 38 single pulses in 37 bursts with at least five $S \geq 20$ time bins. This sample is listed in Table.~\ref{tab:pulses}.

Nevertheless, for the purpose of a catalogue, we still aim to present the properties of the selected sample with as little constraints as possible. Therefore, we present below the results of the overall statistics of this sample (without constraint on $S$)\footnote{Note, however, that the relevance of the data points are still provided by the size of the error bar.} as well as the statistics of this sample with $S \geq 20$. For the purpose of inferring physics from the spectral parameters, only bins with $S \geq 20$ should be used, which, for instance, is done in \citet{Ryde2019}.

Several different models are typically used in the spectral analysis of GRBs, e.g., the cutoff power law (CPL)\footnote{This model is also known as the Comptonised model, abbreviated as COMP due to its theoretical relation to the Comptonised spectral shape.} and the Band function \citep[BAND,][]{Band1993}. The GBM GRB time-resolved catalogue \citep{Yu2016} showed that CPL is preferred over the other frequently used models for a majority (70\%) of bursts, according to the Castor C-Statistic \citep[CSTAT, a modified version of the original Cash statistic derived by][]{Cash1979}. Therefore, for the main analysis below, we fit CPL to all time bins of our 38 pulses. In addition, we also fit BAND to all pulses to allow for a comparison to be made between the models. For each time bin, a Poisson distribution for the source and a Gaussian distribution for the background is used to obtain the likelihood function. This is because the background is estimated from a polynomial fit and the source is not.

We inspected all posteriors of the spectra (2 empirical models for 577 spectra, making up a total of 1,154 corner plots) and checked that {\tt 3ML} signals the fit has converged. We also double-checked that the four independent chains used in the MCMC sampling converged to the same maximum.

In Table.~\ref{tab:pulses}, we list the 38 single pulses from 37 bursts that satisfy all these criteria (Col.~1), together with the detectors used (Col.~2), the source and background intervals (Cols.~3-6), the total number of time bins (Col.~7), and the number of time bins with minimum significance of 20 (Col.~8), and the type of relations for parameter pairs $\alpha$-$E_{\rm p}$, $F$-$E_{\rm p}$, and $F$-$\alpha$ (Cols.~9-11), where $\alpha$ is the low-energy power-law index, $E_{\rm p}$ is the spectral peak, and $F$ is the energy flux. The type of spectral evolution for each pulse is listed in Col.~12. In addition, the Spearman's rank coefficient,{ $r$,   for the paramater relations is also listed in the brackets of Cols.~9-11 next to the type.}

Finally, we note that the models used here are empirical in nature. Physical models can be used and directly compared to each other, but this is out of the scope of the current paper. Note also that model selection is based on prior experience and the statistics (frequentist or Bayesian) cannot identify the true model but only can compare competing ones. While the current study focuses on extracting the parameters of the best model from previous experience, a more standardised study on all kinds of empirical and physical models should be done in the future.

\section{Spectral Results}\label{sect:results}

\begin{deluxetable}{cccccccccccc}
\tabletypesize{\scriptsize}
\tablecaption{GRB name (Col.~1), together with the detectors (Col.~2), and the source (Col.~3) and background intervals (Cols.~4-6) used in the analysis. The number of time bins using Bayesian blocks across the source interval (Col.~7), and the number of time bins with statistical significance at least 20 (Col.~8) are also listed. Columns 9-11 list the type of parameter relations, with the Spearman's rank coefficient, $r$, in the brackets. Column 12 lists the evolutionary trend of the peak energy. The detector in brackets is the brightest one, used for background and Bayesian block fitting. \label{tab:pulses}}
\tablehead{ 
\colhead{GRB} & \colhead{Detectors} & \colhead{$\Delta T_{\rm src}$} & \colhead{$\Delta T_{\rm bkg,1}$} & \colhead{$\Delta T_{\rm bkg,2}$} & \colhead{$\Delta T_{\rm bkg,3}$} & \colhead{$N$} & \colhead{$N_{S \geq 20}$} & \colhead{$\alpha-E_{\rm p}$} & \colhead{$F-E_{\rm p}$} & \colhead{$F-\alpha$} & \colhead{Spectral} \\
 \colhead{} & \colhead{} & \colhead{(s)} & \colhead{(s)} & \colhead{(s)} & \colhead{(s)} & \colhead{} & \colhead{} & \colhead{Type($r$)} & \colhead{Type($r$)} & \colhead{Type($r$)} & \colhead{Evolution} \\
\colhead{(1)} & \colhead{(2)} & \colhead{(3)} & \colhead{(4)} & \colhead{(5)} & \colhead{(6)} & \colhead{(7)} & \colhead{(8)} & \colhead{(9)} & \colhead{(10)} & \colhead{(11)} & \colhead{(12)}
}
\startdata
081009140  & (n3)b1 & 0.-10. & $-25.$-$-5.$ & 15.-30. & 60.-80. & 19 & 16 & -(0.12) & 2(0.64) & 1(0.60) & i.t. \\
081009140  & (n3)b1 & 33.-55. & $-25.$-$-5.$ & 15.-30. & 60.-80. & 13 & 6 & -(0.77) & 3($-0.43$) & 2(0.05) & h.t.s. \\
081125496  & (na)nbb1 & -5.-20. & $-20.$-$-10.$ & 30.-50. & ... & 12 & 6 & 1($-0.69$) & 1($-0.19$) & 1(0.74) & h.t.s. \\
081224887  & n6n7(n9)b1 & 0.-25. & $-25.$-$-5.$ & 30.-60. & ... & 10 & 7 & 1(0.83) & 1(0.83) & 1(0.97) & h.t.s. \\
090530760  & (n1)n2n5b0 & $-1.$-180. & $-25.$-$-10.$ & 200.-250. & ... & 10 & 6 & 1(0.76) & 1(0.95) & 1(0.90) & h.t.s. \\
090620400  & (n6)n7nbb1 & $-1.$-25. & $-25.$-$-10.$ & 30.-45. & ... & 11 & 5 & 2($-0.02$) & 2(0.17) & 1(0.48) & i.t. \\
090626189  & (n0)n1b0 & 30.-39. & $-25.$-$-10.$ & 80.-95. & ... & 15 & 8 & 3($-0.12$) & 1(0.15) & 1(0.84) & i.t. \\
090719063  & n7(n8)b1 & $-1.$-25. & $-25.$-$-10.$ & 35.-50. & ... & 15 & 11 & 2(0.65) & 1(0.71) & 1(0.83) & h.t.s. to i.t. \\
090804940  & n3n4(n5)b0 & $-1.$-15. & $-25.$-$-10.$ & 25.-40. & ... & 14 & 8 & 3($-0.15$) & 2(0.9) & 3($-0.22$) & i.t. \\
090820027  & (n2)n5b0 & 25.-60. & $-20.$-10. & 80.-95. & ... & 25 & 19 & 2(0.53) & 1(0.67) & 1(0.79) & flat to i.t. \\
100122616  & (n6)nab1 & $-5.$-40. & $-20.$-$-10.$ & 50.-80. & ... & 14 & 5 & 2($-0.69$) & 2($-0.83$) & 1(0.78) & i.t. to ? \\
100528075  & n6(n7)nbb1 & $-5.$-60. & $-30.$-$-10.$ & 66.-100. & ... & 16 & 7 & 3($-0.44$) & 3($-0.17$) & 1(0.72) & h.t.s. \\
100612726  & n3n4(n8)b0 & $-2.$-20. & $-30$.-$-5.$ & 25.-100. & ... & 12 & 6 & 1(0.05) & 1(0.03) & 1(0.92) & h.t.s. \\
100707032  & n7(n8)b1 & 0.-30. & $-20.$-$-5.$ & 40.-100. & ... & 19 & 13 & 1(0.58) & 1(0.57) & 1(0.97) & h.t.s. \\
101126198  & (n7)n8nbb1 & $-5.$-40. & $-30.$-$-15.$ & 50.-80. & ... & 15 & 7 & 1($-0.15$) & 2(0.38) & 1(0.54) & flat to h.t.s. \\
110721200  & (n6)n7n9b1 & $-1.$-20. & $-25.$-$-10.$ & 35.-50. & ... & 12 & 9 & 1(0.22) & 1(0.35) & 2(0.51) & h.t.s. to s.t.h. \\
110817191  & n6n7(n9)b1 & $-1.$-11. & $-20.$-$-7.$ & ... & ... & 9 & 5 & 1(0.21) & 1(0.26) & 1(0.98) & h.t.s. \\
110920546  & (n0)n1n3b0 & $-1.$-160. & $-15.$-$-5.$ & 175.-200. & ... & 14 & 10 & 2($-0.86$) & 1(0.88) & 1($-0.71$) & h.t.s. \\
111017657  & (n6)n7n9b1 & $-5.$-20. & $-25.$-$-10.$ & 35.-50. & ... & 13 & 6 & 1($-0.10$) & 1($-0.25$) & 1(0.92) & h.t.s. to s.t.h. \\
120919309  & (n1)n2n5b0 & $-2.$-35. & $-25.$-$-5.$ & 60.-100. & ... & 15 & 6 & 1(0.48) & 1(0.55) & 1(0.88) & i.t. \\
130305486  & n6(n9)nab1 & $-3.$-35. & 50.-70. & ... & ... & 13 & 8 & 2($-0.82$) & 2($-0.44$) & 1(0.81) & h.t.s. to flat \\
130612456  & n6(n7)n8b1 & $-1.$-15. & $-25.$-$-10.$ & 25.-45. & ... & 11 & 6 & 3(0.11) & 2(0.54) & 1(0.86) & flat \\
130614997  & (n0)n1n3b0 & $-1.$-9. & $-25.$-$-10.$ & 20.-45. & ... & 8 & 5 & 3($-0.20$) & 2(0.40) & 1(0.60) & h.t.s. \\
130815660  & (n3)n4n5b0 & $-1.$-47. & $-25.$-$-10.$ & 55.-75. & ... & 13 & 5 & 1($-0.27$) & 1($-0.38$) & 1(0.93) & h.t.s. \\
140508128  & (na)b1 & $-1.$-15. & $-40.$-$-10.$ & 100.-150. & ... & 18 & 11 & 3(0.28) & 2(0.47) & 1(0.91) & i.t. \\
141028455  & (n6)n7n9b1 & 0.-40. & $-30.$-$-10.$ & 50.-100. & ... & 18 & 12 & 2(0.63) & 1(0.37) & 1(0.85) & h.t.s. \\
141205763  & (n2)n5b0 & $-2.$-20. & $-40.$-$-10.$ & 25.-80. & ... & 14 & 5 & 1(0.46) & 2(0.75) & 1(0.75) & i.t. \\
150213001  & n6n7(n8)b1 & $-1.$-10. & $-25.$-$-10.$ & 20.-45. & ... & 24 & 19 & 1(0.03) & 1(0.21) & 1(0.46) & h.t.s. to i.t. \\
150306993  & (n4)b0 & $-1.$-25. & $-25.$-$-10.$ & 35.-55. & ... & 11 & 7 & 2(0.93) & 2(0.83) & 1(0.83) & h.t.s. \\
150314205  & n1(n9)b1 & $-1.$-18. & $-25.$-$-10.$ & 30.-55. & ... & 20 & 14 & 1($-0.23$) & 1($-0.36$) & 1(0.85) & h.t.s. to s.t.h. \\
150510139  & n0n1(n5)b0 & 0.-50. & $-25.$-$-10.$ & 65.-95. & ... & 30 & 16 & 3(0.13) & 1(0.13) & 1(0.81) & s.t.h. to h.t.s. \\
150902733  & (n0)n1n3b0 & $-1.$-25. & $-25.$-$-10.$ & 30.-55. & ... & 22 & 14 & 1(0.13) & 1(0.28) & 1(0.85) & h.t.s. to i.t. to h.t.s. \\
151021791  & n9(na)b1 & $-1.$-10. & $-25.$-$-10.$ & 25.-45. & ... & 10 & 5 & 2(0.86) & 1(0.75) & 1(0.64) & h.t.s. \\
160215773  & n3n4(n5)b0 & 160.-200. & 100.-150. & 250.-300. & ... & 19 & 11 & 2($-0.60$) & 2(0.84) & 1($-0.54$) & i.t. \\
160530667  & n1(n2)n5b0 & $-2.$-25. & $-40.$-$-10.$ & 40.-80. & ... & 22 & 19 & 2(0.73) & 1(0.84) & 1(0.87) & s.t.h. to h.t.s. \\
160910722  & n1n2(n5)b0 & 7.-20. & $-40.$-$-10.$ & 40.-80. & ... & 15 & 14 & 1(0.36) & 1(0.15) & 2(0.86) & h.t.s. \\
161004964  & n3(n4)b0 & $-2.$-25. & $-40.$-$-10.$ & 40.-80. & ... & 11 & 5 & 1(0.20) & 2(0.02) & 1(0.85) & h.t.s. \\
170114917  & n1(n2)nab0 & $-1.$-20. & $-25.$-$-10.$ & 35.-65. & ... & 15 & 9 & 2($-0.34$) & 1($-0.17$) & 1(0.87) & h.t.s. to ? 
\enddata
\end{deluxetable}

\begin{deluxetable}{ccccccccccc}
\tabletypesize{\scriptsize}
\tablecaption{The values of the average and standard deviation of the parameter distributions. For $E_{\rm p}$ and $E_{\rm c}$, only values within the GBM energy range (8~keV--40~MeV) are used in the calculation. \label{tab:parameter_stat}}
\tablehead{ 
\colhead{Model} & \colhead{$\alpha$} & \colhead{$\log_{10}(E_{\rm p}/{\rm keV})$} & \colhead{$\log_{10}(F$/$10^{-6}$~erg$^{-1}$~s$^{-1}$~cm$^{-2}$)} & \colhead{$\log_{10}(E_{\rm c}/{\rm keV})$} & \colhead{$\beta$} 
}
\startdata
CPL & $-1.07\pm0.63$ & $\log_{10}(331)\pm0.53$ & $\log_{10}(1.17)\pm0.77$ & $\log_{10}(457)\pm0.68$ & ... \\
CPL$_{S\geq20}$ & $-0.79\pm0.43$ & $\log_{10}(234)\pm0.44$ & $\log_{10}(2.86)\pm0.44$ & $\log_{10}(206)\pm0.42$ & ... \\
BAND & $-0.31\pm0.84$ & $\log_{10}(170)\pm0.60$ & $\log_{10}(1.39)\pm0.68$ & ... & $-3.18\pm0.66$ \\
BAND$_{S\geq20}$ & $-0.59\pm0.41$ & $\log_{10}(193)\pm0.44$ & $\log_{10}(2.90)\pm0.43$ & ... & $-3.23\pm0.68$ 
\enddata
\end{deluxetable}

\begin{figure*}
\centering

\subfigure{\includegraphics[width=0.45\linewidth]{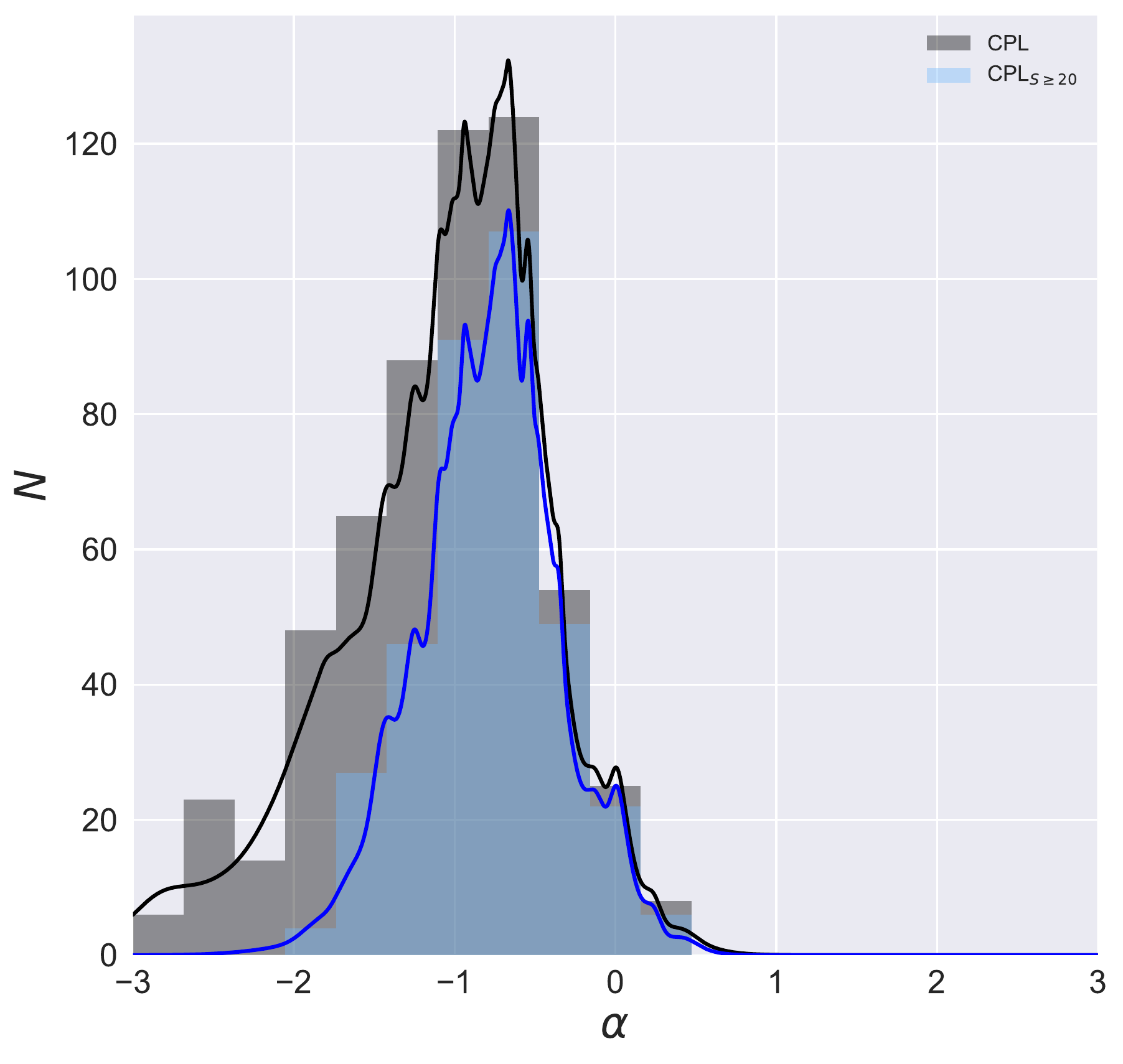}}
\subfigure{\includegraphics[width=0.45\linewidth]{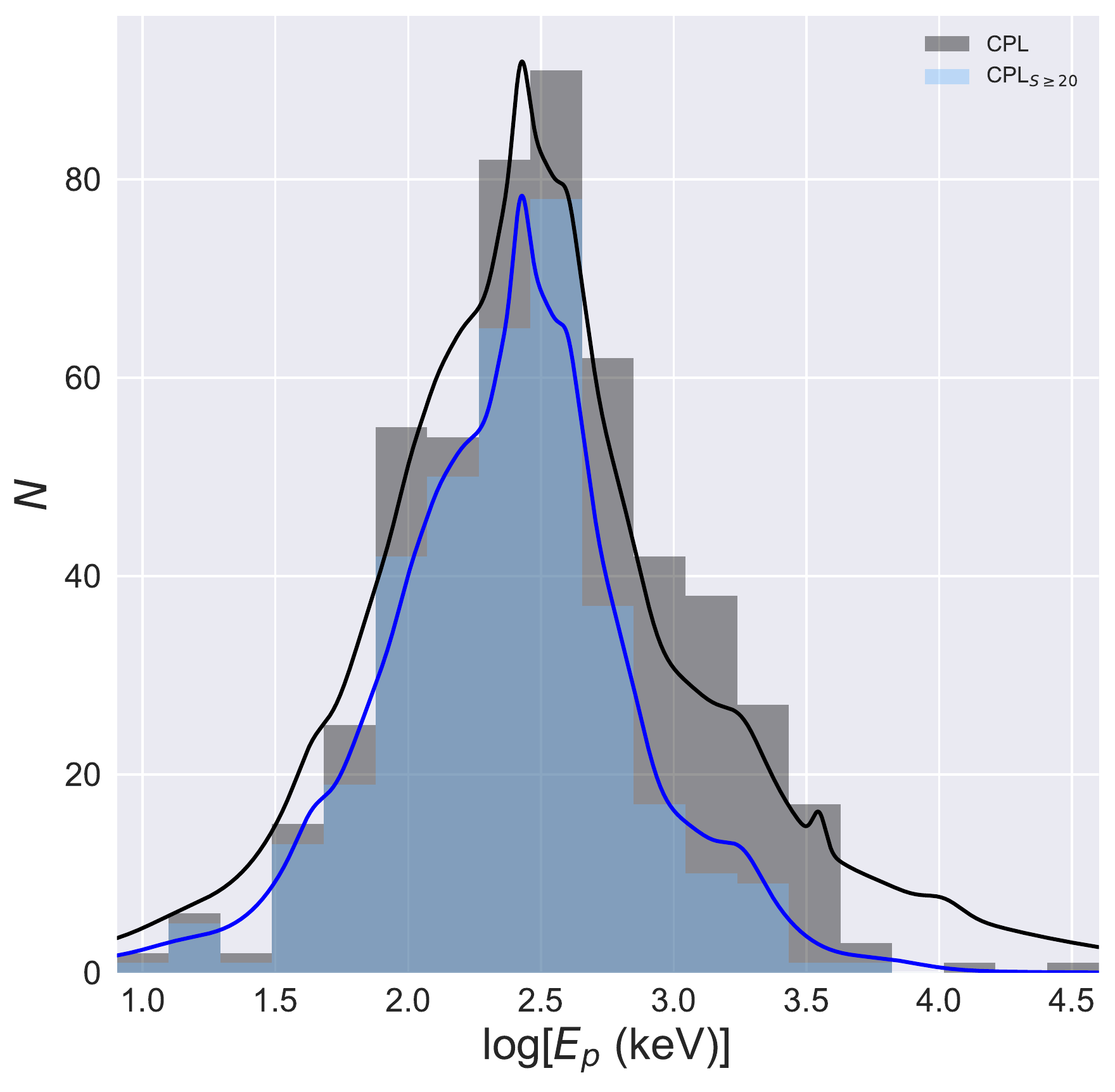}}

\subfigure{\includegraphics[width=0.45\linewidth]{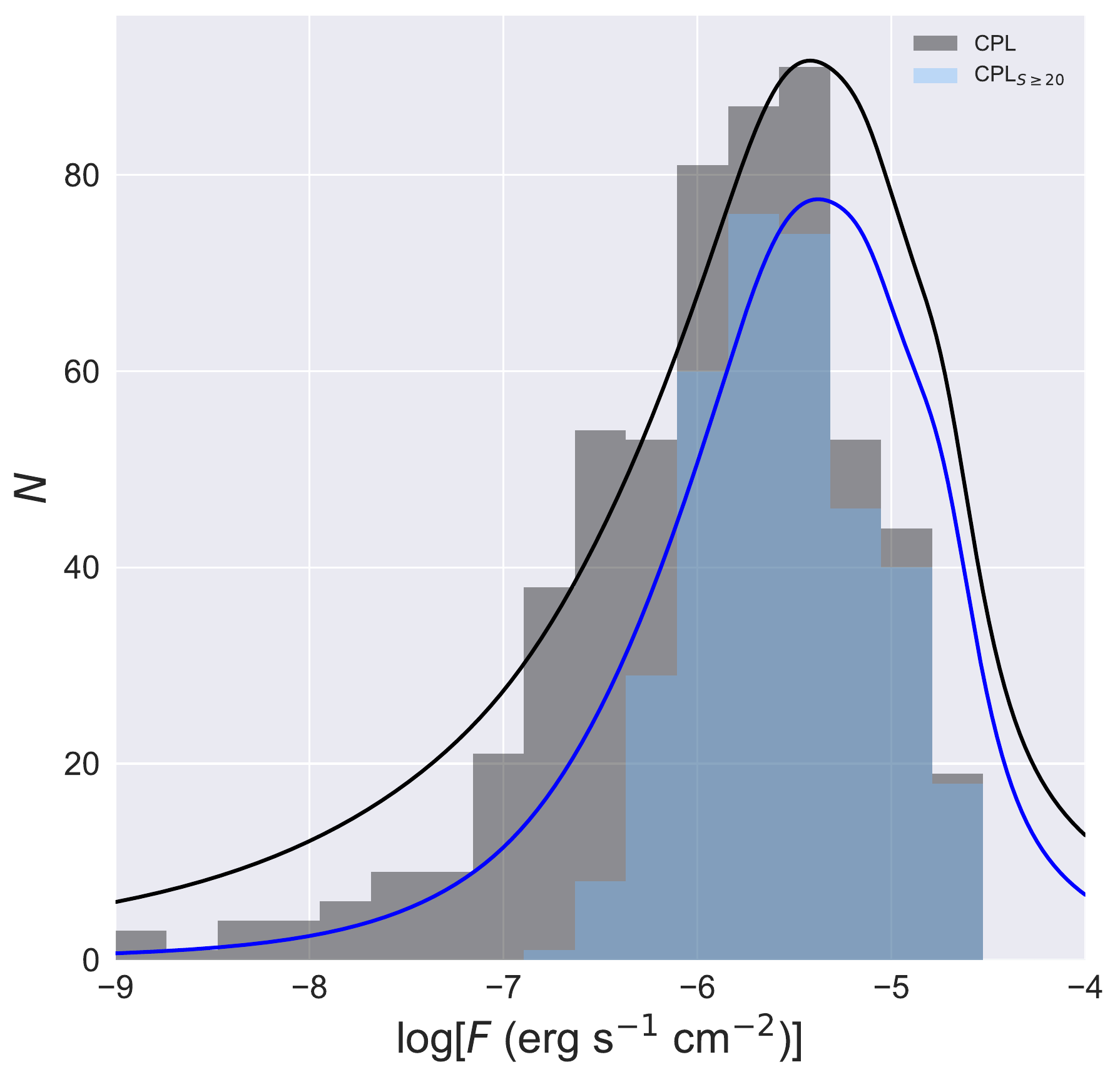}}
\subfigure{\includegraphics[width=0.45\linewidth]{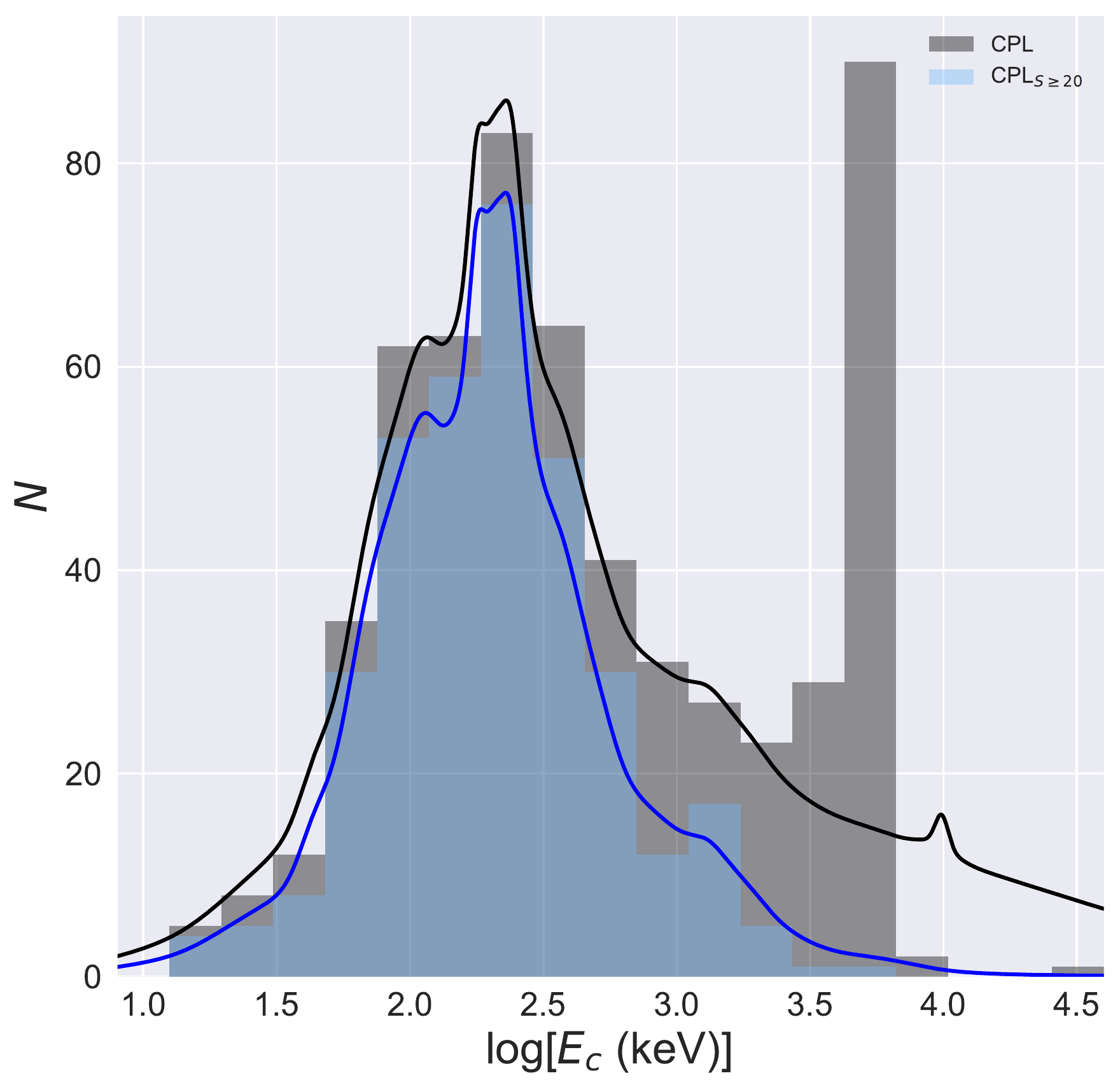}}

\caption{Parameter distributions of the fitted parameters of CPL. Black histogram shows the distributions regardless of significance, and blue histogram shows the distributions with $S \geq 20$. The curves represent the kernel density estimation (KDE) of the distributions, using Gaussian kernels where the standard deviation is set to the larger one of the asymmetrical errors. For $E_{\rm p}$ and $E_{\rm c}$, only the values within the GBM energy range (8~keV-40~MeV) are shown.
\label{fig:CPL_KDE}}
\end{figure*}

\begin{figure*}
\centering

\subfigure{\includegraphics[width=0.45\linewidth]{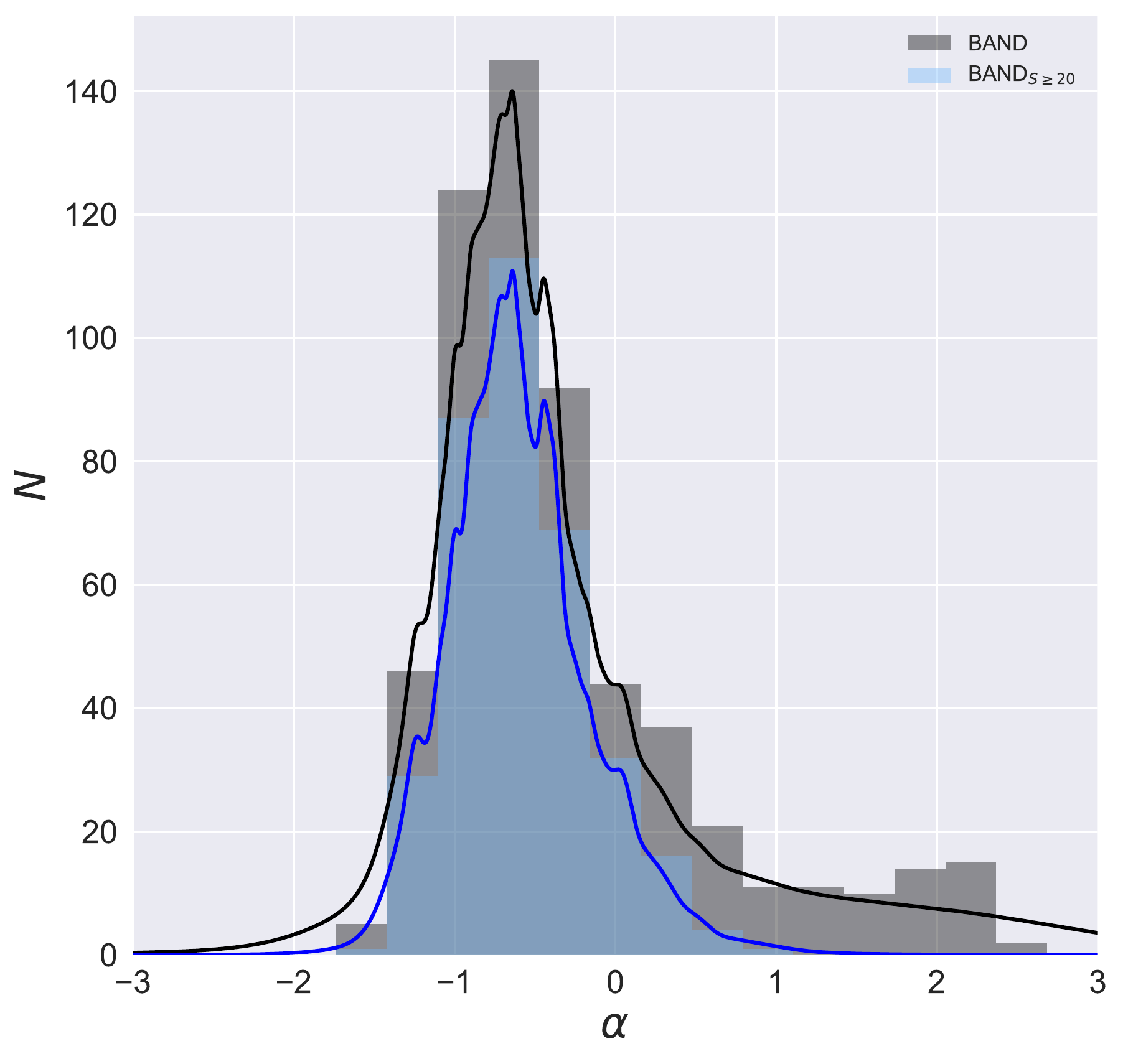}}
\subfigure{\includegraphics[width=0.45\linewidth]{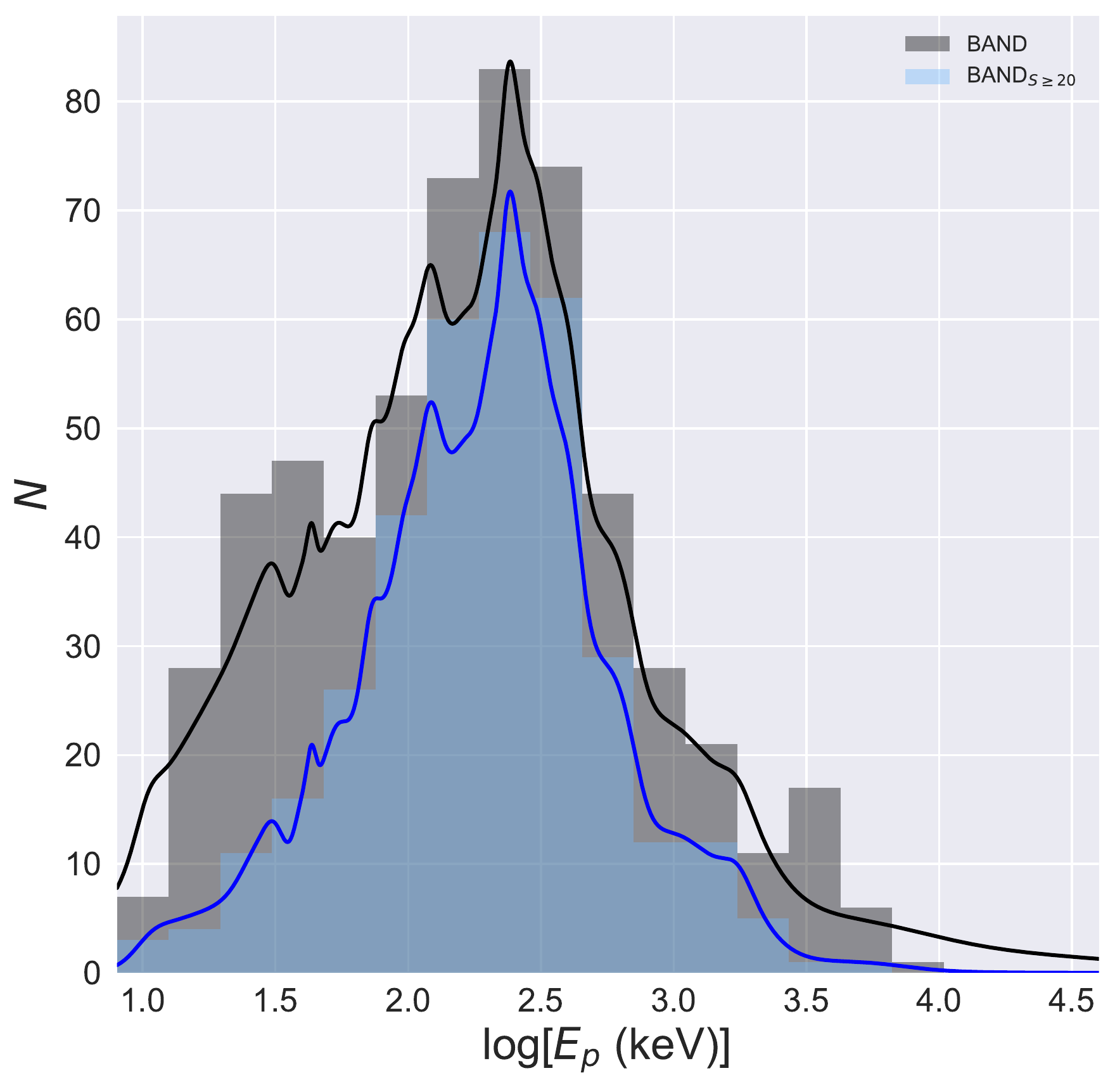}}

\subfigure{\includegraphics[width=0.45\linewidth]{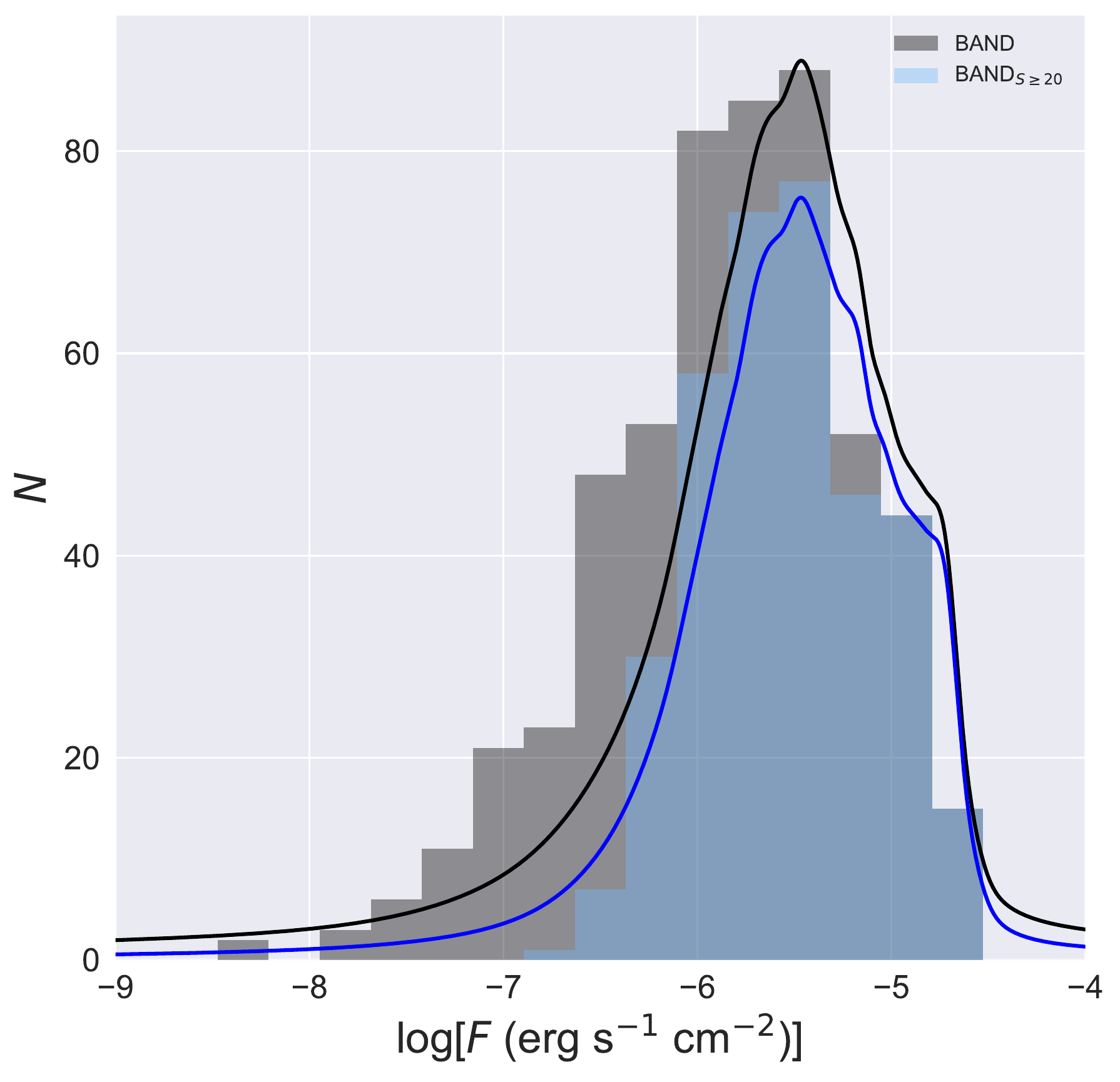}}
\subfigure{\includegraphics[width=0.45\linewidth]{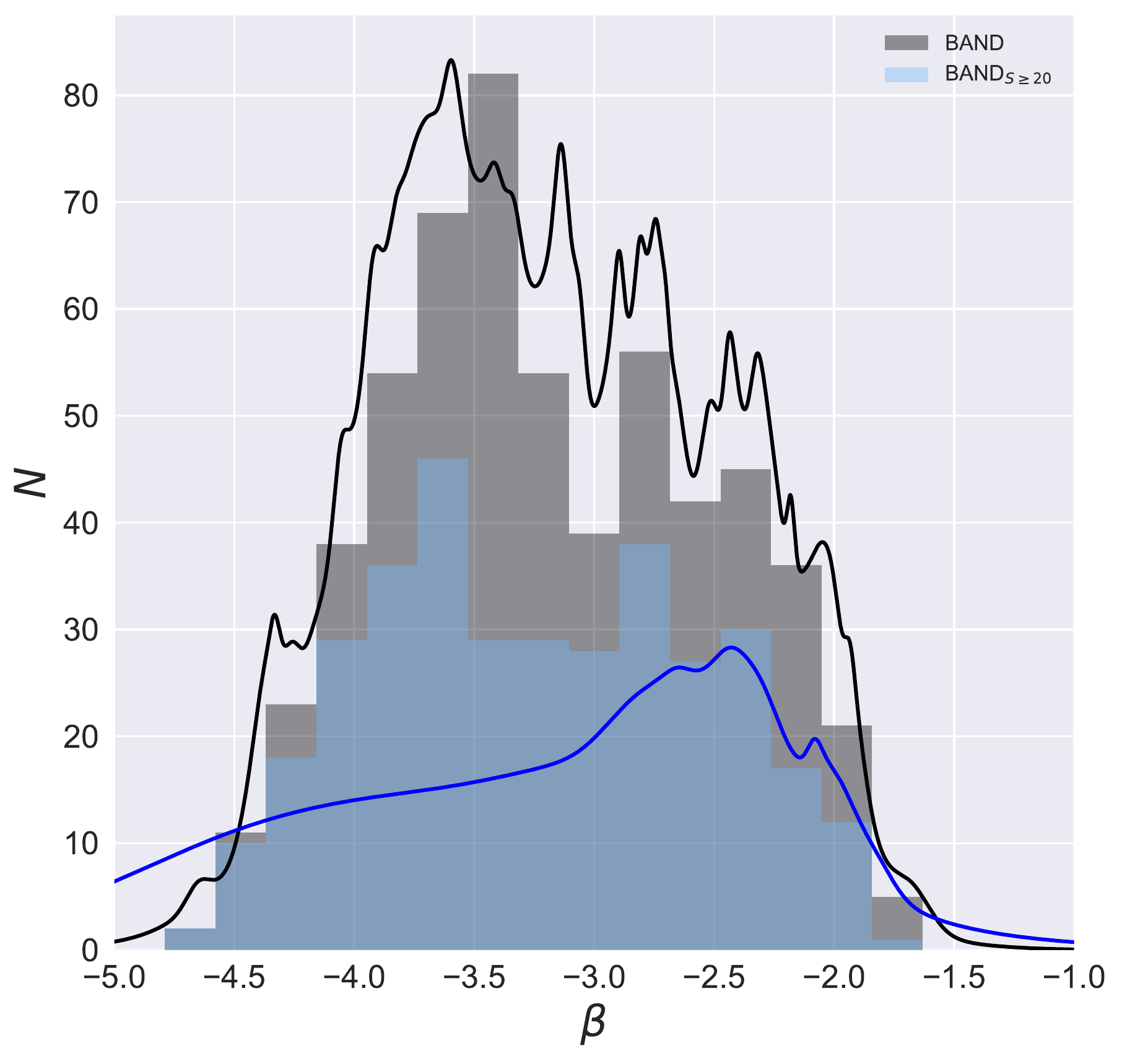}}

\caption{Parameter distributions of the fitted parameters of BAND. Black histogram shows the distributions regardless of significance, and blue histogram shows the distributions with $S \geq 20$. The curves represent the kernel density estimation (KDE) of the distributions, using Gaussian kernels where the standard deviation is set to the larger one of the asymmetrical errors. For $E_{\rm p}$ only the values within the GBM energy range (8~keV-40~MeV) are shown.
\label{fig:BAND_KDE}}
\end{figure*}

The complete fitting results for the CPL and BAND models of all 577 spectra are listed the table of Appendix~\ref{app:tables}. For each pulse, we list the start and stop times of the Bayesian blocks (Cols.~1 and 2), the significance $S$ (Col.~3), the CPL fitted parameters (normalisation $K$ (ph~s$^{-1}$~cm$^{-2}$~keV$^{-1}$), low-energy power-law index $\alpha$, and cutoff energy $E_{\rm c}$, (keV, Cols.~4-6), the derived CPL peak energy $E_{\rm p}$ (keV, Col.~7), the CPL energy flux, $F$ (erg~s$^{-1}$~cm$^{-2}$), (Col.~8), the BAND fitted parameters (normalisation $K_{\rm BAND}$ (ph~s$^{-1}$~cm$^{-2}$~keV$^{-1}$), low-energy power-law index $\alpha_{\rm BAND}$, high-energy power-law index $\beta_{\rm BAND}$, and peak energy $E_{\rm p, BAND}$, (keV, Cols.~9-12), the BAND energy flux, $F_{\rm BAND}$ (erg~s$^{-1}$~cm$^{-2}$), the difference in the deviance information criterion (DIC) between CPL and BAND, $\Delta{\rm DIC}={\rm DIC}_{\rm BAND}-{\rm DIC}_{\rm CPL}$ (Col.~14), and the effective number of parameters of CPL and BAND, $p_{\rm DIC}$ and $p_{\rm DIC, BAND}$ (Cols.~15 and 16).

We also provide the analysis result files in {\tt FITS} format for every time bin, which are available at \dataset[10.5281/zenodo.2601901]{\doi{10.5281/zenodo.2601901}}. They provide complete information of the fits such as the parameter values, covariance matrices, and the statistical information criteria. They can be read readily by {\tt 3ML} to plot the resulting spectra and the posterior probability distributions. The results can be used for further studies of the spectra of these pulses.

\subsection{CPL vs. BAND: which one is ``better''?}

We have fitted the data with the empirical models which have been used as standard models in the field and which have been shown to be compatible with the data \citep[e.g.,][]{Kaneko2006,Goldstein2012,Gruber2014,Yu2016}. In these catalogues, the empirical model fits have also been compared to each other, using the difference in CSTAT. Similarily, in Bayesian statistics, model comparison is done using the so-called information criteria. However, the ``best'' information criterion to use is an active research topic in Bayesian statistics \citep[see, e.g.,][for a recent discussion]{Gelman2014}. In this paper, we compare models by adopting the deviance information criterion \citep[DIC,][]{Spiegelhalter2002}, defined as ${\rm DIC} = -2 \log[p({\rm data}|\hat{\theta})] + 2p_{\rm DIC}$, where $\hat{\theta}$ is the posterior mean of the parameters and $p_{\rm DIC}$ is a term to penalise the more complex model for over-fitting \citep[see, Sect.~3.3 of][]{Gelman2014}.\footnote{The $p_{\rm DIC}$ approaches the total number of parameters of the model when the posterior mean and mode are similar. However, when the posterior is highly skewed, it can become negative. We found that negative $p_{\rm DIC}$ are associated with time bins having low signal-to-noise ratio. In such cases, the value of DIC indeed cannot determine whether a model is preferred. In the extreme example where low signal is present, all model will be performed equally ``bad'', it this case the values of DIC are not trustable.}

The values of the difference between BAND's and CPL's DIC, defined as $\Delta{\rm DIC}={\rm DIC}_{\rm BAND}-{\rm DIC}_{\rm CPL}$, are listed in Col.~14 of Tables~\ref{tab:fitresults} to \ref{tab:fitresults_last}. Since DIC is defined as the negative logarithm of the probability of predicting the observed data given the posterior mean, a positive value of $\Delta{\rm DIC}$ would mean that the CPL is preferred given the observed data (as seen in Col.~14).
%a lot of spectra actually have $|\Delta{\rm DIC}| \lesssim 10$ means that CPL model fit the data well.

However, just like any other statistical measures, attempting to summarise the multi-dimensional posterior distribution in just one number can often be misleading. In some cases, we see that $|\Delta{\rm DIC}|$ can be as large as hundreds of thousand. It is of course very dangerous to blindly believe such a number and claim that one model is exceedingly better than the other. Thus, we need to check the values of $p_{\rm DIC}$ for both models (Cols.~15 and 16).
%As mentioned, this is a term for penalising over-complex models and statistically can be interpreted as the effective number of parameters \citep[see, Sect.~3.3 of][]{Gelman2014}. Its value can vary across spectra even for a single model, however, if one finds its value highly negative, this means that the posterior mode and mean are very far away from each other. In other words, the posterior is highly skewed and/or multi-modal, which is a signal of a bad fit. 
We checked that in almost all of the cases for which the $\Delta{\rm DIC}$ is highly negative, so is $p_{\rm DIC, BAND}$. We found that when ${\rm DIC}_{\rm BAND} < {\rm DIC}_{\rm CPL}$, $p_{\rm DIC, BAND} \ll 0$ in most cases.

The posterior corner plot contains the 2-dimensional probability density maps for each pair of parameters. The marginal probability for each parameter is also computed by the integral of the conditional probability over all but the desired parameter. An acceptable fit is indicated when the probability density map is centred within the prior limits. When the probability density of the normalisation increases towards zero, upper limits can only be inferred. A check to the posteriors of those Band fits with highly negative $p_{\rm DIC, BAND}$ reveals that the normalisation is often small, indicating that the addition of an extra power-law segment has a negative impact to the fit.

%Alternatively, the goodness-of-fit of a physical model in Bayesian statistics can be assessed though the so-called method of posterior predictive check \citep[see, e.g.,][]{Burgess2018b}.

%In this paper we made use of the conclusion from previous catalogues and assumed that the CPL and BAND are good empirical descriptions that are compatible with the observed spectra. 
%%%%%%%%
%Alternative models have been used, both empirical, such as the smoothly broken power-law complicated \citep{Ryde1999}, and composite models,  (e.g., \citet{Guiriec2013}; see also \S \ref{sect:intro}) and physical models synchrotron \citep[e.g.][]{Baring} and  photospheric models \citep{Ghirlanda2003, Ryde2004}.
%%%%%%%%%%%
%} 

%Since the CPL and BAND are not nested, the {  mathematical} interpretation of the low-energy power-law slope and peak energy of CPL and BAND are indeed not the same. Therefore, we adopt the principle that when trying to explain time-resolved parameter evolution or relation, it is better

In this study, we use the same empirical model throughout the whole burst for consistency. However, when there is enough data in the high-energy range (higher than a few 100 keV where the spectral peak usually resides), the Band function might be preferred as indicated by the value of $\Delta{\rm DIC}$. This usually occurs around the peak time in the light curve. 
%The fact that CPL and BAND do not have identical results shows that empirical models are, at best, approximate descriptions of the actual physical process. Nevertheless, empirical model fitting 
Although the models used in the current study are empirical, they are useful in extracting spectral parameters and their evolution, which can give an indication of the physical model underlying the emission. Such investigations can thus motivate physical models to be fit directly to the data.

In summary, we found that the cutoff power-law model is the preferred model, since it systematically has a lower DIC value. In addition, the resulting parameters for the CPL fits are constrained within the prior ranges more often than for the Band function fits. This result is consistent with previous GBM spectral catalogues. However, we note that the preference of the exponential cutoff model could be due to the lack of photon counts at high energy in the GBM energy channels.

\subsection{Parameter Distributions}\label{subsubsec:paradist}

Figures~\ref{fig:CPL_KDE} and \ref{fig:BAND_KDE} show the overall parameter distributions, including $\alpha$, $E_{\rm c}$, and $\beta$, $E_{\rm p}$ and the derived parameters, $E_{\rm p}$ (for CPL) and $F$. The average values and standard deviations of the distributions are listed in Table~\ref{tab:parameter_stat}. 

Since the errors of the fitted spectral parameters could not be taken into consideration in the histograms, we performed kernel density estimation (KDE) on individual parameter distributions. A Gaussian kernel is chosen. In order to be conservative, the standard deviation of the Gaussian kernel is set equal to the larger one of the asymmetrical errors. The KDEs are overlaid on Figs.~\ref{fig:CPL_KDE} and \ref{fig:BAND_KDE}.

The average values of $\alpha$, $E_{\rm p}$, and $F$ distributions for CPL and BAND agree within 1-$\sigma$ for $S\geq20$. Since the data and analysis conditions of the current study is different from previous catalogues \citep[e.g.,][]{Gruber2014,Yu2016}, the parameter distributions shown here should not be treated as a one-to-one direct comparison. Nevertheless, we still find that the distributions of $\alpha$ and $E_{\rm p}$ are in agreement with previous time-resolved catalogues \citep[e.g., see Table~2 and Fig.~3 of][]{Yu2016}. Therefore, the frequentist and Bayesian analysis give consistent results. The distribution of $\beta$ that we obtained has lower values than that of \citet{Yu2016}, who did not distinguish between single and composite pulses (c.f. lower right panel of Fig.~3 therein). This indicates that single pulses are in general softer, and that the higher values of $\beta$ might be a result of overlapping spectra from composite pulses which contain photons from various emission sites and times. 

It is observed that the majority of the low-significance data points in the $\alpha$-$E_{\rm c}$ plot have $\alpha < -2$ and $E_{\rm c} \sim 5$~MeV, which is reflected by an unexpected peak at 5~MeV in the $E_{\rm c}$ histogram. 
%It is necessary to investigate whether this peak could be real or artificial. 
First, we noticed that when plotting the $E_{\rm c}$ distribution of time bins with $S \geq 20$ only, the peak at 5~MeV completely disappeares. Second, these are spectra not from one particular burst but from either the beginning or the end of multiple bursts. Third, we also noticed that noise dominates at energies $\gtrsim 1$~MeV, resulting in overall lower significance for the time bin. Last but not least, this peak does not show up in the KDE. This implies the errors on those values of $E_{\rm c}$ are very large, which means that the Bayesian inference struggled to find a cutoff point. We therefore repeated the spectral analysis on these time bins using a simple power law, and found that they are indeed well fit by a single power law. Since noise dominates at energy $E \gtrsim 1$~MeV for these spectra, the value of $E_{\rm c}$ cannot go beyond 5~MeV and is highly uncertain as indicated by the error bars. This indicates that for low-significance time bins, the spectrum can be sufficiently described by a single power law and a spectral break is not necessary.

%We further demonstrate this point by Fig.~\ref{fig:compare} which shows a plot of four CPLs with different $E_{\rm c}$ together with a simple power law with the same spectral index as the CPLs ($-2.5$ in this example). We found that at $E_{\rm c} \sim 1$-5~MeV the CPL mimics the simple power law within half an order of magnitude. Mathematically, as $E_{\rm c}$ becomes larger, the CPL mimics more of a simple power law. 

%The distribution of $E_{\rm c}$ show an unexpected peak at 5~MeV. This peak is indeed artificial and caused by the limited data quality above $E\sim 1$~MeV in the BGOs. An inspection to each spectrum reveals that these are all very low-significance time bins ($S<10$). An analysis of this is in Sect.~\ref{subsubsec:global}. In short, the spectrum is trying to mimic a simple power law with a cutoff, and the value of 5~MeV is a combined effect of BGO noise and the observing energy window.

\begin{figure}
 \includegraphics[width=\columnwidth]{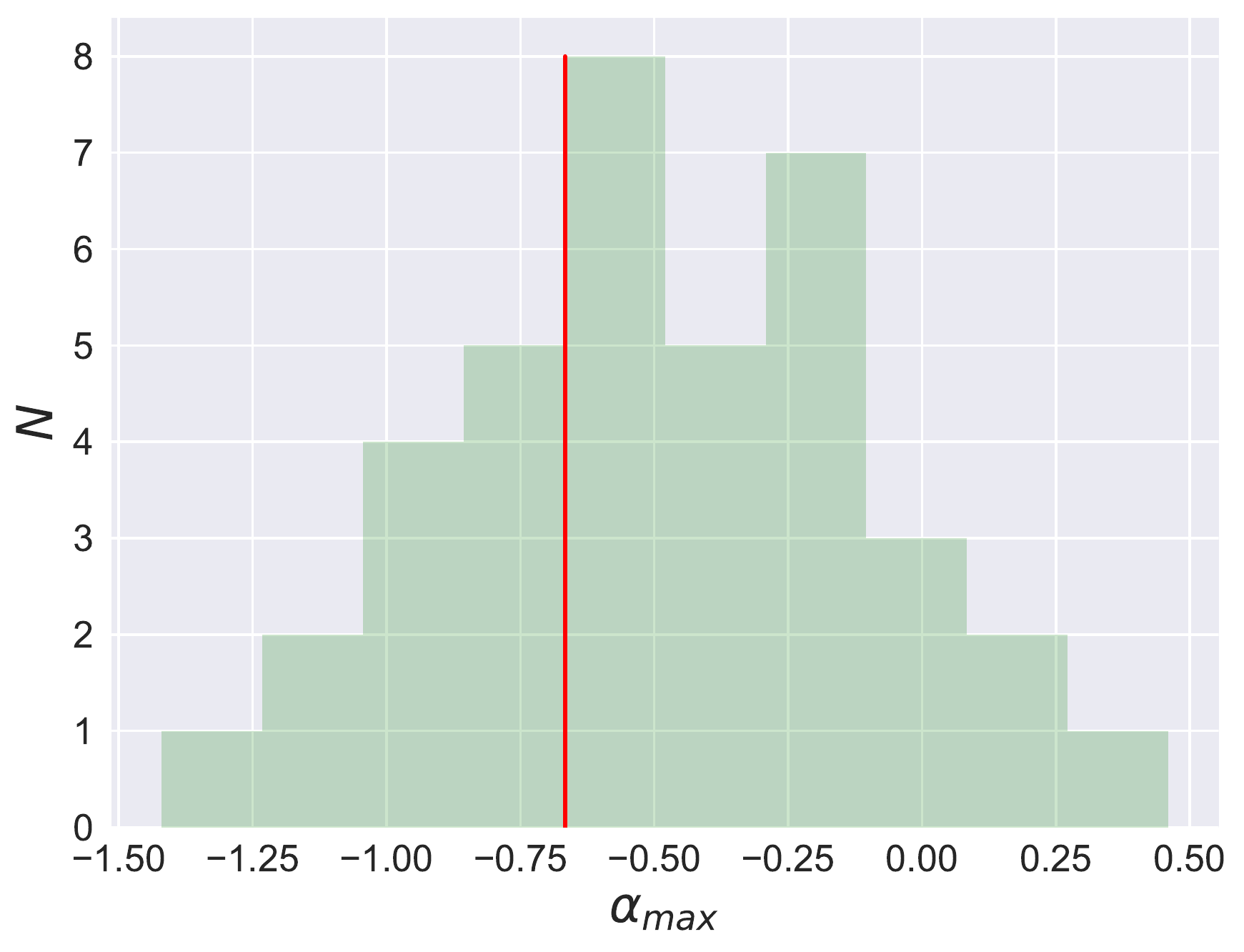}
 \caption{Histogram of the maximal value of $\alpha$ in each of the 38 pulses in the sample. The red line indicated the line-of-death for the synchrotron interpretation for individual pulses, assuming that the same emission mechanism operates throughout the pulse. {60 \% of the pulses have $\alpha_{\rm max}$ (within a $1\sigma$ lower limit of the error) that is incompatible with synchrotron emission.}
 \label{fig:new_histogram}}
\end{figure}

In Figs.~\ref{fig:CPL_KDE} and \ref{fig:BAND_KDE}, the $\alpha$--distributions contain all analysed time-bins. A consequence of this is that individual bursts contribute to the distribution with a varying number of bins. This leads to a bias towards bursts with many time bins. In order to avoid such a bias, one can instead interpret the distribution containing only one bin per burst. Furthermore, 
the best bin to characterise the emission mechanism during a pulse is the bin containing the 
largest value of $\alpha$ indexes in each pulse/burst. The reason for this is that physical models typically have a limit to how hard the spectra are allowed to get. Therefore, if one single bin violates such a limit the corresponding emission model is rejected by the data. This is, of course, under the assumption that a single emission mechanism is responsible for the full duration.
%Such a distribution then gives an indication of how large fraction of the bursts (and not time-bins) that are consistent or not with an emission model
We, therefore, identify the largest value of $\alpha$ in each of the 38 pulses in the sample:
$\alpha_{\rm max} \equiv \max (\alpha(t))$. We present their histogram in Figure~\ref{fig:new_histogram} in which we also plot the $\alpha = -2/3$-line, which is the "line-of-death" for synchrotron emission. 
In order to calculate the fraction of $\alpha_{\rm max}$-values that are incompatible with synchrotron emission, i.e., the fraction of bins lying to the right side of the red line, we identify the cases for which the 1$\sigma$ lower limit of the $\alpha_{\rm max}$ is larger than $-2/3$. 
We find that a majority of the pulses (60 \%) are inconsistent with synchrotron emission, using this criterion. This fraction is significantly larger than what is found by applying the line-of-death to the full distribution of $\alpha$-values \citep[][see also Ghirlanda et al. 2002]{Preece1998}. %A similar conclusion was drawn by \cite{Ghirlanda2002} who found that in 44 \% of cases $\alpha$ violates synchrotron emission around the emission peak. 
%However, if the emission mechanism is assumed to vary during a pulse, then the distribution in Figure~\ref{fig:para_histogram} is the relevant one.

\subsection{Spectral Evolution}\label{subsubsec:specevo}

In Appendix \ref{app:plotsevolution}, we show the CPL and BAND parameter evolutions across the duration of each pulse, with color scale from light blue (start) to deep blue (end) showing temporal evolution and the light curve overlaid (Figs.~\ref{fig:evolution_group1} to \ref{fig:evolution_group10}).
%Figures~\ref{fig:evolution_group1} to \ref{fig:evolution_group10} shows the evolution of the spectral parameters of CPL and BAND across the bursting duration of the pulses from pale to deep colors. 
Data points with red, orange, yellow, and no circles indicate statistical significance $S\geq20$, $20>S\geq15$, $15>S\geq10$, and $S<10$, respectively. Many of the low-significance data points are not constrained, as seen from the huge negative-side error bars.

%In the left panels, the evolution of the power-law indices, $\alpha$ and $\beta$, are shown in blue ($\alpha$ of CPL), green ($\alpha_{\rm BAND}$ of BAND), and purple ($\beta_{\rm BAND}$ of BAND); in the middle panels, the evolution of the peak energy $E_{\rm p}$ are shown in blue ($E_{\rm p}$ of CPL) and green ($E_{\rm p, BAND}$ of BAND); and in the right panels, the evolution of the flux $F$ are shown in blue ($F$ of CPL) and green ($F_{\rm BAND}$ of BAND).

It is observed that the values of the low-energy spectral indices of CPL and BAND, $\alpha$ and $\alpha_{\rm BAND}$, are approximately equal to within errors and track each other during the main emission periods of the pulses (which are also the most significant time bins, indicated by red circles). As discussed in Sect.~\ref{subsubsec:paradist}, $\alpha_{\rm BAND}$ tends to have slightly higher values than $\alpha$, and $\beta_{\rm BAND}$ usually have lower values than $-3$. %In the cases how many cases? when $\alpha$ and $\alpha_{\rm BAND}$ differ significantly ($\alpha_{\rm BAND}$ becomes much harder with values $\approx -1$-0 or above, e.g., the first few data points of GRB081009140 and the rising phase of GRB150213001, and some data points of a few other pulses with low significance), $\beta_{\rm BAND}$ also becomes much harder ($\approx -3$--$-2$ or above). 

In most pulses the evolution of $\alpha$ exhibits a variation that appears to track the  variation in the light curve. This is most pronounced around the pulse peak, where the time bins also have the  highest significance. In some cases there is a slight temporal shift between the $\alpha$-variation and the light curve. These observed properties  are similar to earlier findings by, for instance,  \citet{Crider97, Ghirlanda2002, Lloyd2002, BasakRao2014} and are further discussed in \S \ref{sec:Fa}.

For the behavior of the peak energy, it is obvious that in almost all time bins $E_{\rm p}$ and $E_{\rm p,BAND}$ are well within half an order of magnitude. It is noticed that $E_{\rm p,BAND} \lesssim E_{\rm p}$ during time bins with $S \geq 20$. Combining with the observation that $\alpha_{\rm BAND} > \alpha > \beta_{\rm BAND}$, this suggests that BAND is trying to fit the spectrum by mimicking the curvature below the CPL's peak energy using two power-law segments. This can also explain the hard BAND spectrum during low-significance time bins: the overshot of $\beta_{\rm BAND}$ at high energies is tolerated by the noisy time bins.

The evolution of $E_{\rm p}$ is observed to exhibit various trends (Col.~11 of Table \ref{tab:pulses}). We found that 16 exhibit pure hard-to-soft (h.t.s.) evolution (42\%), while 8 exhibit pure intensity tracking (i.t.) evolution (21\%). Seven pulses change from either h.t.s. or flat to i.t. or soft-to-hard (s.t.h.) evolution. \citet{Lu2012} studied simulated GRB pulses and claimed that an i.t. evolving pulse can be composed by multiple h.t.s. evolving pulses. %According to \citet{Lu2012}, while these pulses could be intrinsically i.t., they might also actually be composite pulses. 
Four cases cannot be classified into the above categories: GRB150510139 and GRB 160530667 exhibit s.t.h. to h.t.s. evolution; GRB100122616 i.t. to unclassified, and GRB170114917 h.t.s. to unclassified (marked by a ``?'') during part of the pulse.

The calculated energy fluxes for CPL and BAND, $F$ and $F_{\rm BAND}$, agree very well for every spectrum and they basically track the photon light curve. During low-significance time bins, $F_{\rm BAND}$ is always larger than $F$, which could be explained by the aforementioned harder BAND spectrum.

\subsection{Global Parameter Relations}\label{subsubsec:global}

\begin{figure*}
\centering
\subfigure{\includegraphics[width=0.45\linewidth]{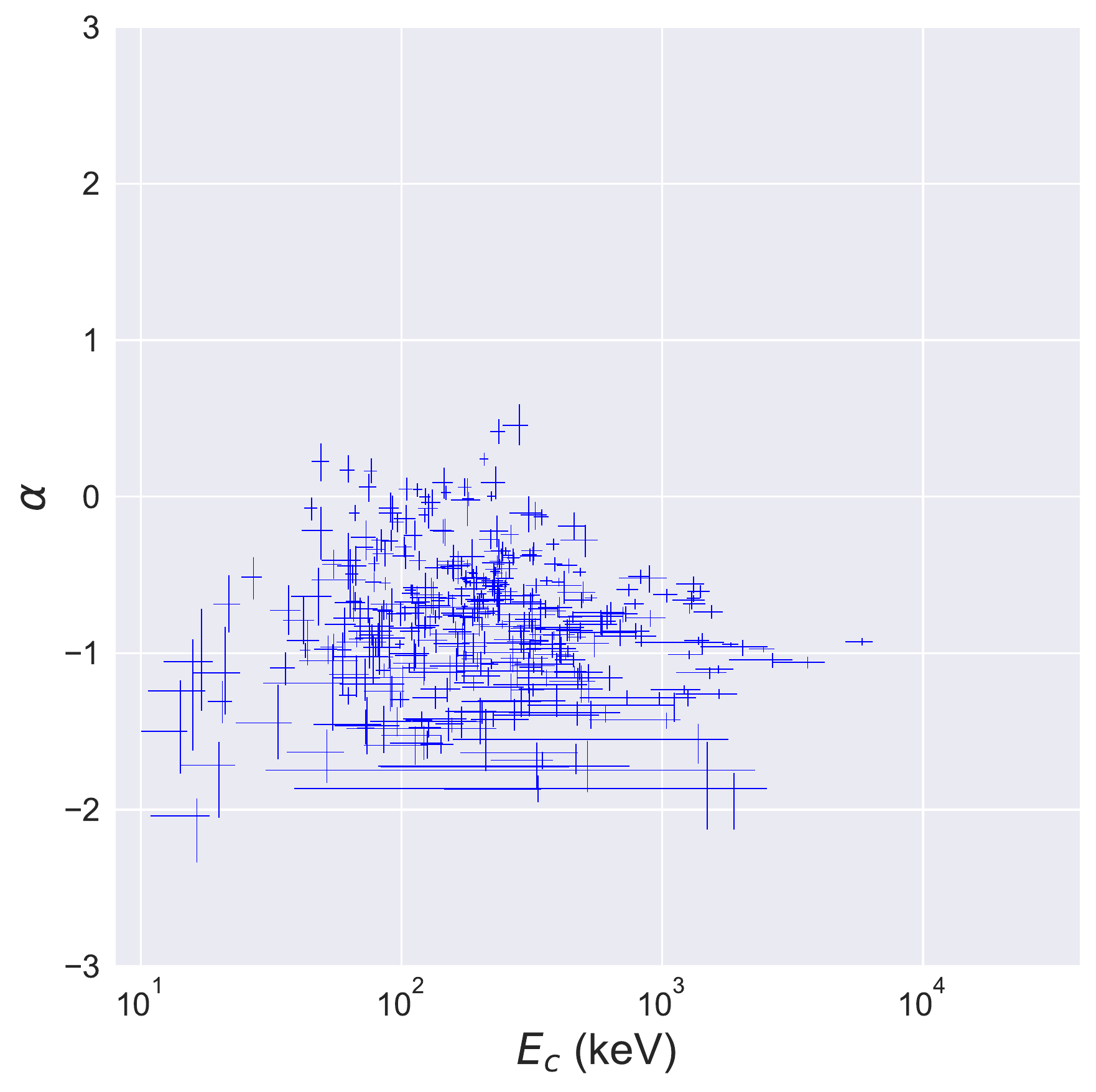}}
\subfigure{\includegraphics[width=0.45\linewidth]{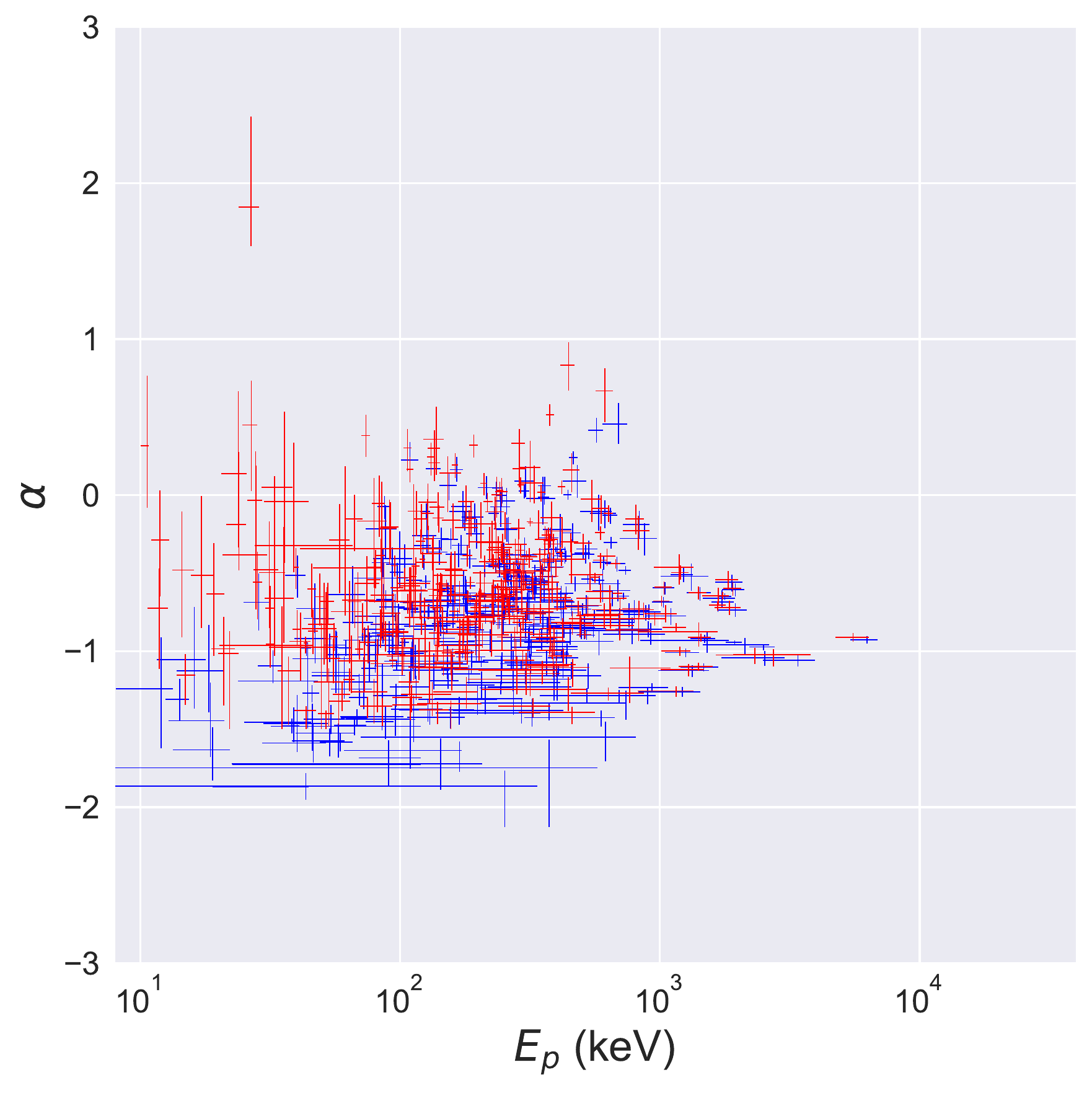}}
\subfigure{\includegraphics[width=0.45\linewidth]{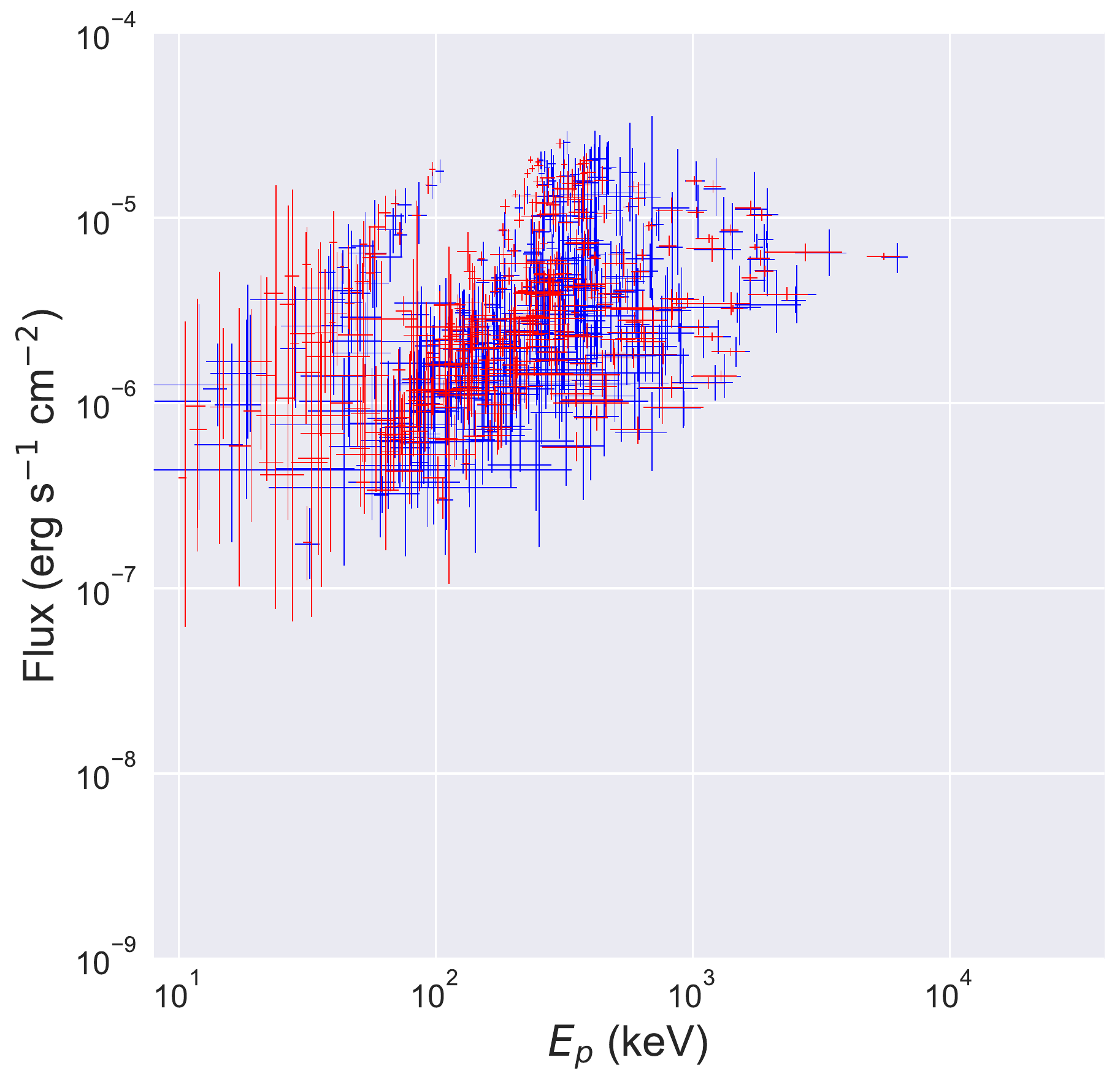}}
\subfigure{\includegraphics[width=0.45\linewidth]{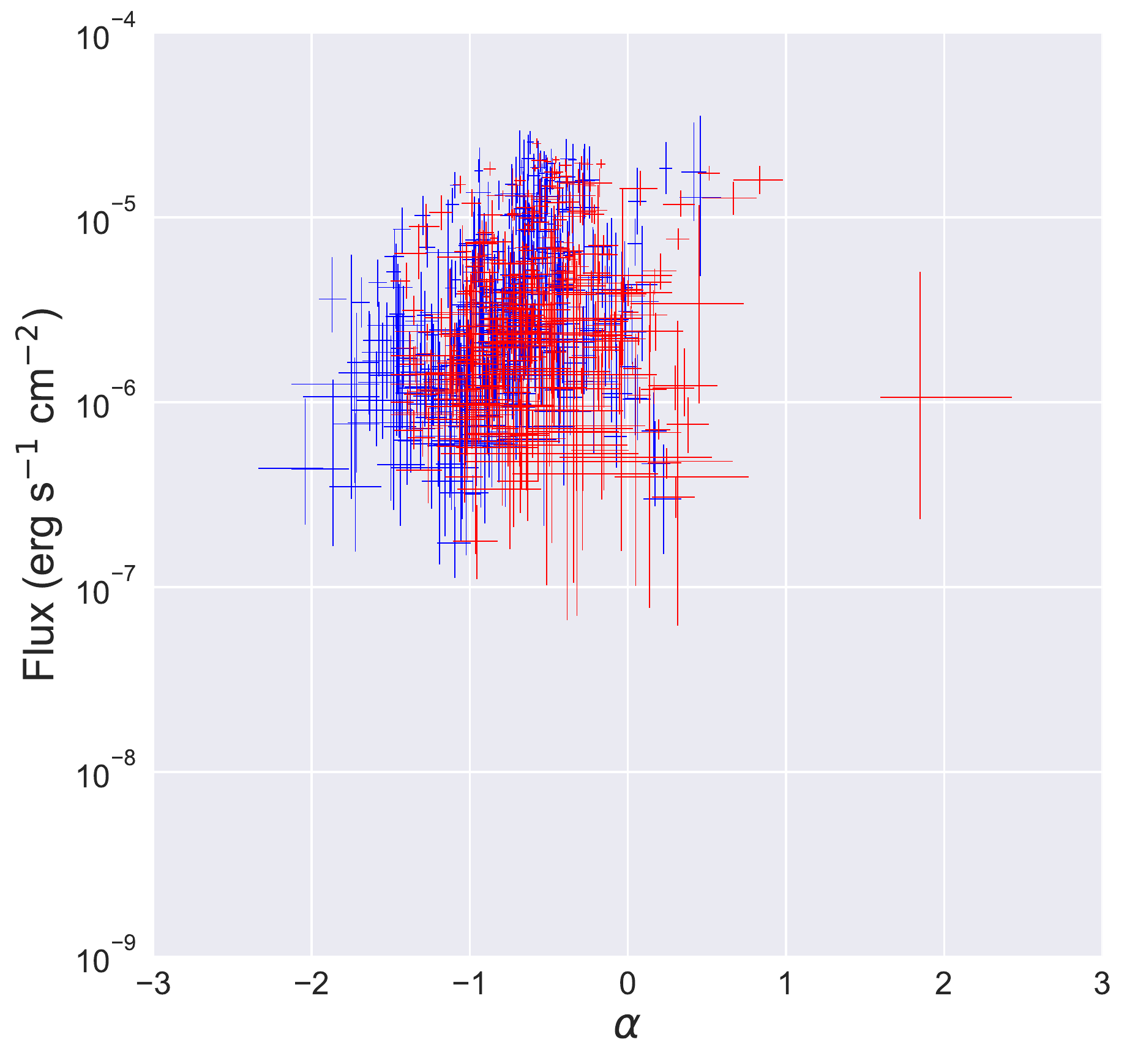}}
\subfigure{\includegraphics[width=0.45\linewidth]{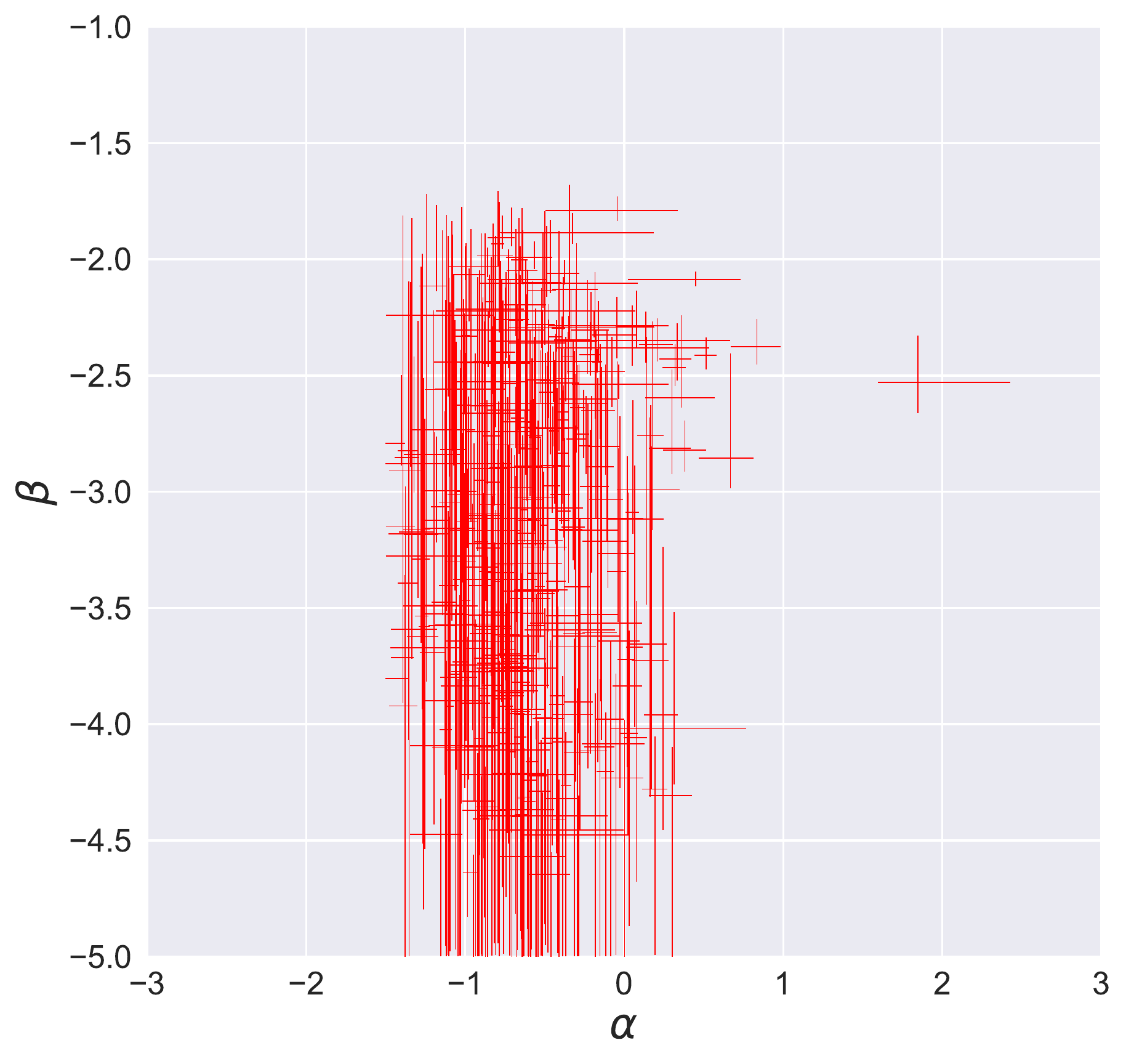}}
\caption{Global relations of the fitted parameters within the GBM energy range (8~keV-40~MeV) with statistical significance $S\geq20$. Blue data points are for CPL and red for BAND.
\label{fig:correlations_ALL}}
\end{figure*}

%\begin{figure}
% \includegraphics[width=\columnwidth]{compare.pdf}
% \caption{A synthetic simple power law (black solid line) with $K=1$~ph~s$^{-1}$~cm$^{-2}$~keV$^{-1}$ and $\alpha=-2.5$ and four synthetic CPL spectra with $K=1$~ph~s$^{-1}$~cm$^{-2}$~keV$^{-1}$, $\alpha=-2.5$, and various $E_{\rm c}$ (50~keV, 500~keV, 5~MeV, and 40~MeV: blue, green, red, and purple, respectively). The vertical black dashed line indicates $E = 5$~MeV. The figure is plotted for the GBM energy range, 8~keV-40~MeV.
% \label{fig:compare}}
%\end{figure}

%The global relation plots of the fitted parameters could help to inspect the quality of the spectral fits. 

Figure~\ref{fig:correlations_ALL} shows the overall parameter relations {  within the GBM energy range (8~keV-40~MeV) with statistical significance $S\geq20$,} in five panels of the parameter pairs: $\alpha$--$E_{\rm c}$ for CPL (upper left panel), $\alpha$--$E_{\rm p}$ for CPL and BAND (upper right), $F$--$E_{\rm p}$ for CPL and BAND (middle left), $F$--$\alpha$ for CPL and BAND (middle right), and  $\beta$--$\alpha$ for BAND (bottom). It is observed that the distributions of parameters for CPL and BAND show no obvious difference nor any global relation.

%In the first glance, it seems that the distribution of $\alpha$ for CPL (blue data points) and BAND (green data points) are very different. However, it is important to notice that the very soft and very hard values of $\alpha$ actually came from spectra with low significance, as shown also in Fig.~\ref{fig:para_histogram}. This indicates that those unexpectedly soft or hard values of the low-energy power-law slope might be results of poorly constrained {\fb are they removed now?} fits due to low photon statistics. 

For statistical significance $S\geq20$, all except one $\alpha$-values are between $-2$ and 1 to within 1-$\sigma$ uncertainty for both CPL and BAND, which are typically observed in studies of GRB prompt spectra. We note that  that $\beta$ can be very negative ($\sim -3$ or below). There are two possible reasons for this. Either the BAND model is trying to mimic a cutoff in the high-energy spectrum, or the poor count statistics at high energies prevents a determination of $\beta$ \citep[see, e.g.,][]{Kaneko2006,Goldstein2012,Gruber2014,Yu2016}. The threshold of $F$ for the high-significance data points is $\sim 10^{-7}$--$10^{-6}$~erg~s$^{-1}$~cm$^{-2}$. 

%In the imaginary case where the observed data is perfect without any error bar, the value of $E_{\rm c}$ will be further shifted to outside of the energy range of the BGOs.

The peak energy, $E_{\rm p}$ for BAND is a fitted parameter, while that for CPL is calculated from $E_{\rm p} = (\alpha+2)E_{\rm c}$. Notice that when $\alpha$ has lower values than $-2$ or when $\beta_{\rm BAND}$ has higher values than $-2$, $E_{\rm p}$ becomes negative and thus there is no peak in the $\nu F_\nu$ spectrum. 
%The values of $E_{\rm p}$ for the high-significance data points are found to cluster between 50~keV to 5~MeV for both models.

\subsection{Individual Parameter Relations}\label{subsubsec:individual}

%%%%%%%%

%{\color{red}  %We plot the observed $\alpha$-$E_{\rm p}$, $F$-$E_{\rm p}$, and $F$-$\alpha$ relations in Figures~\ref{fig:correlation_group1} to \ref{fig:correlation_group10} (\S~\ref{subsubsec:individual}).

%Notice that in the $E_{\rm p}$ plots of some bursts, the number of data points are less than the number of time bins. This is because the peak energy of CPL is given by $E_{\rm p}=(2+\alpha)E_{\rm c}$. Thus, for $\alpha<-2$ the $\nu F_\nu$-spectrum is monotonically decreasing and no peak can be found in the observed energy window. This happens, for instance, in the second episode of GRB081009.
%}

%%%%%%%%%

Relations over individual pulses are of greatest interest since they carry the information closest to the physics of the emission. We, therefore, provide the relation plots of the 38  pulses in our sample in  Appendix \ref{app:correlationplots}. In Figs.~\ref{fig:correlation_group1} to \ref{fig:correlation_group10}, the relation between $\alpha$ and $E_{\rm p}$ are shown in the left-hand panels; the relation between energy flux $F$ and $E_{\rm p}$, i.e., the Golenetskii correlation  \citep{Golenetskii1983}, are shown in the middle panels; and, finally, the relation between $F$ and $\alpha$ are shown in the right-hand panels. 

Below, we will discuss the appearances of the temporal tracks in the relation planes by visual inspection. We consider all time bins with significance $S>10$ (i.e., yellow, orange, and red data points) that are fitted with a cutoff power law function.

%Below, we summarise the morphological appearance of the correlations.

\subsubsection{$\alpha$-$E_{\rm p}$--Relation}

The $\alpha$-$E_{\rm p}$ relations show three main types of behaviours. The most common behaviour is a non-monotonic relation, with a clear break. This occurs in 17 pulses. The break either occurs at the maximal $\alpha$-value (e.g., GRB081125), or at the minimum $E_{\rm p}$ (e.g., GRB150314). Another common behaviour (12 pulses) is a monotonic, straight line in the linear-log plots  \citep[see, also ][]{Crider97}. Of these bursts, 6 have a positive relation (e.g., GRB090719) and 6 have a negative relation (e.g., GRB130305), {  even if GRB090620 has a weaker correlation.} The third behaviour (7 bursts) is given by pulses in which the $E_{\rm p}$ does not vary much, while $\alpha$ does vary more significantly. This leads to a vertical relation, or a weakly negative relation (e.g., GRB100528). In one of these cases (GRB090804), though, there is only little variation in $\alpha$ as well, it even being consistent with a constant at around $\alpha \sim -0.5$.
In Table \ref{tab:pulses}, all bursts are assigned to one of these three groups, 1, 2, and 3, respectively. The remaining two pulses (both in GRB081009) do not show any clear trend. For the second pulse in GRB081009, the reason is that in most of the high-significance time bins there is no $E_{\rm p}$  (the $\nu F_\nu$ spectrum is monotonically decreasing) leaving only a few data points left for the relation.
 
 {  To quantify the relations, we calculate the Spearman's rank coefficient, $r$, which is also provided in Table \ref{tab:pulses}. In general, values over 0.7 indicate strong correlations. However, only for a few pulses (8 cases) $r > 0.7$. A large majority of the pulses (20 cases) have weak correlations as indicated by the $r$--value being below 0.4.}
 
 It is noteworthy that among all the $E_{\rm p}$-$\alpha$--relations, only three pulses have a relation that follows what is expected for synchrotron emission \citep[][their Fig. 5]{Lloyd2000}, namely GRB120919, GRB130815 and GRB141205, see further discussion in \citet{Ryde2019}.

\subsubsection{$F$-$E_{\rm p}$--Relation}

Turning over to the Golenetskii relation, again three main different types of relations are revealed \citep[see, also ][]{Borgonovo2001, Firmani2009, Ghirlanda2010}. The most common behaviour (in 23 pulses) is a non-monotonic relation with a distinct break and having power-law segments (e.g., GRB160530). The break typically occurs at the flux peak of the pulse, that is, the relation is different during the rise phase and the decay phase of the pulse. Another common behaviour has a relation described by a single power law (in 13 of the pulses). Of these, 11 pulses have a positive relation (e.g., GRB090804) and in 2 cases it is negative (e.g., GRB130305). Finally, in two cases there are no clear trends. These pulses are from the second episode of GRB081009 (again mainly due to fact that many of the $E_{\rm p}$ are not determined) and from GRB100528. In Table \ref{tab:pulses}, these three groups, are denoted by 1, 2, and 3, respectively. 

{  For these relations, we also calculate the Spearman's rank coefficient (provided Table \ref{tab:pulses}). Again, only a few pulses (11 cases) have strong correlations ($r > 0.7$), while a large majority of the pulses (18 cases) have weak correlations ($r < 0.4$).}

GRB090804 is an interesting case in which the Golenetskii-relation is prominent, but both $E_{\rm p}$-$\alpha$ and the $F$-$\alpha$ relations are very weak. Such a behaviour is, however, an exception. We also note that \citet{Guiriec2013,Guiriec2015a,Guiriec2015b,Guiriec2016a,Guiriec2016b} have shown that, in their three-component model, a correlation between the energy flux  and the $\nu F_\nu$ peak energy  manifests itself, for one of the non-thermal components, {  even in GRBs where the Golenetskii-relation is not valid.}

\subsubsection{$F$-$\alpha$--Relation}
\label{sec:Fa}

{  Finally, the $F$-$\alpha$--relations differ clearly from the two first relations, by it having a much more homogeneous behaviour. In nearly all cases the relation is very similar, with a linear relation appearing in the semi-log plots. Of these, 32 pulses show a positive and 2 pulses show a negative relations (e.g., GRB110920).} In only three bursts there is non-monotonic relation with a break, albeit being weak (GRBs 081009 [second episode], 110721, 160910). In the last case, GRB090804 the relation is weak, since there is only little variation in both the parameters. Again, this classification is shown in Table \ref{tab:pulses}
by group 1 (34 pulses), group 2 (three pulses), and group 3 (one pulse), respectively. 

{  To quantify the observed correlations, we again calculate the Spearman's rank coefficient (Table \ref{tab:pulses}). Indeed, for a large majority of the pulses (28 cases) $r > 0.7$, and of these, 8 have very strong correlations, with values over 0.9. There are only two pulses which have weak correlations ($r < 0.4$).}

We note that in the cases where the variation in $E_{\rm p}$ is small, it is only $F$ and $\alpha$ that are correlated. An example is GRB100528 for which the Golenetskii correlation is very weak, but the $F$-$\alpha$ is very clear.

%The purpose of this catalogue is to identify trends and the variety of behaviours of the relation between parameters. 
The fact that the relation between $F$ and $\alpha$ has a similar behaviour for a majority of the pulses, instigates searches for  possible functional relations between the parameters, for use in physical interpretations of the underlying mechanisms. With such an goal in mind, we have interpreted the $F$-$\alpha$ relation in the context of photospheric models in Ryde et al. (2019). 

%In summary, the $F$-$\alpha$ stands out among the three relations with a majority having a similar behaviour. Indeed,

\section{Summary and Conclusion}\label{sec:summary}

In summary, {  we have defined a sample of 38 single pulses} from 37 GRBs out of 2,050 {\it Fermi}/GBM detected bursts. These pulses all have more than 5 highly significant time bins, which allows  time-resolved spectroscopy to be performed and the spectral evolution to be investigated. A total of 577 time-resolved spectra were obtained and their spectral properties investigated using a fully Bayesian method. The time bins were selected using the Bayesian block method \citep{Scargle2013} in contrast to the signal-to-noise ratio method, employed in the previous time-resolved GRB spectral catalogue \citep{Yu2016}. A new statistical measure of the data significance \citep{Vianello2018} was also used to indicate various significance level. 

We confirm the finding in previous catalogues that the cutoff power-law function is better than the Band function for most bursts, when considering the number of degrees of freedom. {  In the current study, we found that, among the frequently used empirical functions, a consistent description of the time-resolved spectra of GRB pulses could be achieved by using the power-law function with an exponential cutoff.}

The distributions of the low-energy power-law slope and peak energy of the $\nu F_\nu$ spectra from the highest-significance time bins are consistent to previous results, while the distribution of the high-energy slope, when using a Band function instead of a cutoff power-law, has a lower value than that of \citet{Yu2016}. The latter study did not distinguish between single and composite pulses, which thus indicates that the high-energy slope observed in composite pulses might not be intrinsic in nature, but an effect of spectral evolution.

In contrast to previous catalogues, we also investigate the distribution of the maximal value of $\alpha$ in each pulse. Assuming that one and the same emission mechanism operates through out the pulse, we show that majority of the pulses (60\%) are inconsistent with synchrotron emission, solely based on the line-of-death of $\alpha = -2/3$.

Finally, we found that a majority of the pulses have a congruent, monotonic behaviour between the low-energy power-law index $\alpha$ and the energy flux $F$, which is largely independent of the flux variation in the light curve. This parameter correlation is studied in detail in a separate paper \citep{Ryde2019}.

%As a remark for different methods to perform spectral analysis of GRBs, this work, along with all the existing GRB spectral catalogues, uses empirical photon models for spectral fitting using the forward-folding technique. While the empirical parameter behaviours give valuable evidence for the intrinsic process(es), physical model fitting should be performed in order to directly obtain physical parameters from the data. Only a few studies have attempted to fit physical models and only a handful of bursts are studied in this way \citep[e.g.,][]{Burgess2011,Zhang2016}. 
%{\color{red} Iyyani2011 Ahlgren2015, Ahlgren2019 Ryde2017}

%%%%%%%
%% Needed in the conclusions?
%\citet{Burgess2018} combined localisation and spectroscopy of GRBs within a Bayesian framework, enabling simultaneous fitting of location and spectral parameters. This method can eliminate the uncertainty propagated to the spectral parameters from using a pre-defined, fixed spectral template for localisation. Future studies of GRB emission mechanism should consider this method in order to obtain the most reliable results.
%%%%%%

%% If you wish to include an acknowledgments section in your paper,
%% separate it off from the body of the text using the \acknowledgments
%% command.
\acknowledgments

We wish to thank Damien B\'egu\'e, J.~Michael Burgess, Liang Li, Daniel Mortlock, \& Asaf Pe'er for contributions in different parts of the project. We also thank the anonymous referee for useful comments on the manuscript. This research made use of the High Energy Astrophysics Science Archive Research Center Online Service HEASARC at the NASA/Goddard Space Flight Center. We acknowledge support from the Swedish National Space Agency and the Swedish Research Council (Vetenskapsr{\aa}det). FR is supported by the  G\"oran Gustafsson Foundation for Research in Natural Sciences and Medicine.

%% To help institutions obtain information on the effectiveness of their 
%% telescopes the AAS Journals has created a group of keywords for telescope 
%% facilities.
%
%% Following the acknowledgments section, use the following syntax and the
%% \facility{} or \facilities{} macros to list the keywords of facilities used 
%% in the research for the paper.  Each keyword is check against the master 
%% list during copy editing.  Individual instruments can be provided in 
%% parentheses, after the keyword, but they are not verified.

\vspace{5mm}

\facilities{{\it Fermi}/GBM}

%% Similar to \facility{}, there is the optional \software command to allow 
%% authors a place to specify which programs were used during the creation of 
%% the manusscript. Authors should list each code and include either a
%% citation or url to the code inside ()s when available.

\software{{\tt 3ML} \citep{Vianello2015}}

%% Appendix material should be preceded with a single \appendix command.
%% There should be a \section command for each appendix. Mark appendix
%% subsections with the same markup you use in the main body of the paper.

%% Each Appendix (indicated with \section) will be lettered A, B, C, etc.
%% The equation counter will reset when it encounters the \appendix
%% command and will number appendix equations (A1), (A2), etc. The
%% Figure and Table counter will not reset.

\appendix

\section{Plots of the evolutions}
\label{app:plotsevolution}

\begin{figure*}
\centering

\subfigure{\includegraphics[width=0.3\linewidth]{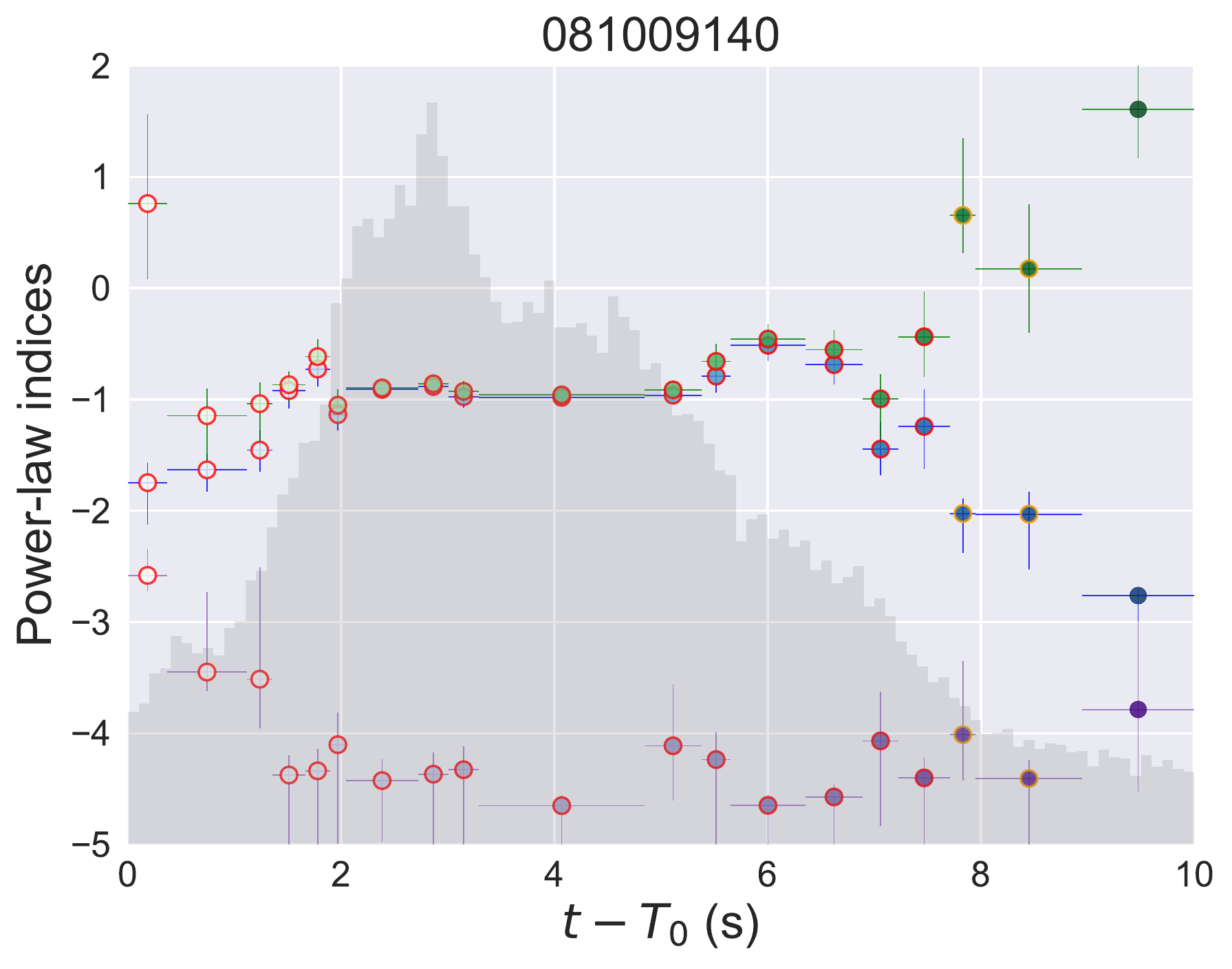}}
\subfigure{\includegraphics[width=0.3\linewidth]{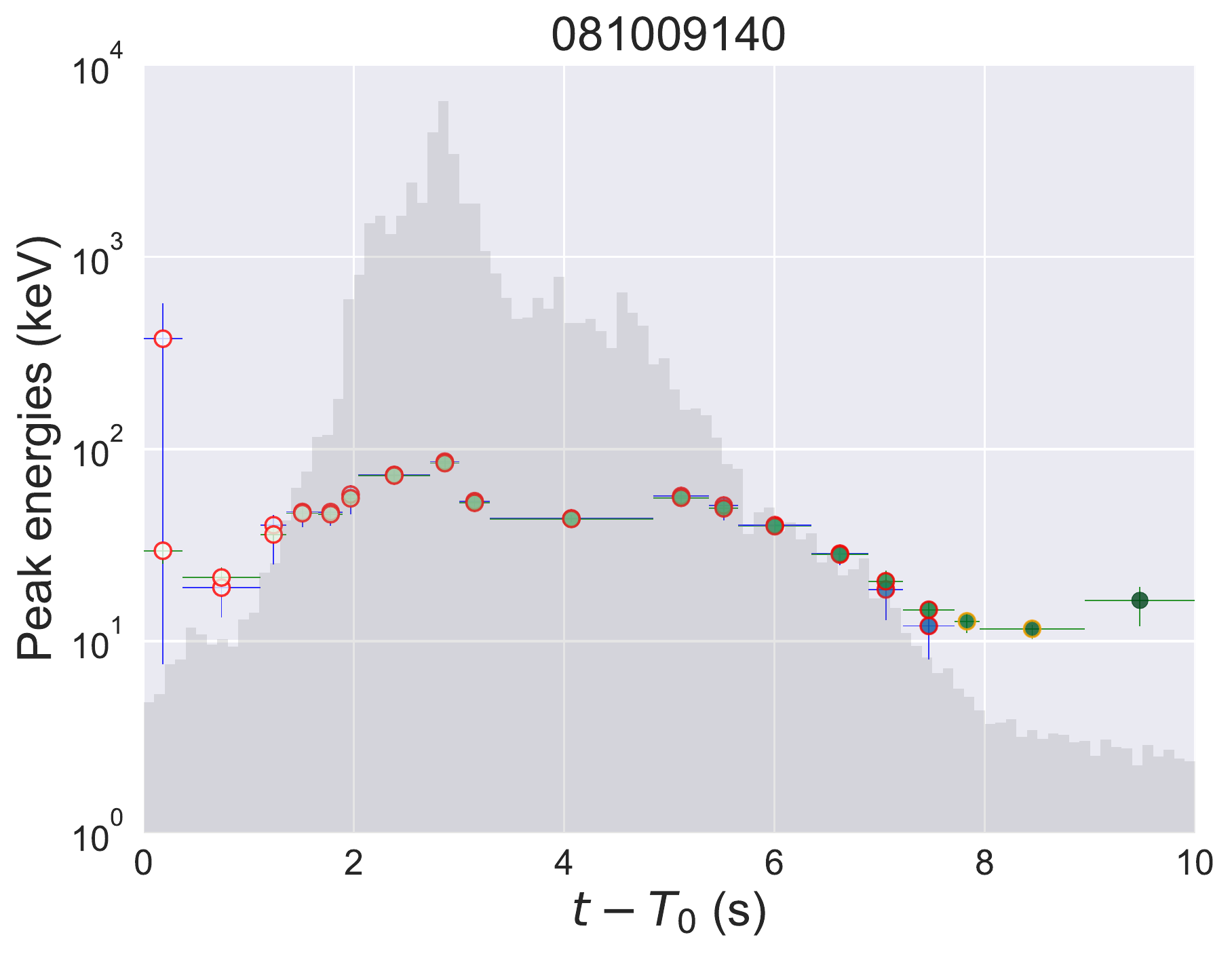}}
\subfigure{\includegraphics[width=0.3\linewidth]{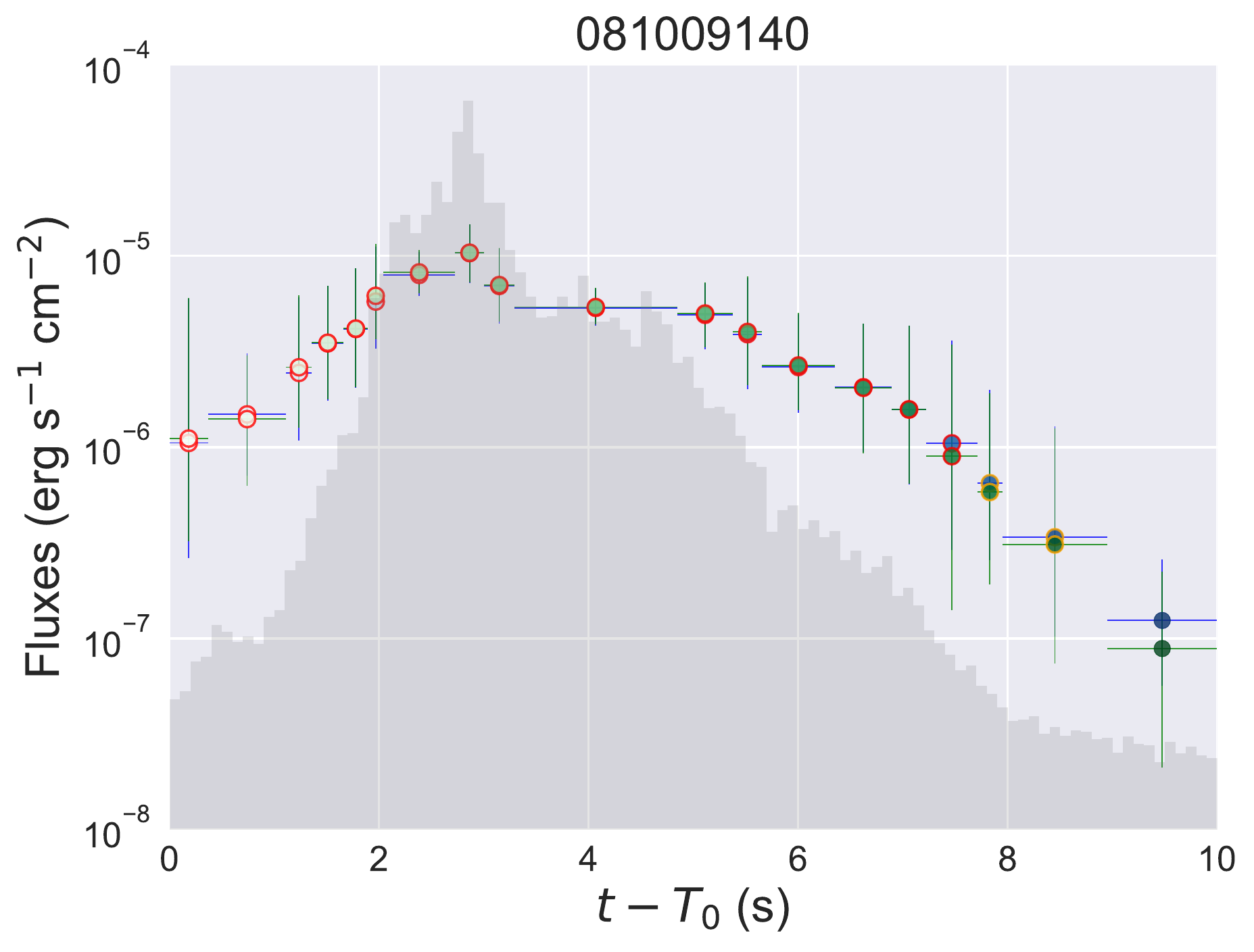}}

\subfigure{\includegraphics[width=0.3\linewidth]{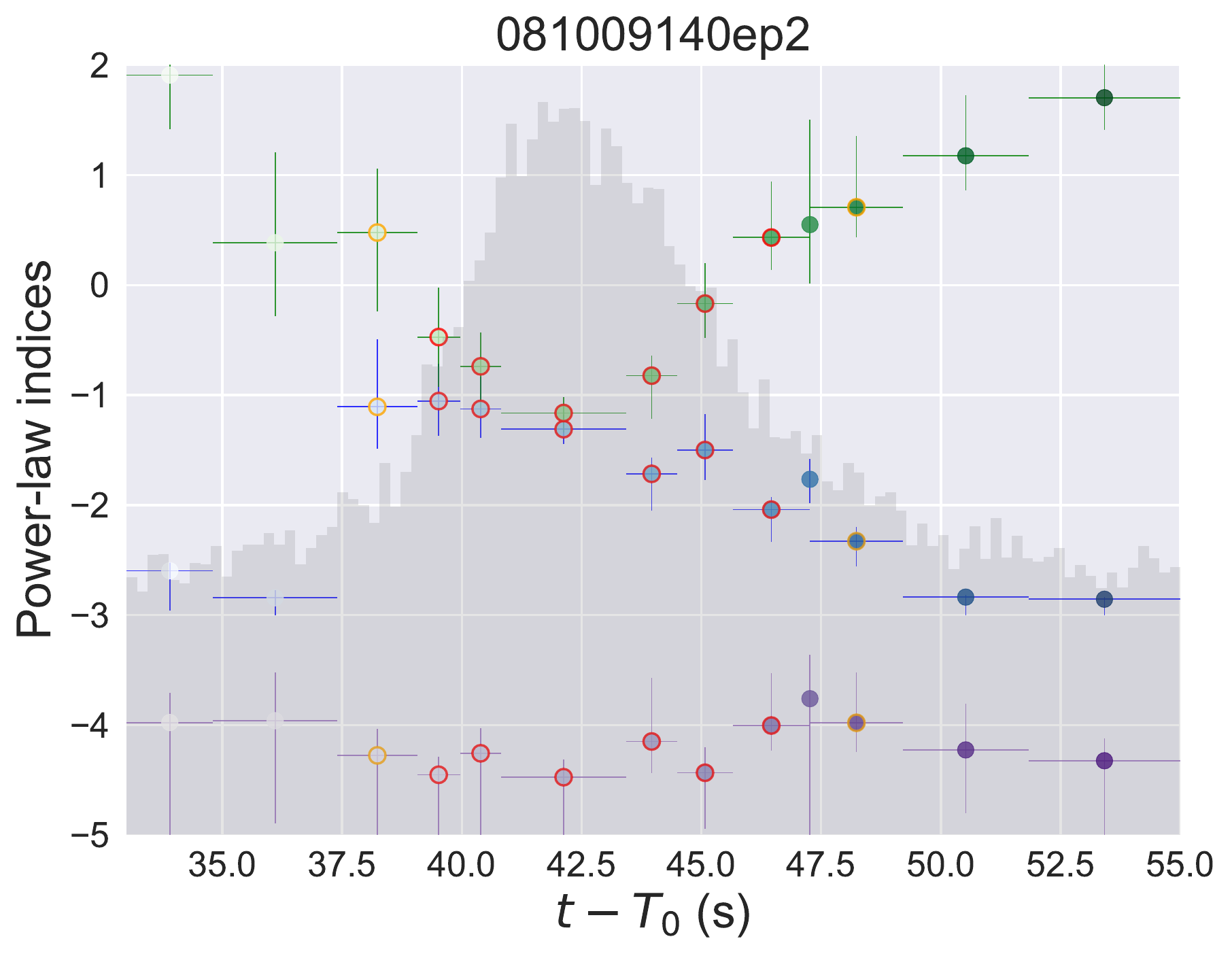}}
\subfigure{\includegraphics[width=0.3\linewidth]{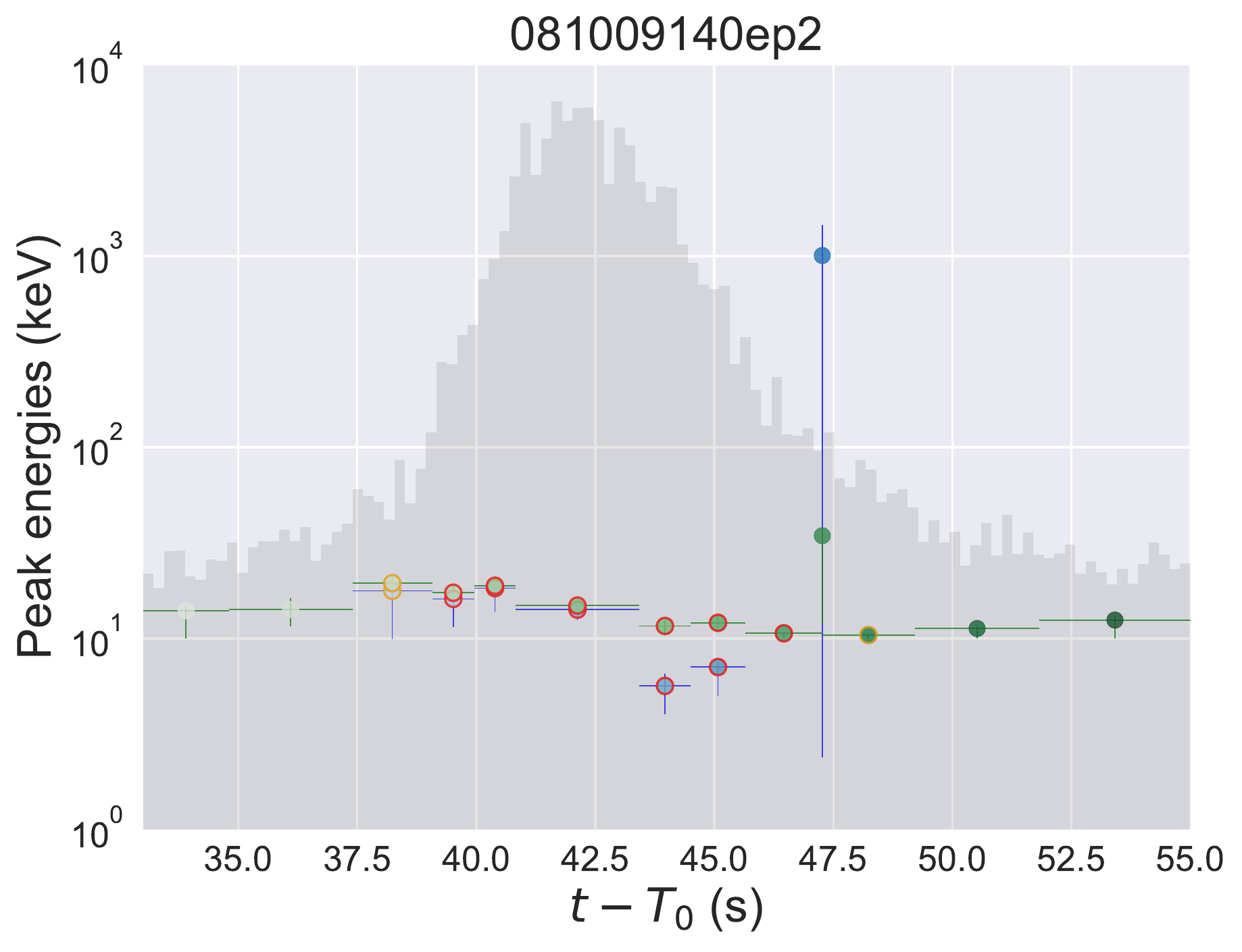}}
\subfigure{\includegraphics[width=0.3\linewidth]{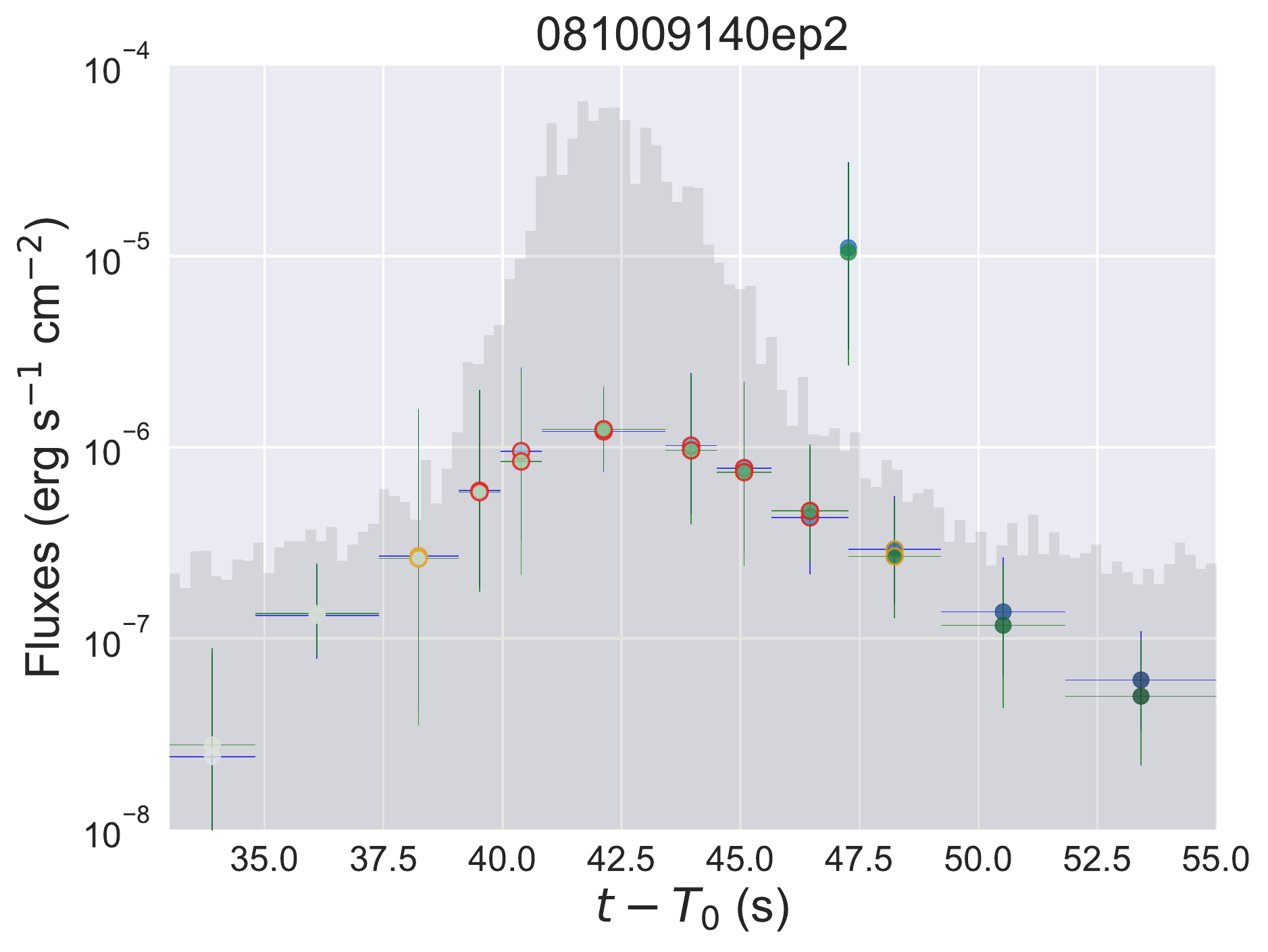}}

\subfigure{\includegraphics[width=0.3\linewidth]{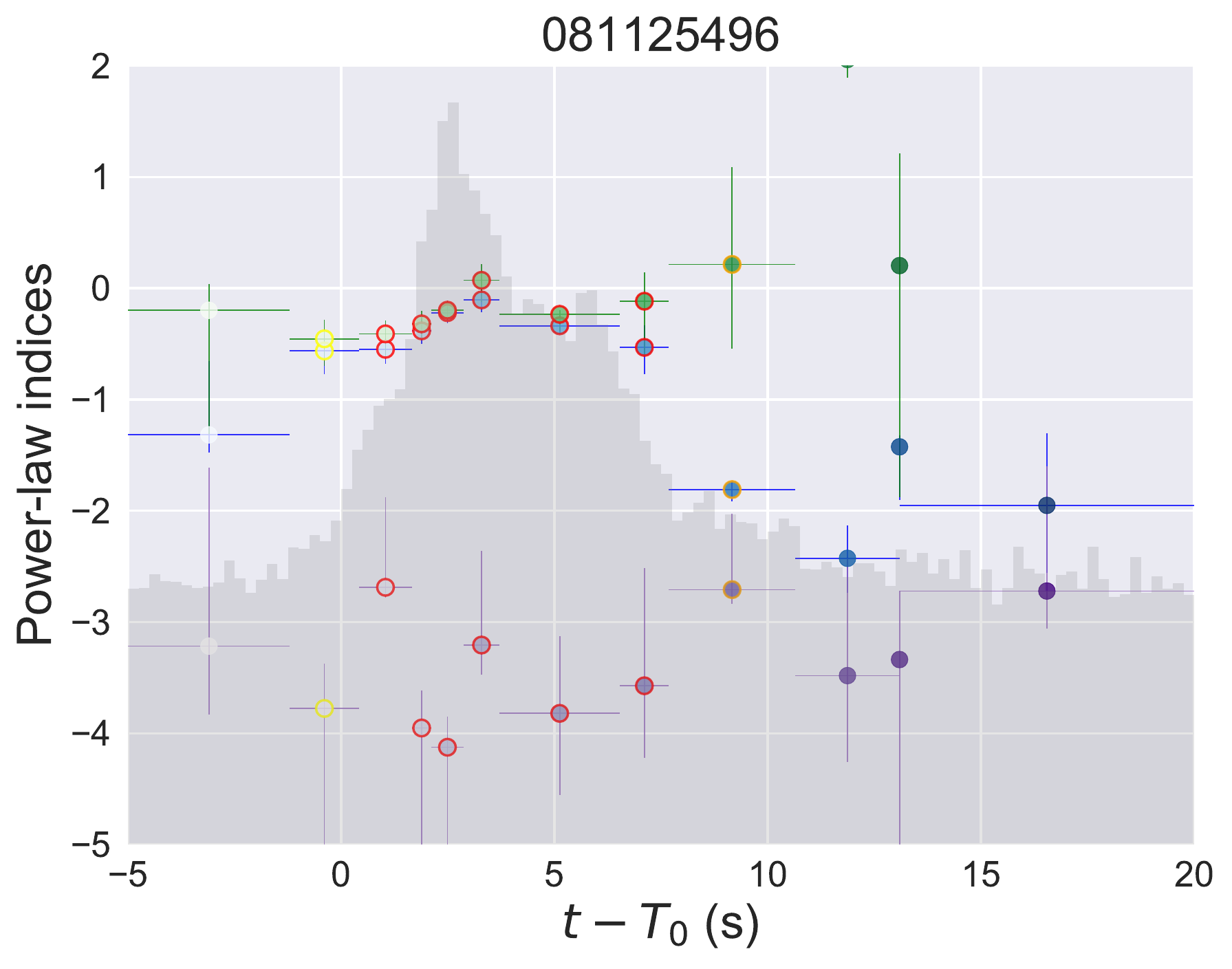}}
\subfigure{\includegraphics[width=0.3\linewidth]{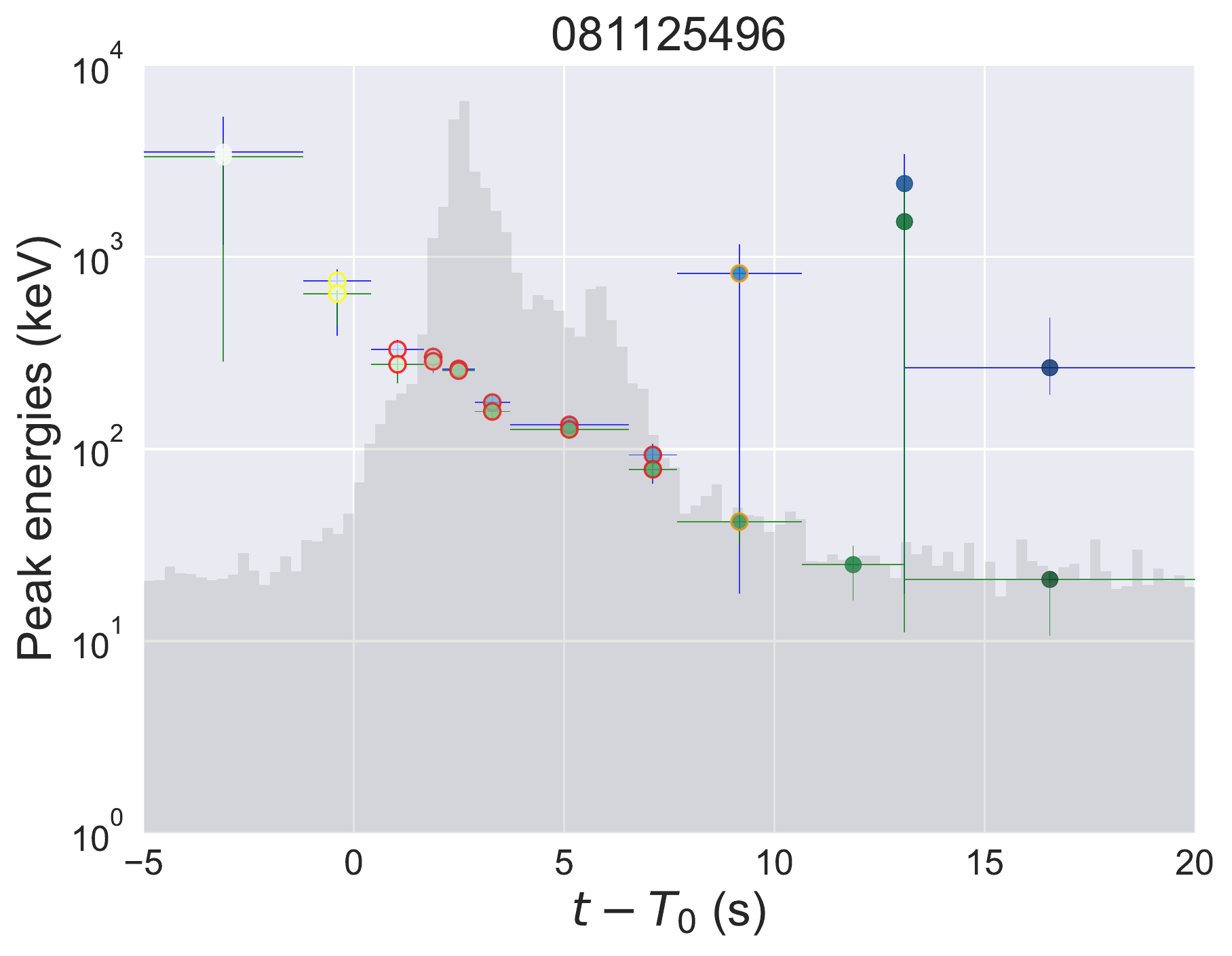}}
\subfigure{\includegraphics[width=0.3\linewidth]{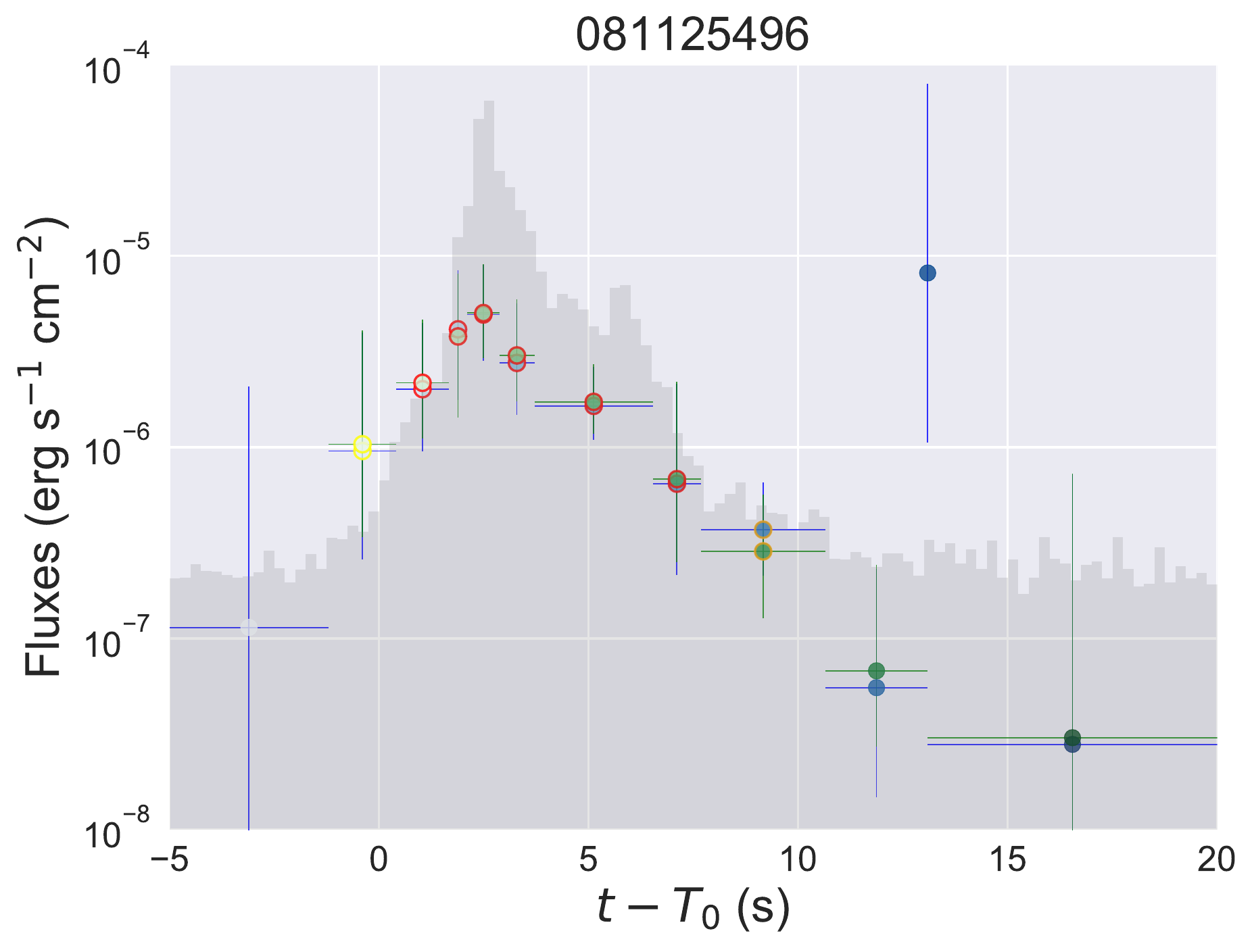}}

\subfigure{\includegraphics[width=0.3\linewidth]{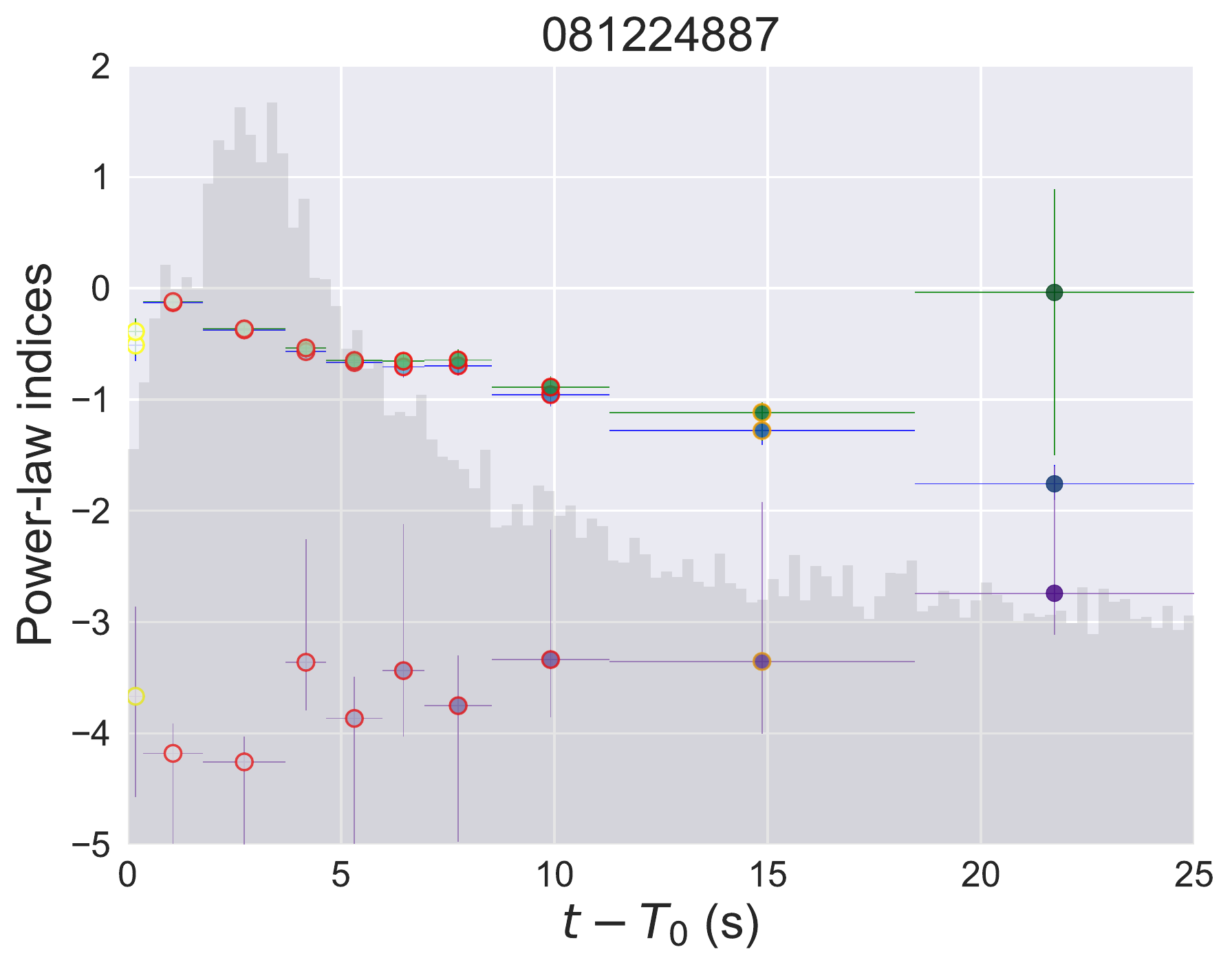}}
\subfigure{\includegraphics[width=0.3\linewidth]{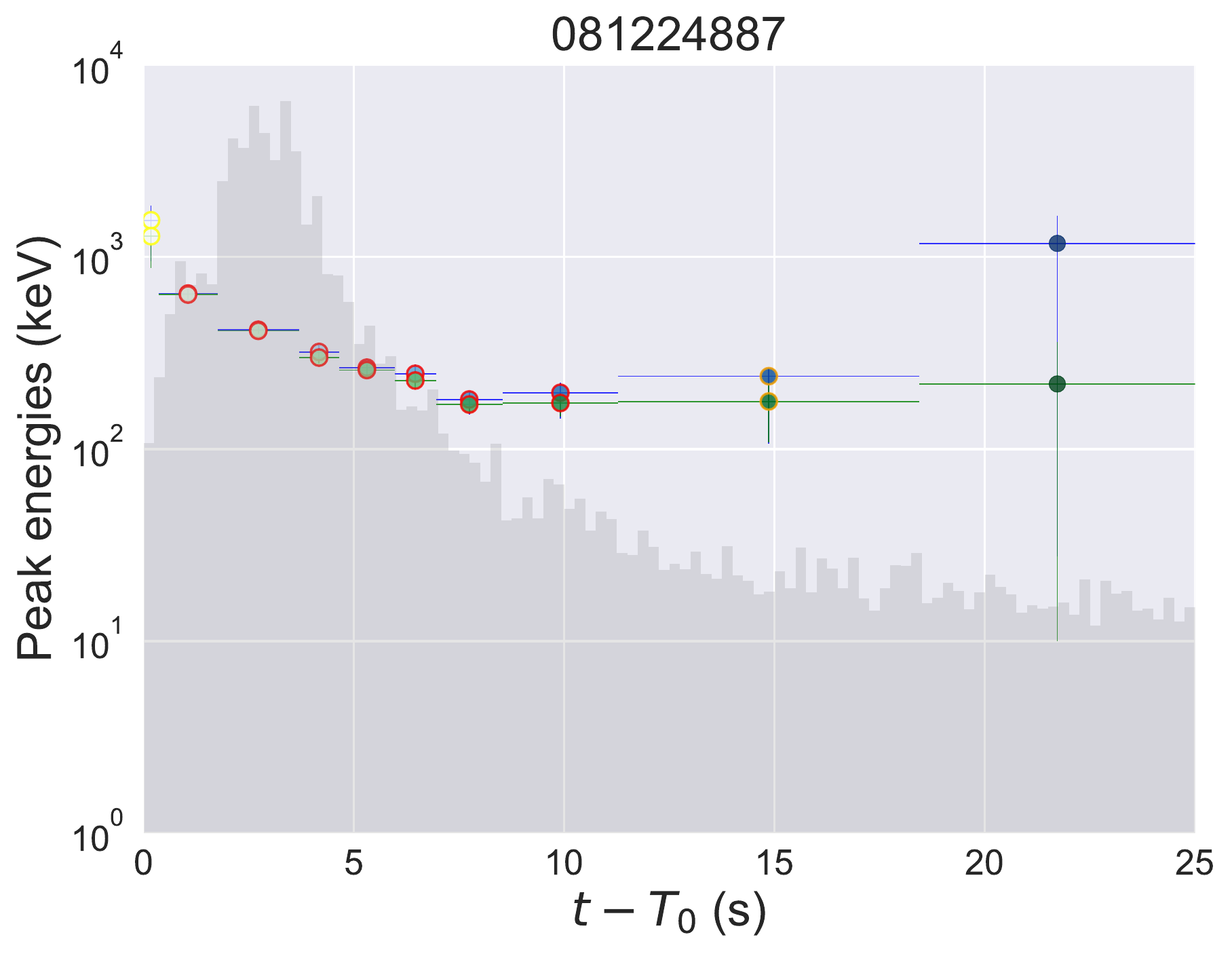}}
\subfigure{\includegraphics[width=0.3\linewidth]{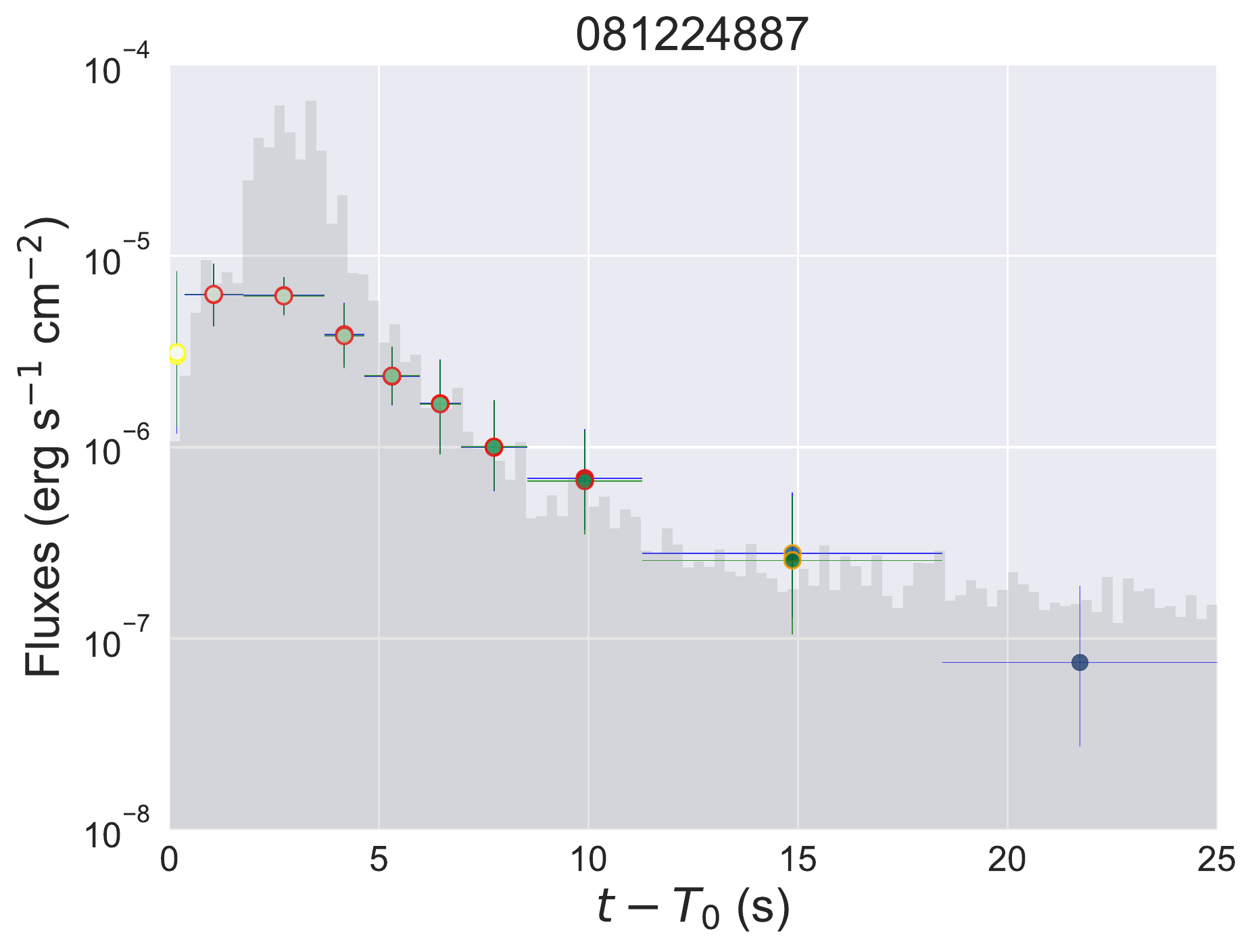}}

\caption{Left panels: temporal evolution of $\alpha$ (blue), $\alpha_{\rm BAND}$ (green), and $\beta_{\rm BAND}$ (purple). Middle panels: temporal evolution of $E_{\rm p}$ (blue) and $E_{\rm p, BAND}$ (green). Right panels: temporal evolution of $F$ (blue) and $F_{\rm BAND}$ (green). Light curves are overlaid in grey colour. Data points with red, orange, yellow, and no circles indicate statistical significance $S\geq20$, $20>S\geq15$, $15>S\geq10$, and $S<10$, respectively. Color scale from light blue (start) to deep blue (end) shows temporal evolution. Many of the low-significance data points are marginally or not constrained, as seen from the huge negative-side error bars.
\label{fig:evolution_group1}}
\end{figure*}

\begin{figure*}
\centering

\subfigure{\includegraphics[width=0.3\linewidth]{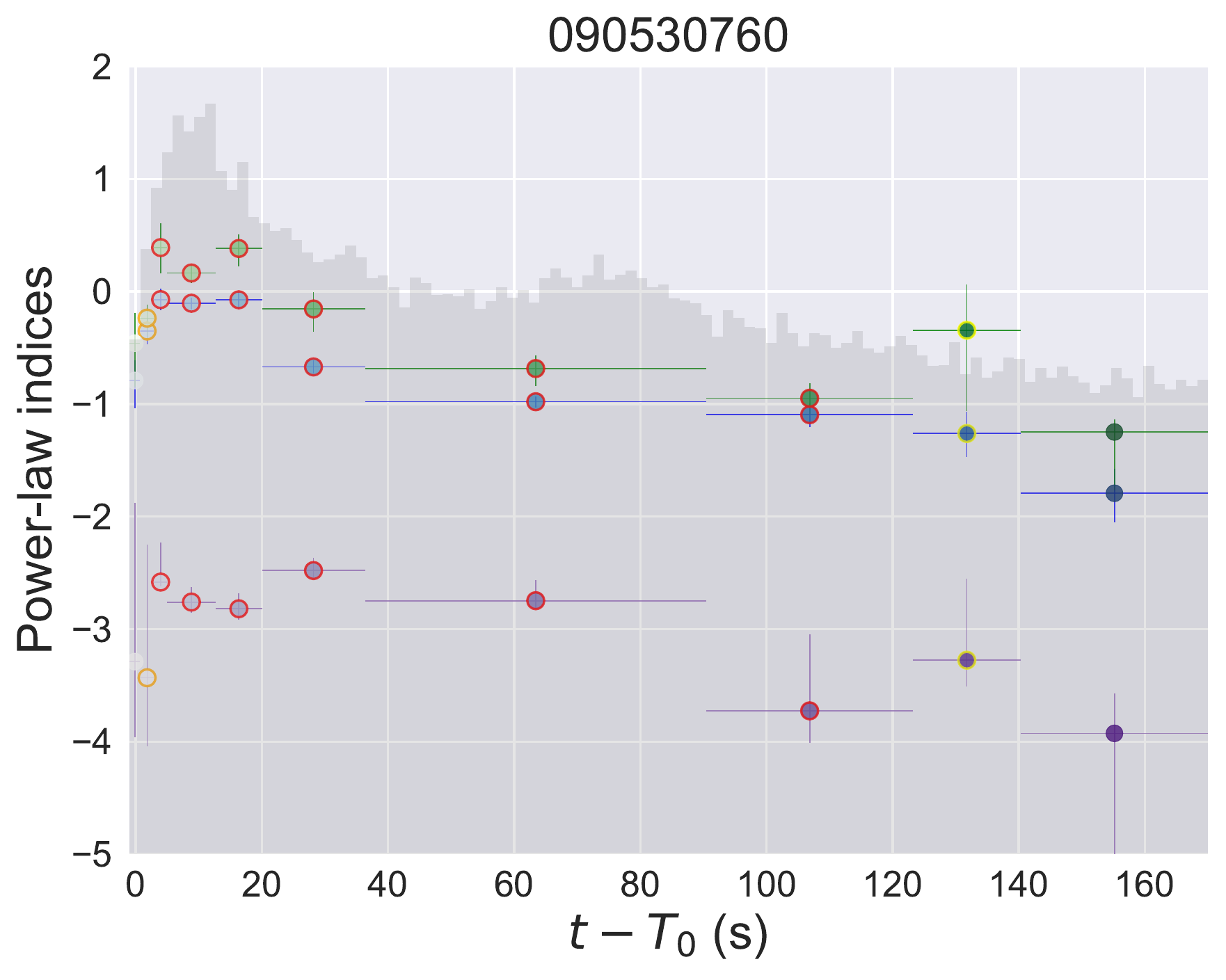}}
\subfigure{\includegraphics[width=0.3\linewidth]{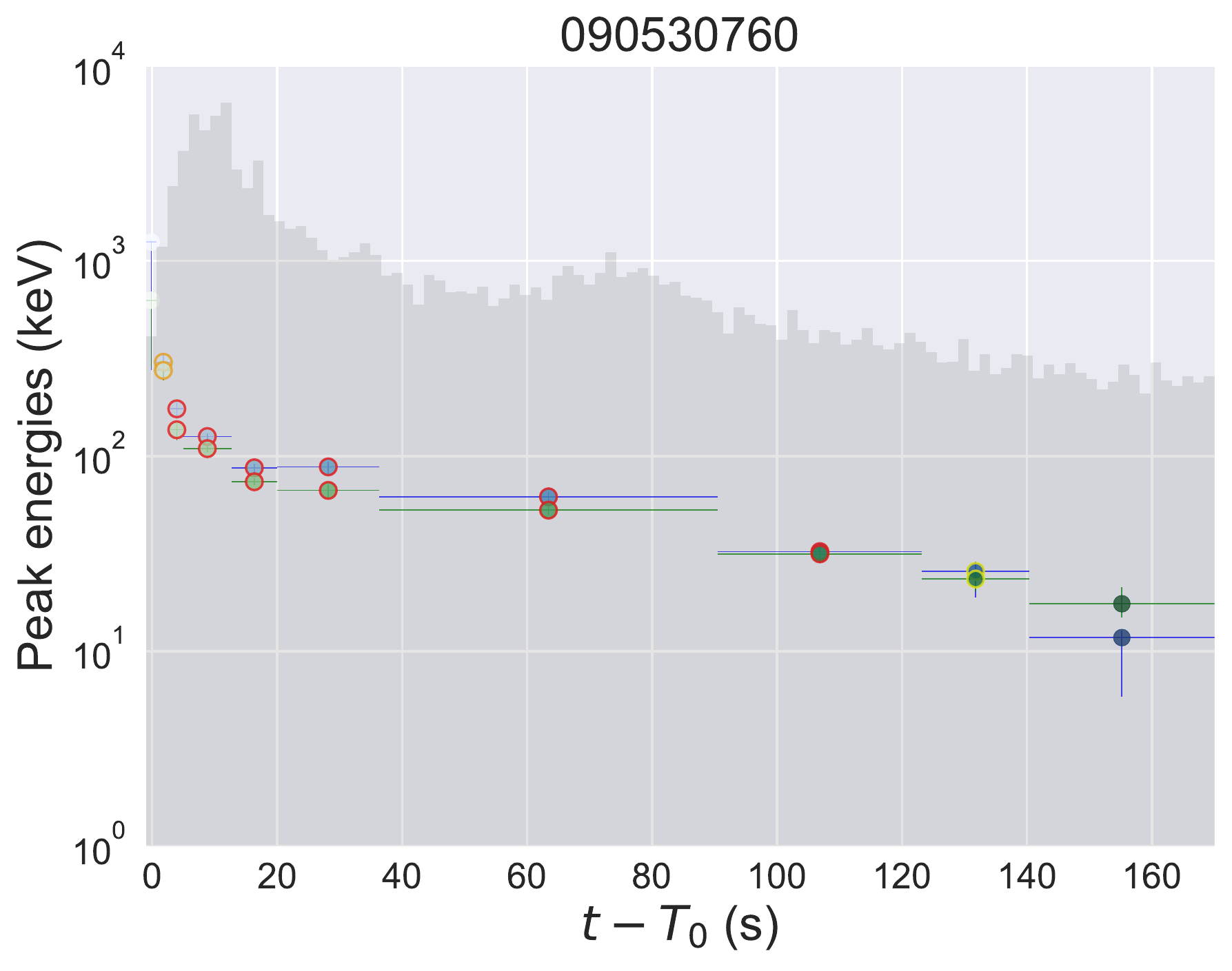}}
\subfigure{\includegraphics[width=0.3\linewidth]{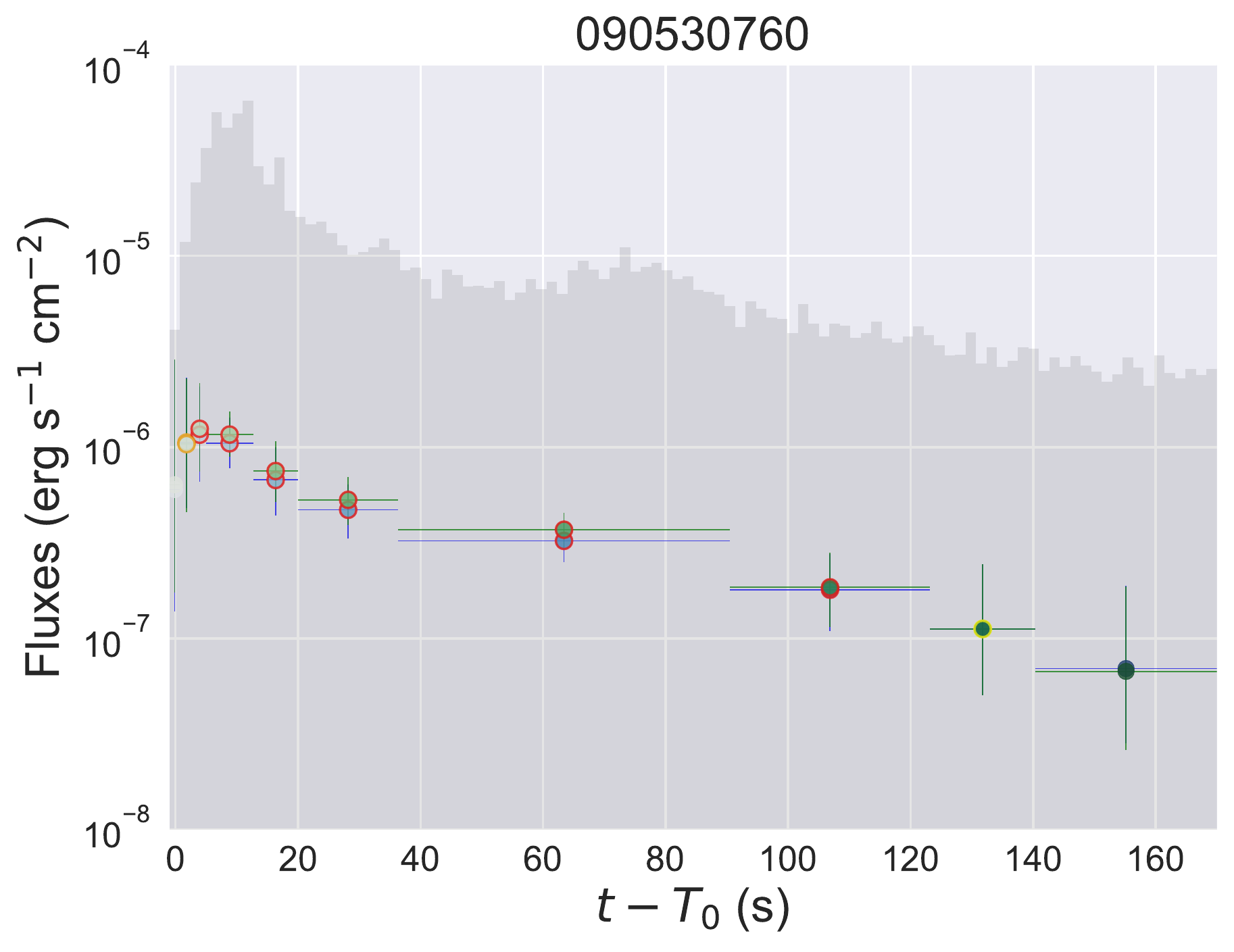}}

\subfigure{\includegraphics[width=0.3\linewidth]{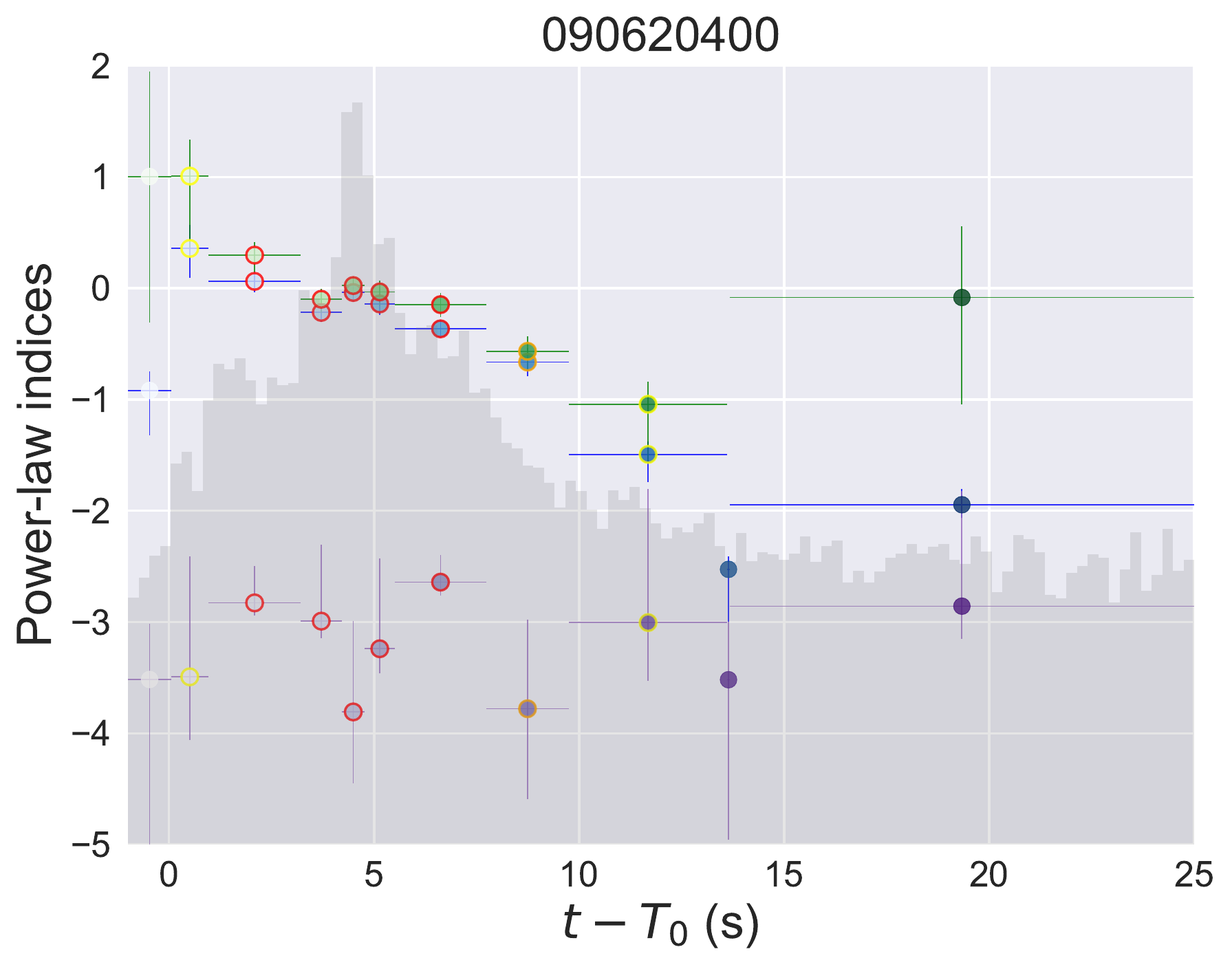}}
\subfigure{\includegraphics[width=0.3\linewidth]{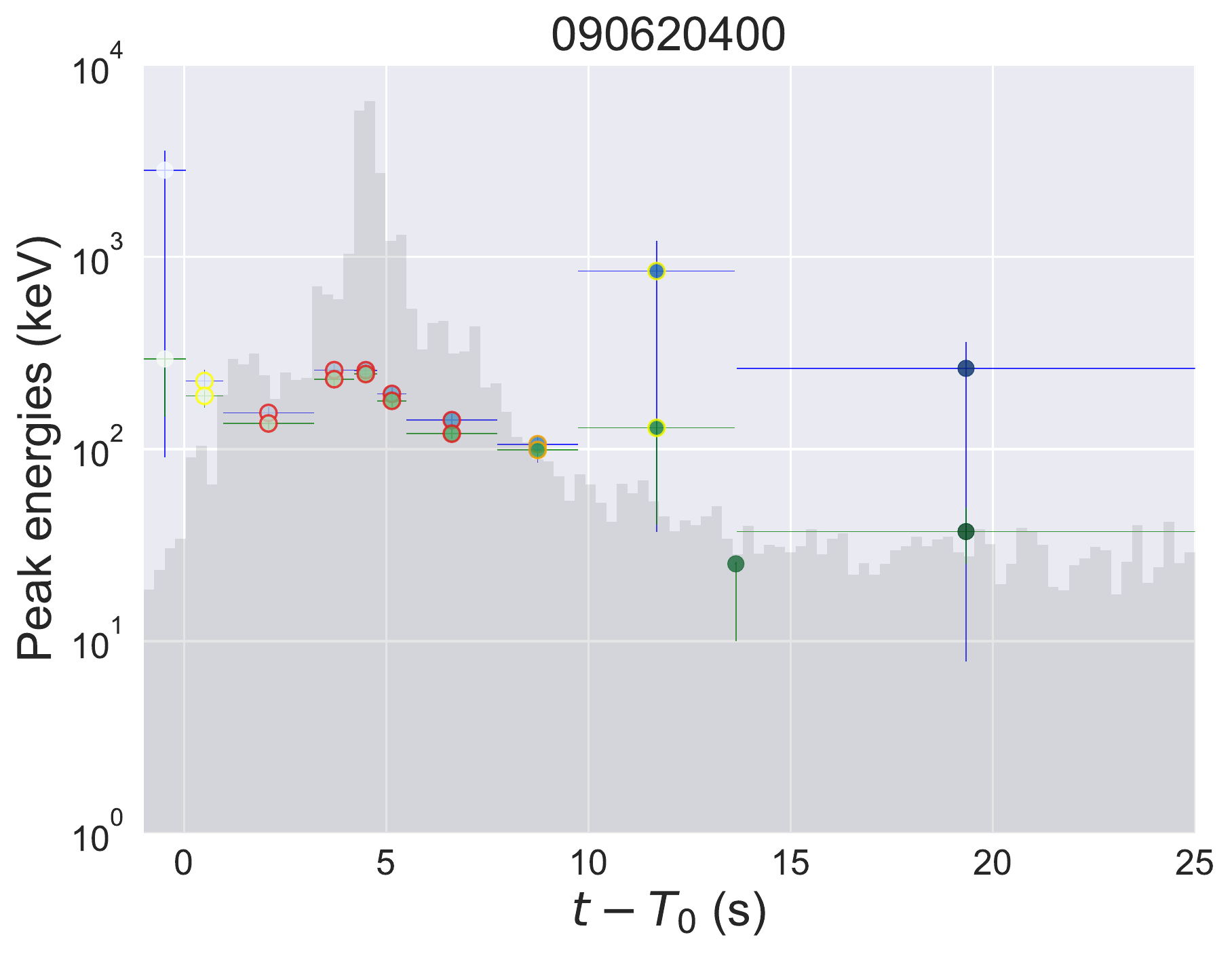}}
\subfigure{\includegraphics[width=0.3\linewidth]{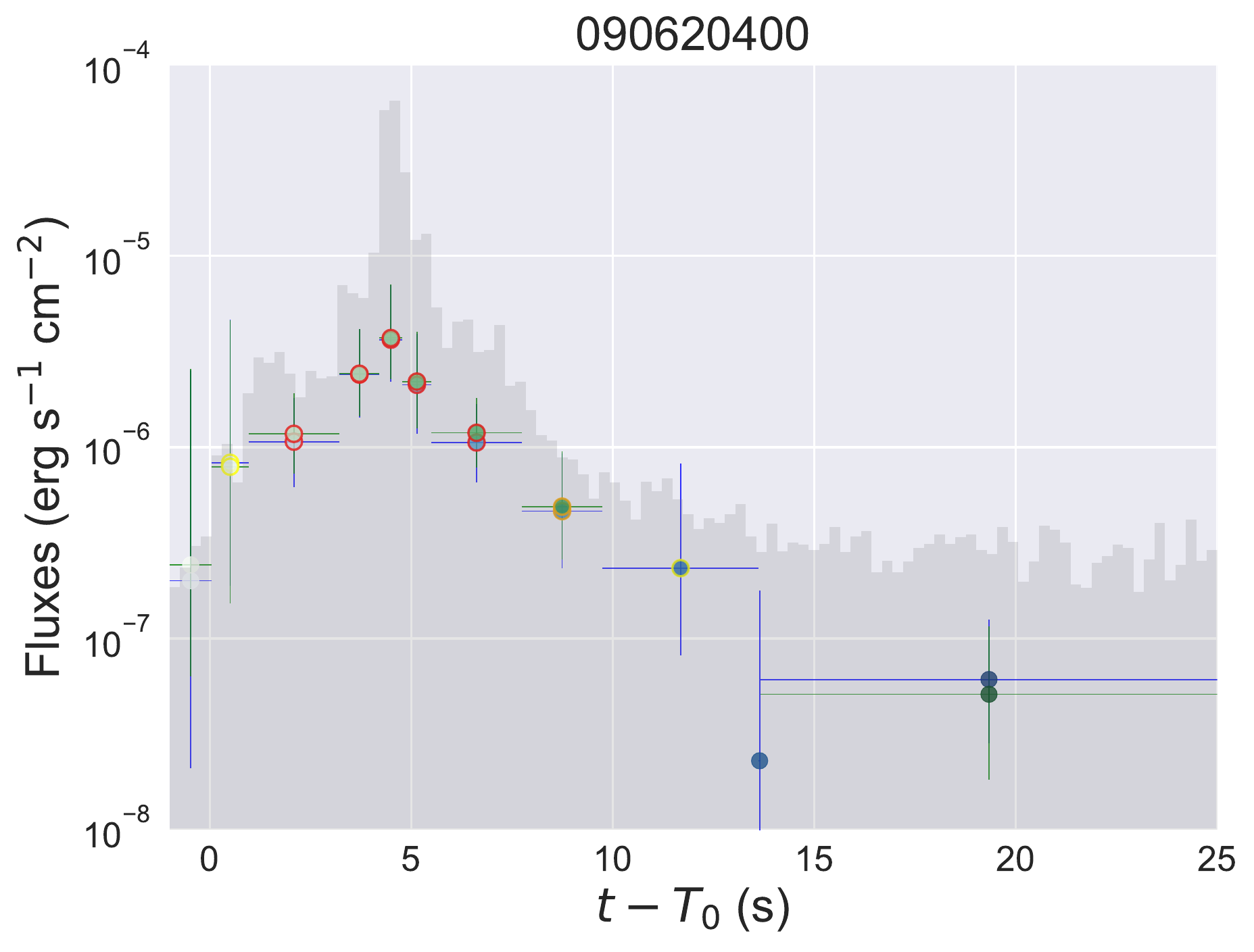}}

\subfigure{\includegraphics[width=0.3\linewidth]{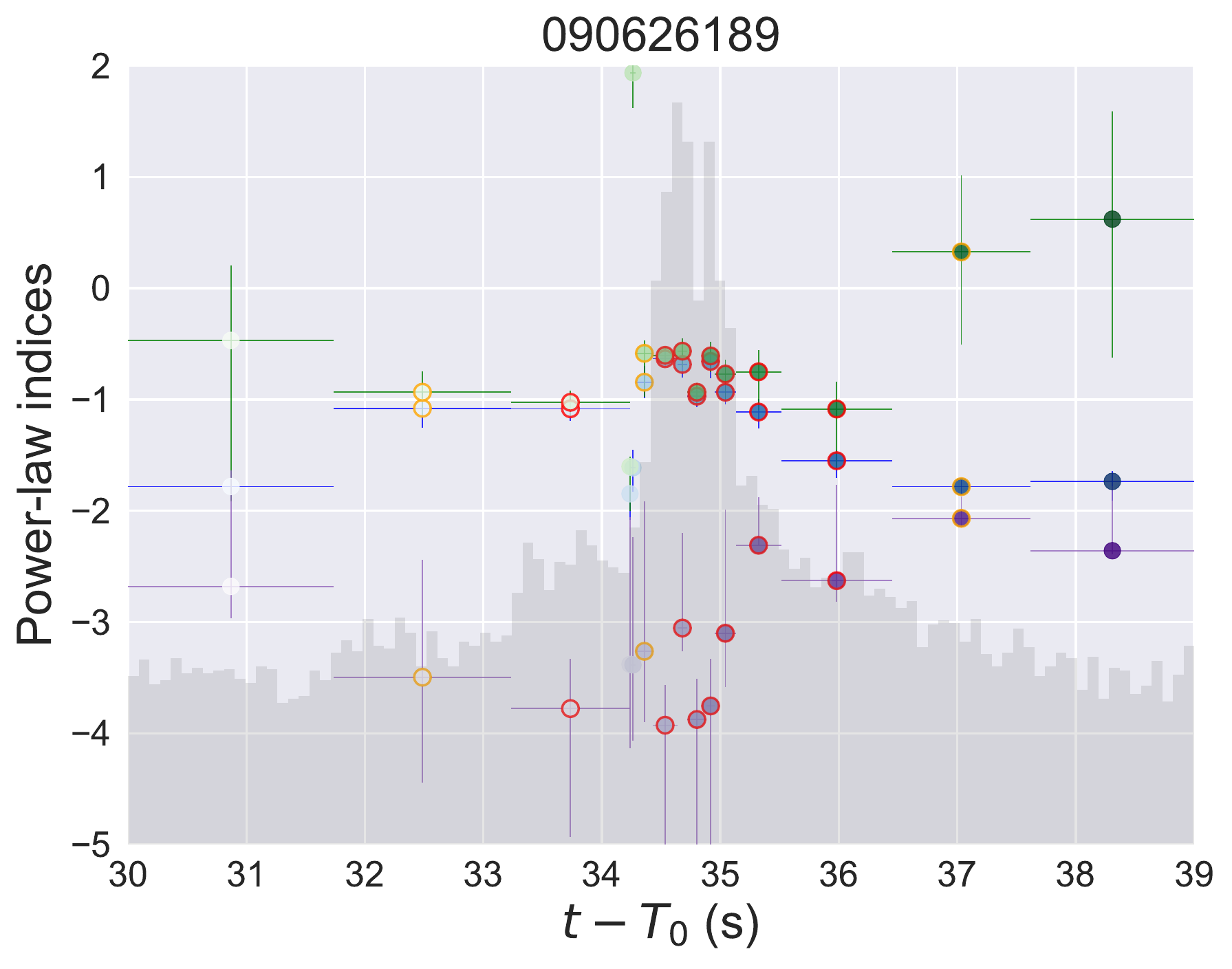}}
\subfigure{\includegraphics[width=0.3\linewidth]{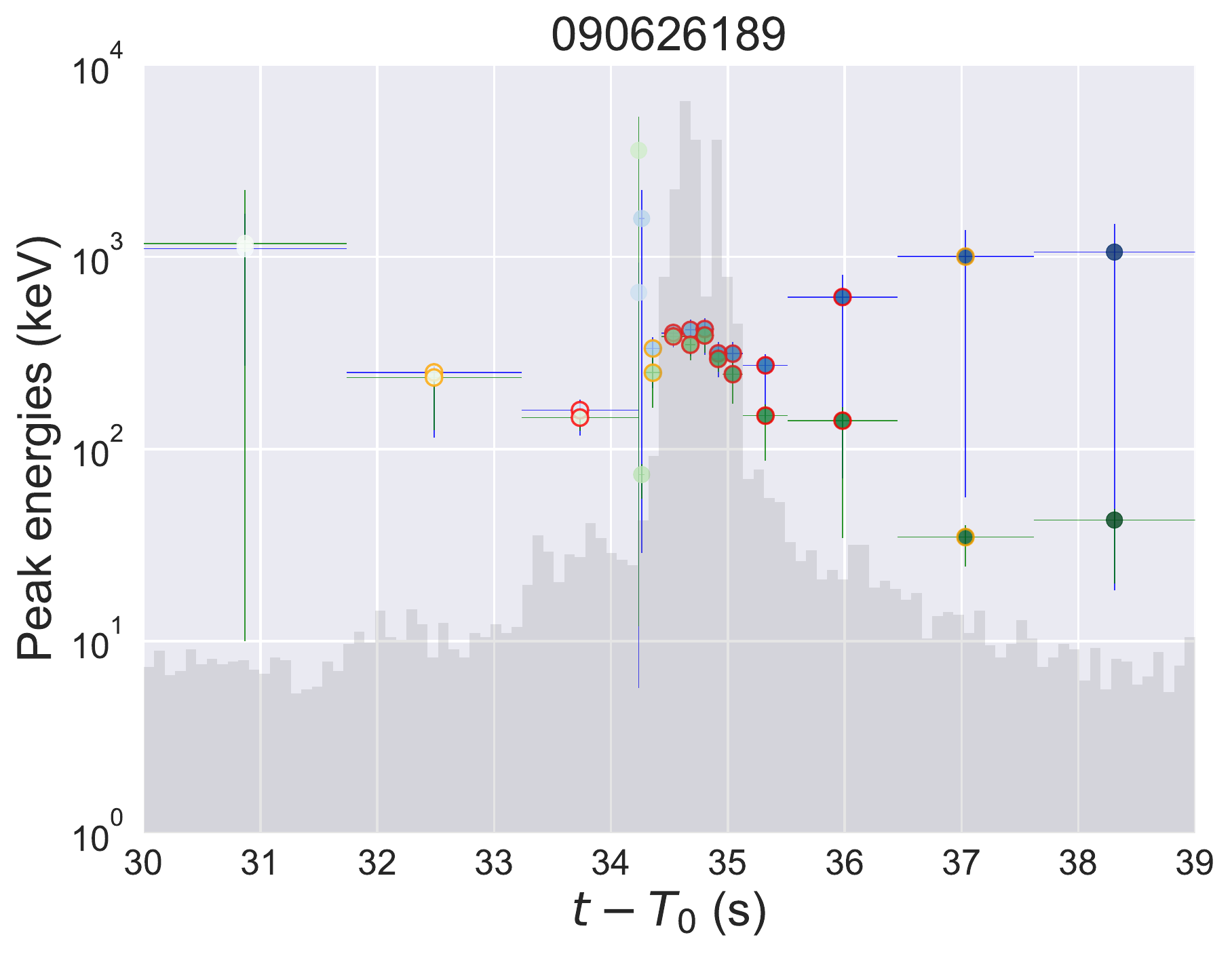}}
\subfigure{\includegraphics[width=0.3\linewidth]{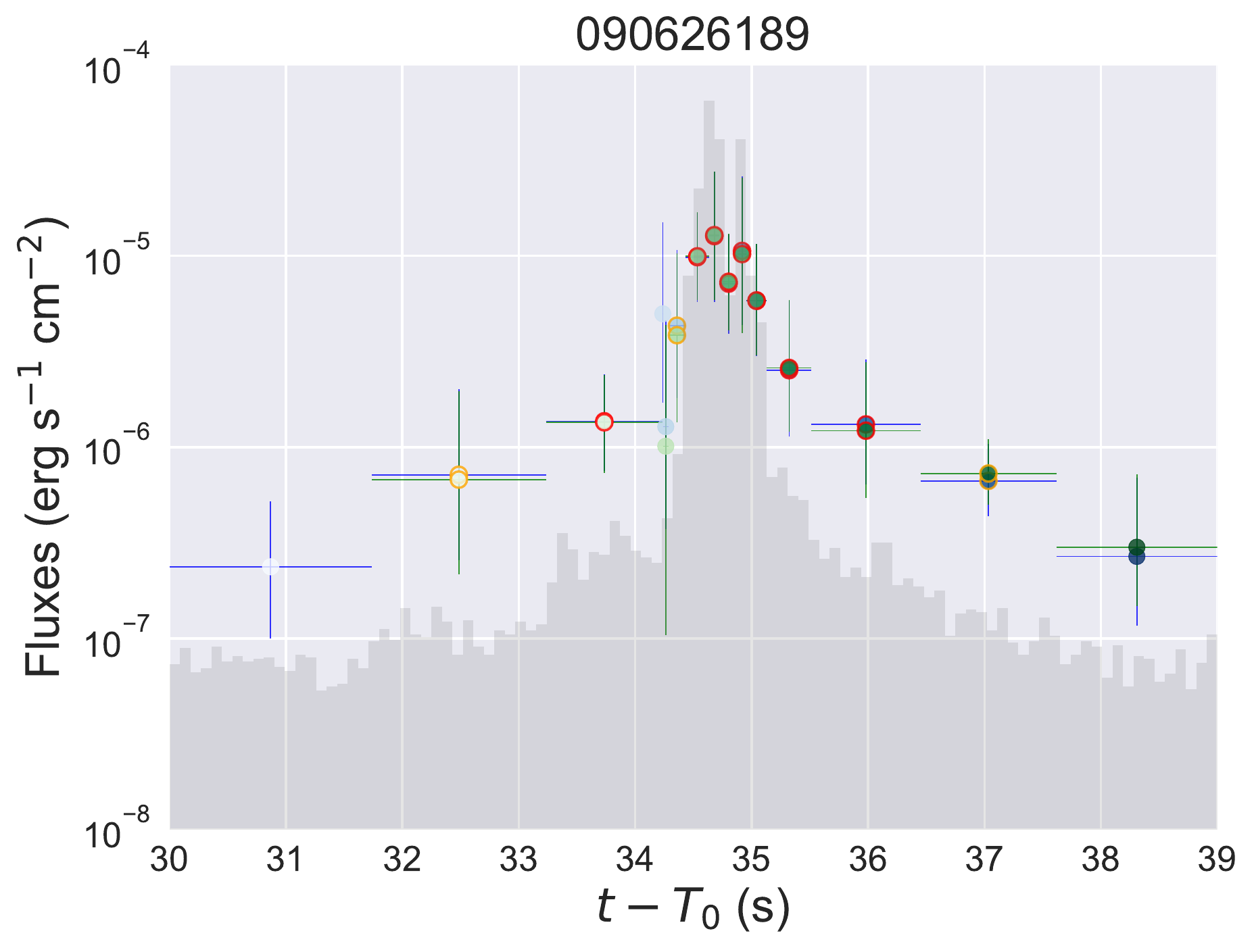}}

\subfigure{\includegraphics[width=0.3\linewidth]{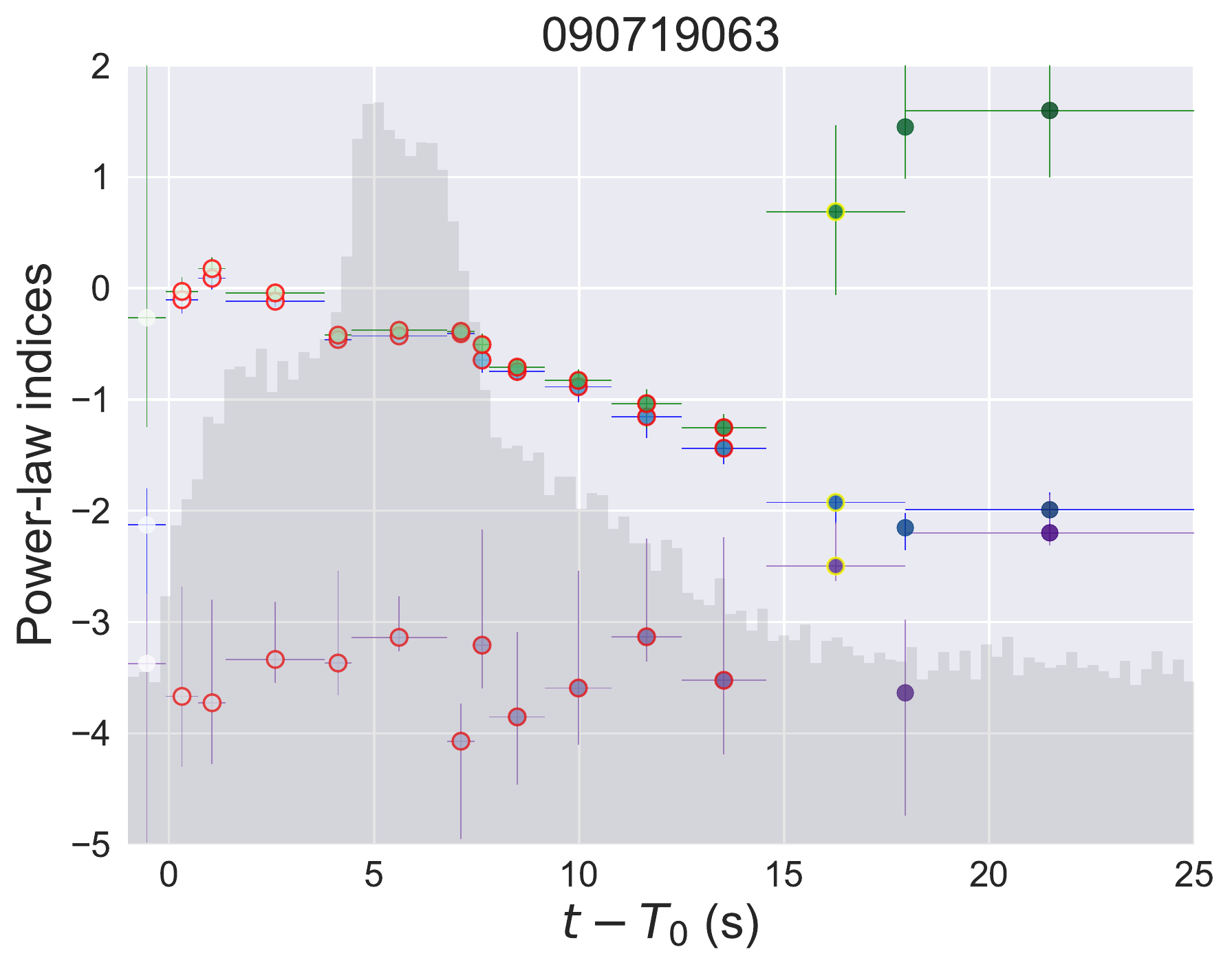}}
\subfigure{\includegraphics[width=0.3\linewidth]{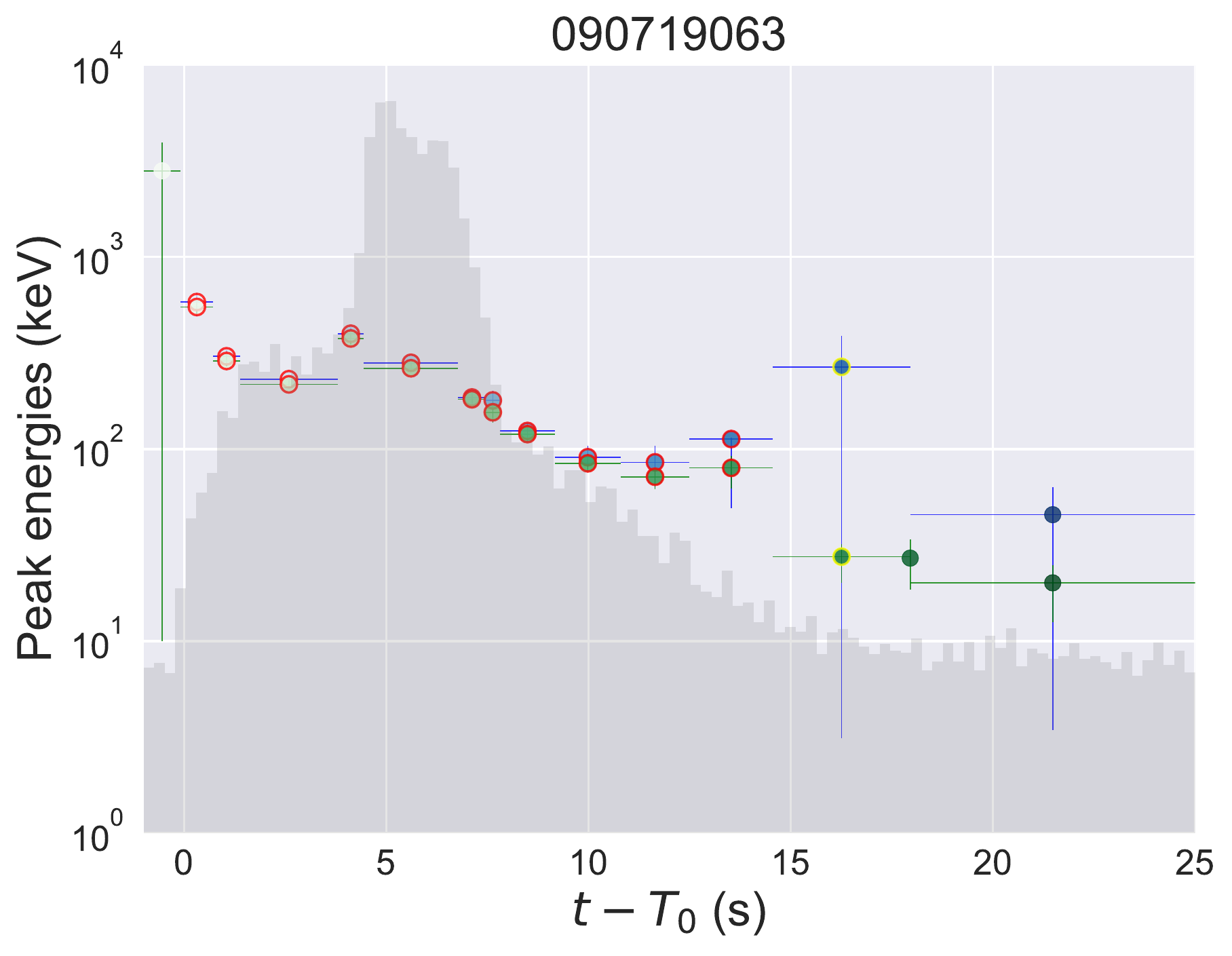}}
\subfigure{\includegraphics[width=0.3\linewidth]{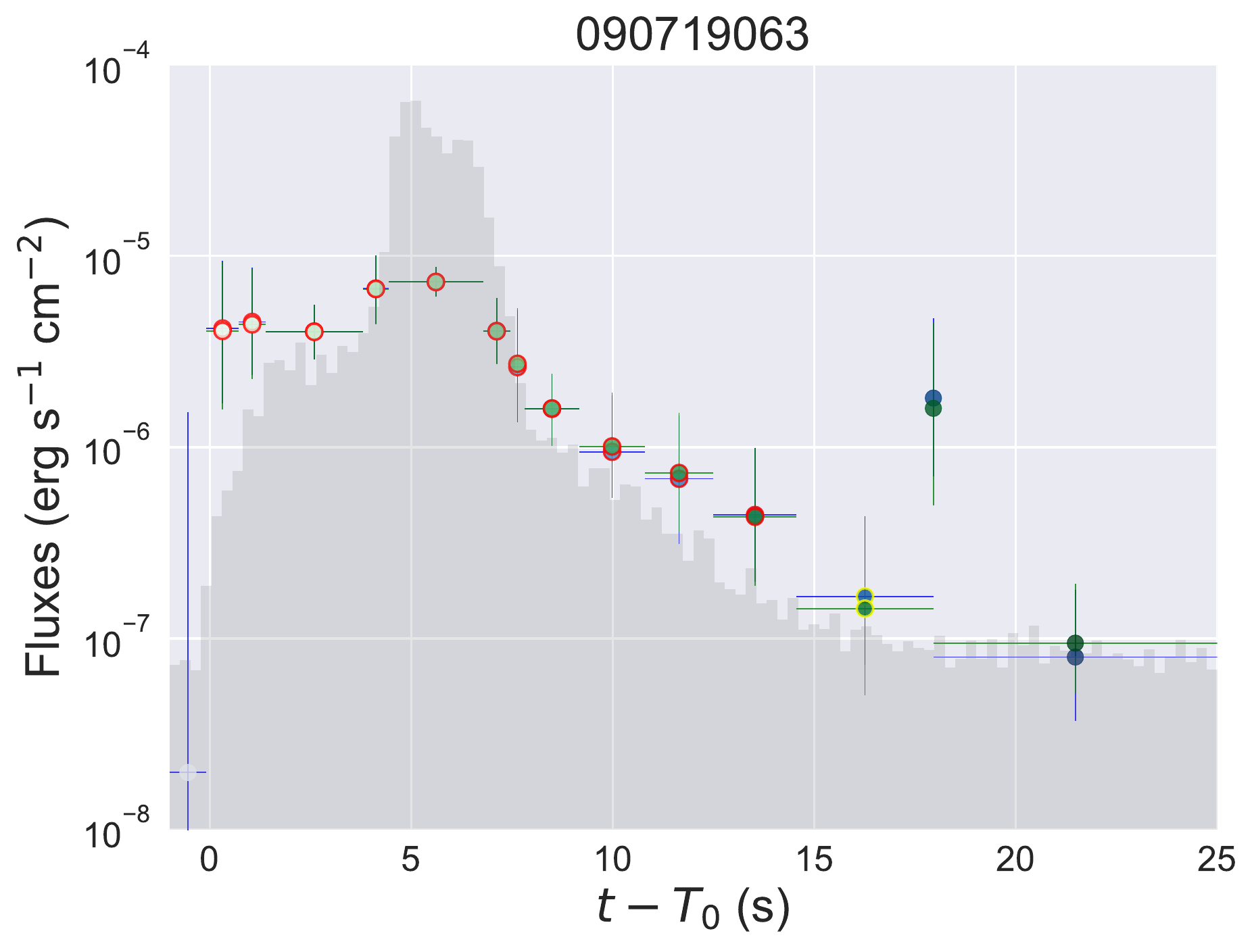}}

\caption{Same as Fig.~\ref{fig:evolution_group1}.
\label{fig:evolution_group2}}
\end{figure*}

\begin{figure*}
\centering

\subfigure{\includegraphics[width=0.3\linewidth]{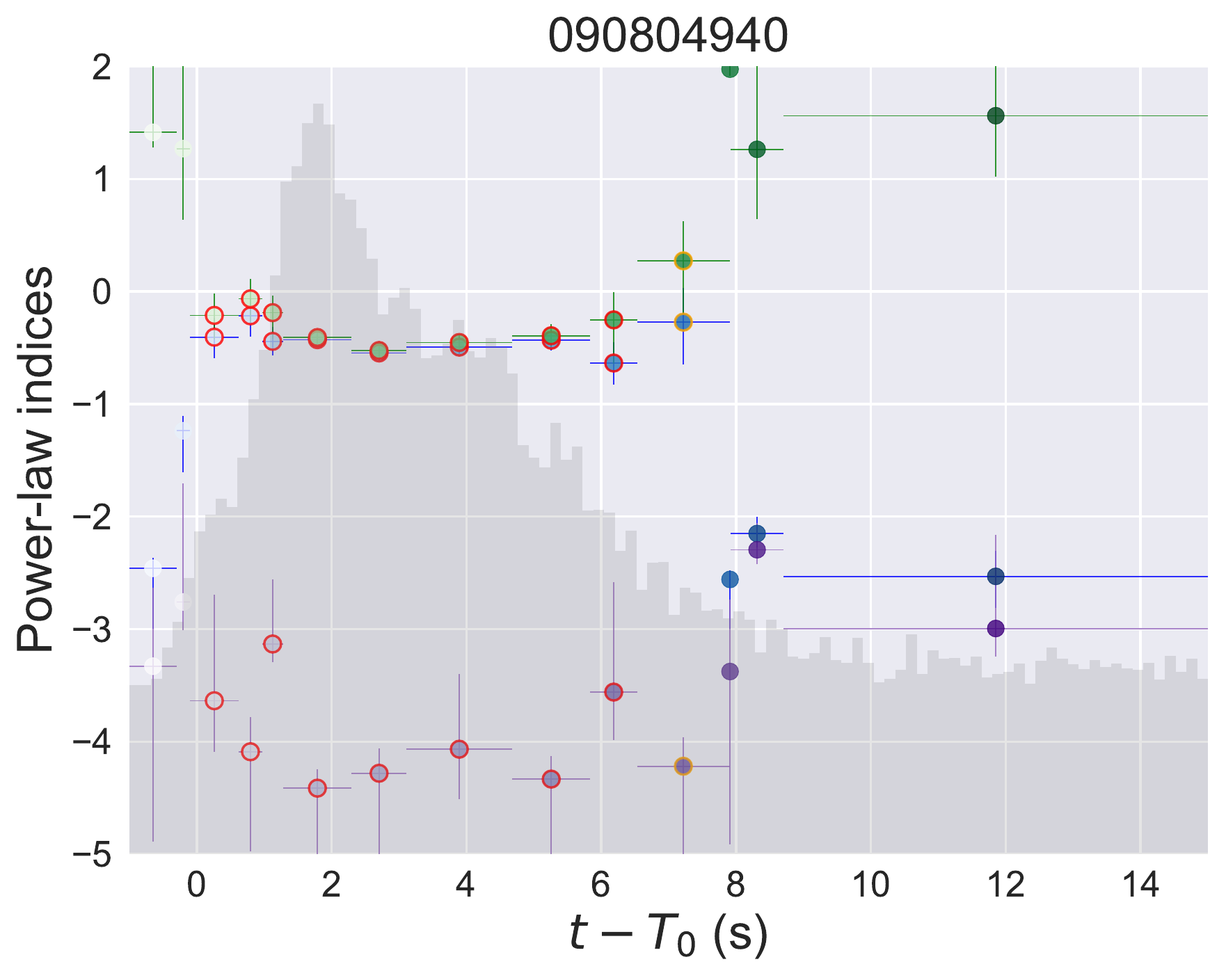}}
\subfigure{\includegraphics[width=0.3\linewidth]{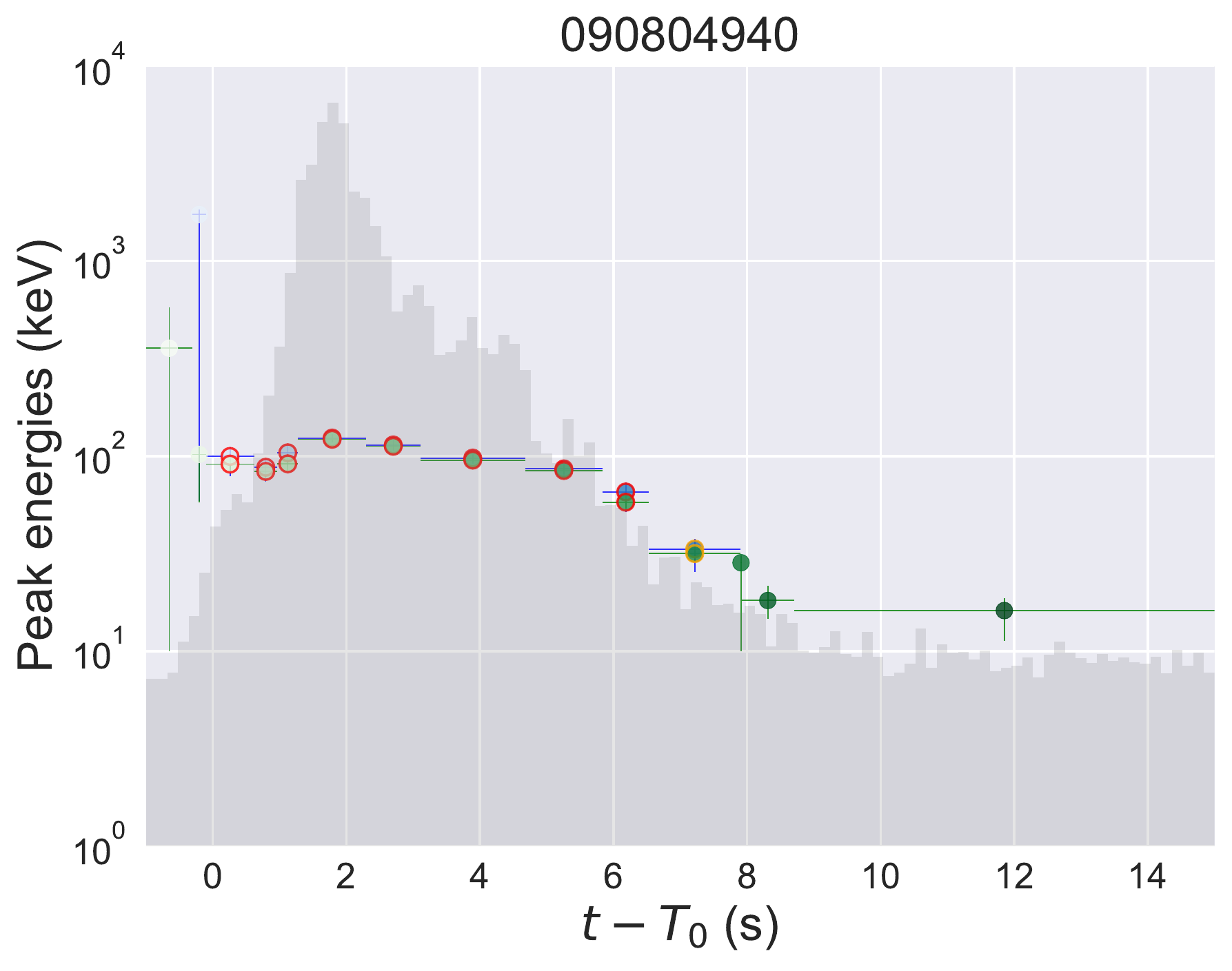}}
\subfigure{\includegraphics[width=0.3\linewidth]{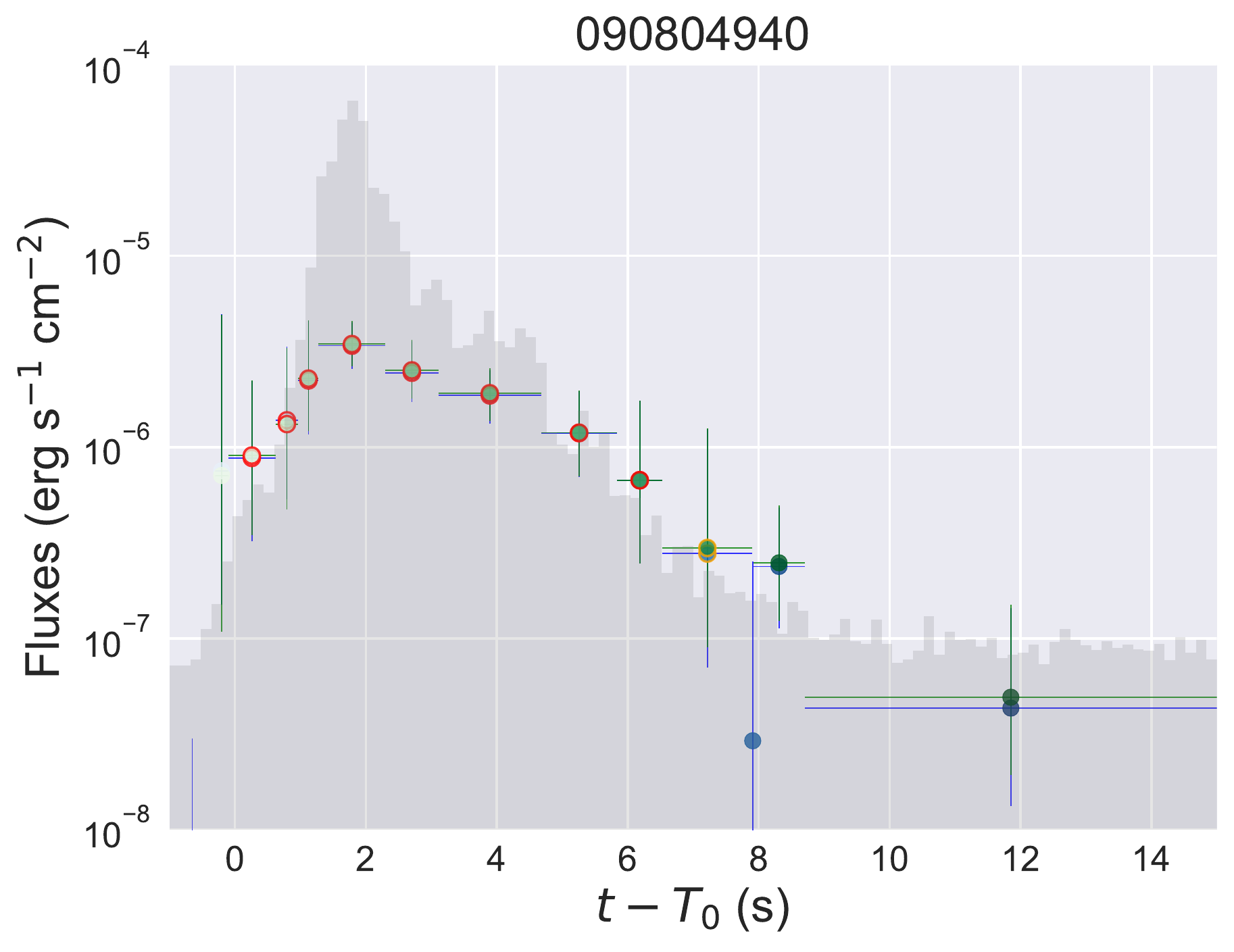}}

\subfigure{\includegraphics[width=0.3\linewidth]{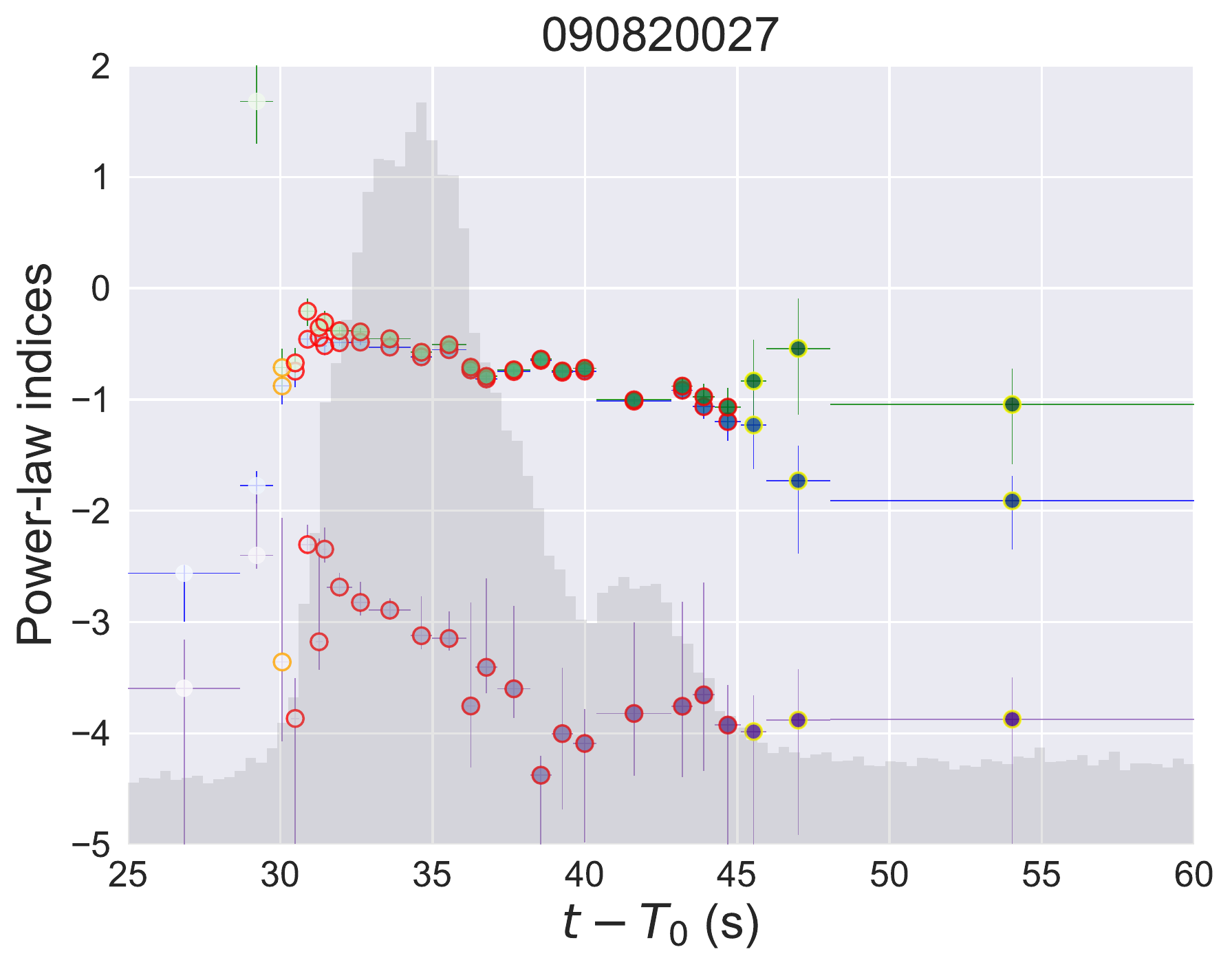}}
\subfigure{\includegraphics[width=0.3\linewidth]{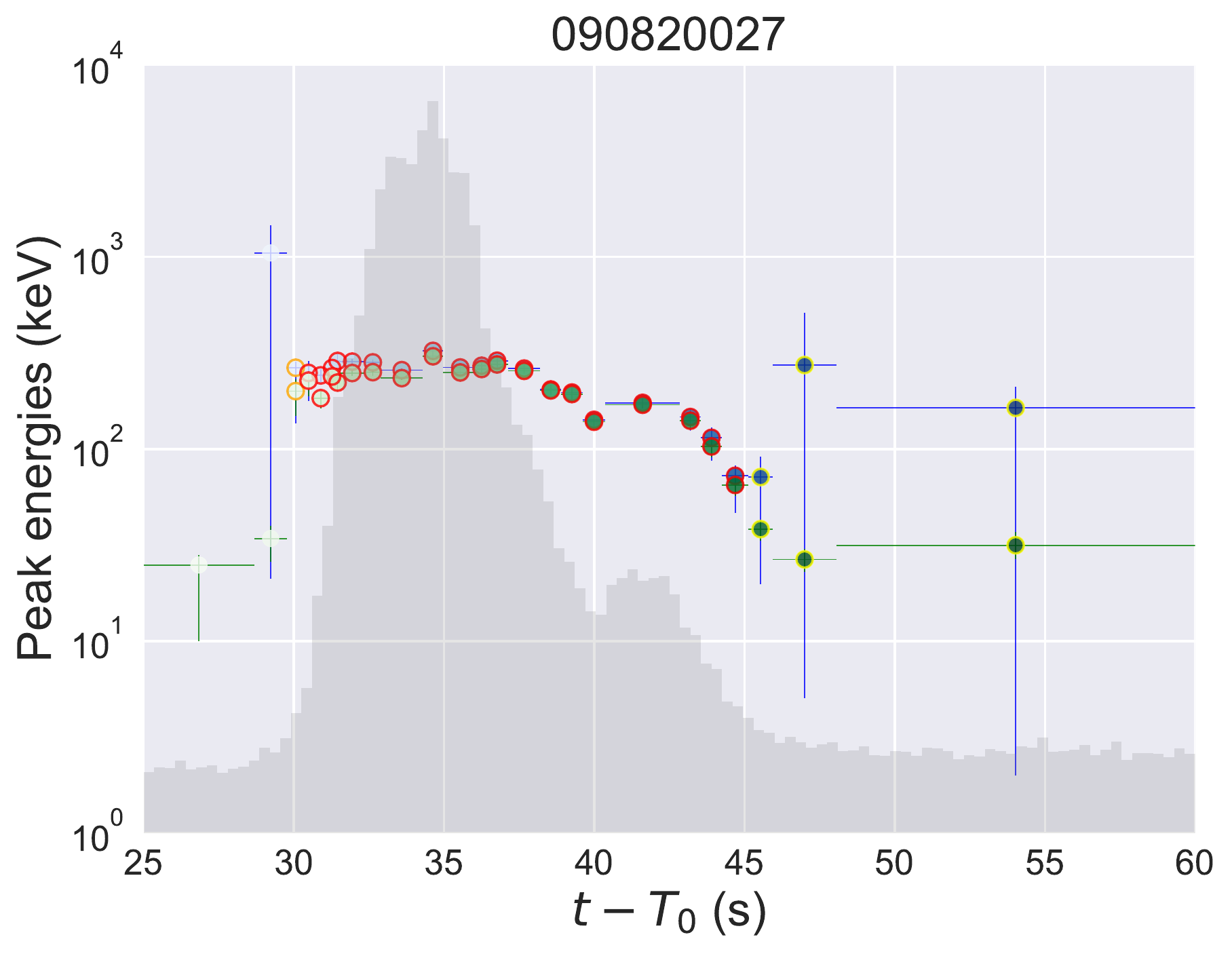}}
\subfigure{\includegraphics[width=0.3\linewidth]{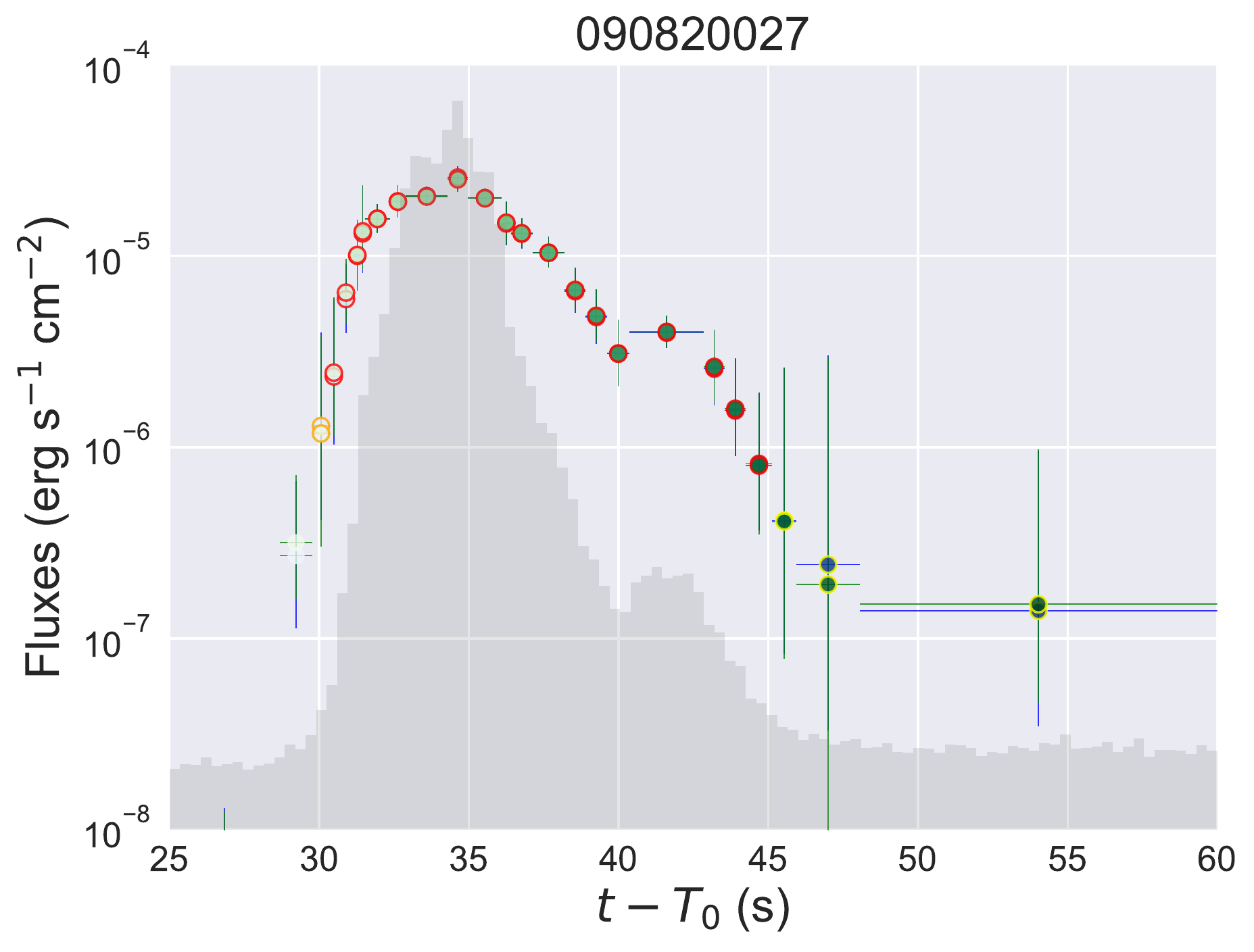}}

\subfigure{\includegraphics[width=0.3\linewidth]{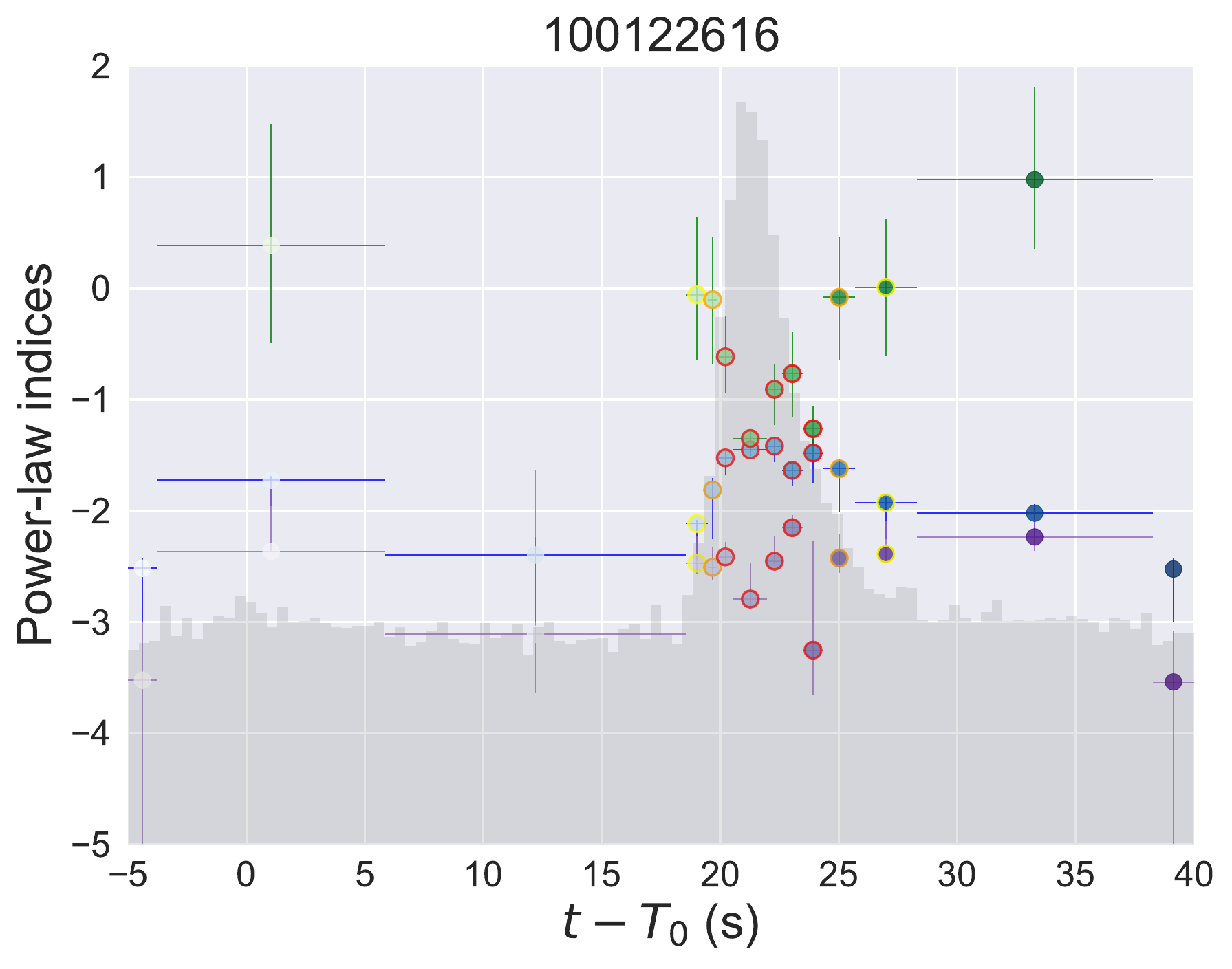}}
\subfigure{\includegraphics[width=0.3\linewidth]{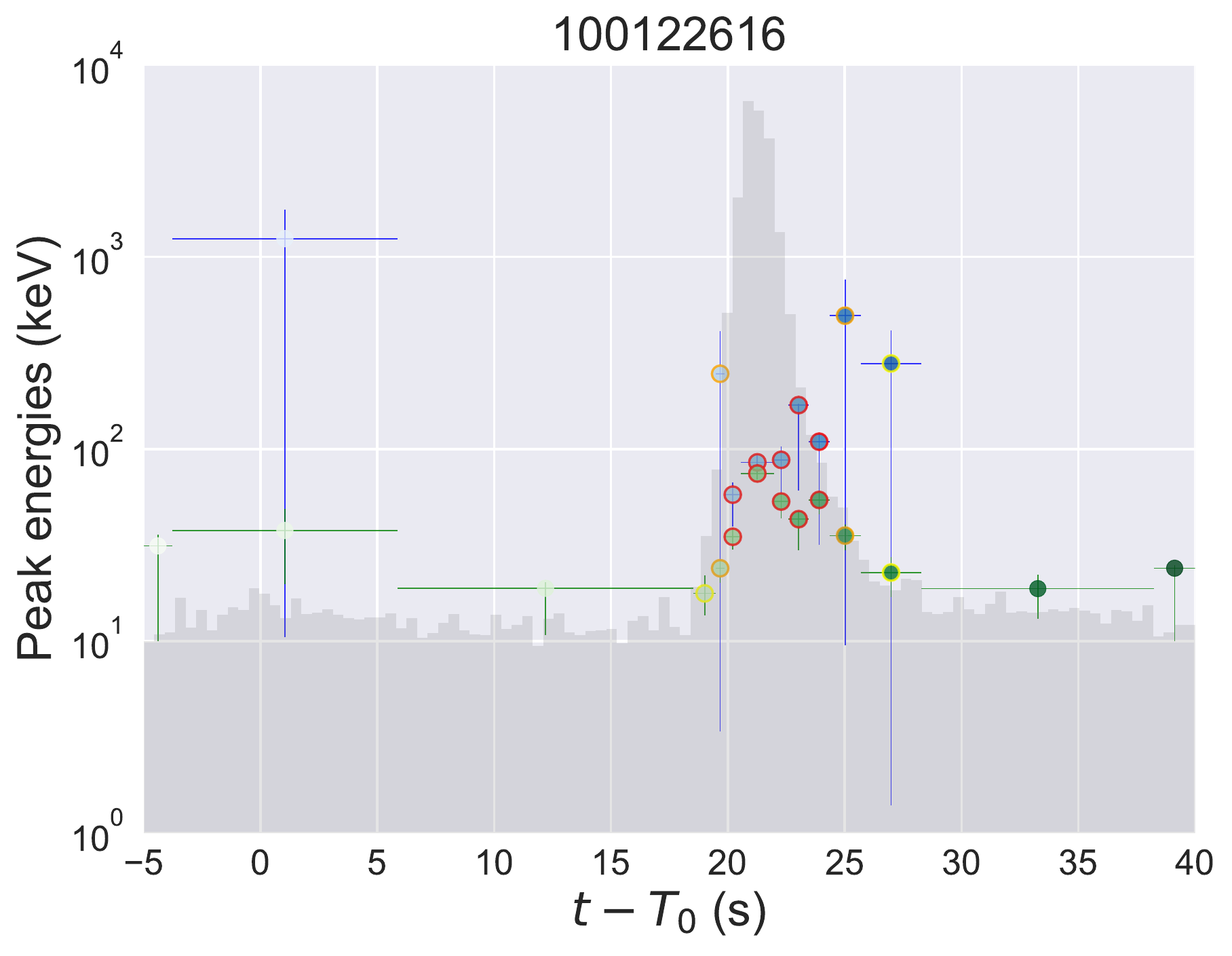}}
\subfigure{\includegraphics[width=0.3\linewidth]{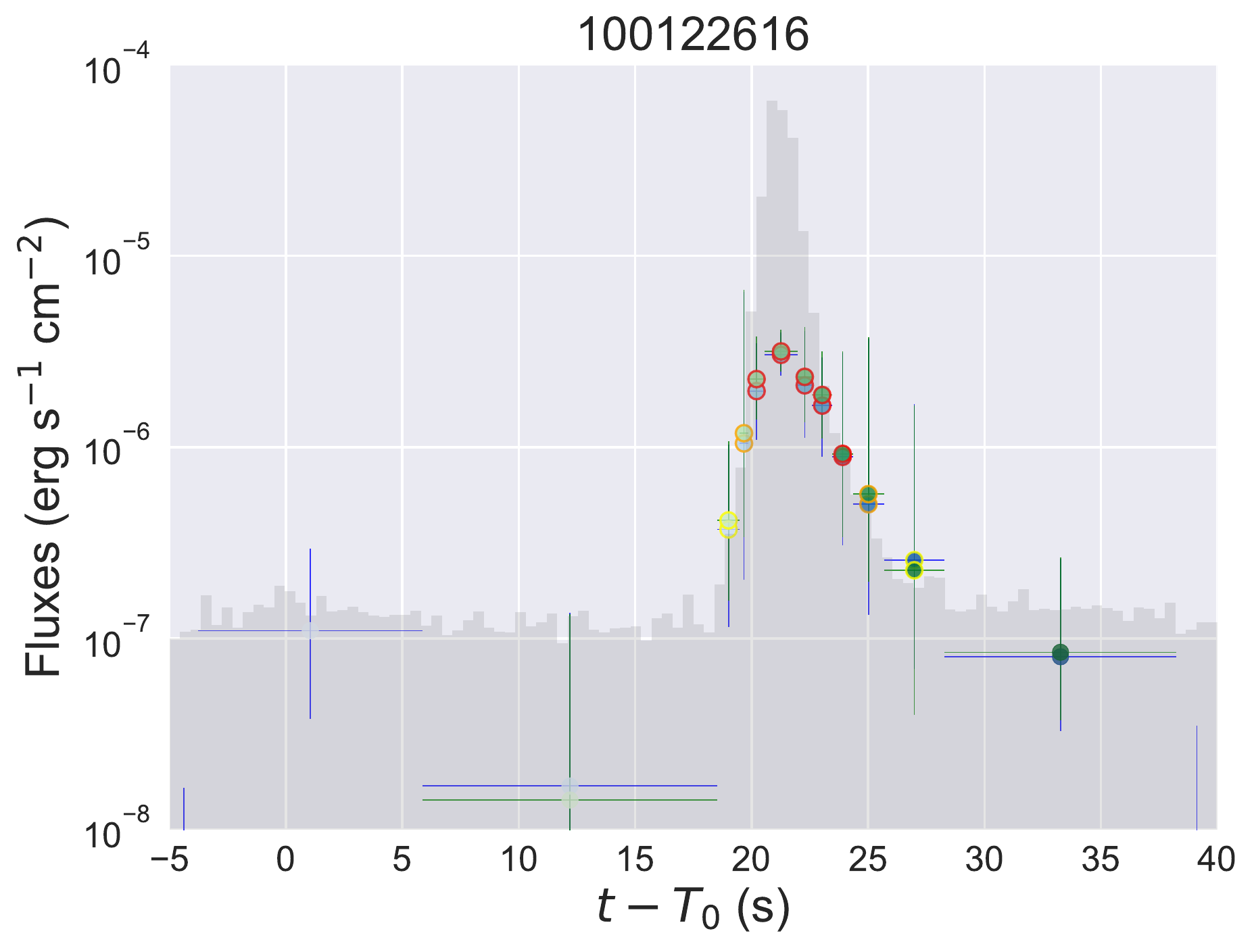}}

\subfigure{\includegraphics[width=0.3\linewidth]{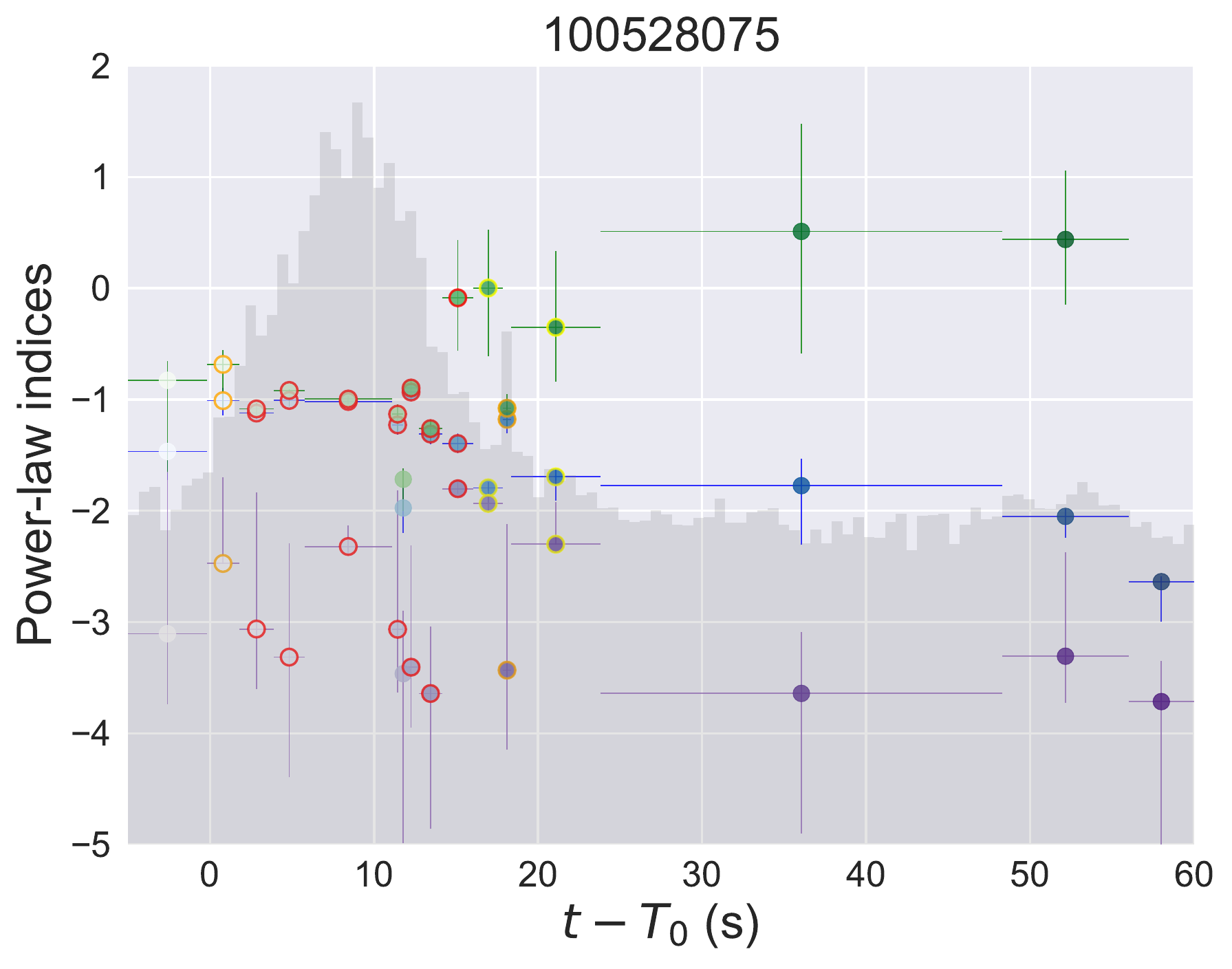}}
\subfigure{\includegraphics[width=0.3\linewidth]{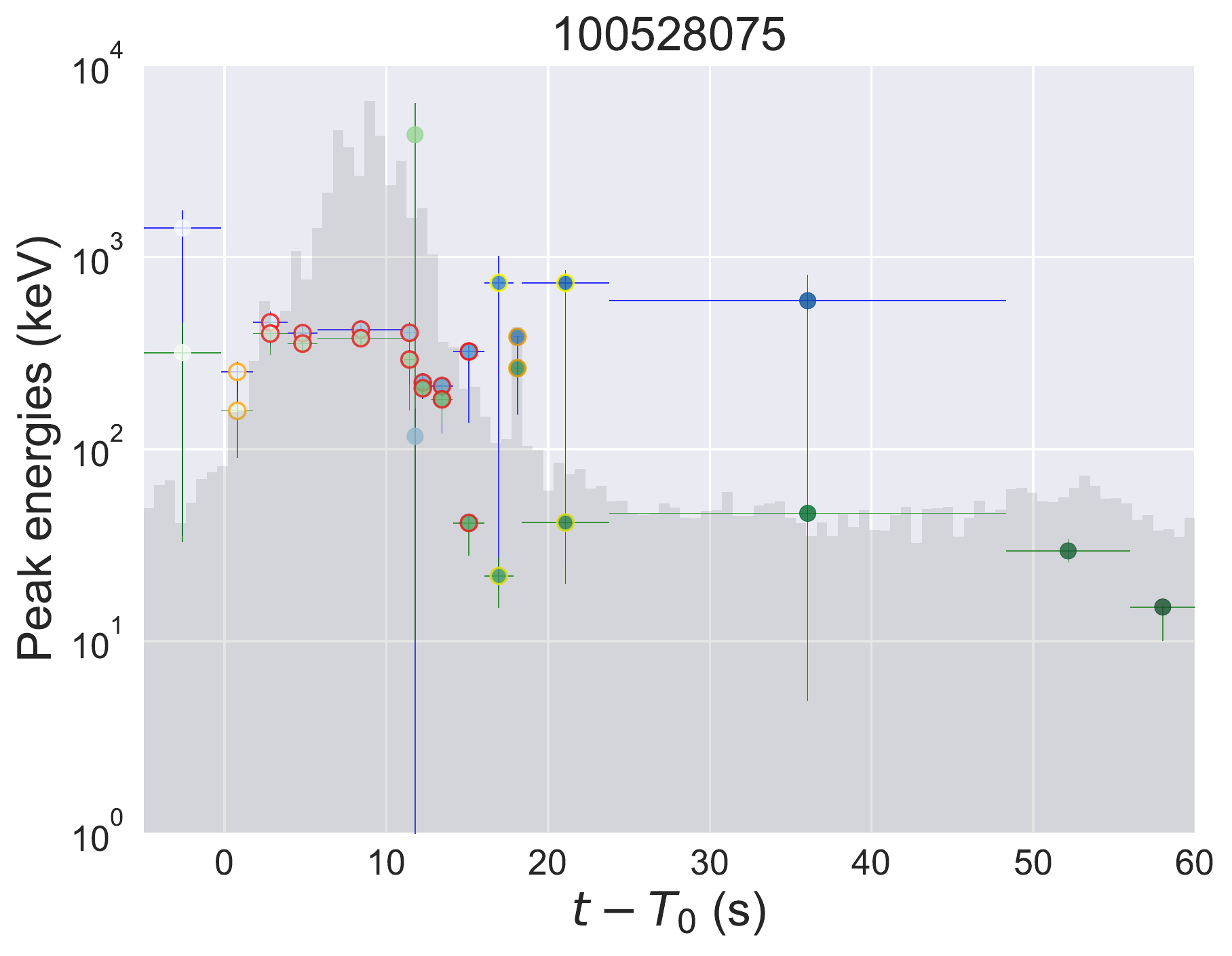}}
\subfigure{\includegraphics[width=0.3\linewidth]{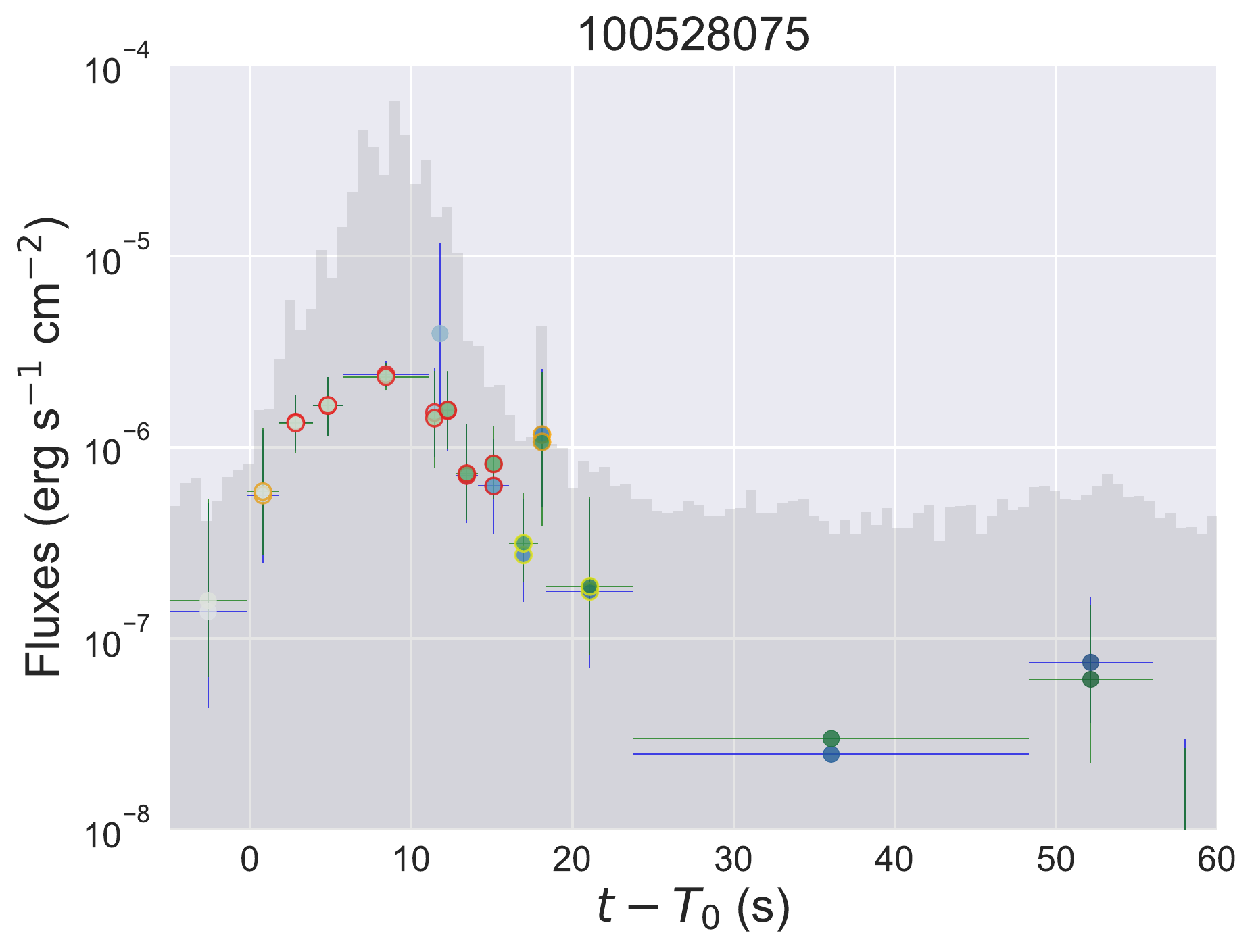}}

\caption{Same as Fig.~\ref{fig:evolution_group1}.
\label{fig:evolution_group3}}
\end{figure*}

\begin{figure*}
\centering

\subfigure{\includegraphics[width=0.3\linewidth]{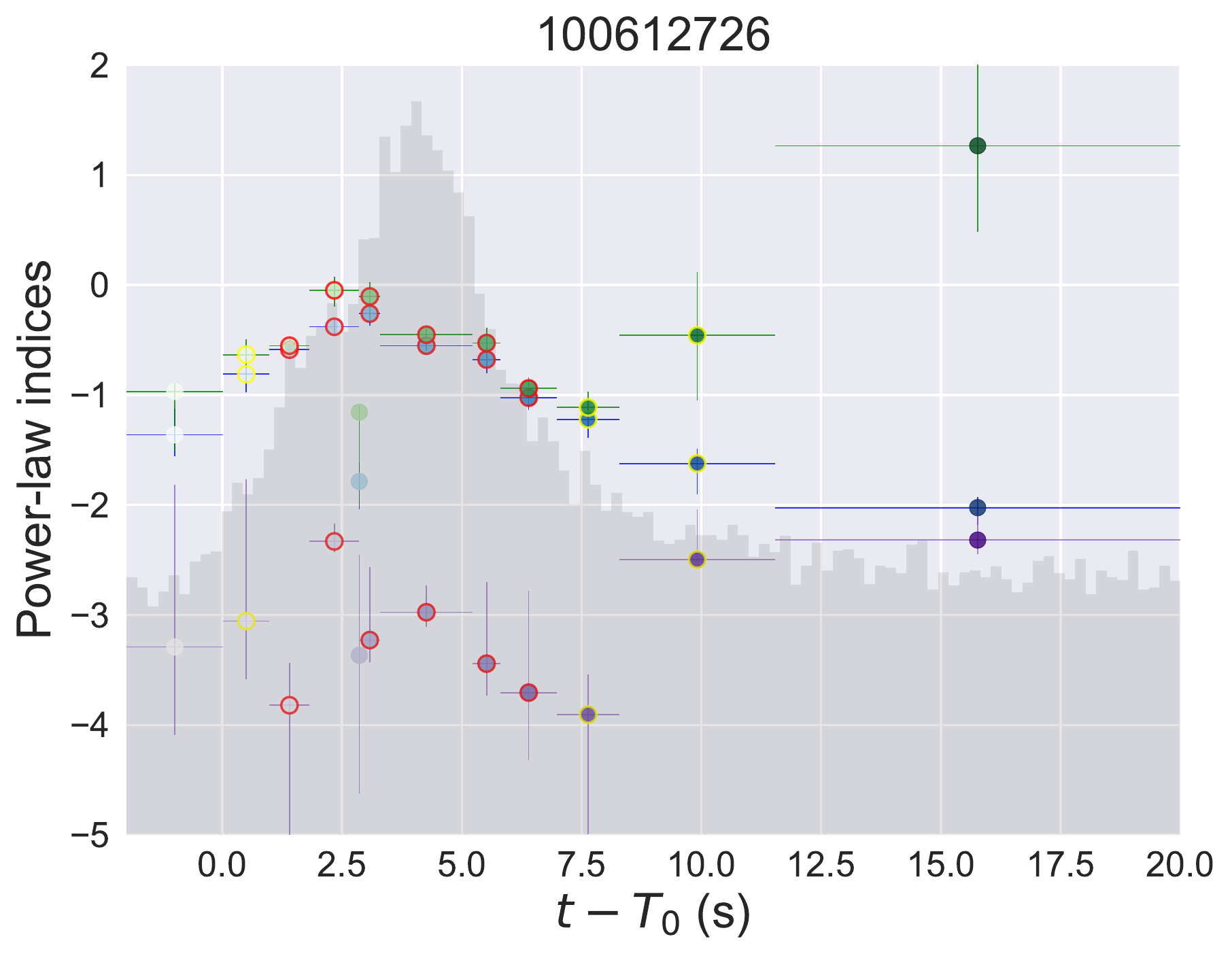}}
\subfigure{\includegraphics[width=0.3\linewidth]{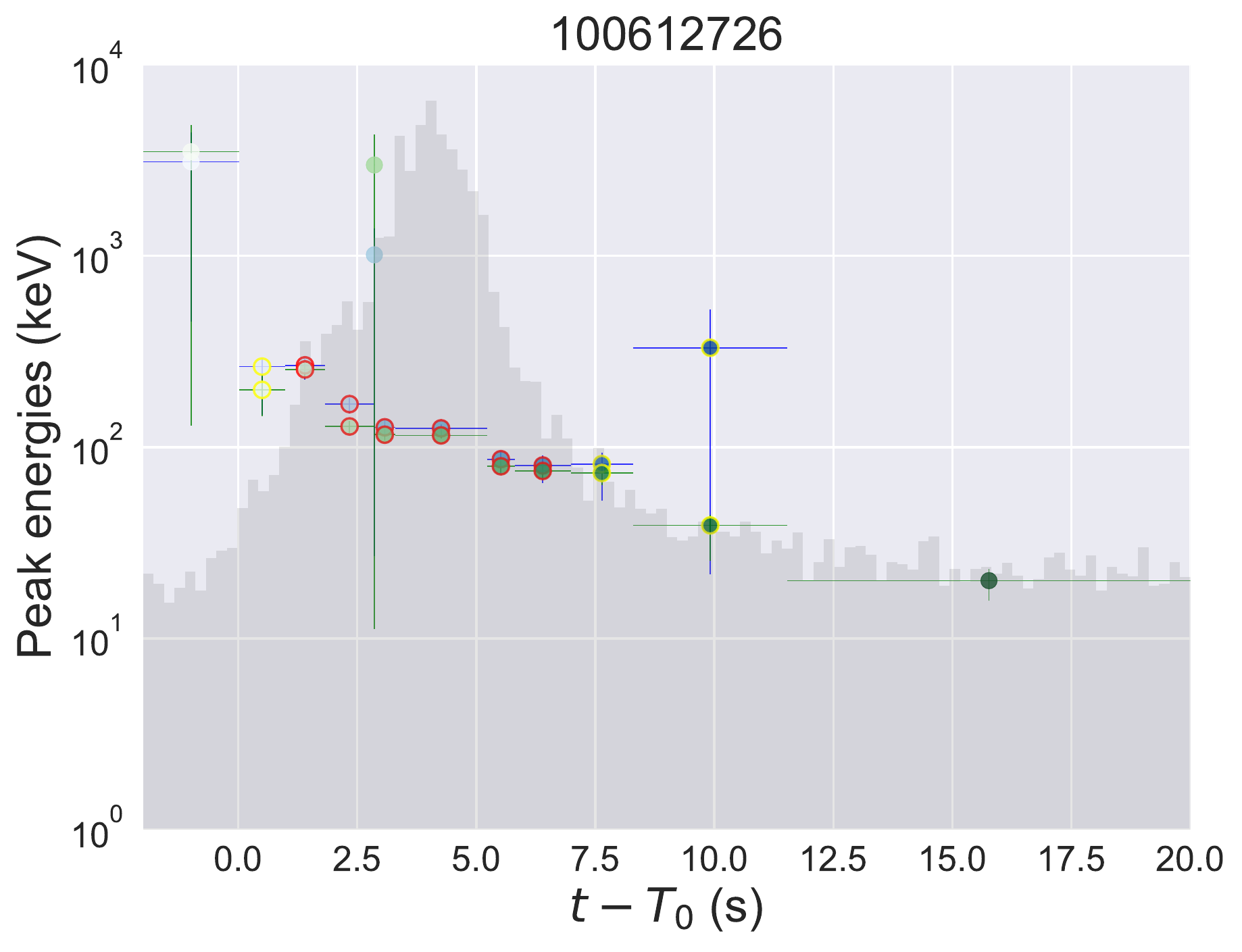}}
\subfigure{\includegraphics[width=0.3\linewidth]{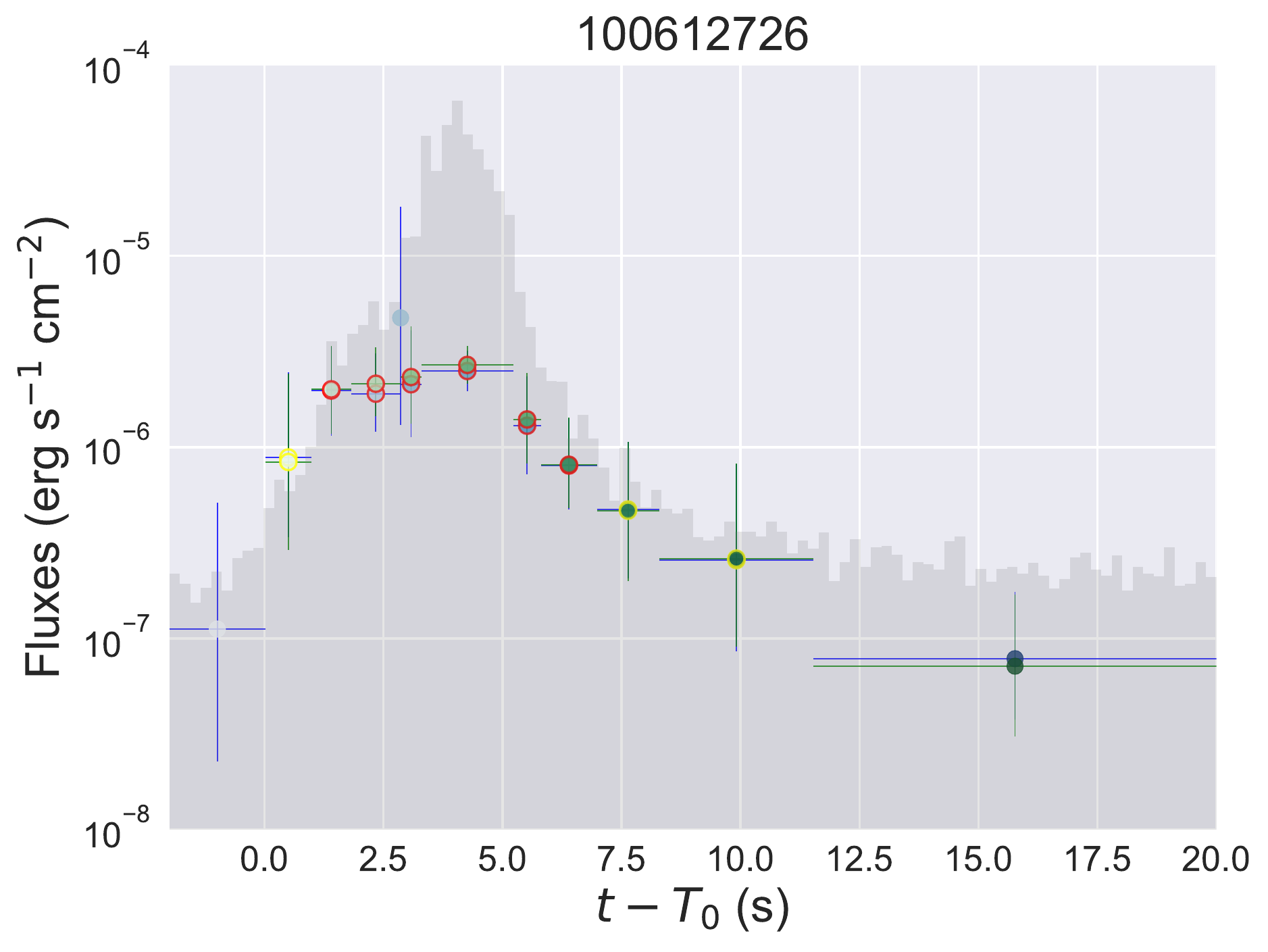}}

\subfigure{\includegraphics[width=0.3\linewidth]{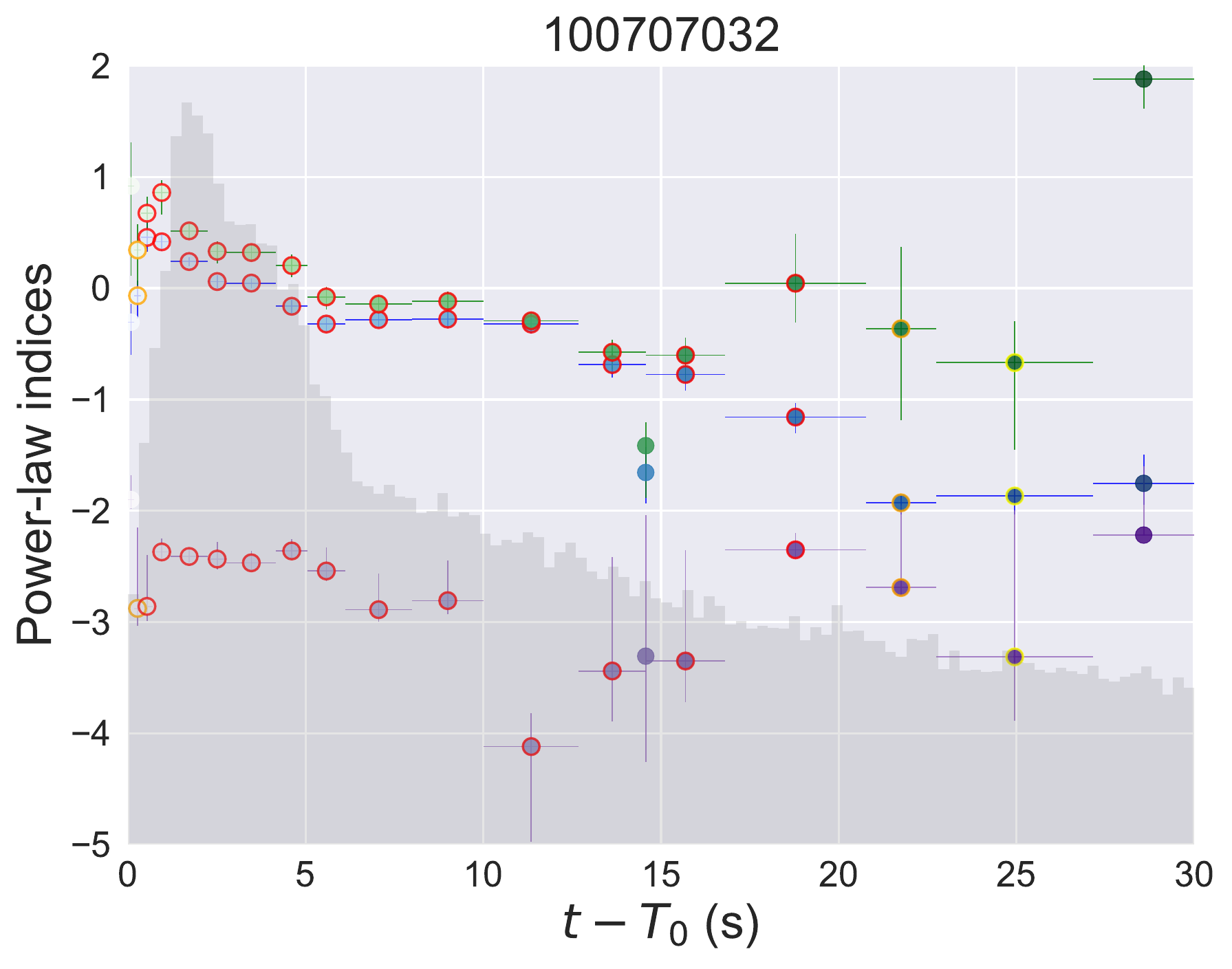}}
\subfigure{\includegraphics[width=0.3\linewidth]{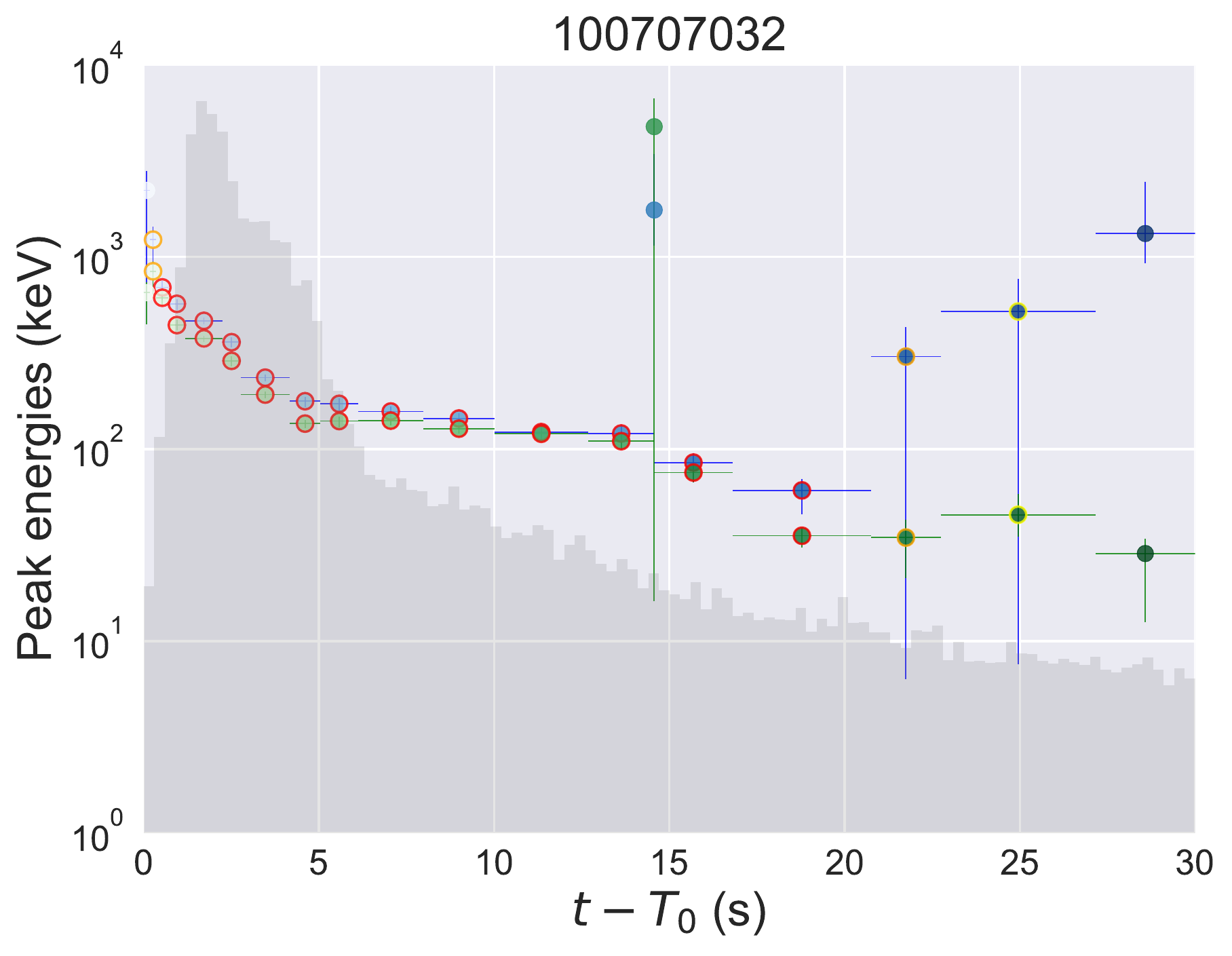}}
\subfigure{\includegraphics[width=0.3\linewidth]{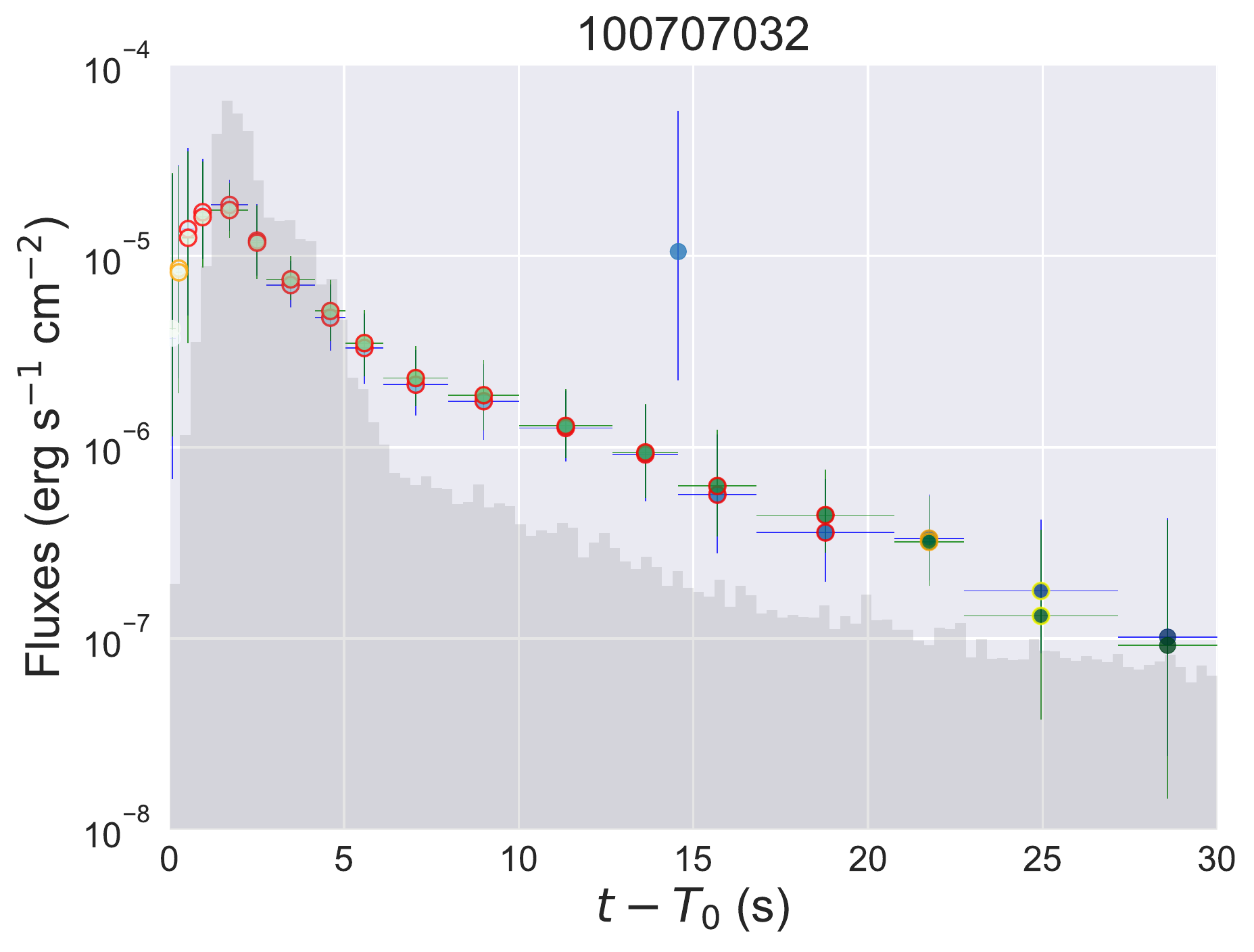}}

\subfigure{\includegraphics[width=0.3\linewidth]{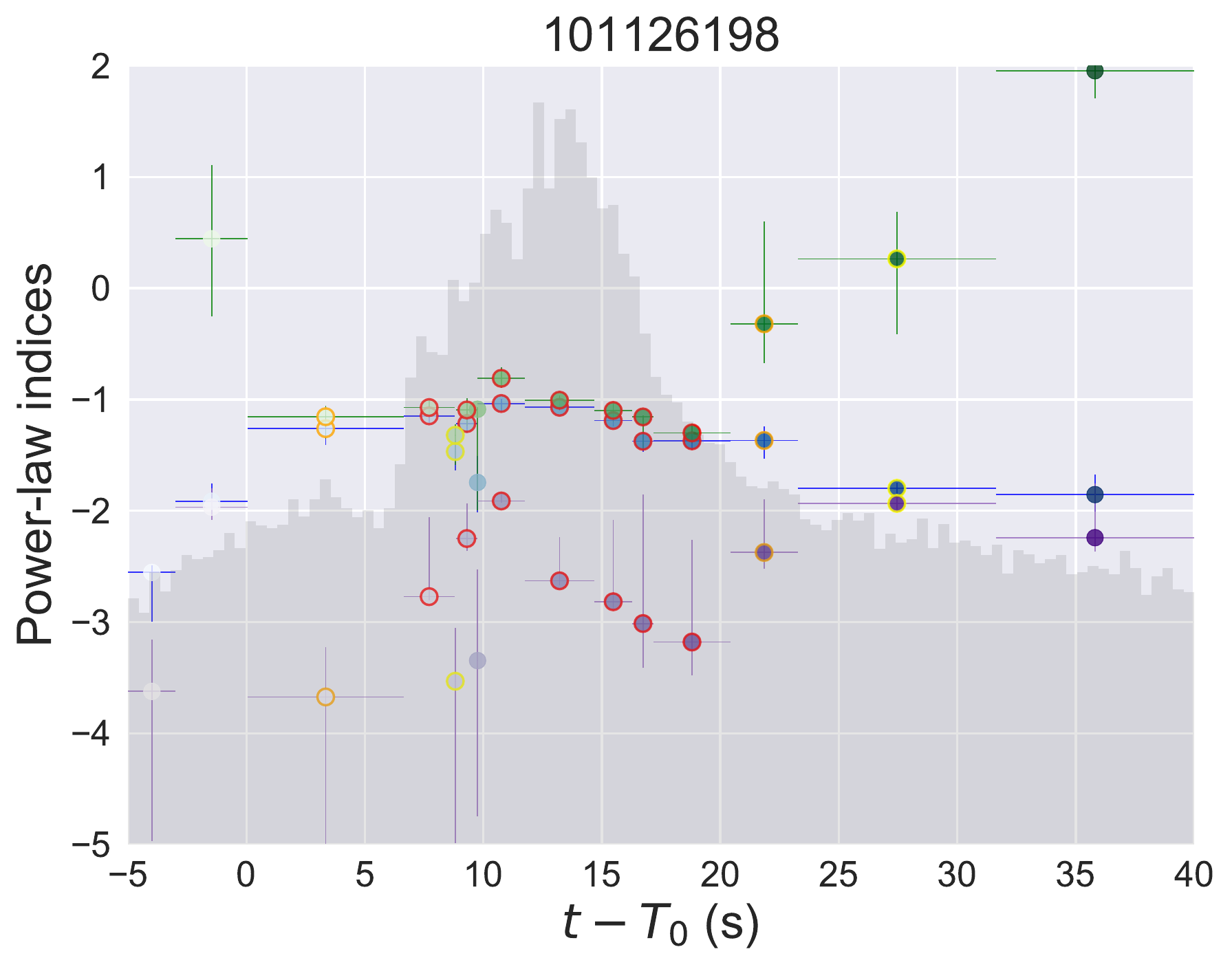}}
\subfigure{\includegraphics[width=0.3\linewidth]{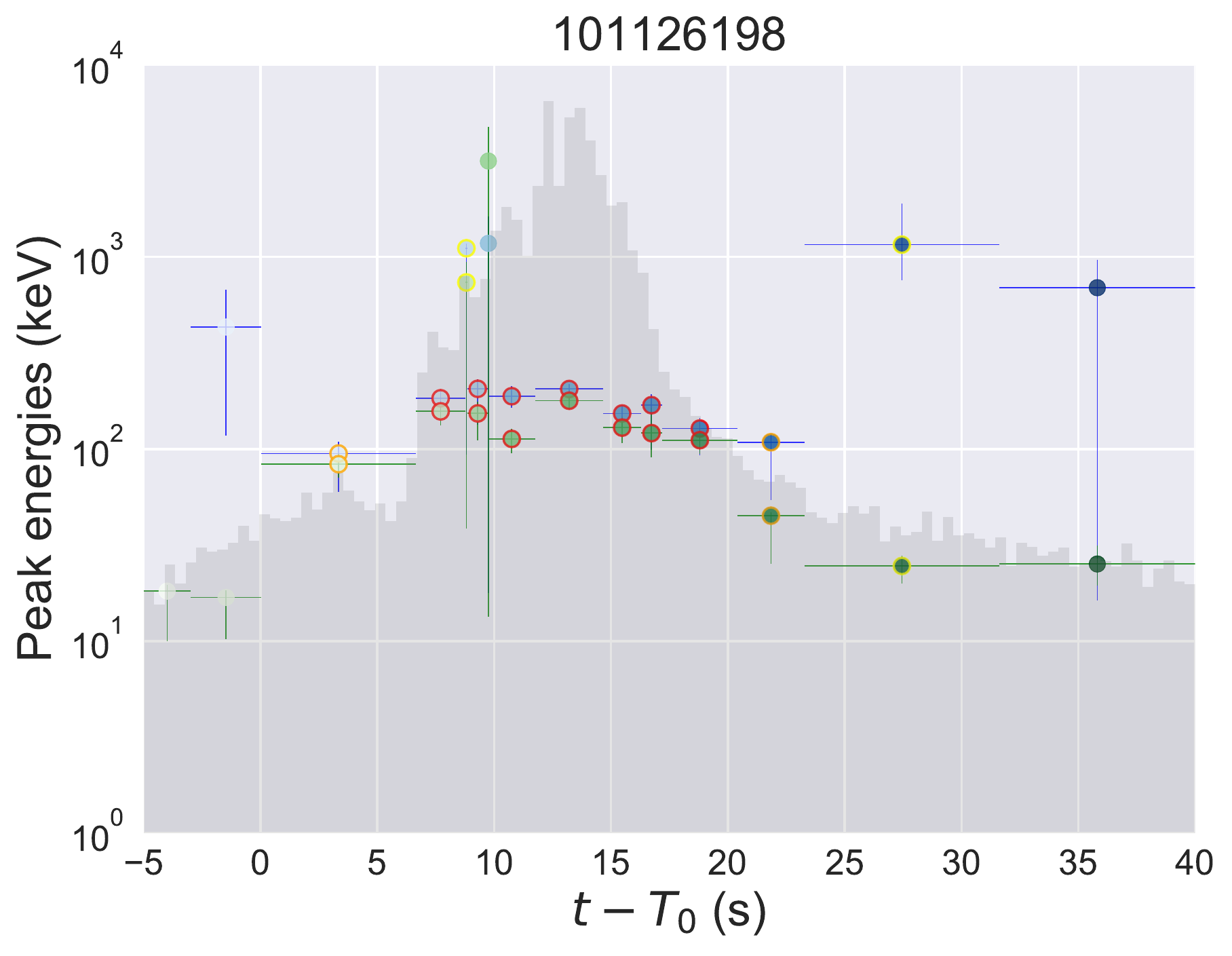}}
\subfigure{\includegraphics[width=0.3\linewidth]{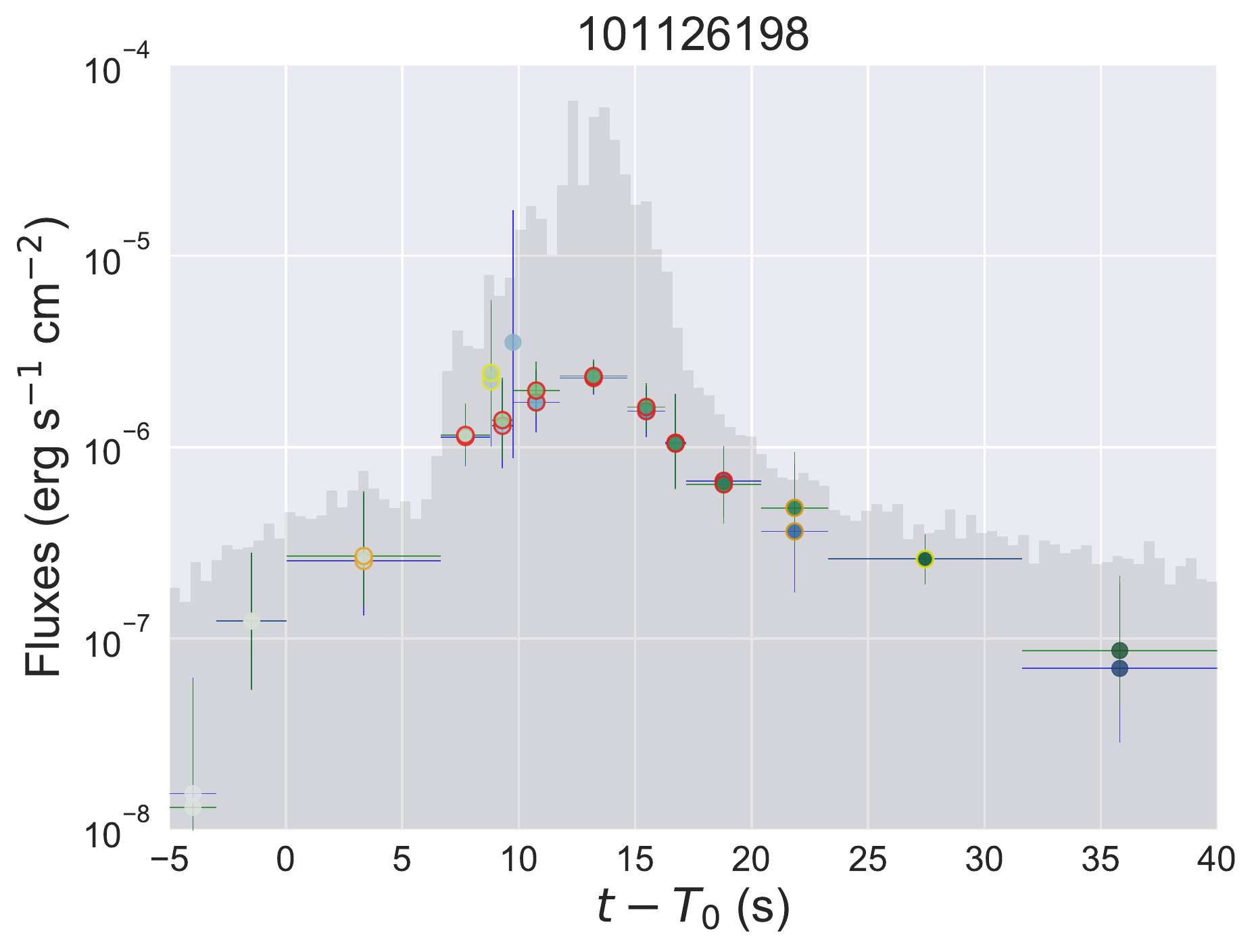}}

\subfigure{\includegraphics[width=0.3\linewidth]{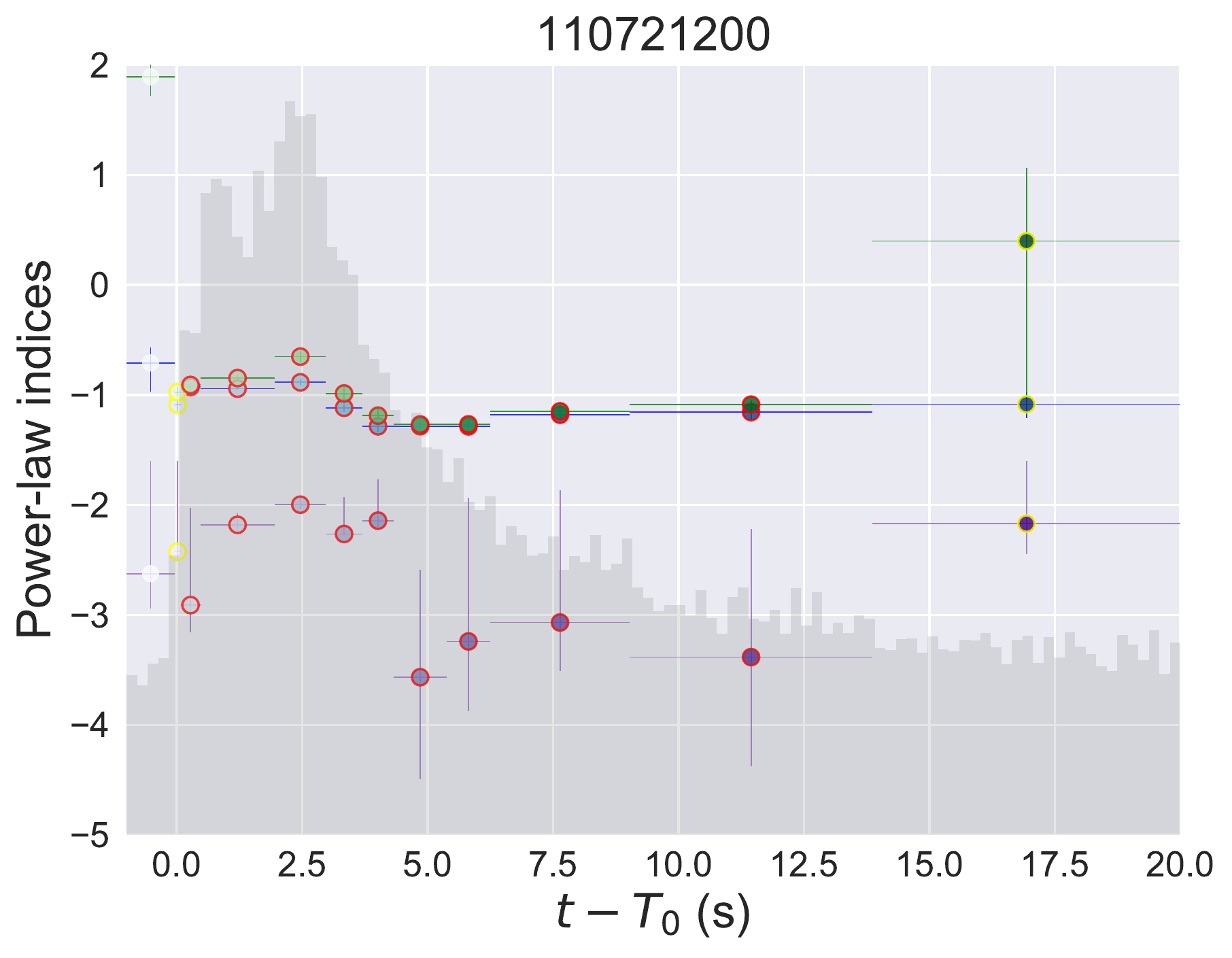}}
\subfigure{\includegraphics[width=0.3\linewidth]{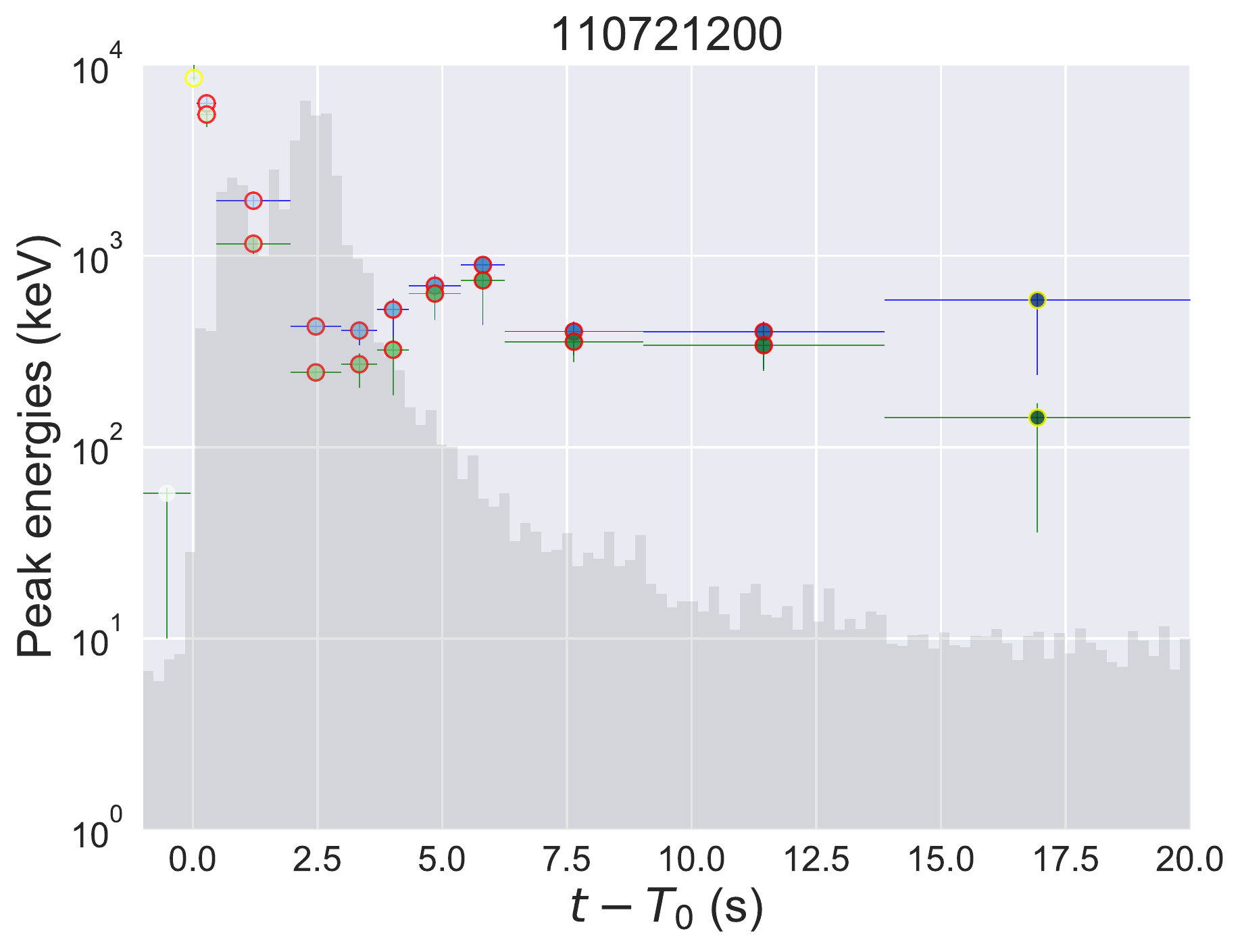}}
\subfigure{\includegraphics[width=0.3\linewidth]{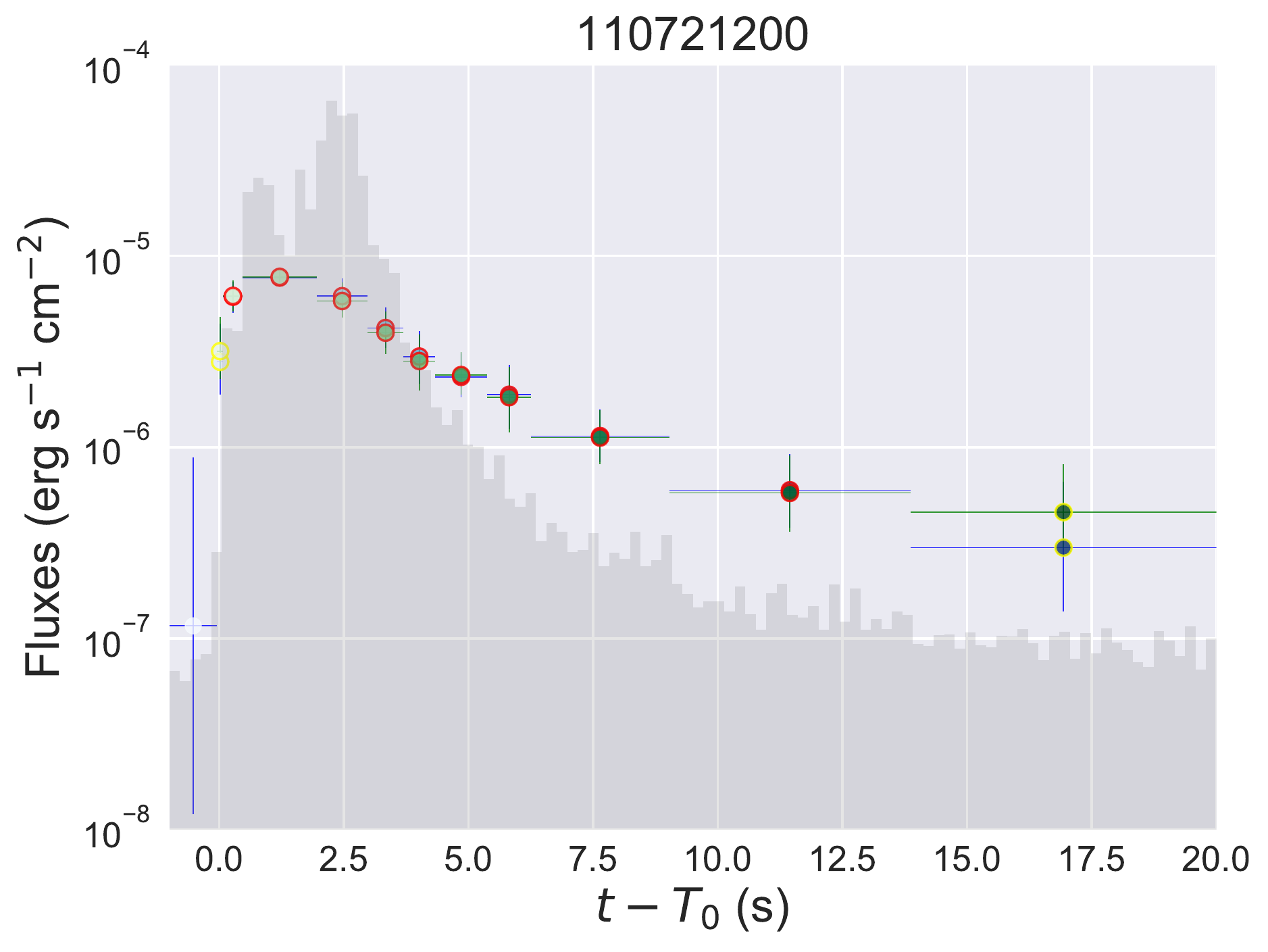}}

\caption{Same as Fig.~\ref{fig:evolution_group1}.
\label{fig:evolution_group4}}
\end{figure*}

\begin{figure*}
\centering

\subfigure{\includegraphics[width=0.3\linewidth]{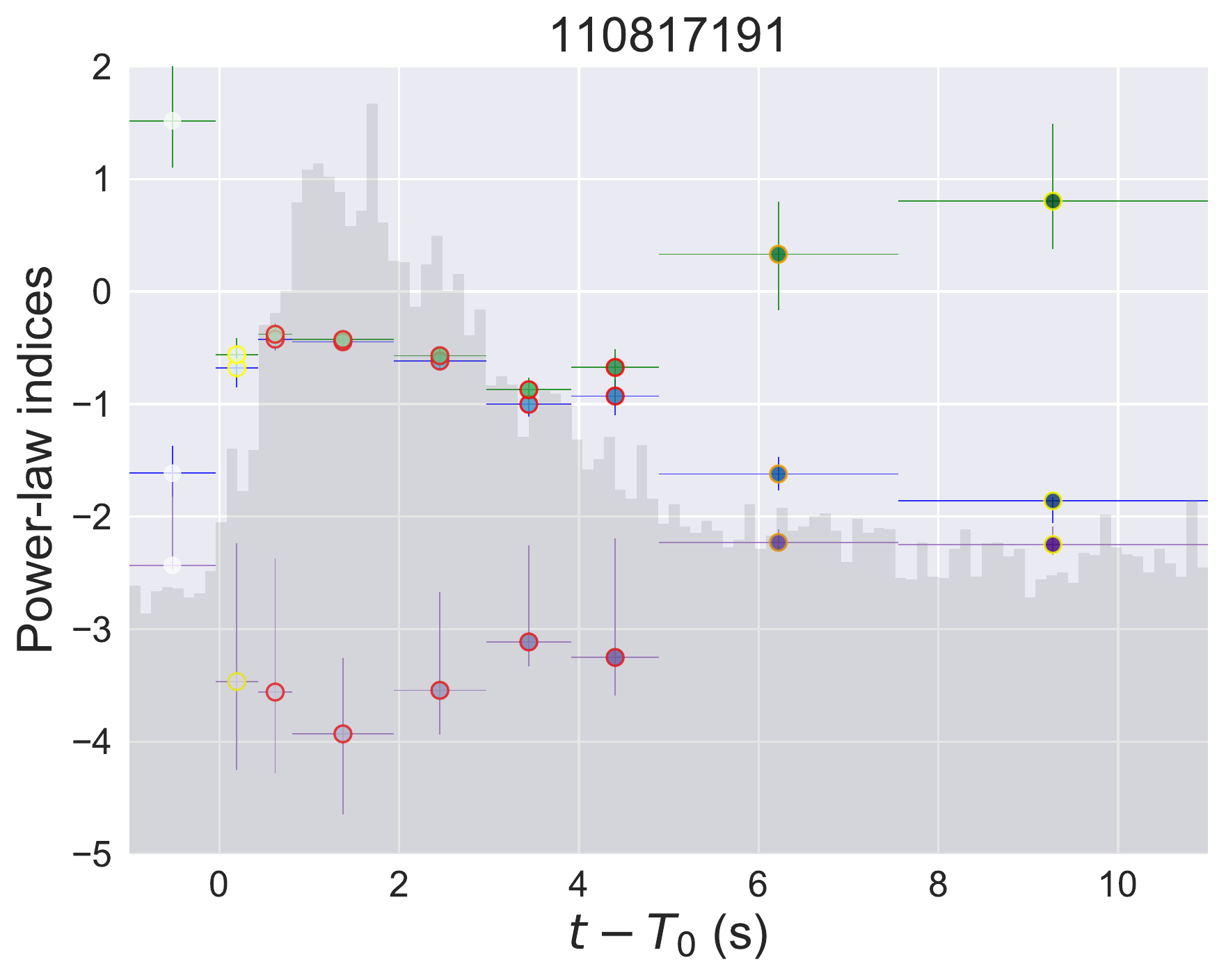}}
\subfigure{\includegraphics[width=0.3\linewidth]{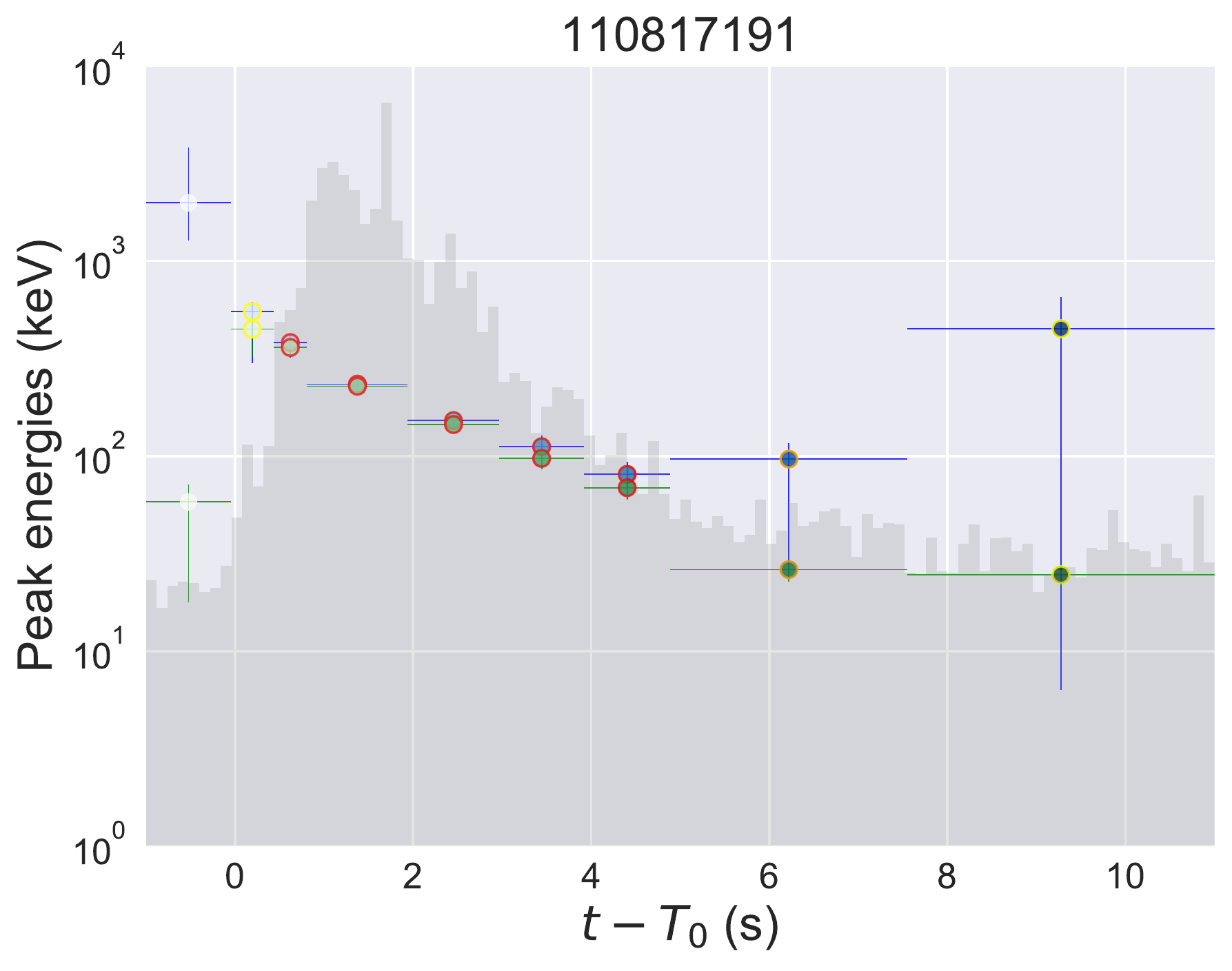}}
\subfigure{\includegraphics[width=0.3\linewidth]{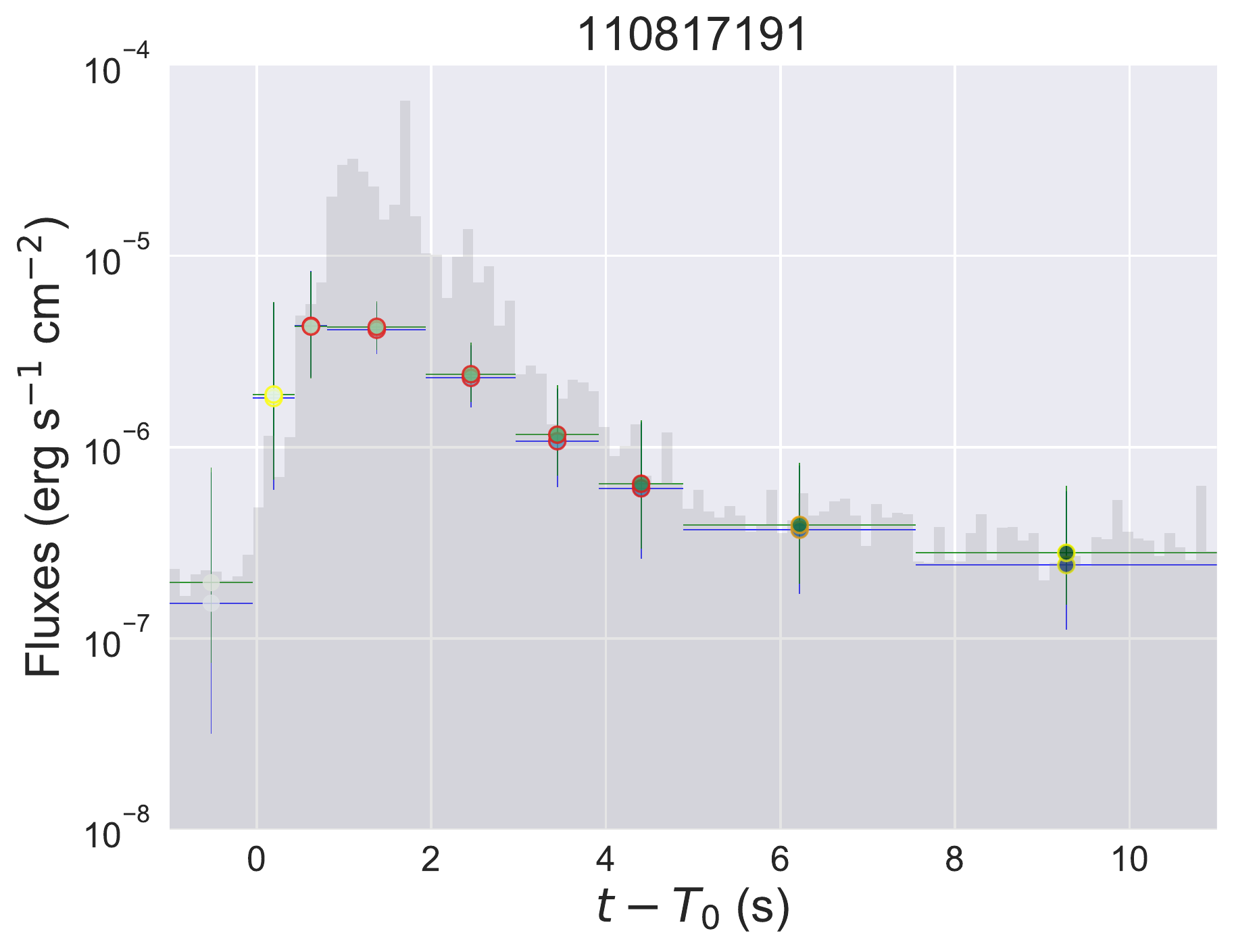}}

\subfigure{\includegraphics[width=0.3\linewidth]{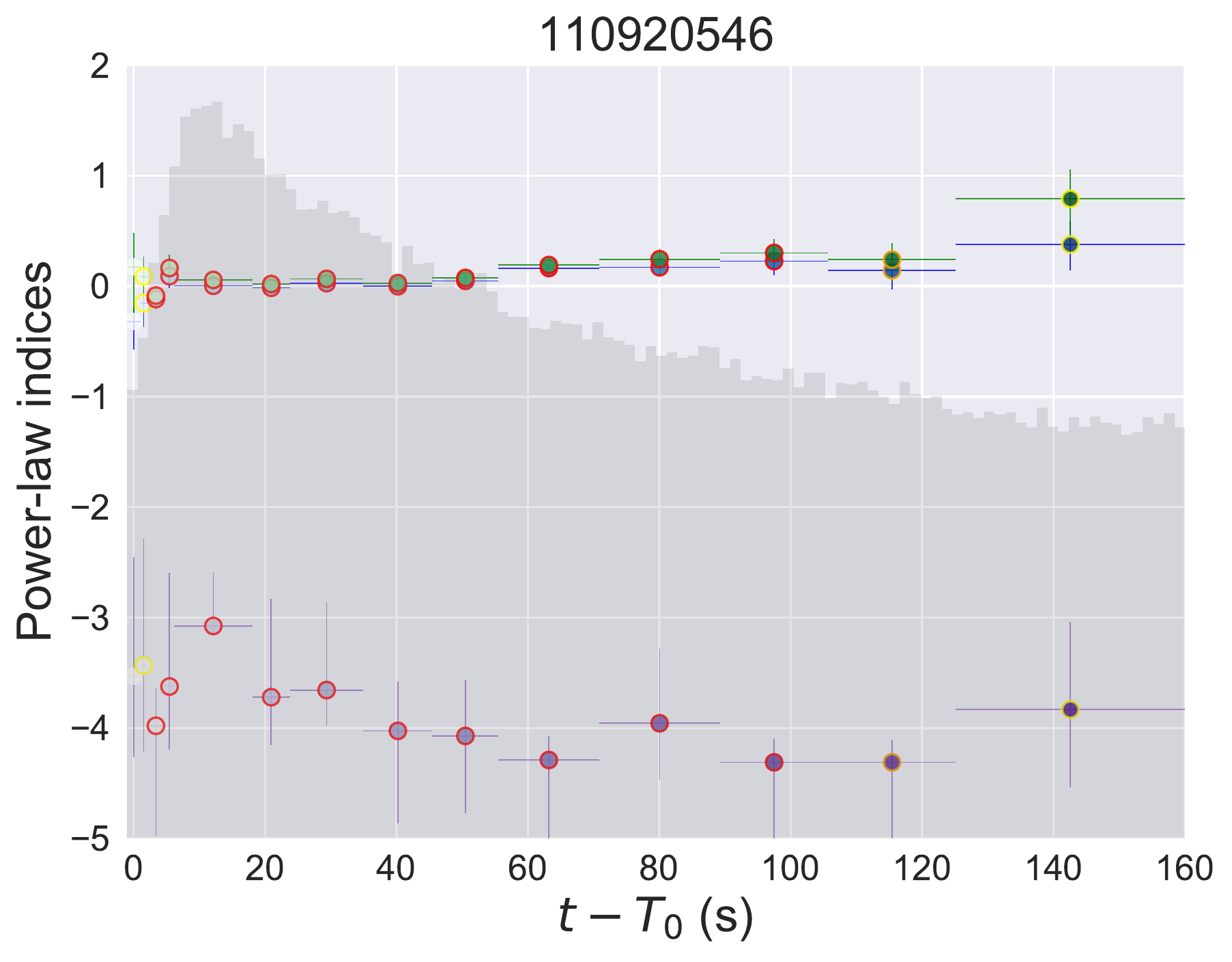}}
\subfigure{\includegraphics[width=0.3\linewidth]{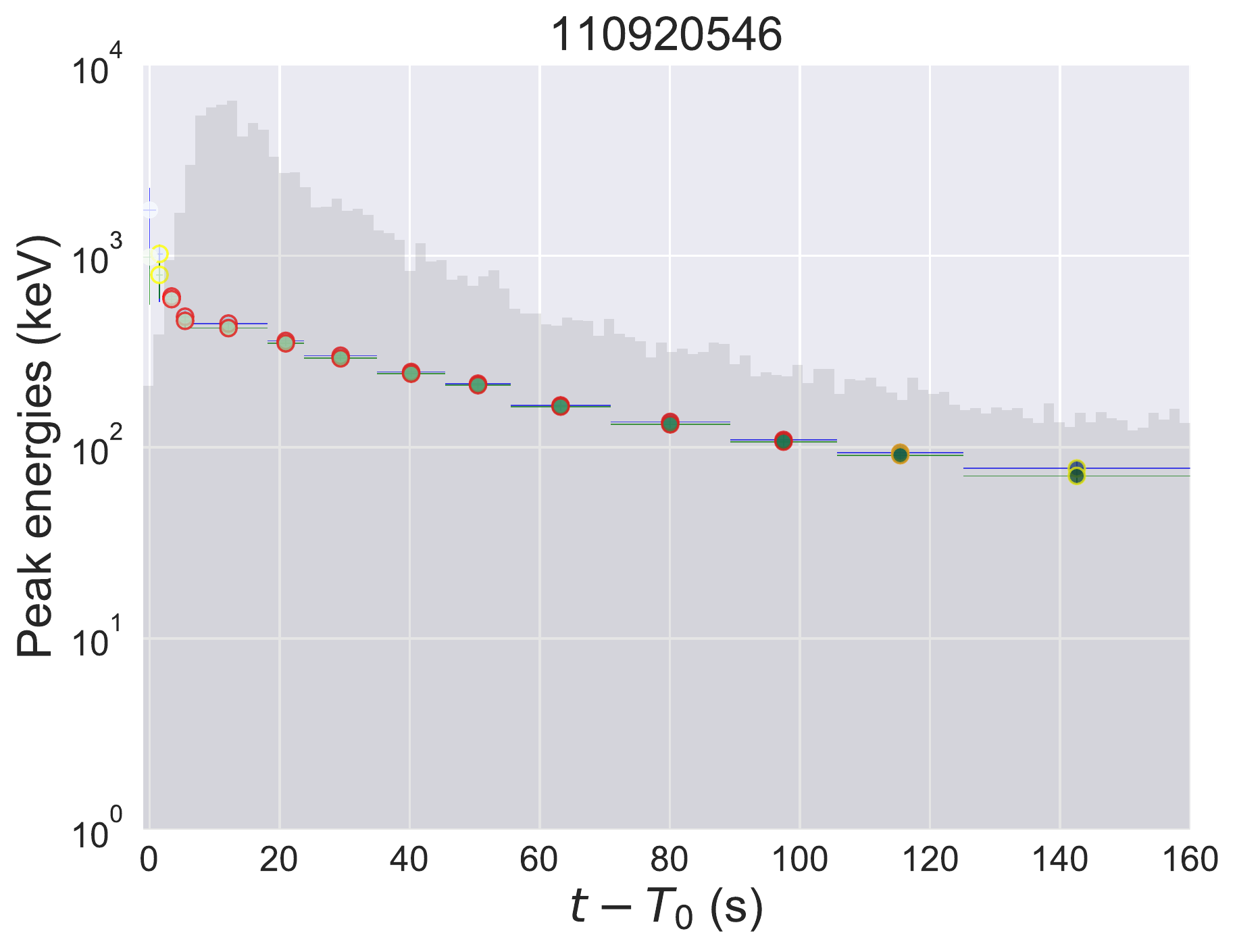}}
\subfigure{\includegraphics[width=0.3\linewidth]{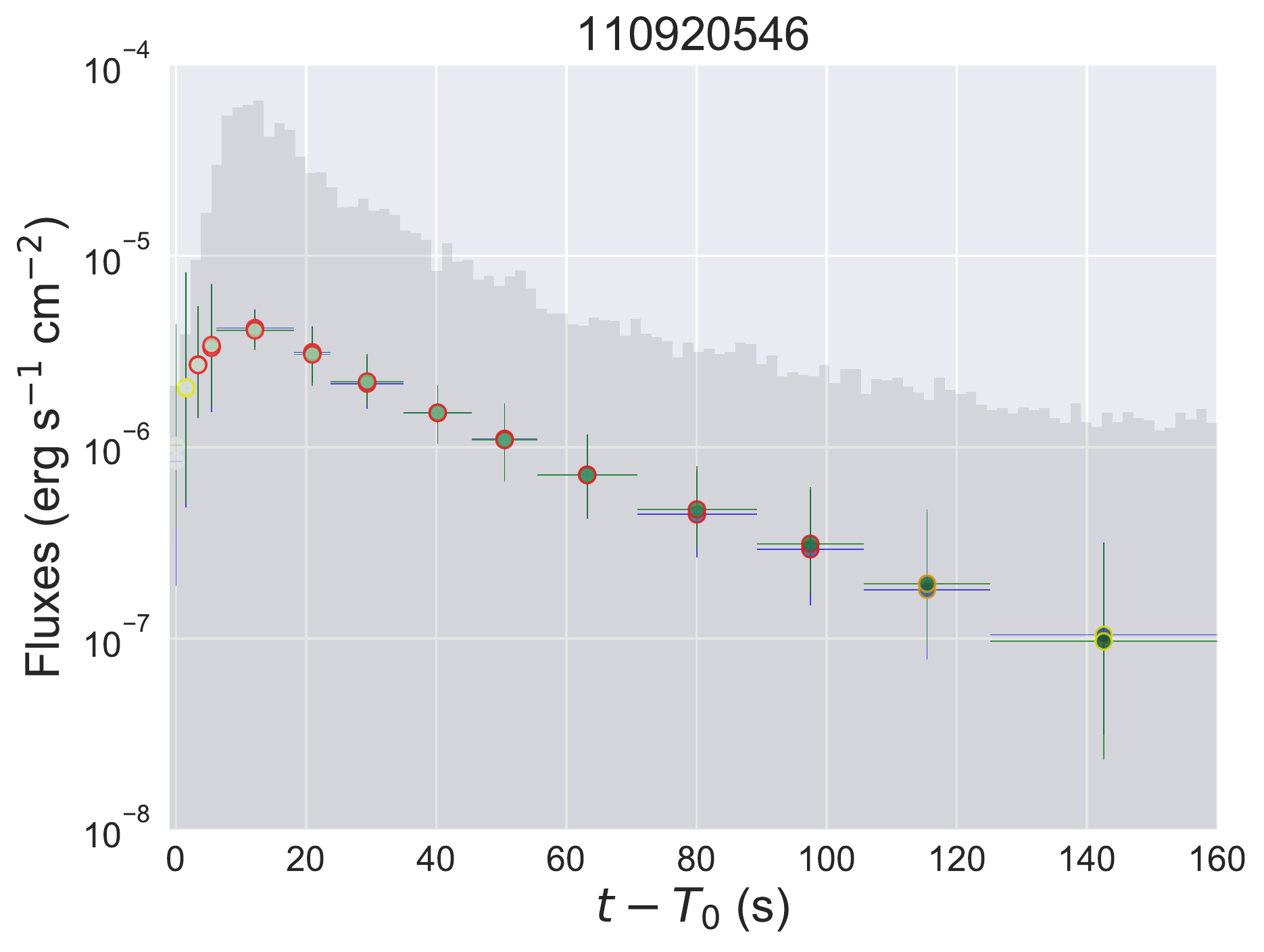}}

\subfigure{\includegraphics[width=0.3\linewidth]{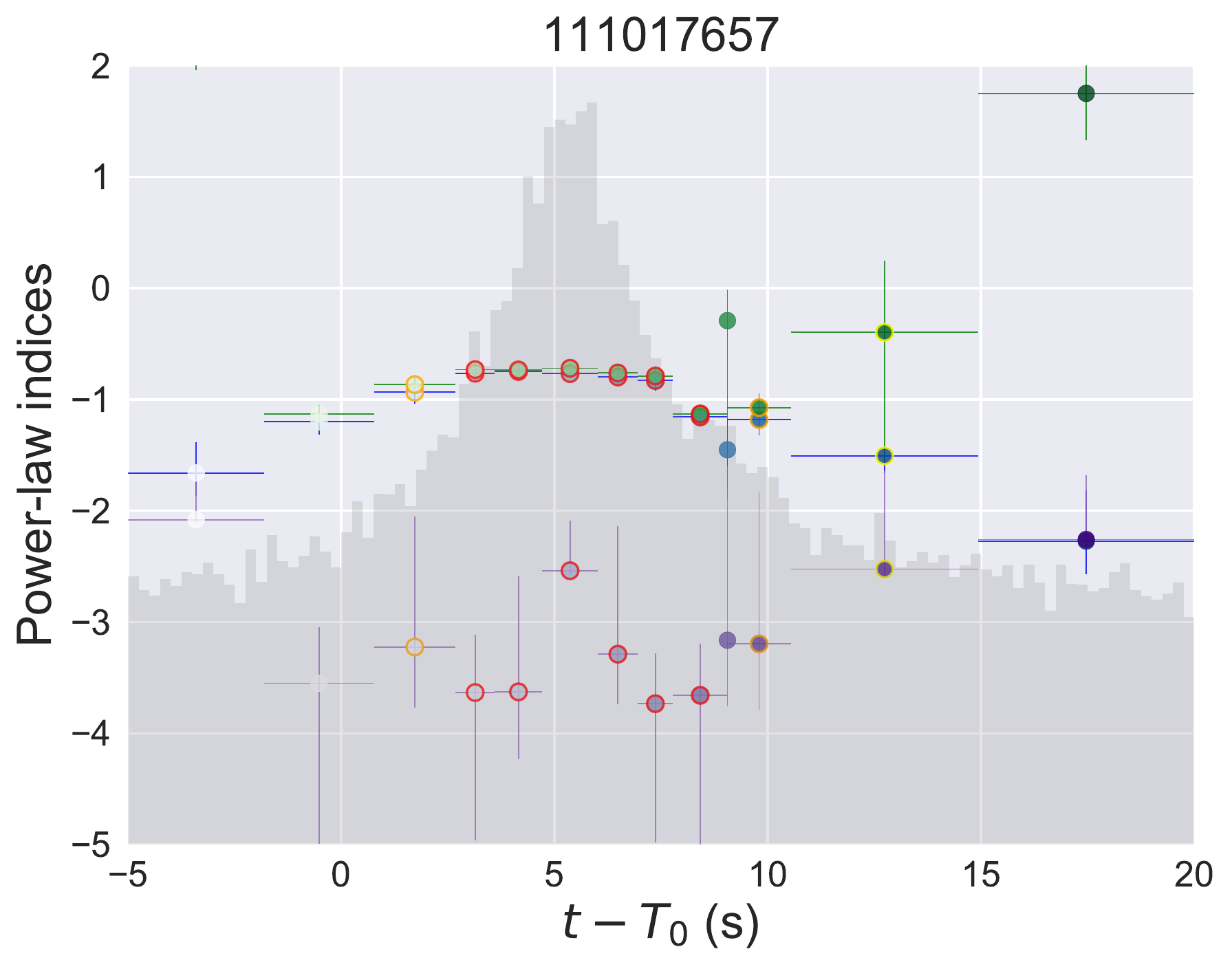}}
\subfigure{\includegraphics[width=0.3\linewidth]{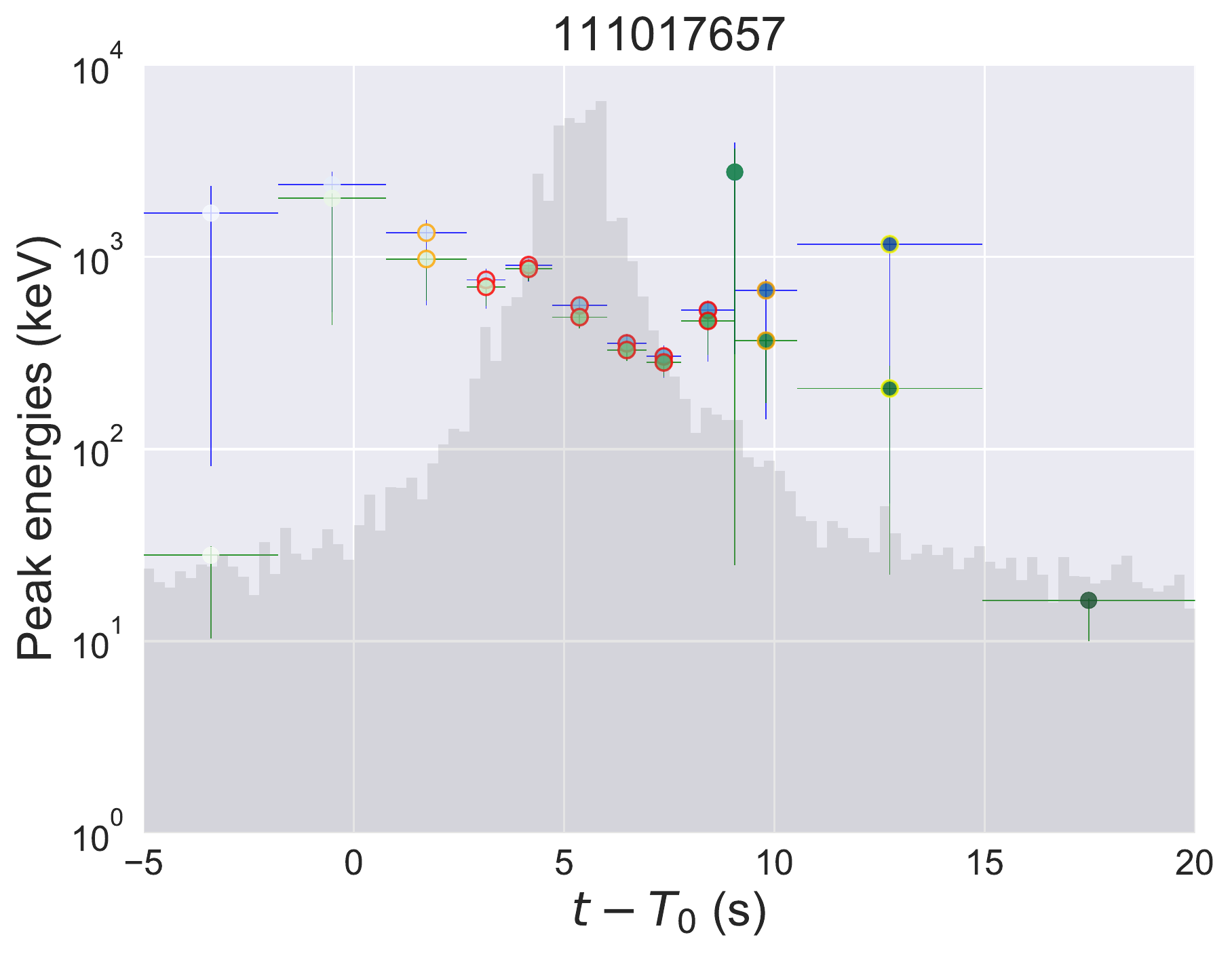}}
\subfigure{\includegraphics[width=0.3\linewidth]{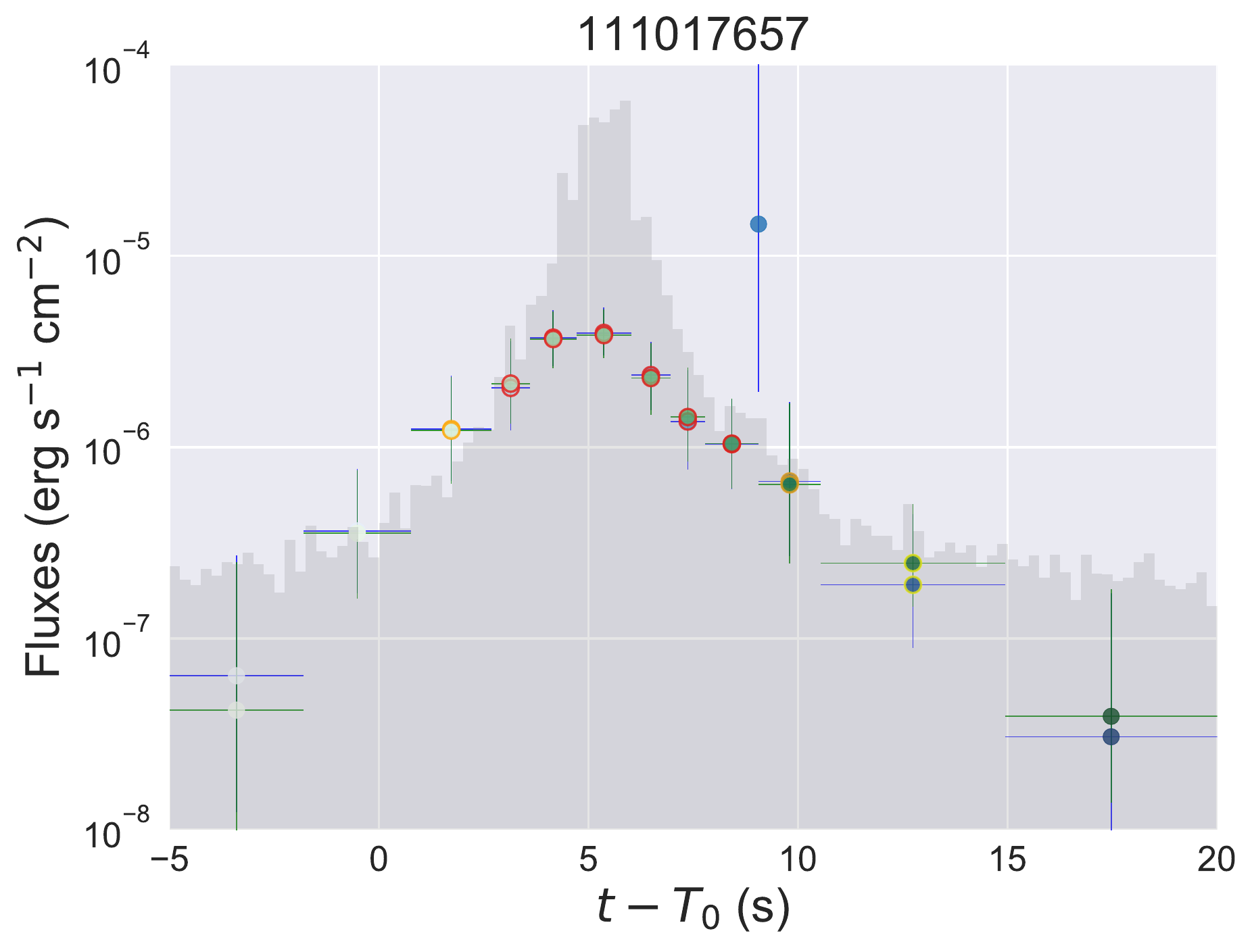}}

\subfigure{\includegraphics[width=0.3\linewidth]{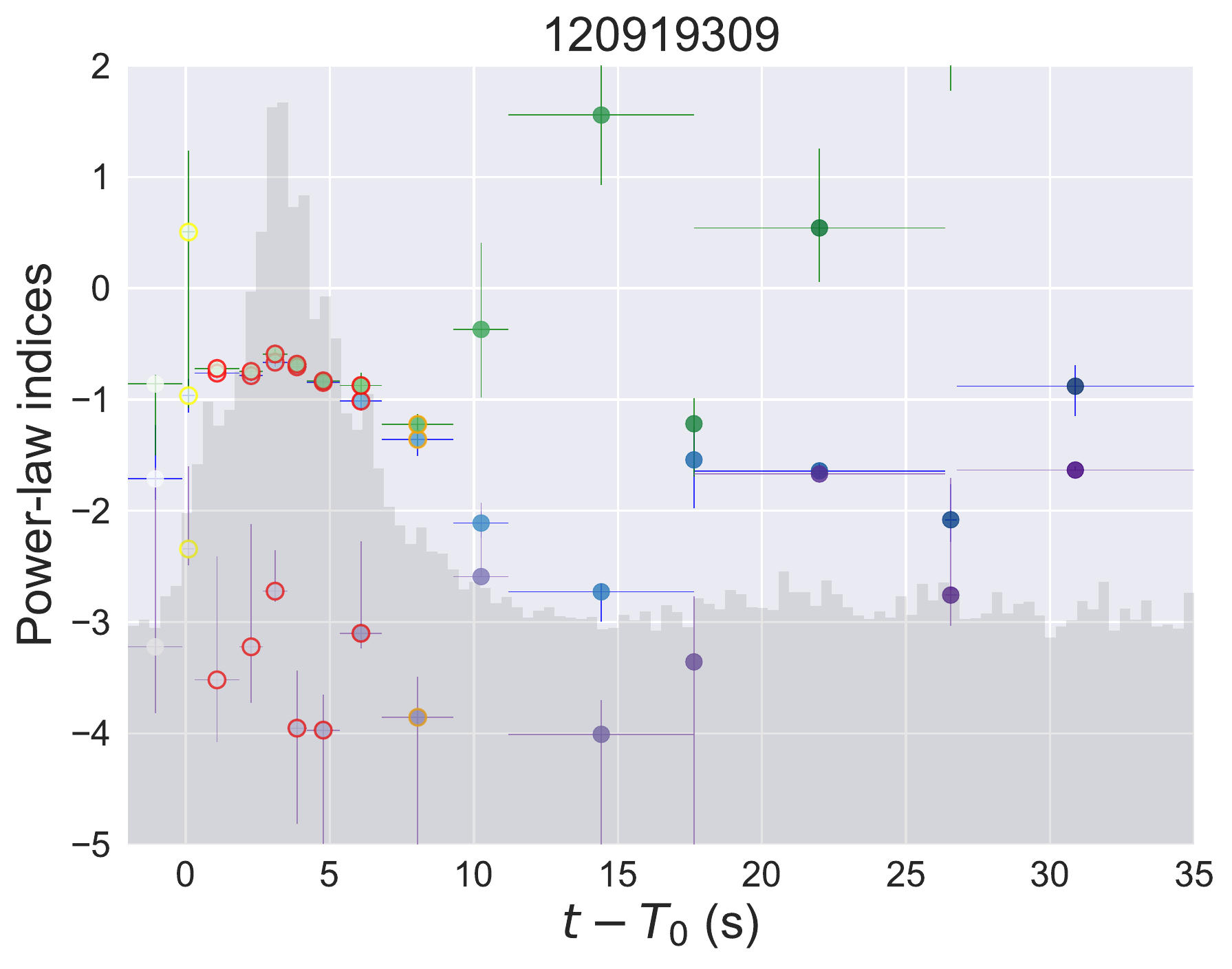}}
\subfigure{\includegraphics[width=0.3\linewidth]{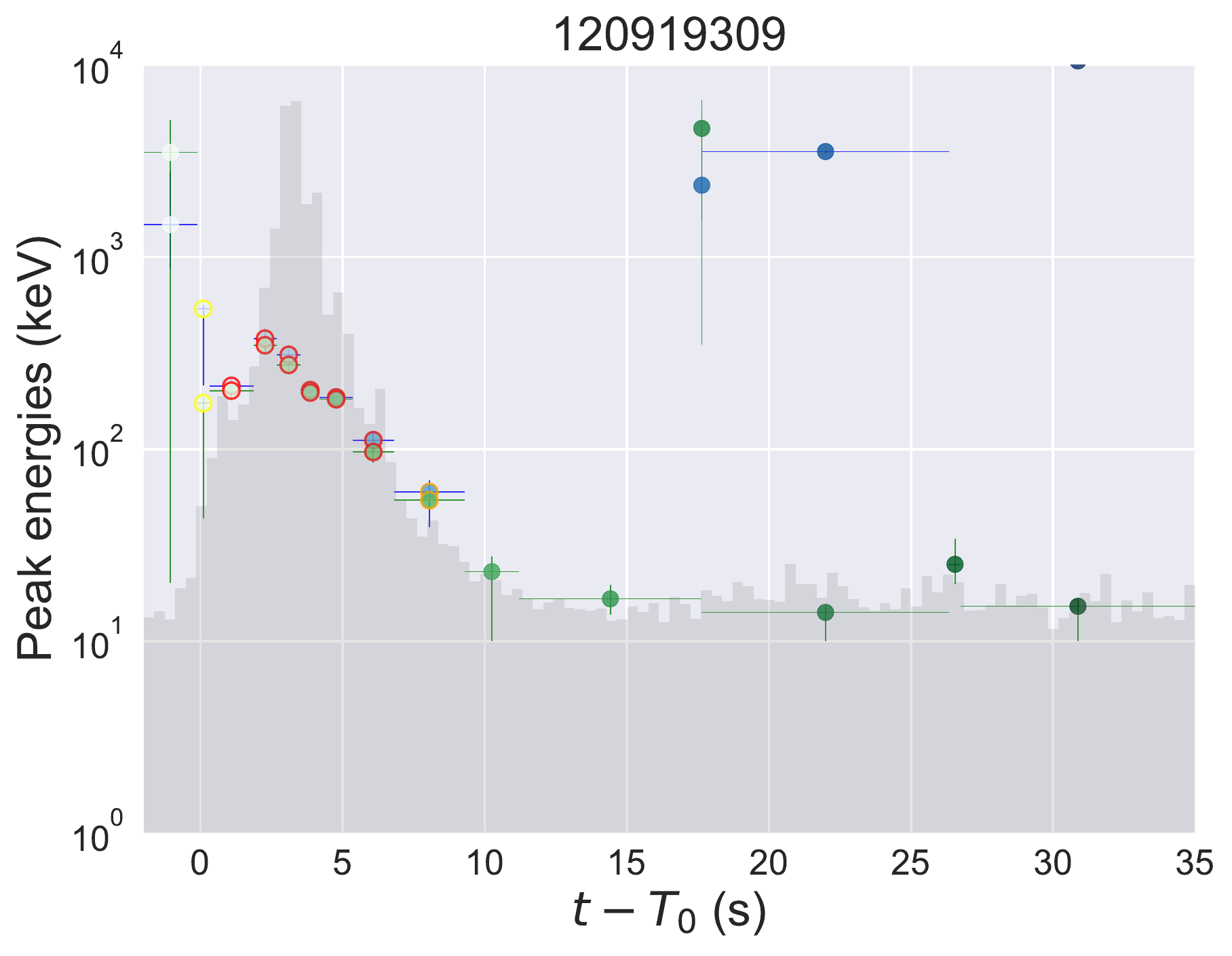}}
\subfigure{\includegraphics[width=0.3\linewidth]{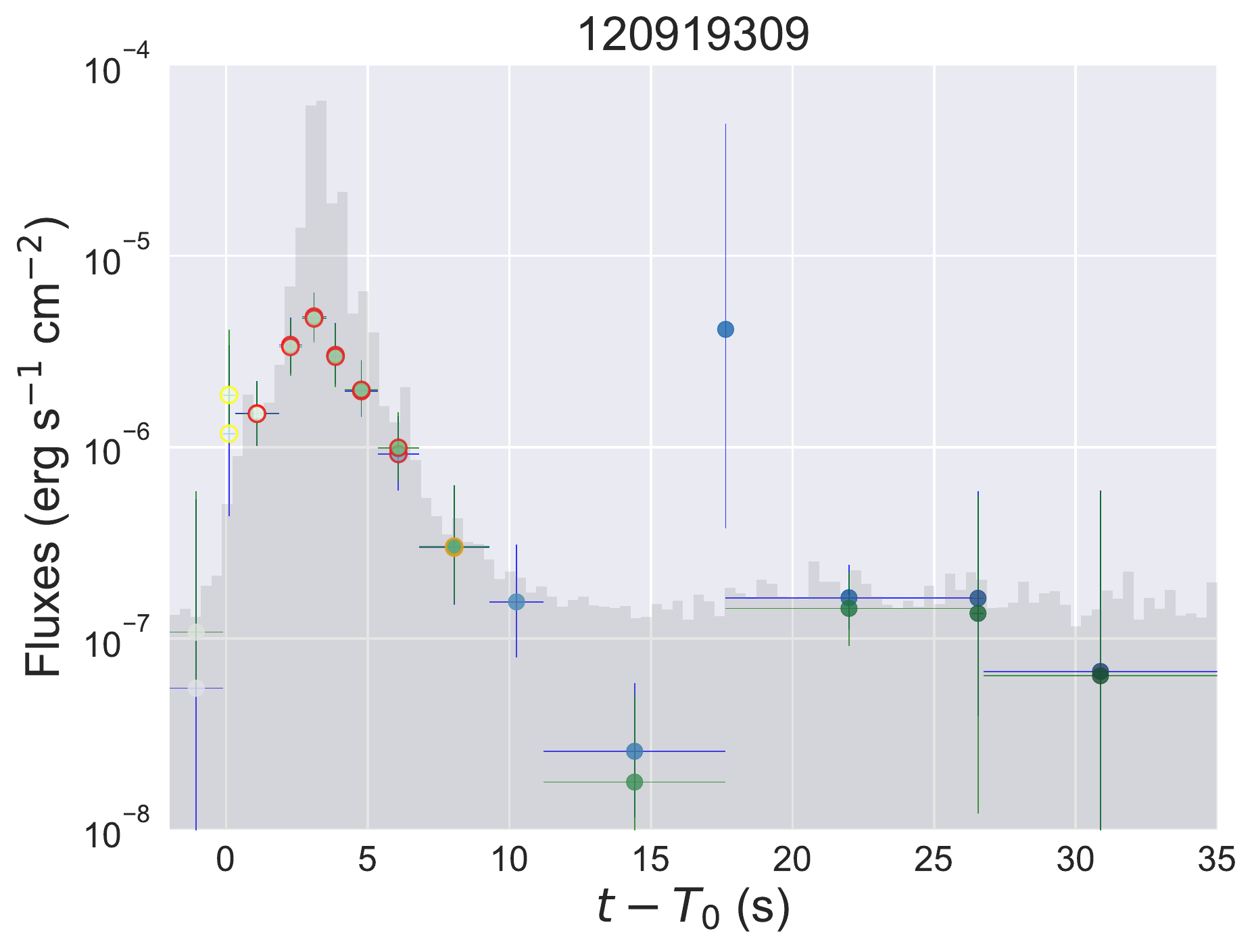}}

\caption{Same as Fig.~\ref{fig:evolution_group1}.
\label{fig:evolution_group5}}
\end{figure*}

\begin{figure*}
\centering

\subfigure{\includegraphics[width=0.3\linewidth]{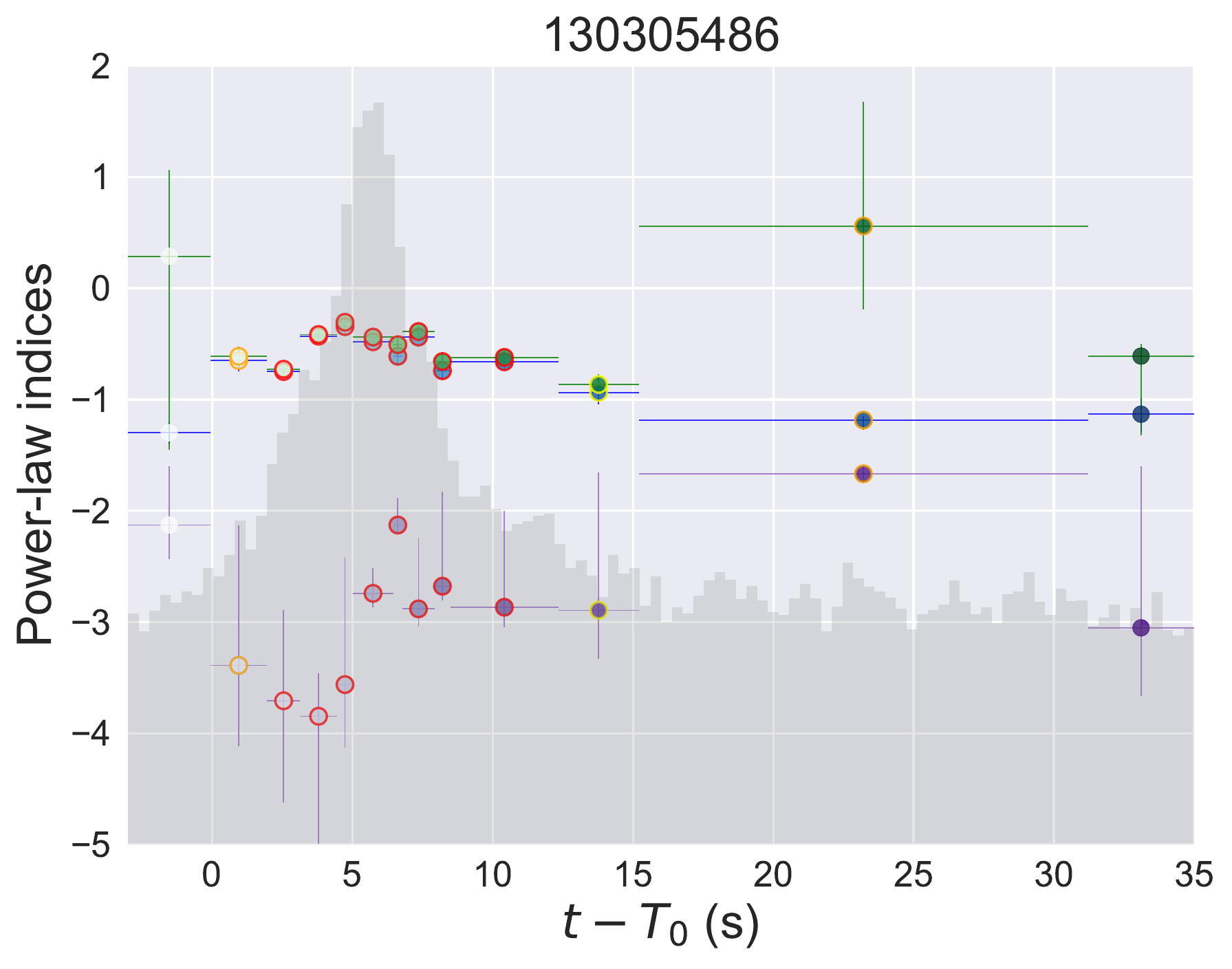}}
\subfigure{\includegraphics[width=0.3\linewidth]{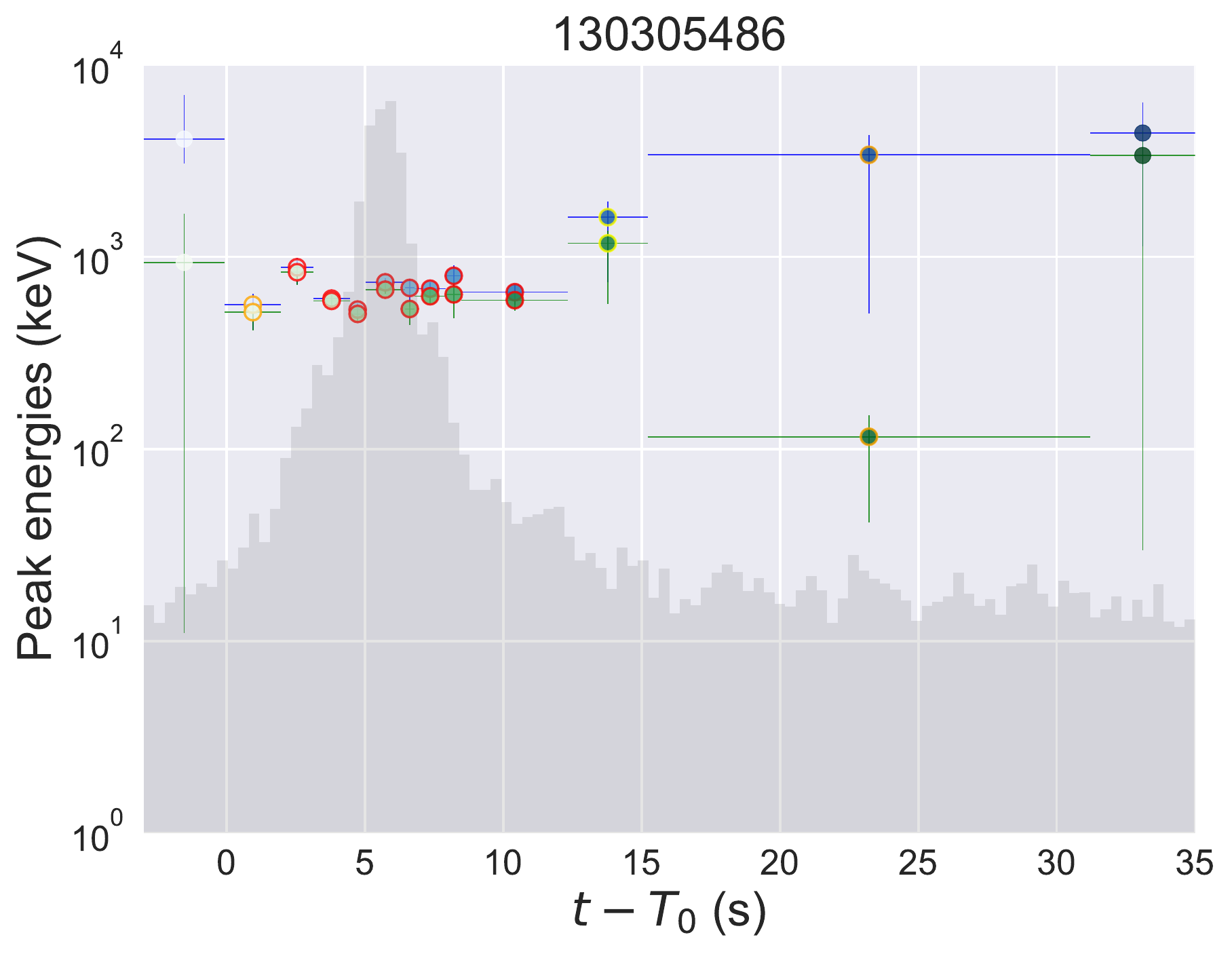}}
\subfigure{\includegraphics[width=0.3\linewidth]{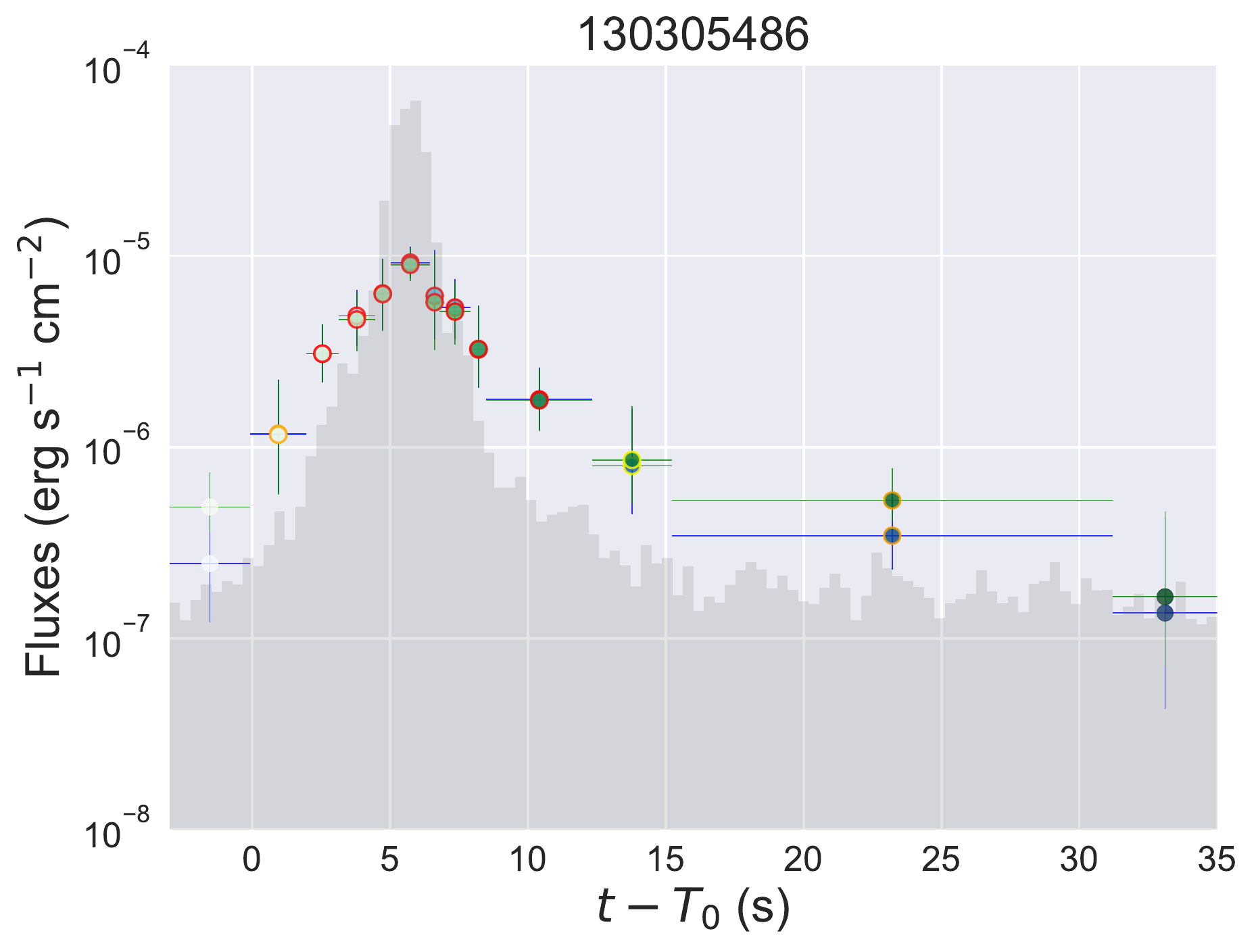}}

\subfigure{\includegraphics[width=0.3\linewidth]{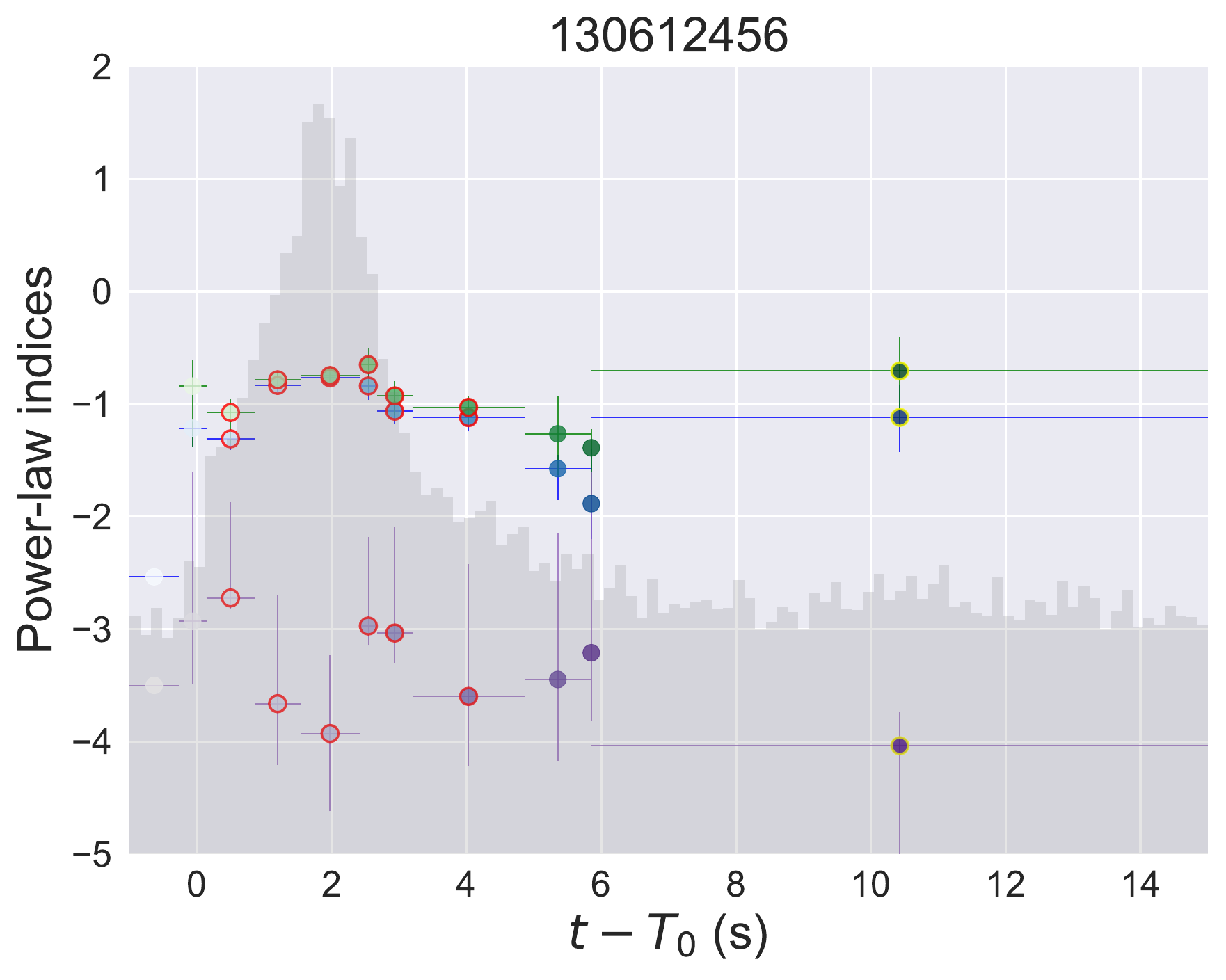}}
\subfigure{\includegraphics[width=0.3\linewidth]{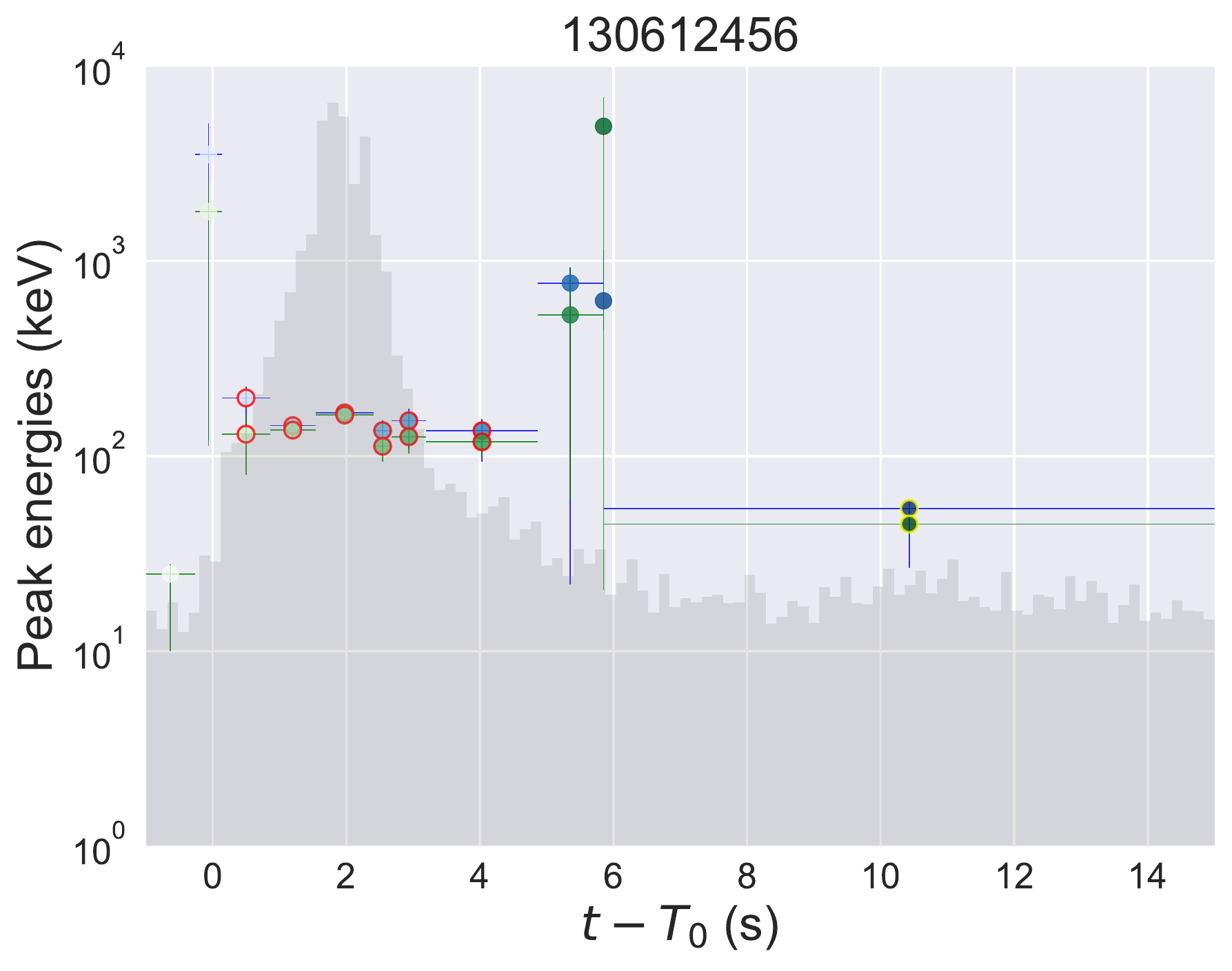}}
\subfigure{\includegraphics[width=0.3\linewidth]{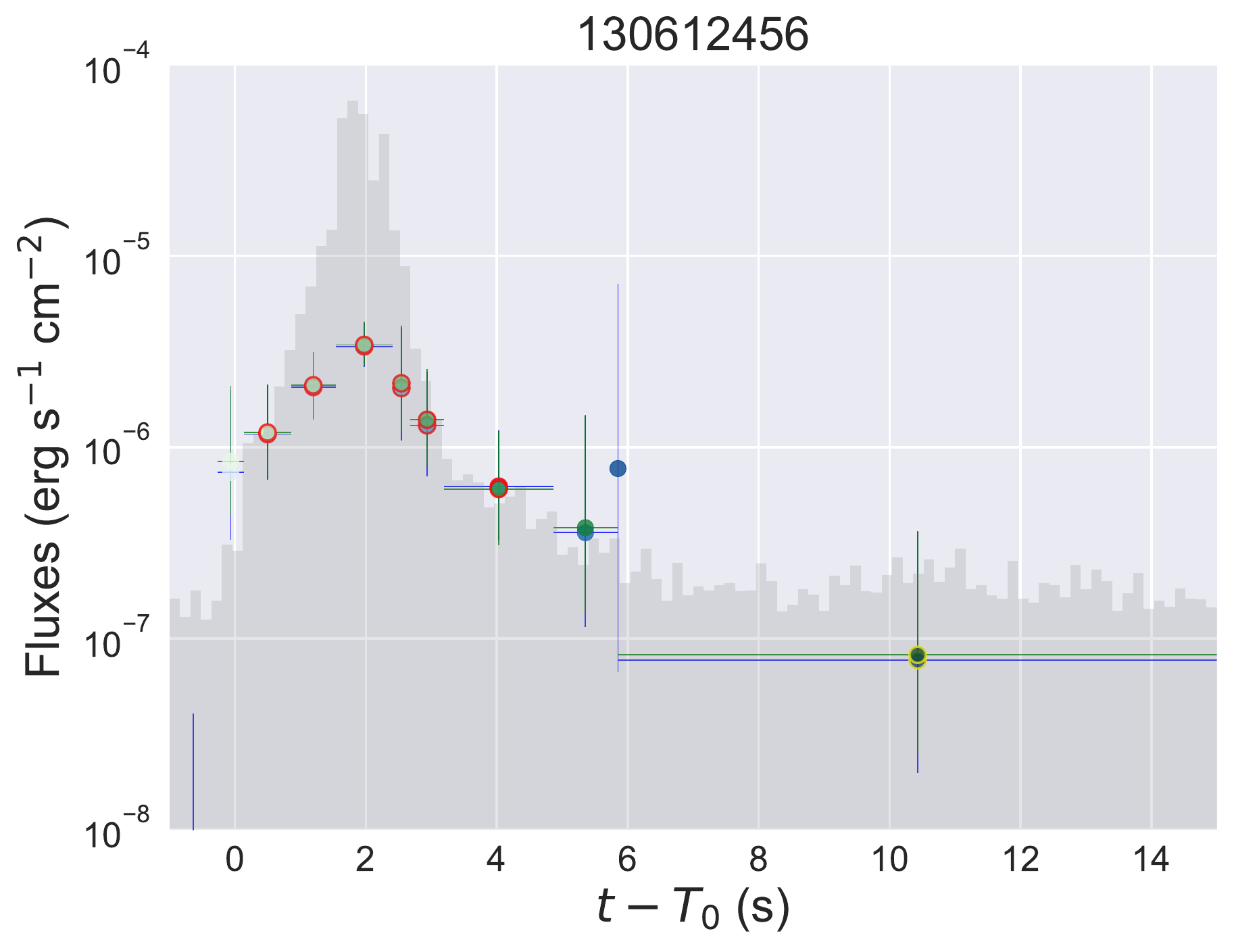}}

\subfigure{\includegraphics[width=0.3\linewidth]{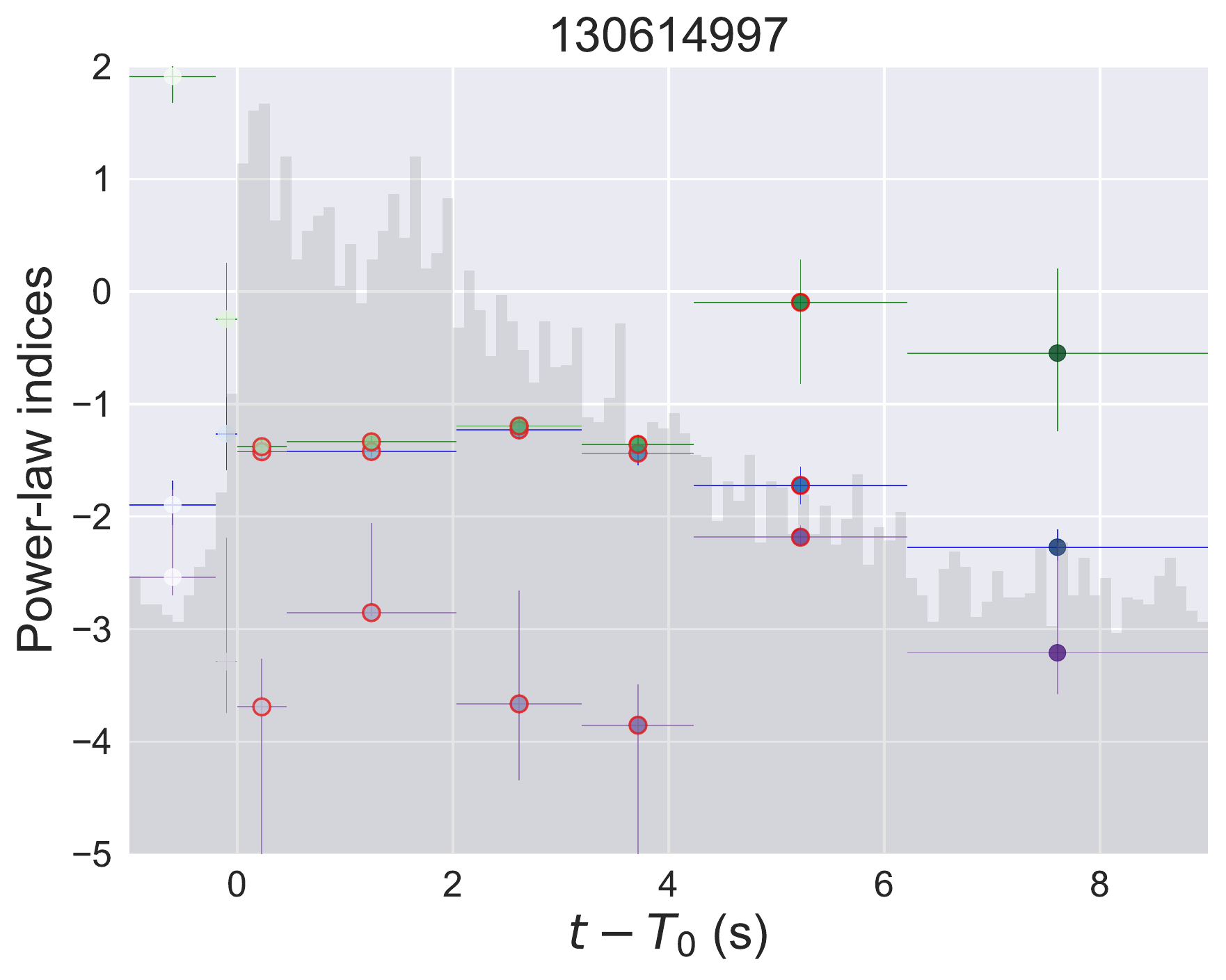}}
\subfigure{\includegraphics[width=0.3\linewidth]{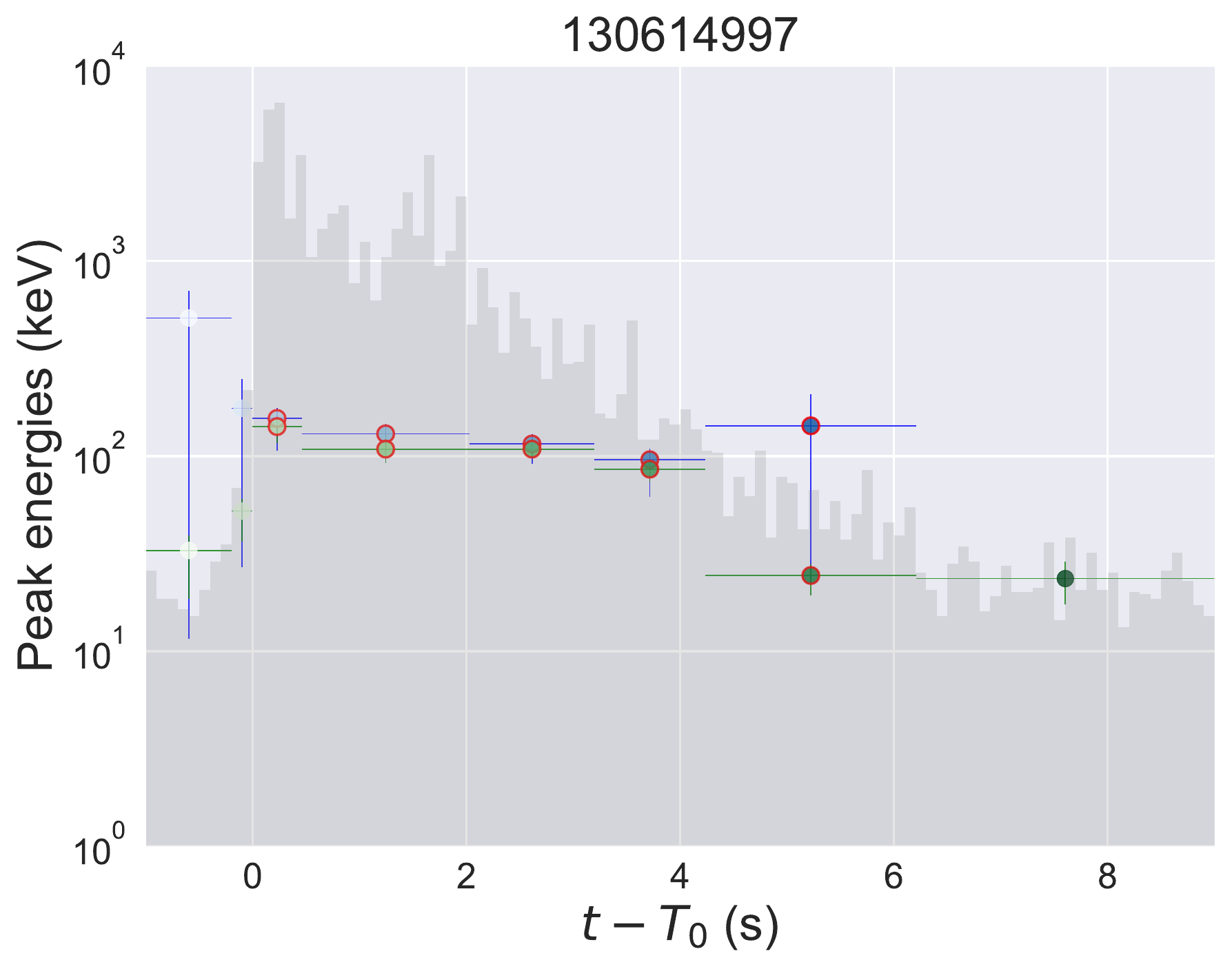}}
\subfigure{\includegraphics[width=0.3\linewidth]{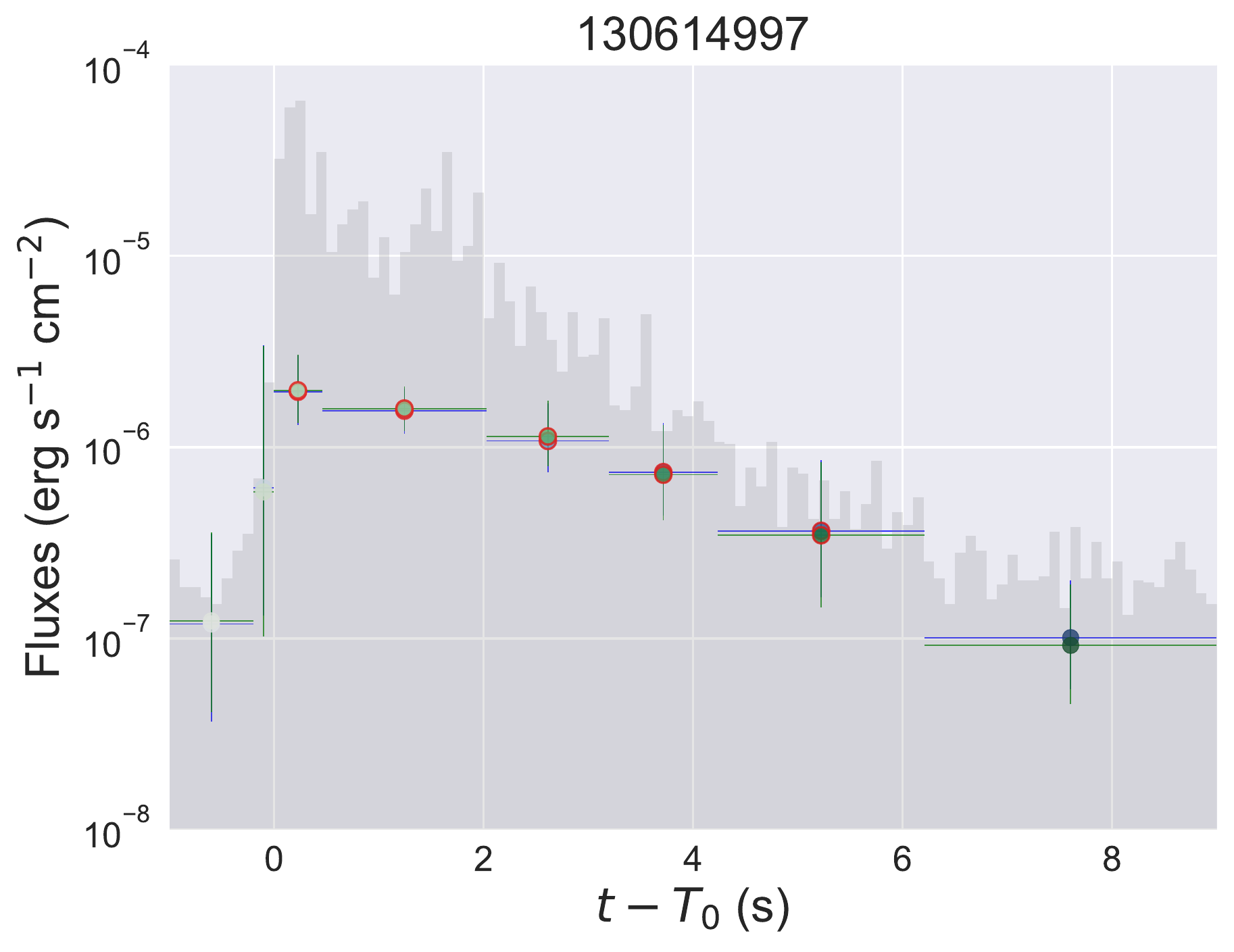}}

\subfigure{\includegraphics[width=0.3\linewidth]{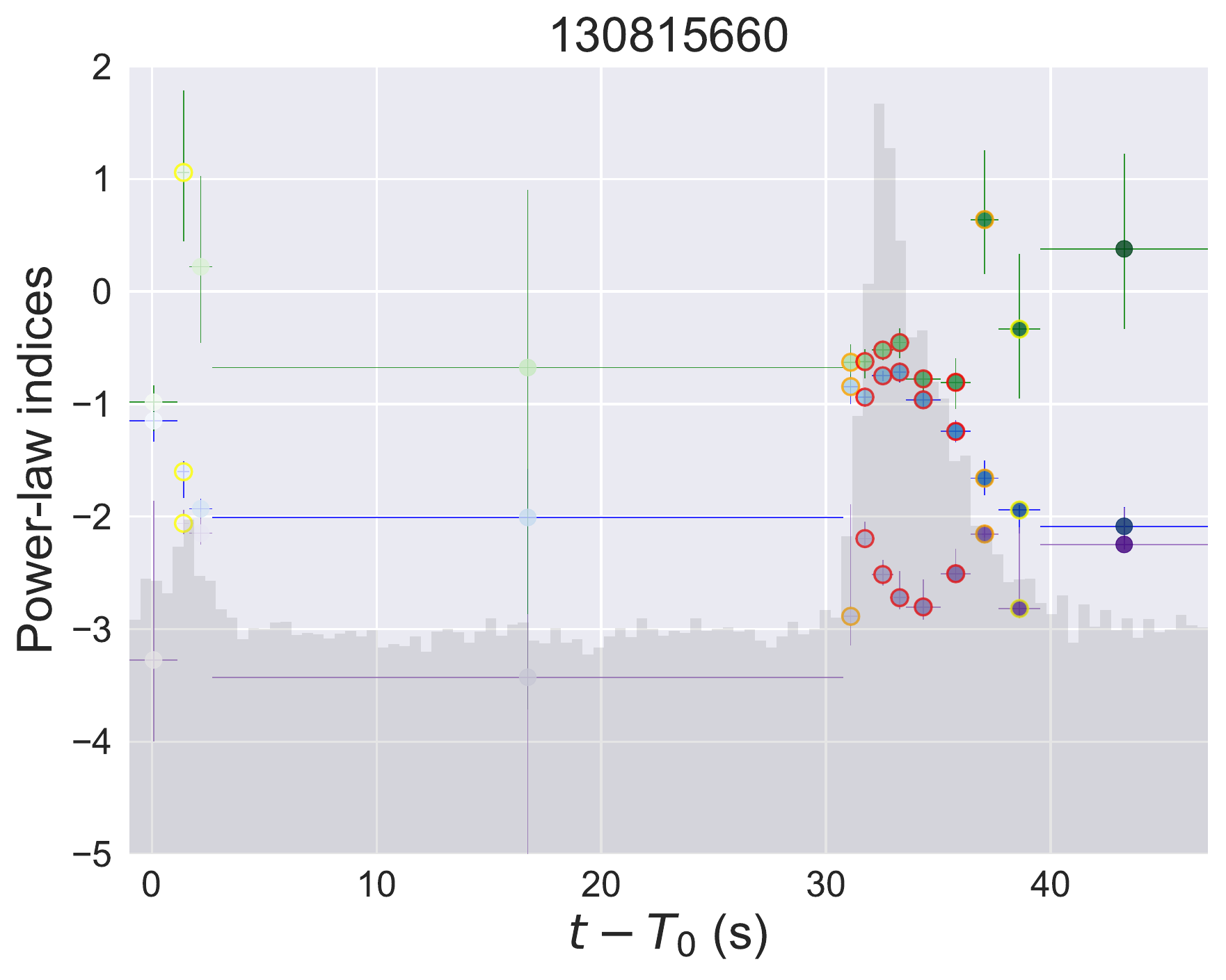}}
\subfigure{\includegraphics[width=0.3\linewidth]{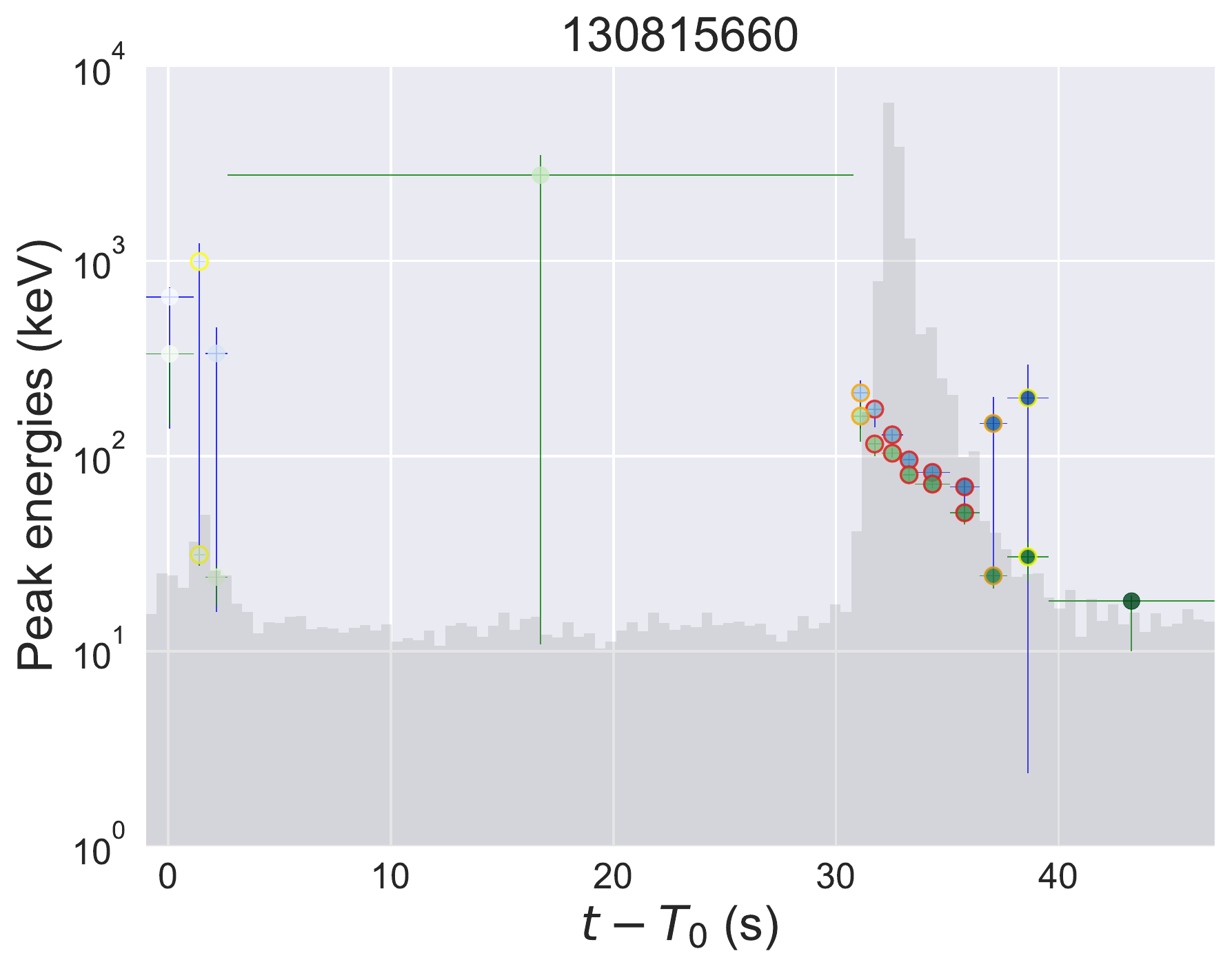}}
\subfigure{\includegraphics[width=0.3\linewidth]{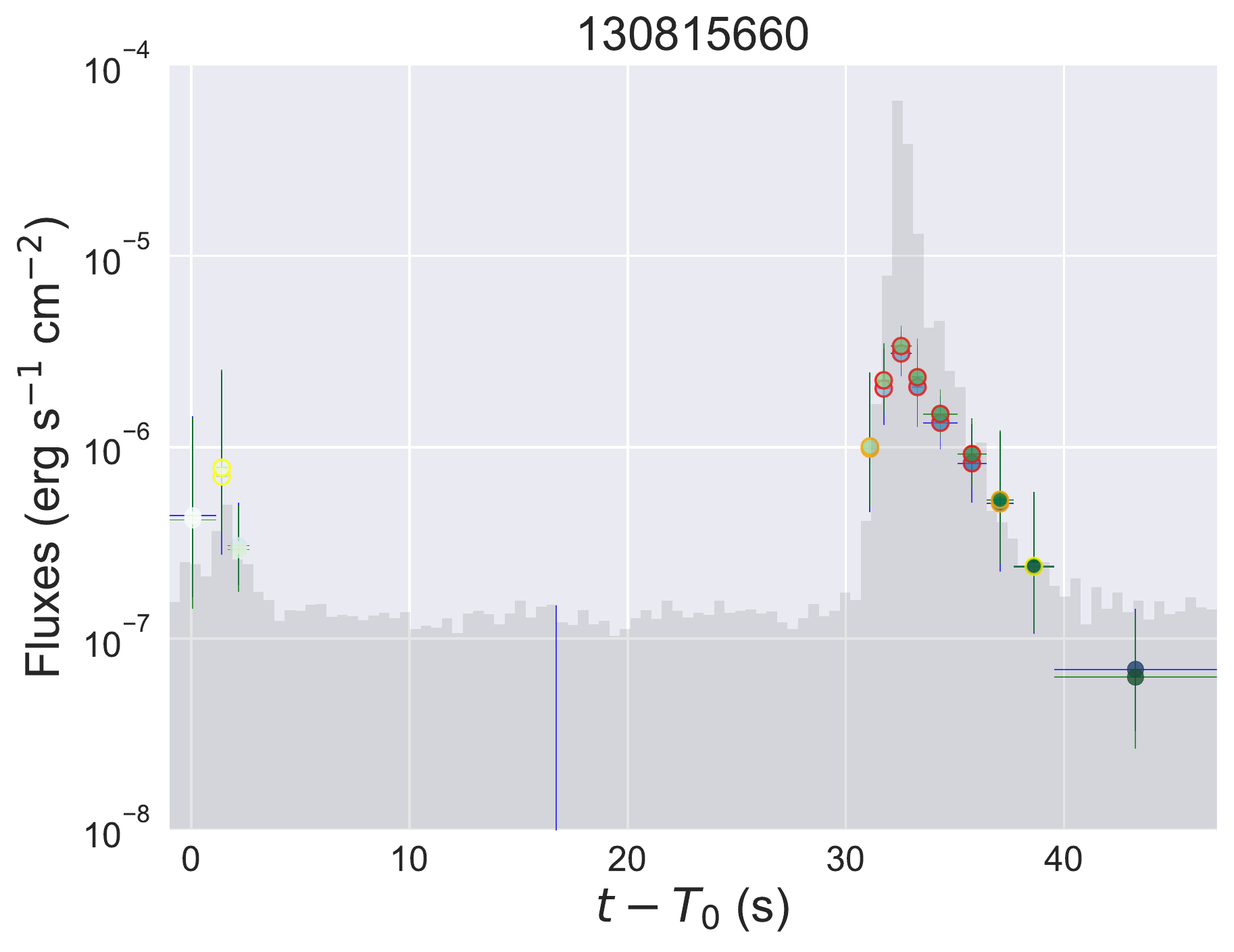}}

\caption{Same as Fig.~\ref{fig:evolution_group1}.
\label{fig:evolution_group6}}
\end{figure*}

\begin{figure*}
\centering

\subfigure{\includegraphics[width=0.3\linewidth]{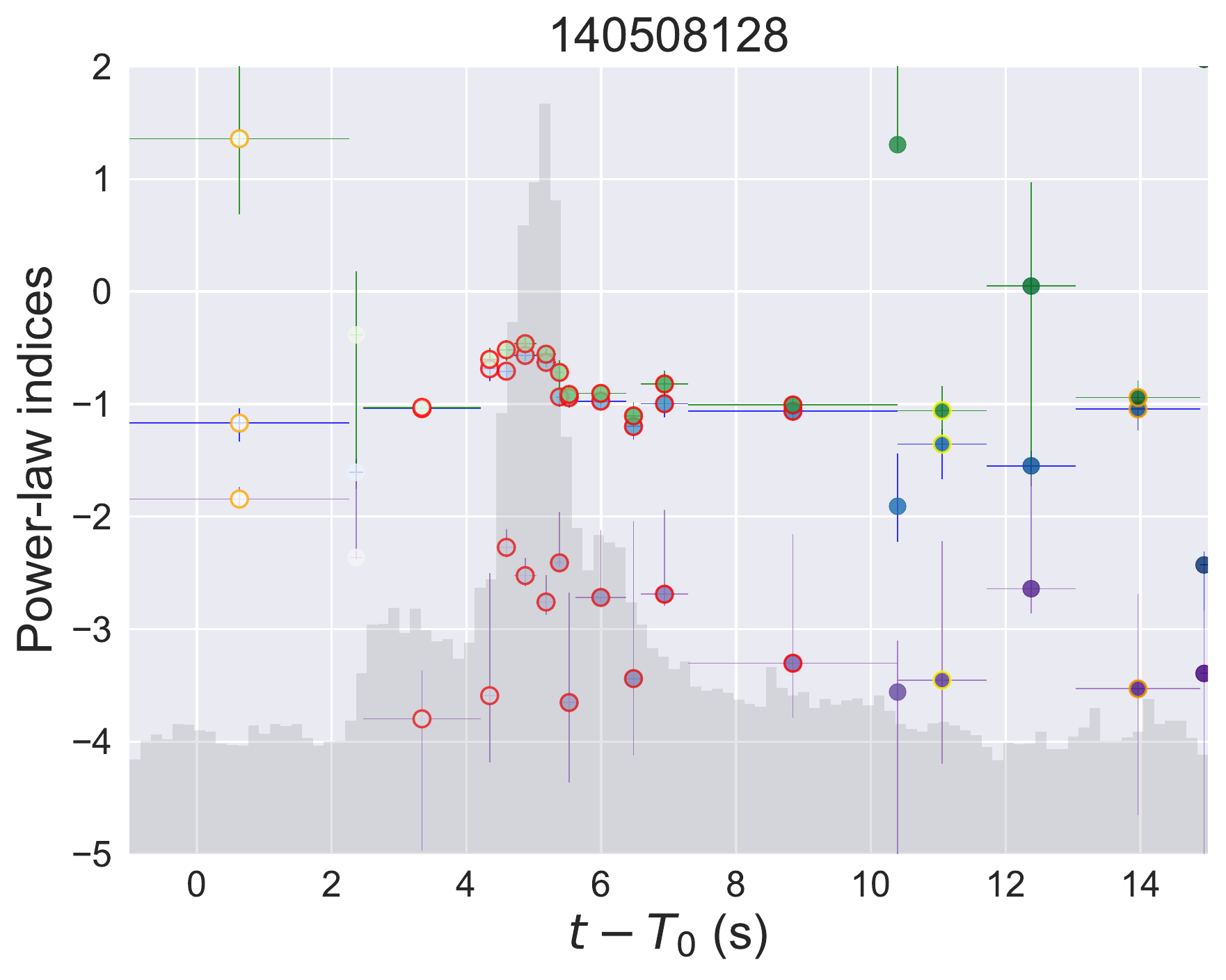}}
\subfigure{\includegraphics[width=0.3\linewidth]{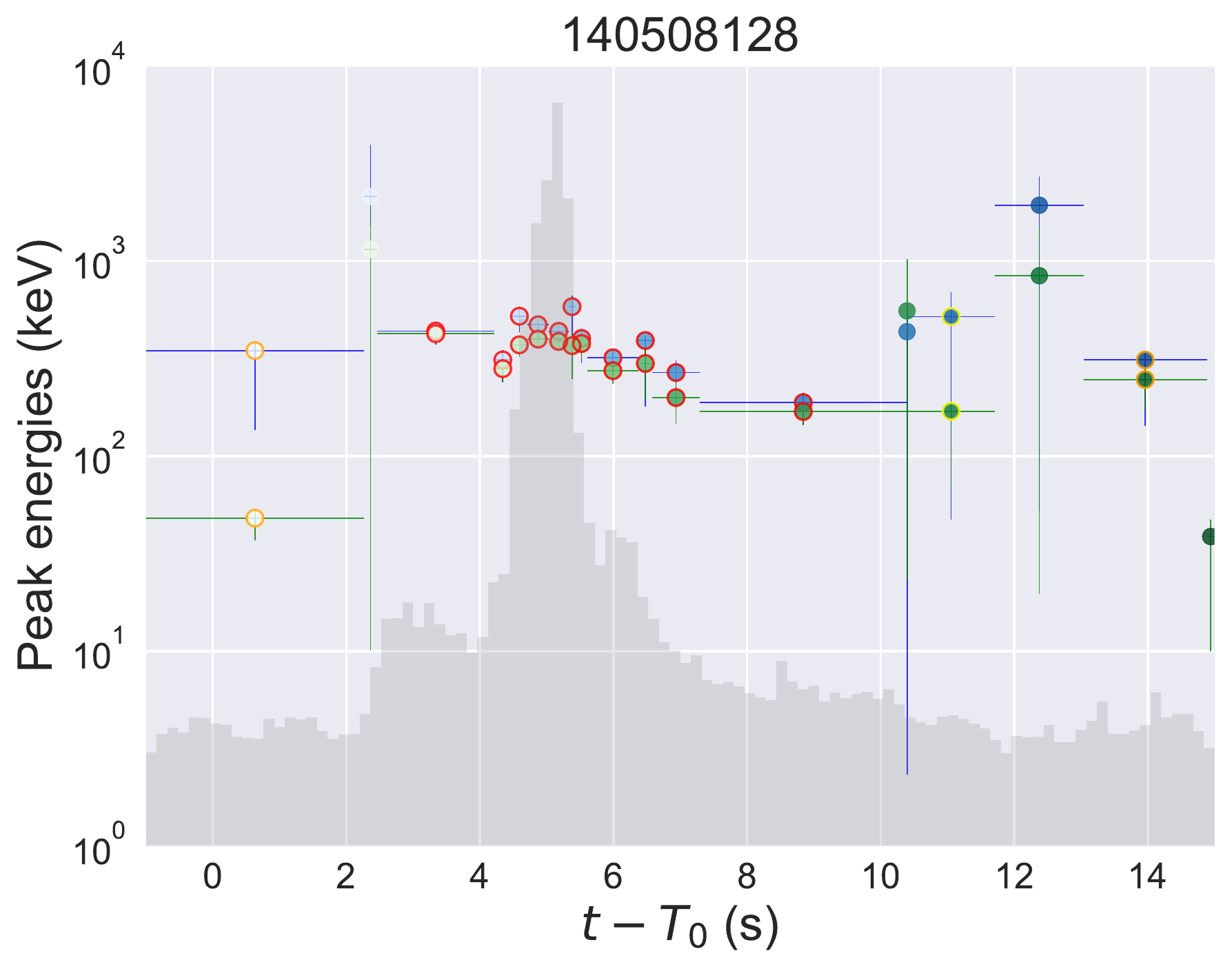}}
\subfigure{\includegraphics[width=0.3\linewidth]{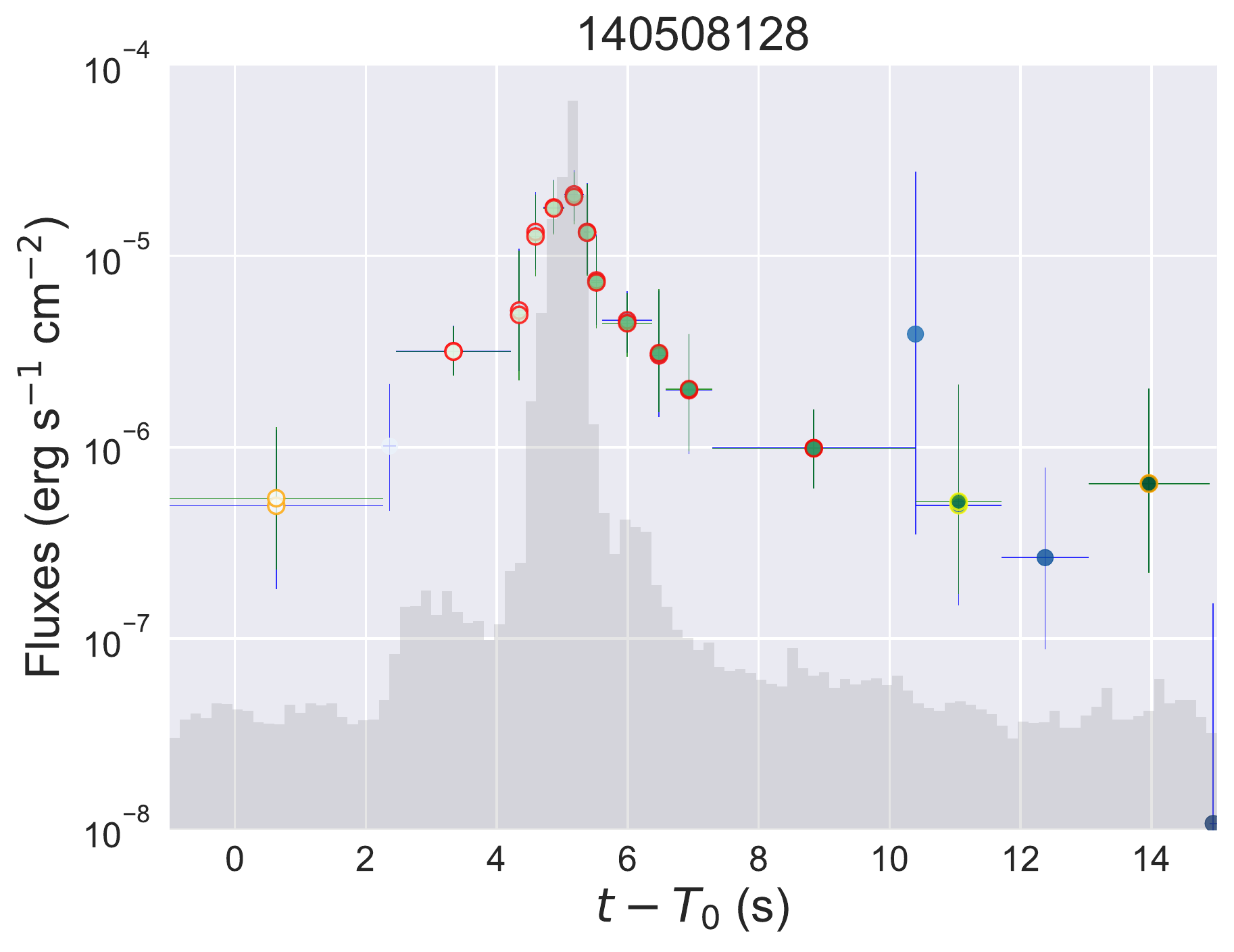}}

\subfigure{\includegraphics[width=0.3\linewidth]{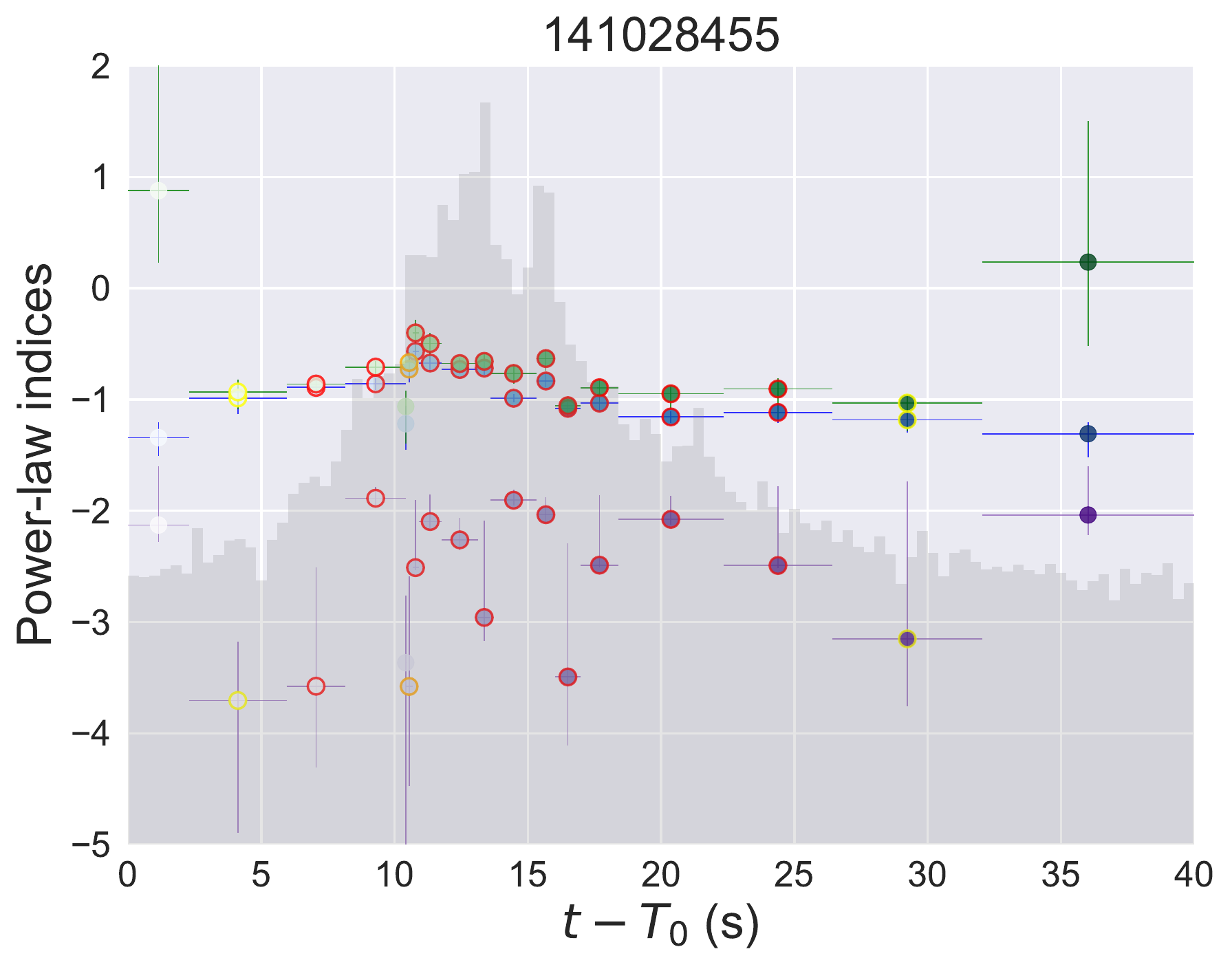}}
\subfigure{\includegraphics[width=0.3\linewidth]{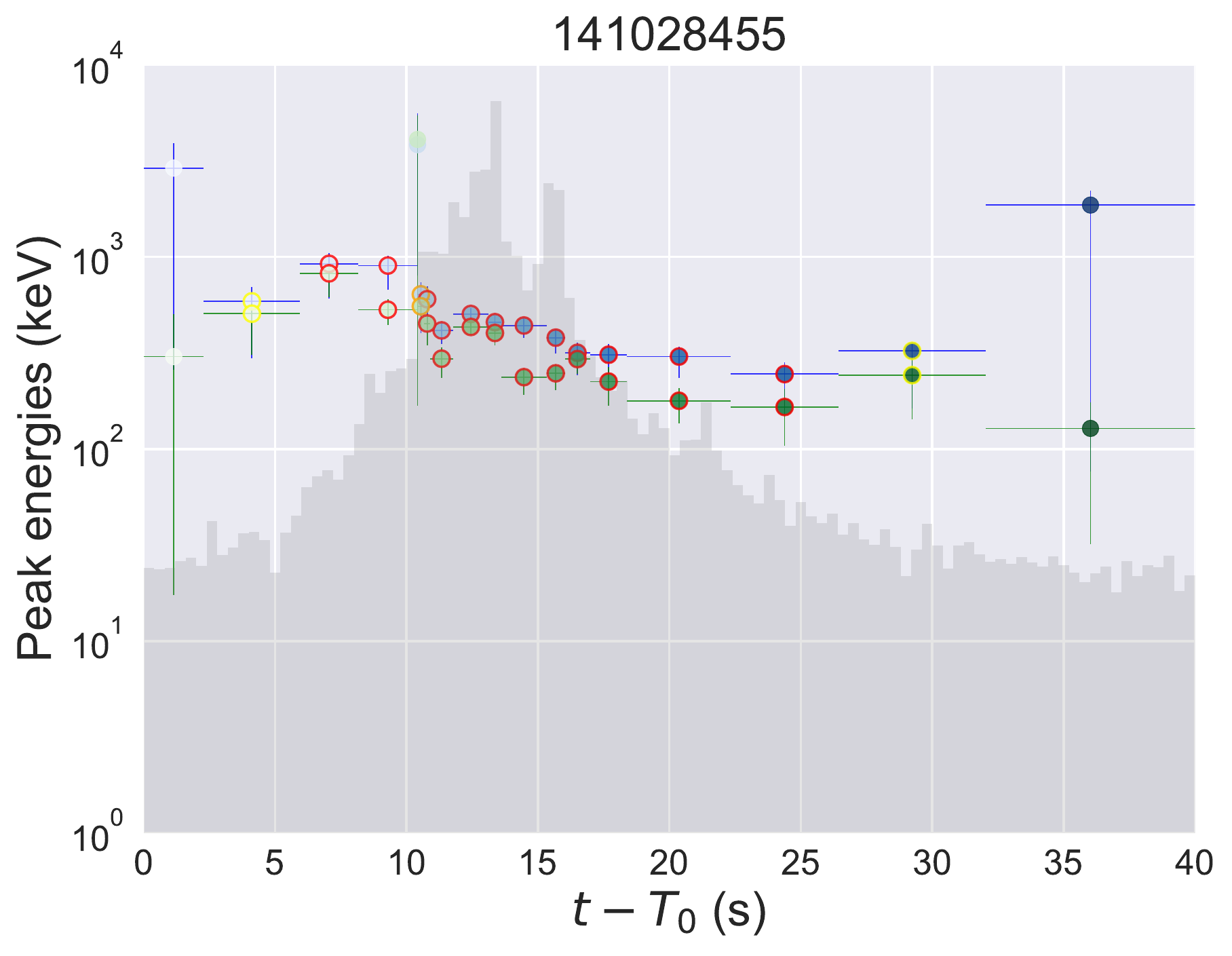}}
\subfigure{\includegraphics[width=0.3\linewidth]{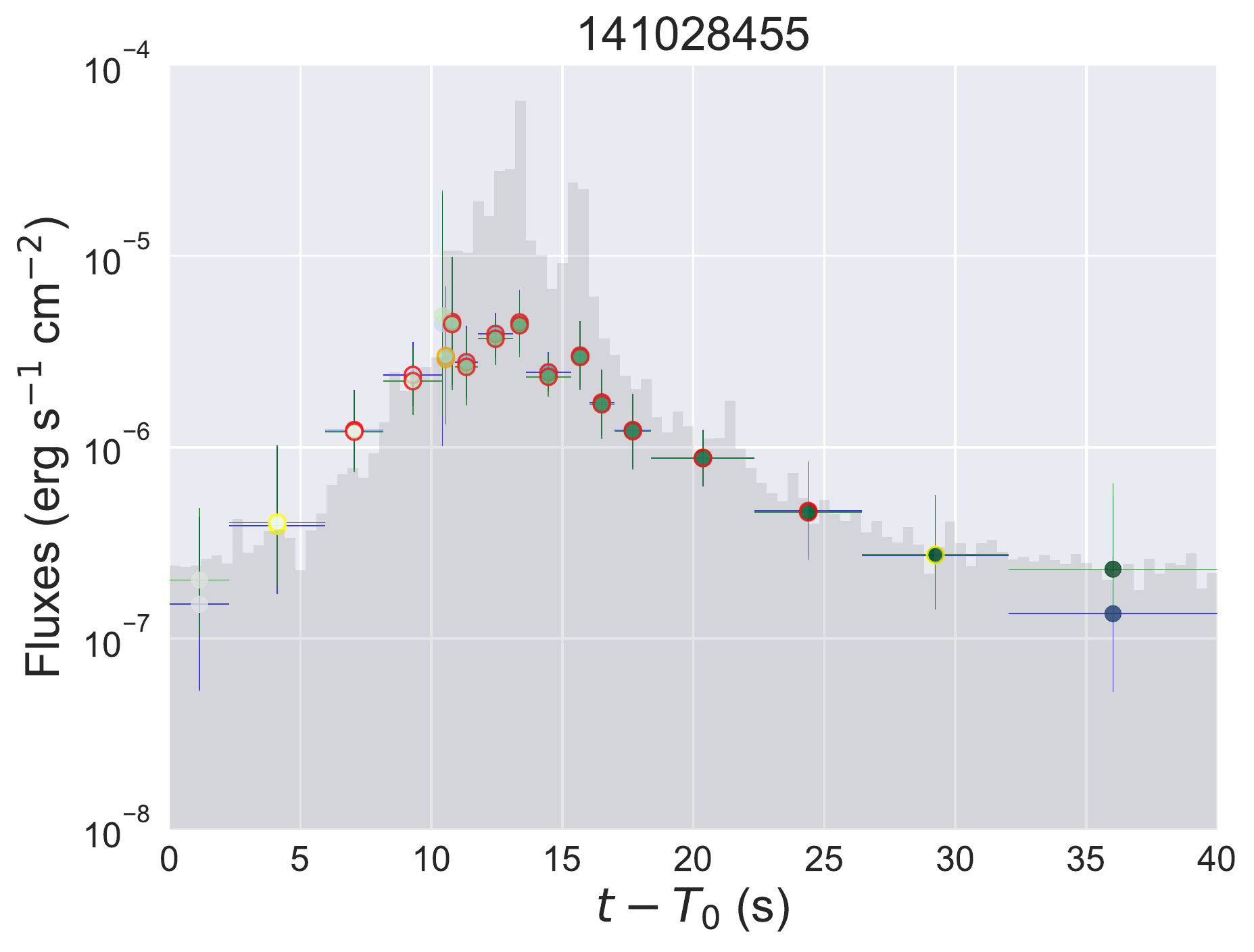}}

\subfigure{\includegraphics[width=0.3\linewidth]{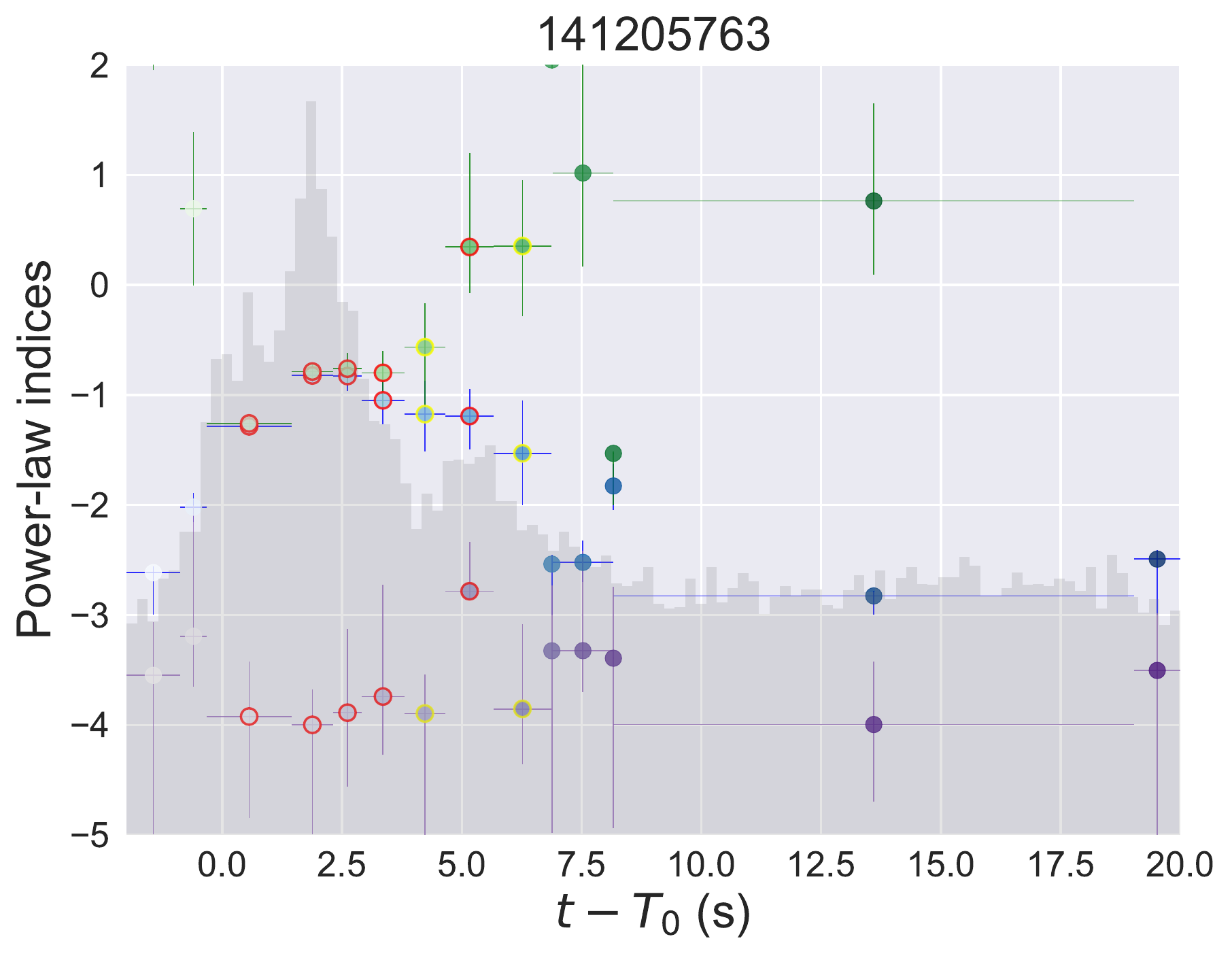}}
\subfigure{\includegraphics[width=0.3\linewidth]{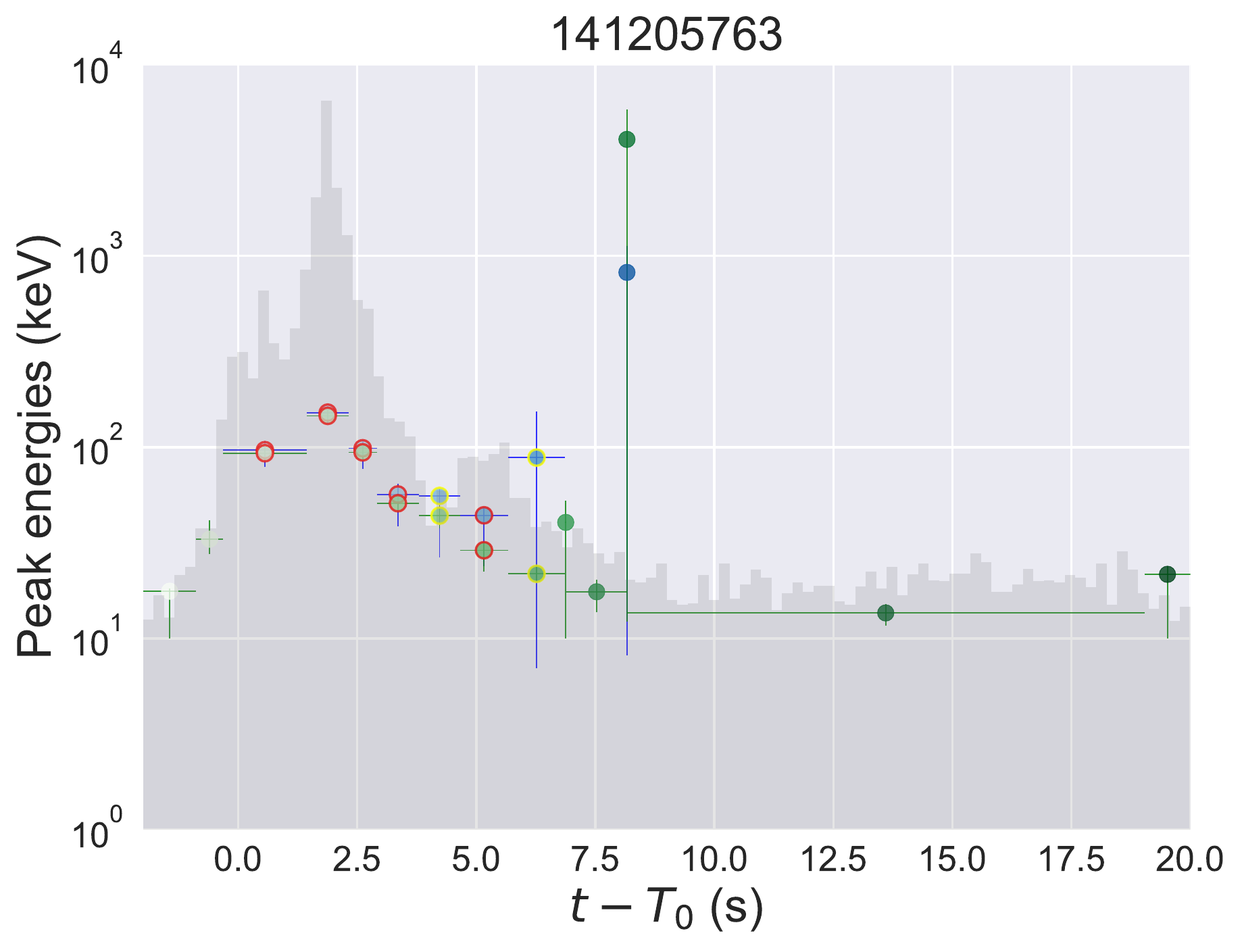}}
\subfigure{\includegraphics[width=0.3\linewidth]{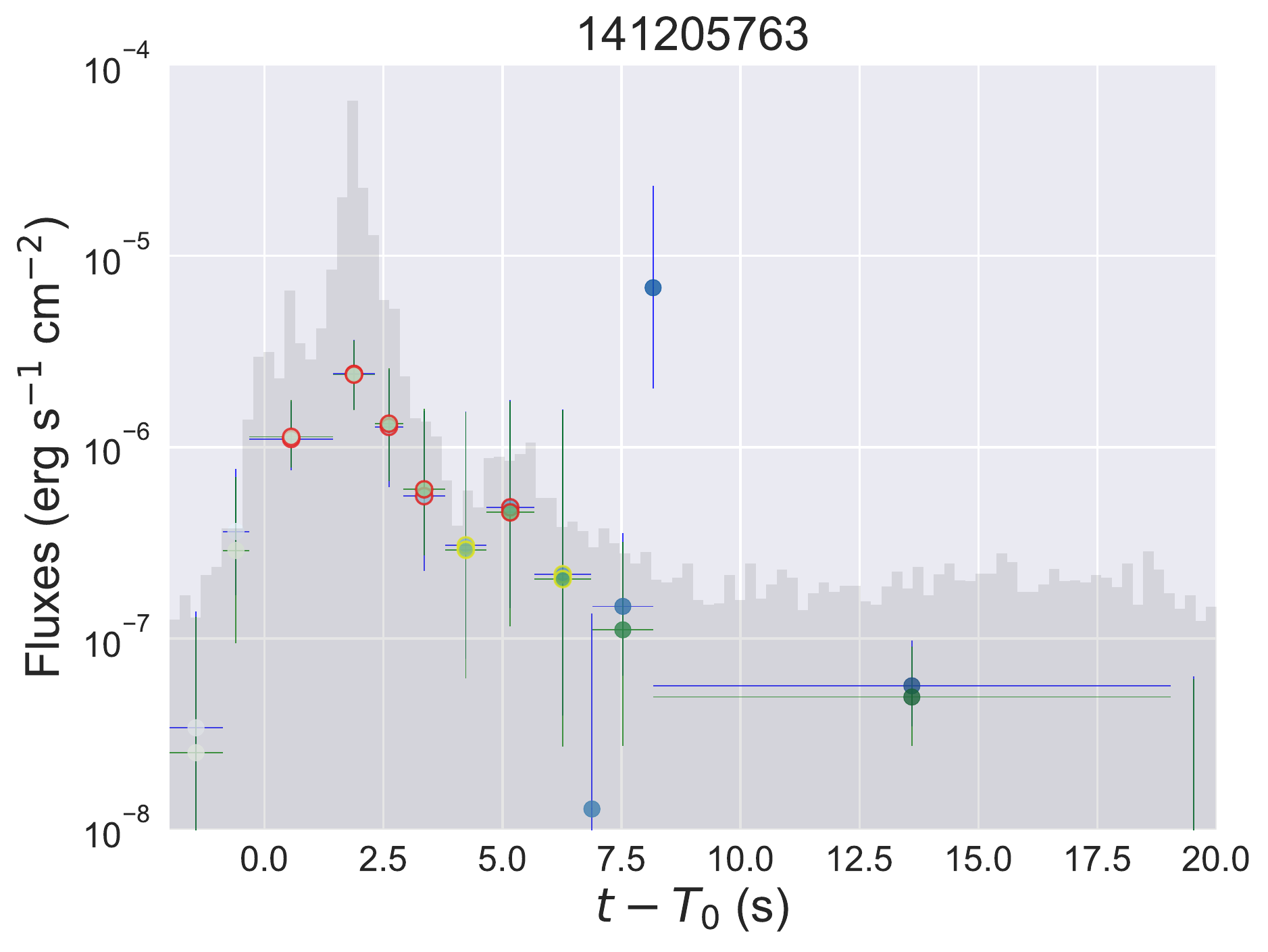}}

\subfigure{\includegraphics[width=0.3\linewidth]{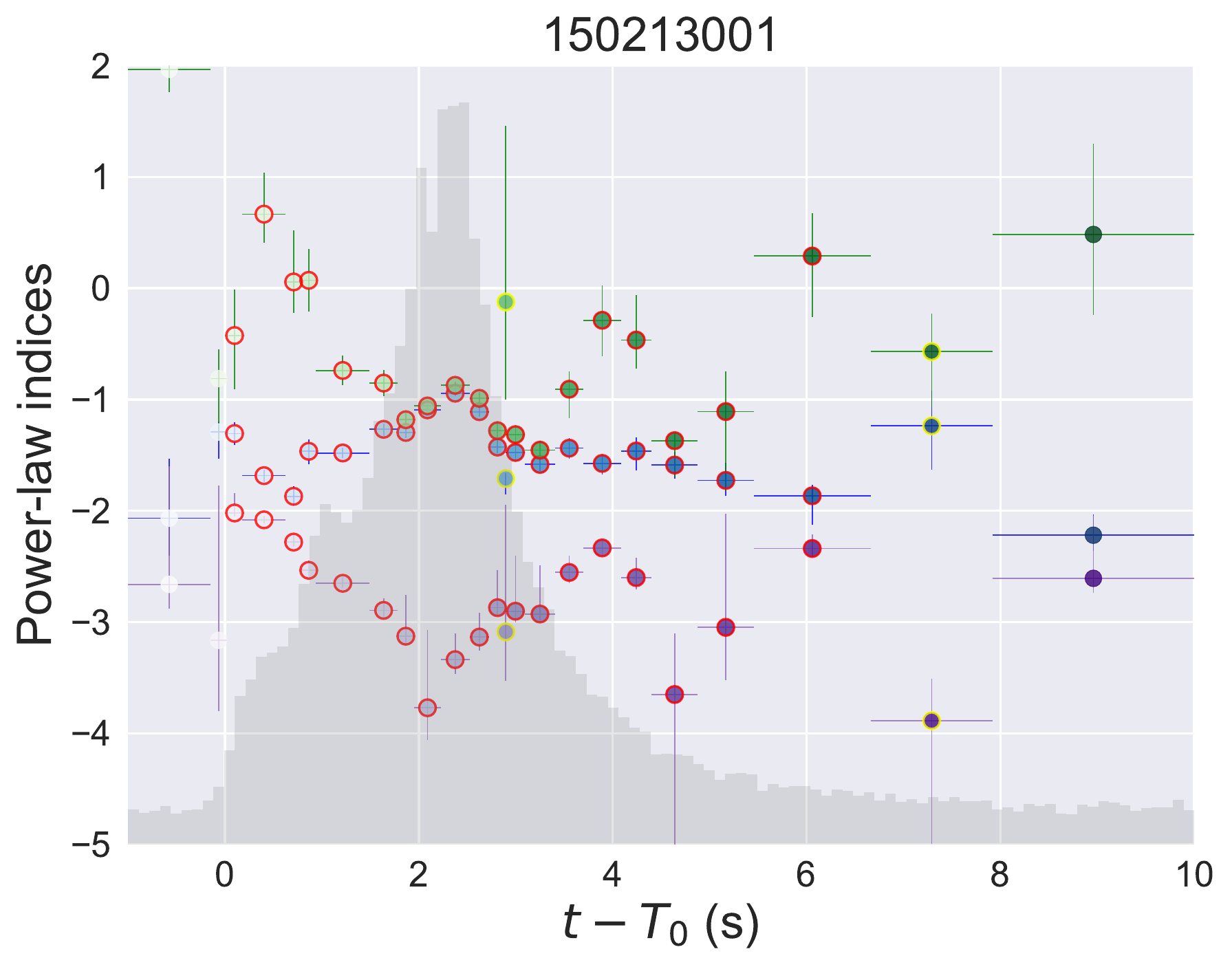}}
\subfigure{\includegraphics[width=0.3\linewidth]{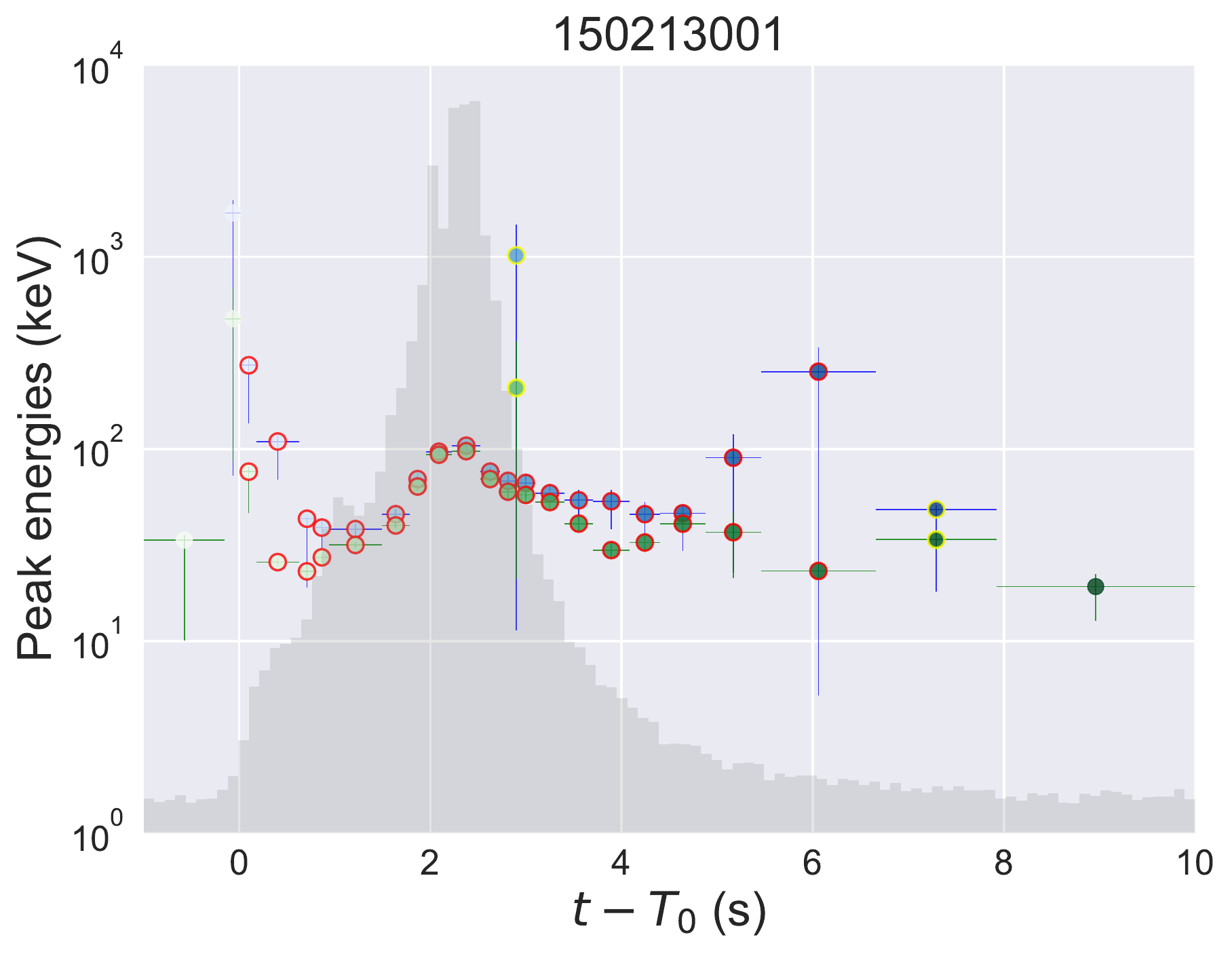}}
\subfigure{\includegraphics[width=0.3\linewidth]{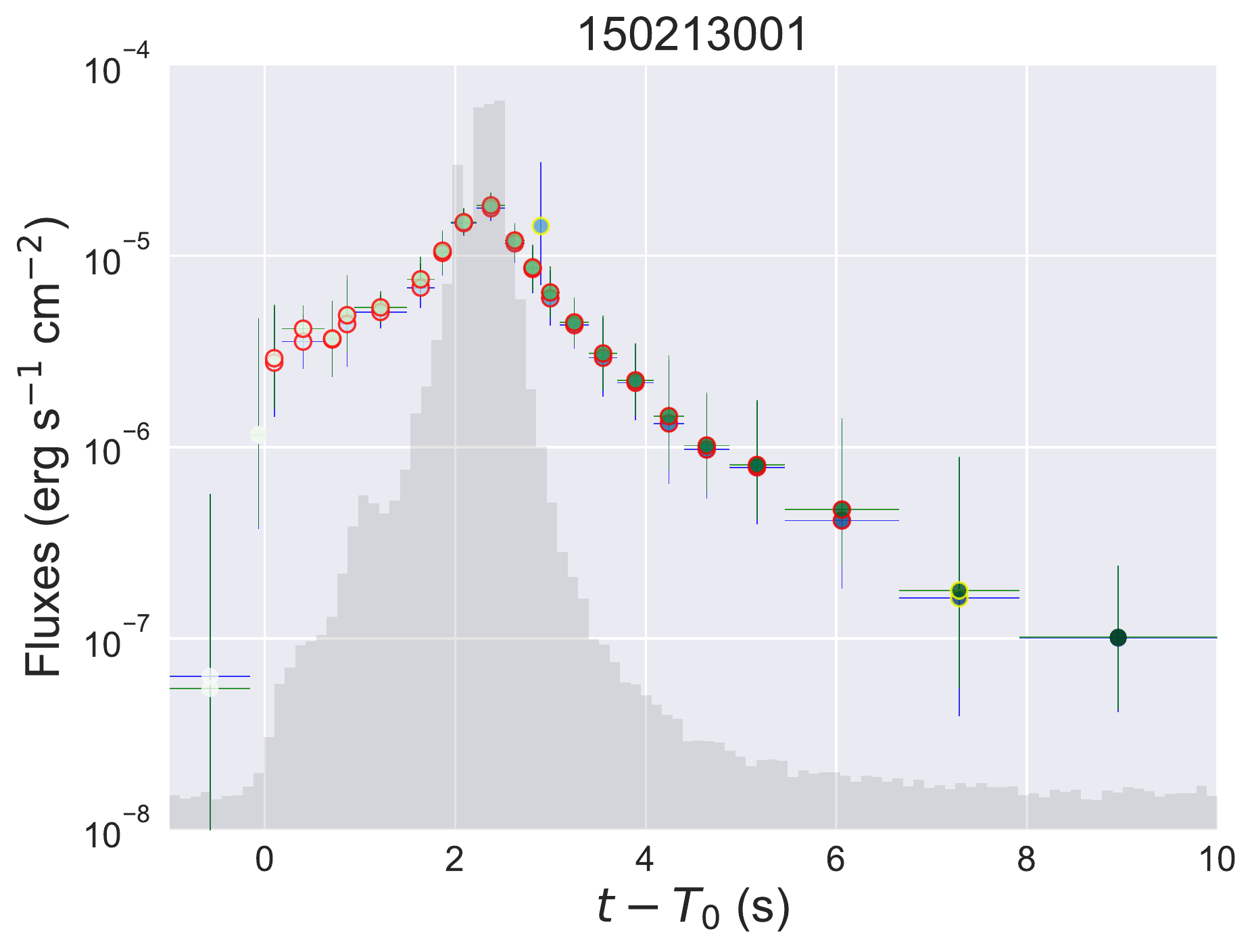}}

\caption{Same as Fig.~\ref{fig:evolution_group1}.
\label{fig:evolution_group7}}
\end{figure*}

\begin{figure*}
\centering

\subfigure{\includegraphics[width=0.3\linewidth]{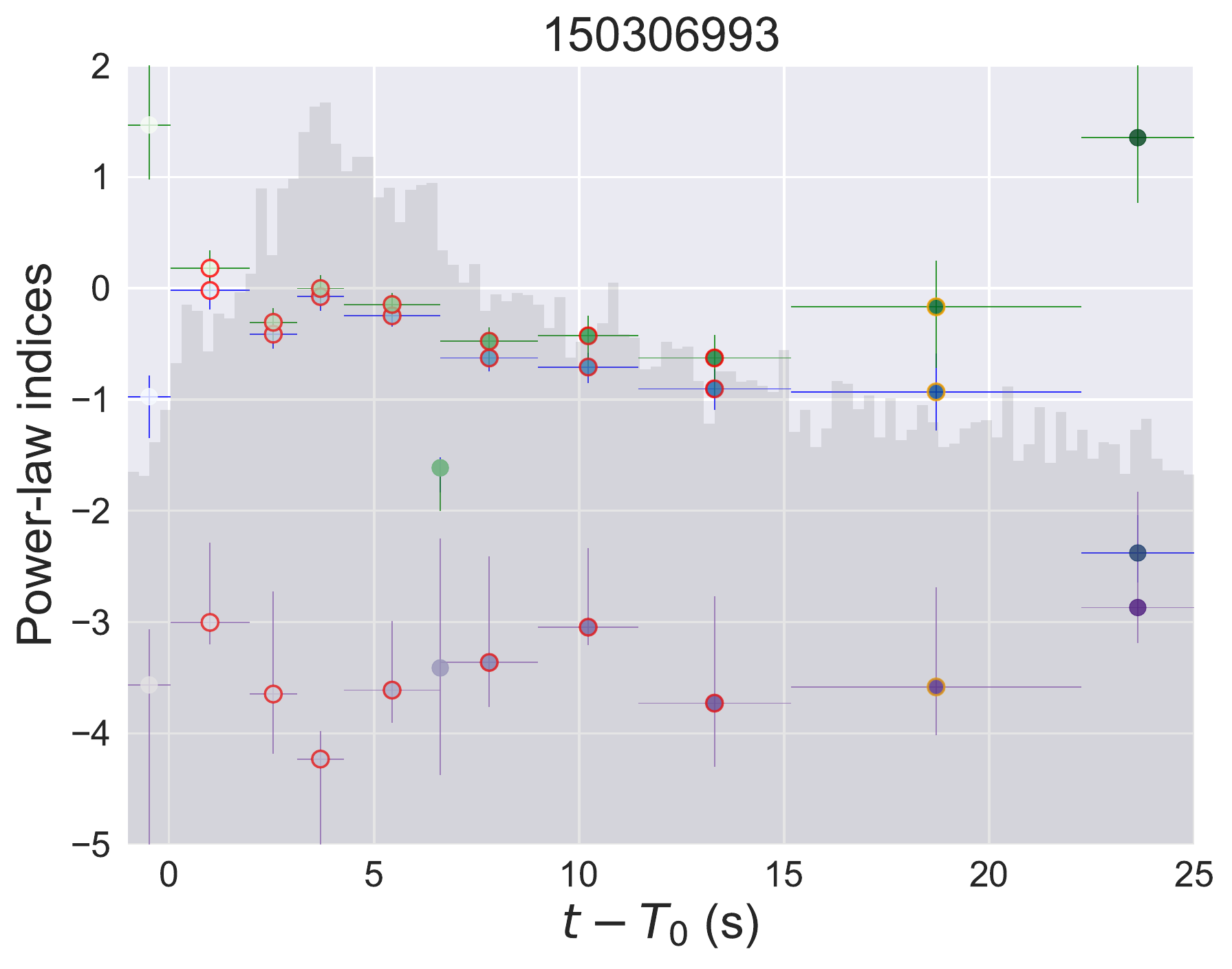}}
\subfigure{\includegraphics[width=0.3\linewidth]{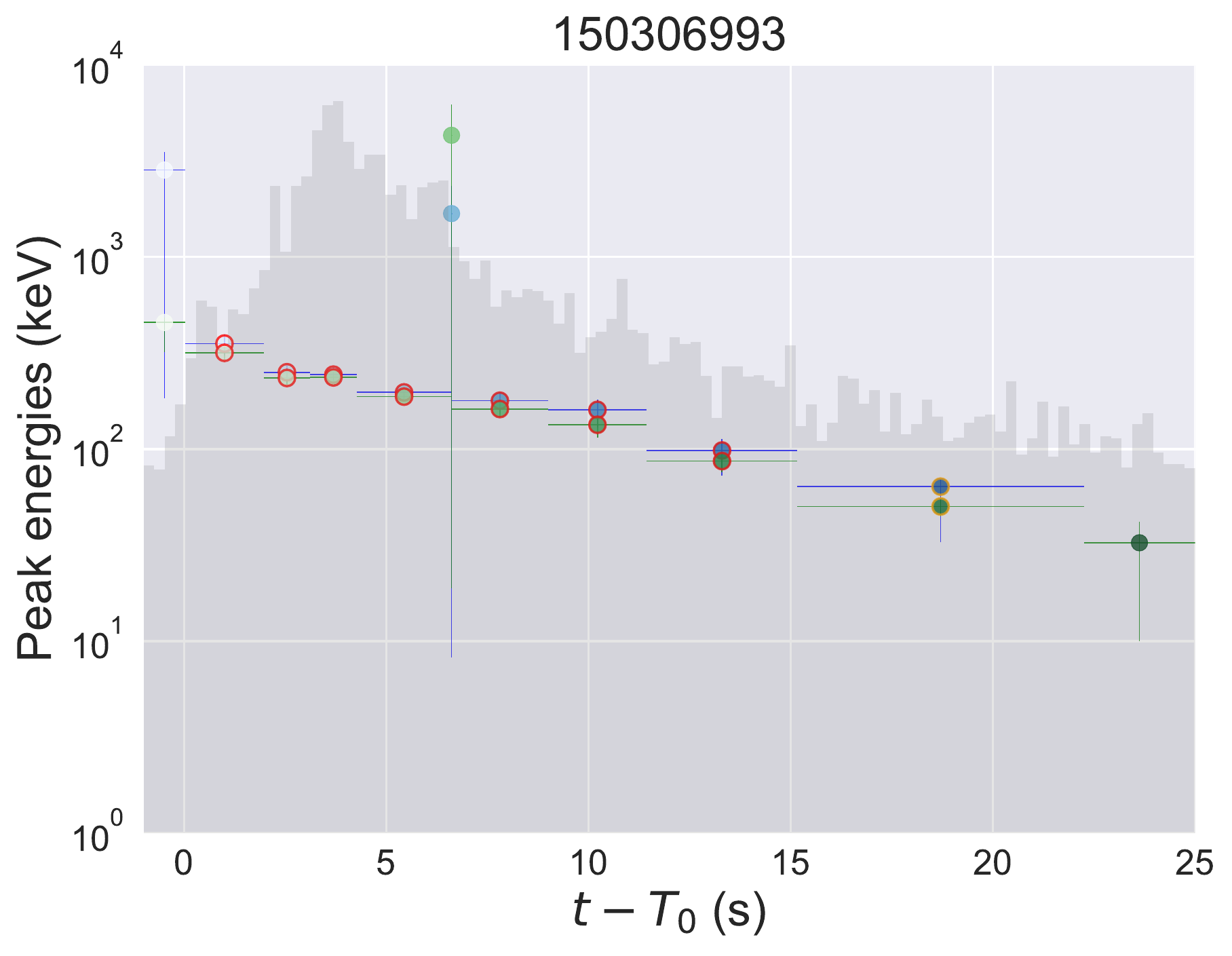}}
\subfigure{\includegraphics[width=0.3\linewidth]{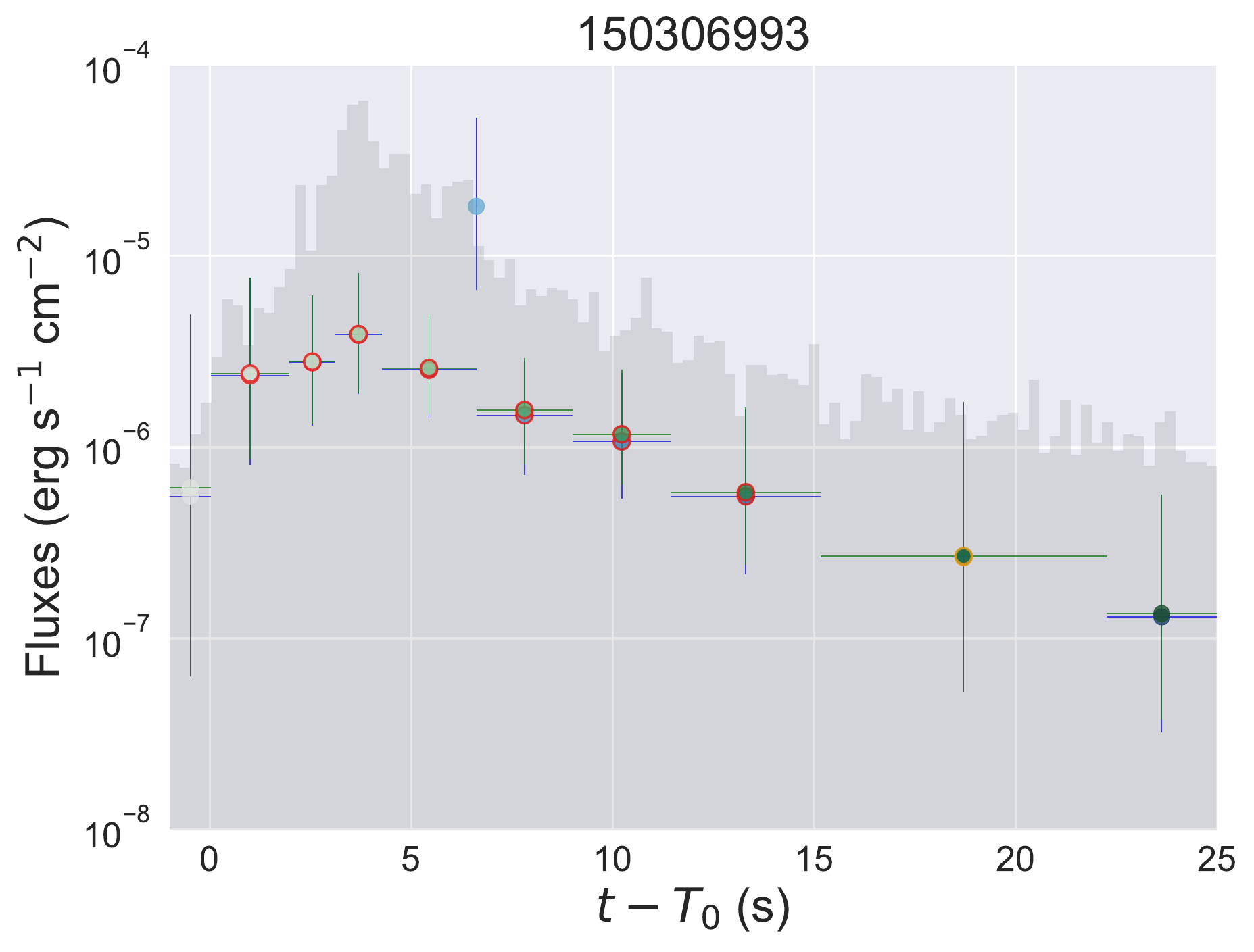}}

\subfigure{\includegraphics[width=0.3\linewidth]{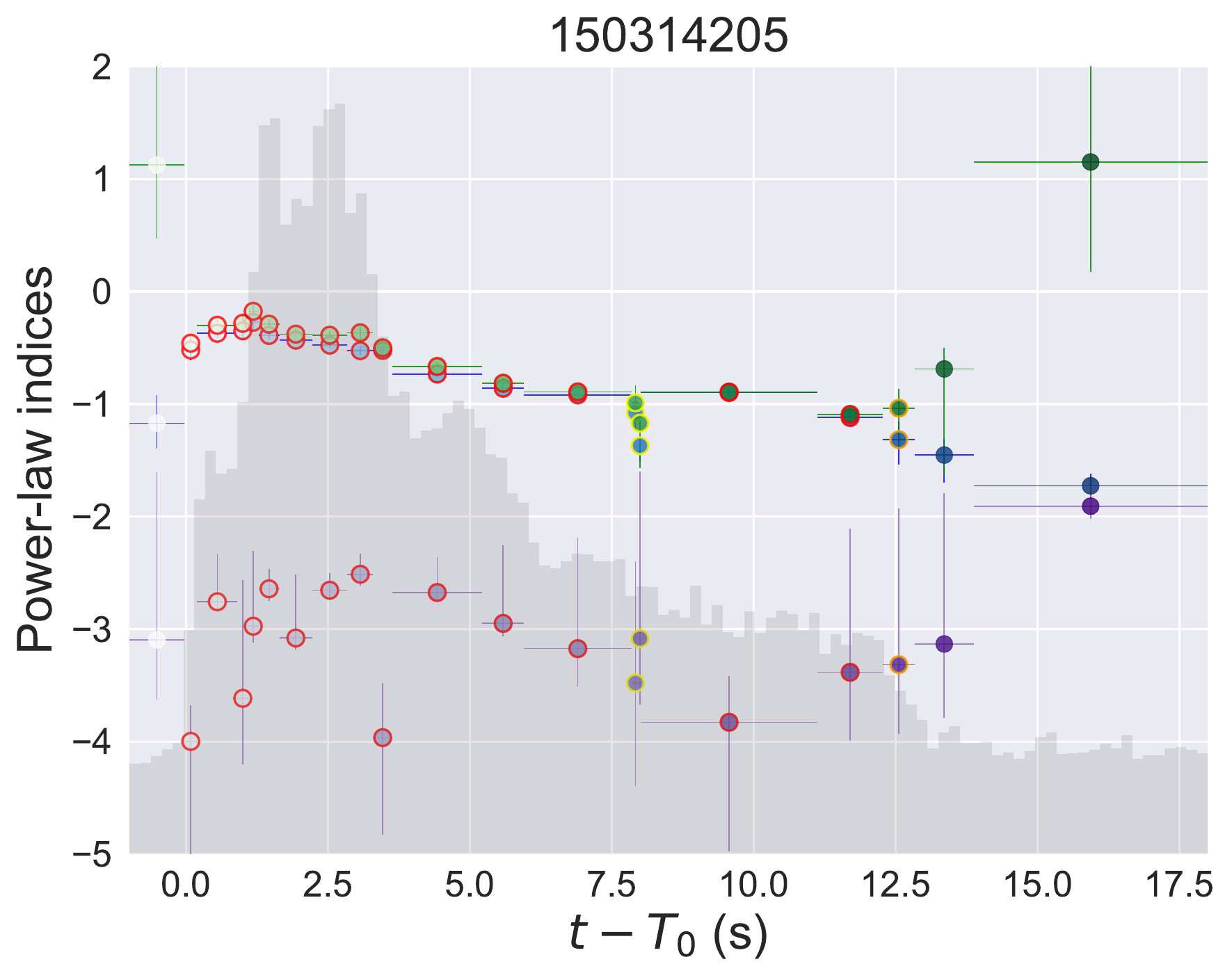}}
\subfigure{\includegraphics[width=0.3\linewidth]{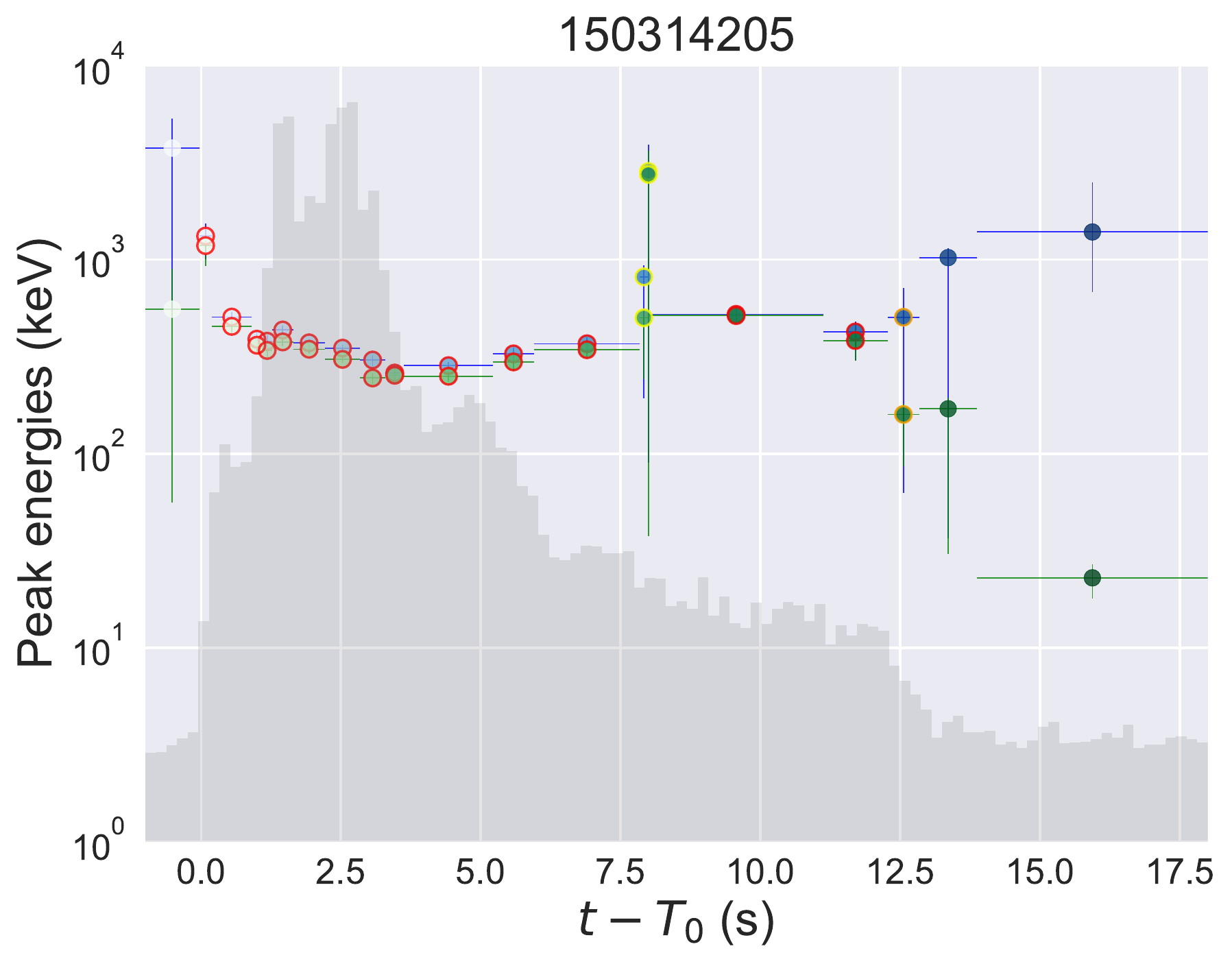}}
\subfigure{\includegraphics[width=0.3\linewidth]{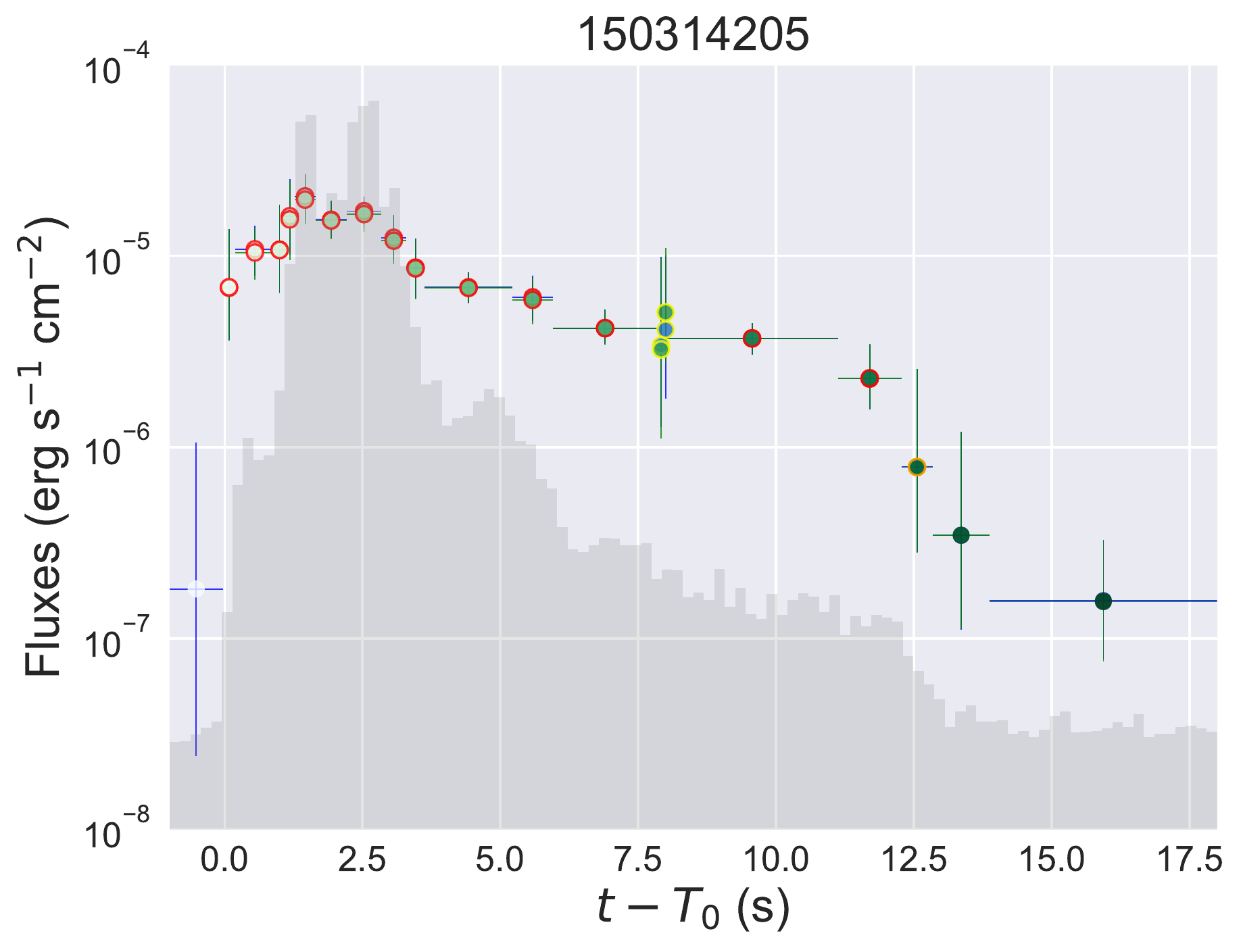}}

\subfigure{\includegraphics[width=0.3\linewidth]{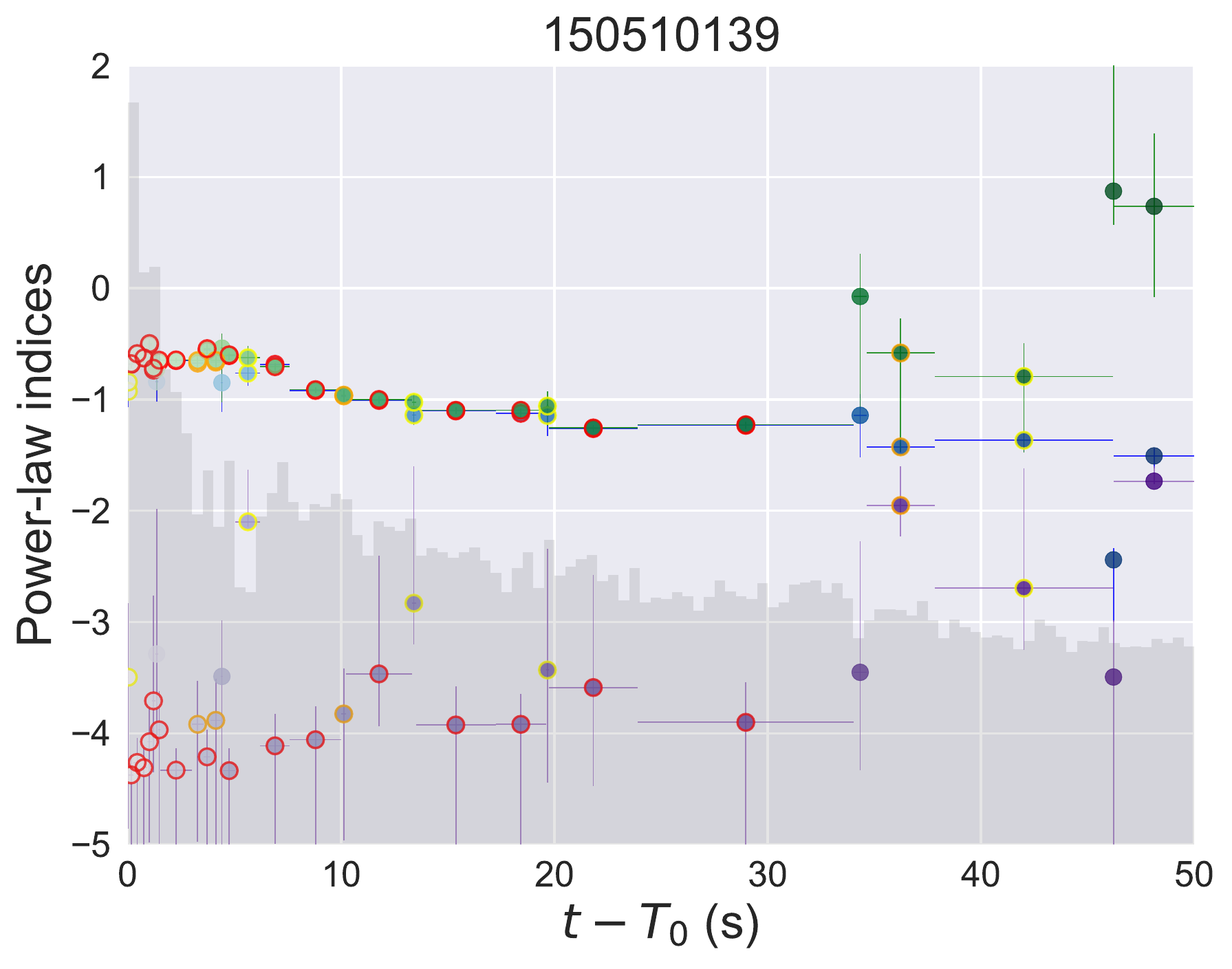}}
\subfigure{\includegraphics[width=0.3\linewidth]{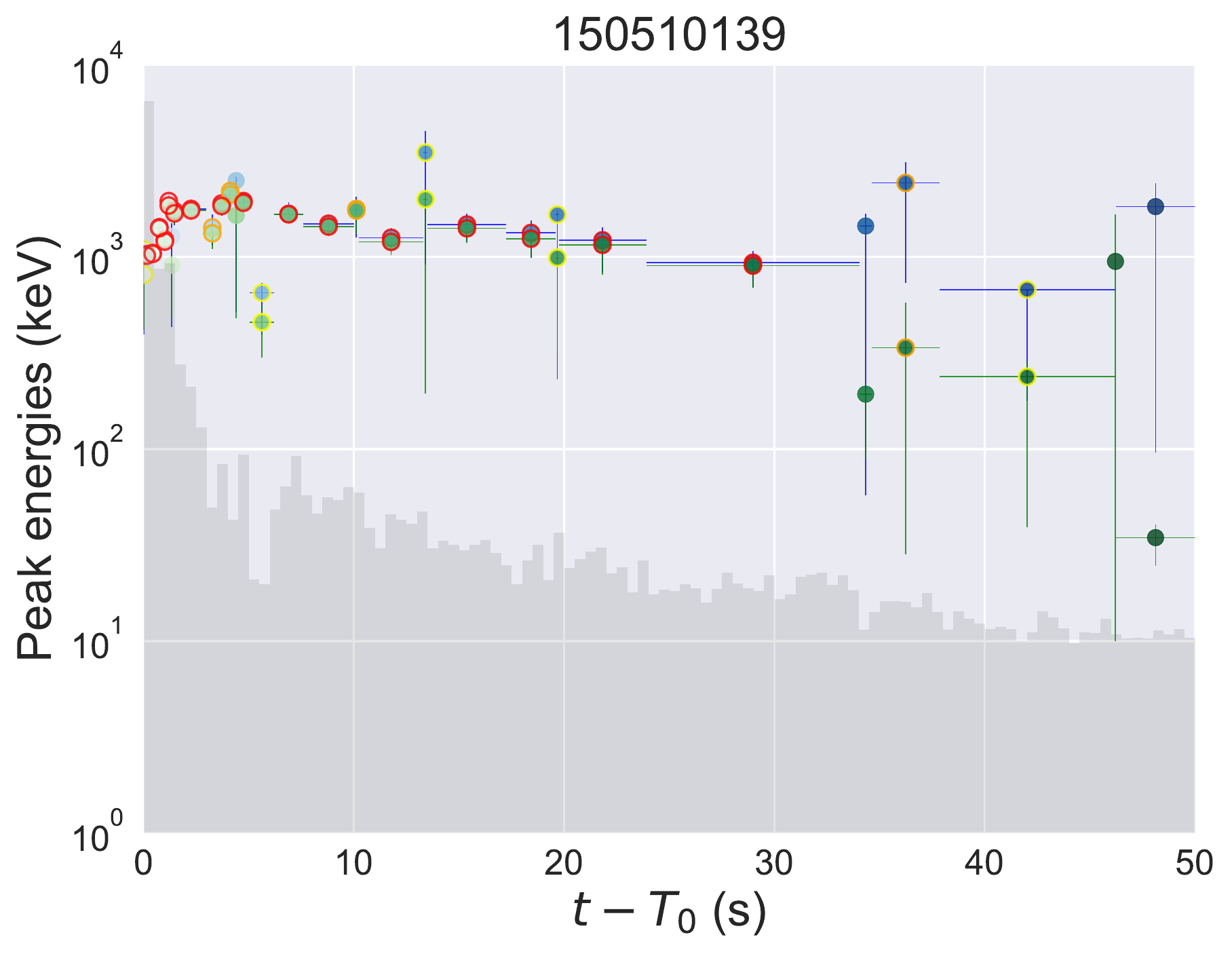}}
\subfigure{\includegraphics[width=0.3\linewidth]{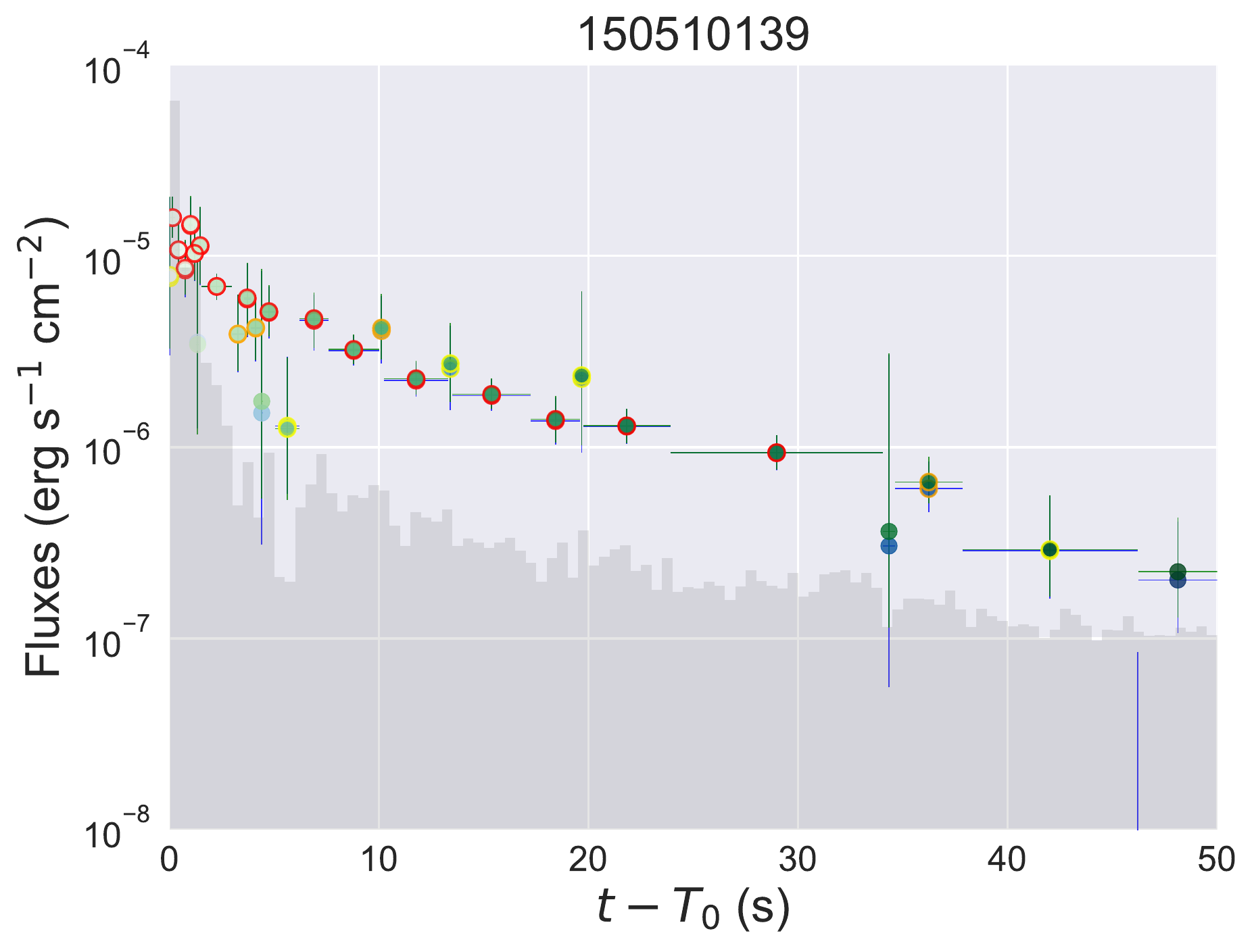}}

\subfigure{\includegraphics[width=0.3\linewidth]{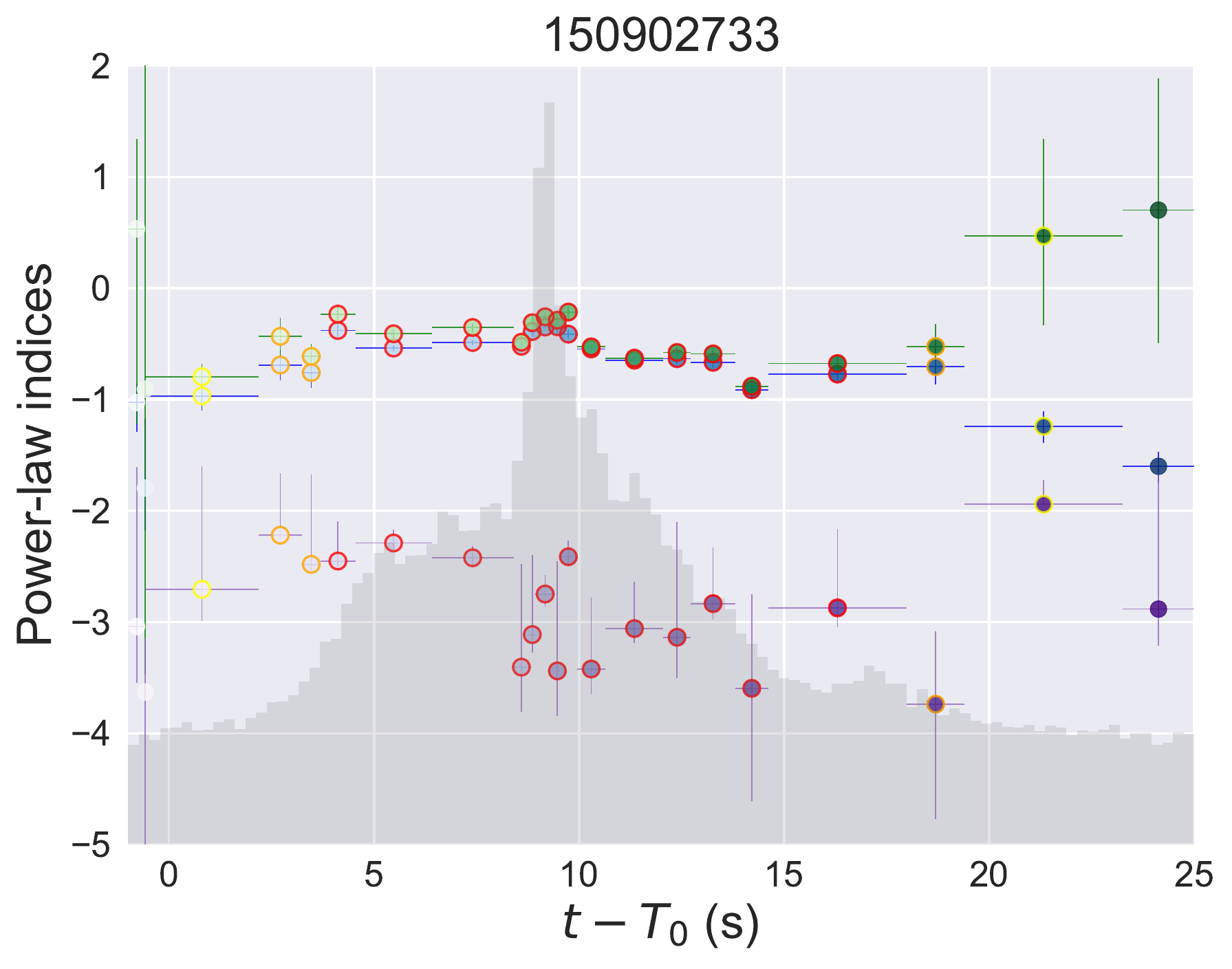}}
\subfigure{\includegraphics[width=0.3\linewidth]{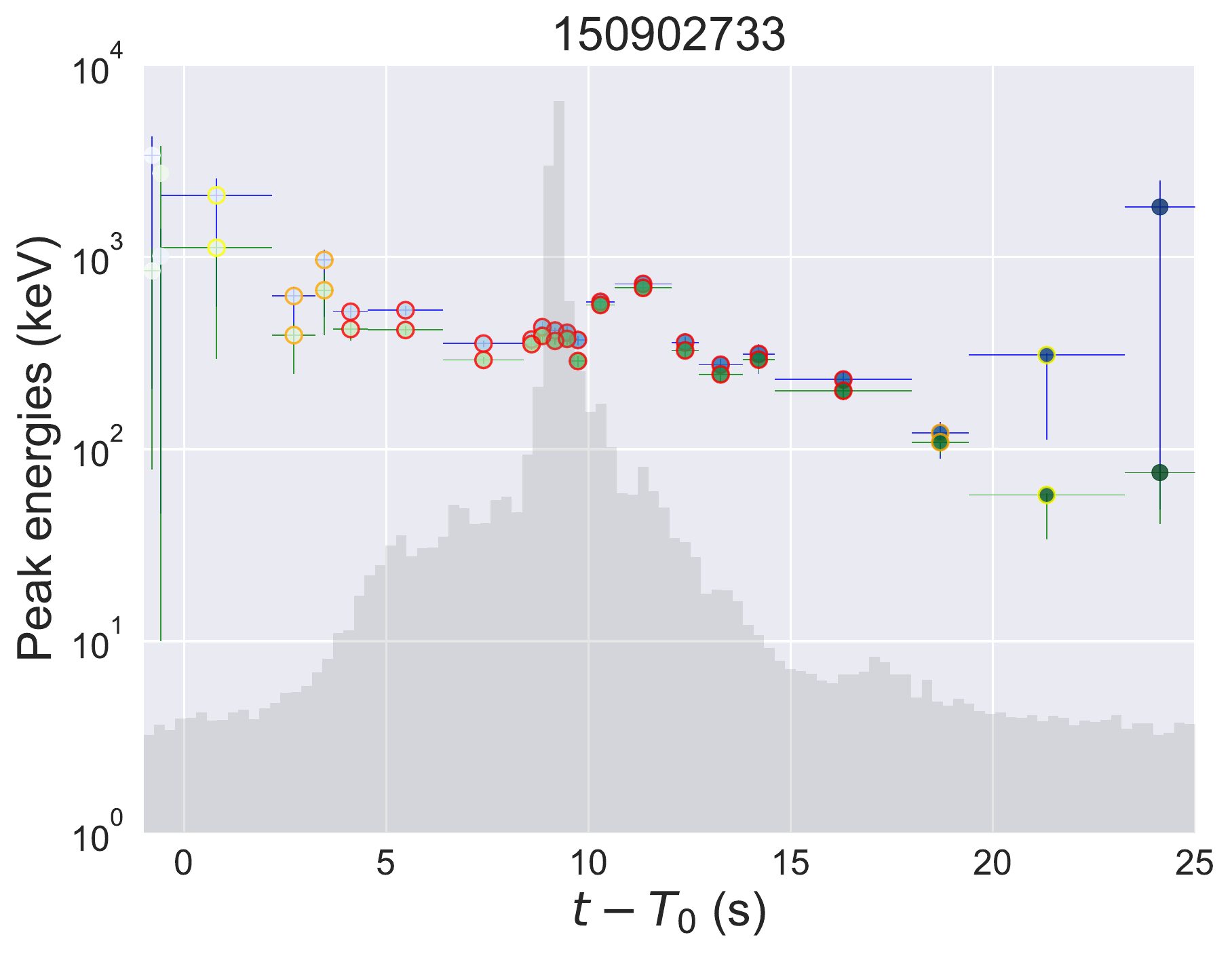}}
\subfigure{\includegraphics[width=0.3\linewidth]{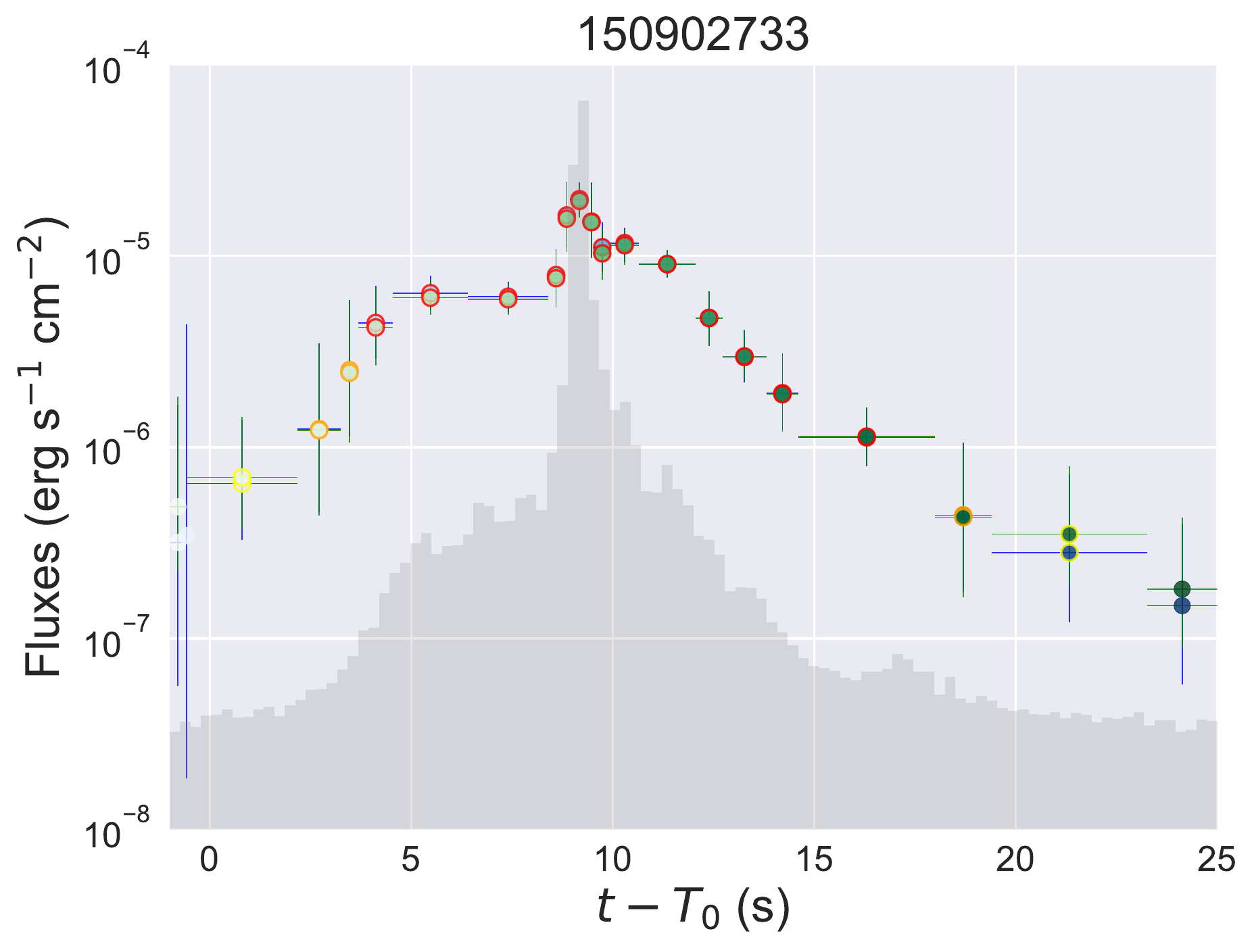}}

\caption{Same as Fig.~\ref{fig:evolution_group1}.
\label{fig:evolution_group8}}
\end{figure*}

\begin{figure*}
\centering

\subfigure{\includegraphics[width=0.3\linewidth]{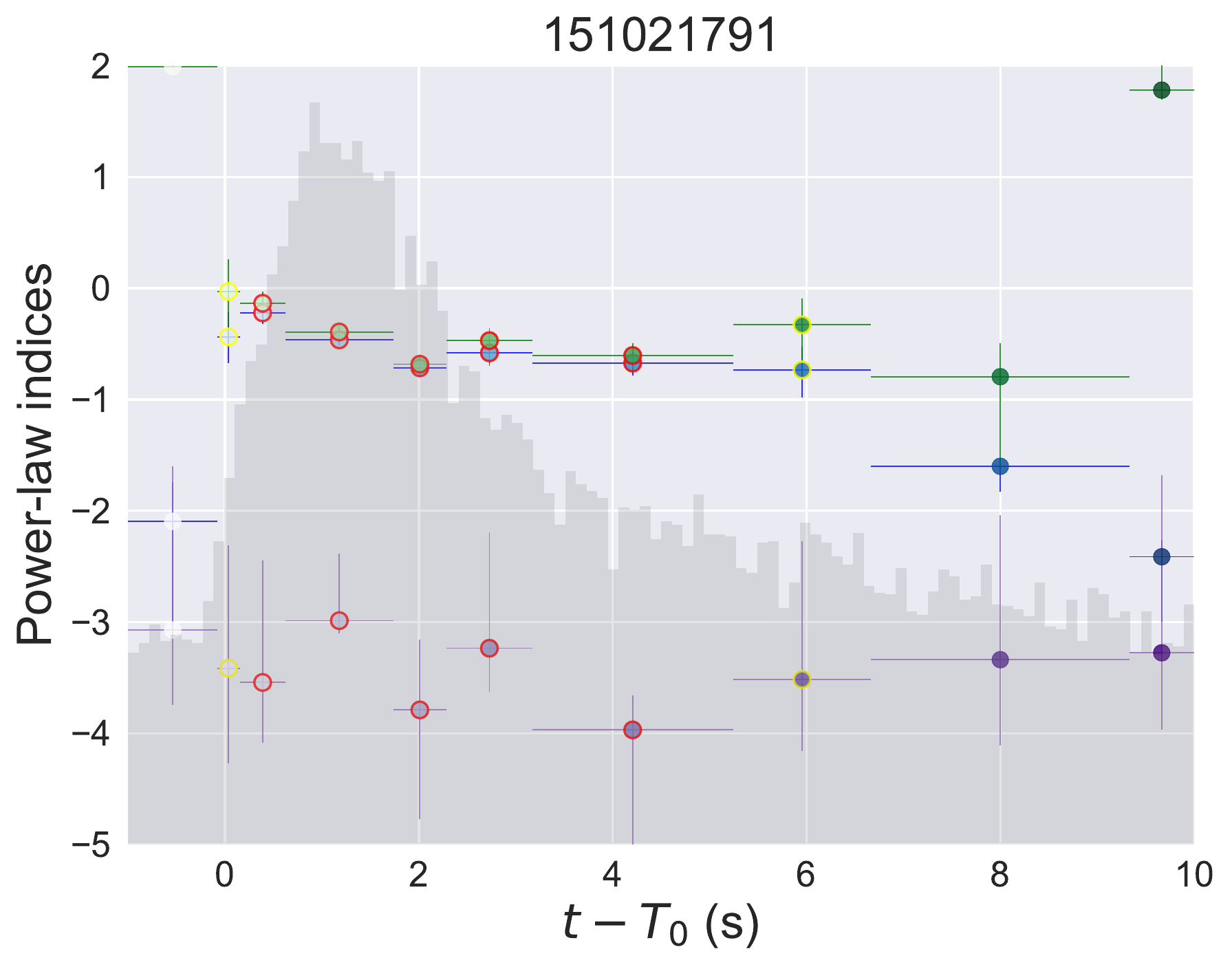}}
\subfigure{\includegraphics[width=0.3\linewidth]{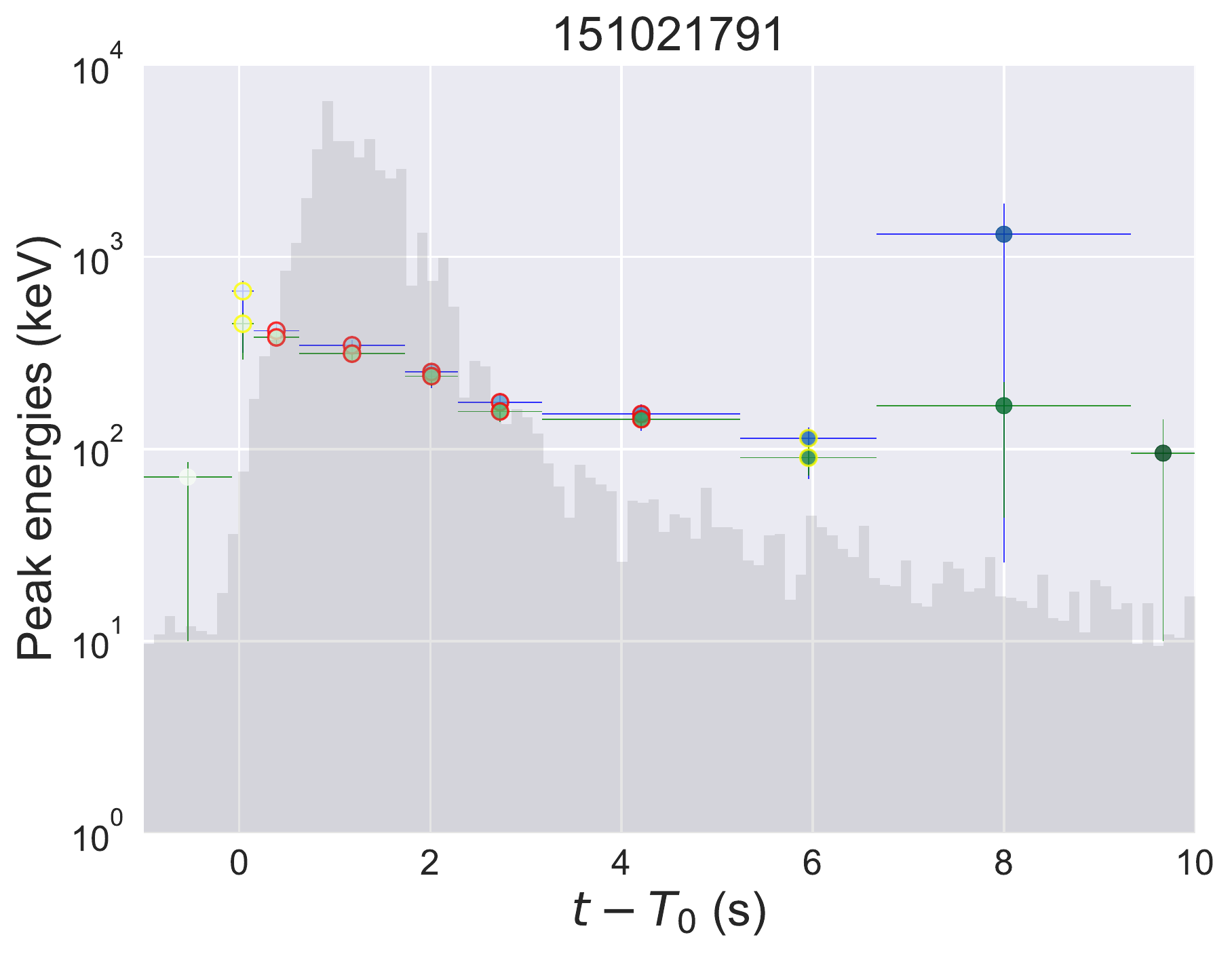}}
\subfigure{\includegraphics[width=0.3\linewidth]{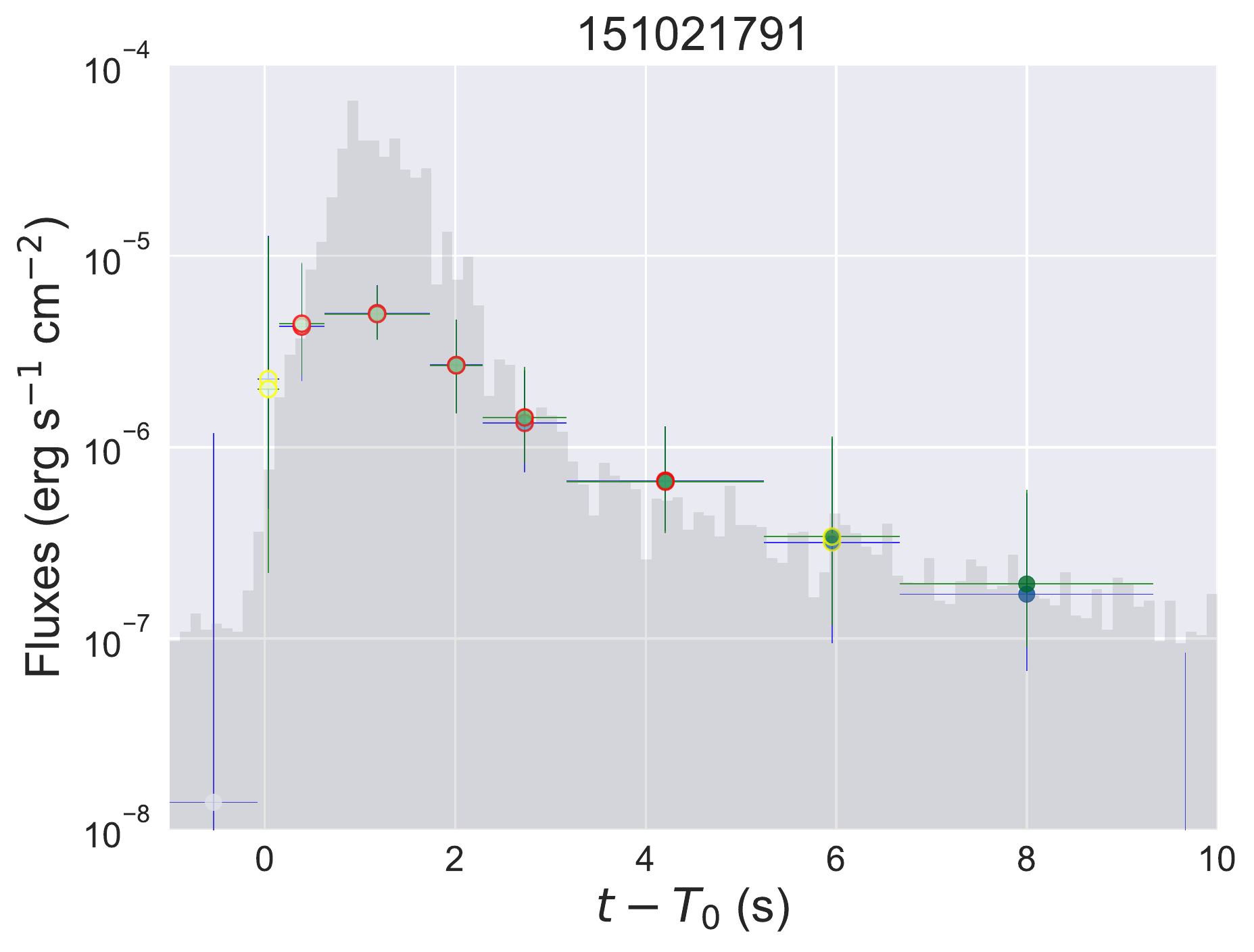}}

\subfigure{\includegraphics[width=0.3\linewidth]{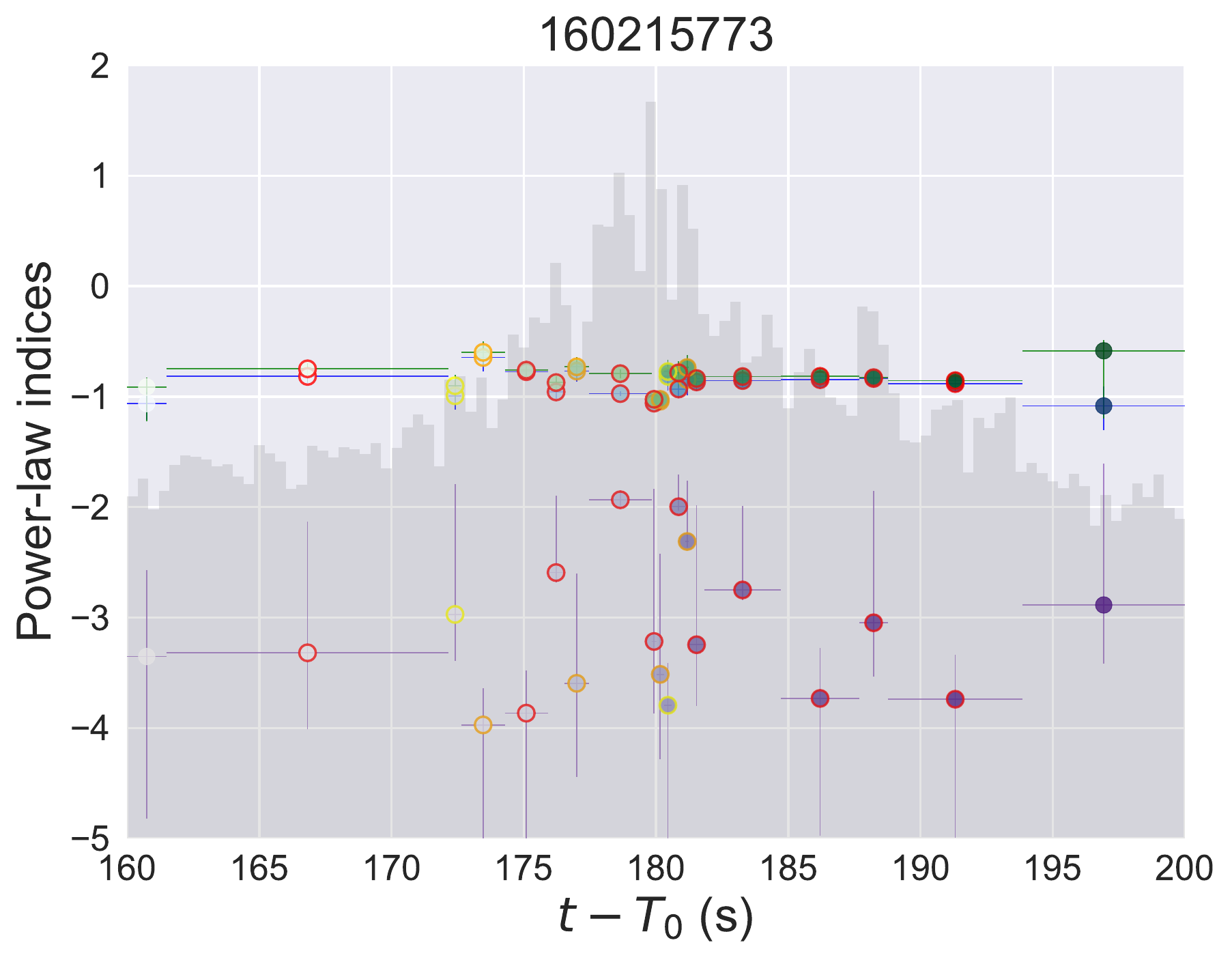}}
\subfigure{\includegraphics[width=0.3\linewidth]{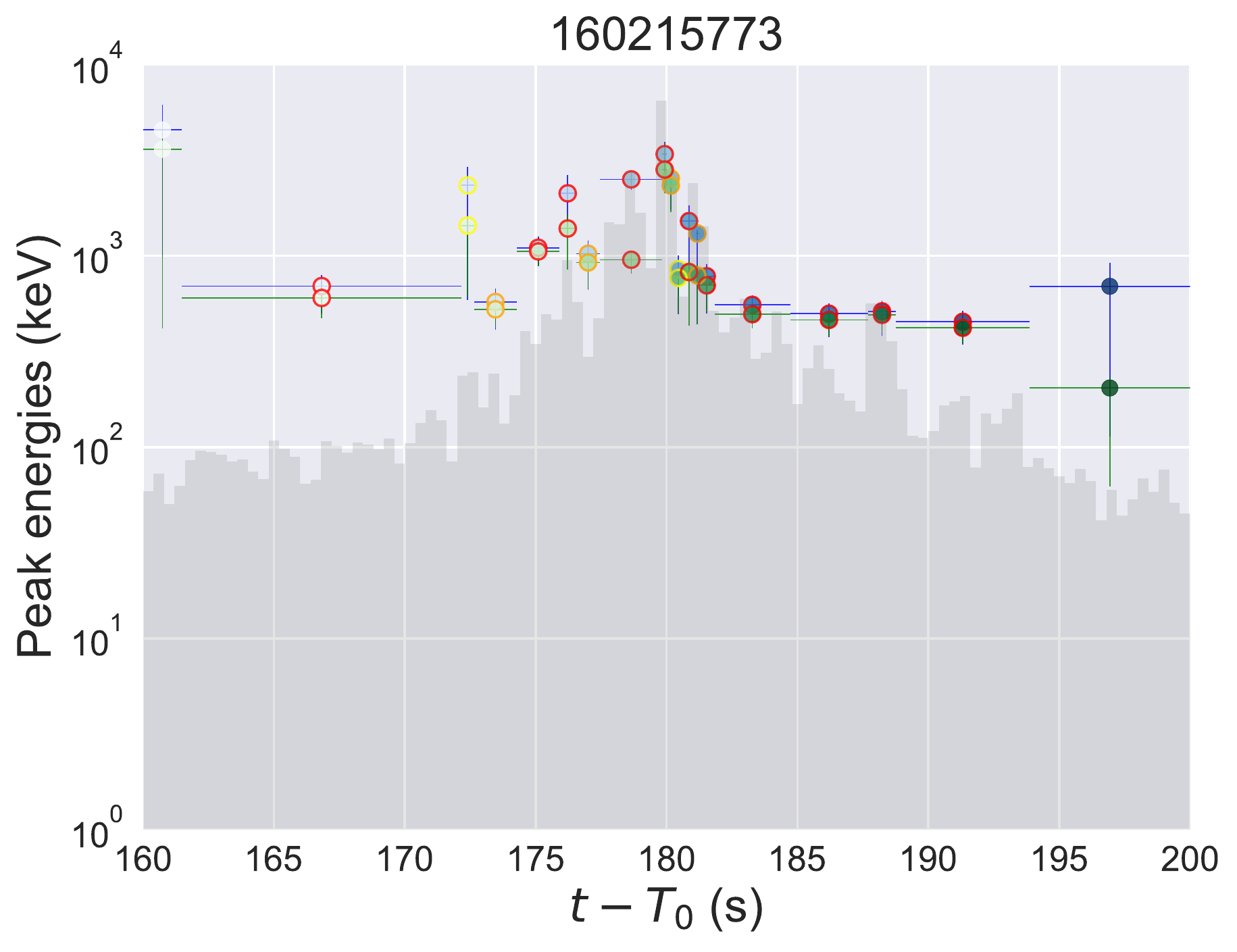}}
\subfigure{\includegraphics[width=0.3\linewidth]{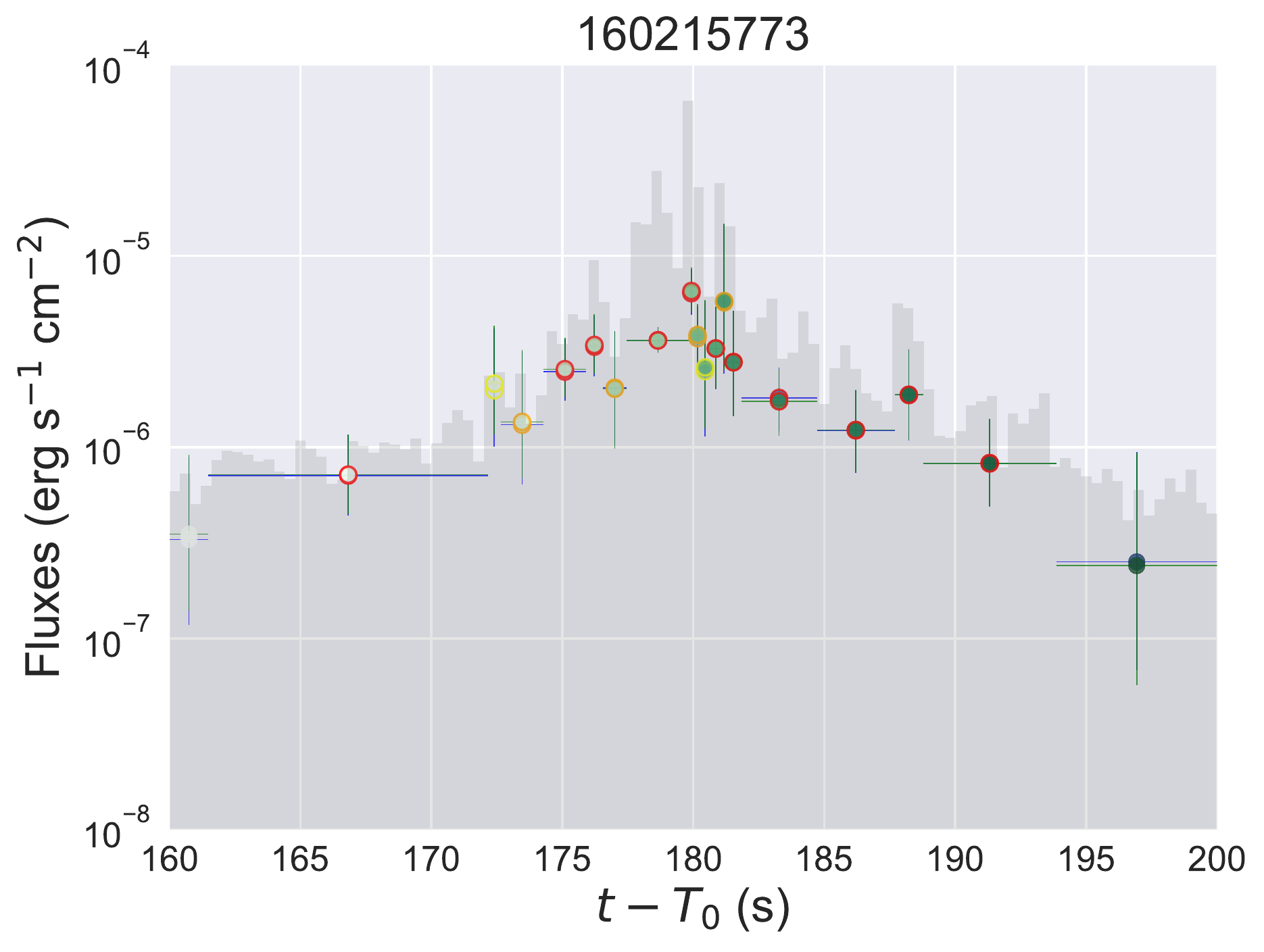}}

\subfigure{\includegraphics[width=0.3\linewidth]{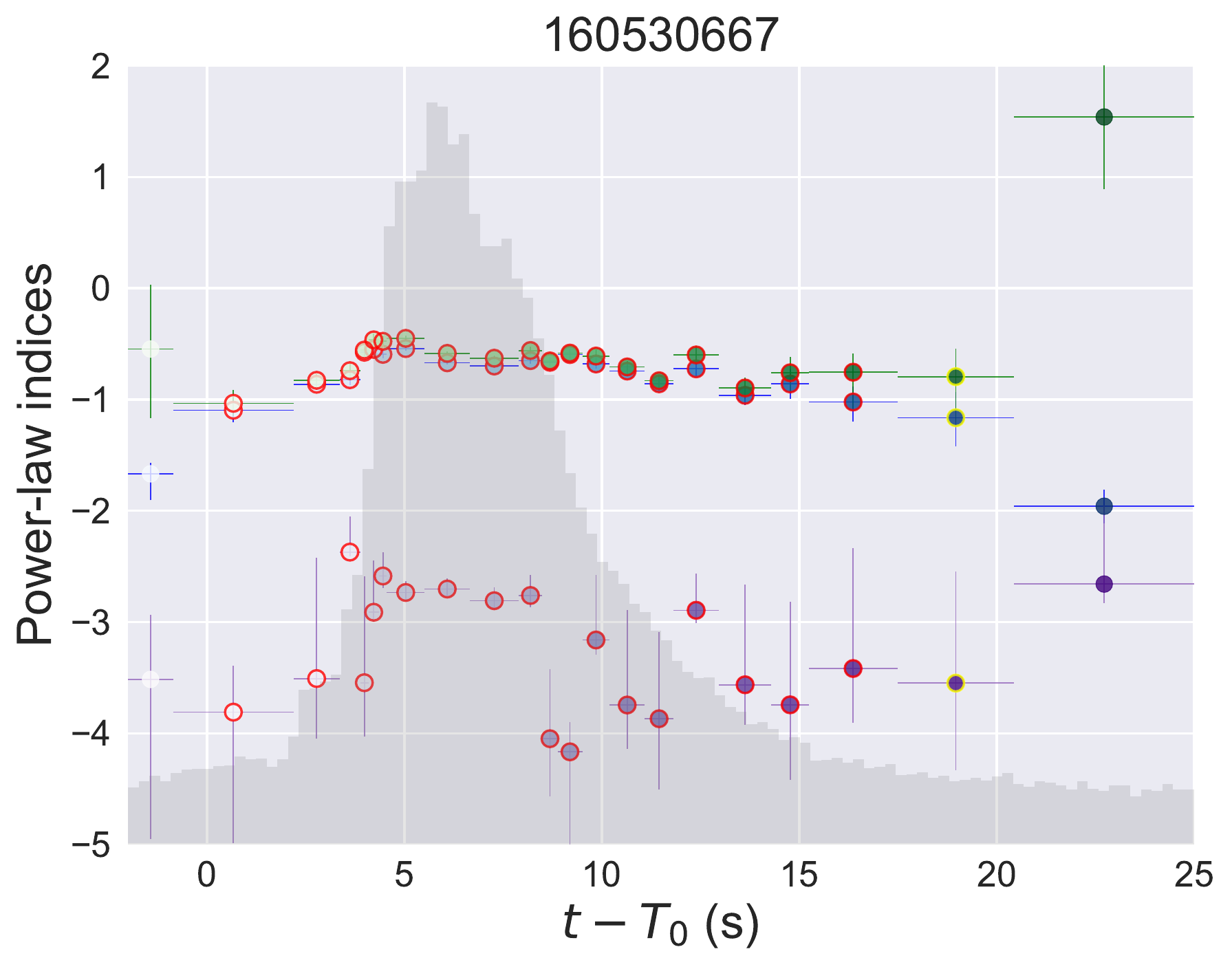}}
\subfigure{\includegraphics[width=0.3\linewidth]{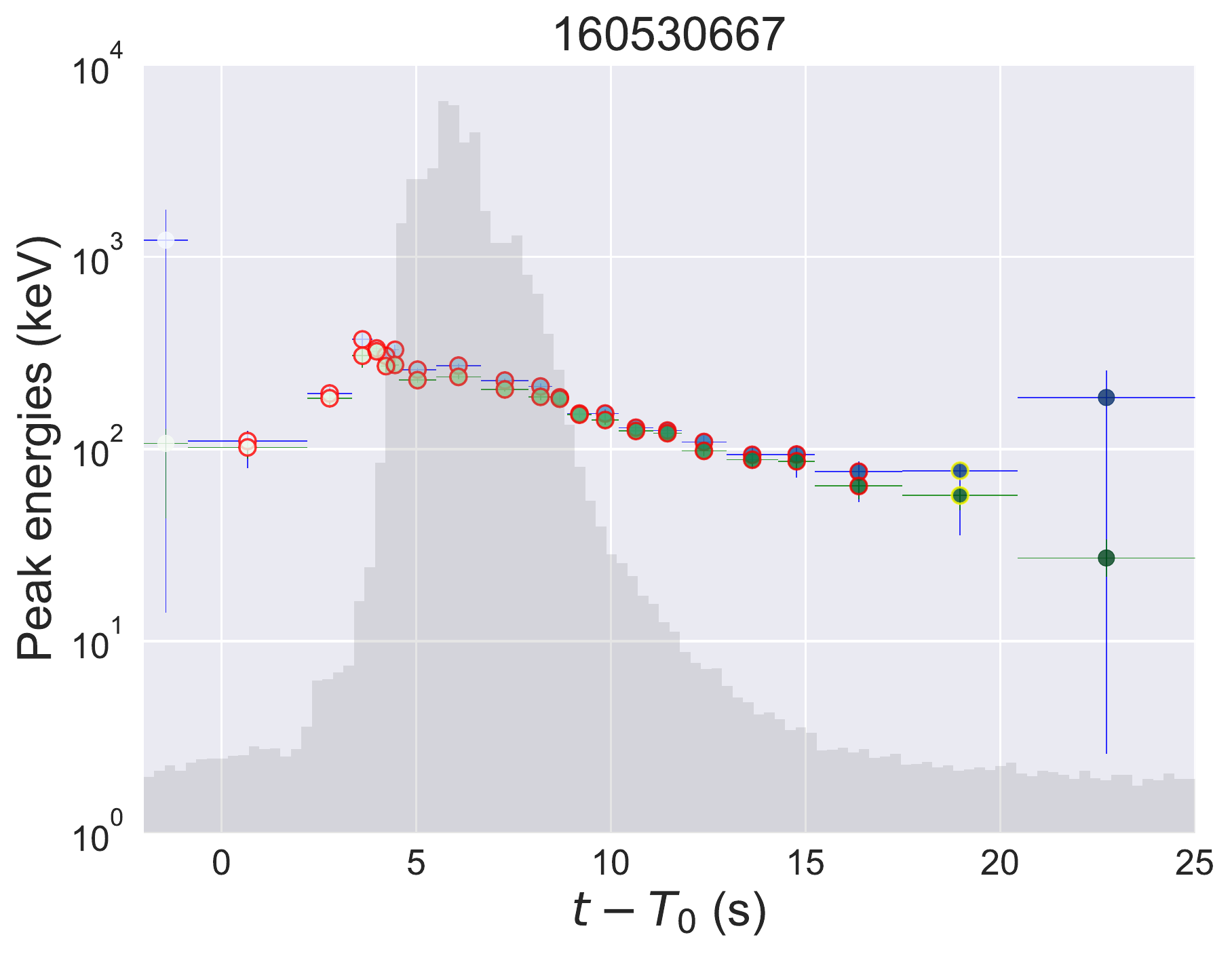}}
\subfigure{\includegraphics[width=0.3\linewidth]{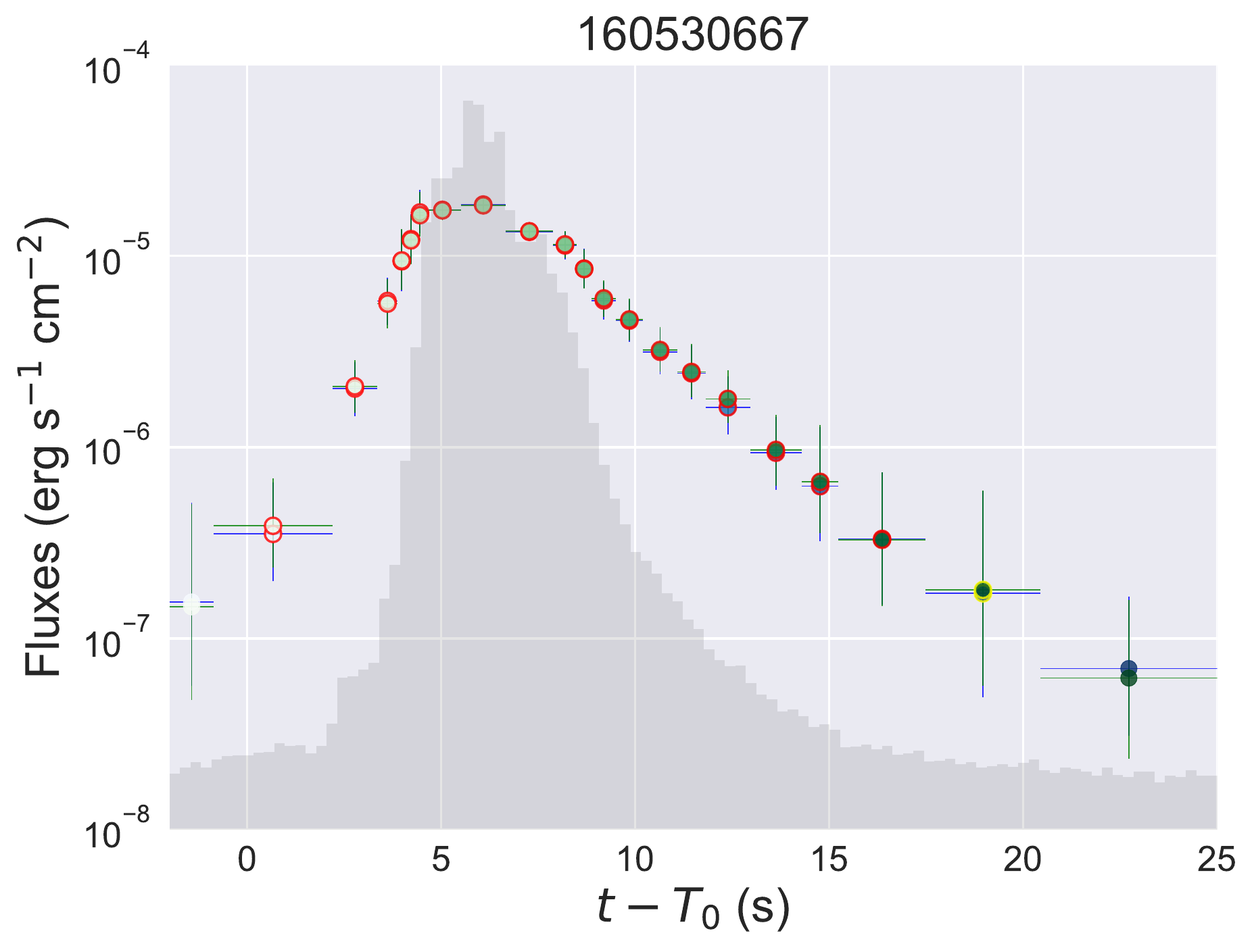}}

\subfigure{\includegraphics[width=0.3\linewidth]{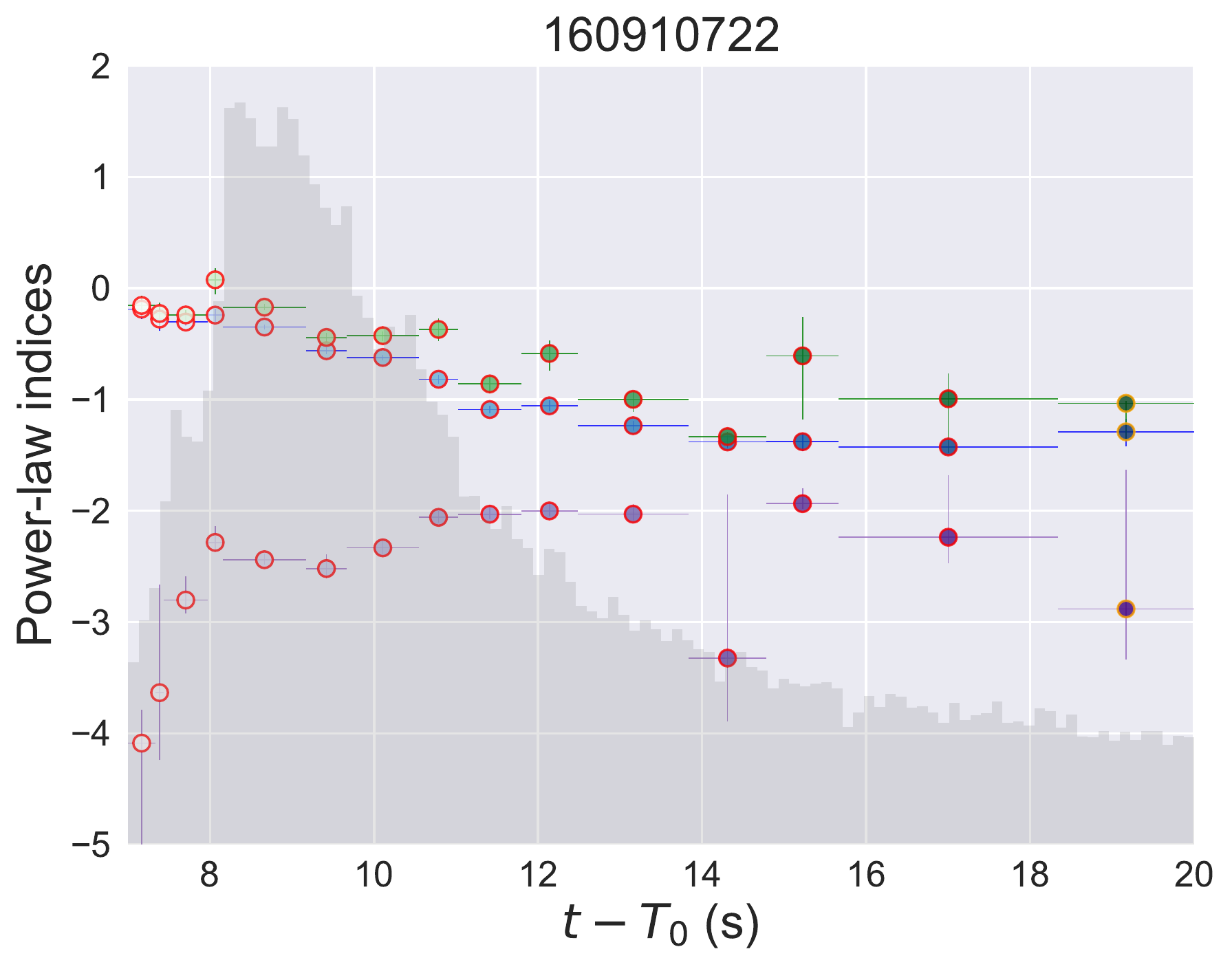}}
\subfigure{\includegraphics[width=0.3\linewidth]{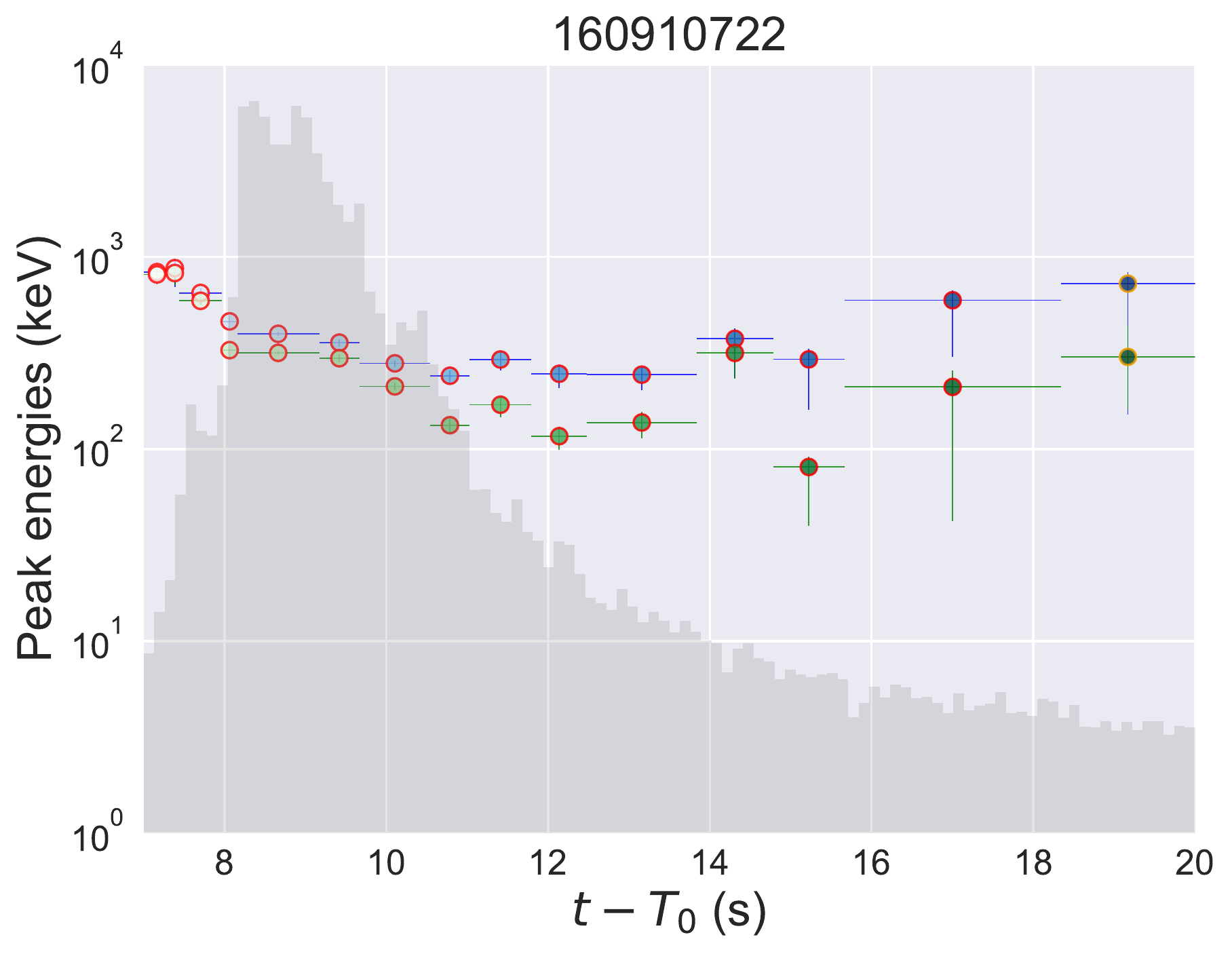}}
\subfigure{\includegraphics[width=0.3\linewidth]{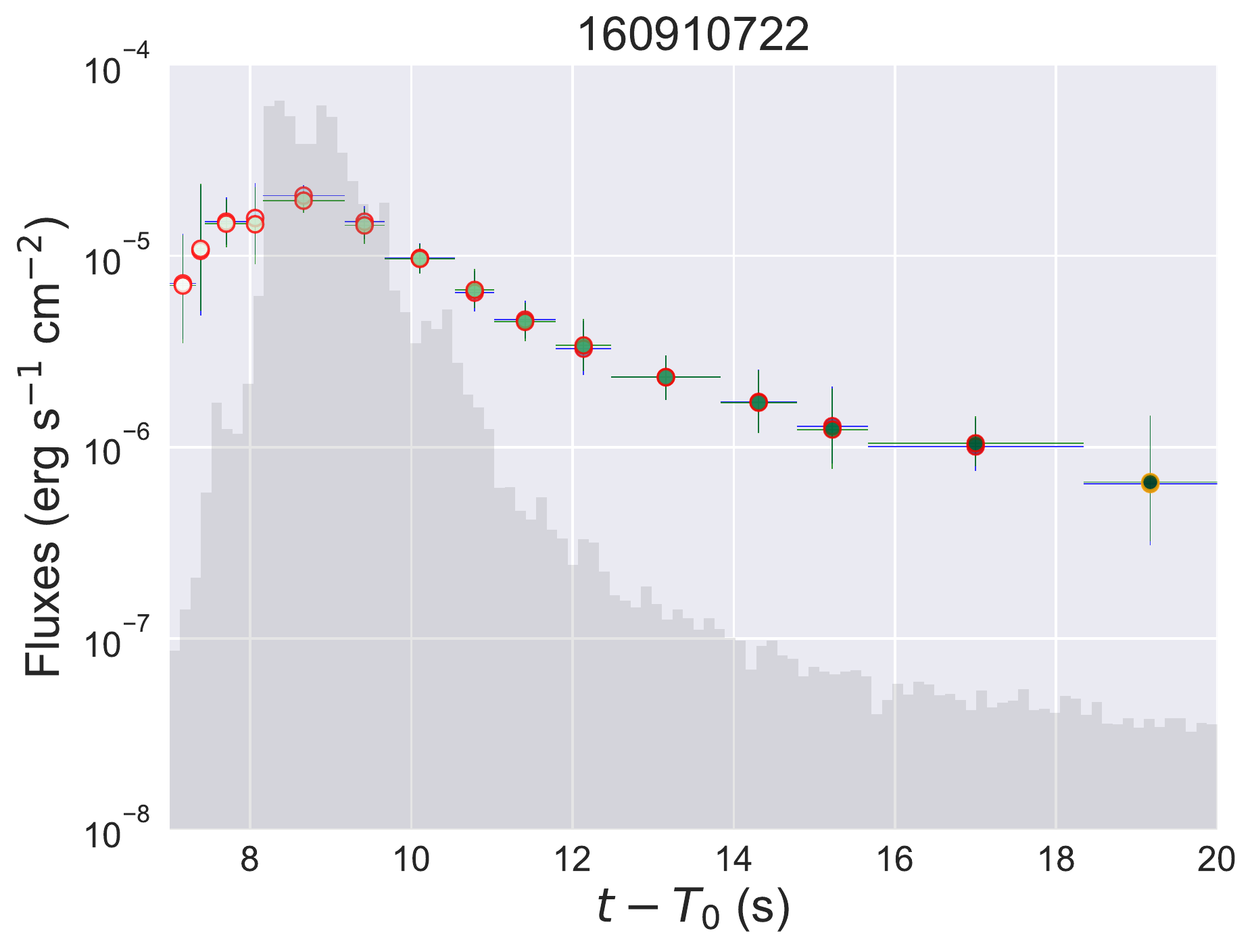}}

\caption{Same as Fig.~\ref{fig:evolution_group1}.
\label{fig:evolution_group9}}
\end{figure*}

\begin{figure*}
\centering

\subfigure{\includegraphics[width=0.3\linewidth]{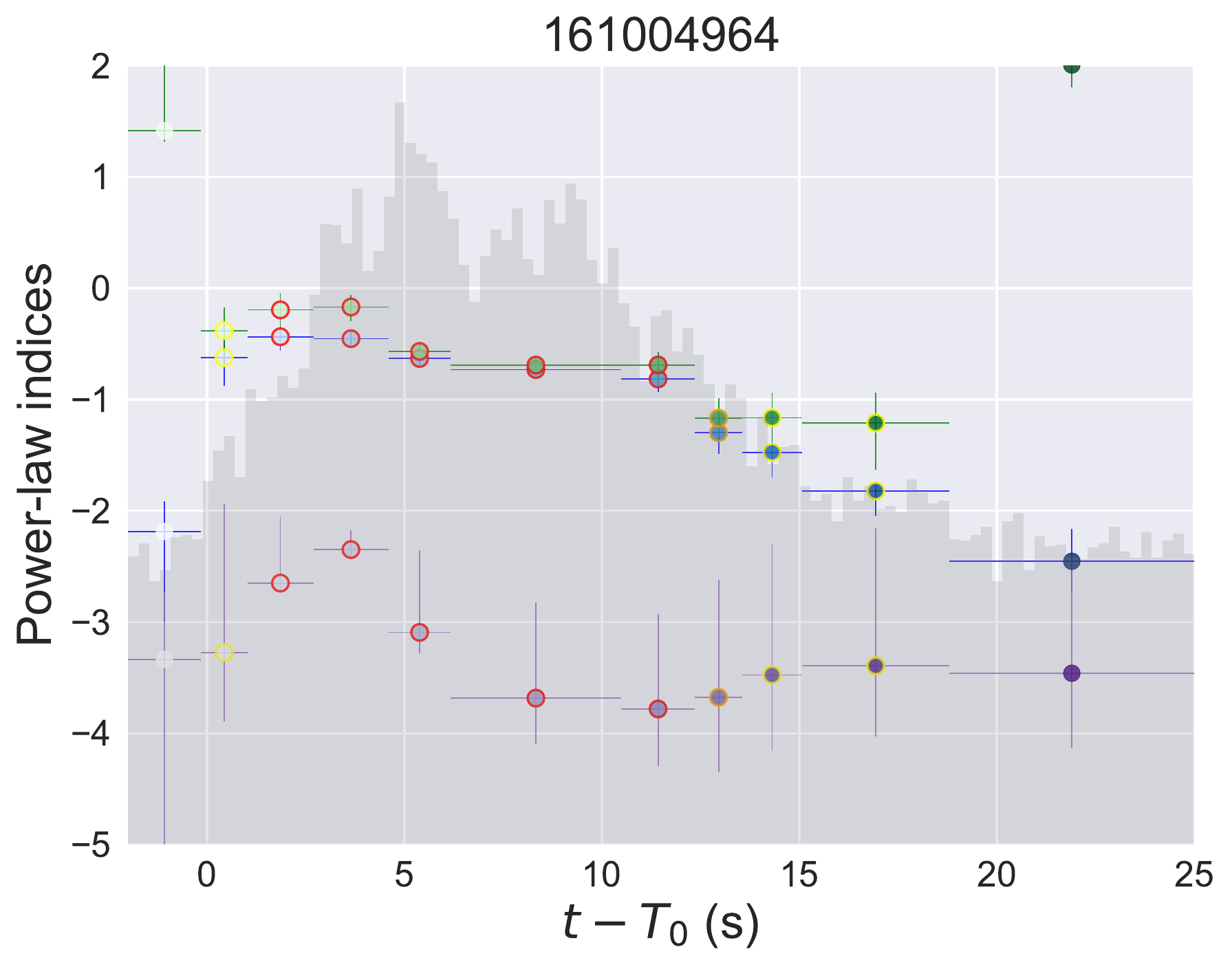}}
\subfigure{\includegraphics[width=0.3\linewidth]{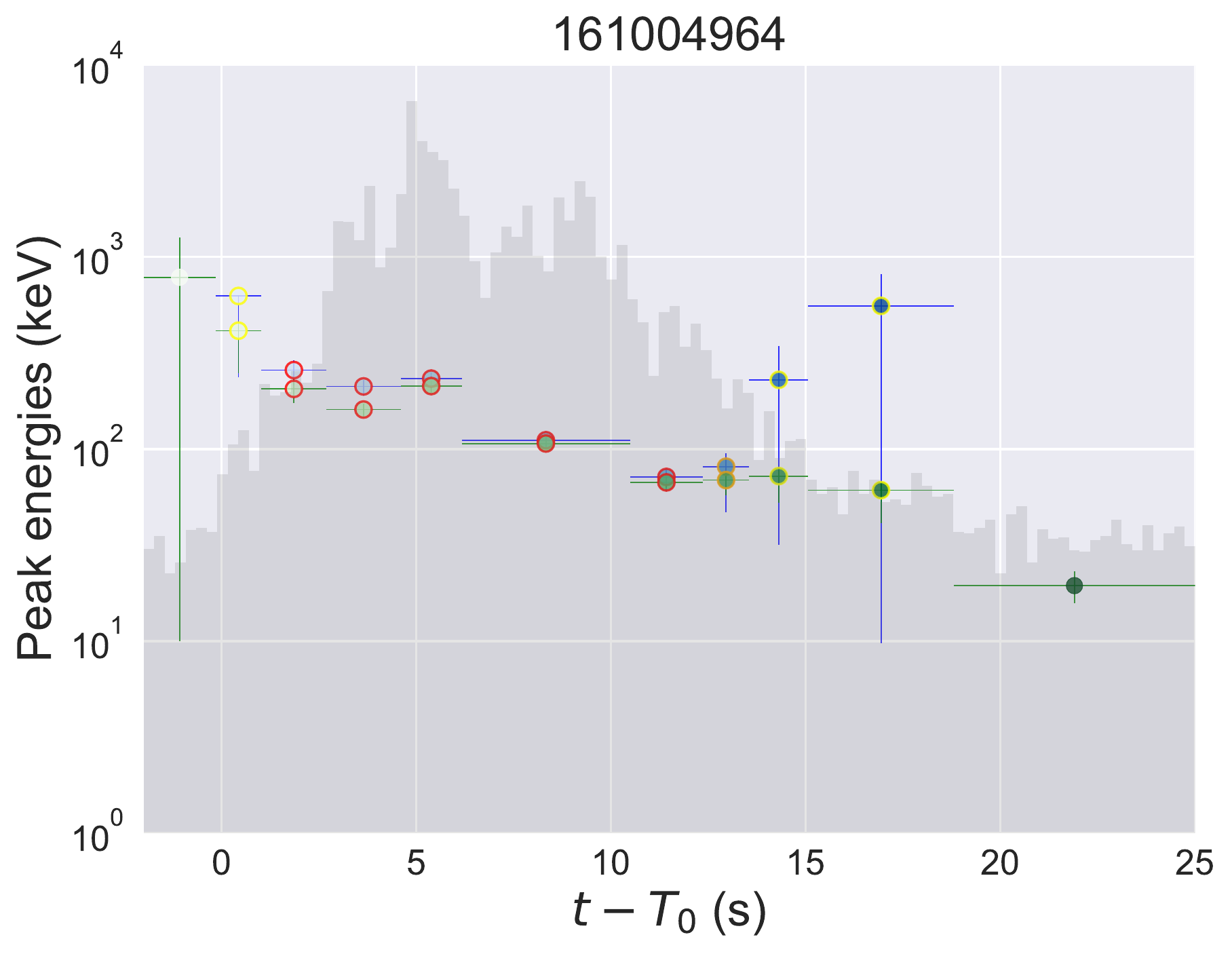}}
\subfigure{\includegraphics[width=0.3\linewidth]{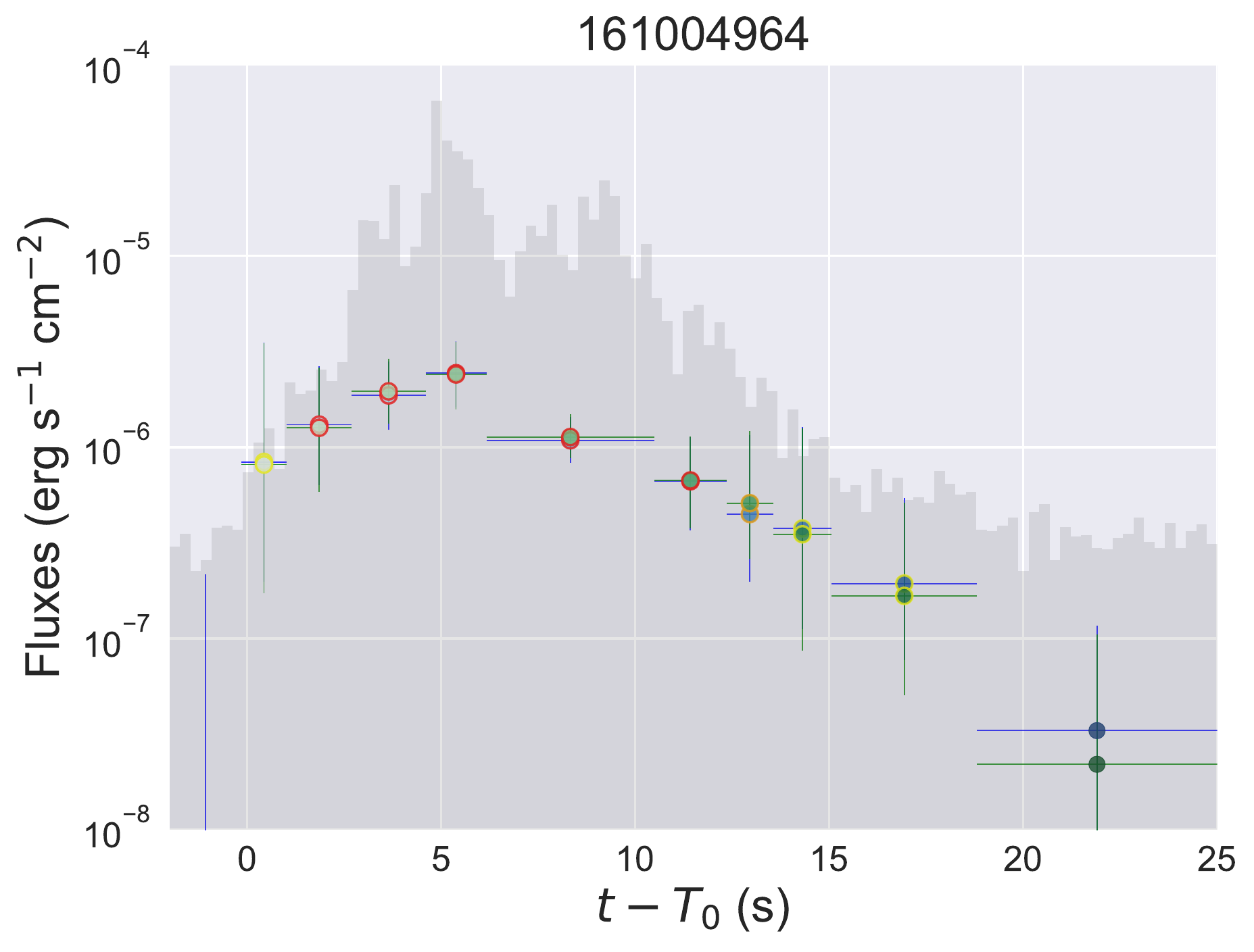}}

\subfigure{\includegraphics[width=0.3\linewidth]{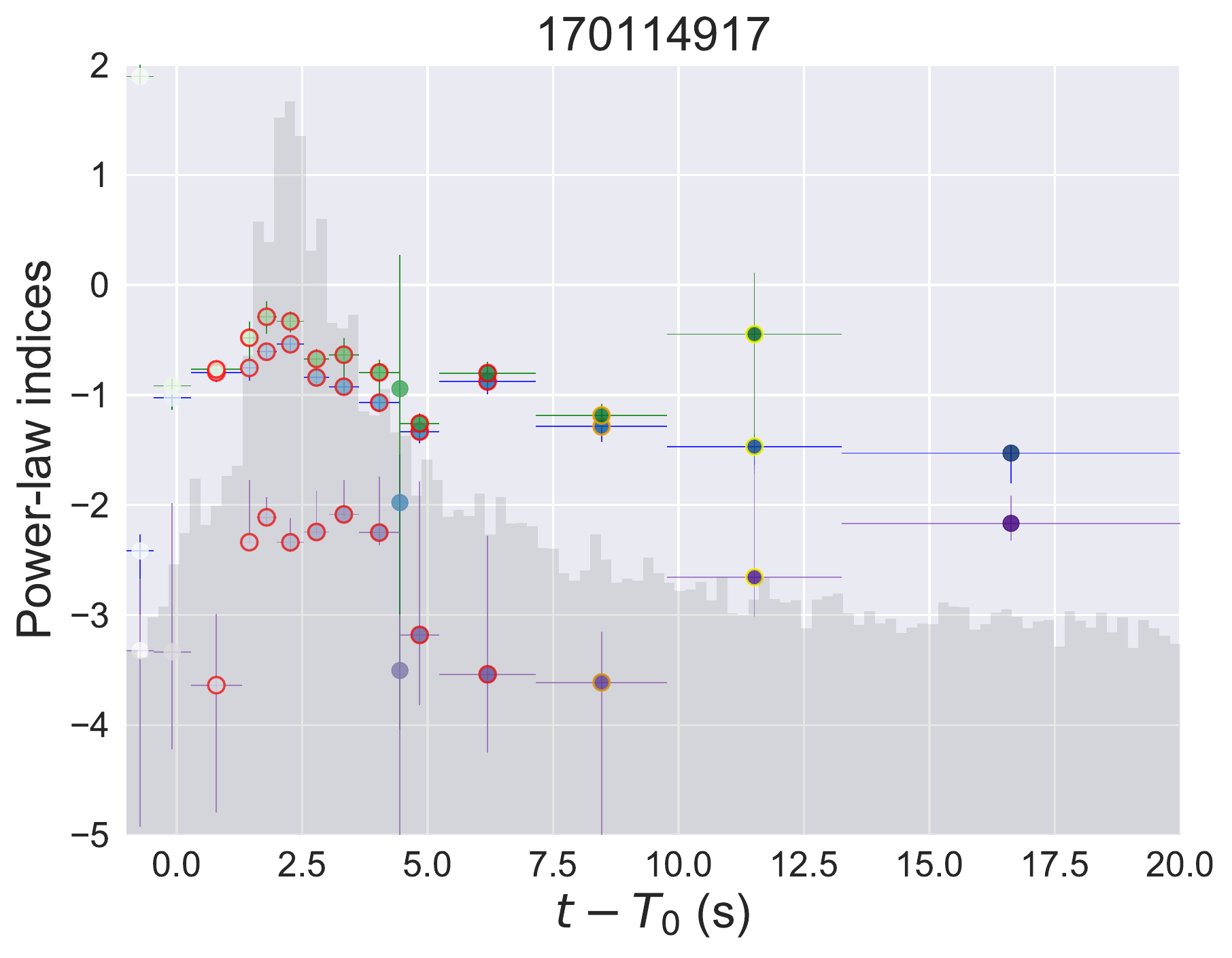}}
\subfigure{\includegraphics[width=0.3\linewidth]{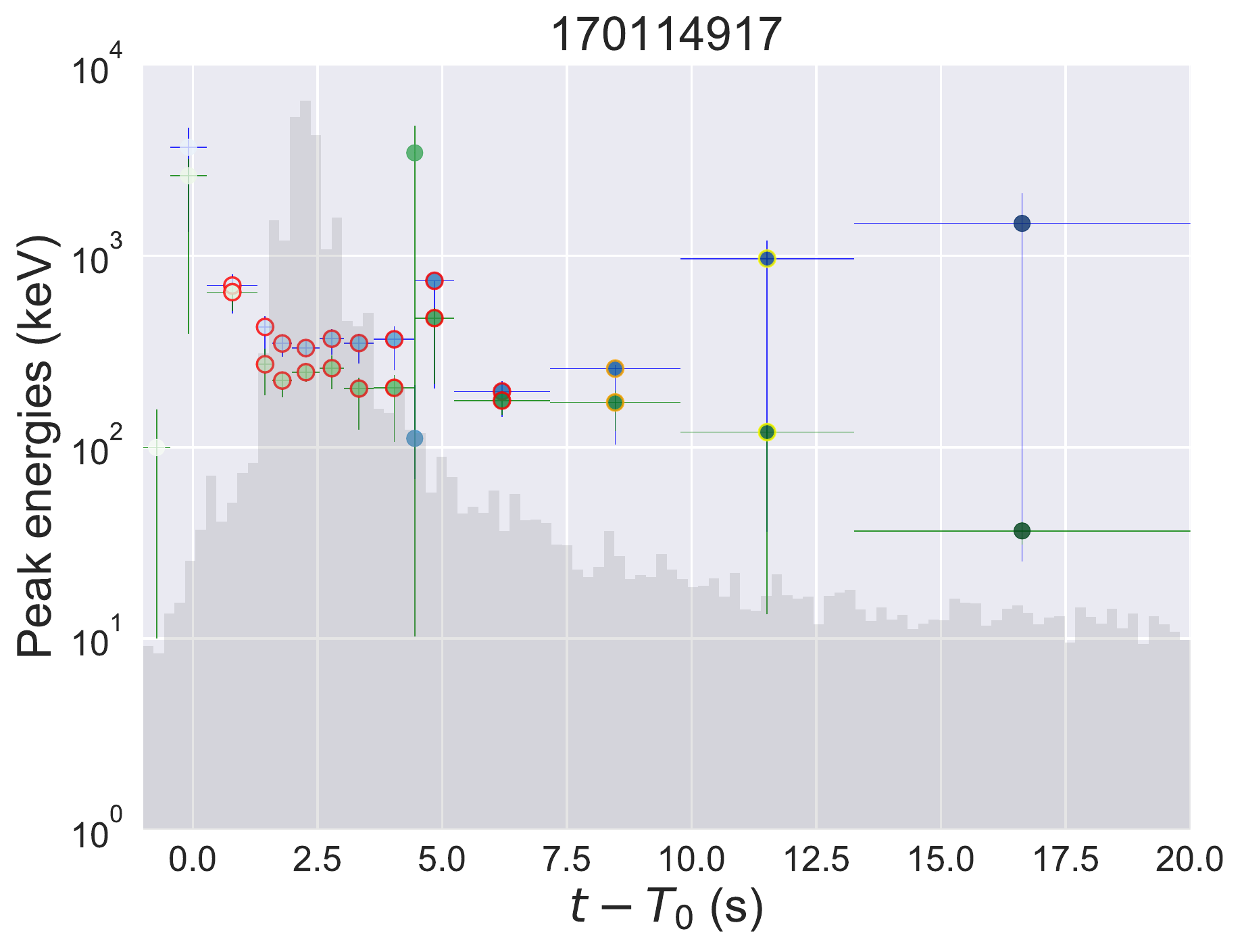}}
\subfigure{\includegraphics[width=0.3\linewidth]{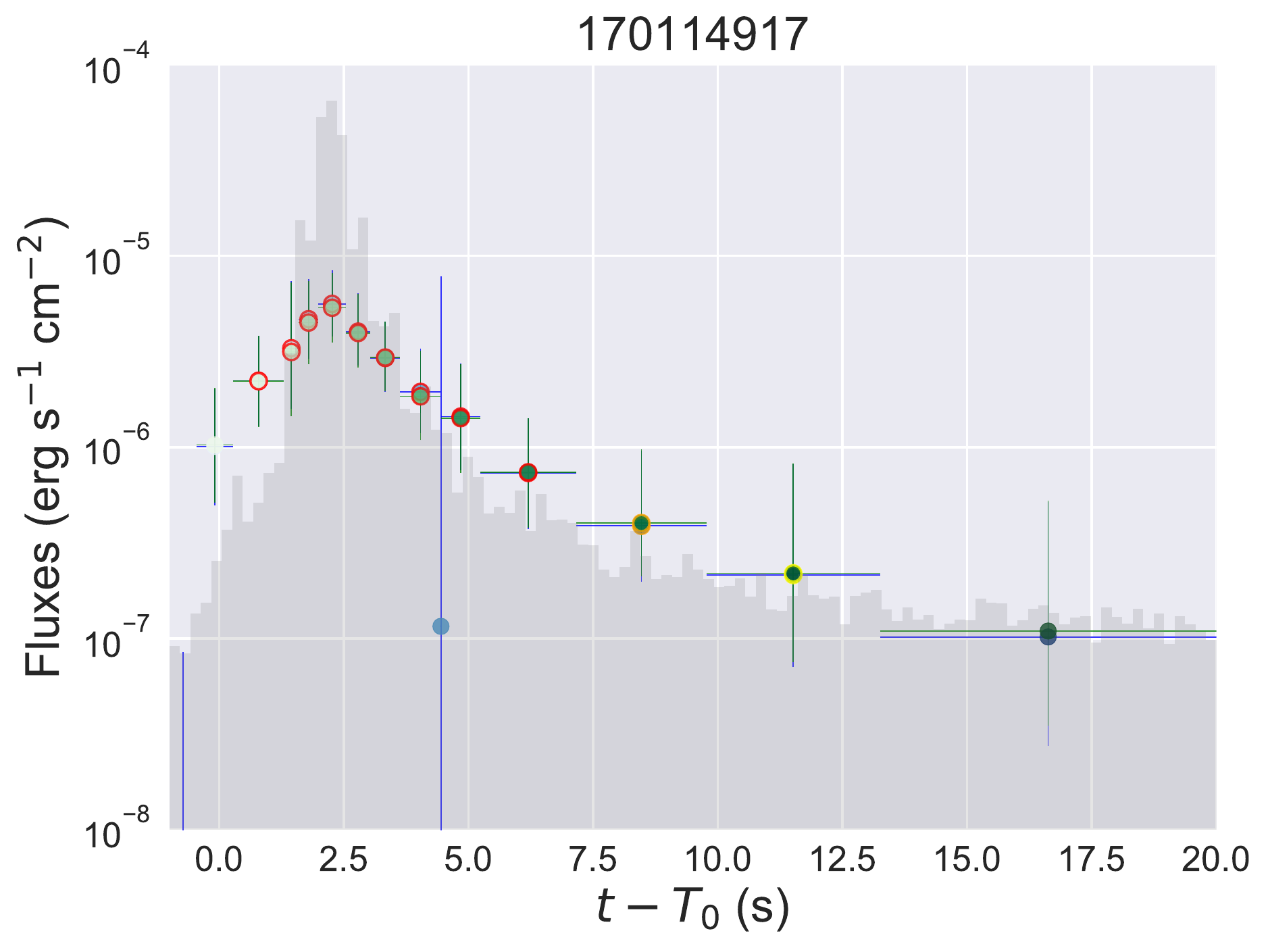}}

\caption{Same as Fig.~\ref{fig:evolution_group1}.
\label{fig:evolution_group10}}
\end{figure*}

\section{Plots of the correlations}
\label{app:correlationplots}

\begin{figure*}
\centering

\subfigure{\includegraphics[width=0.3\linewidth]{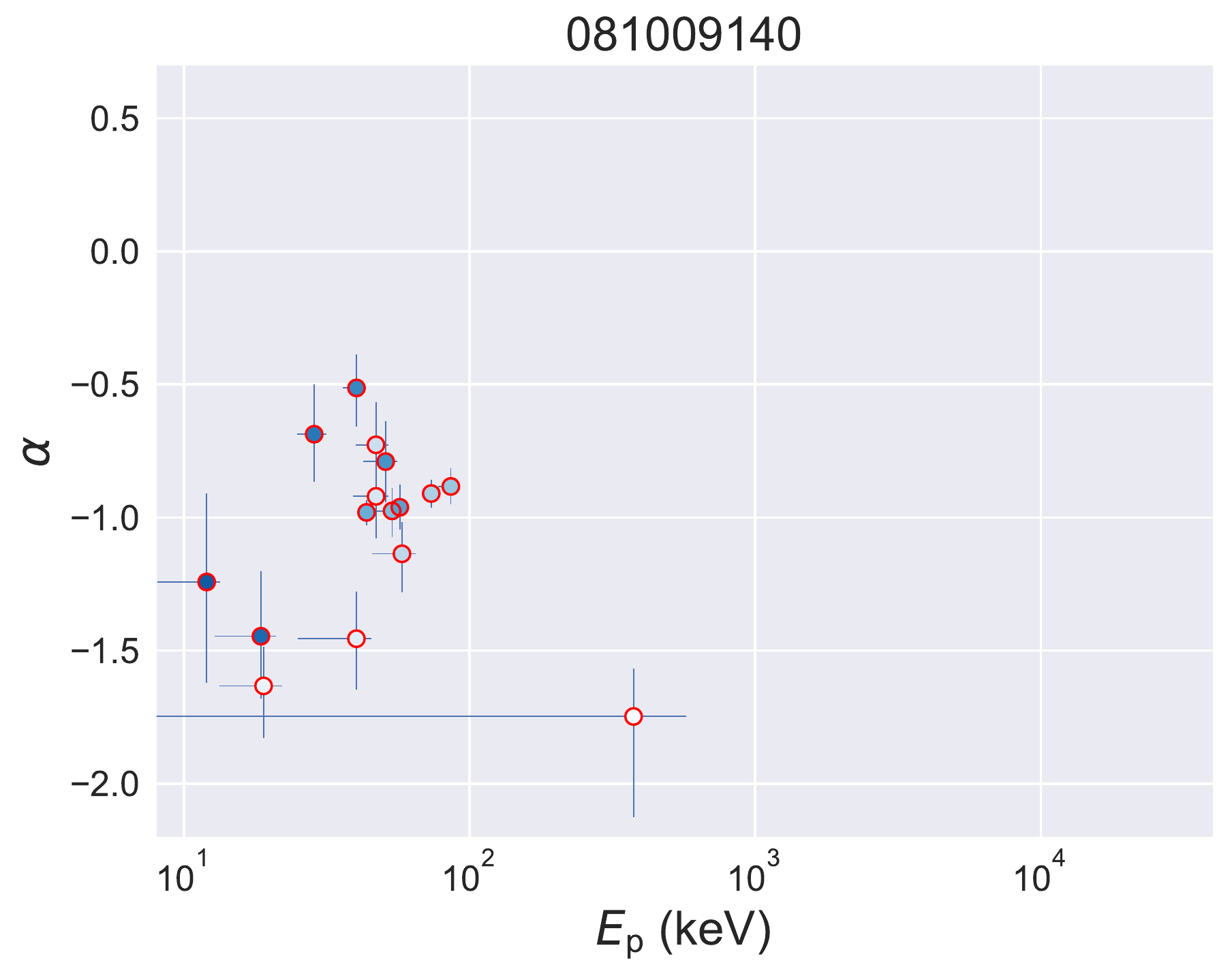}}
\subfigure{\includegraphics[width=0.3\linewidth]{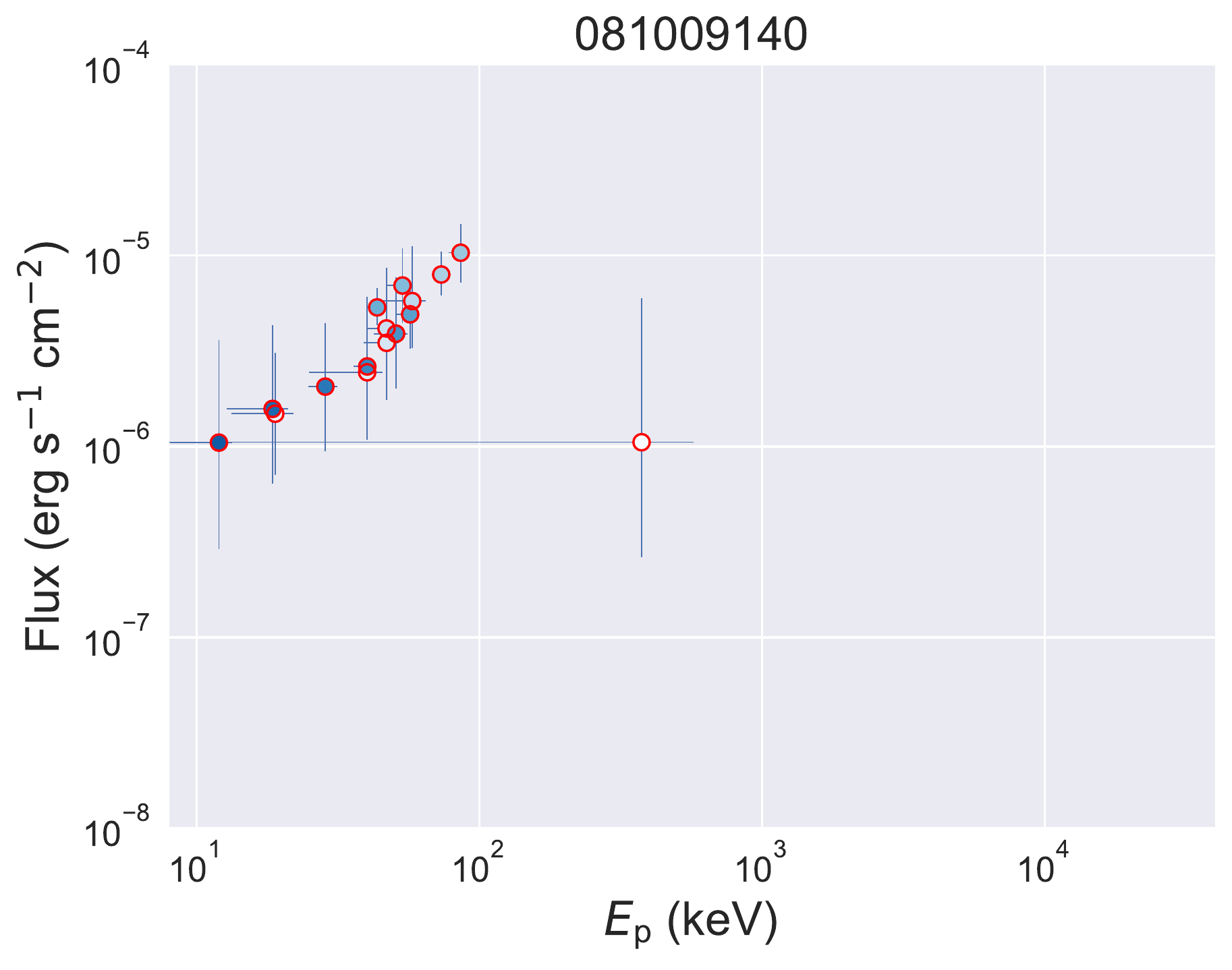}}
\subfigure{\includegraphics[width=0.3\linewidth]{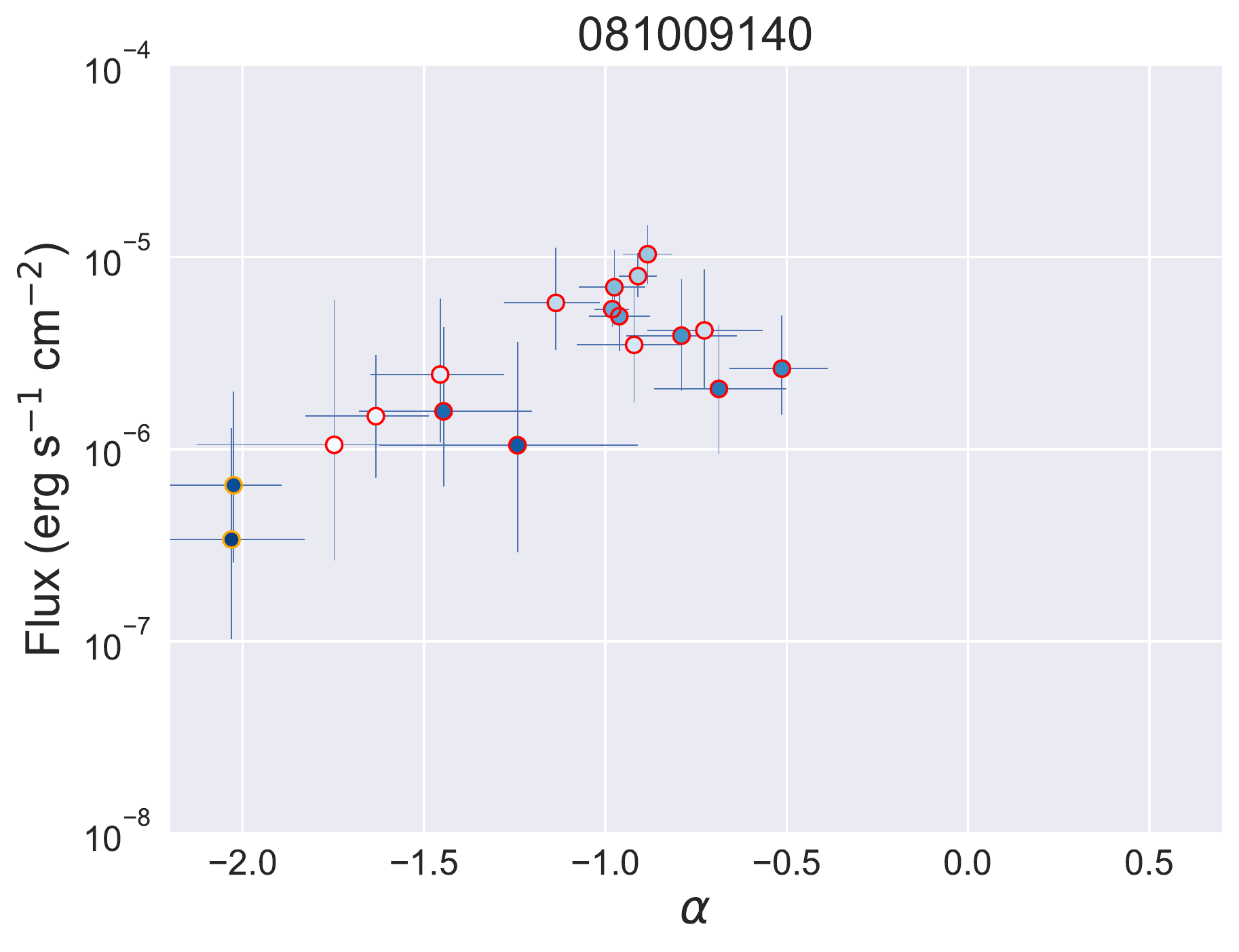}}

\subfigure{\includegraphics[width=0.3\linewidth]{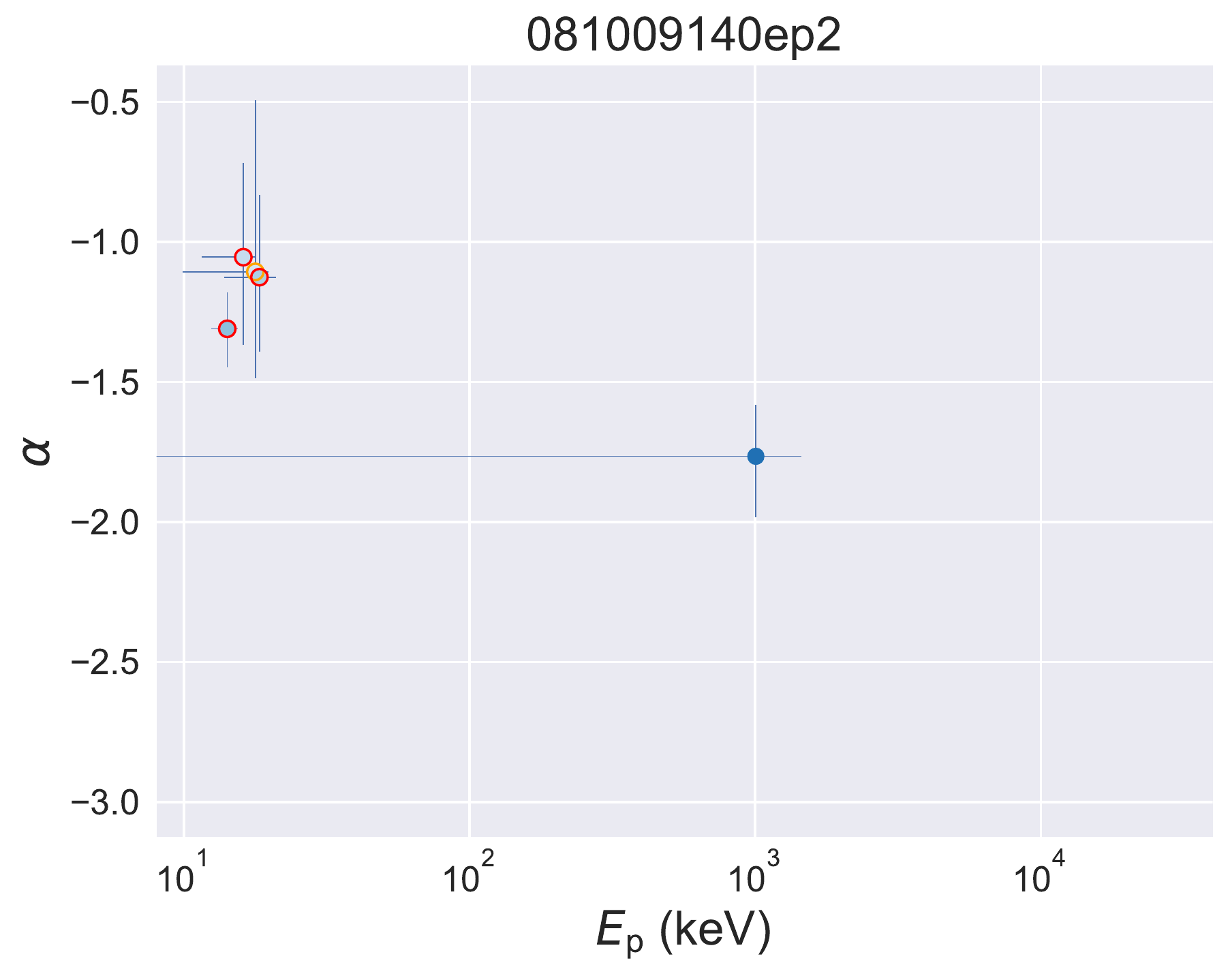}}
\subfigure{\includegraphics[width=0.3\linewidth]{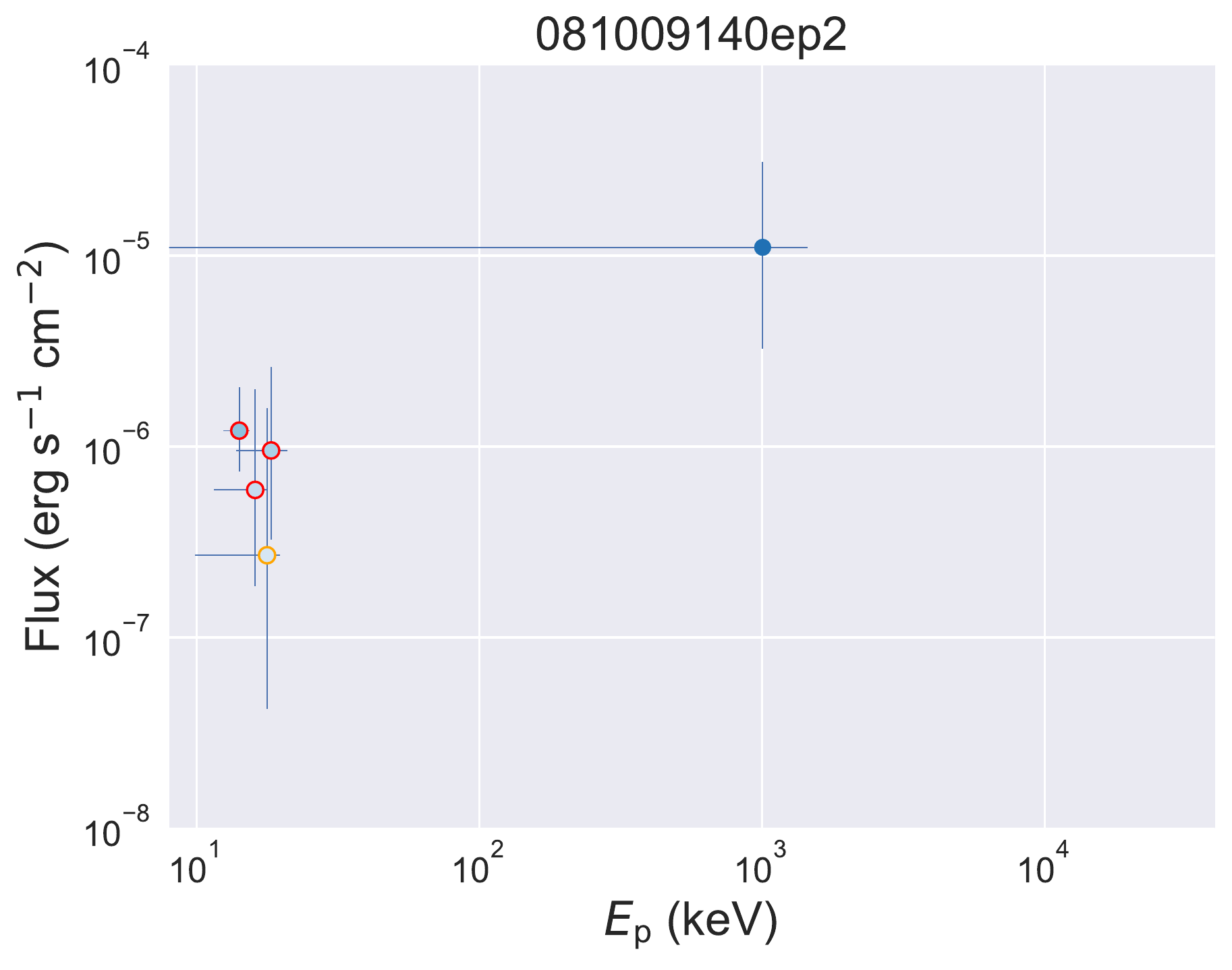}}
\subfigure{\includegraphics[width=0.3\linewidth]{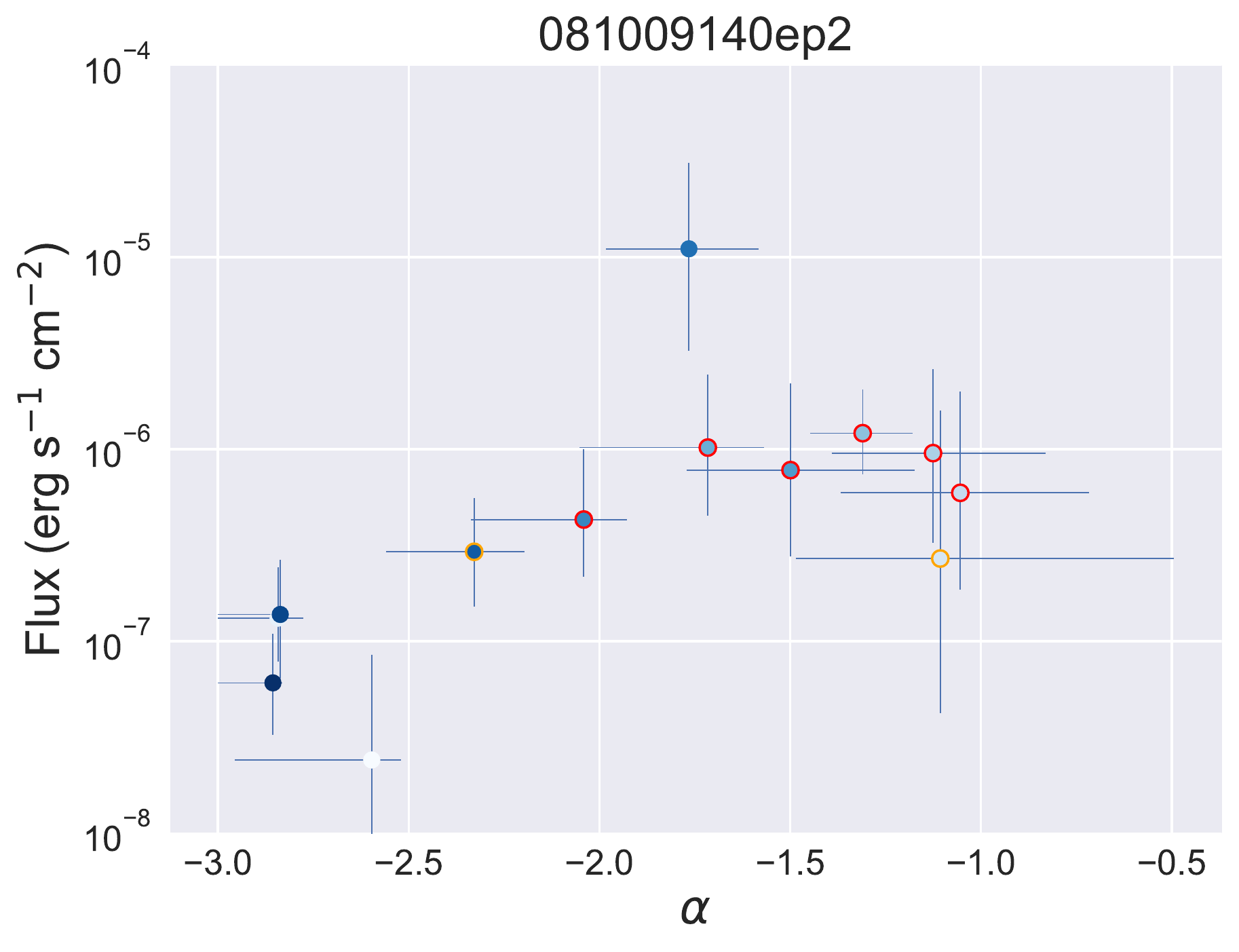}}

\subfigure{\includegraphics[width=0.3\linewidth]{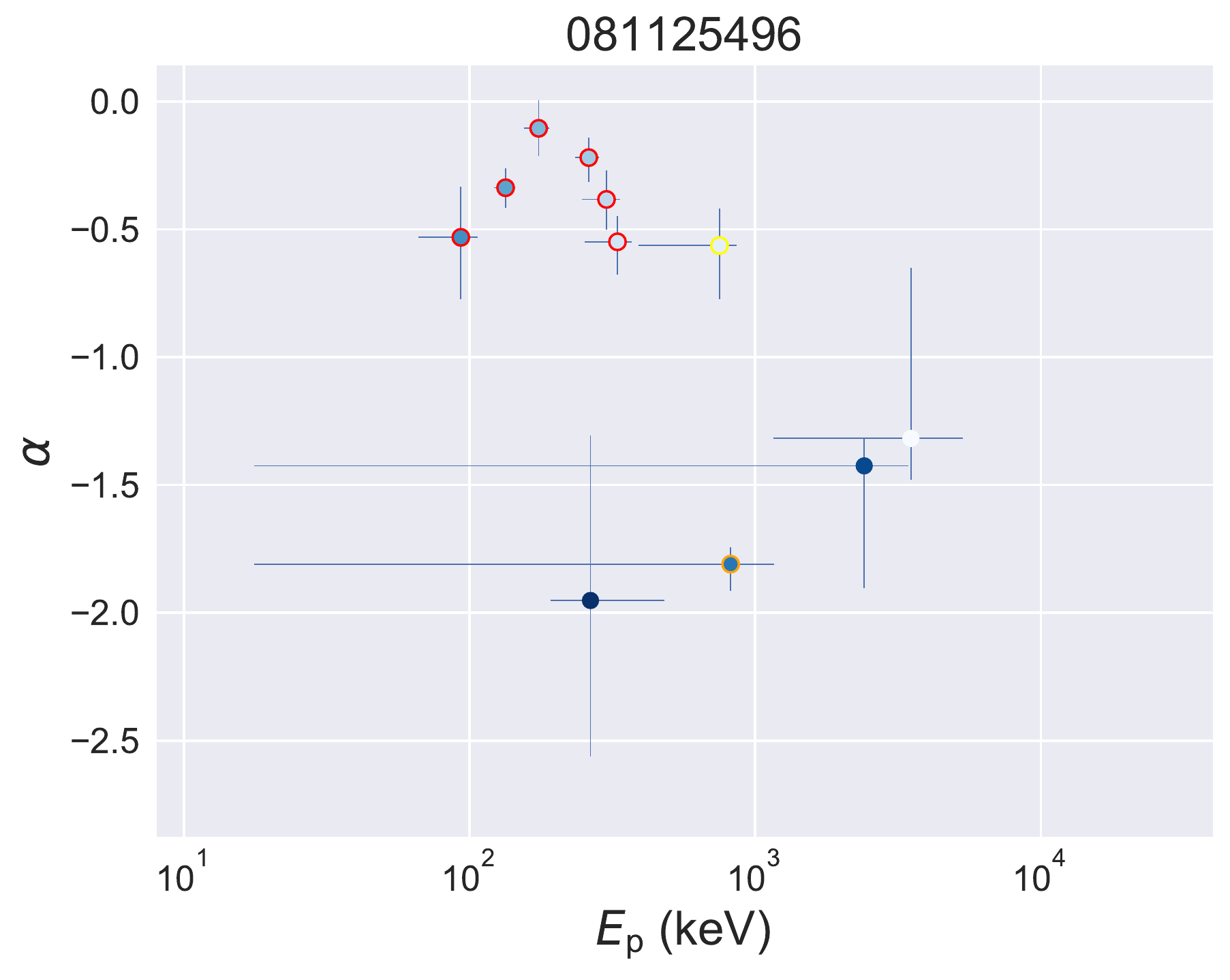}}
\subfigure{\includegraphics[width=0.3\linewidth]{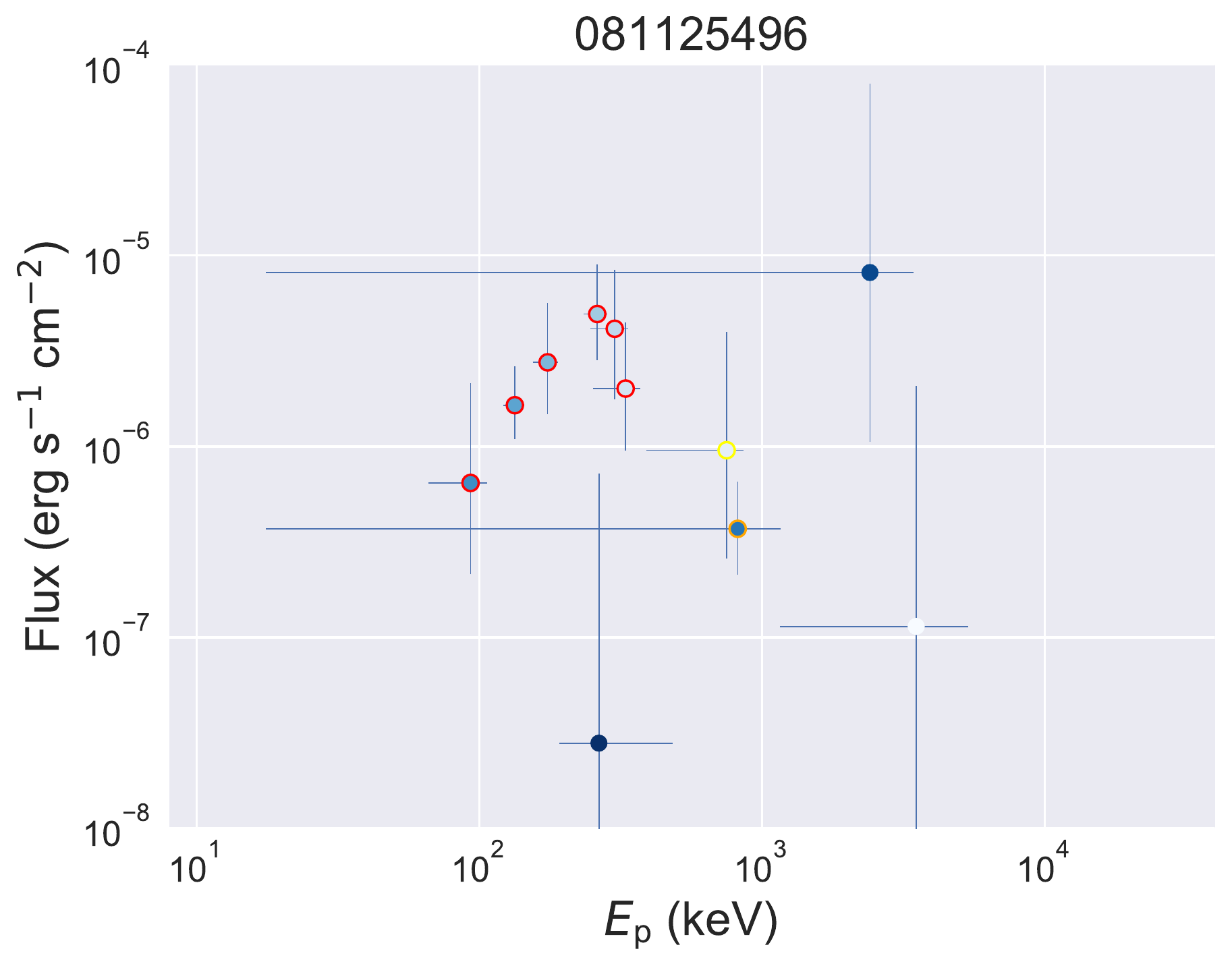}}
\subfigure{\includegraphics[width=0.3\linewidth]{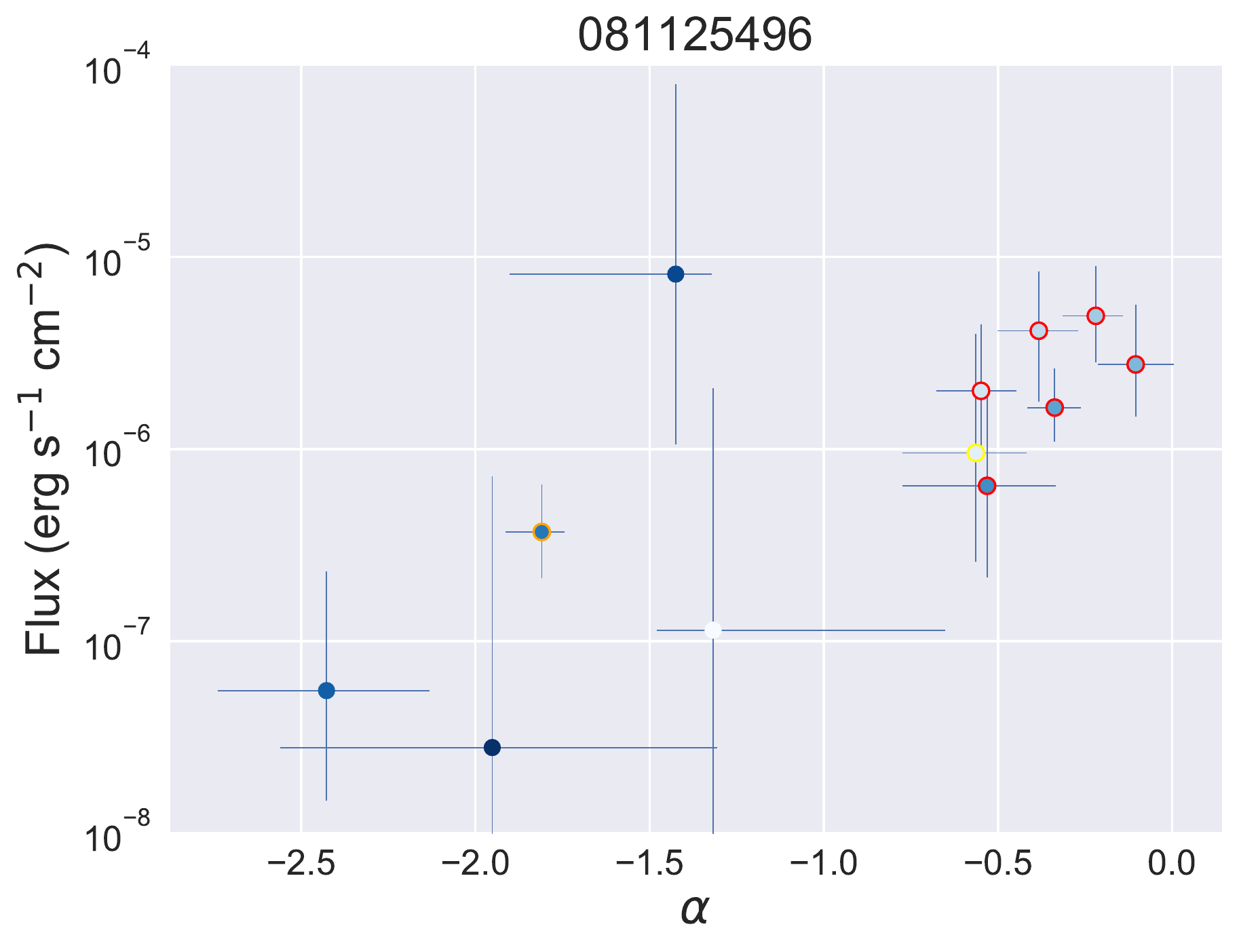}}

\subfigure{\includegraphics[width=0.3\linewidth]{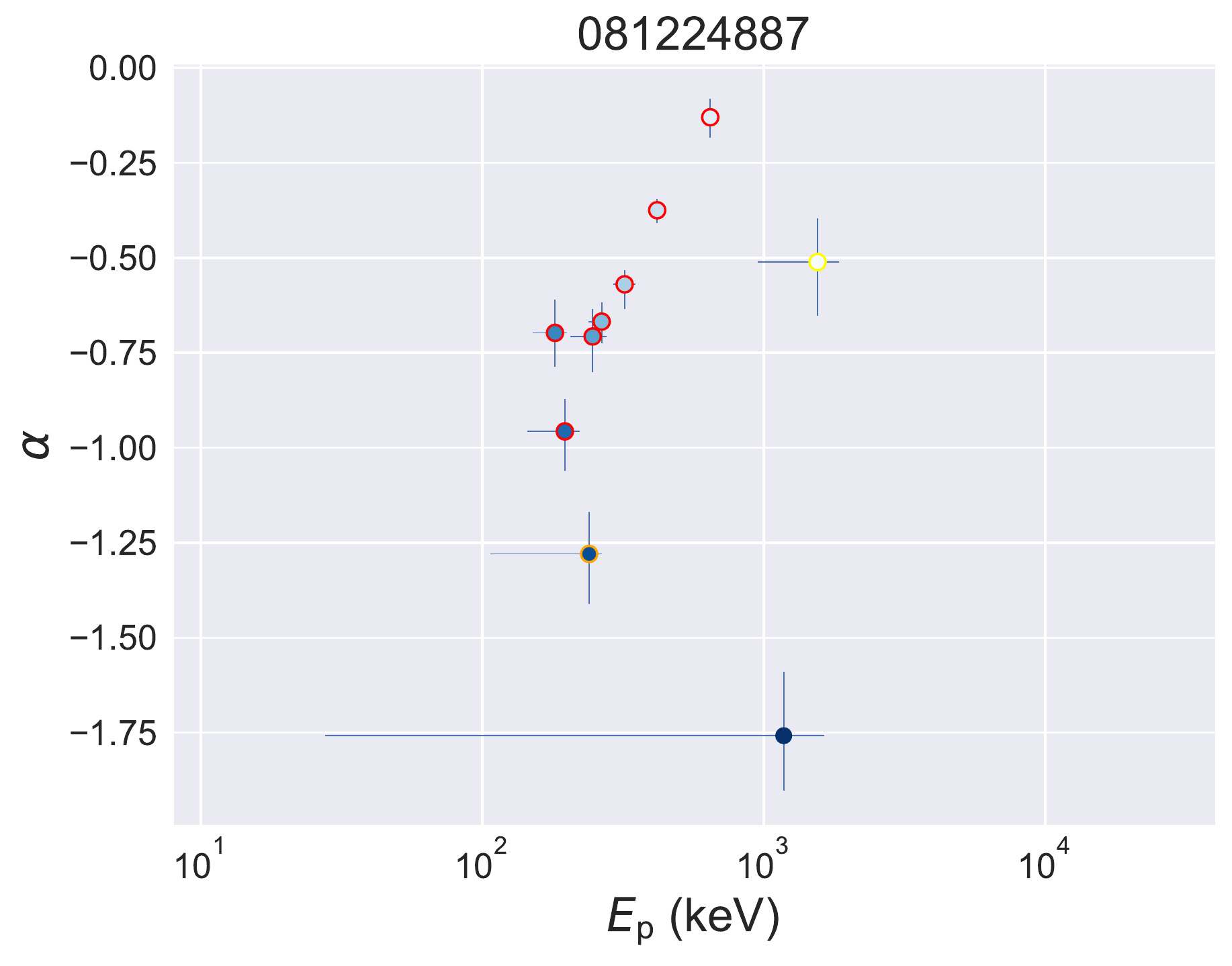}}
\subfigure{\includegraphics[width=0.3\linewidth]{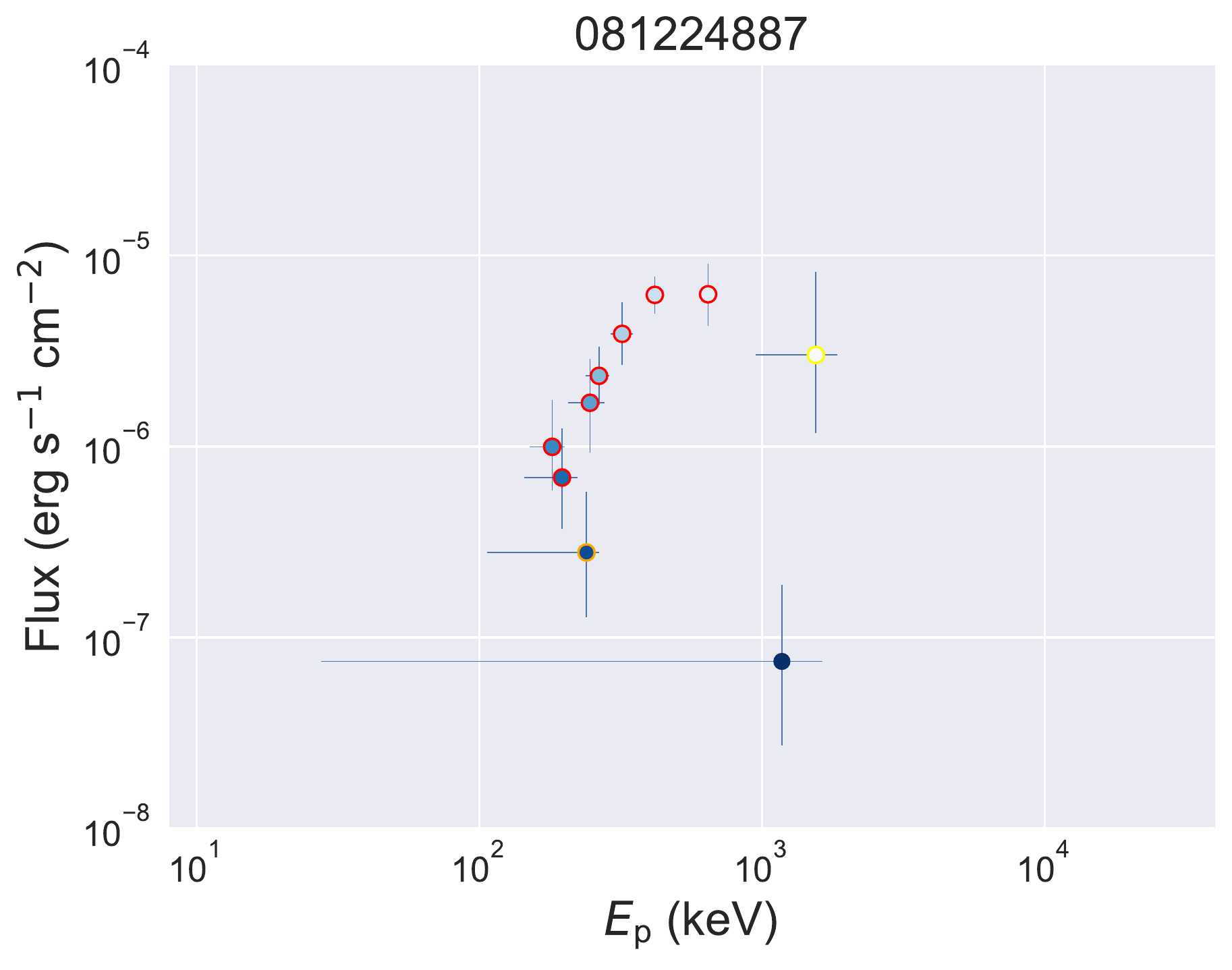}}
\subfigure{\includegraphics[width=0.3\linewidth]{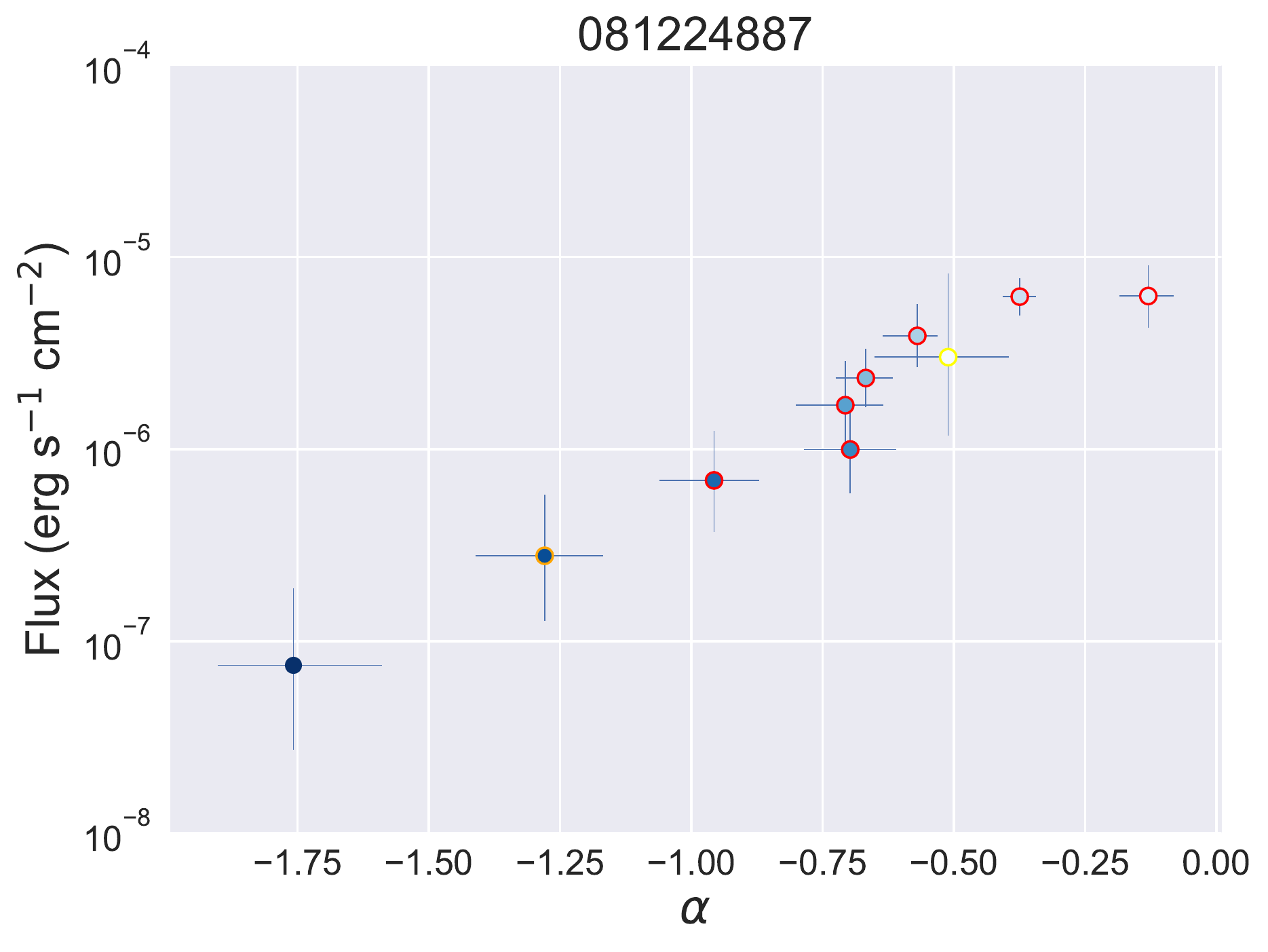}}

\caption{Left panels: relation of $E_{\rm p}$-$\alpha$. Middle panels: relation of $F$-$E_{\rm p}$. Right panels: relation of $F$-$\alpha$. Data points with red, orange, yellow, and no circles indicate statistical significance $S\geq20$, $20>S\geq15$, $15>S\geq10$, and $S<10$, respectively. Color scale from light blue (start) to deep blue (end) shows temporal evolution. Many of the low-significance data points are marginally or not constrained, as seen from the huge negative-side error bars.
\label{fig:correlation_group1}}
\end{figure*}

\begin{figure*}
\centering

\subfigure{\includegraphics[width=0.3\linewidth]{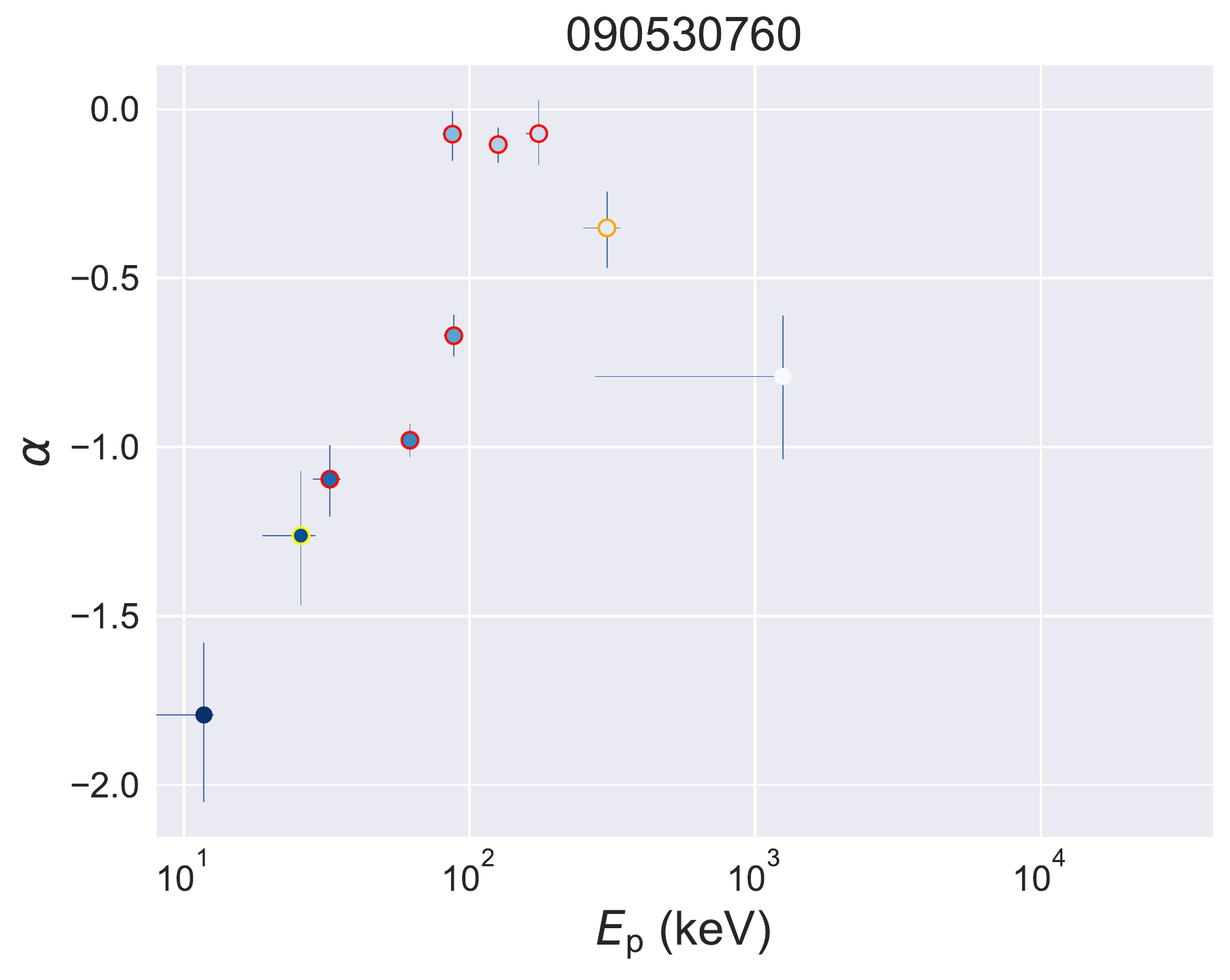}}
\subfigure{\includegraphics[width=0.3\linewidth]{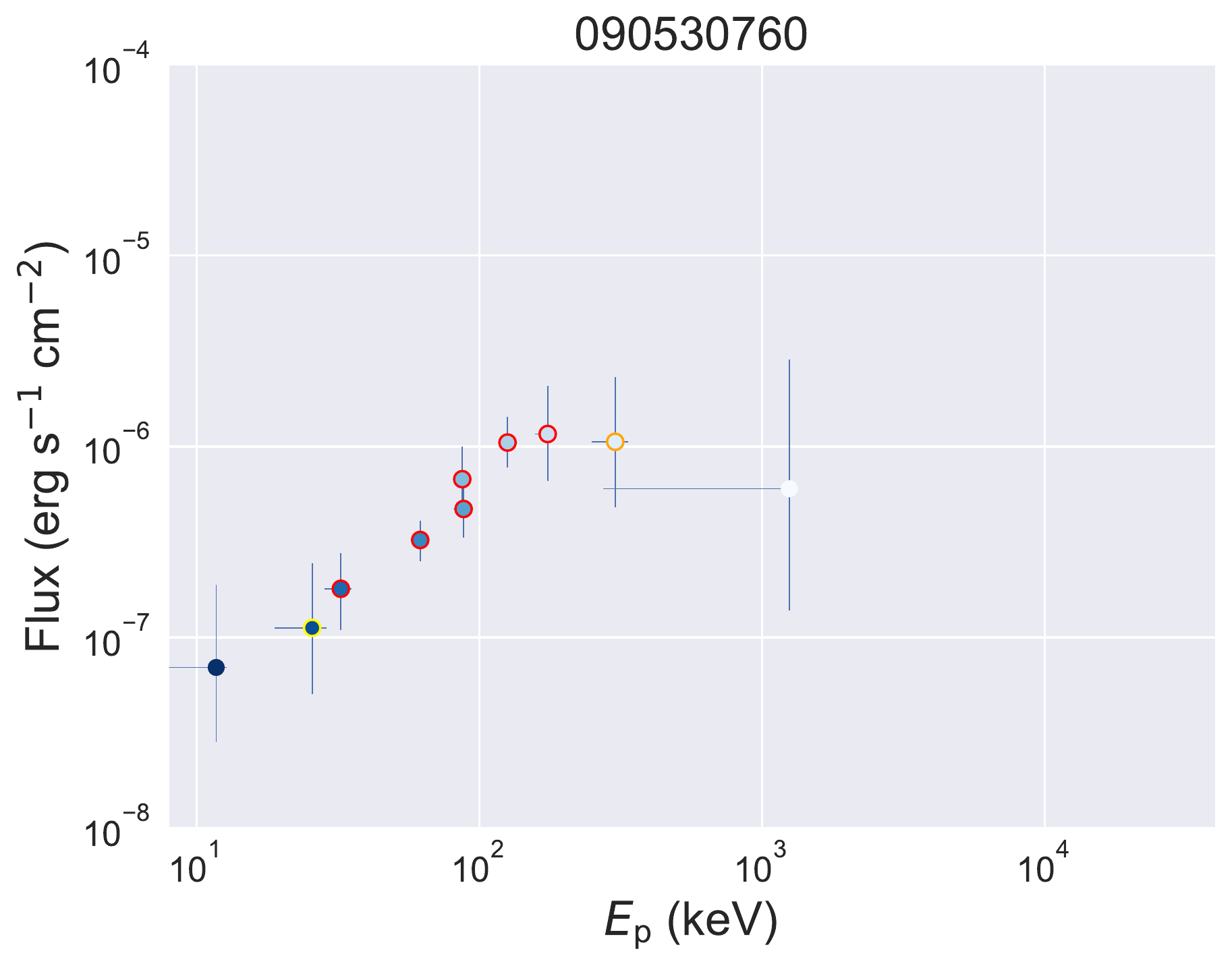}}
\subfigure{\includegraphics[width=0.3\linewidth]{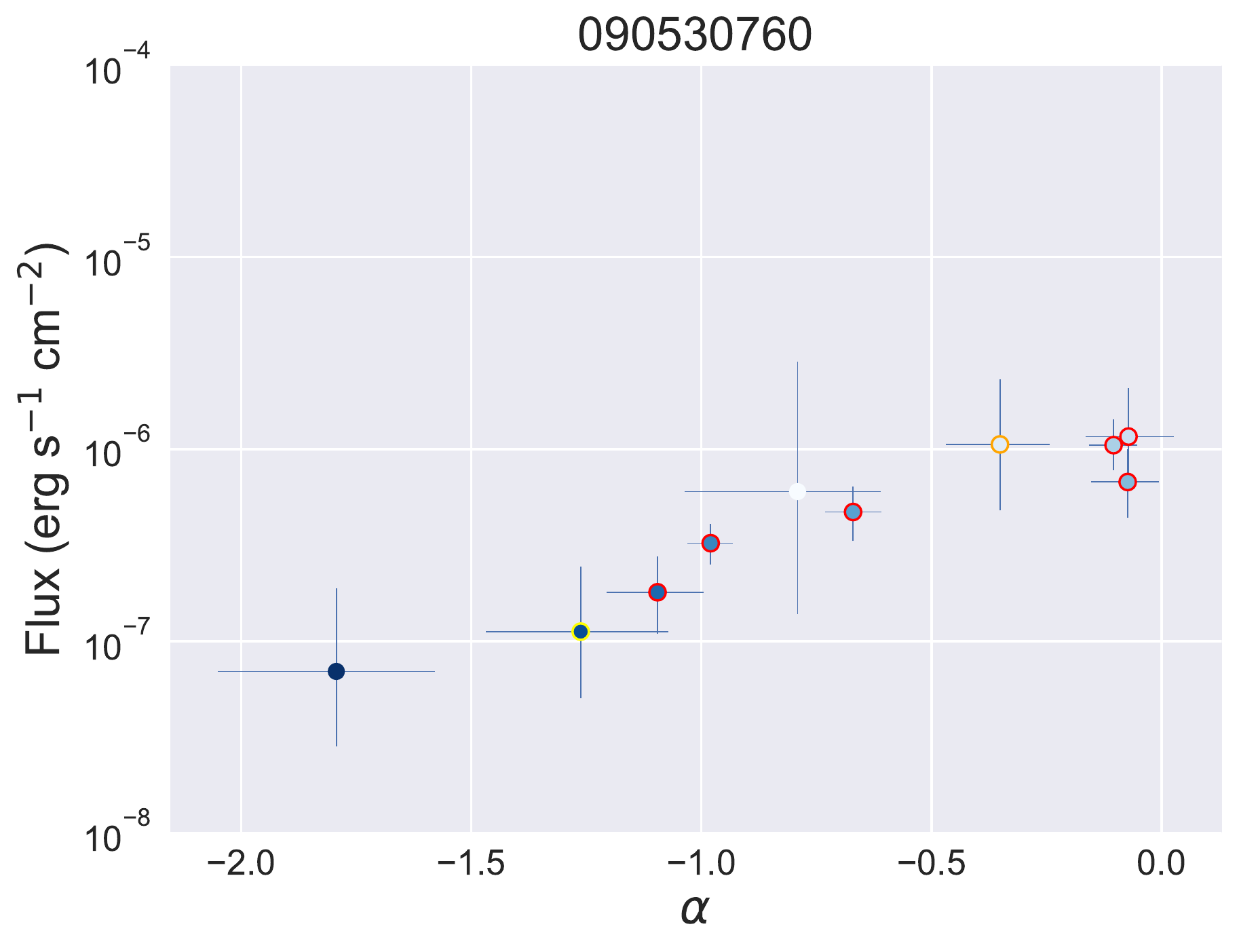}}

\subfigure{\includegraphics[width=0.3\linewidth]{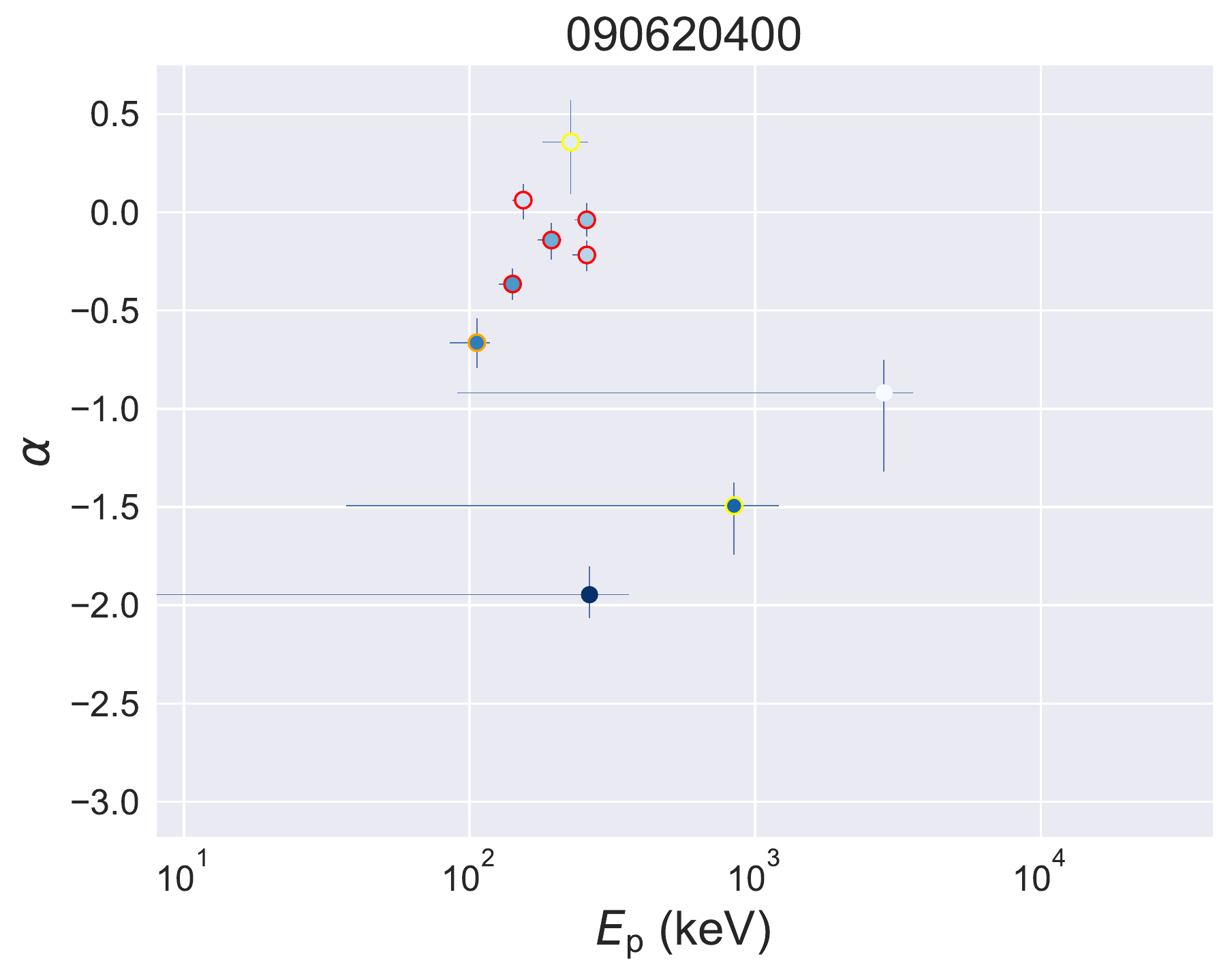}}
\subfigure{\includegraphics[width=0.3\linewidth]{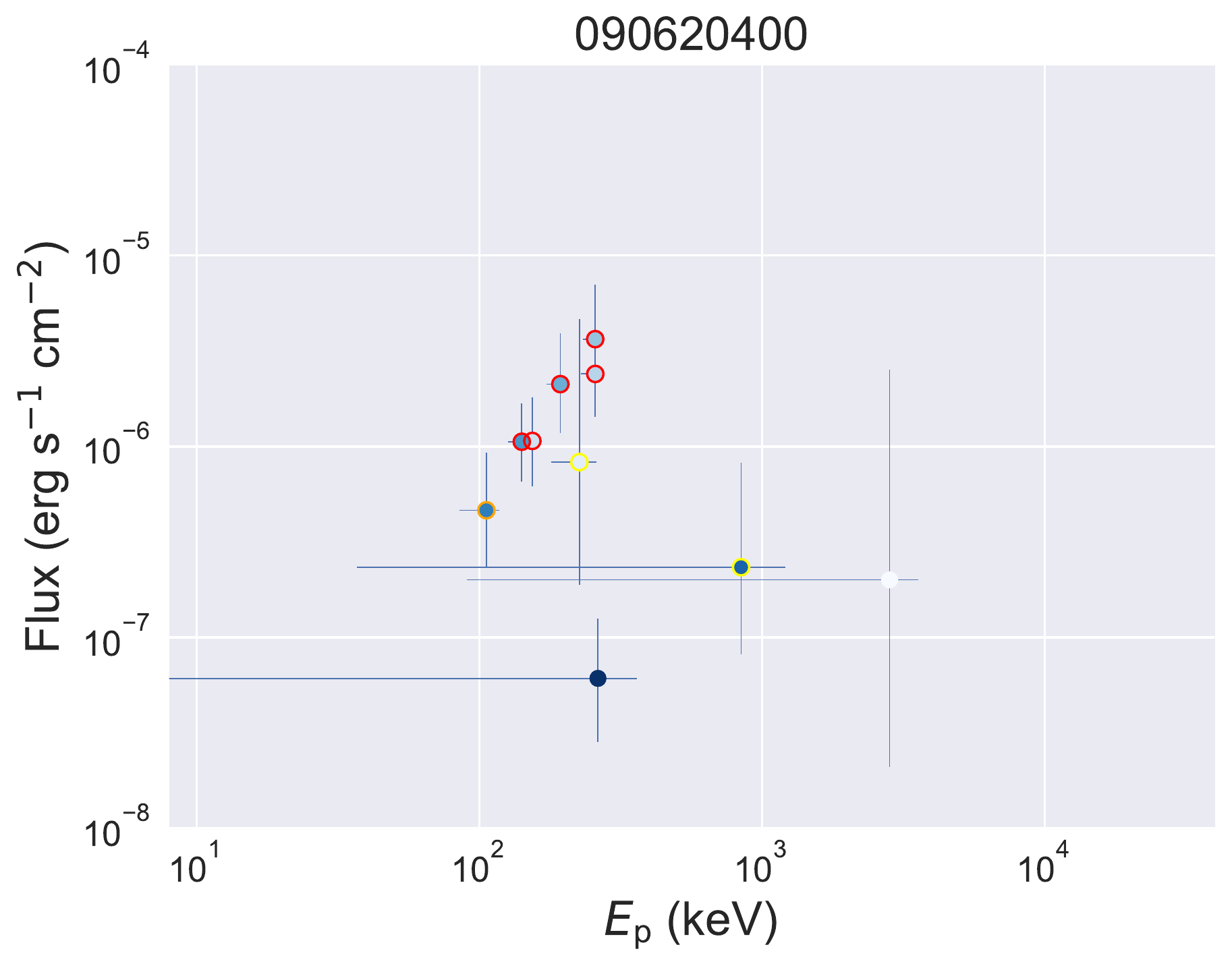}}
\subfigure{\includegraphics[width=0.3\linewidth]{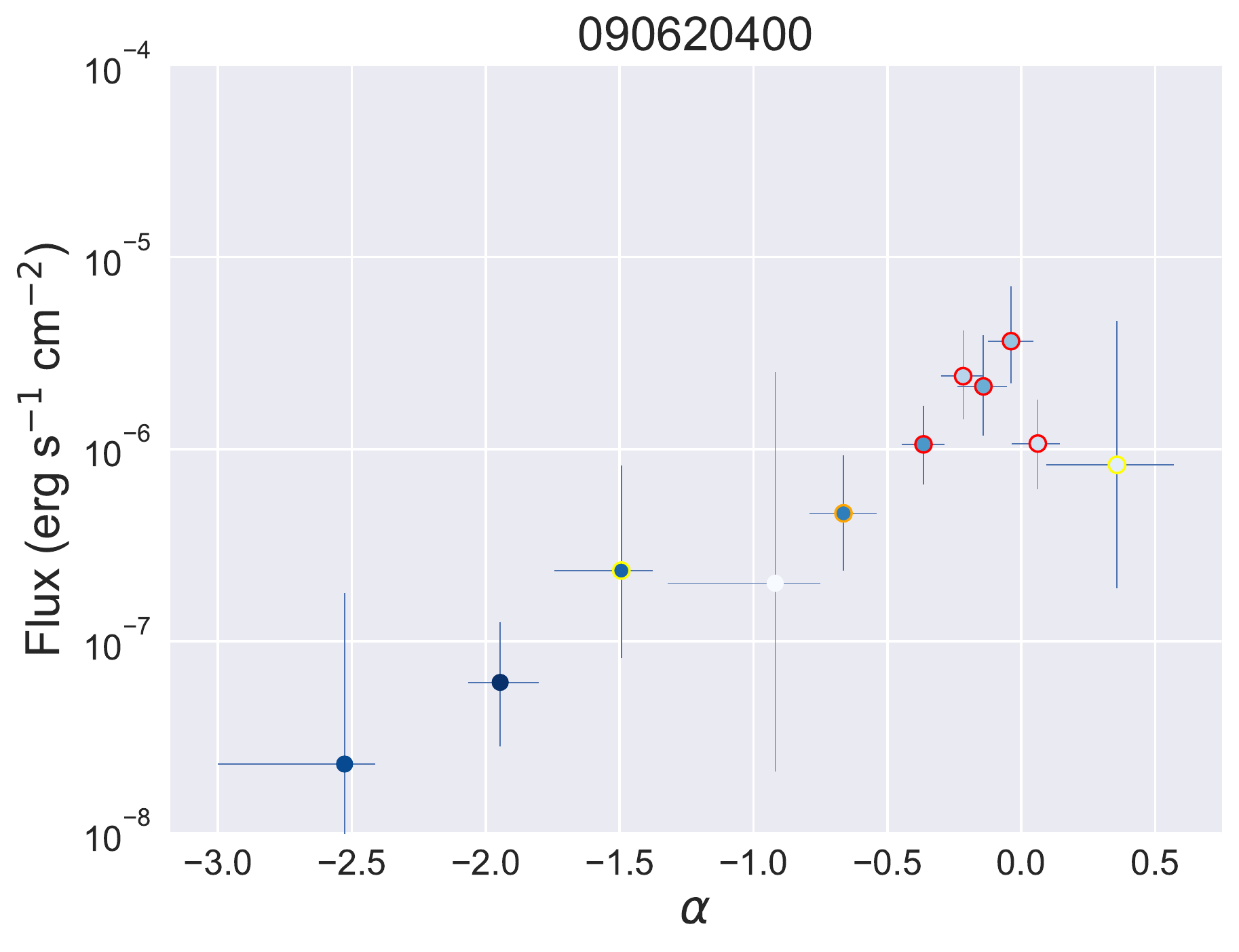}}

\subfigure{\includegraphics[width=0.3\linewidth]{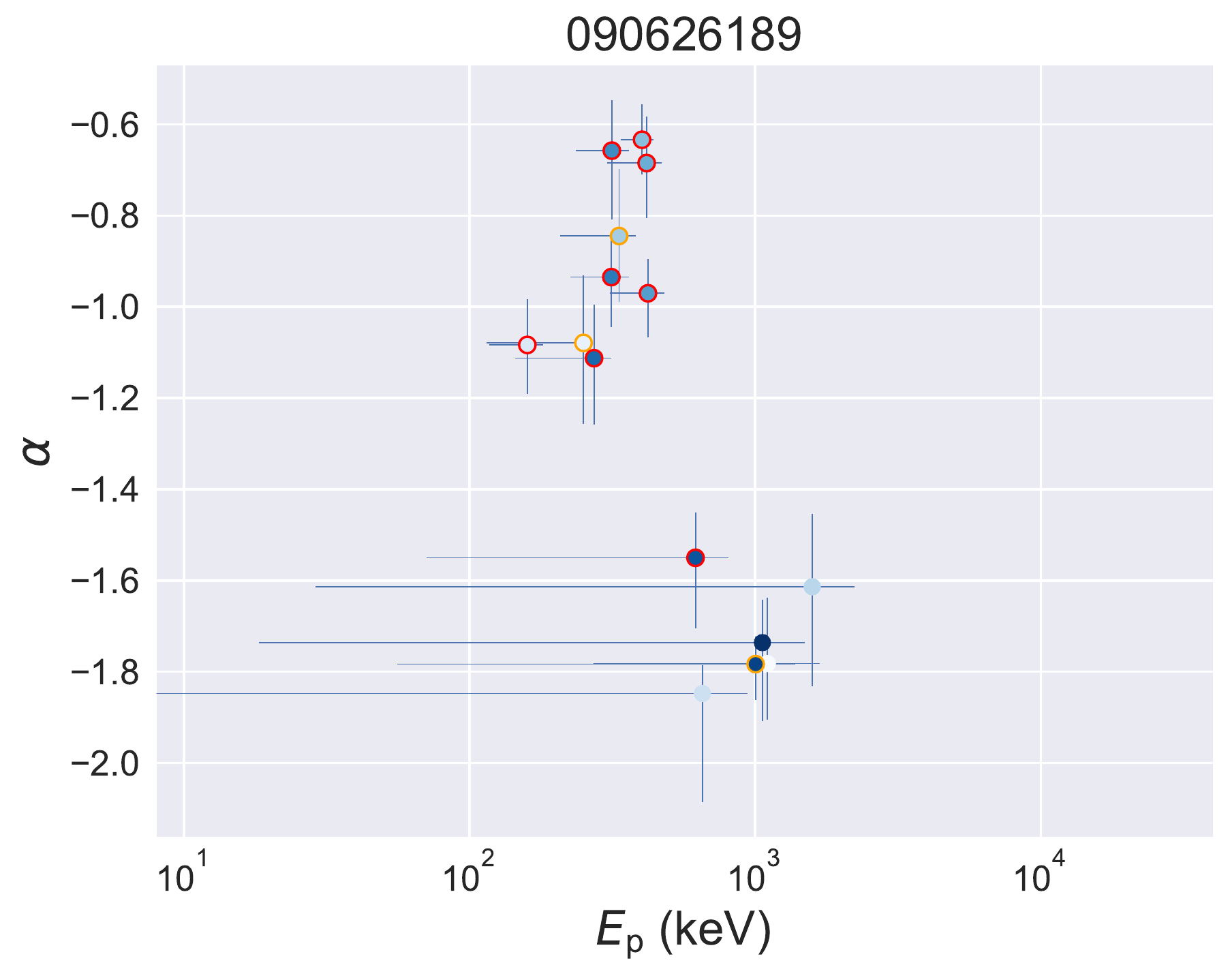}}
\subfigure{\includegraphics[width=0.3\linewidth]{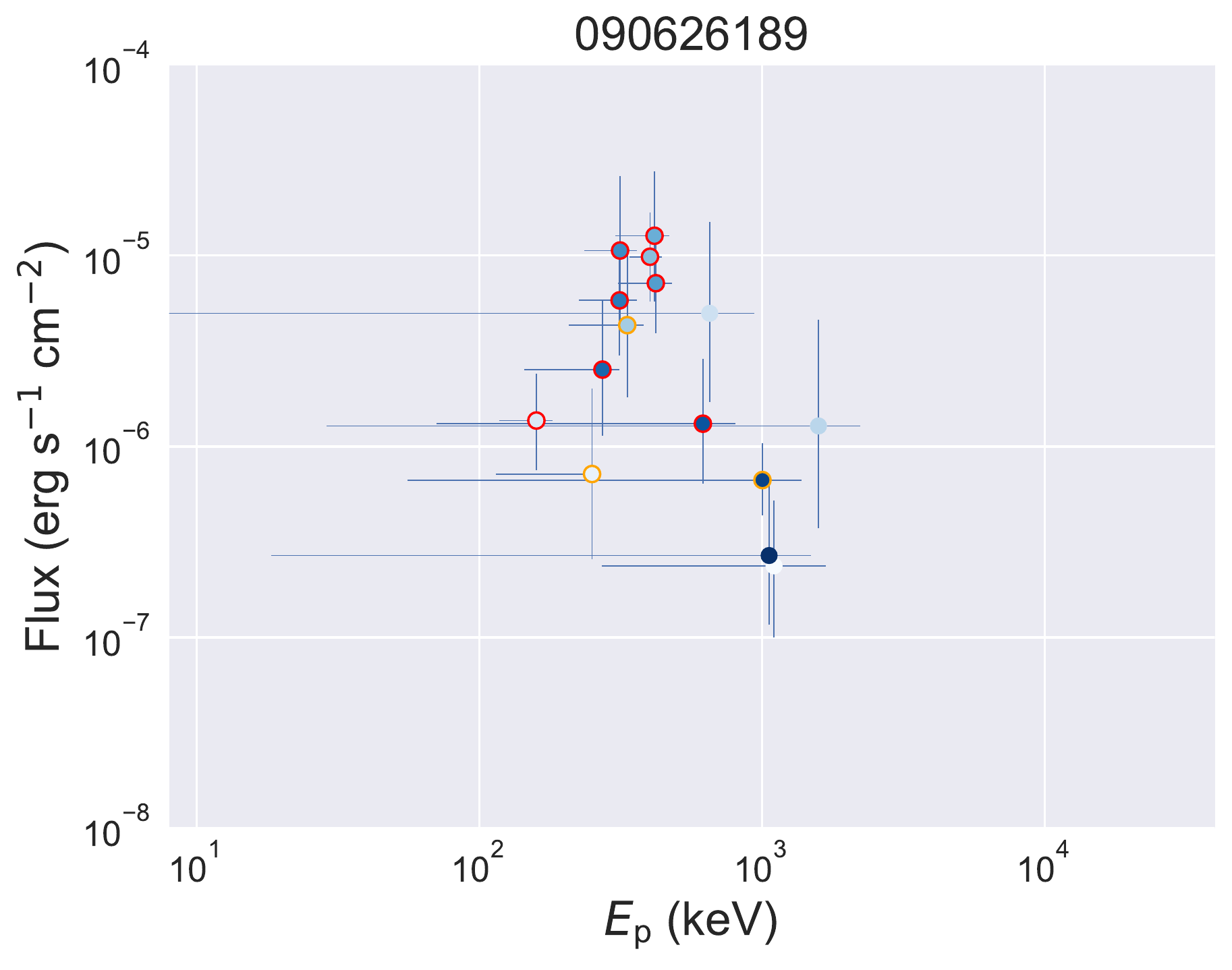}}
\subfigure{\includegraphics[width=0.3\linewidth]{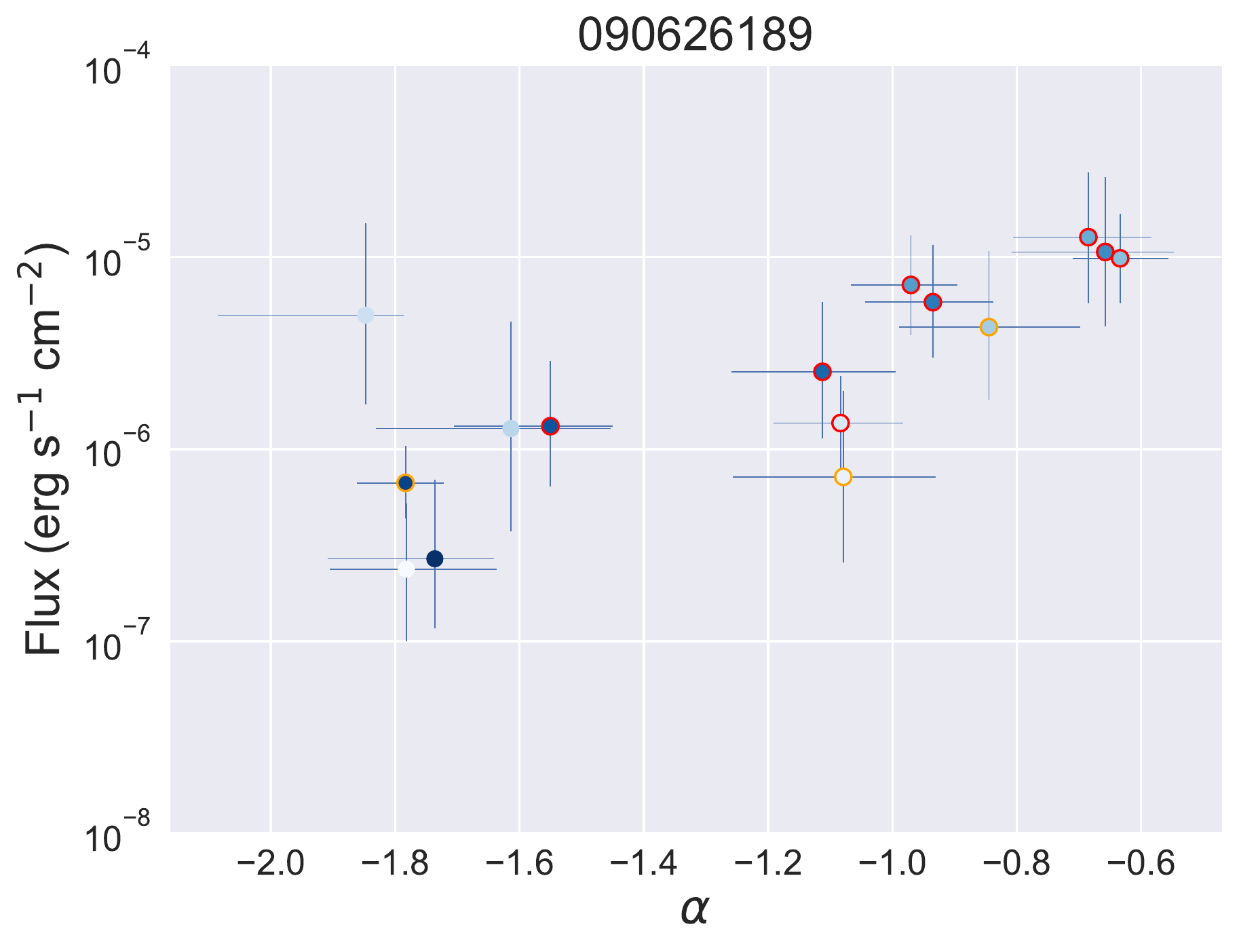}}

\subfigure{\includegraphics[width=0.3\linewidth]{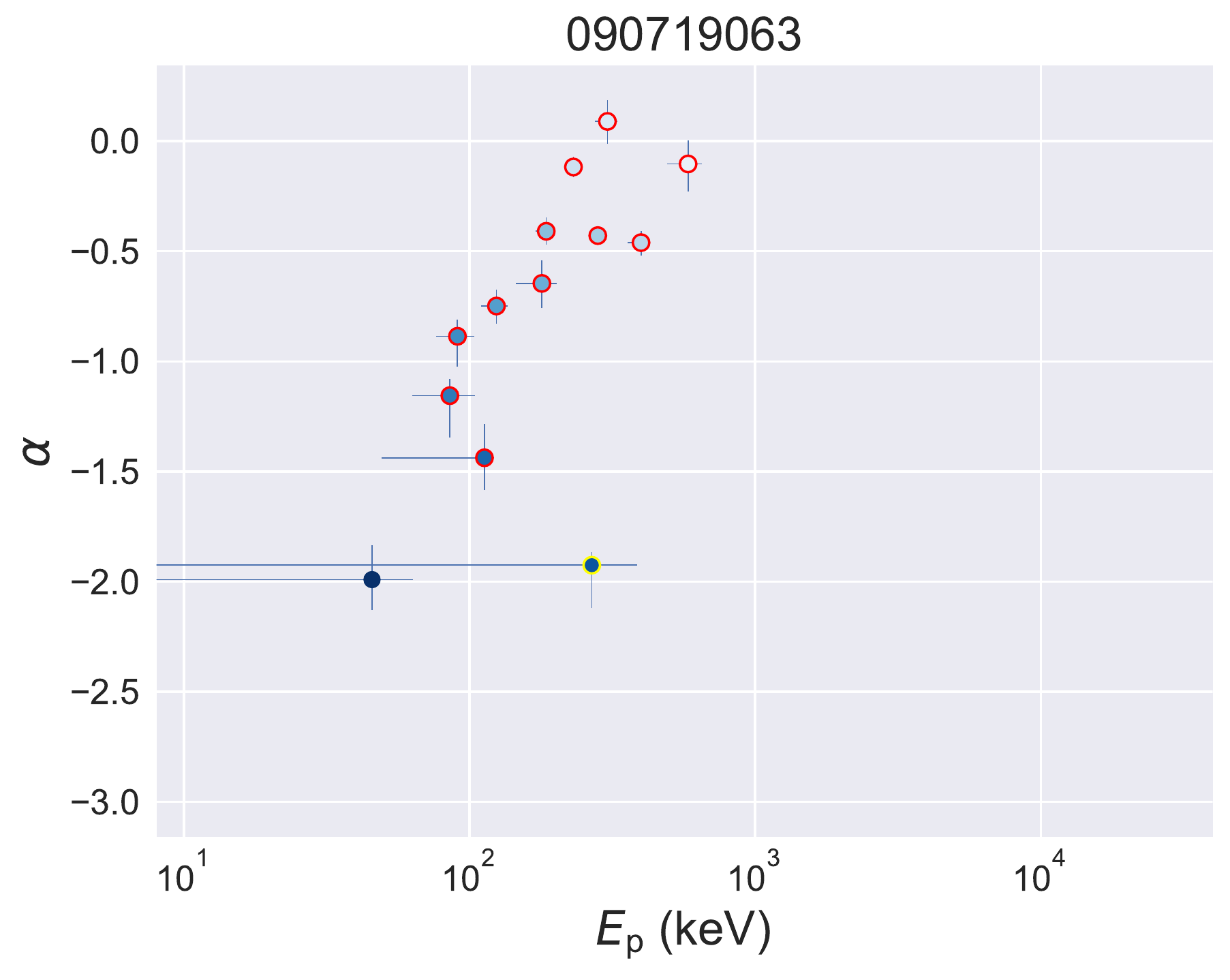}}
\subfigure{\includegraphics[width=0.3\linewidth]{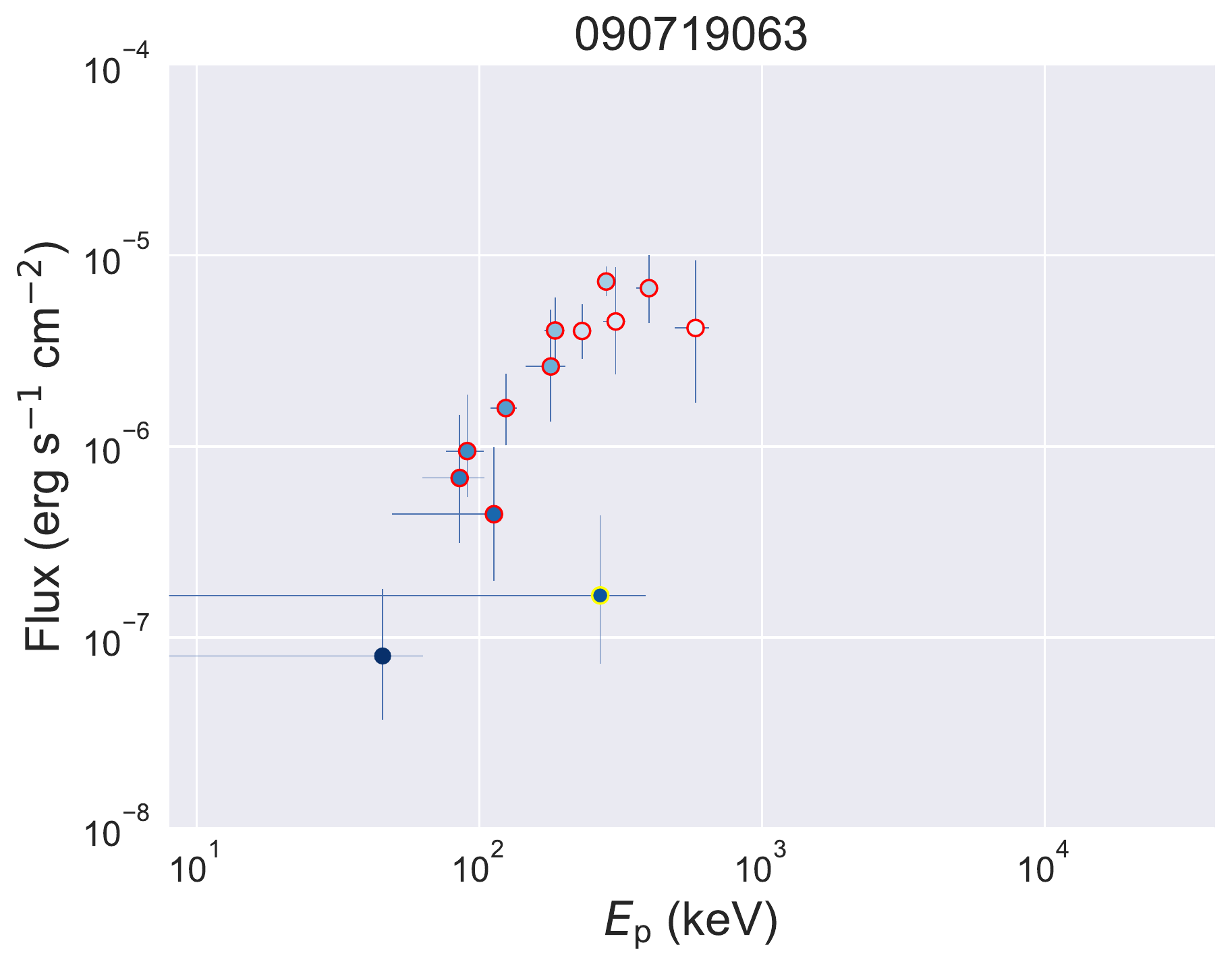}}
\subfigure{\includegraphics[width=0.3\linewidth]{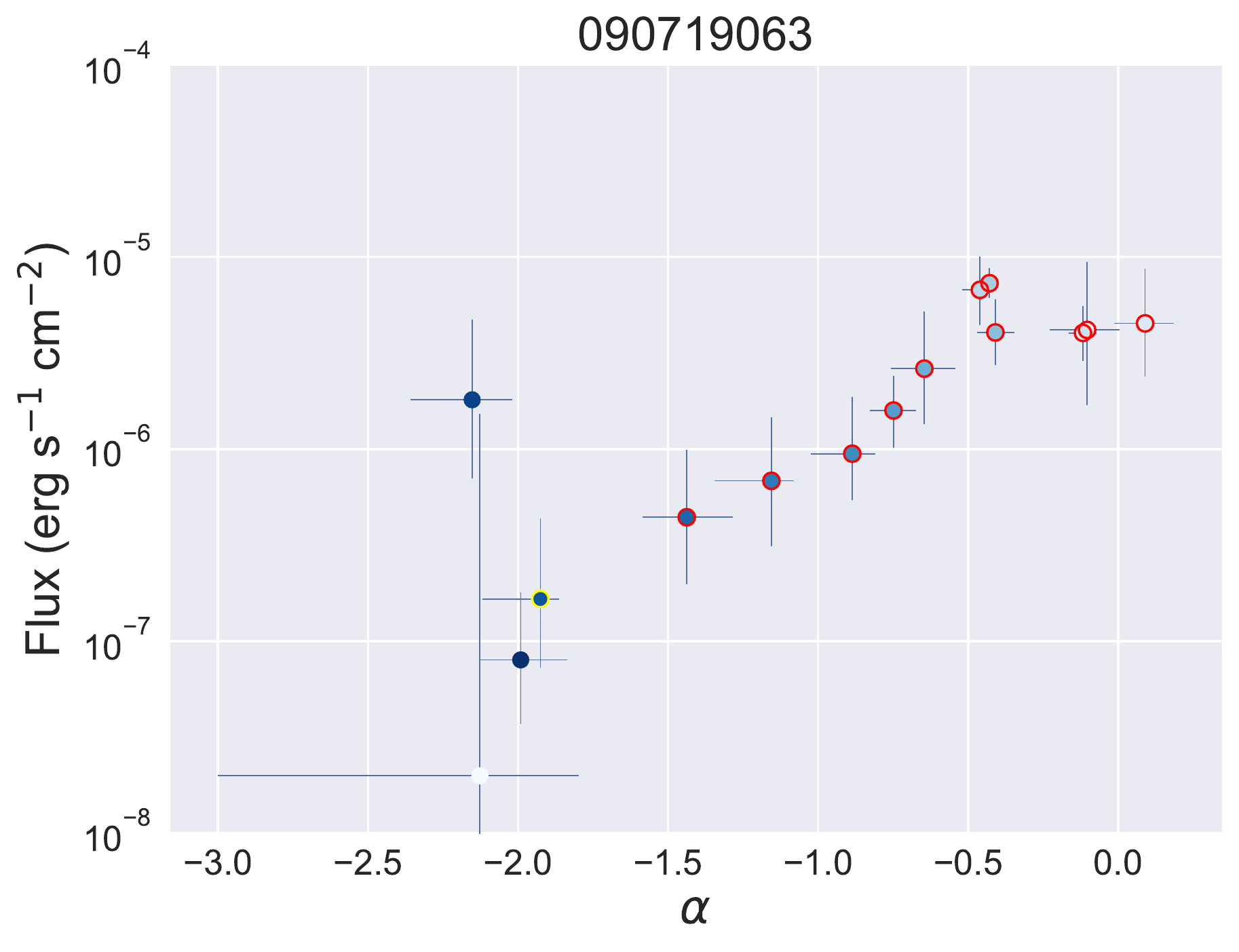}}

\caption{Same as Fig.~\ref{fig:correlation_group1}.
\label{fig:correlation_group2}}
\end{figure*}

\begin{figure*}

\subfigure{\includegraphics[width=0.3\linewidth]{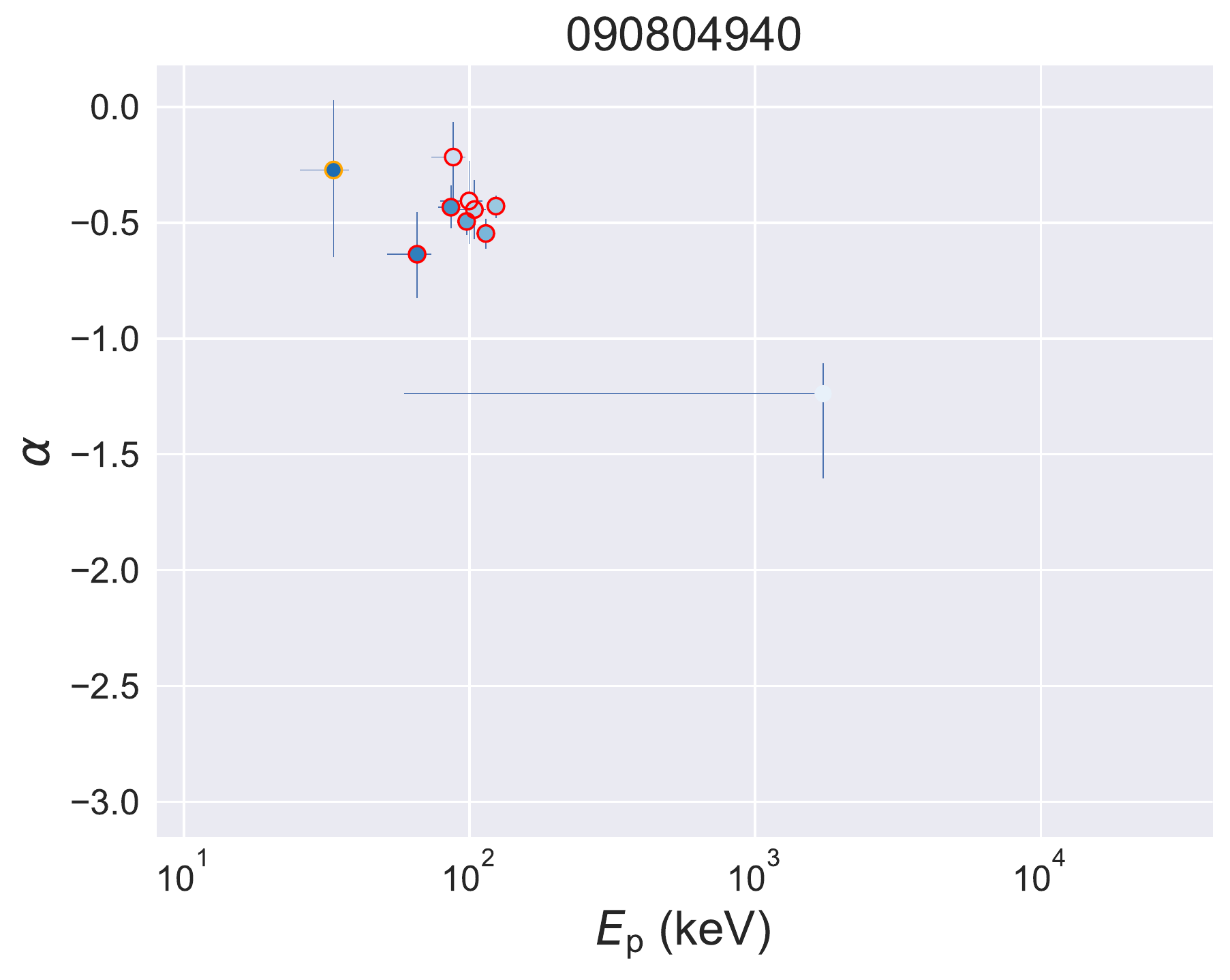}}
\subfigure{\includegraphics[width=0.3\linewidth]{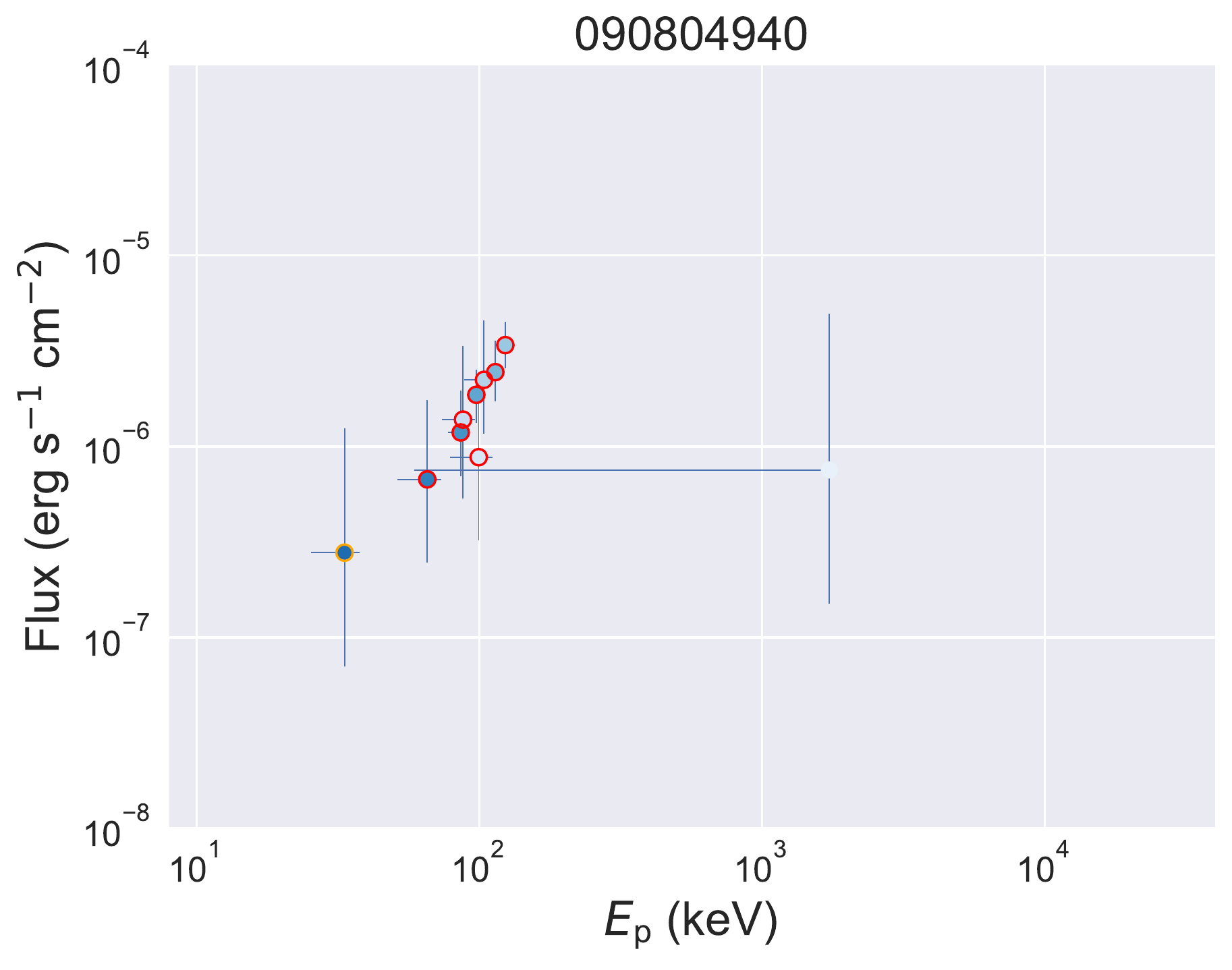}}
\subfigure{\includegraphics[width=0.3\linewidth]{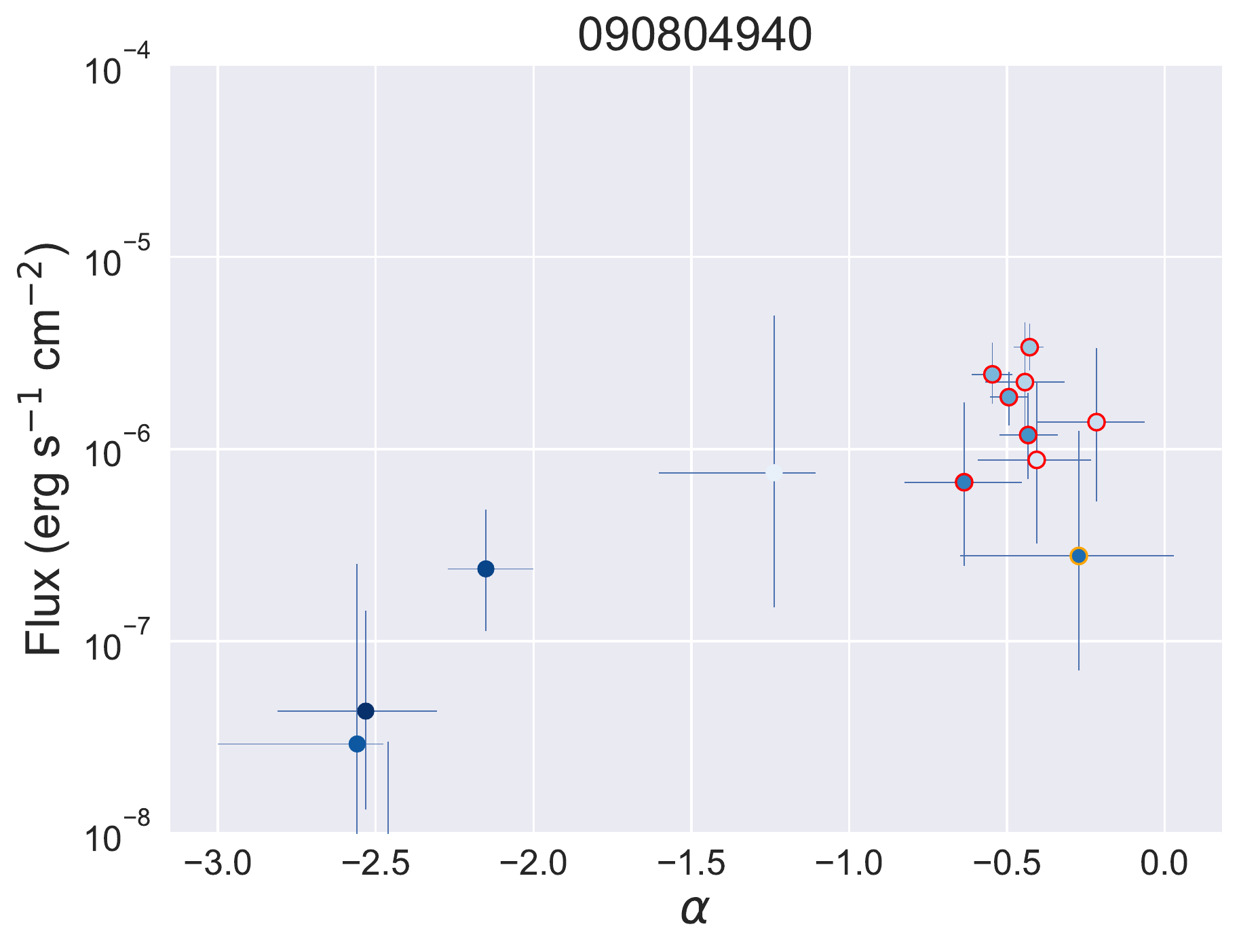}}

\subfigure{\includegraphics[width=0.3\linewidth]{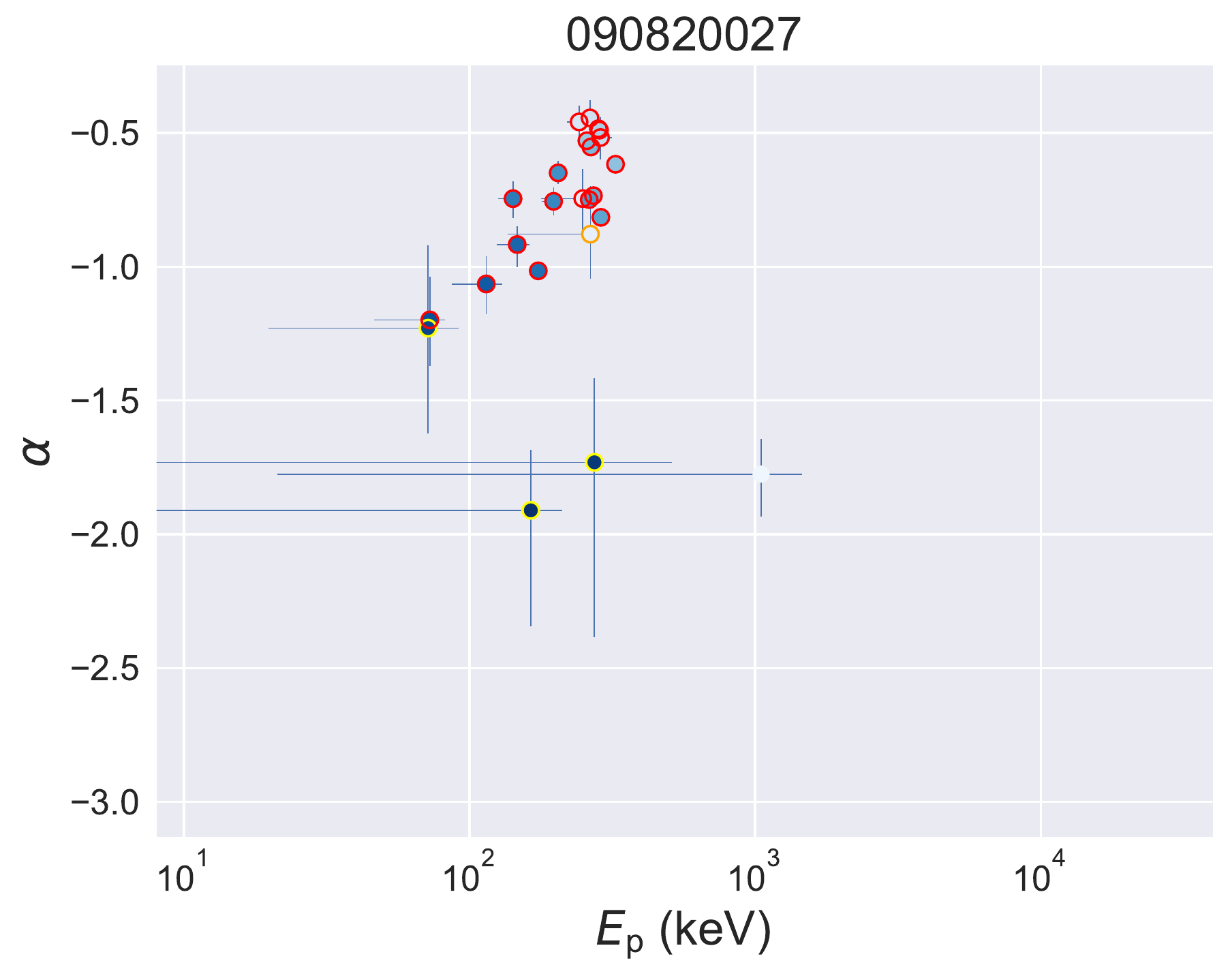}}
\subfigure{\includegraphics[width=0.3\linewidth]{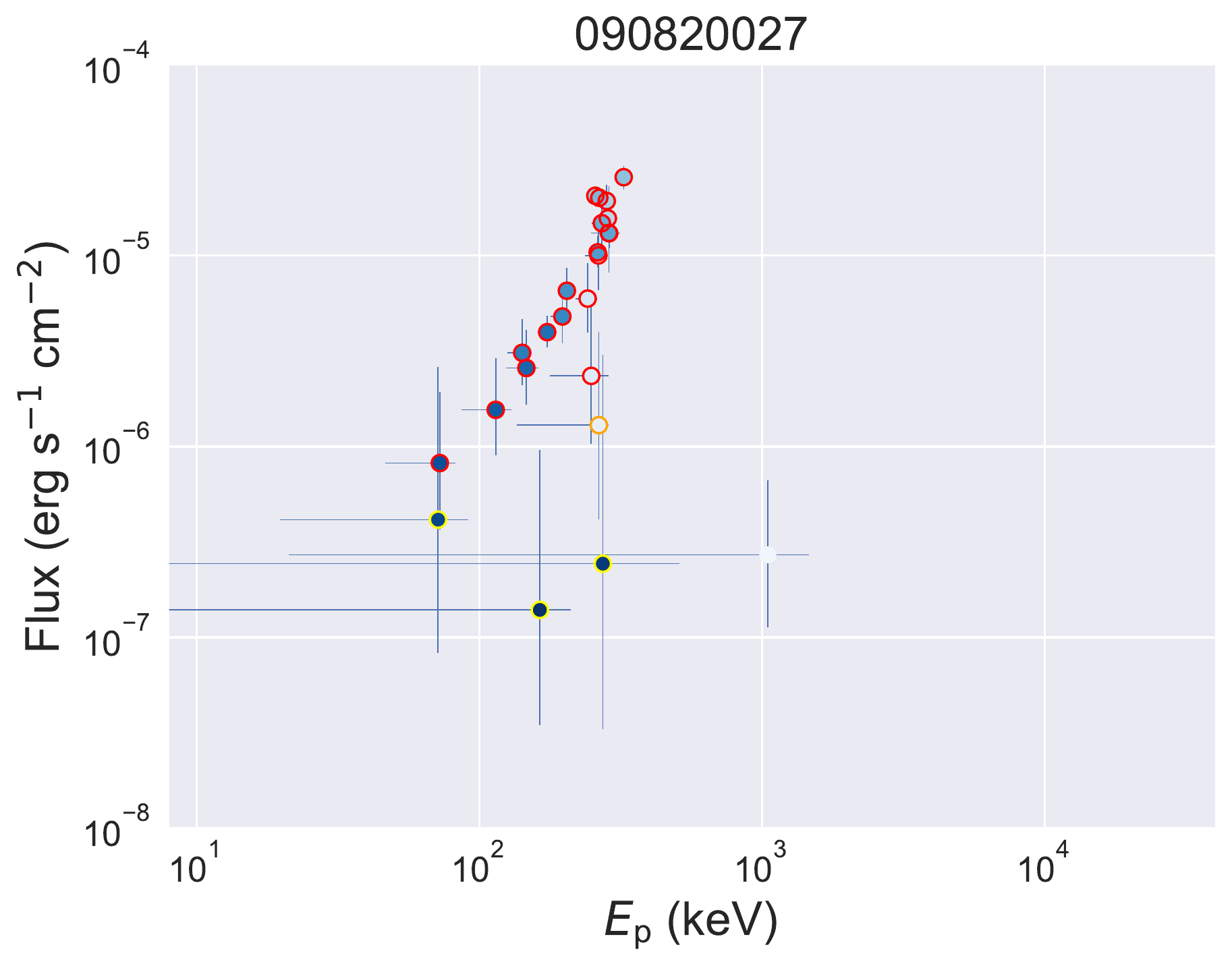}}
\subfigure{\includegraphics[width=0.3\linewidth]{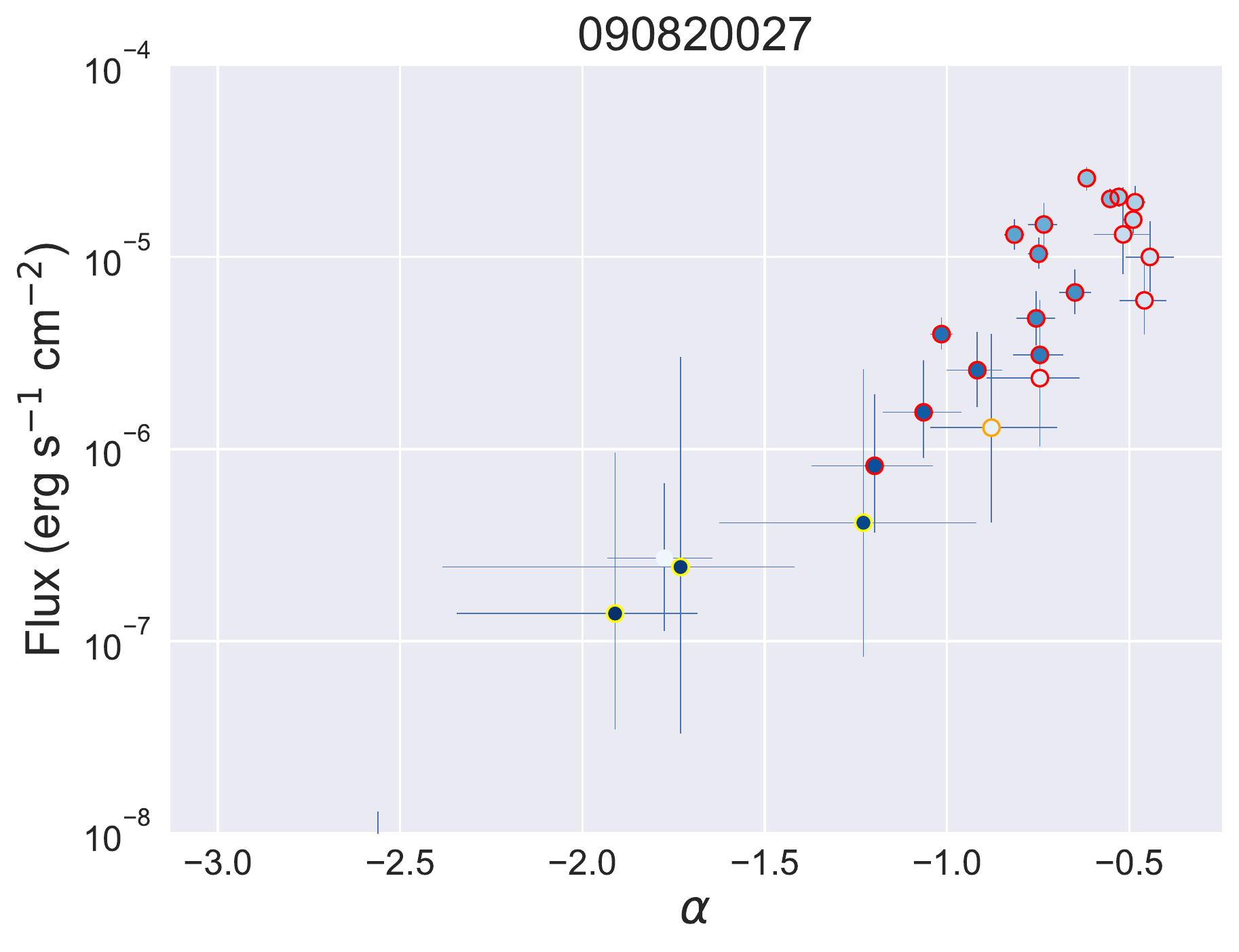}}

\subfigure{\includegraphics[width=0.3\linewidth]{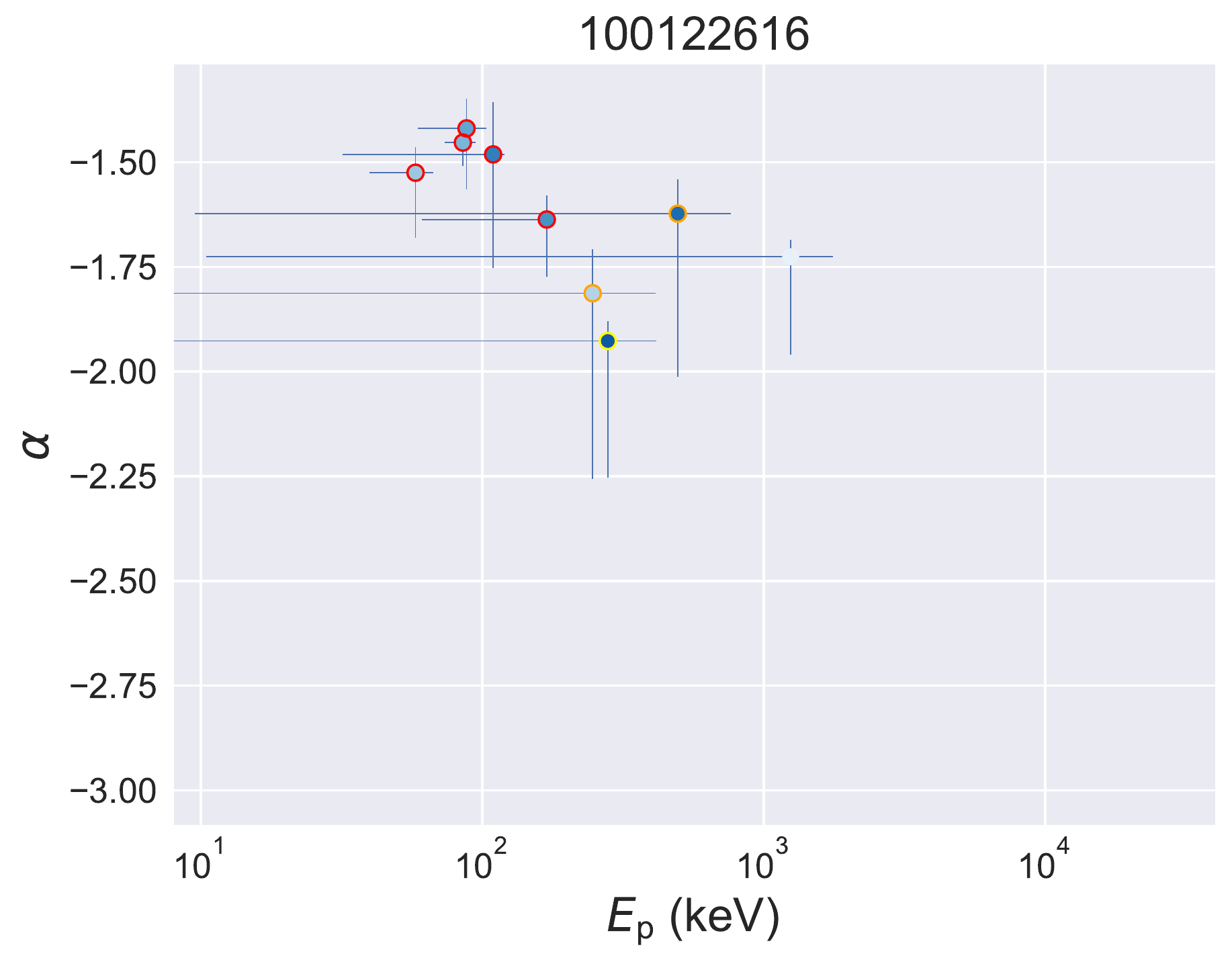}}
\subfigure{\includegraphics[width=0.3\linewidth]{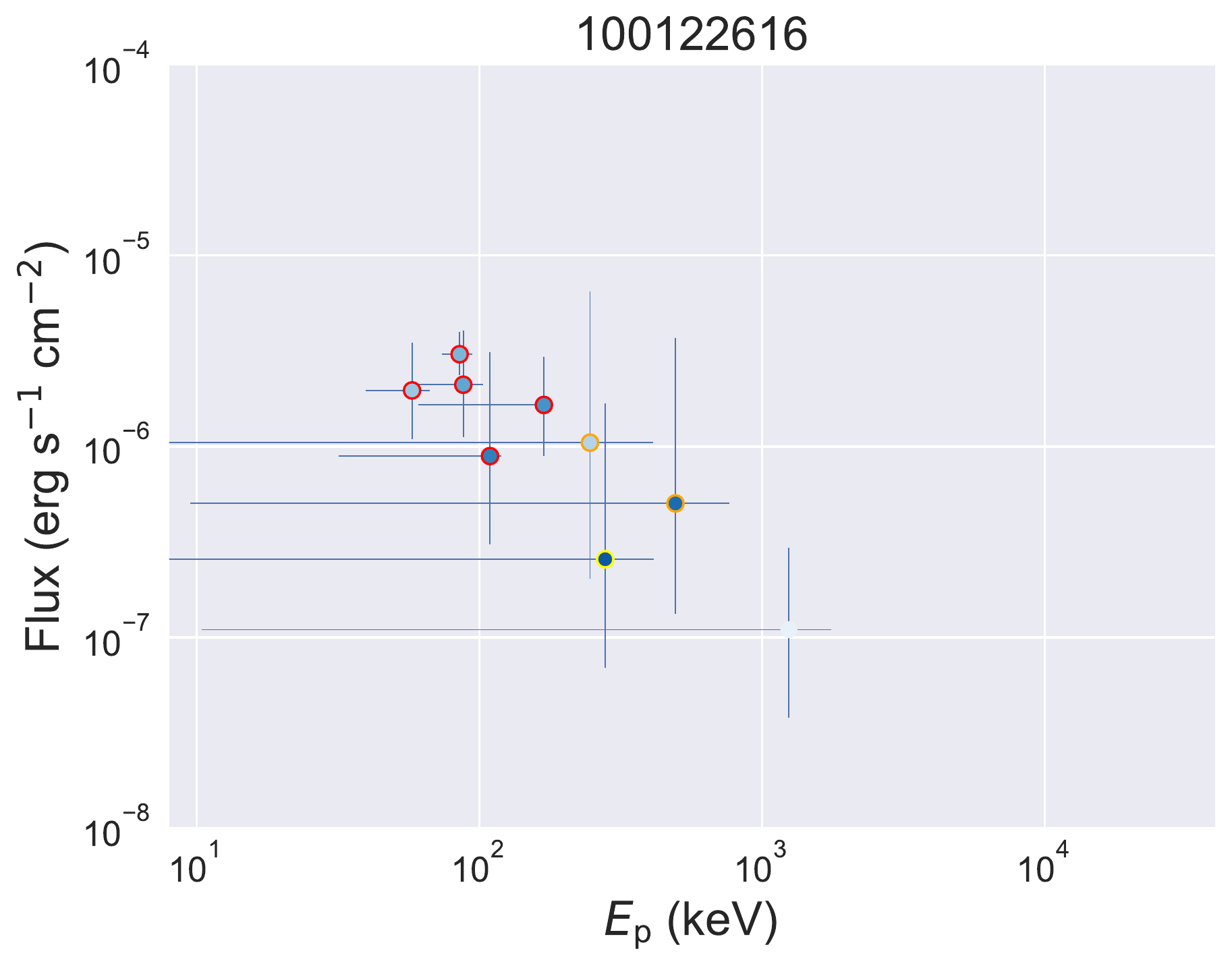}}
\subfigure{\includegraphics[width=0.3\linewidth]{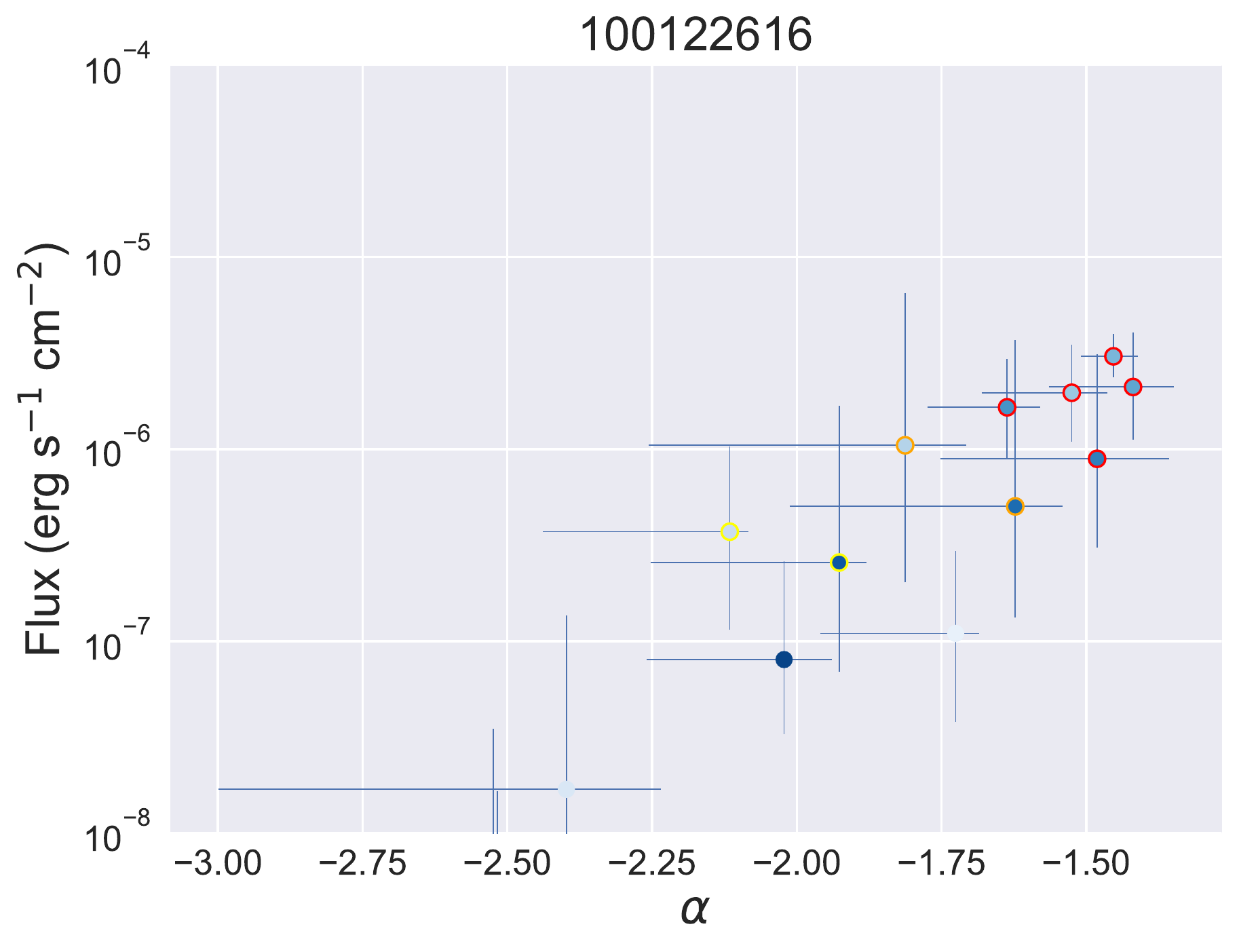}}

\subfigure{\includegraphics[width=0.3\linewidth]{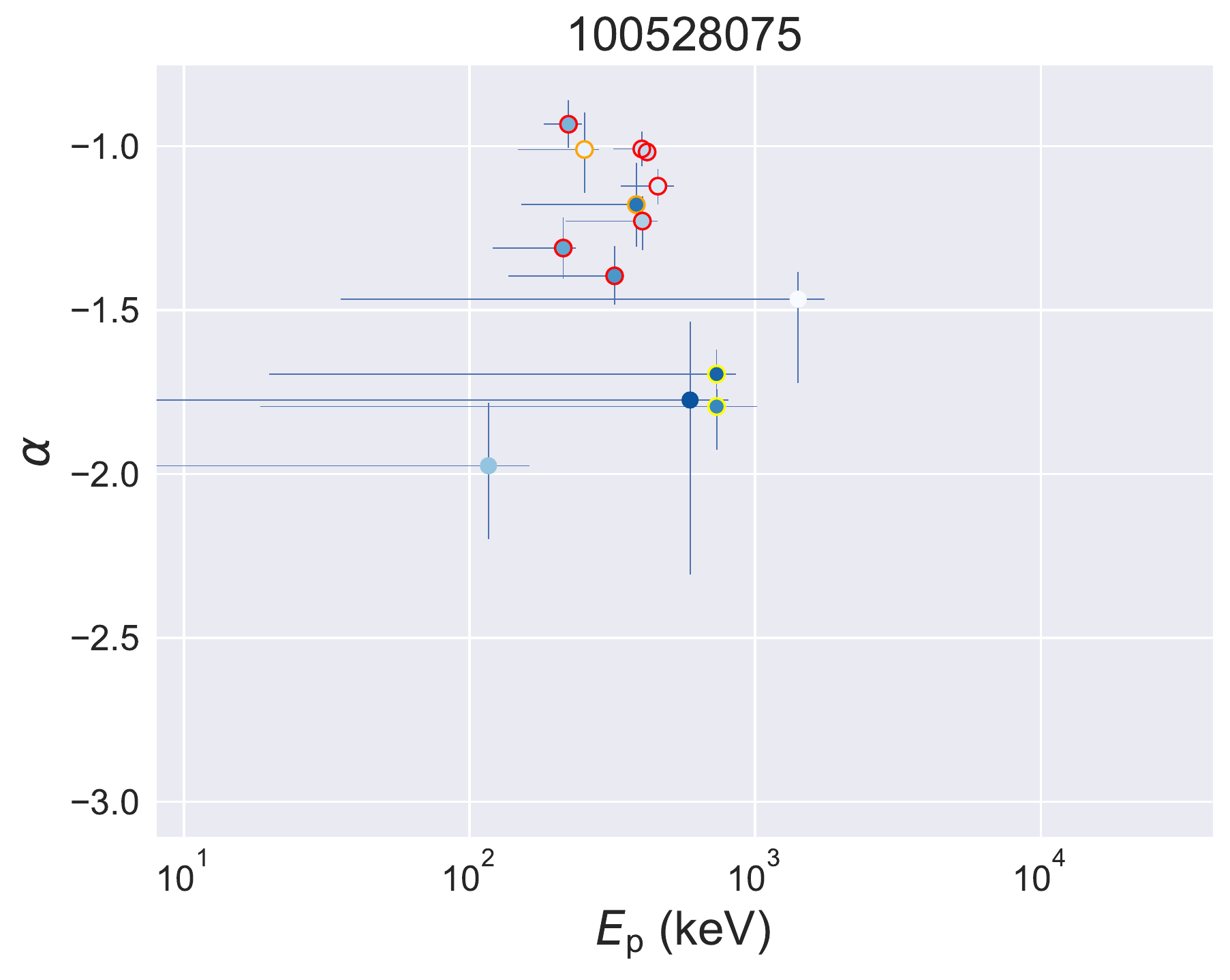}}
\subfigure{\includegraphics[width=0.3\linewidth]{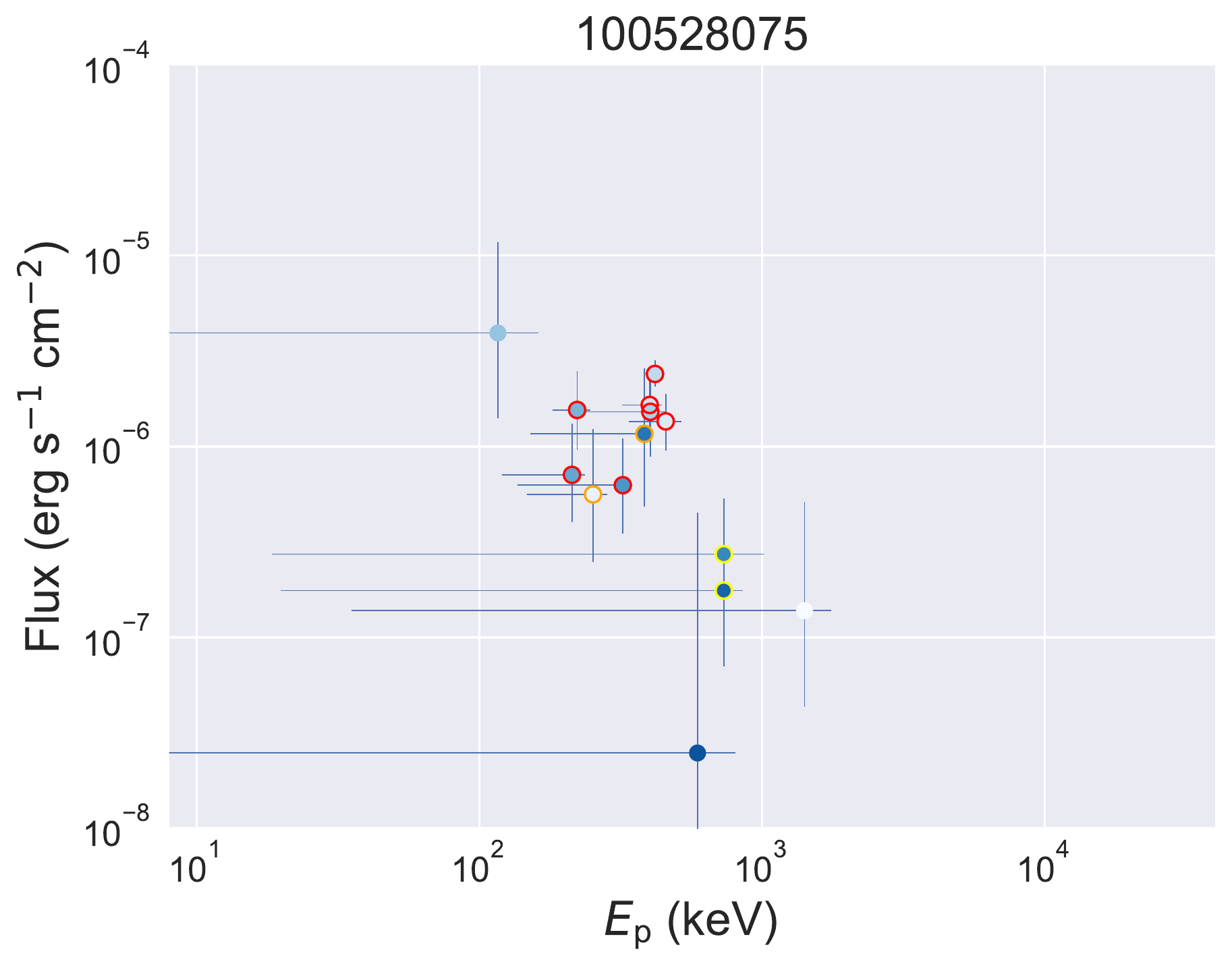}}
\subfigure{\includegraphics[width=0.3\linewidth]{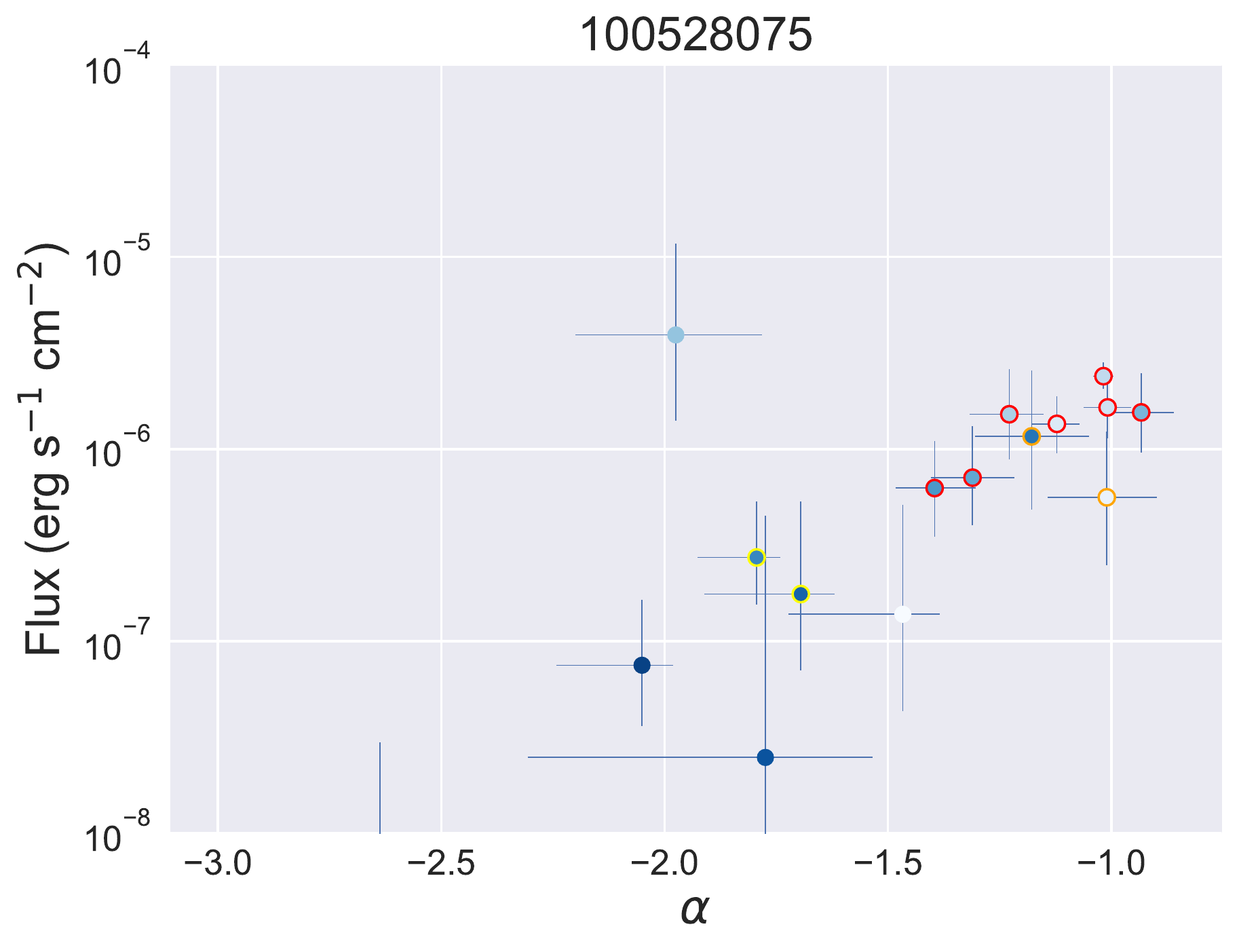}}

\caption{Same as Fig.~\ref{fig:correlation_group1}.
\label{fig:correlation_group3}}
\end{figure*}

\begin{figure*}

\subfigure{\includegraphics[width=0.3\linewidth]{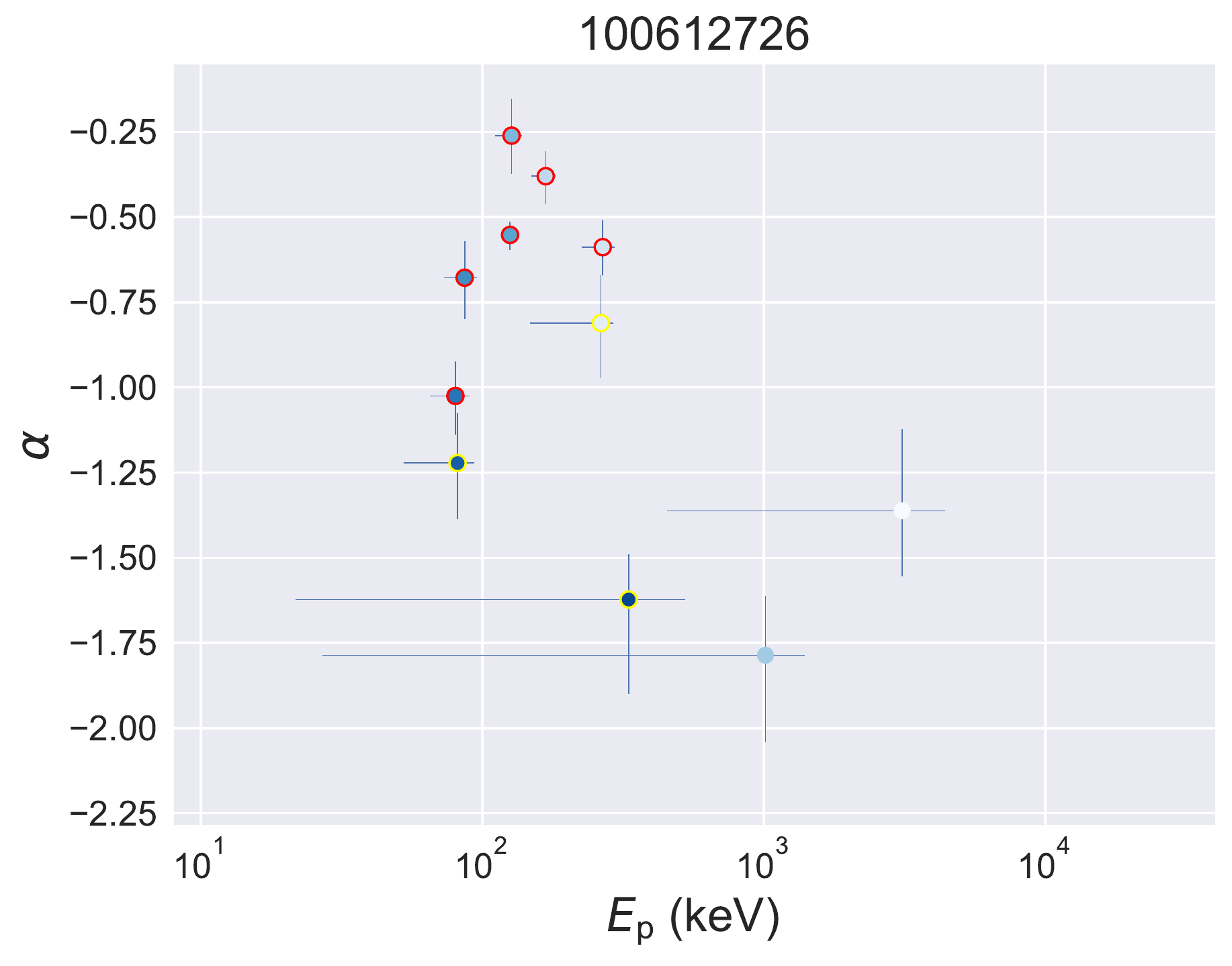}}
\subfigure{\includegraphics[width=0.3\linewidth]{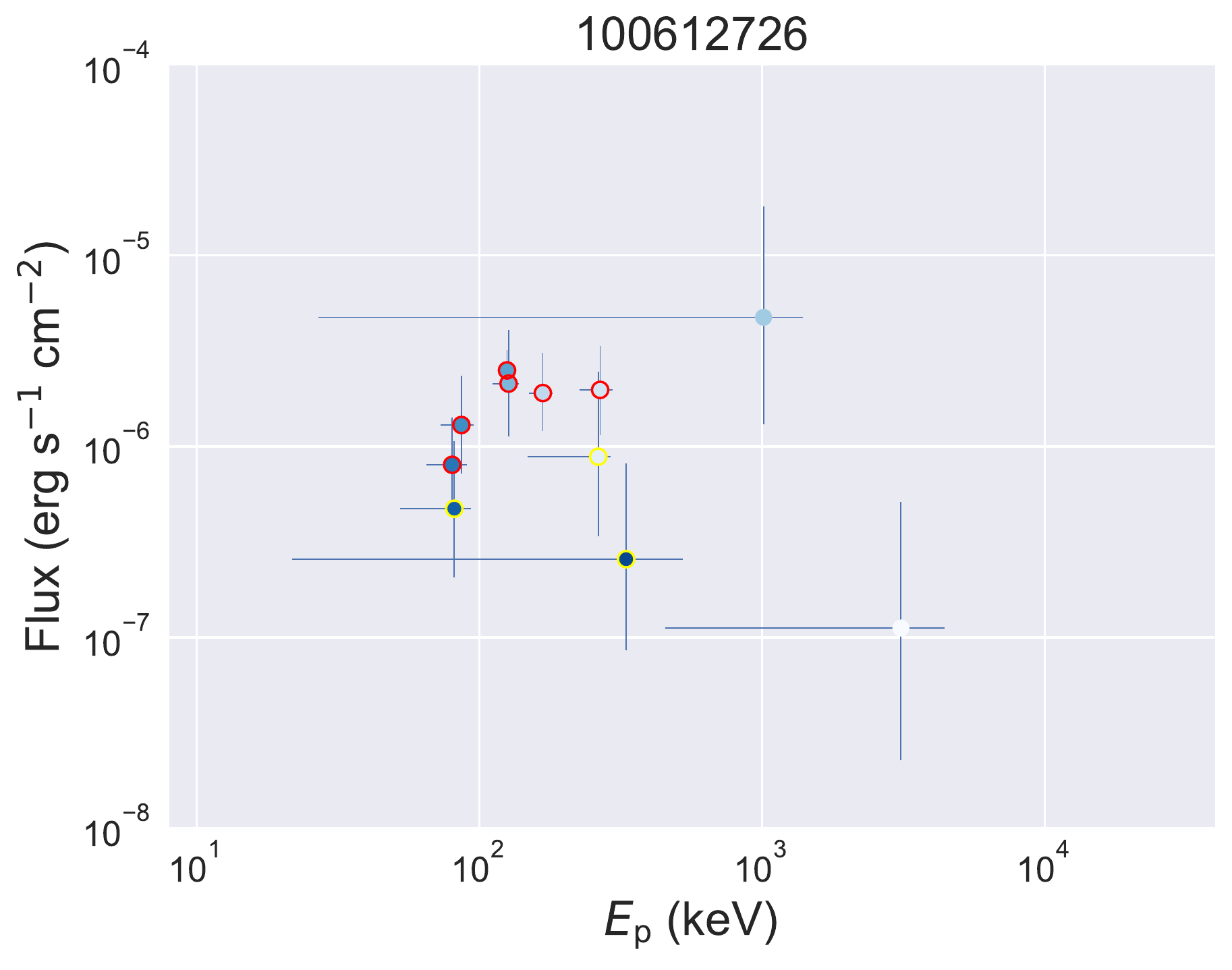}}
\subfigure{\includegraphics[width=0.3\linewidth]{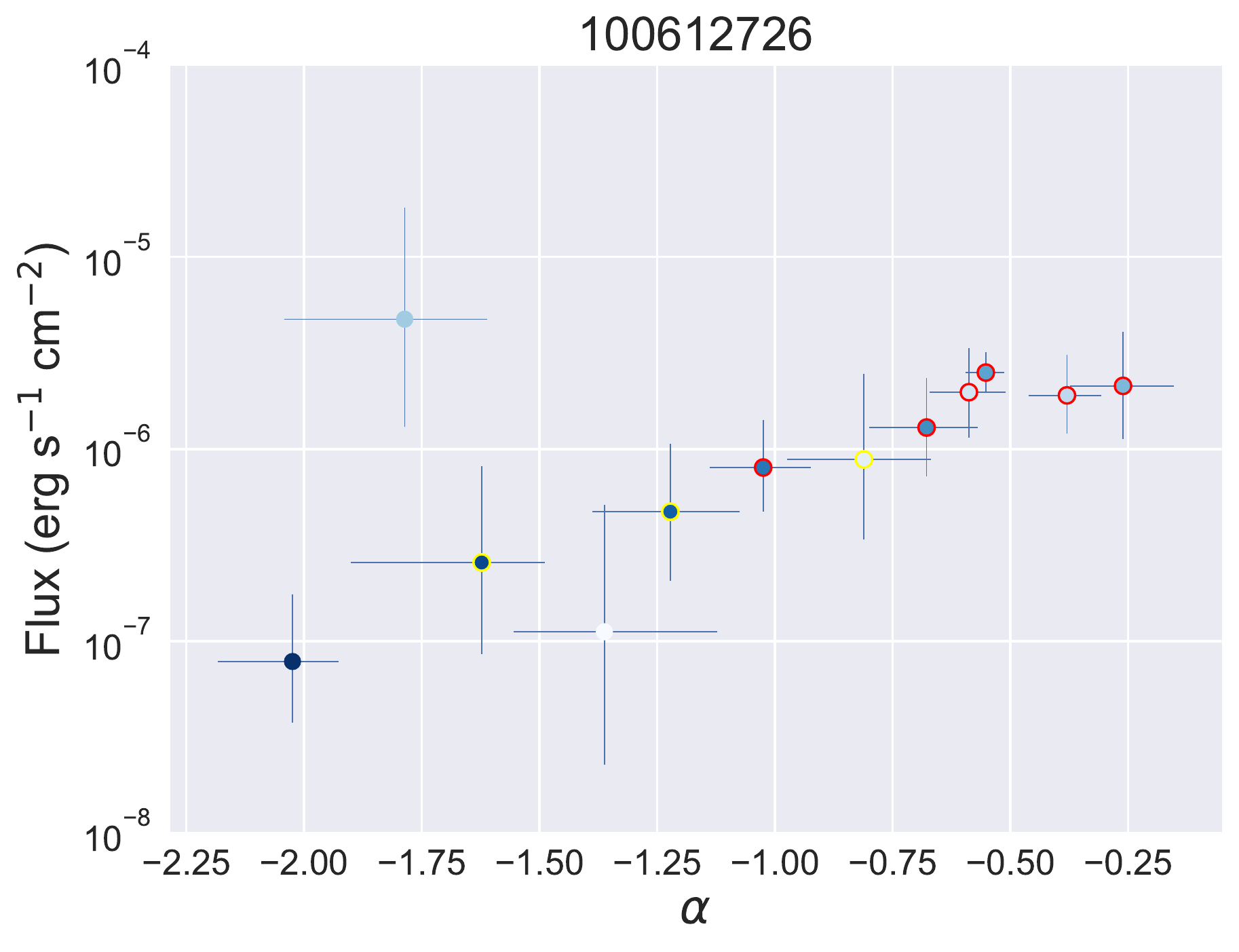}}

\subfigure{\includegraphics[width=0.3\linewidth]{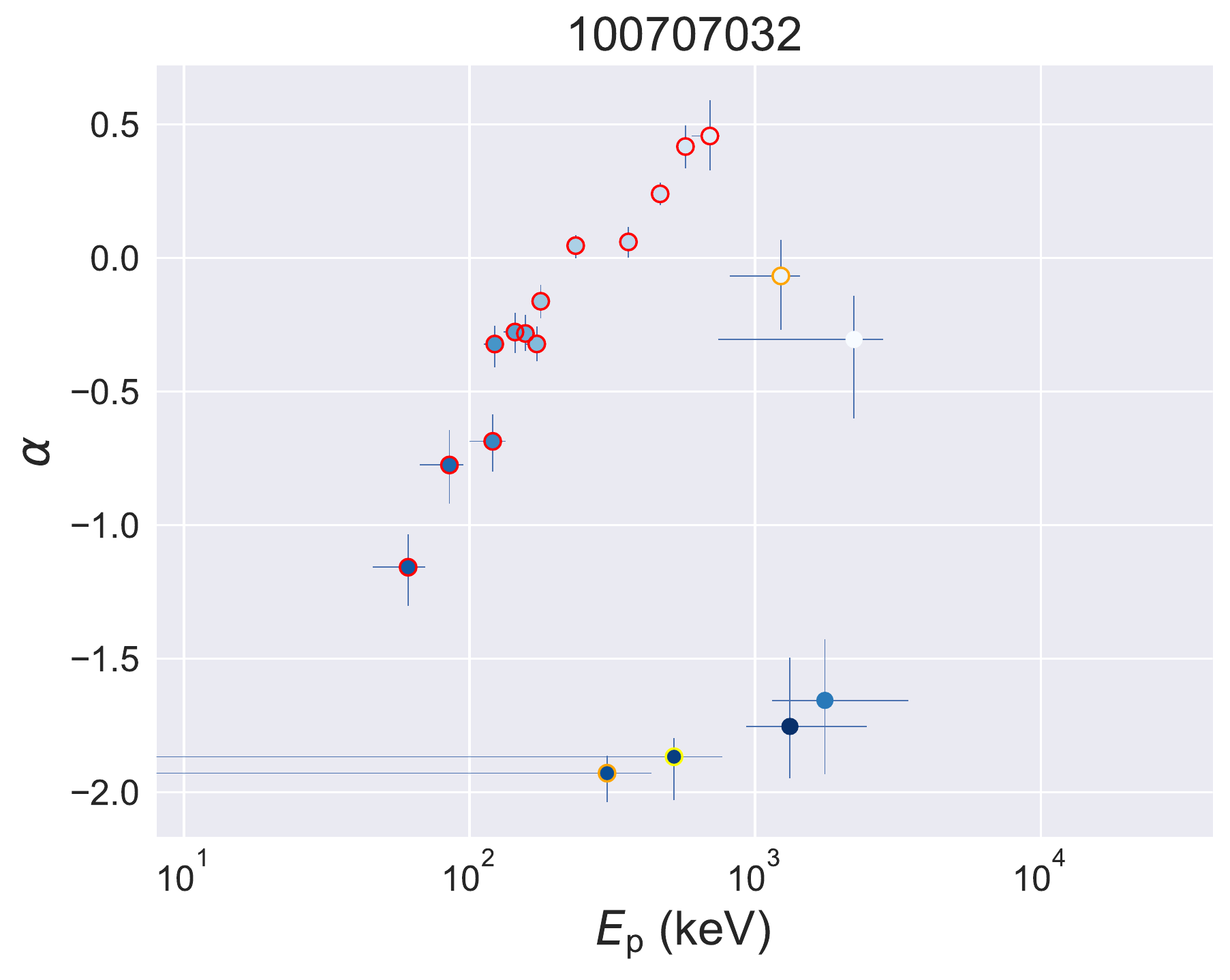}}
\subfigure{\includegraphics[width=0.3\linewidth]{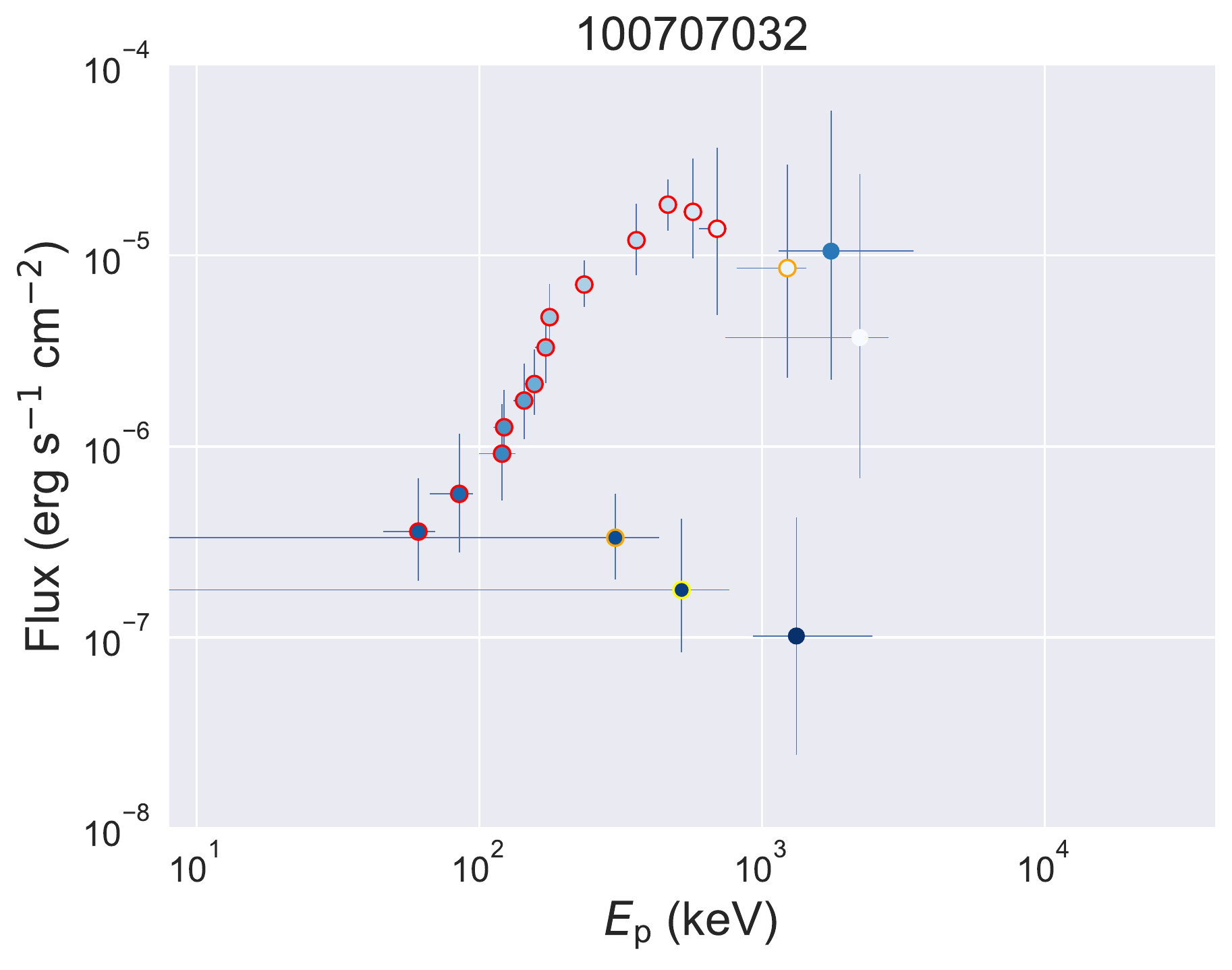}}
\subfigure{\includegraphics[width=0.3\linewidth]{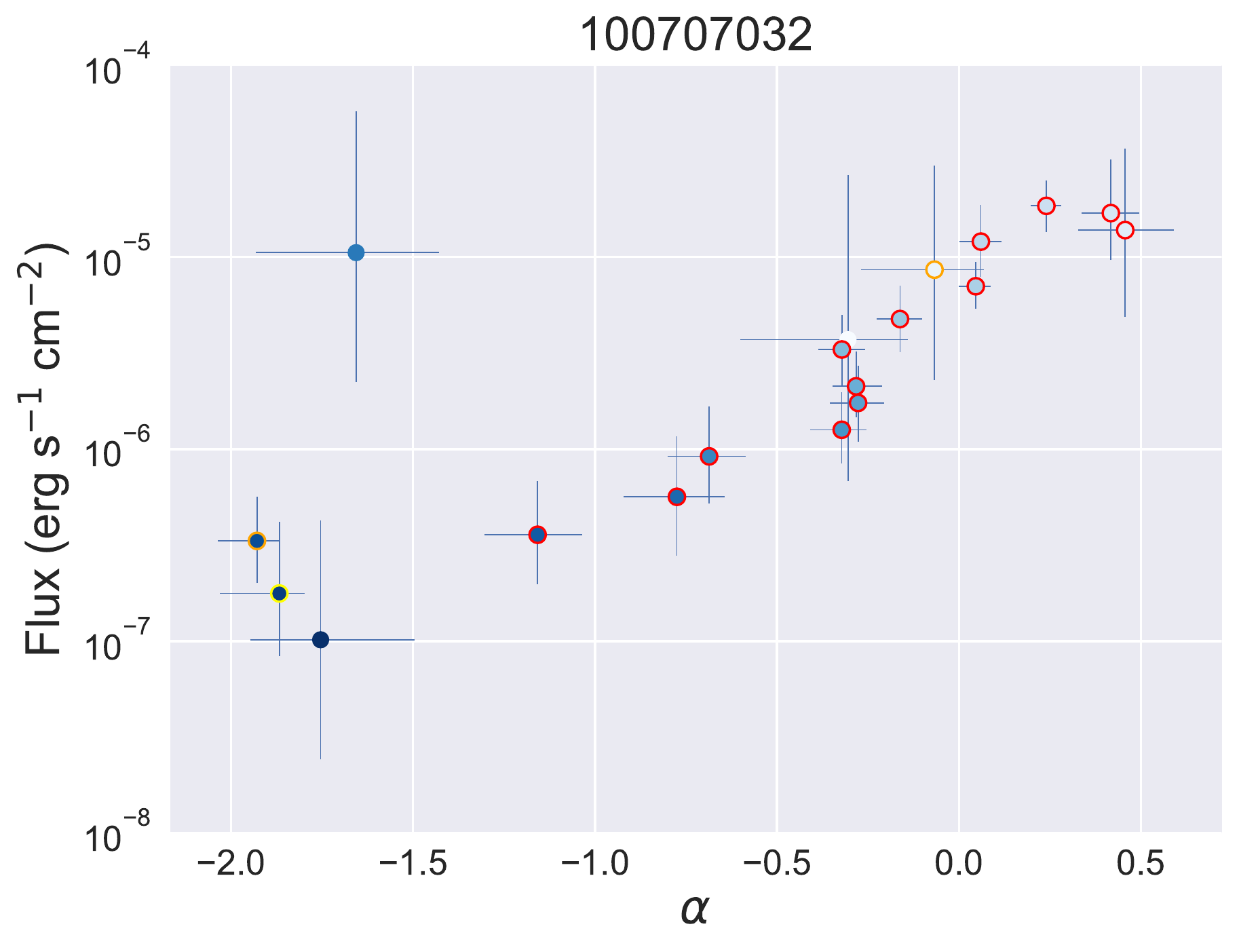}}

\subfigure{\includegraphics[width=0.3\linewidth]{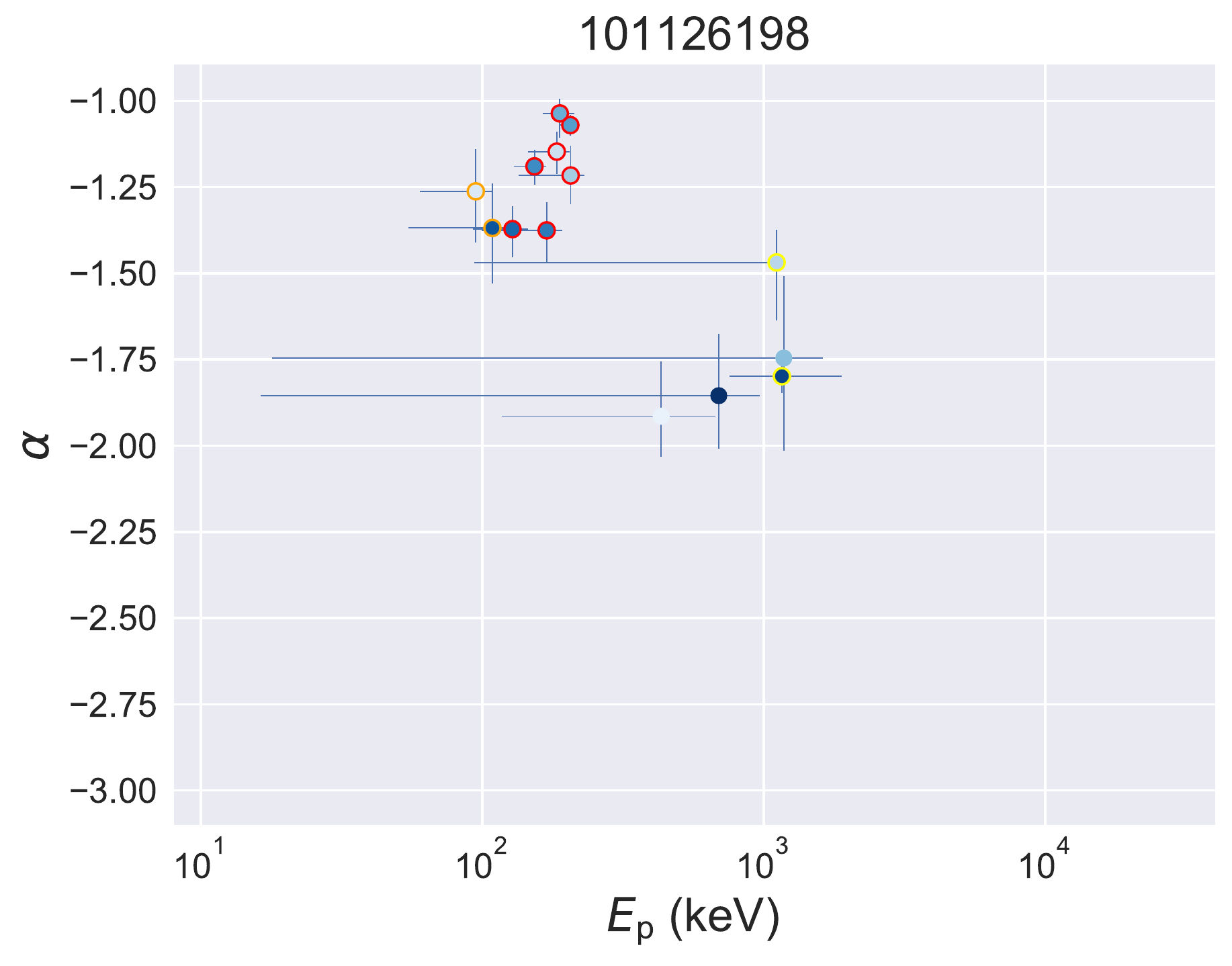}}
\subfigure{\includegraphics[width=0.3\linewidth]{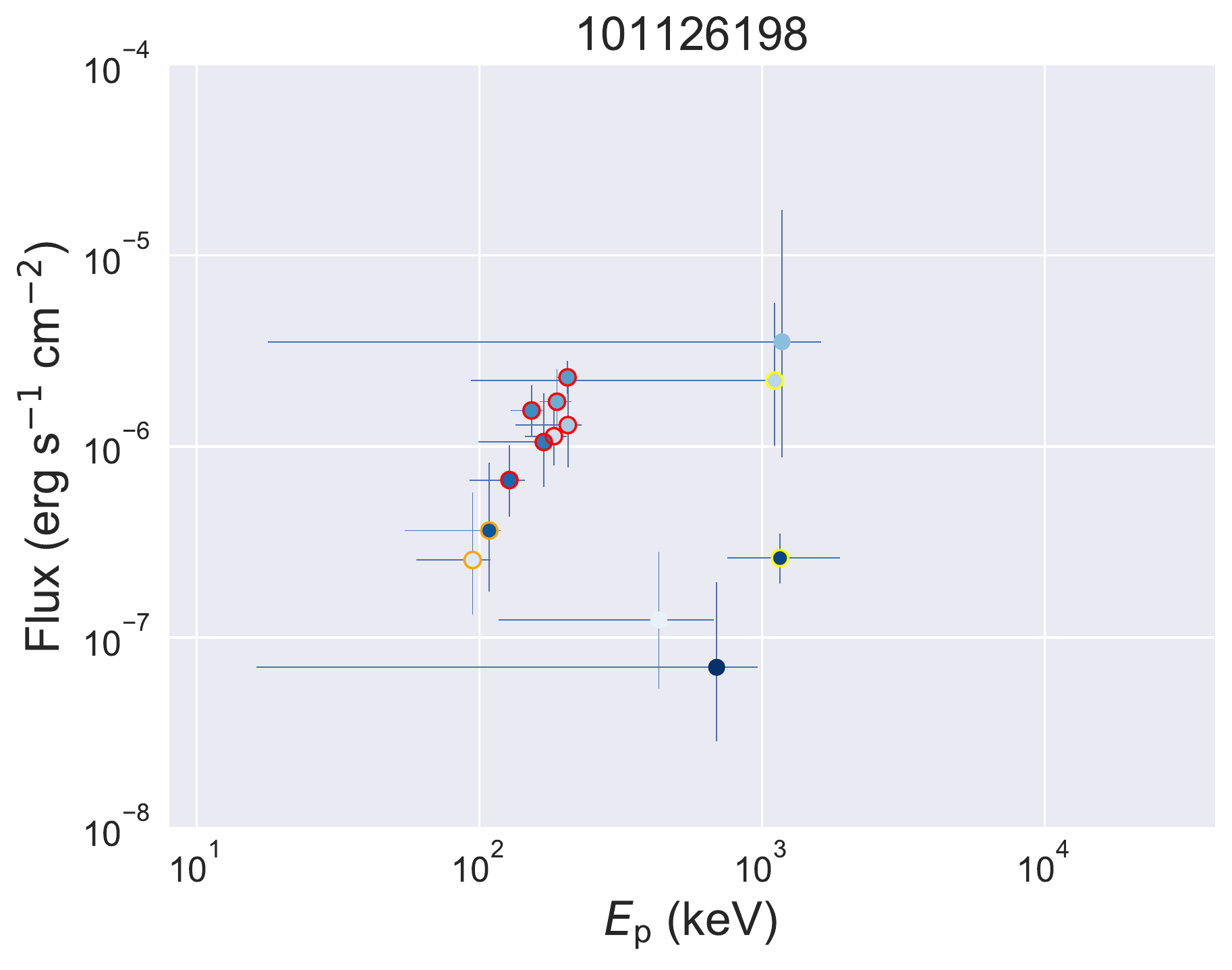}}
\subfigure{\includegraphics[width=0.3\linewidth]{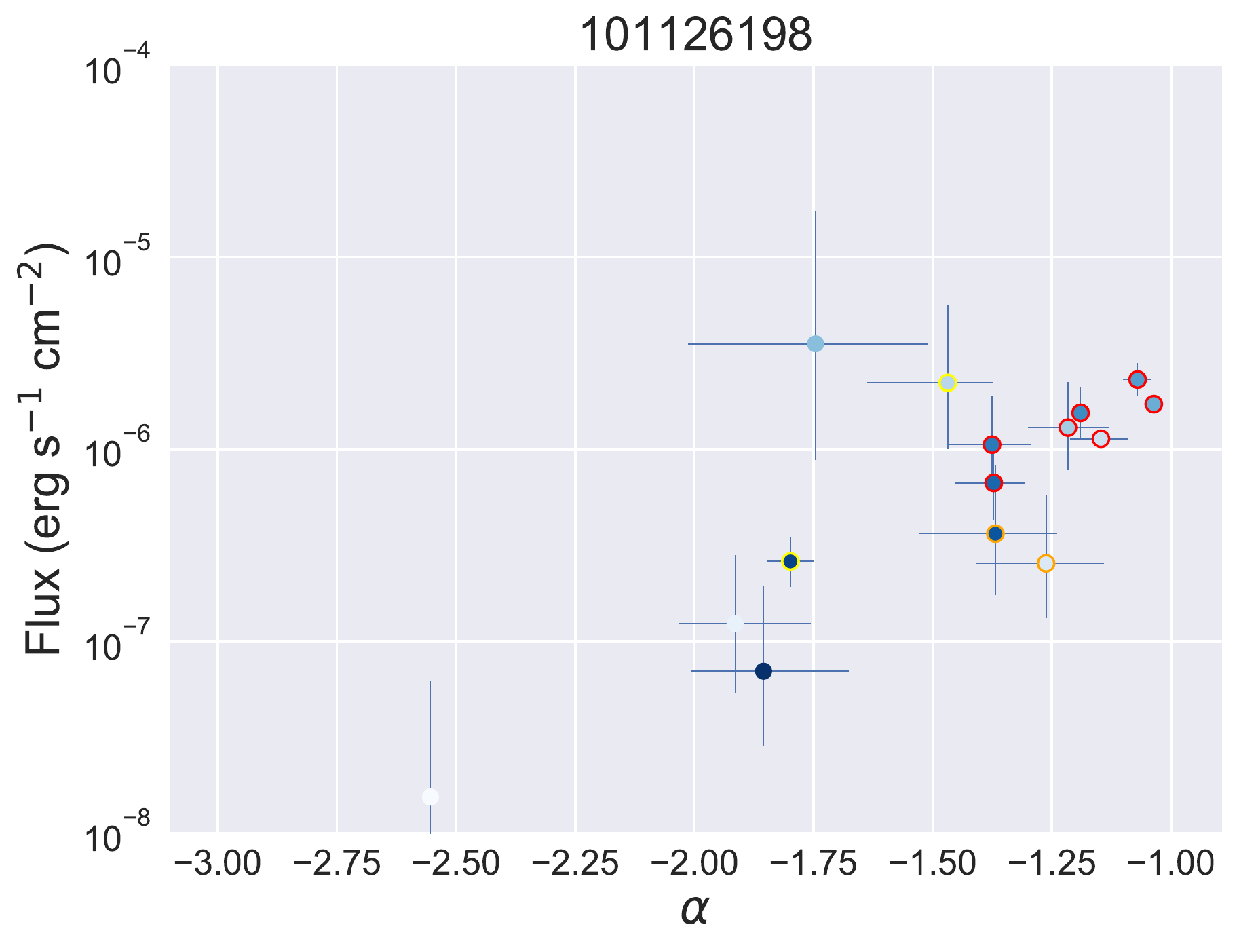}}

\subfigure{\includegraphics[width=0.3\linewidth]{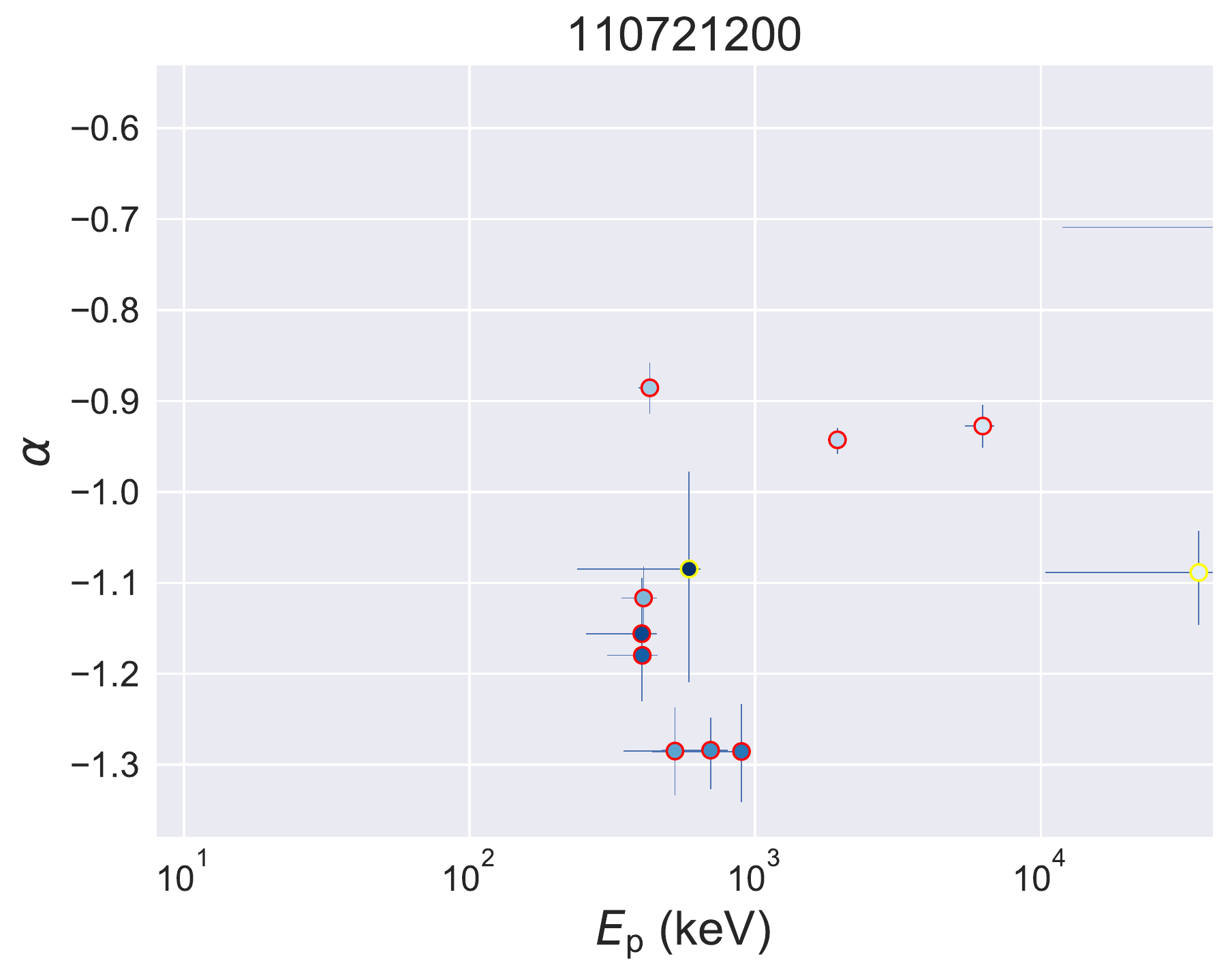}}
\subfigure{\includegraphics[width=0.3\linewidth]{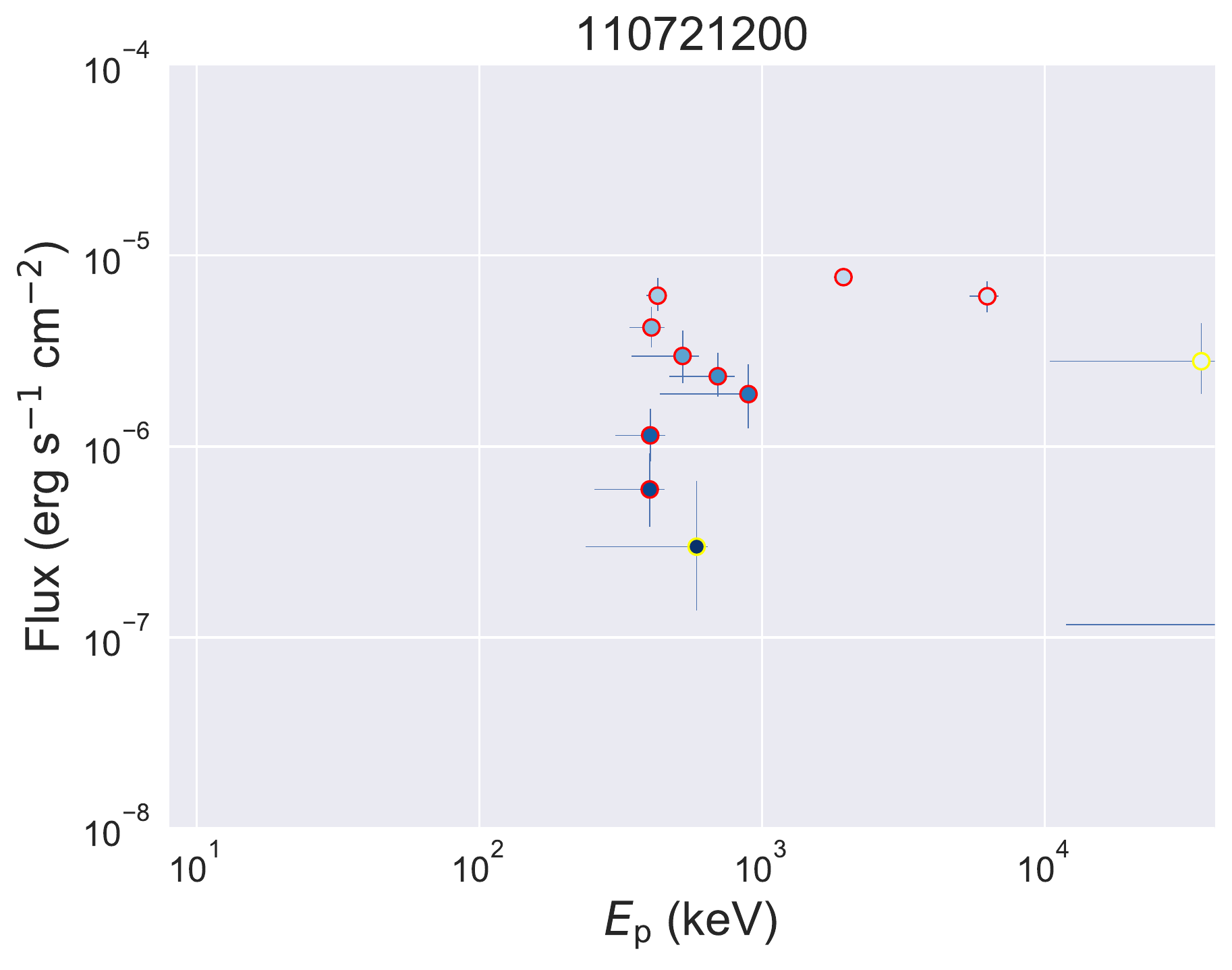}}
\subfigure{\includegraphics[width=0.3\linewidth]{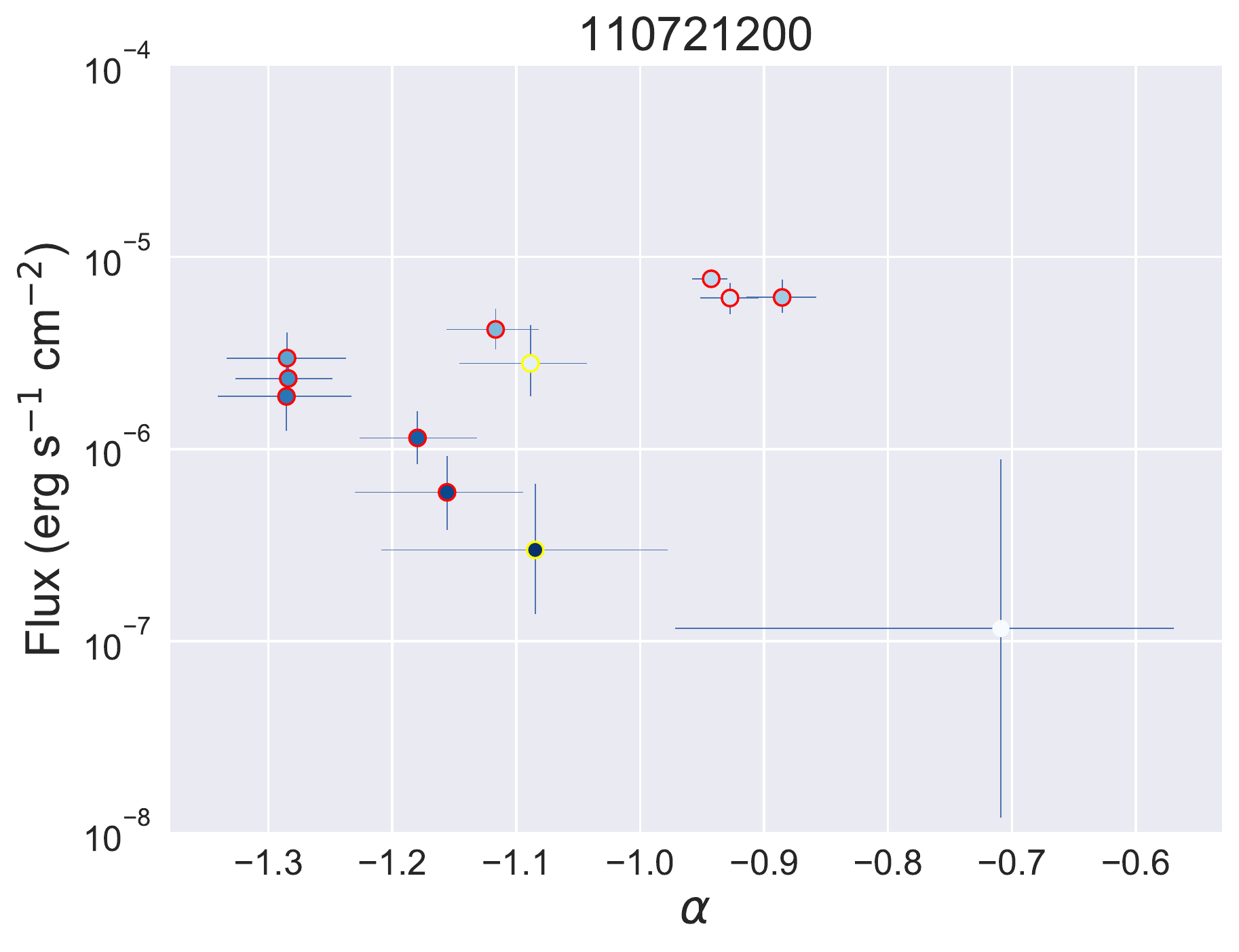}}

\caption{Same as Fig.~\ref{fig:correlation_group1}.
\label{fig:correlation_group4}}
\end{figure*}

\begin{figure*}

\subfigure{\includegraphics[width=0.3\linewidth]{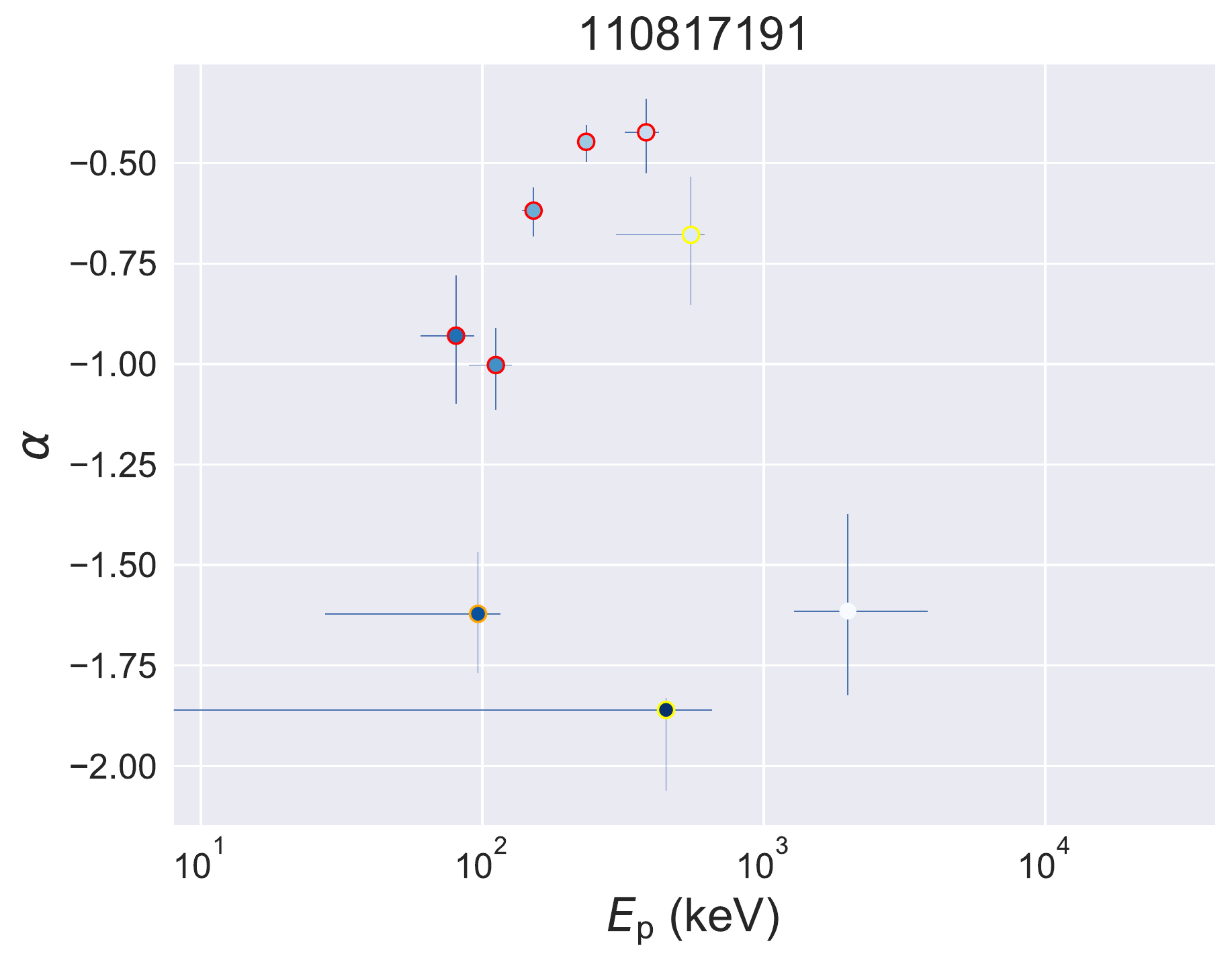}}
\subfigure{\includegraphics[width=0.3\linewidth]{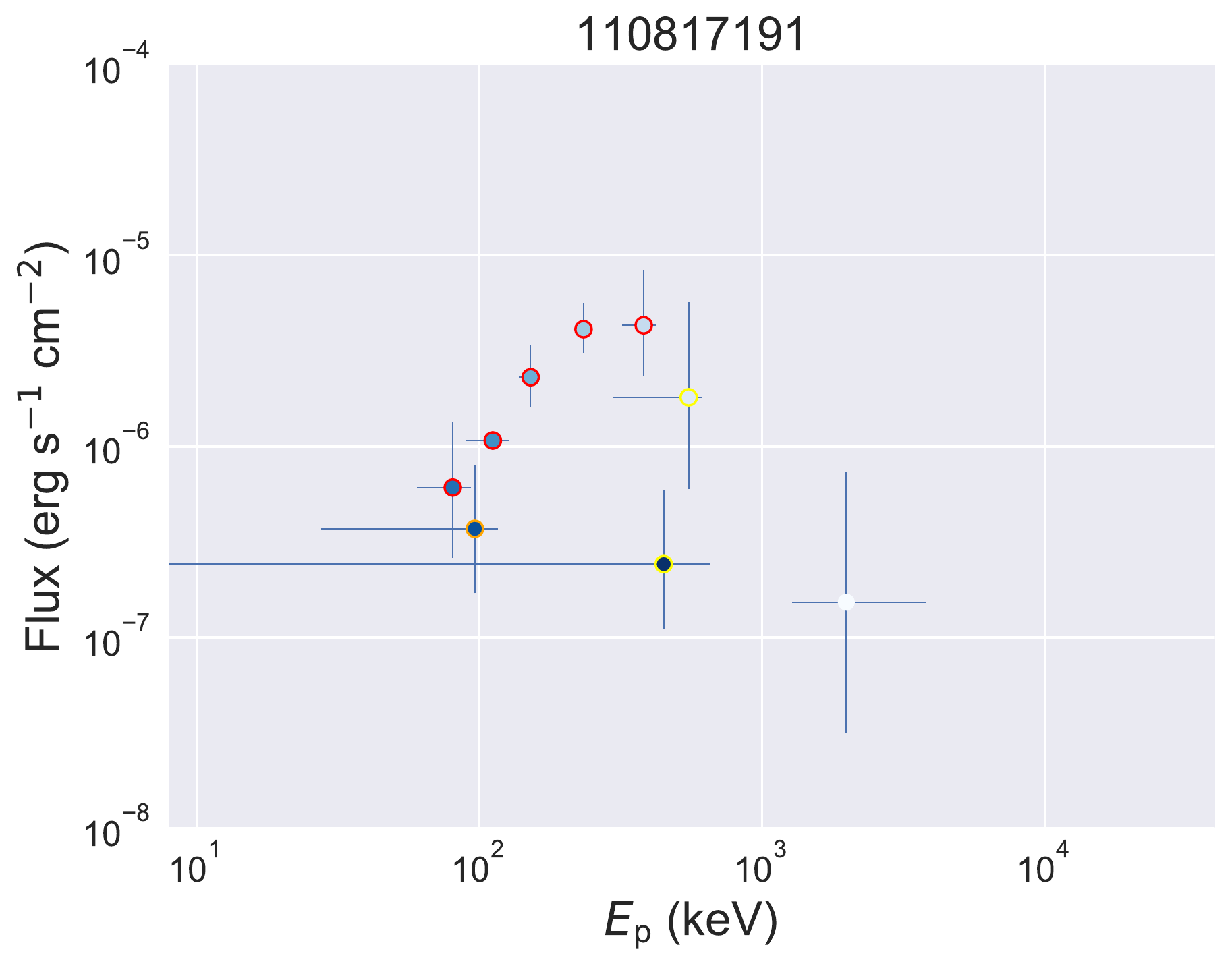}}
\subfigure{\includegraphics[width=0.3\linewidth]{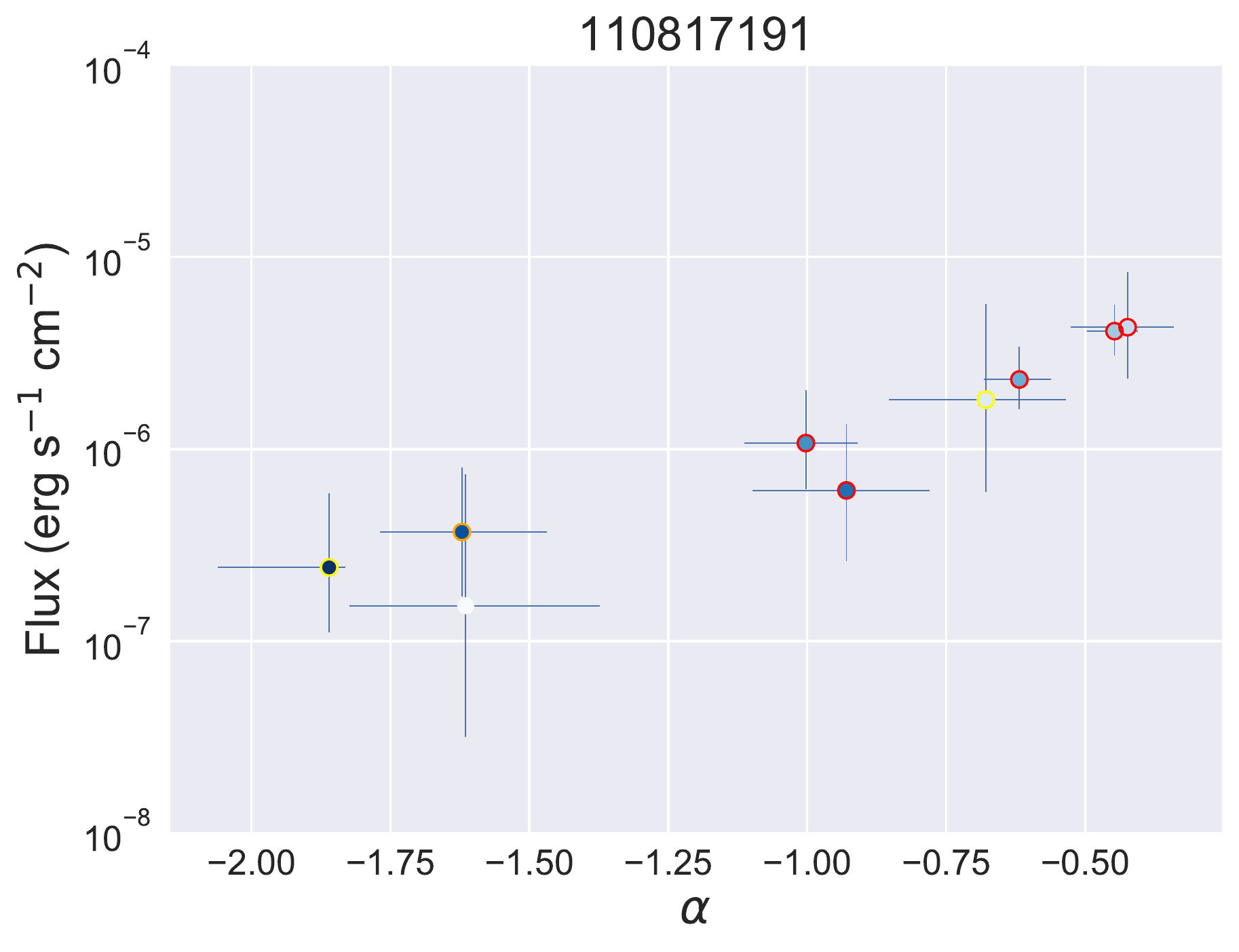}}

\subfigure{\includegraphics[width=0.3\linewidth]{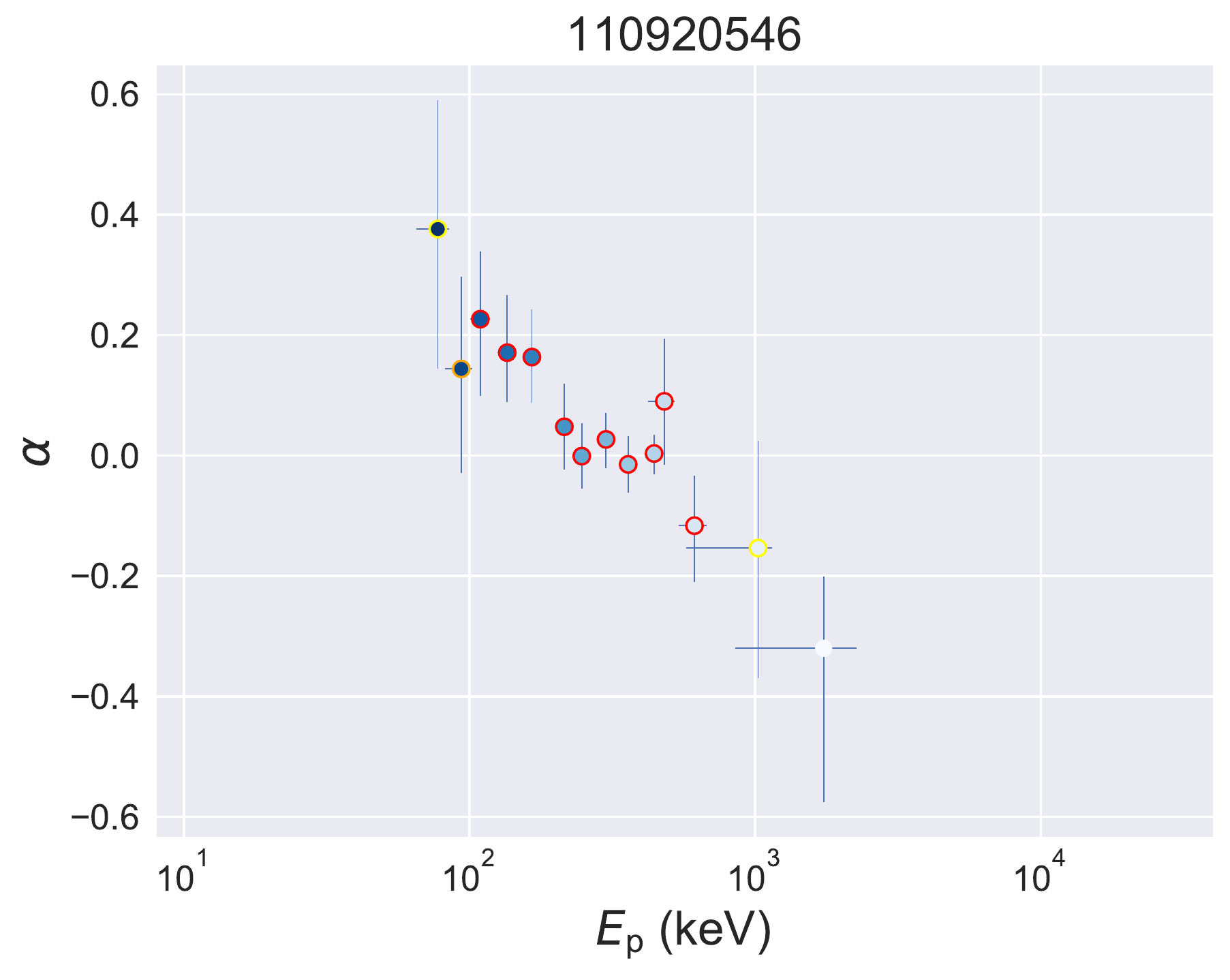}}
\subfigure{\includegraphics[width=0.3\linewidth]{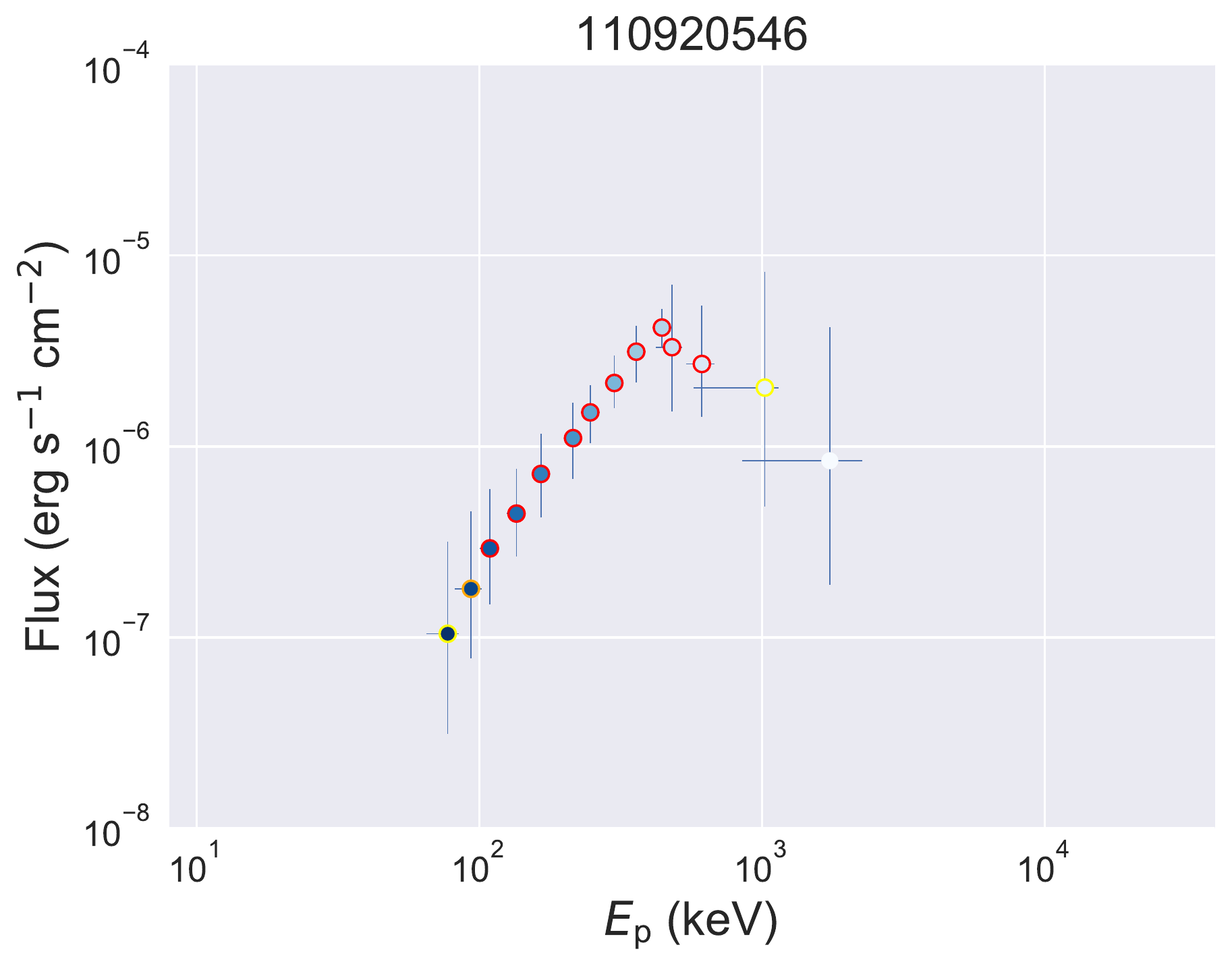}}
\subfigure{\includegraphics[width=0.3\linewidth]{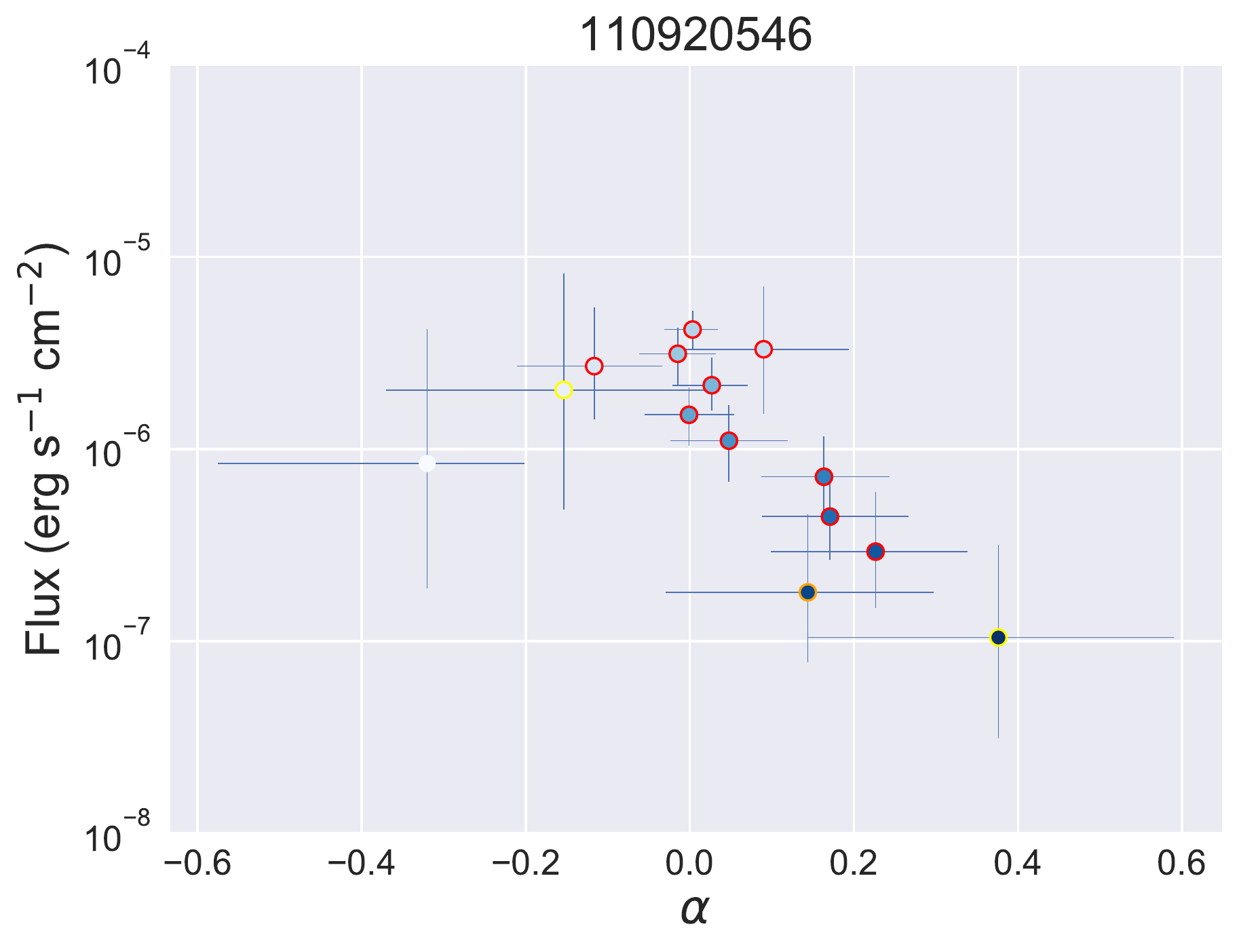}}

\subfigure{\includegraphics[width=0.3\linewidth]{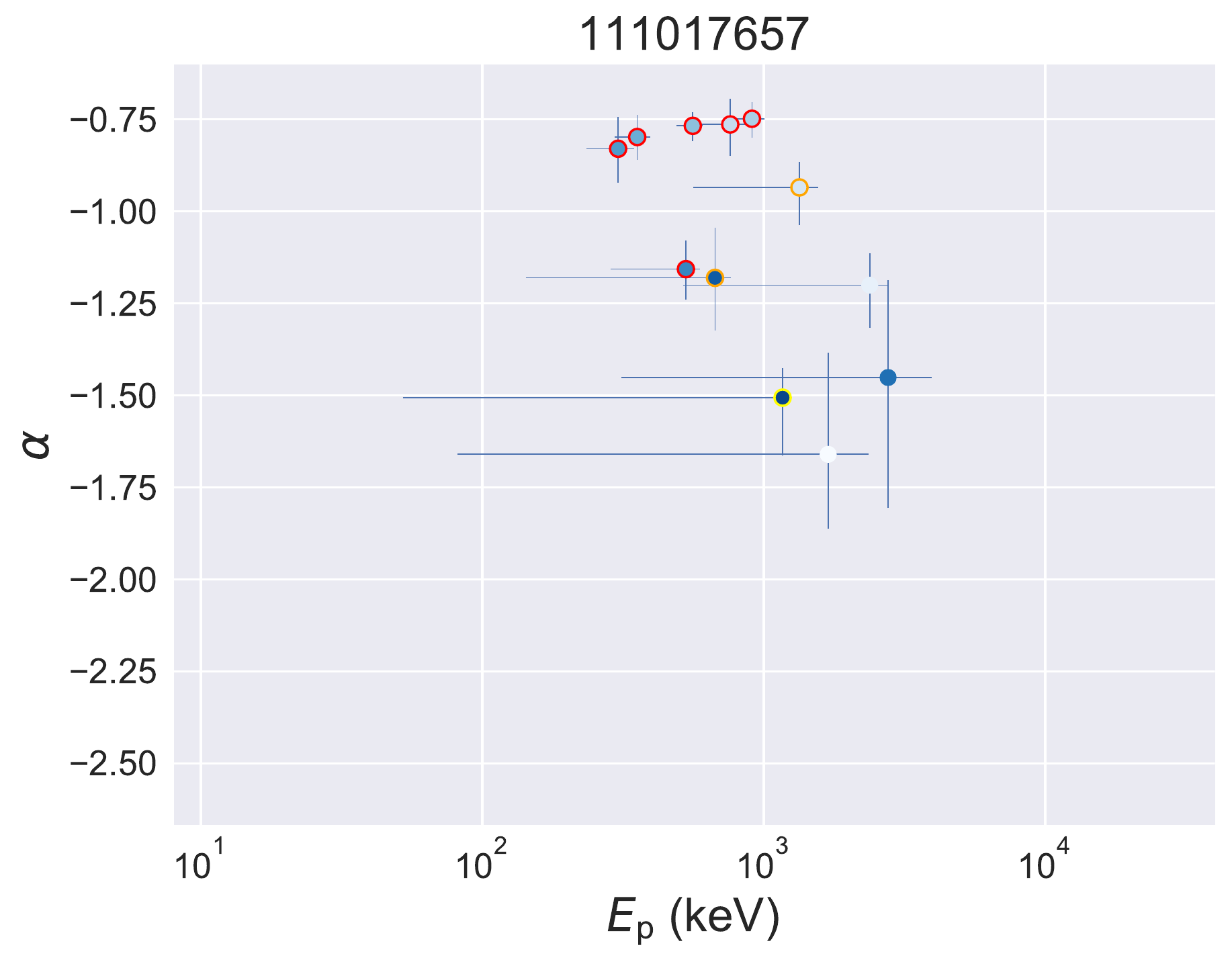}}
\subfigure{\includegraphics[width=0.3\linewidth]{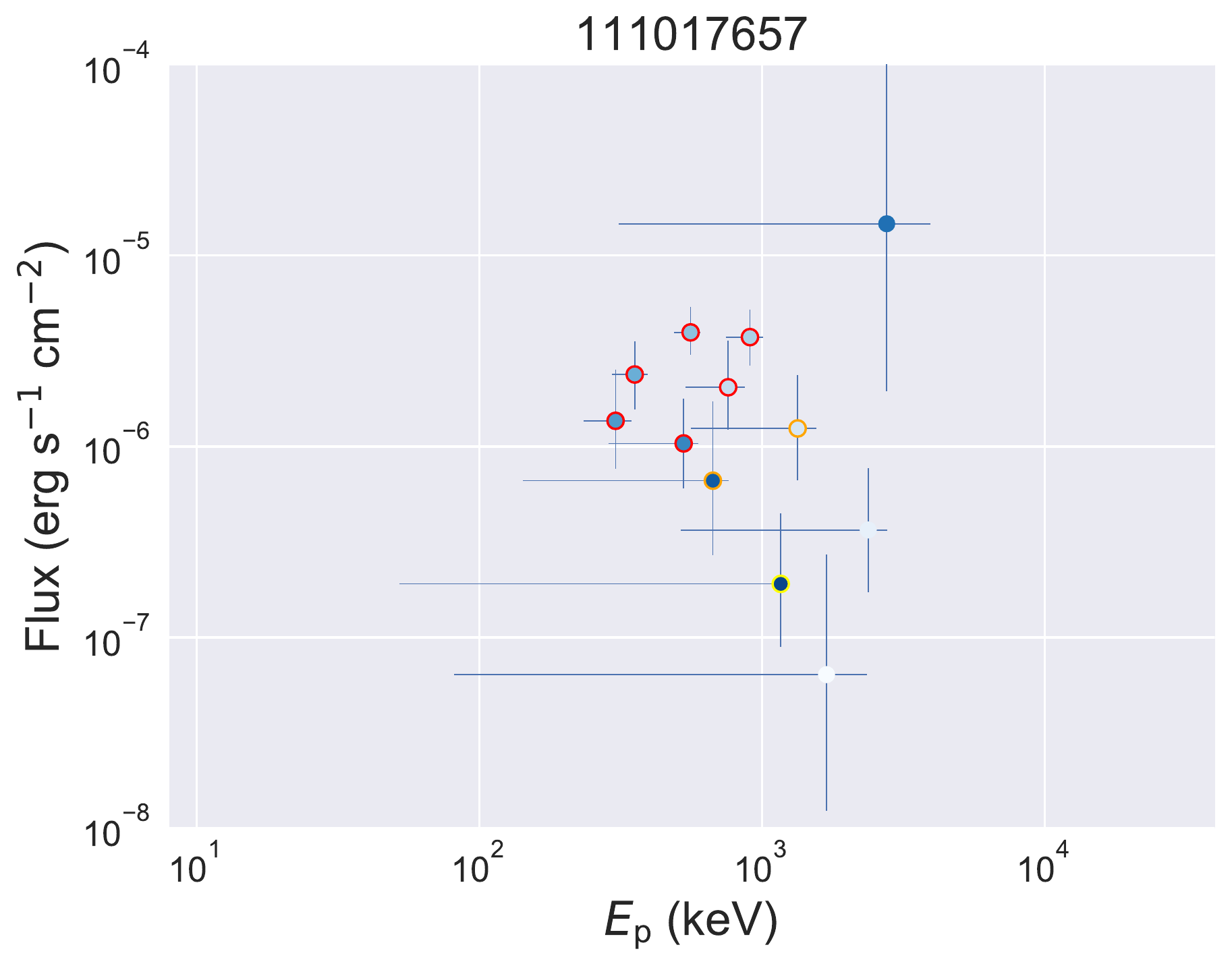}}
\subfigure{\includegraphics[width=0.3\linewidth]{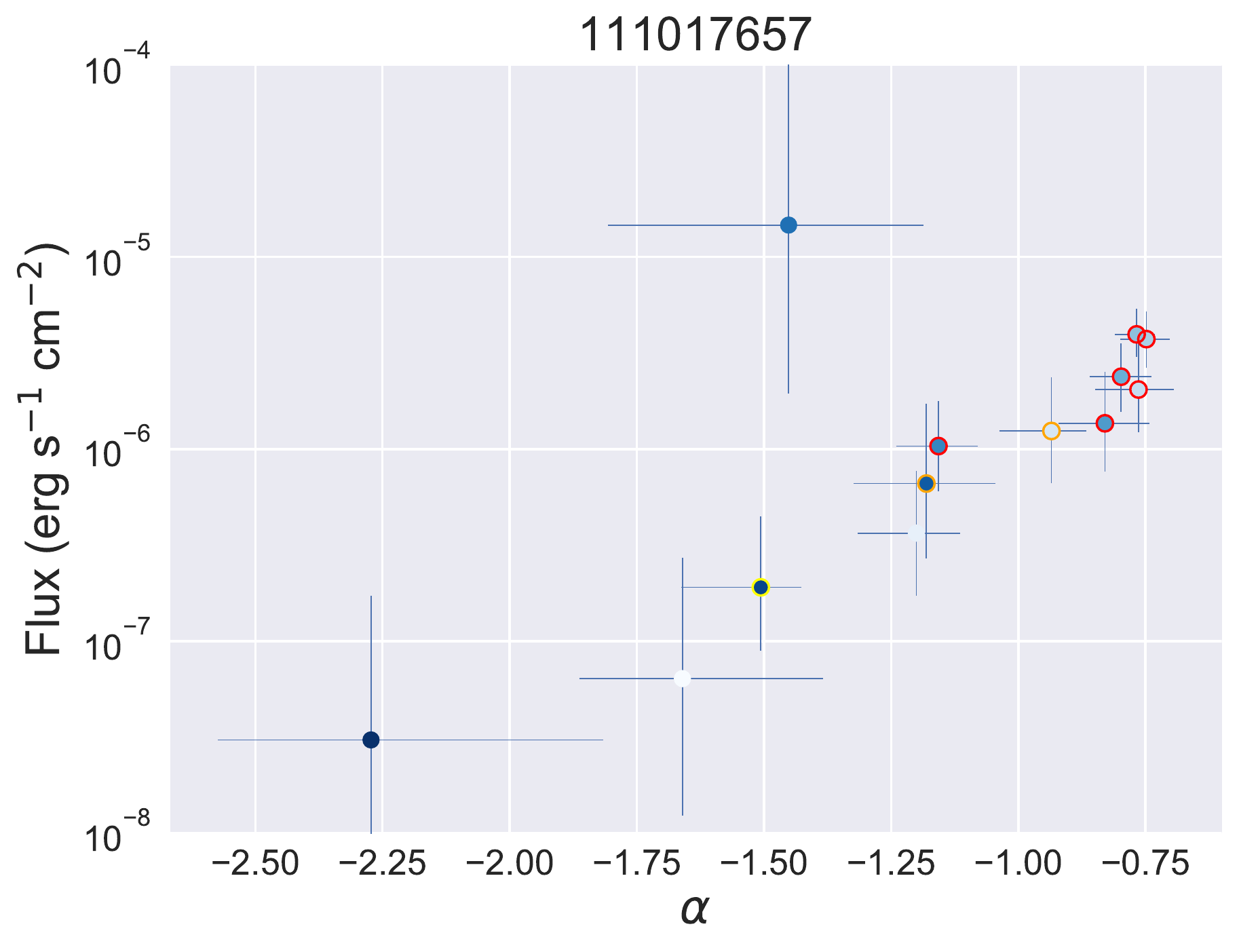}}

\subfigure{\includegraphics[width=0.3\linewidth]{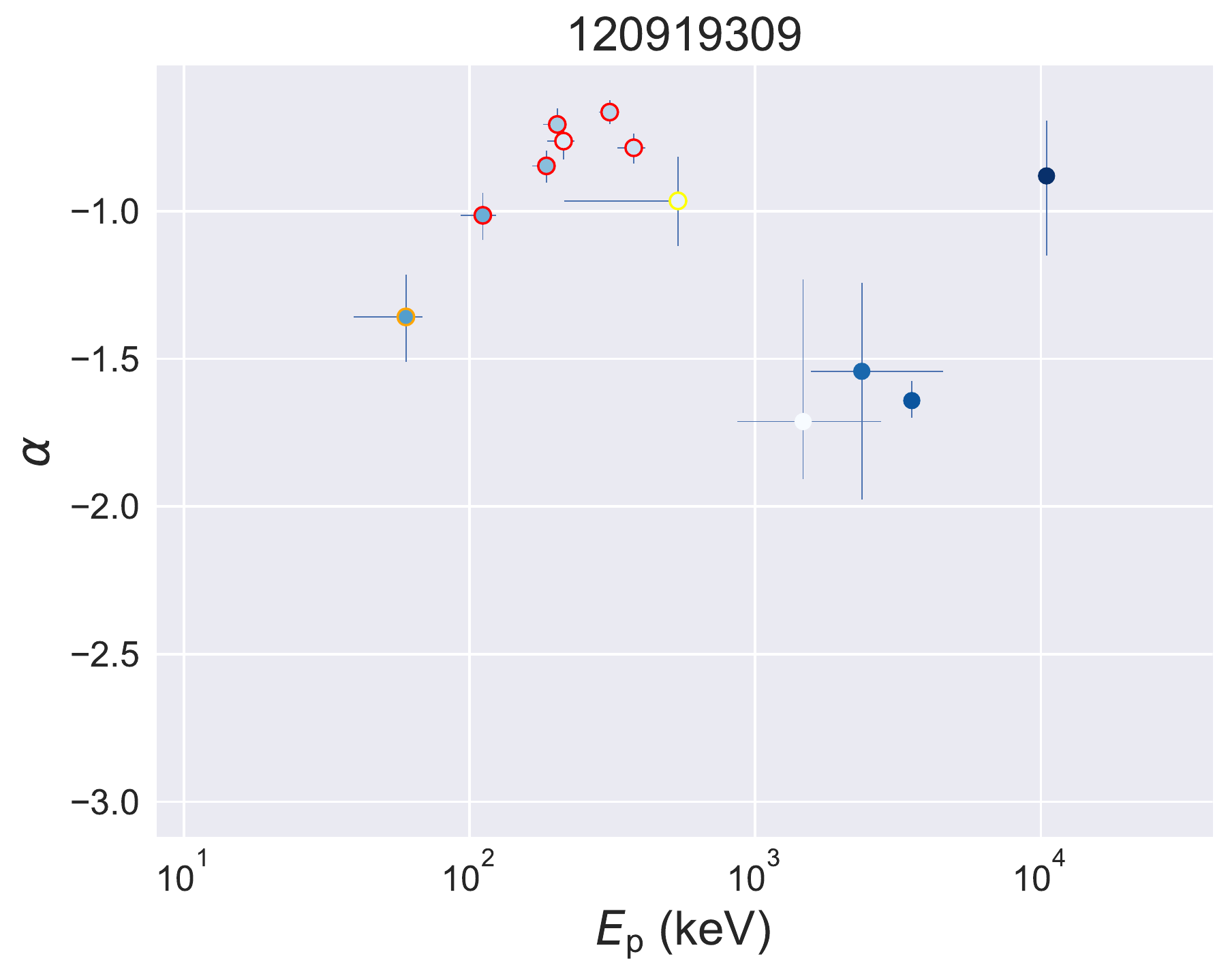}}
\subfigure{\includegraphics[width=0.3\linewidth]{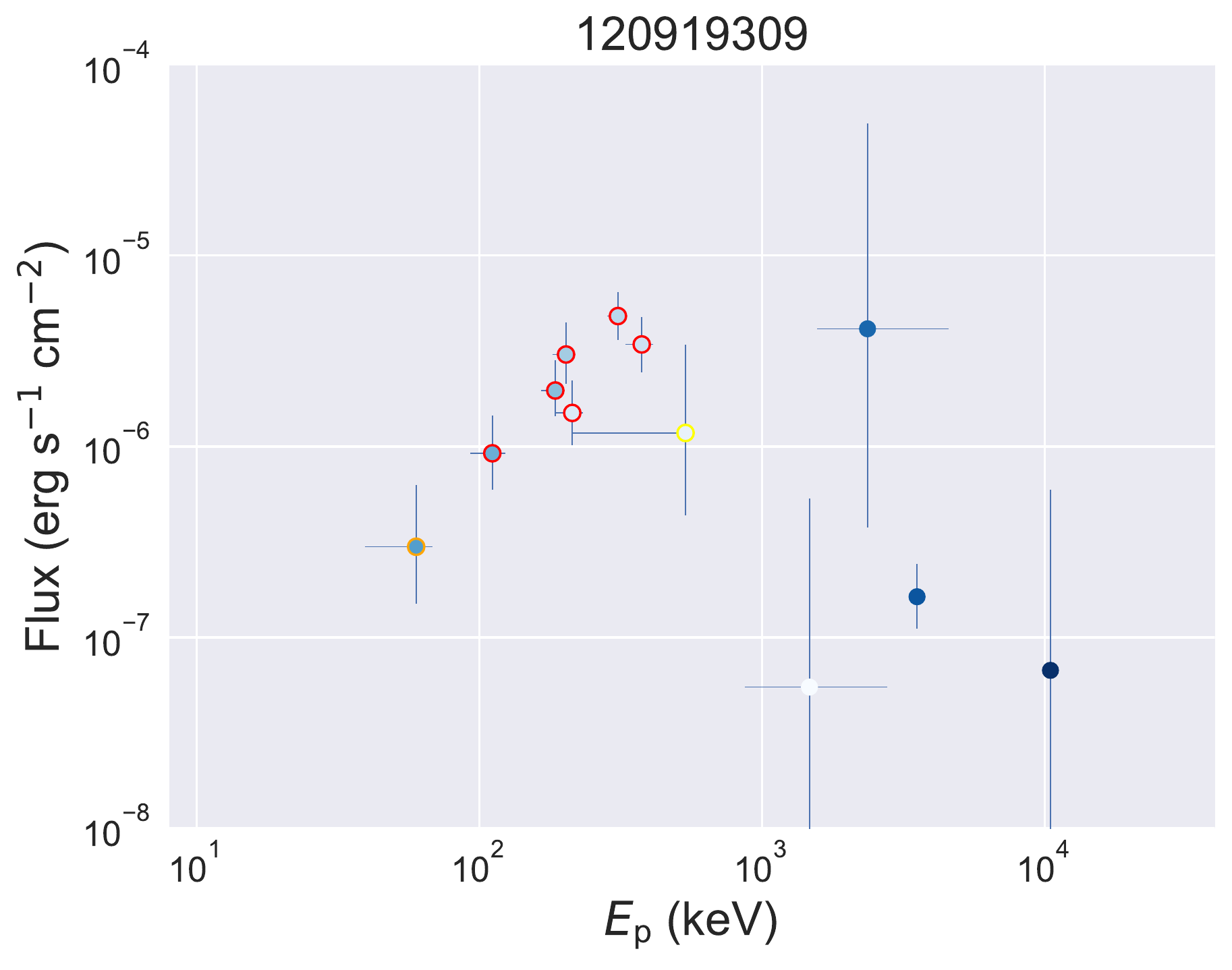}}
\subfigure{\includegraphics[width=0.3\linewidth]{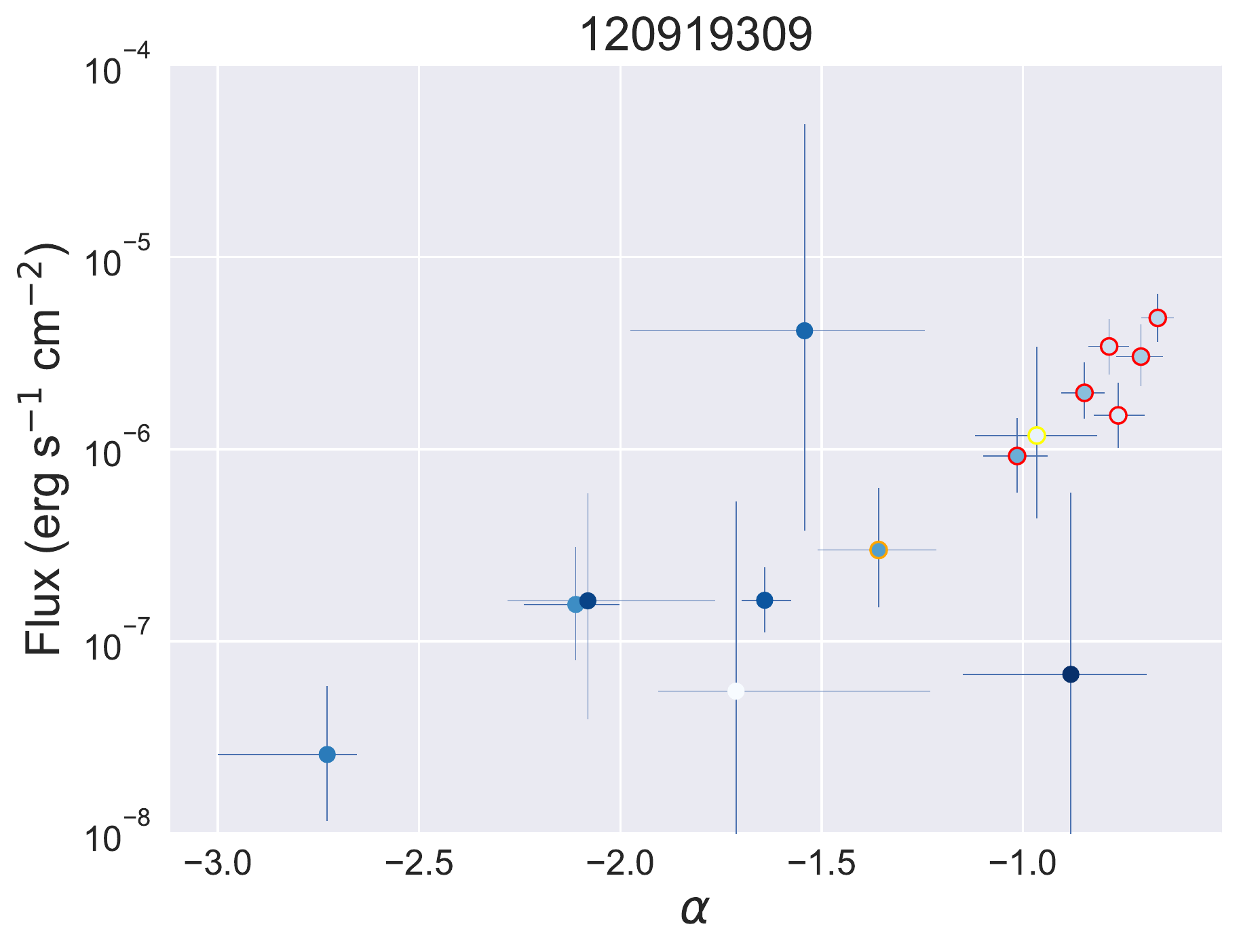}}

\caption{Same as Fig.~\ref{fig:correlation_group1}.
\label{fig:correlation_group5}}
\end{figure*}

\begin{figure*}

\subfigure{\includegraphics[width=0.3\linewidth]{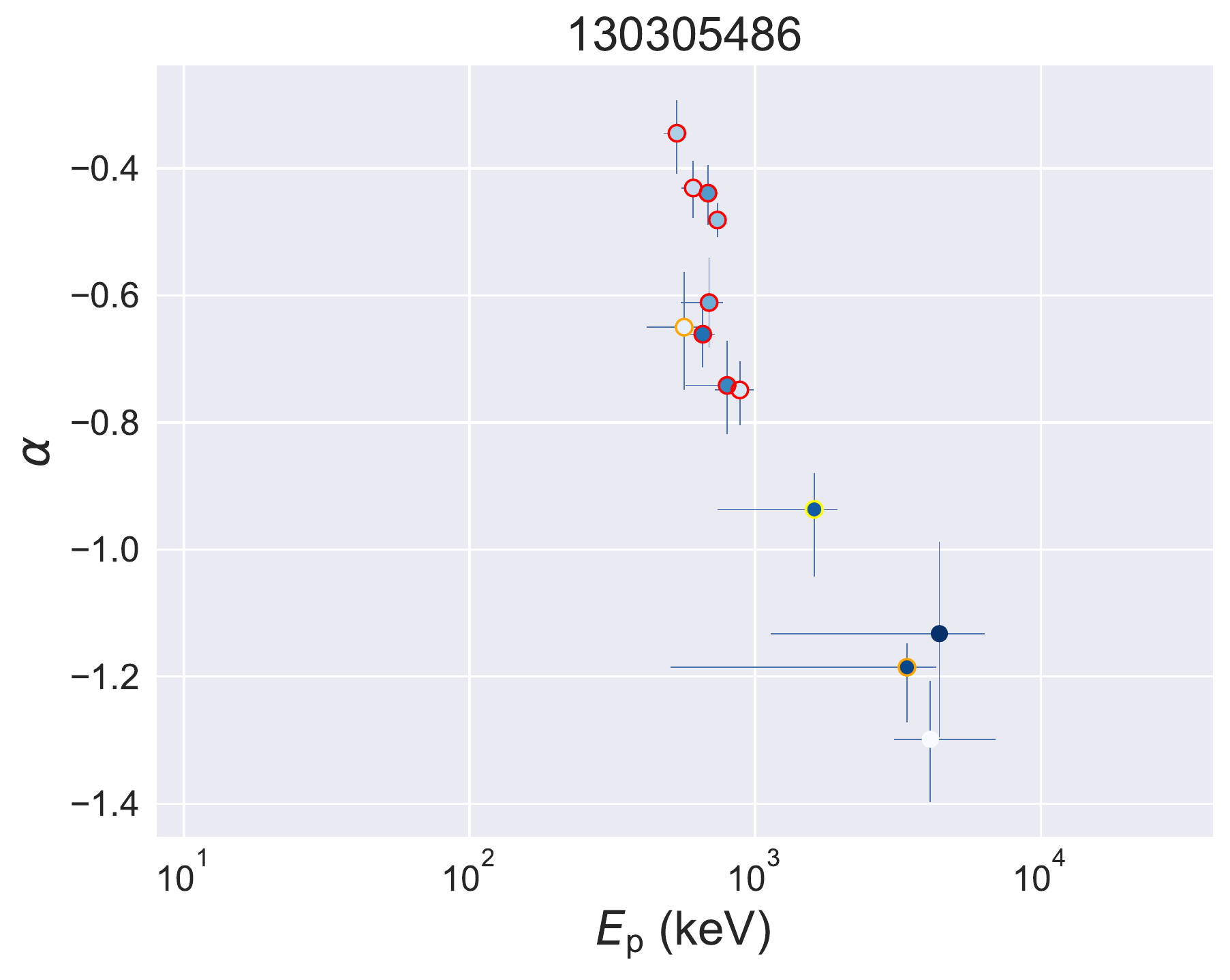}}
\subfigure{\includegraphics[width=0.3\linewidth]{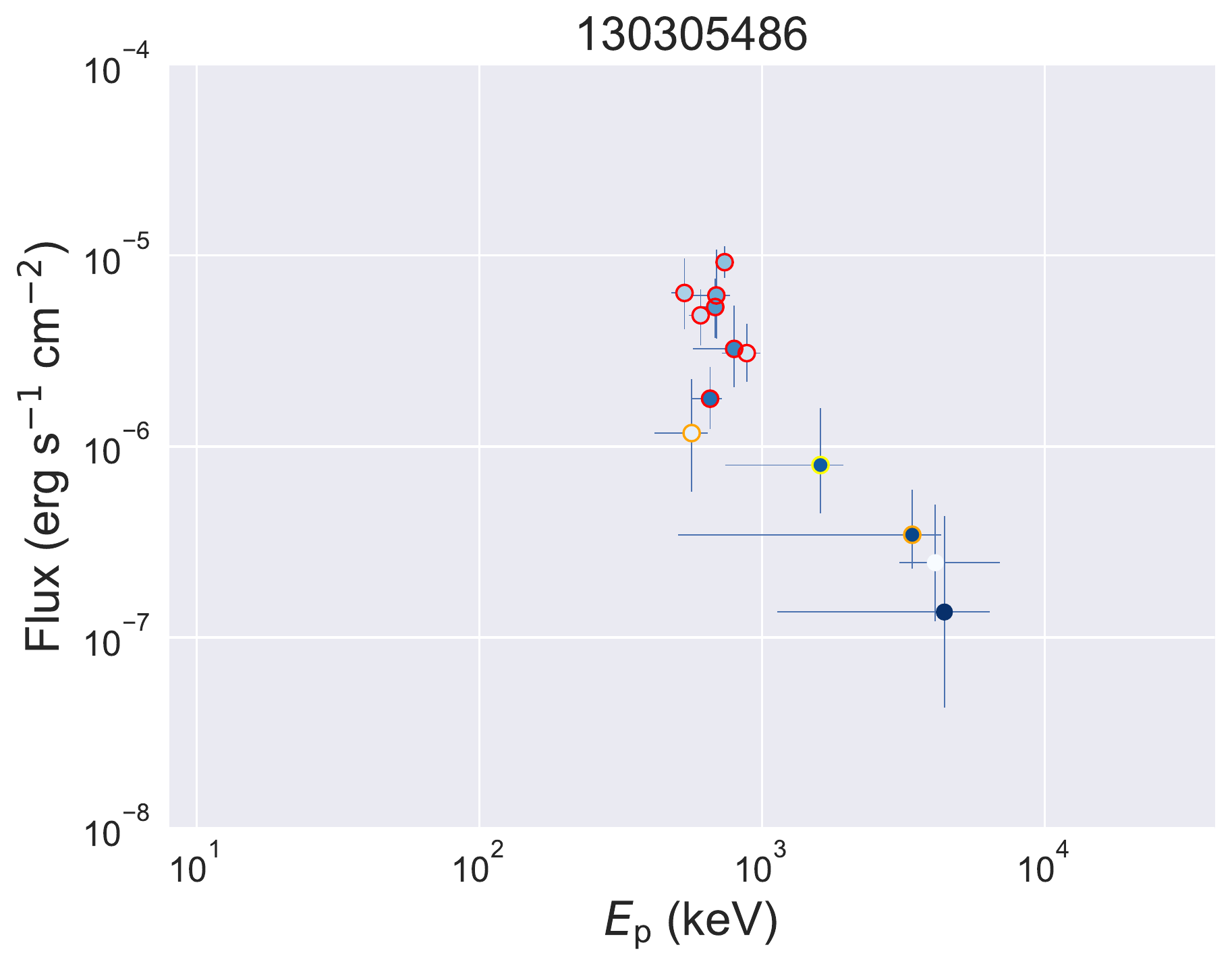}}
\subfigure{\includegraphics[width=0.3\linewidth]{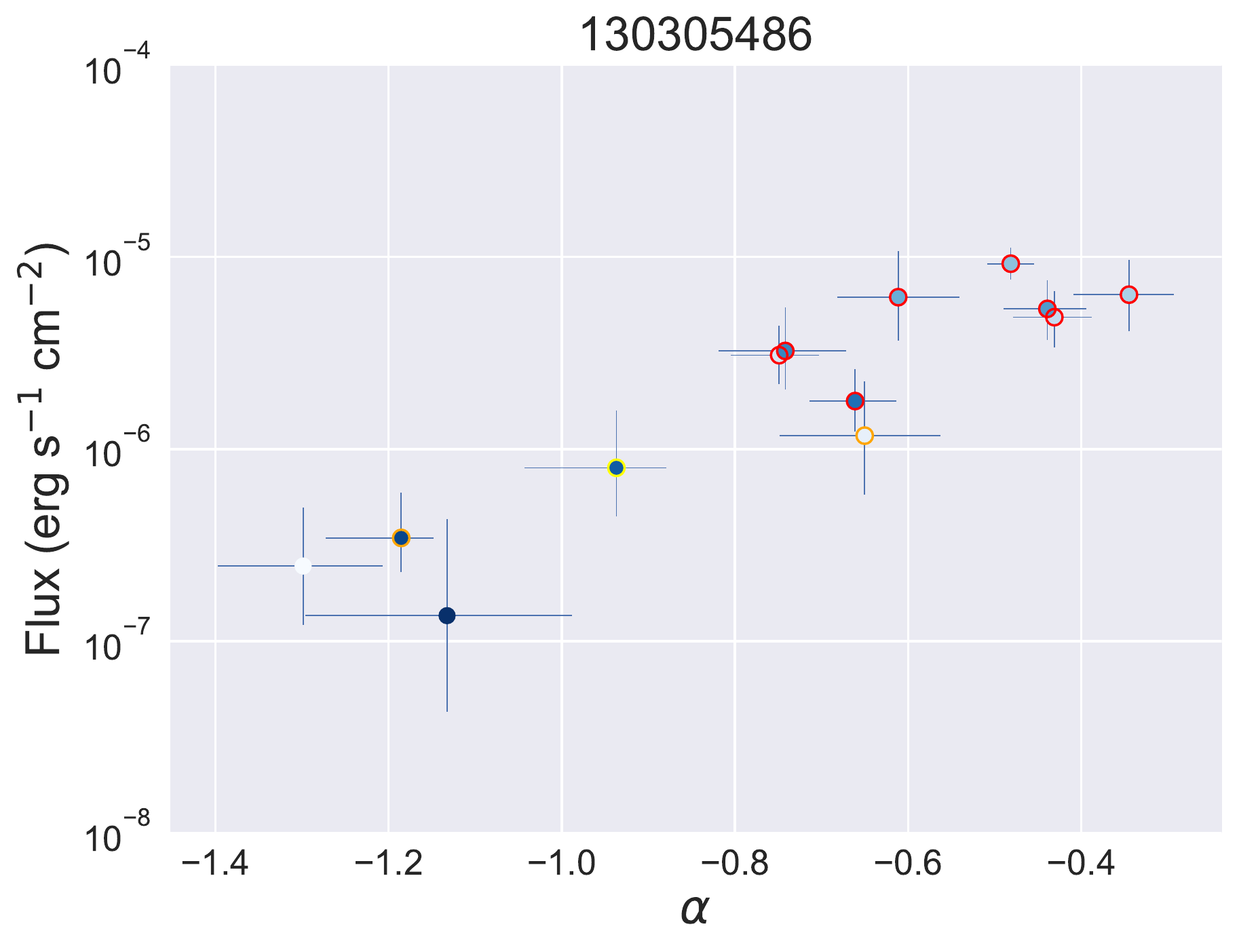}}

\subfigure{\includegraphics[width=0.3\linewidth]{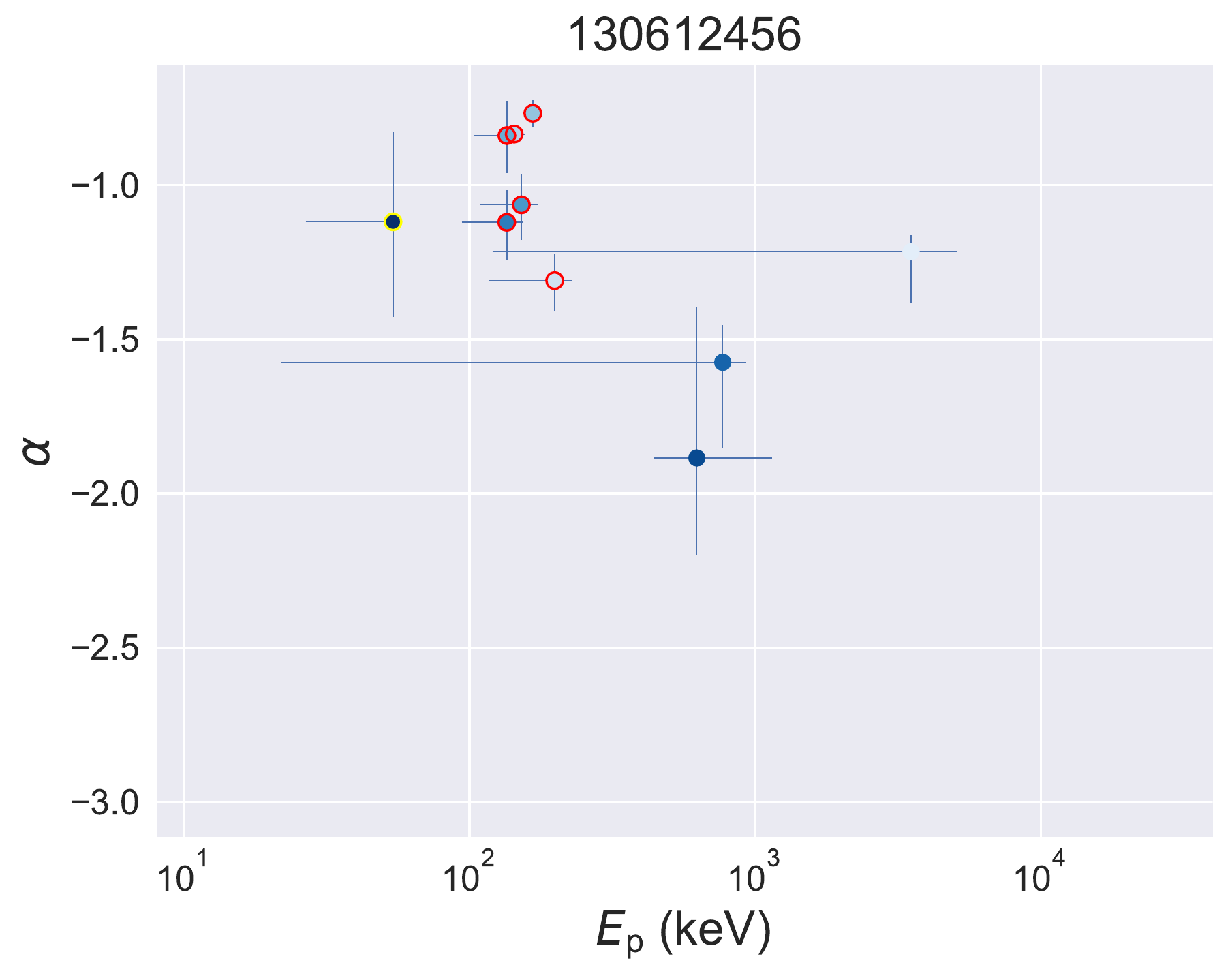}}
\subfigure{\includegraphics[width=0.3\linewidth]{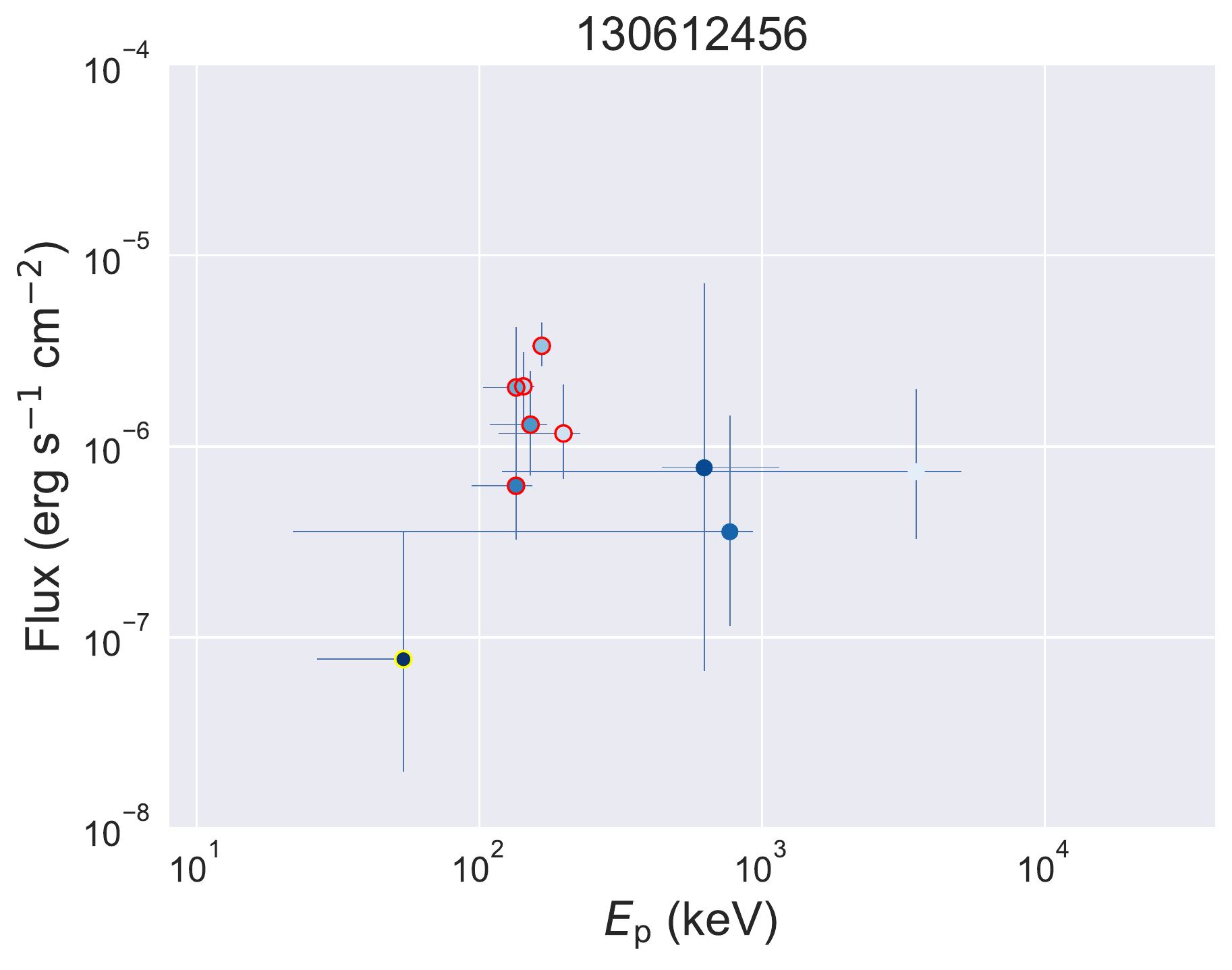}}
\subfigure{\includegraphics[width=0.3\linewidth]{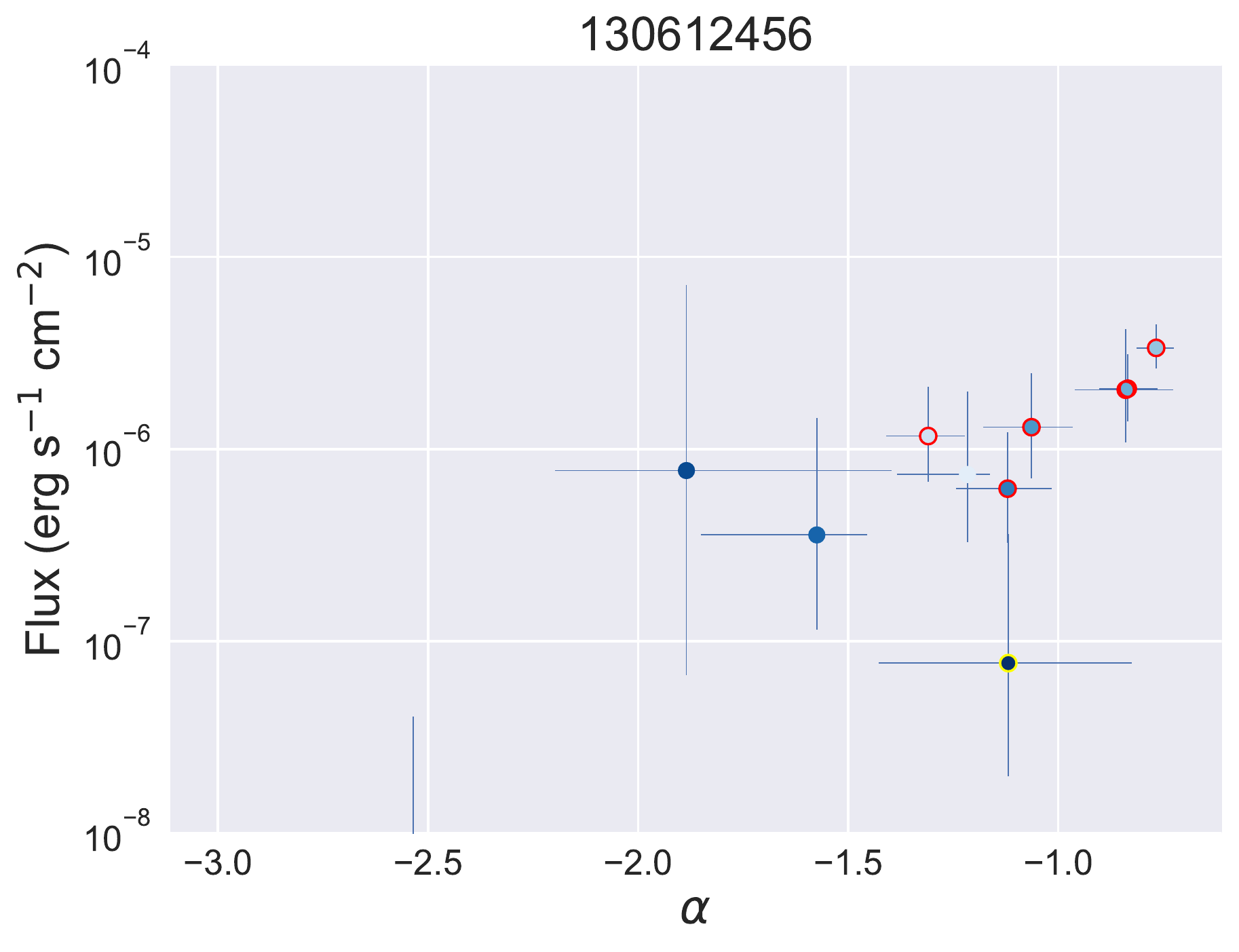}}

\subfigure{\includegraphics[width=0.3\linewidth]{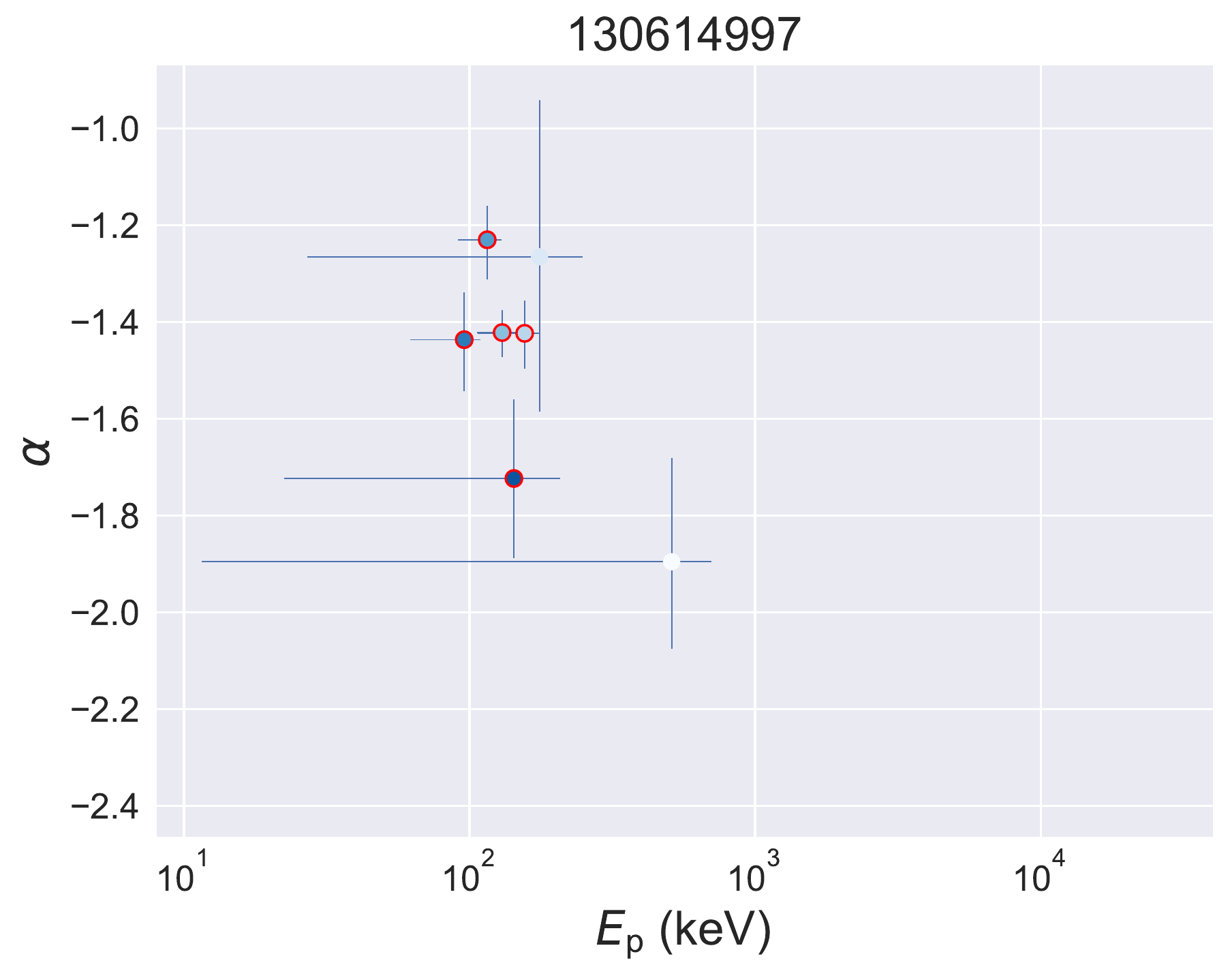}}
\subfigure{\includegraphics[width=0.3\linewidth]{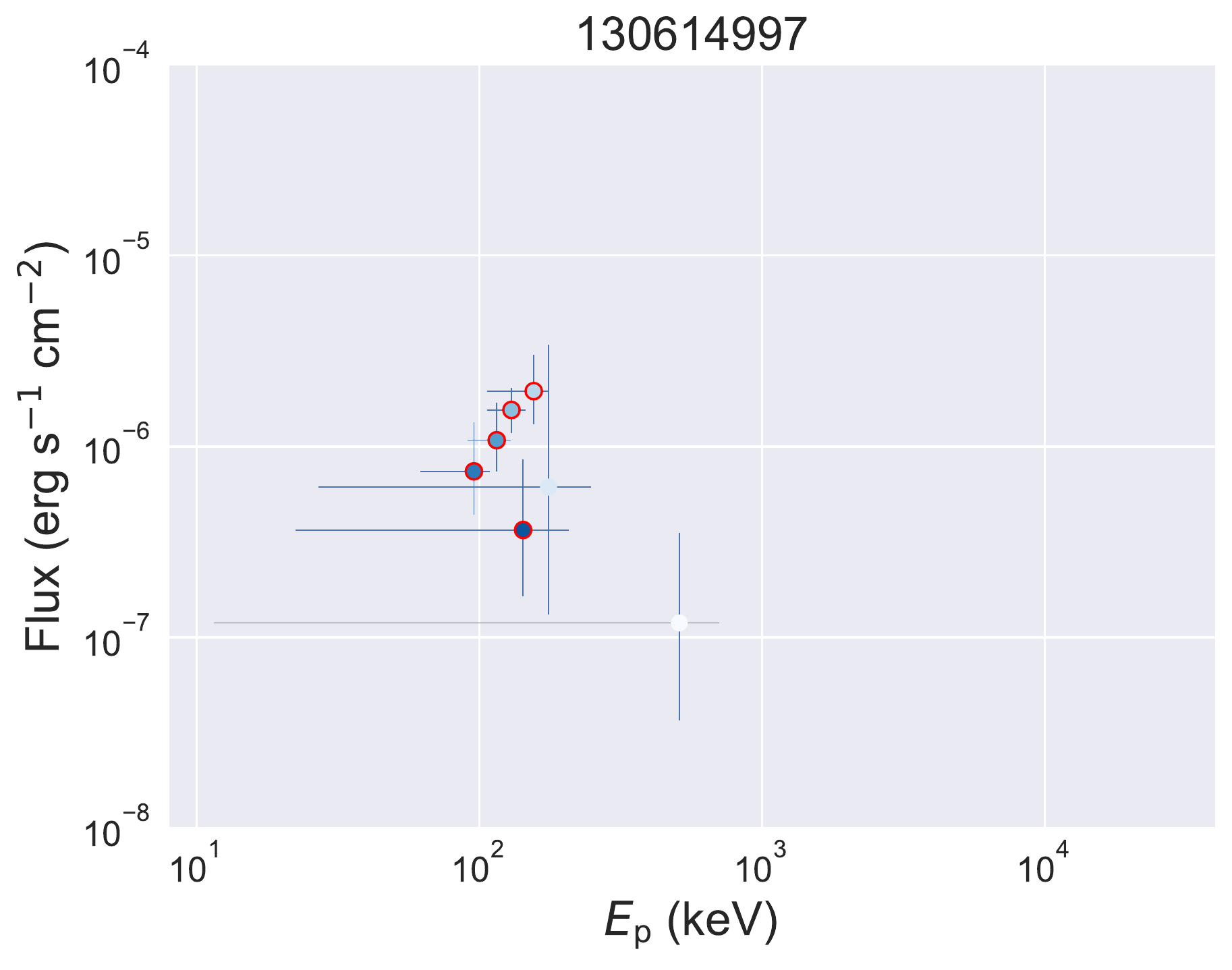}}
\subfigure{\includegraphics[width=0.3\linewidth]{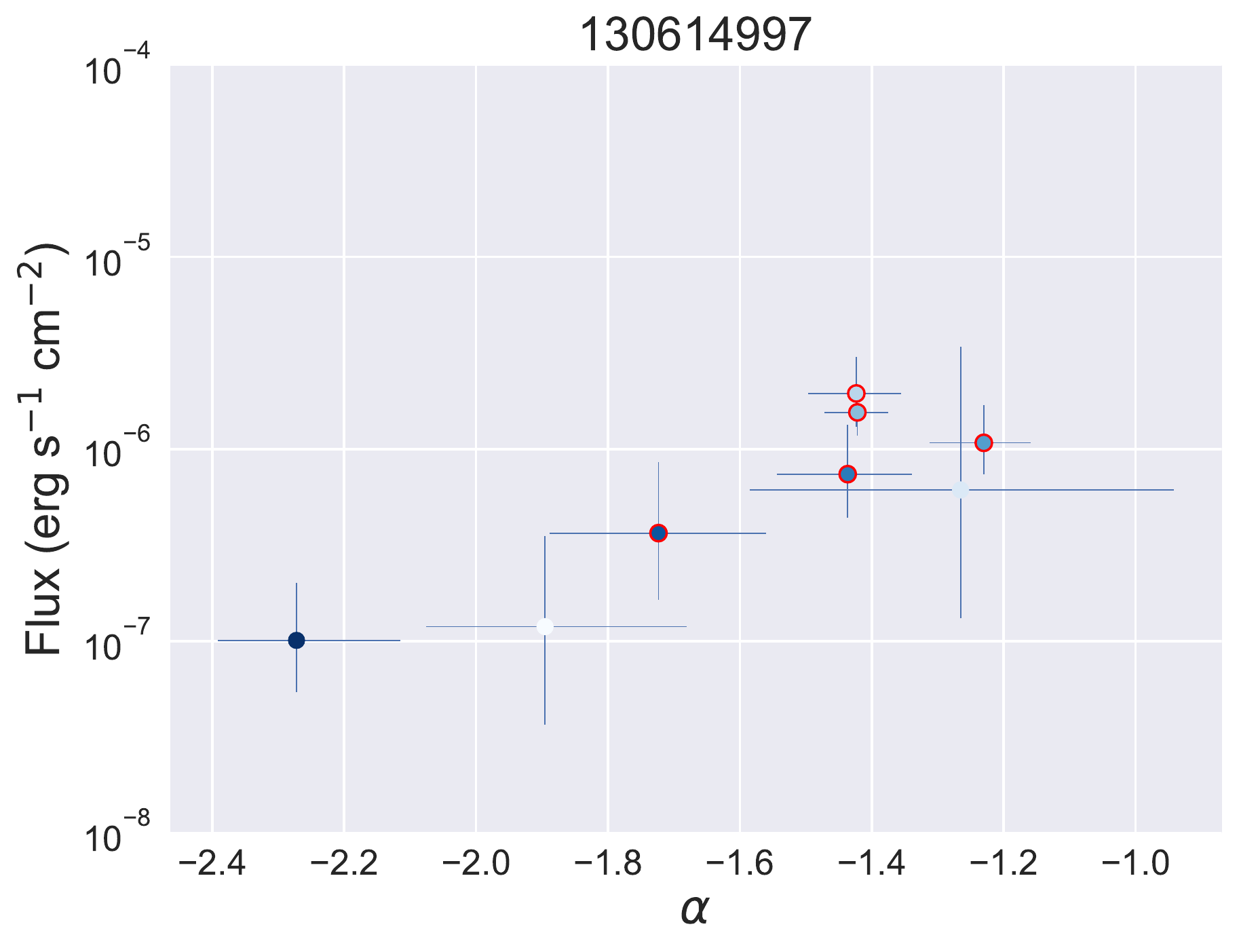}}

\subfigure{\includegraphics[width=0.3\linewidth]{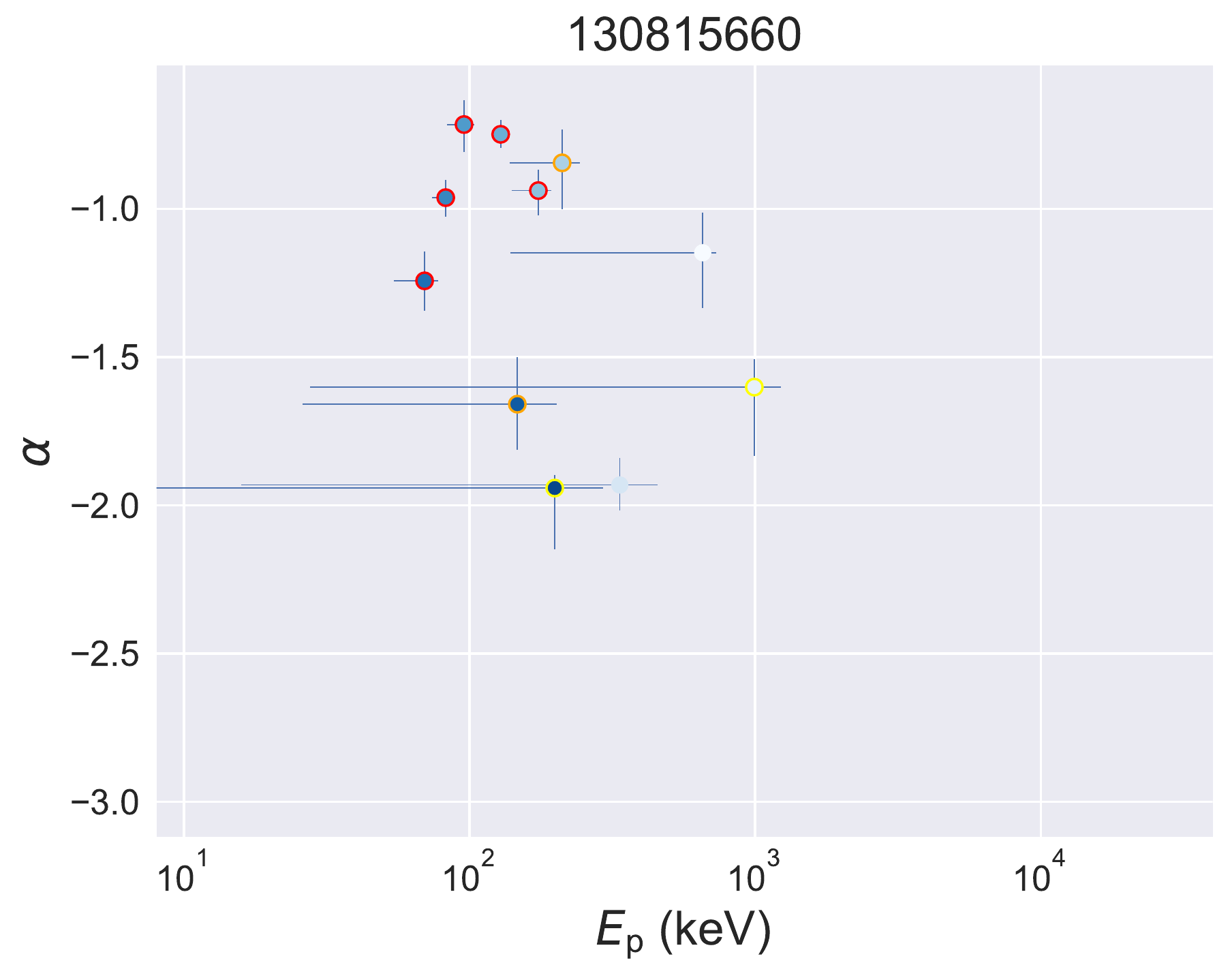}}
\subfigure{\includegraphics[width=0.3\linewidth]{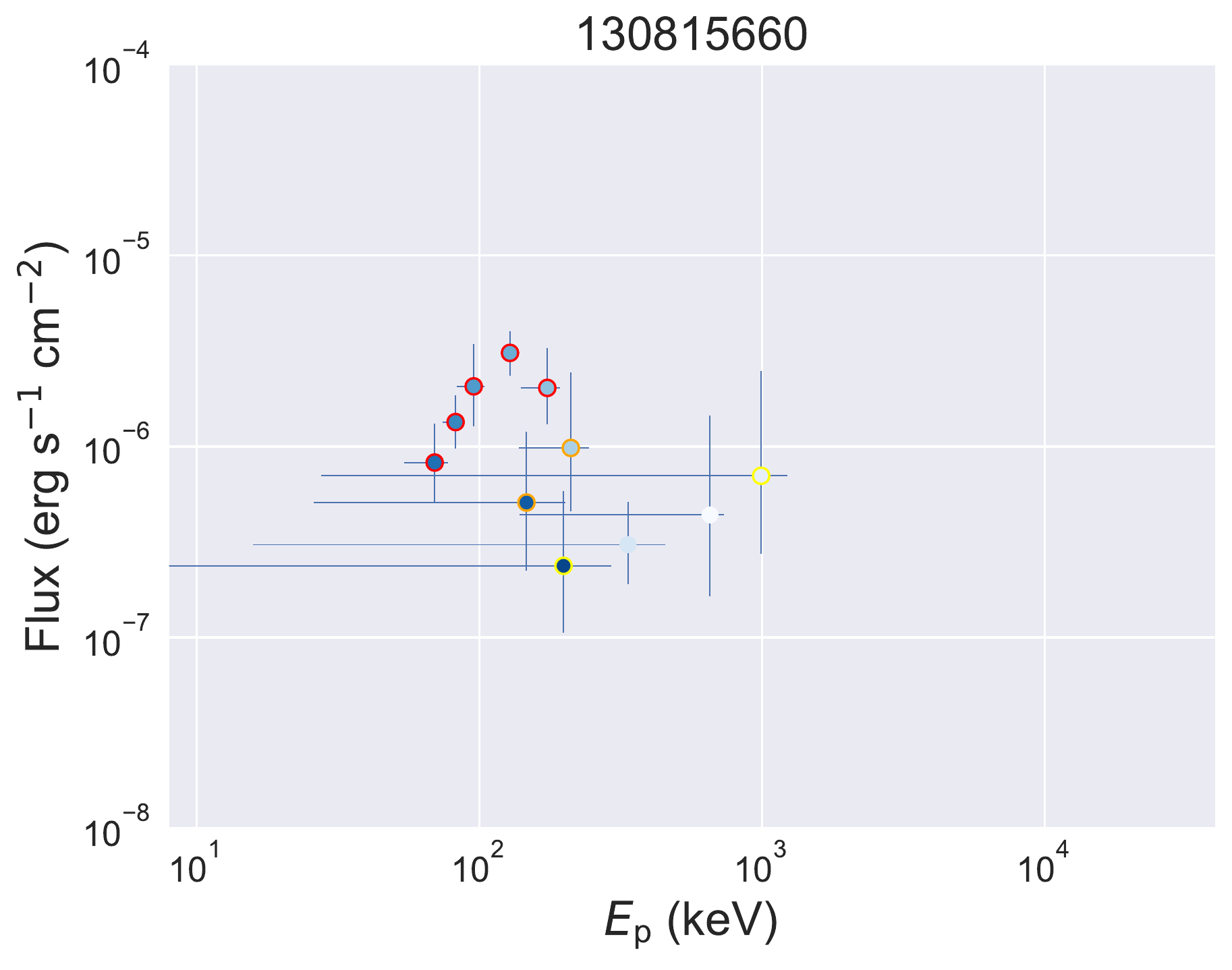}}
\subfigure{\includegraphics[width=0.3\linewidth]{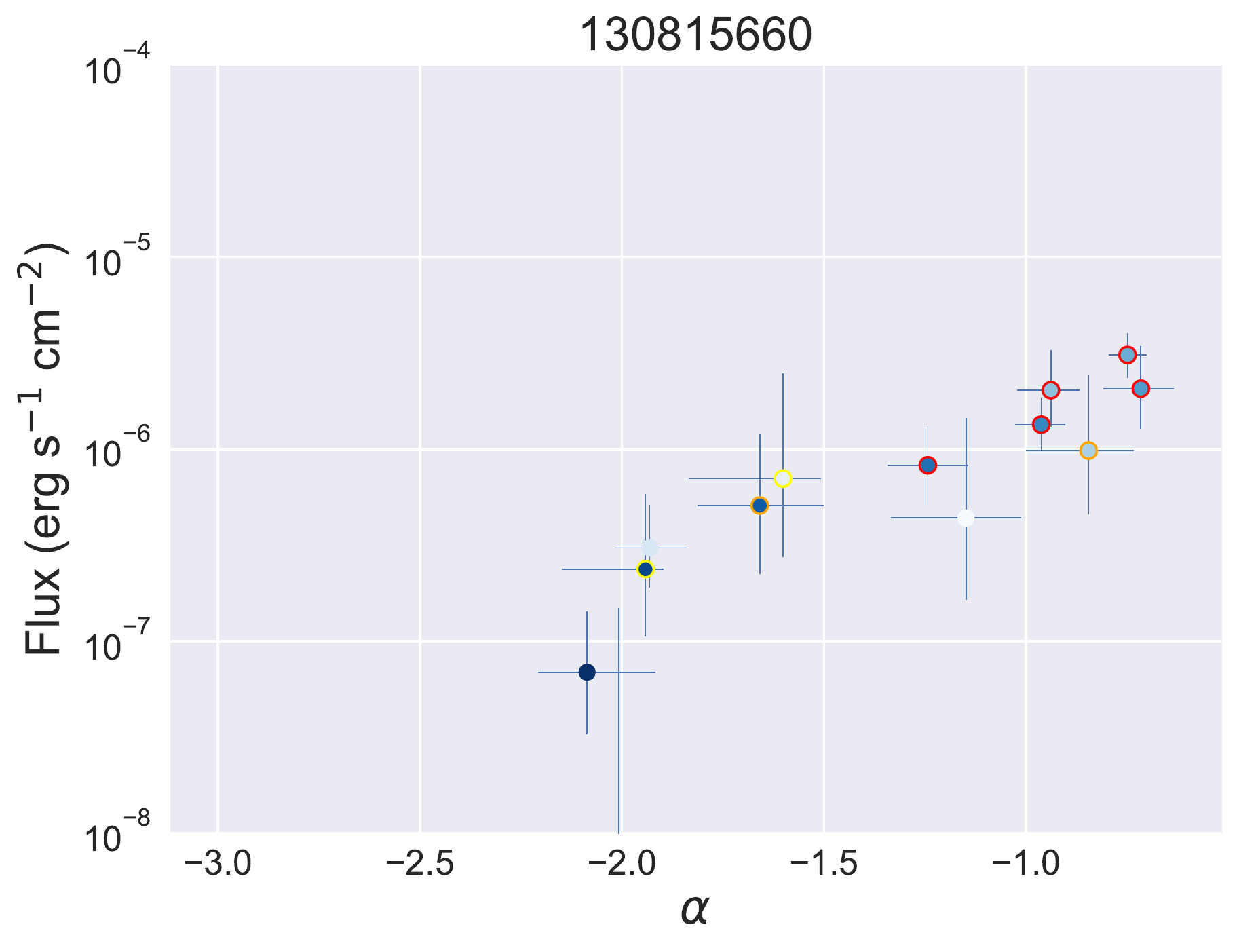}}

\caption{Same as Fig.~\ref{fig:correlation_group1}.
\label{fig:correlation_group6}}
\end{figure*}

\begin{figure*}

\subfigure{\includegraphics[width=0.3\linewidth]{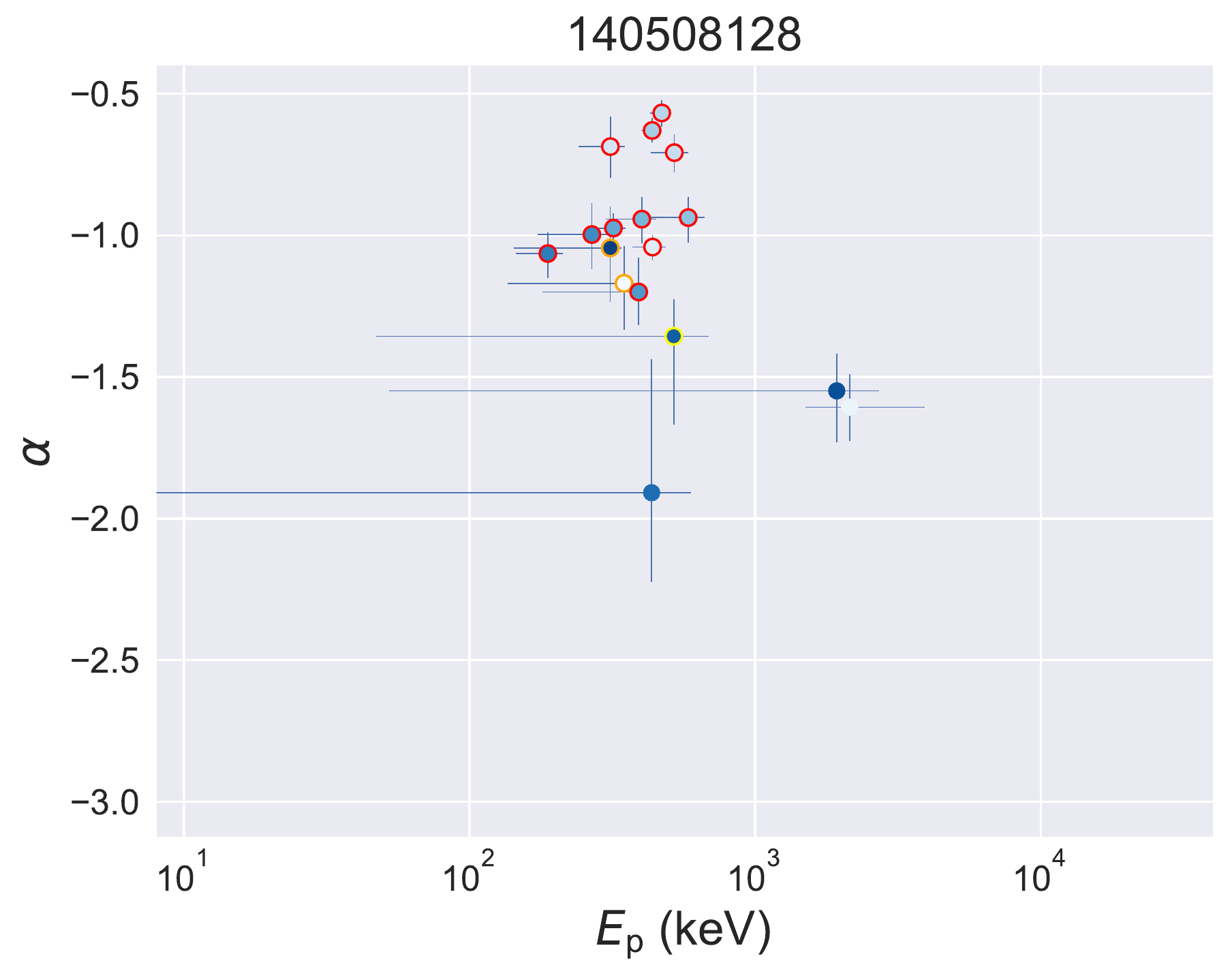}}
\subfigure{\includegraphics[width=0.3\linewidth]{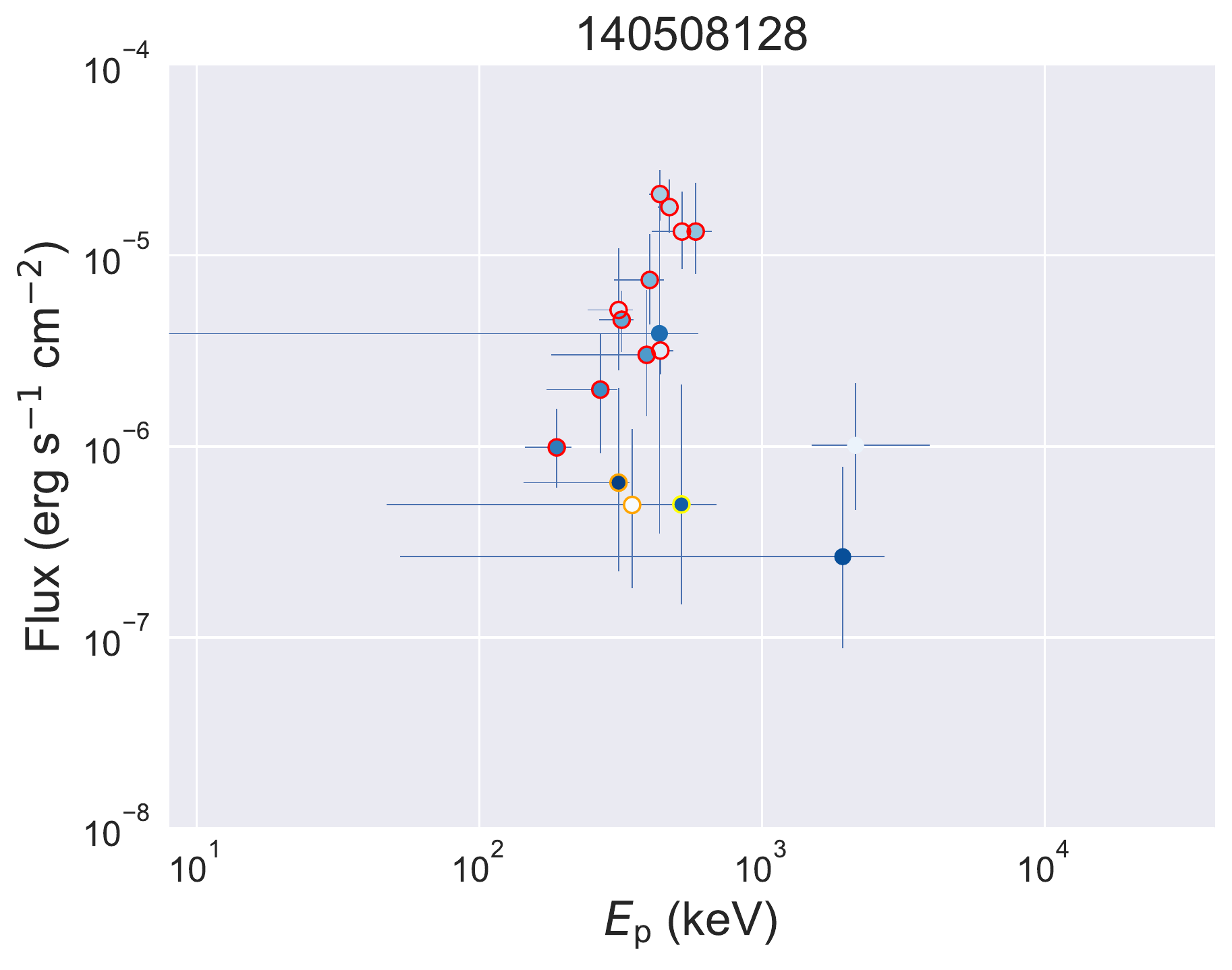}}
\subfigure{\includegraphics[width=0.3\linewidth]{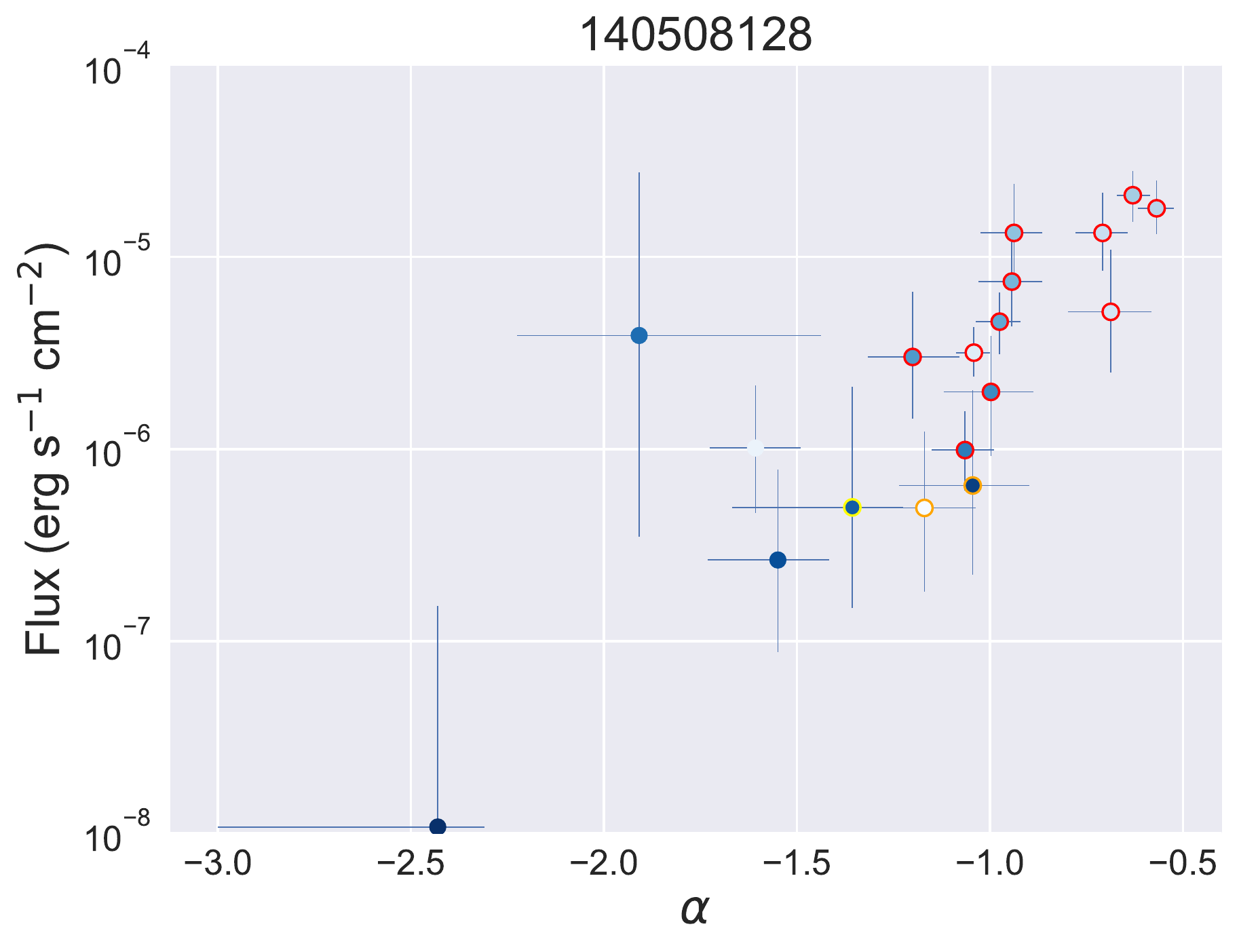}}

\subfigure{\includegraphics[width=0.3\linewidth]{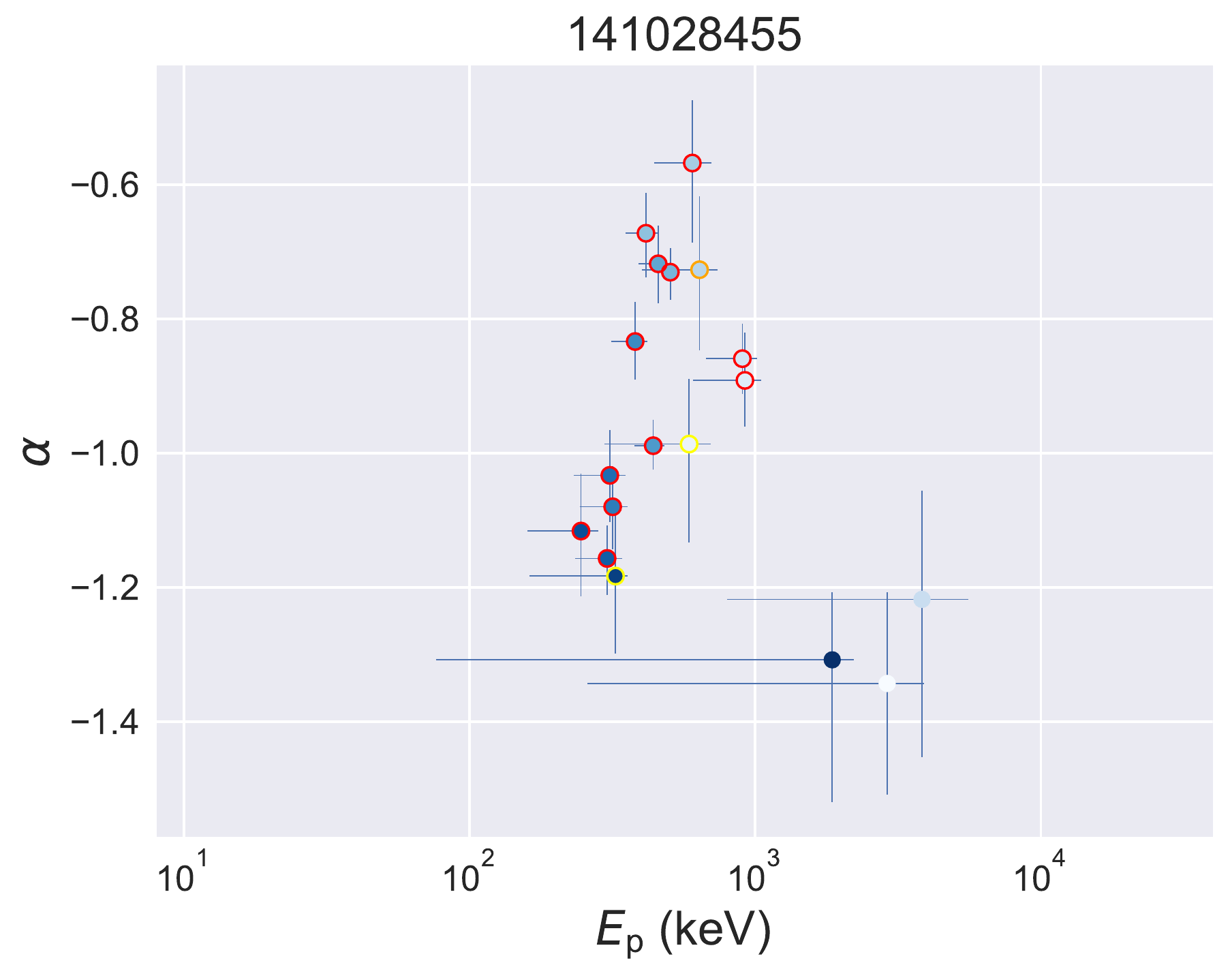}}
\subfigure{\includegraphics[width=0.3\linewidth]{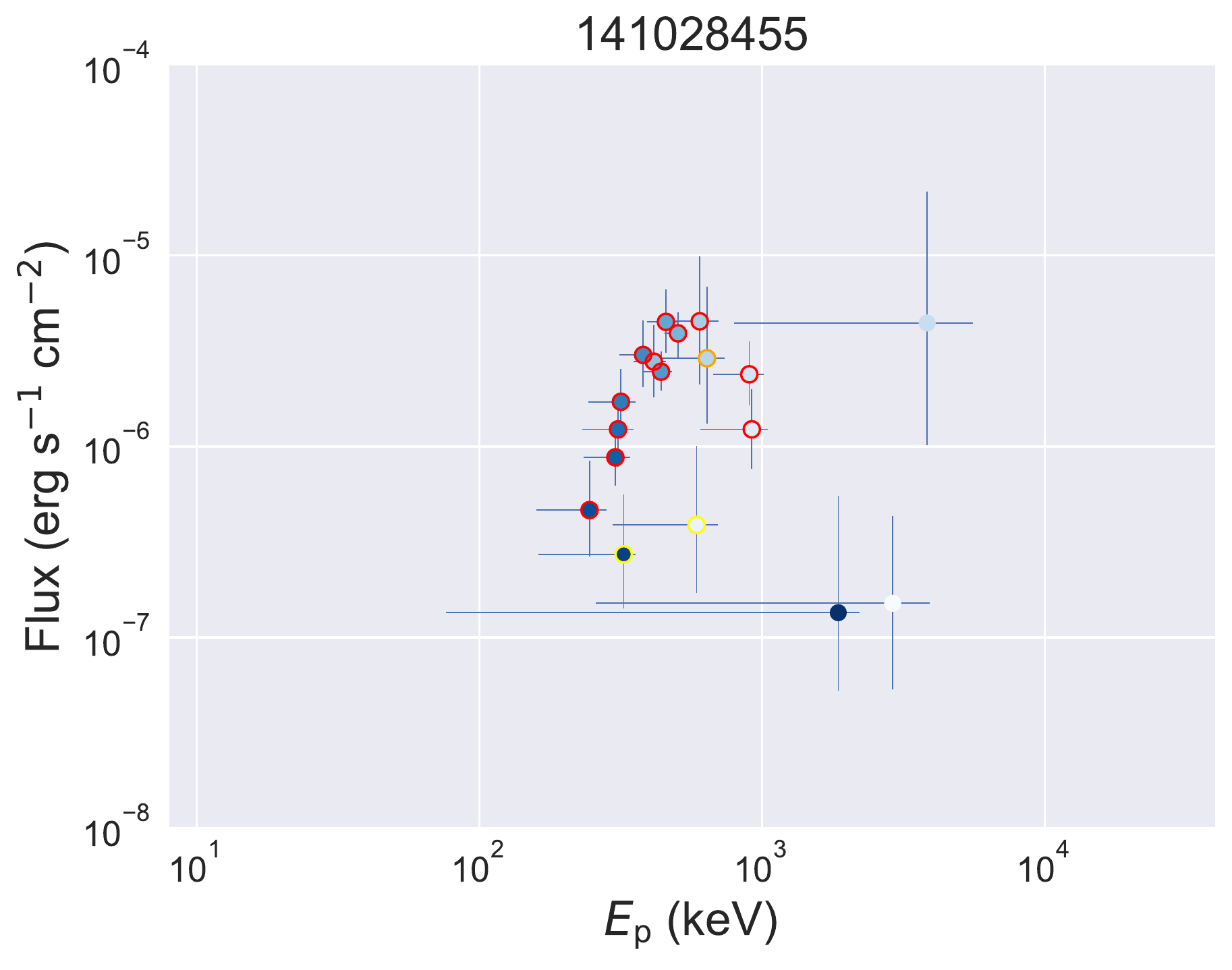}}
\subfigure{\includegraphics[width=0.3\linewidth]{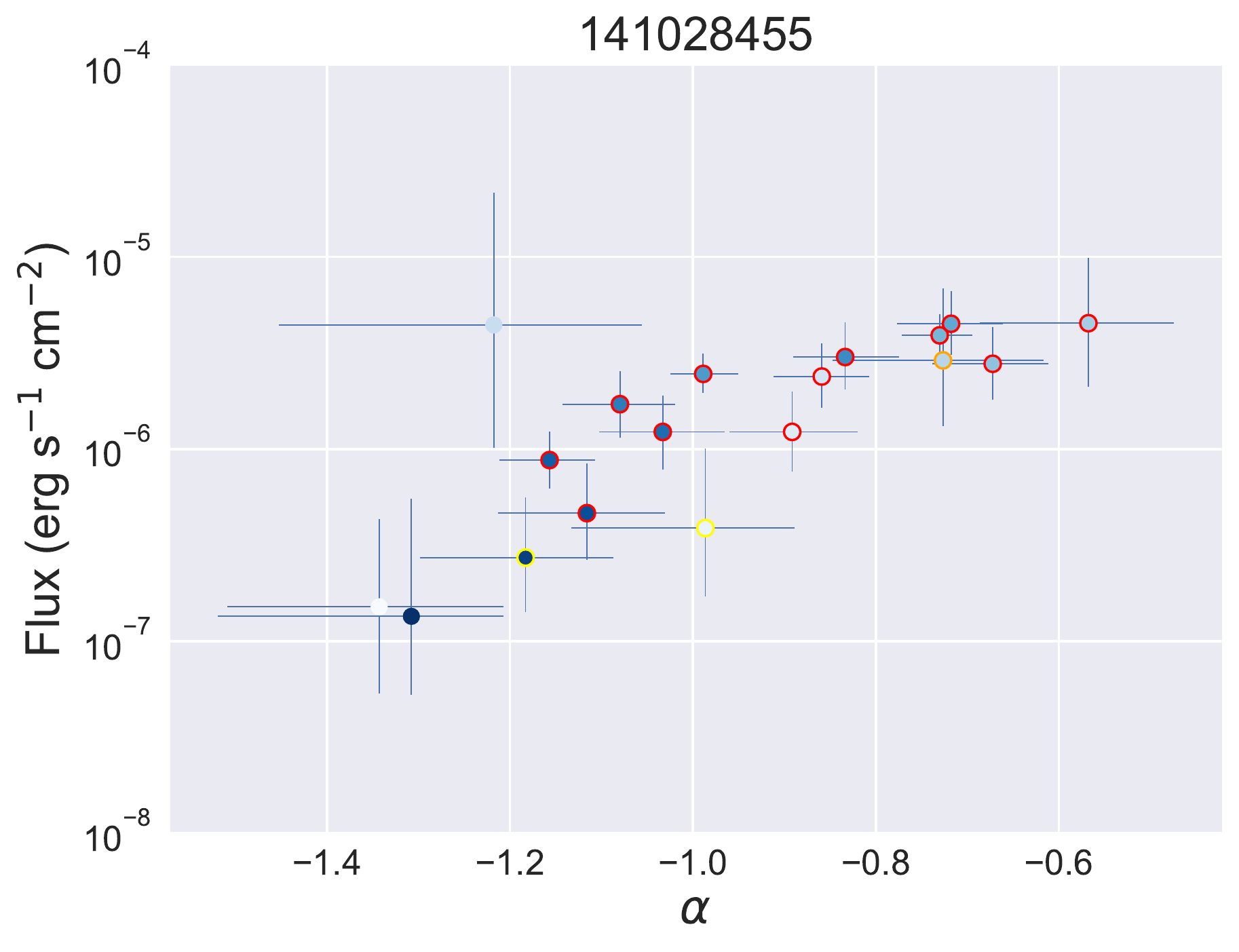}}

\subfigure{\includegraphics[width=0.3\linewidth]{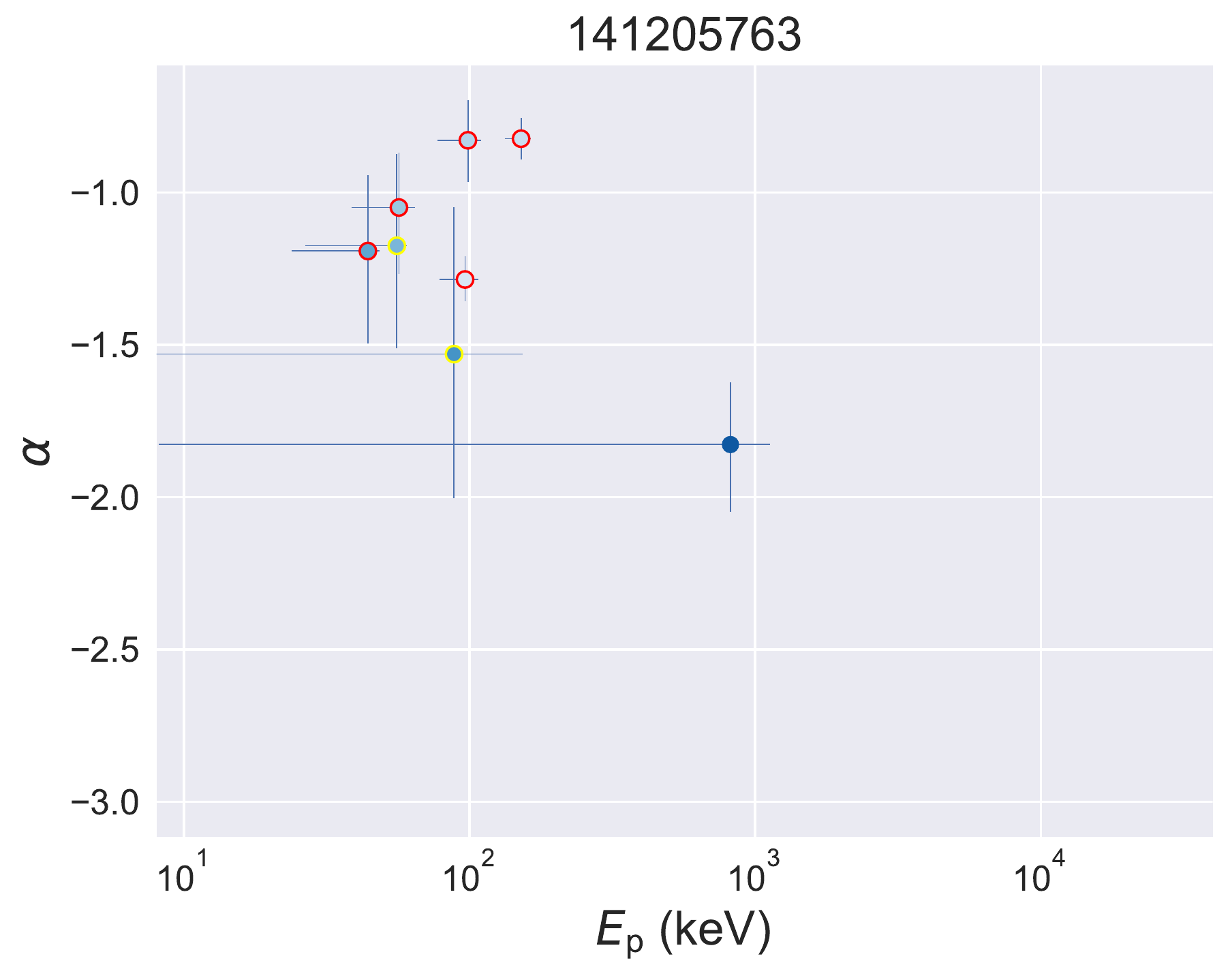}}
\subfigure{\includegraphics[width=0.3\linewidth]{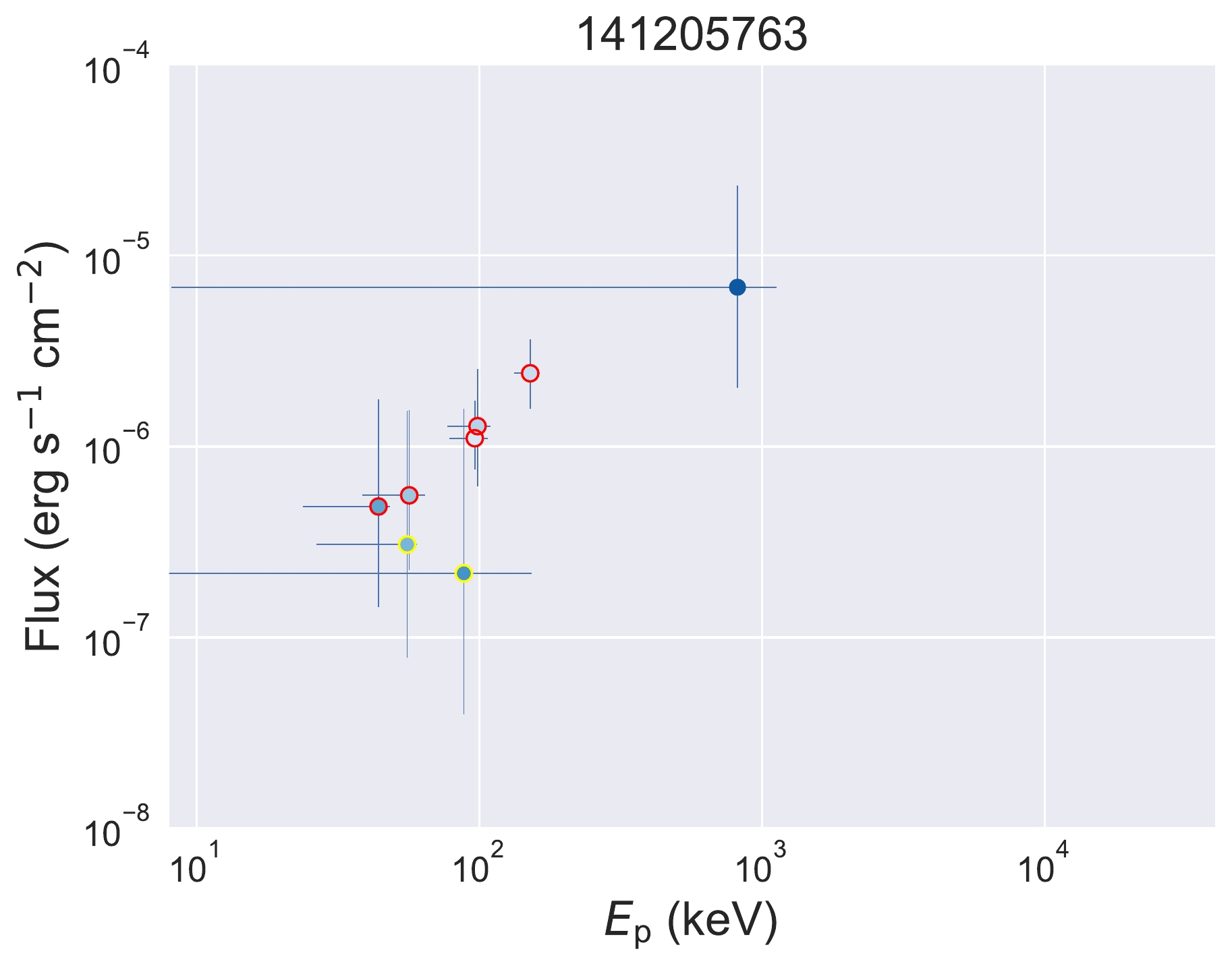}}
\subfigure{\includegraphics[width=0.3\linewidth]{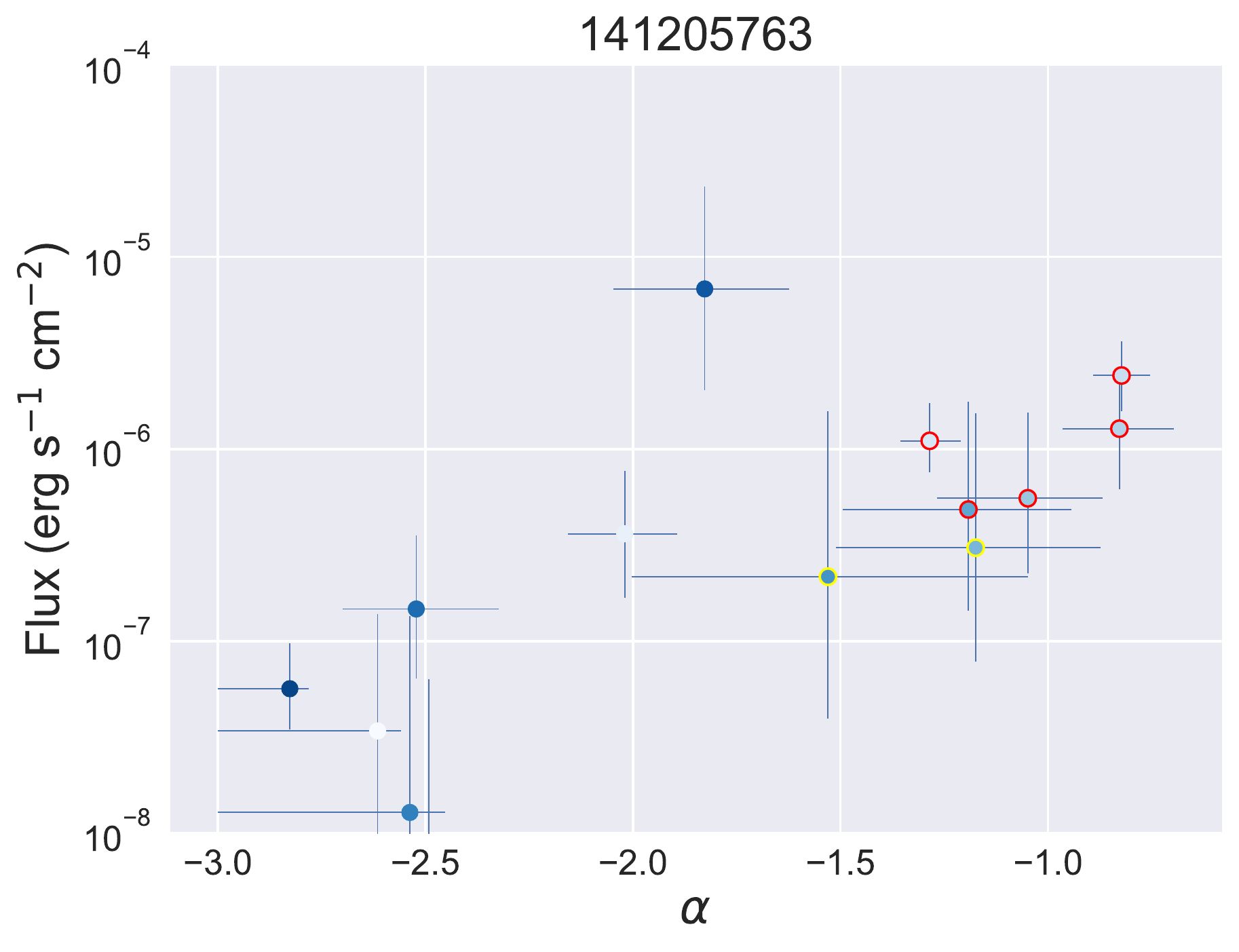}}

\subfigure{\includegraphics[width=0.3\linewidth]{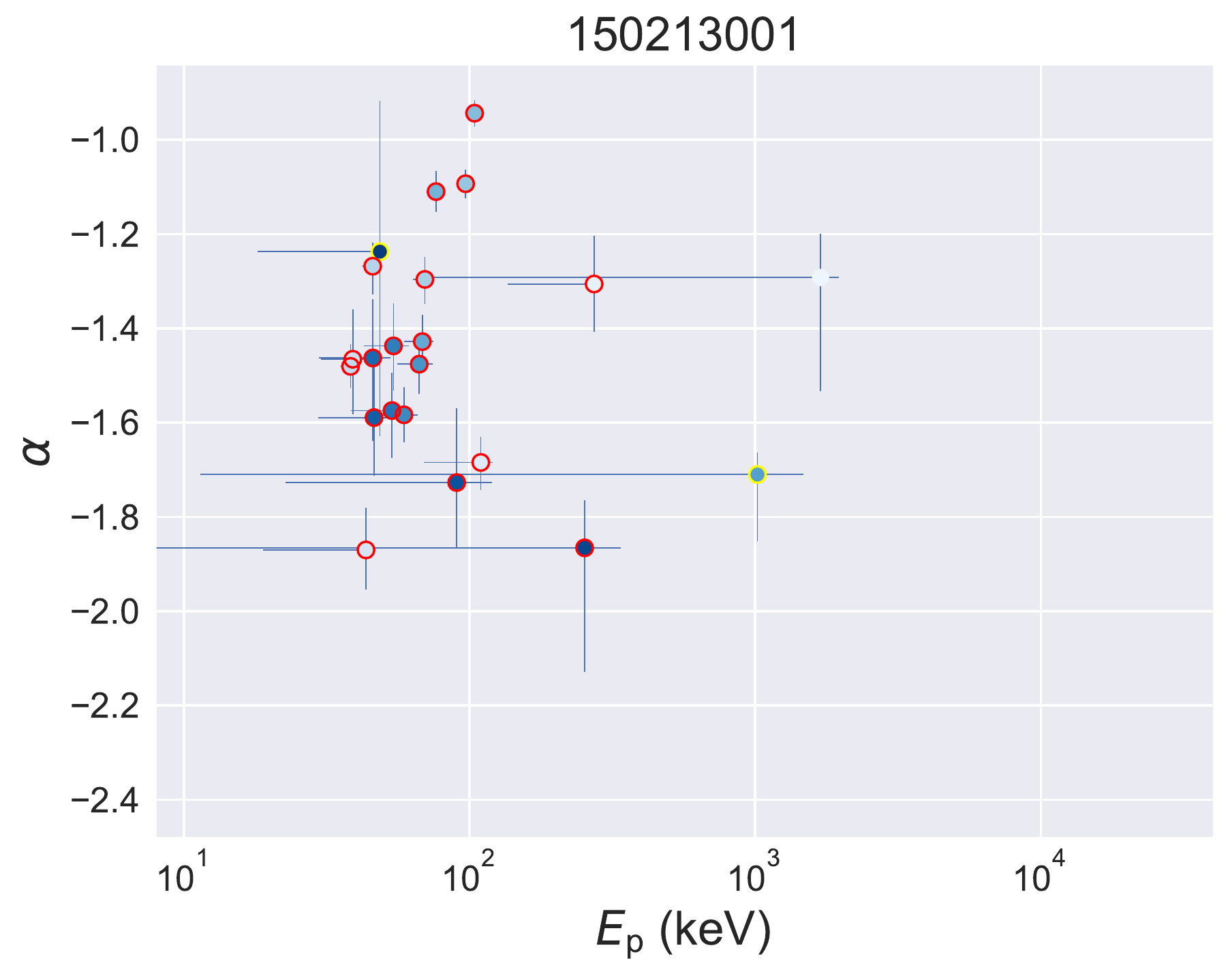}}
\subfigure{\includegraphics[width=0.3\linewidth]{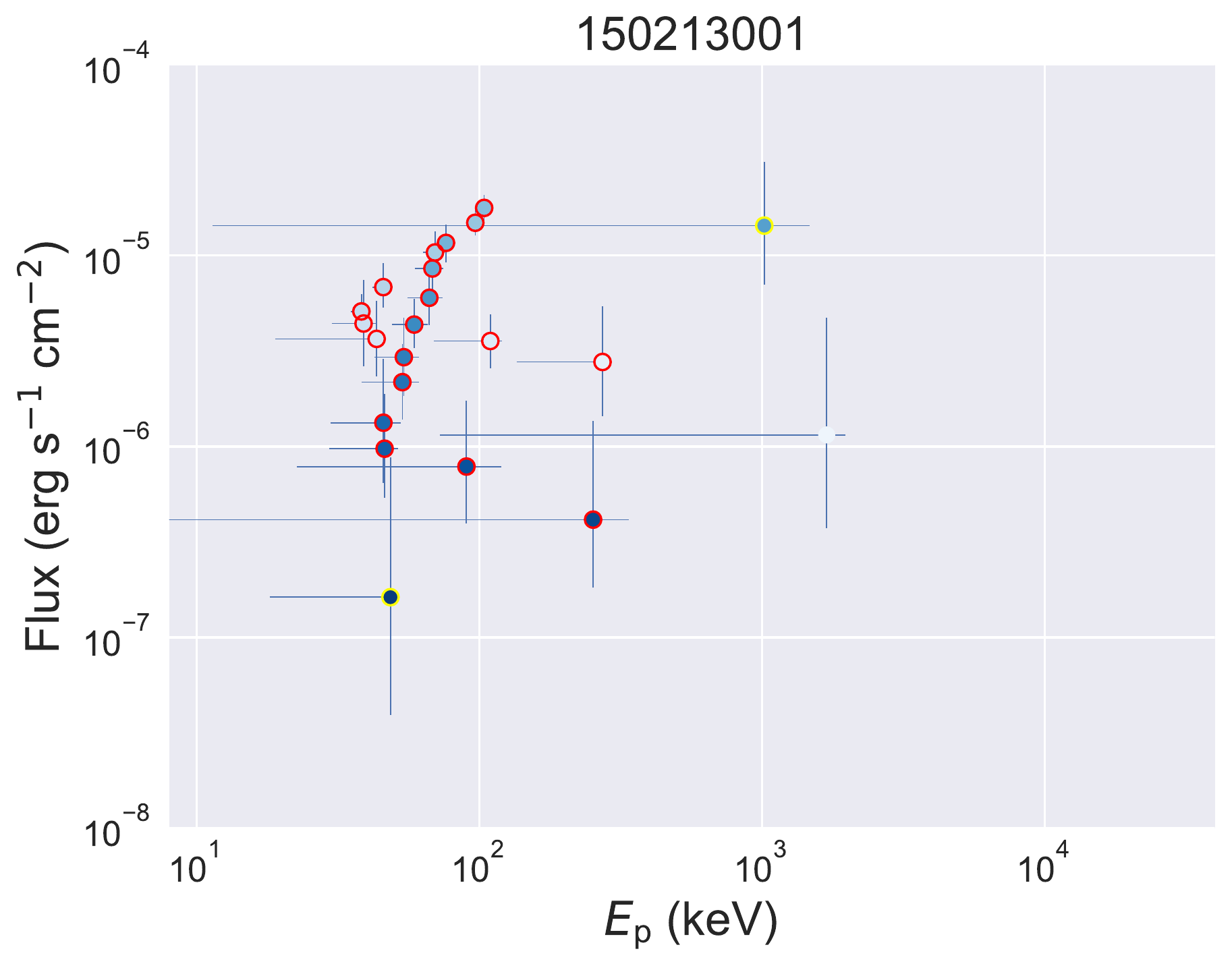}}
\subfigure{\includegraphics[width=0.3\linewidth]{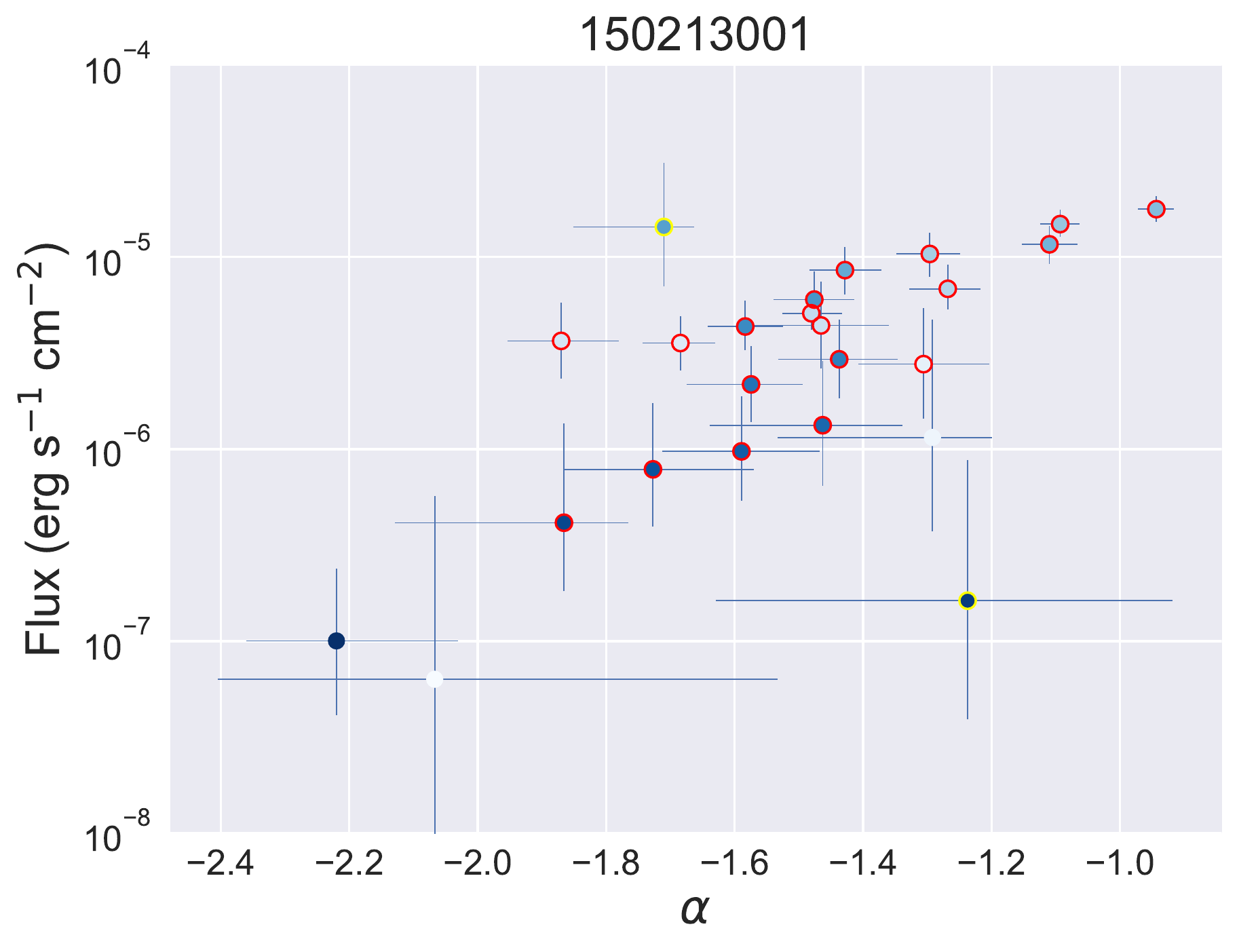}}

\caption{Same as Fig.~\ref{fig:correlation_group1}.
\label{fig:correlation_group7}}
\end{figure*}

\begin{figure*}

\subfigure{\includegraphics[width=0.3\linewidth]{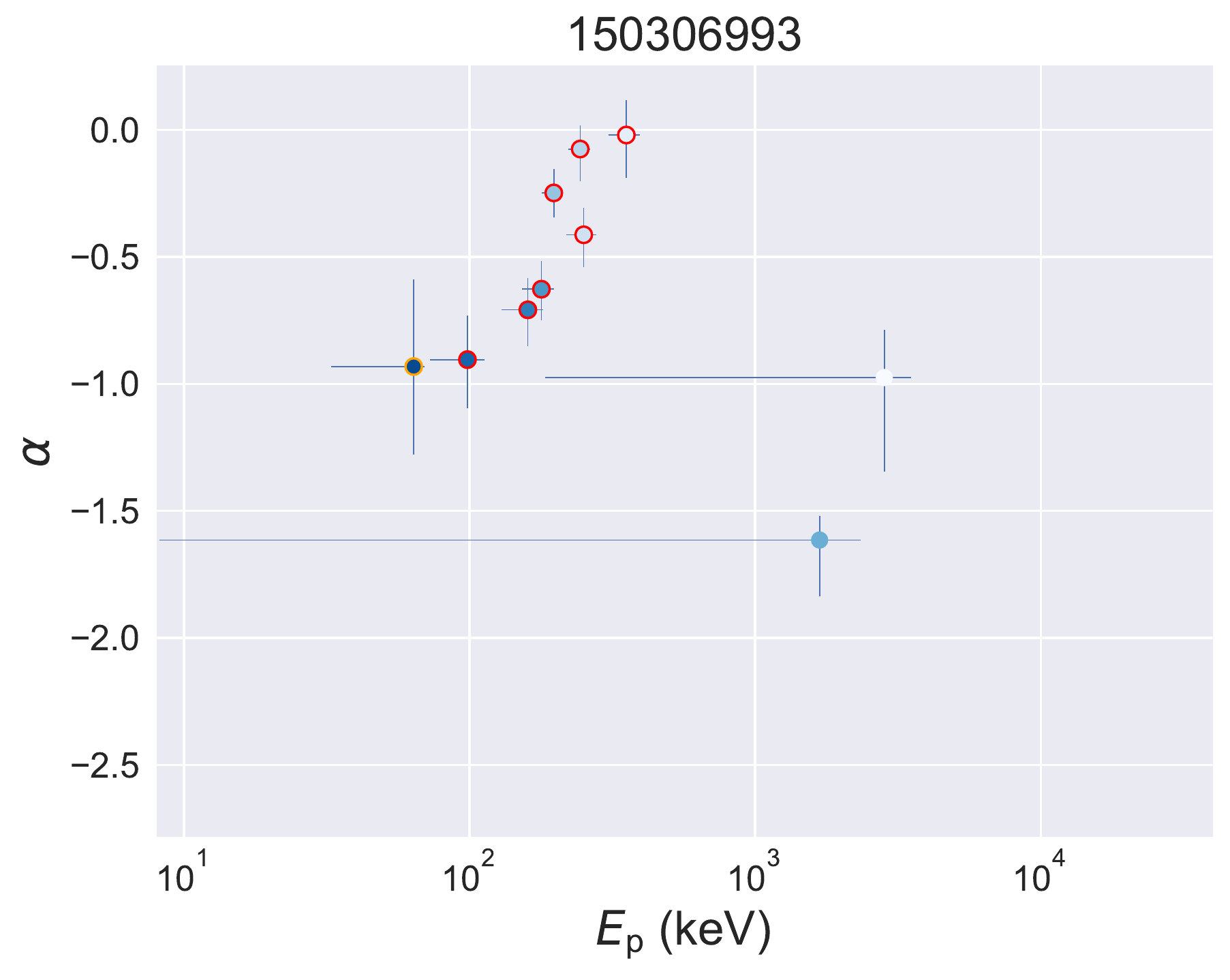}}
\subfigure{\includegraphics[width=0.3\linewidth]{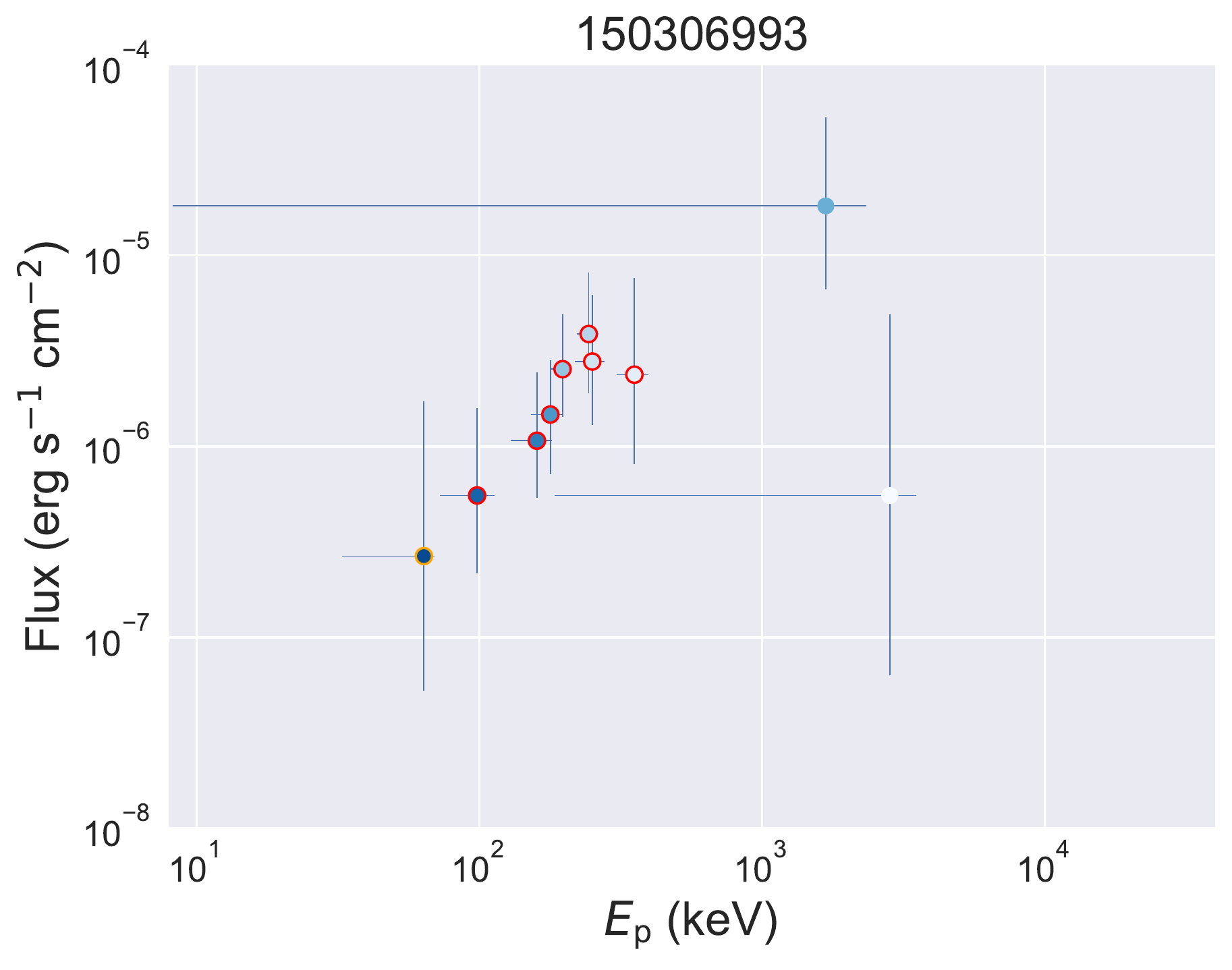}}
\subfigure{\includegraphics[width=0.3\linewidth]{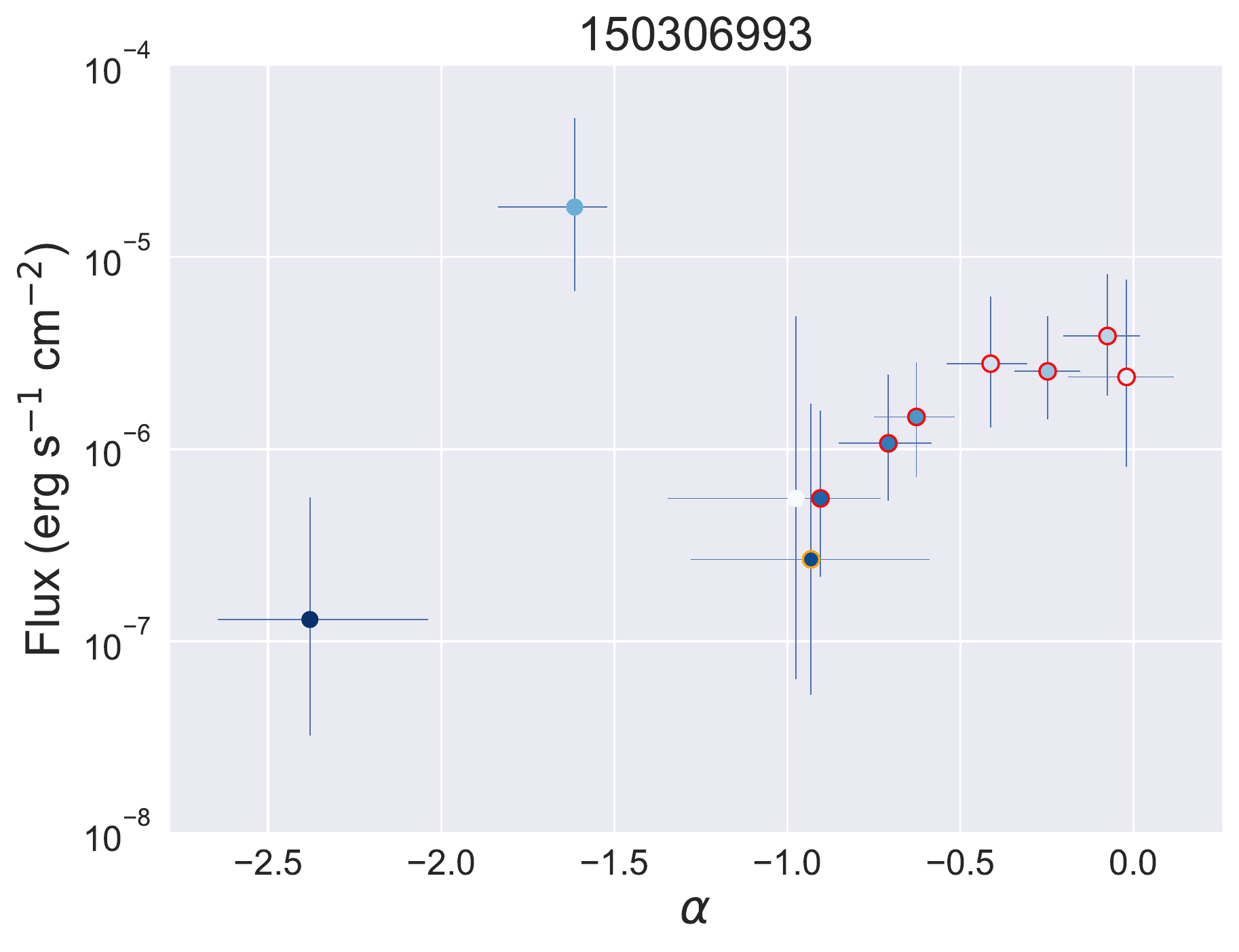}}

\subfigure{\includegraphics[width=0.3\linewidth]{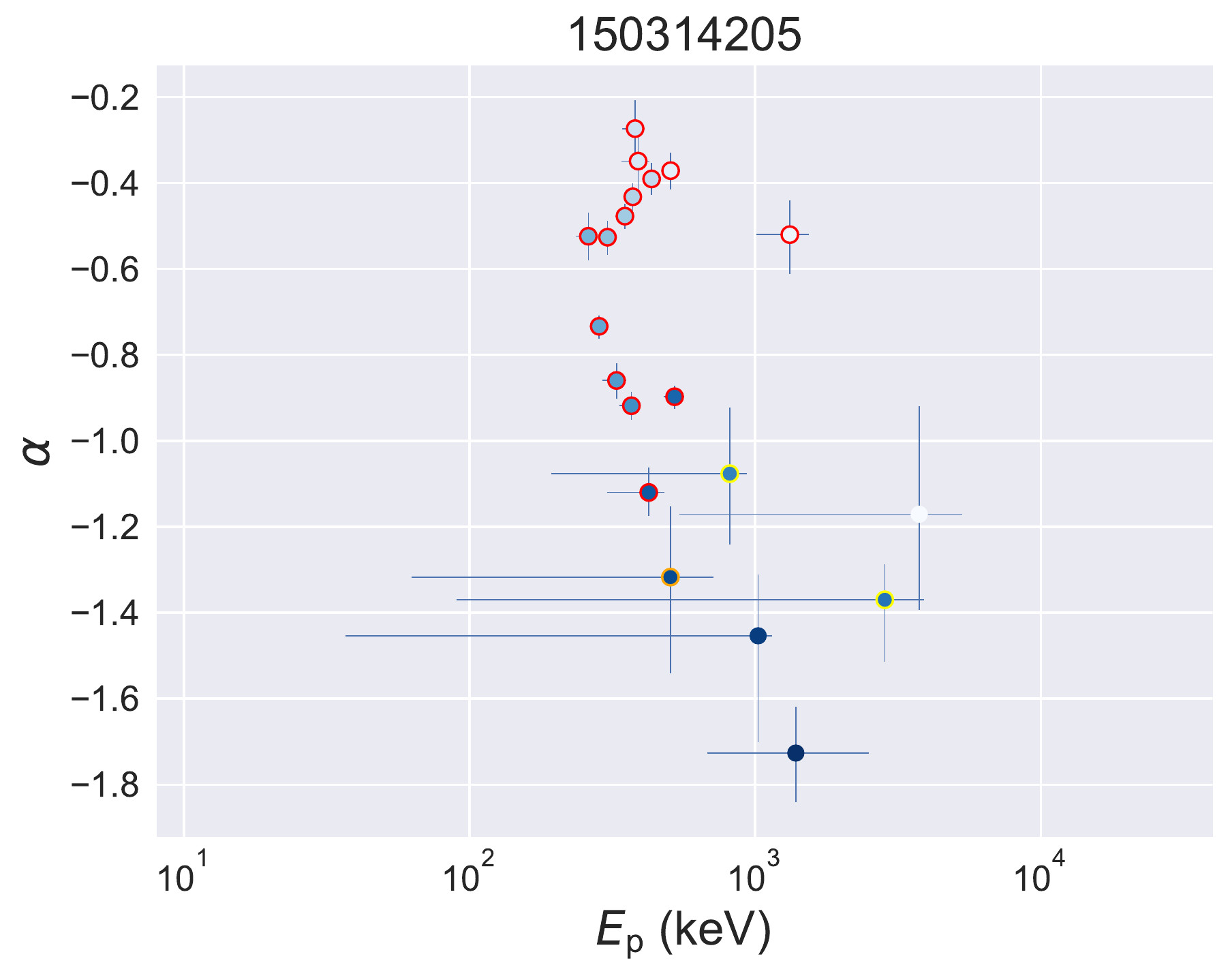}}
\subfigure{\includegraphics[width=0.3\linewidth]{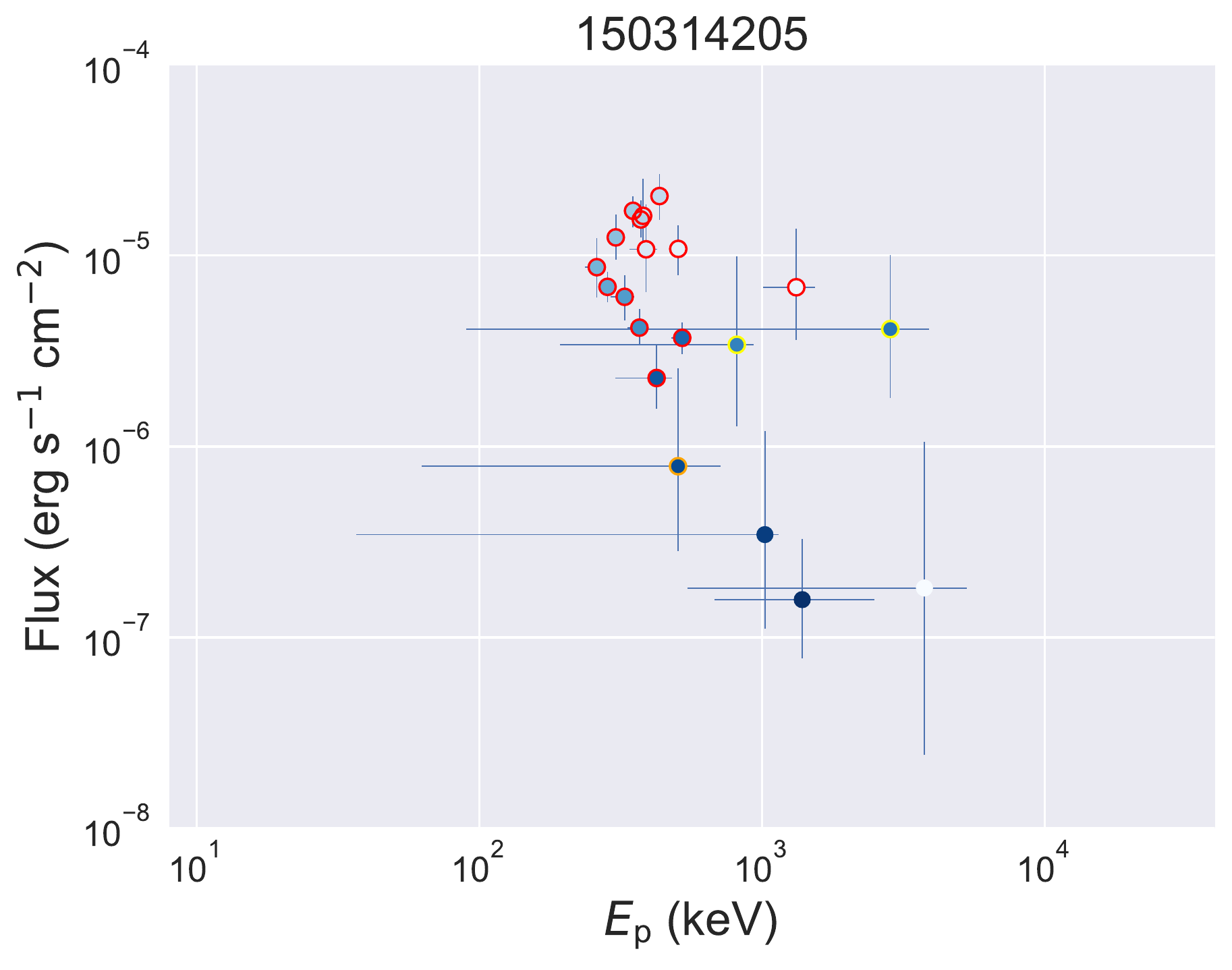}}
\subfigure{\includegraphics[width=0.3\linewidth]{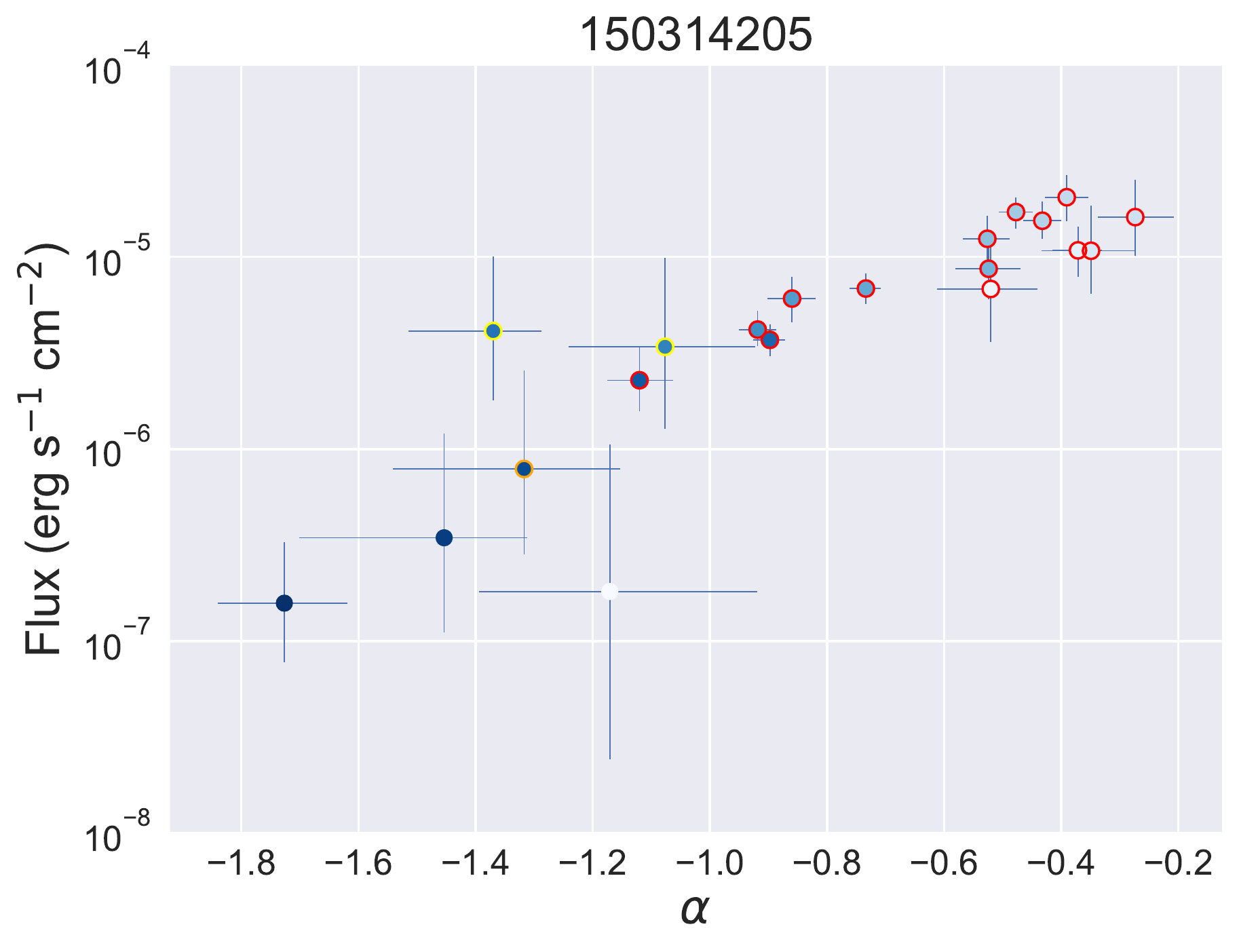}}

\subfigure{\includegraphics[width=0.3\linewidth]{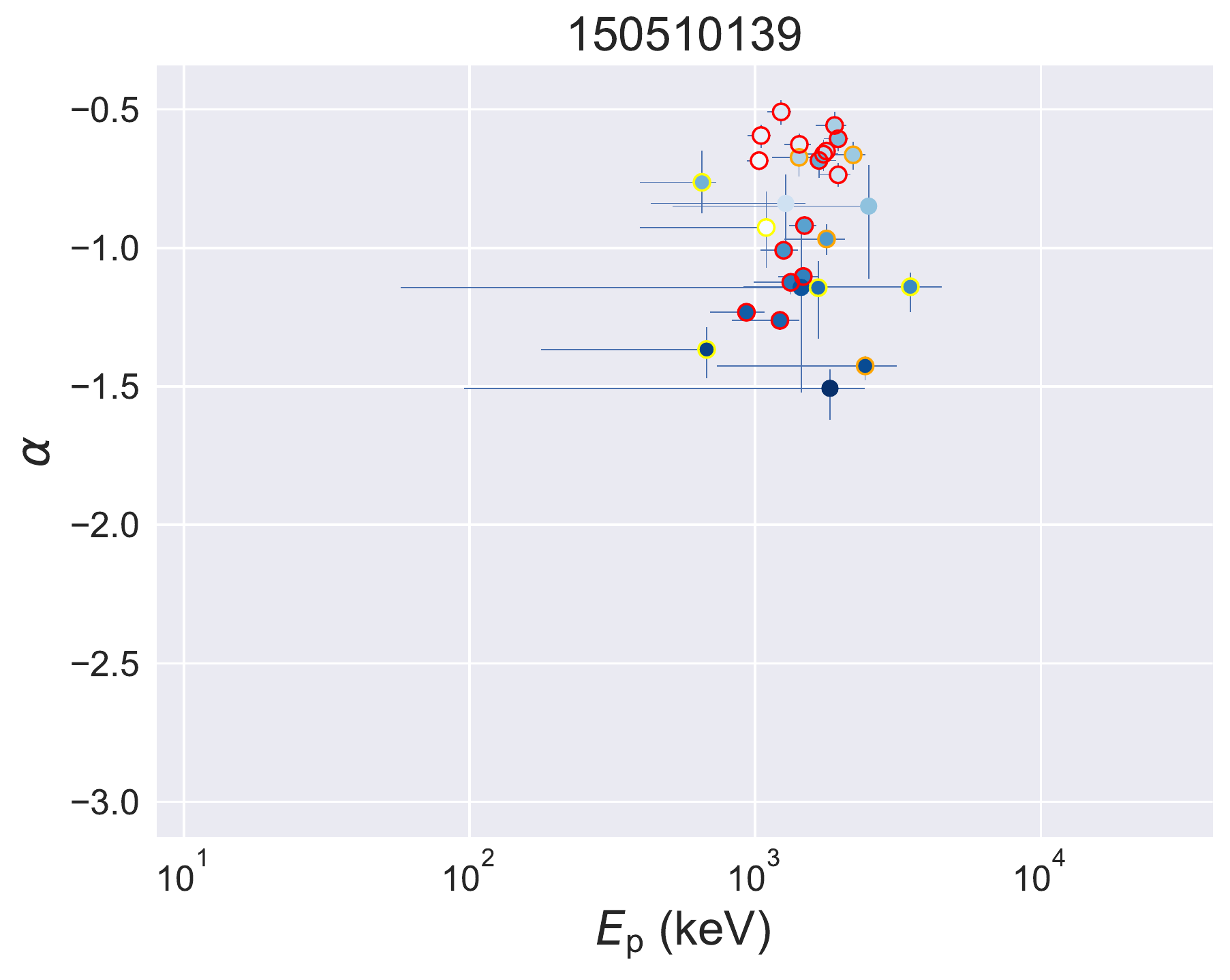}}
\subfigure{\includegraphics[width=0.3\linewidth]{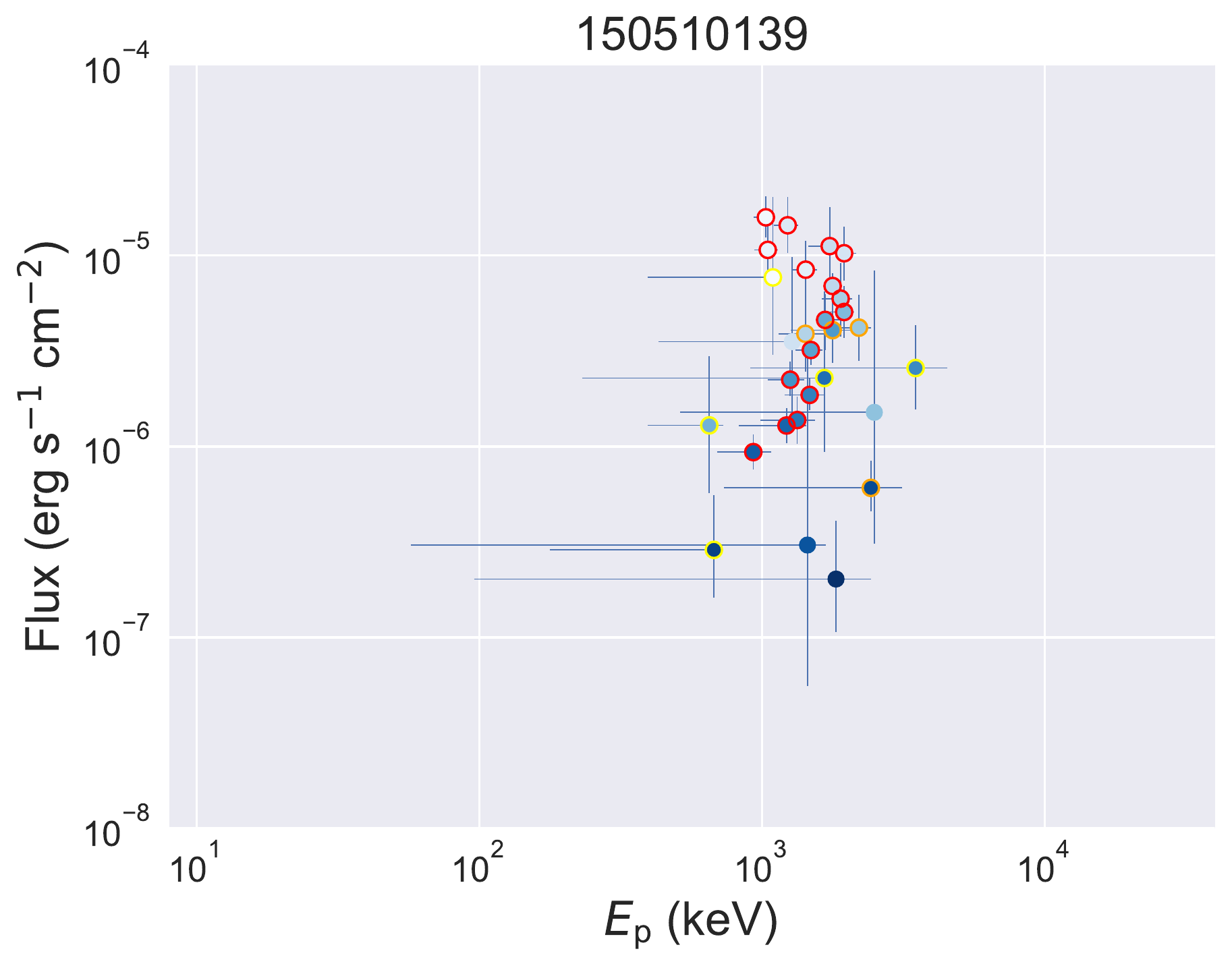}}
\subfigure{\includegraphics[width=0.3\linewidth]{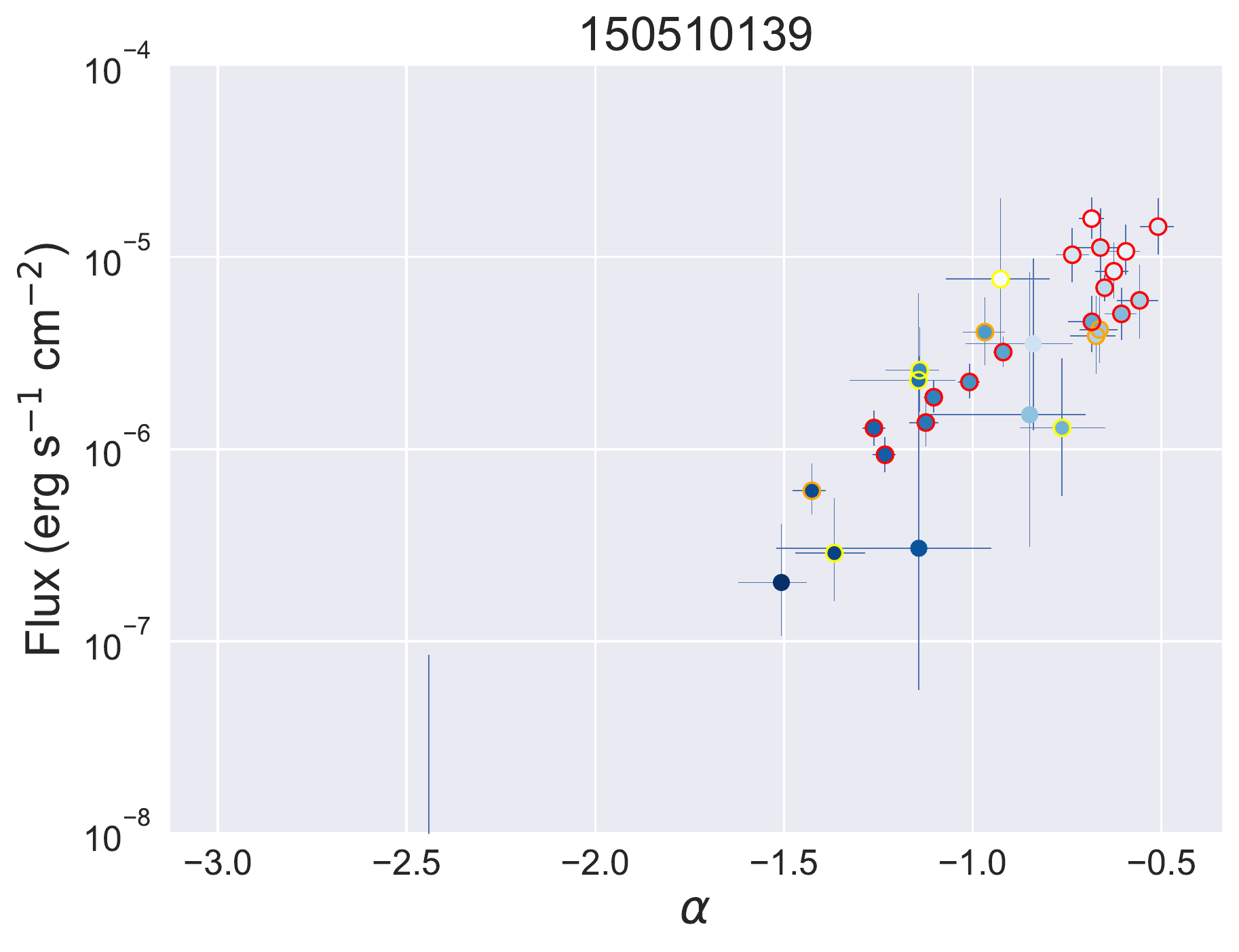}}

\subfigure{\includegraphics[width=0.3\linewidth]{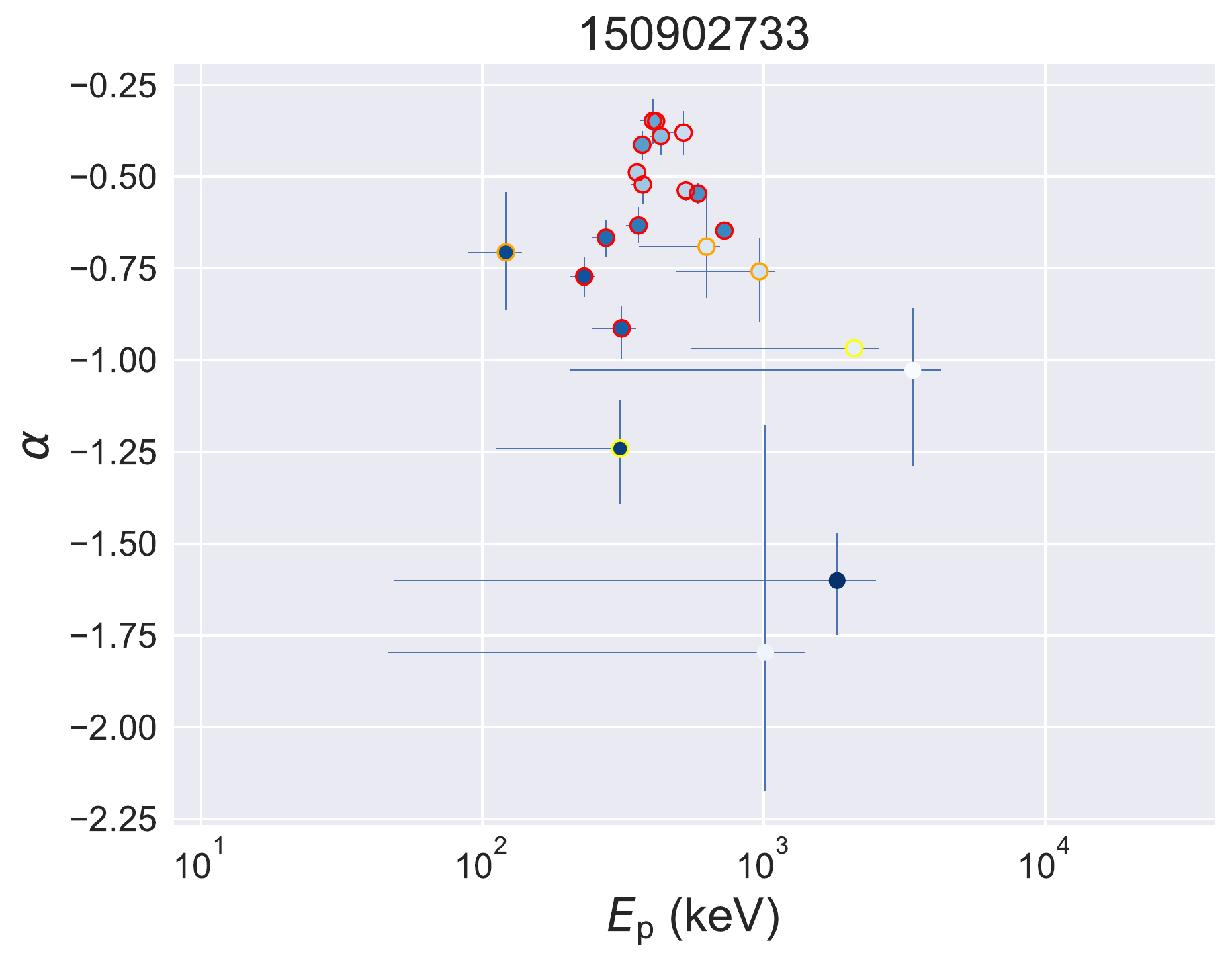}}
\subfigure{\includegraphics[width=0.3\linewidth]{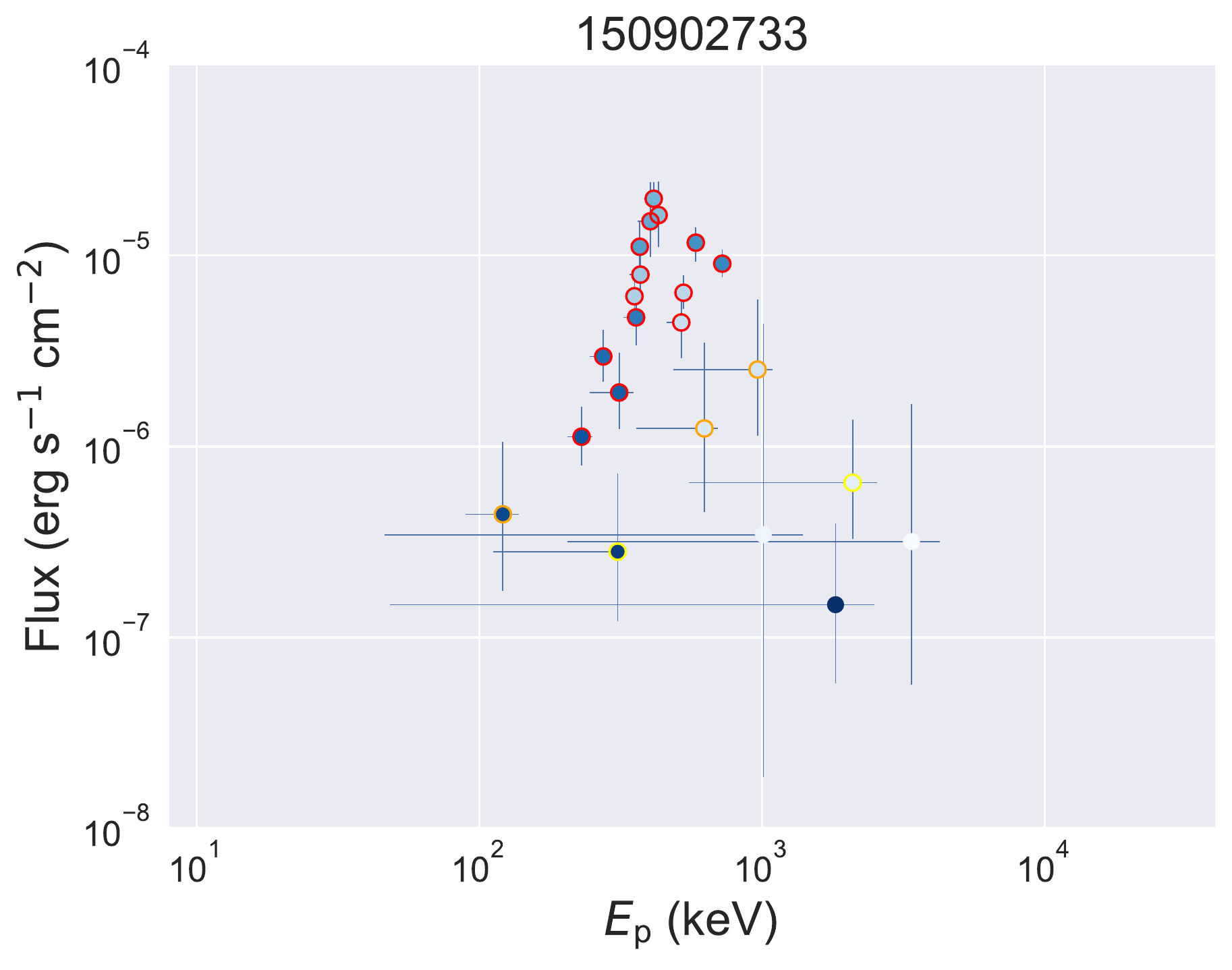}}
\subfigure{\includegraphics[width=0.3\linewidth]{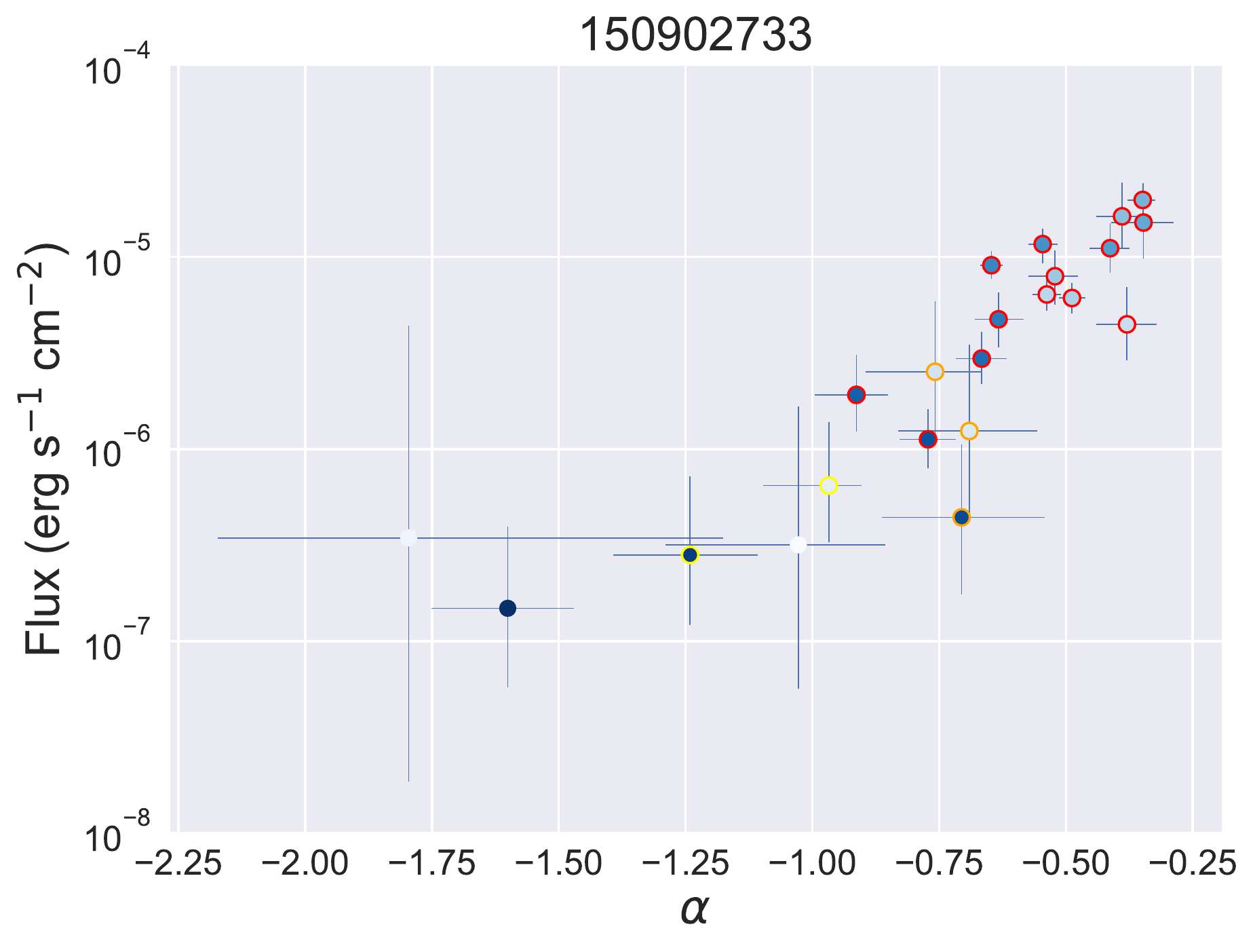}}

\caption{Same as Fig.~\ref{fig:correlation_group1}.
\label{fig:correlation_group8}}
\end{figure*}

\begin{figure*}

\subfigure{\includegraphics[width=0.3\linewidth]{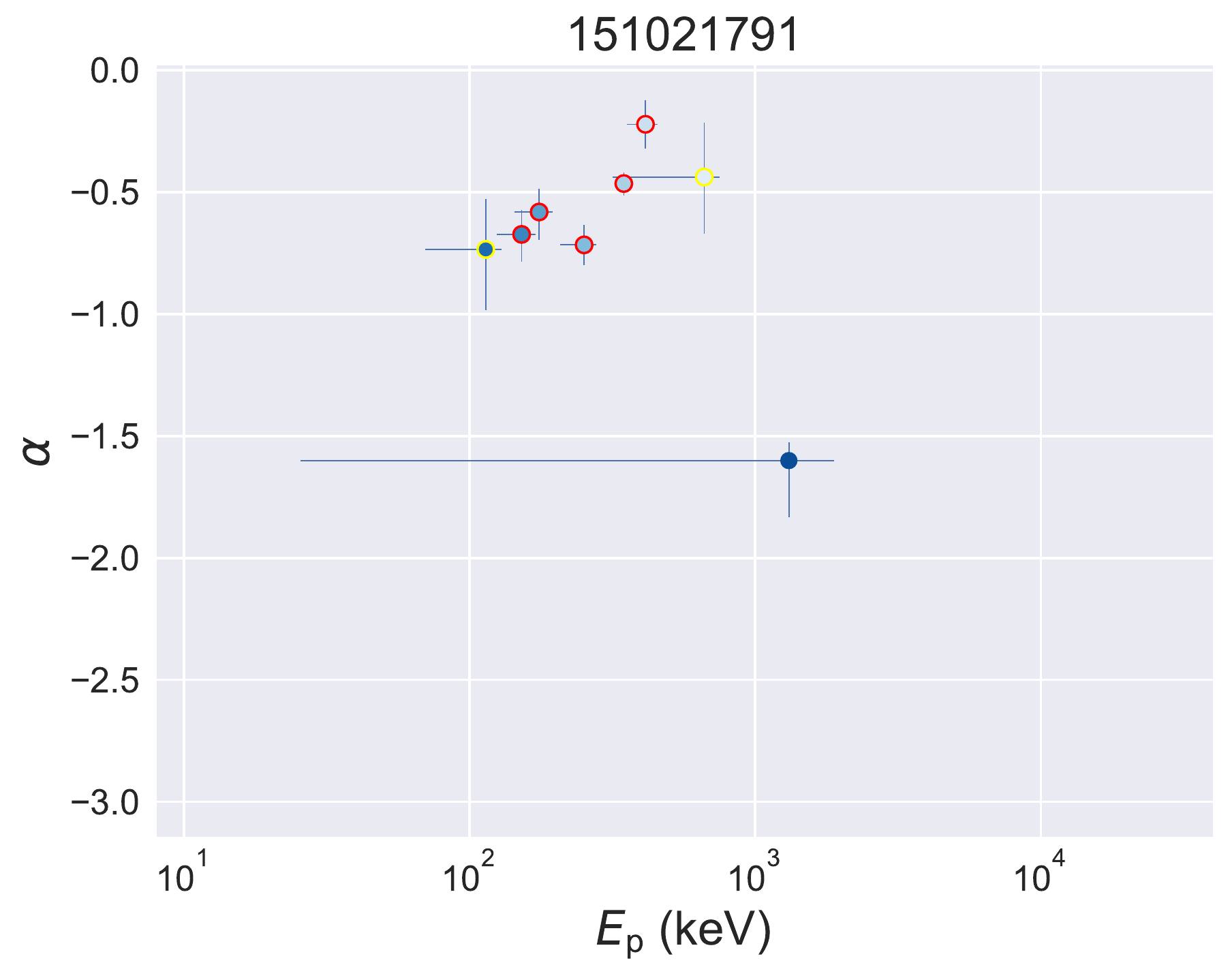}}
\subfigure{\includegraphics[width=0.3\linewidth]{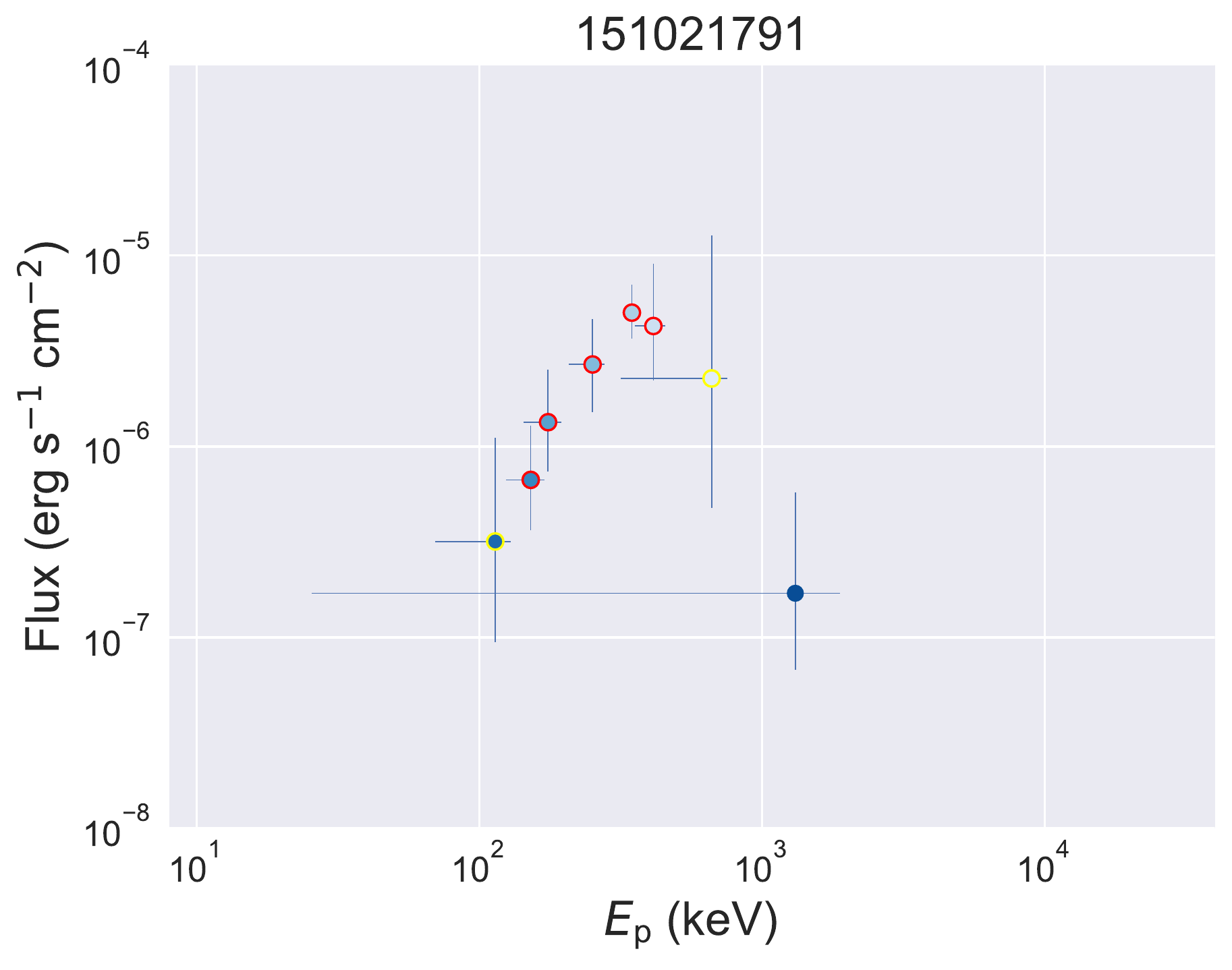}}
\subfigure{\includegraphics[width=0.3\linewidth]{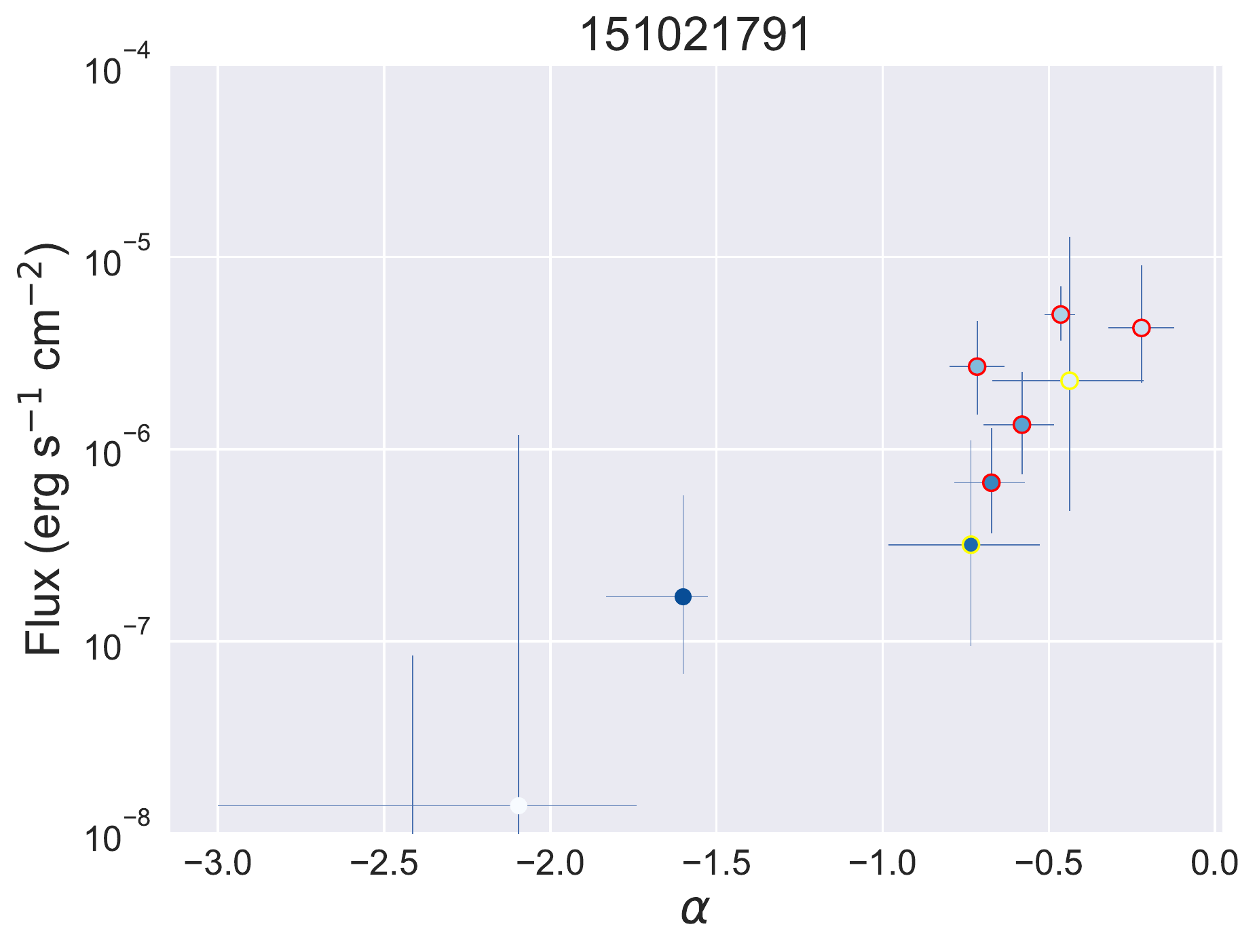}}

\subfigure{\includegraphics[width=0.3\linewidth]{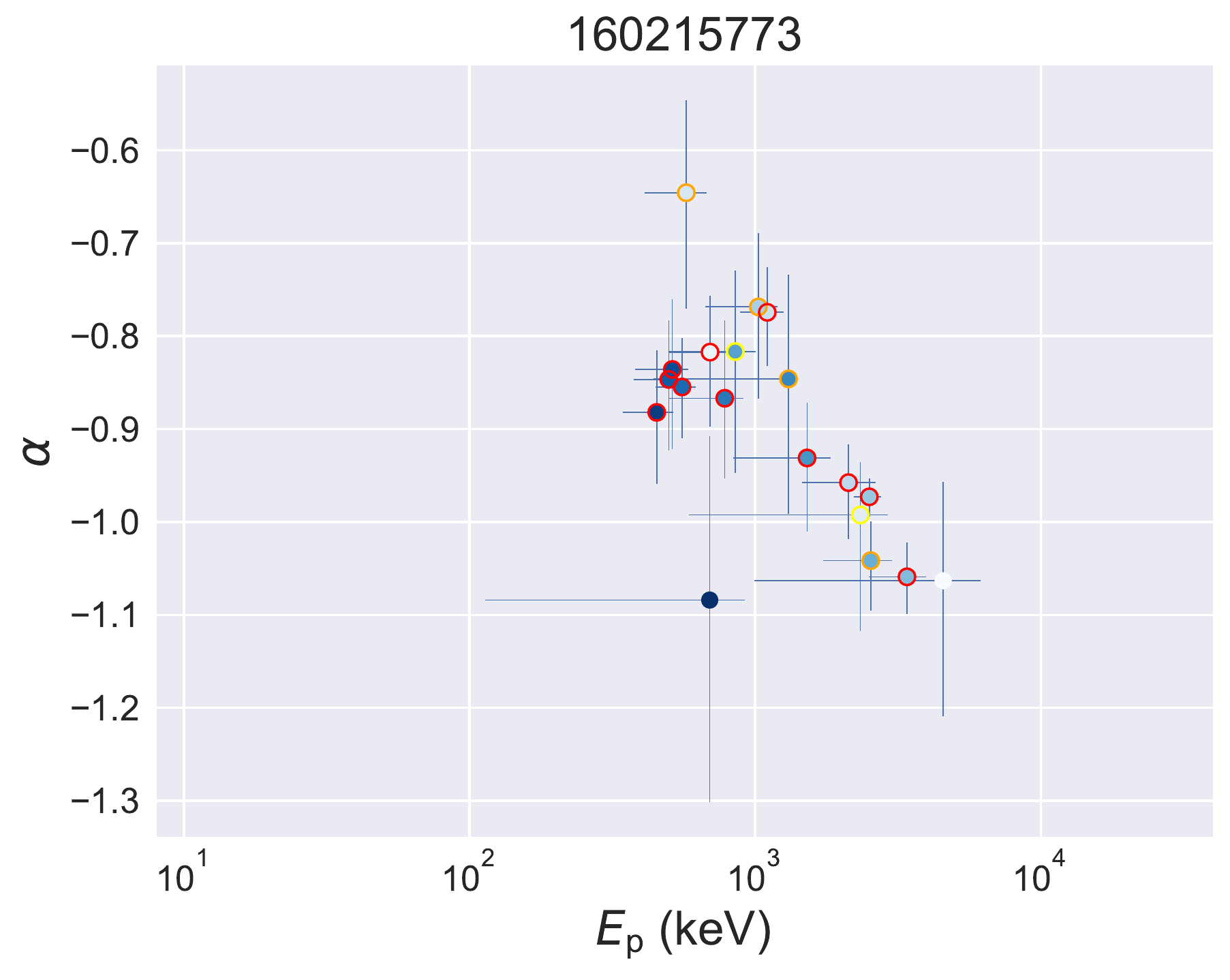}}
\subfigure{\includegraphics[width=0.3\linewidth]{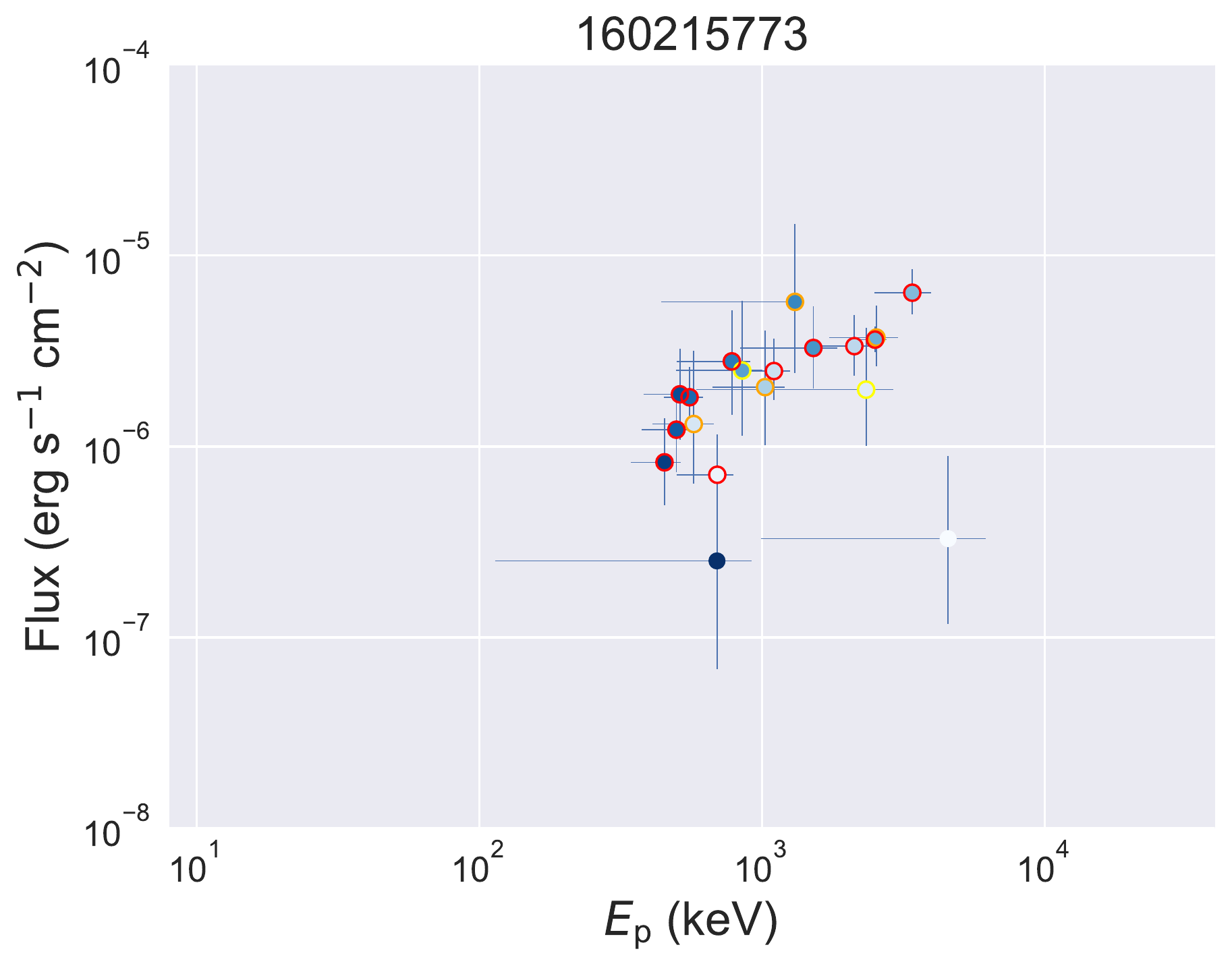}}
\subfigure{\includegraphics[width=0.3\linewidth]{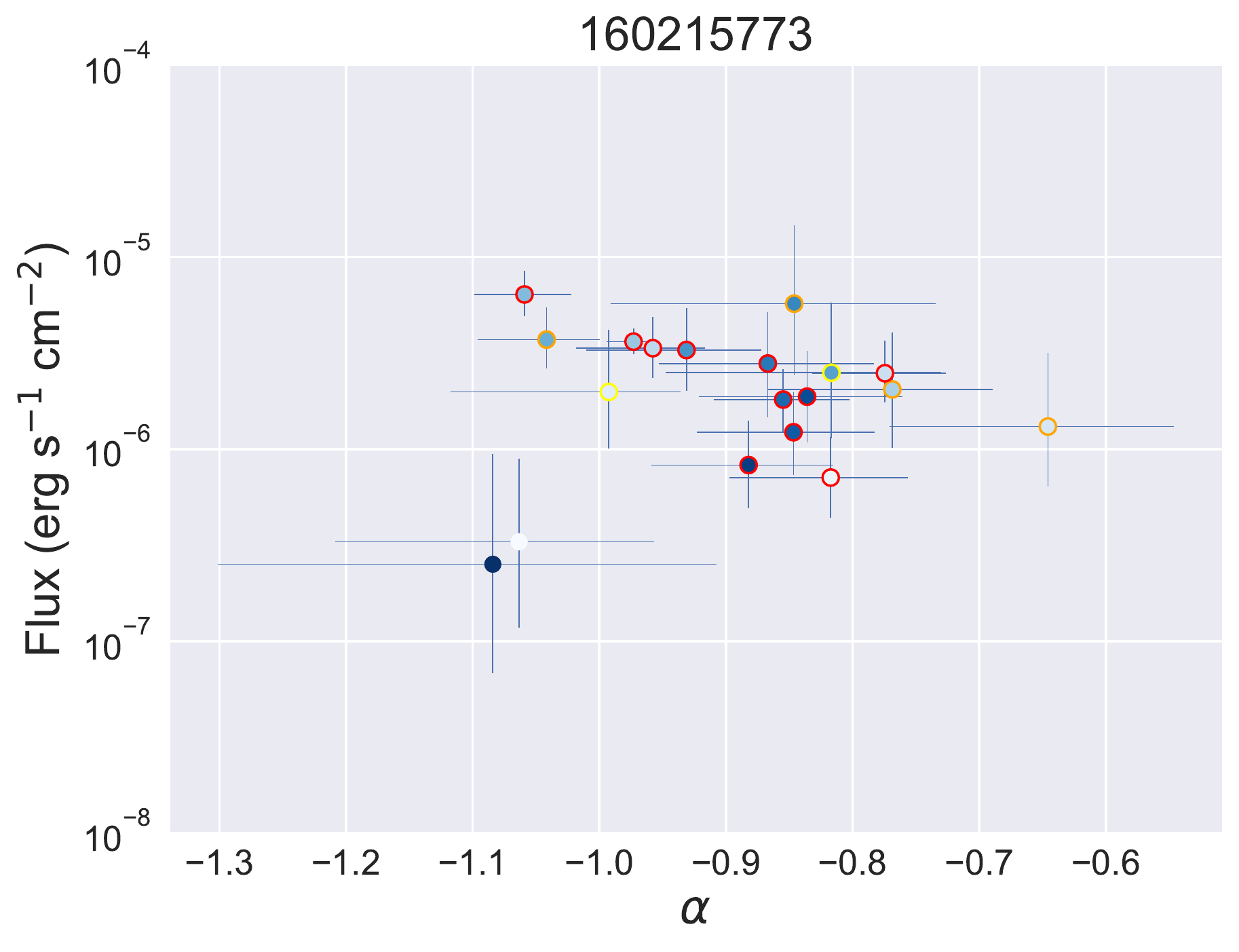}}

\subfigure{\includegraphics[width=0.3\linewidth]{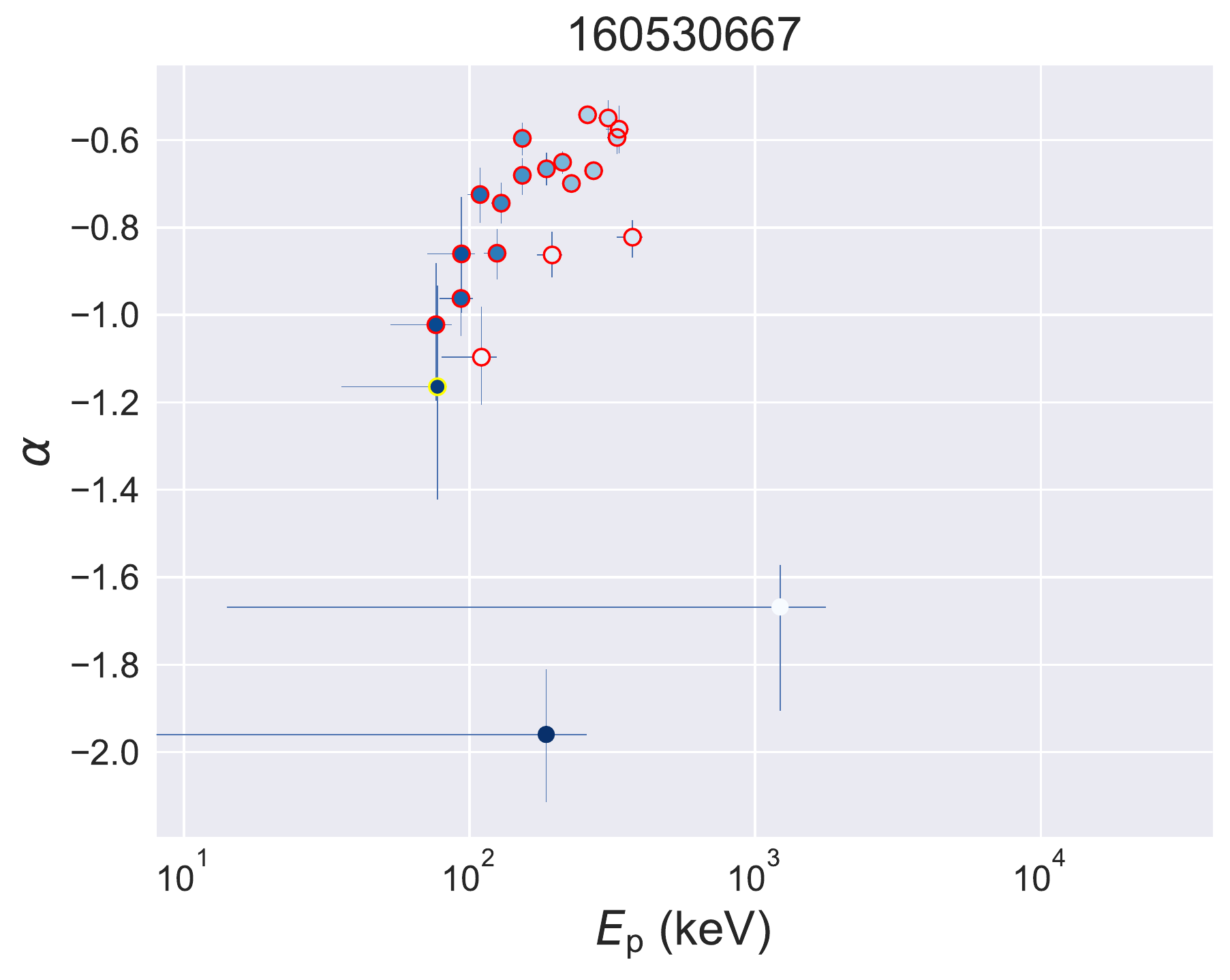}}
\subfigure{\includegraphics[width=0.3\linewidth]{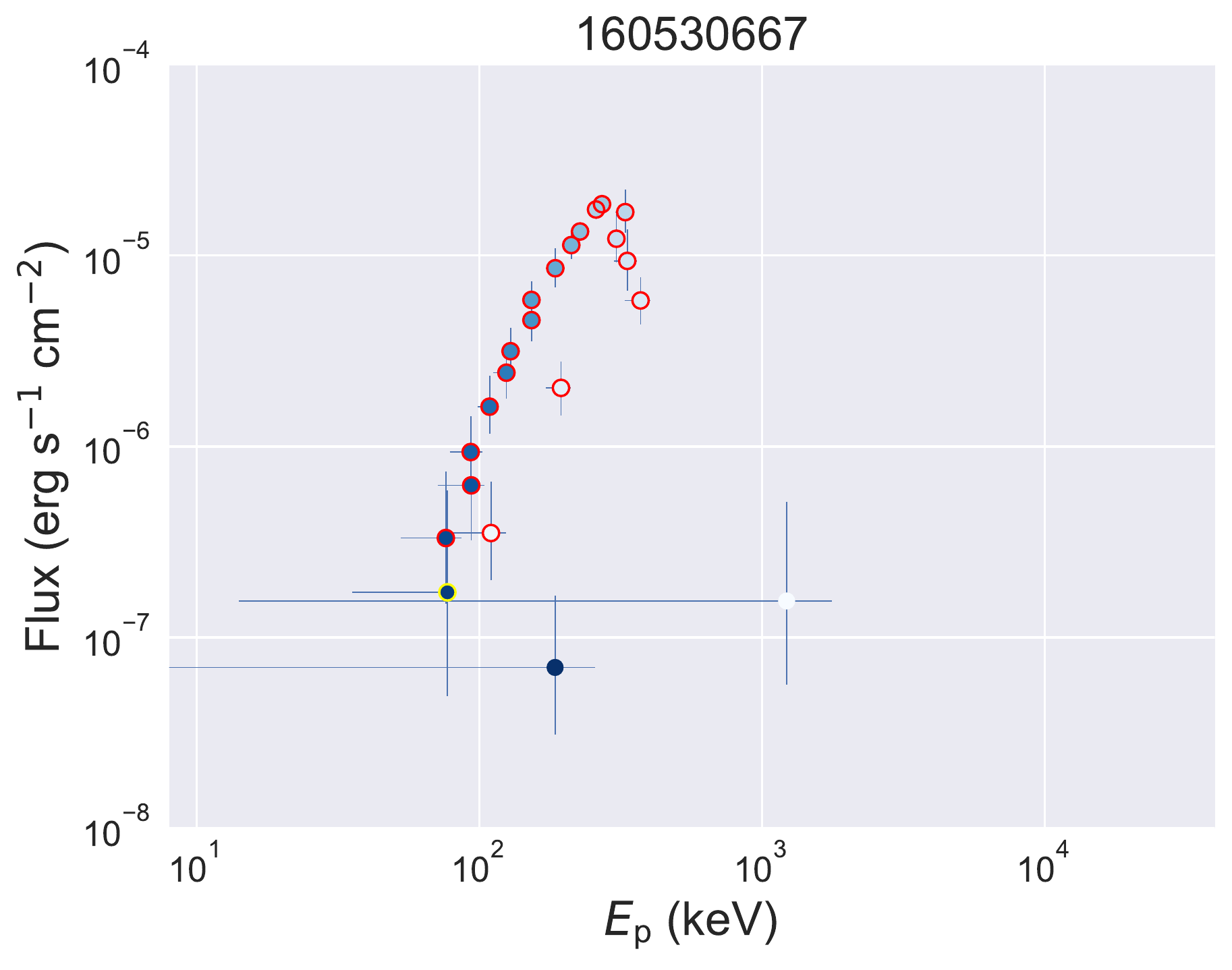}}
\subfigure{\includegraphics[width=0.3\linewidth]{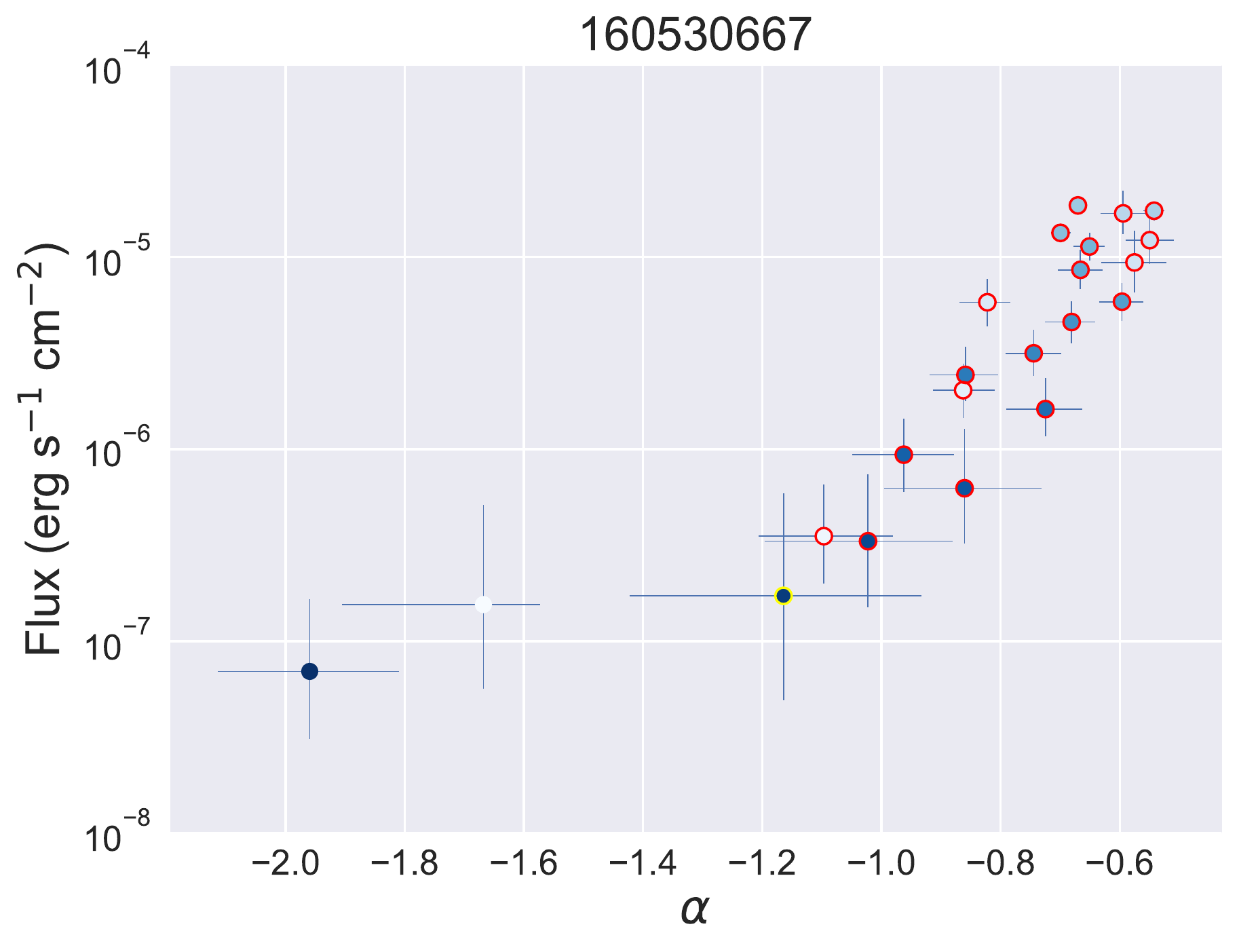}}

\subfigure{\includegraphics[width=0.3\linewidth]{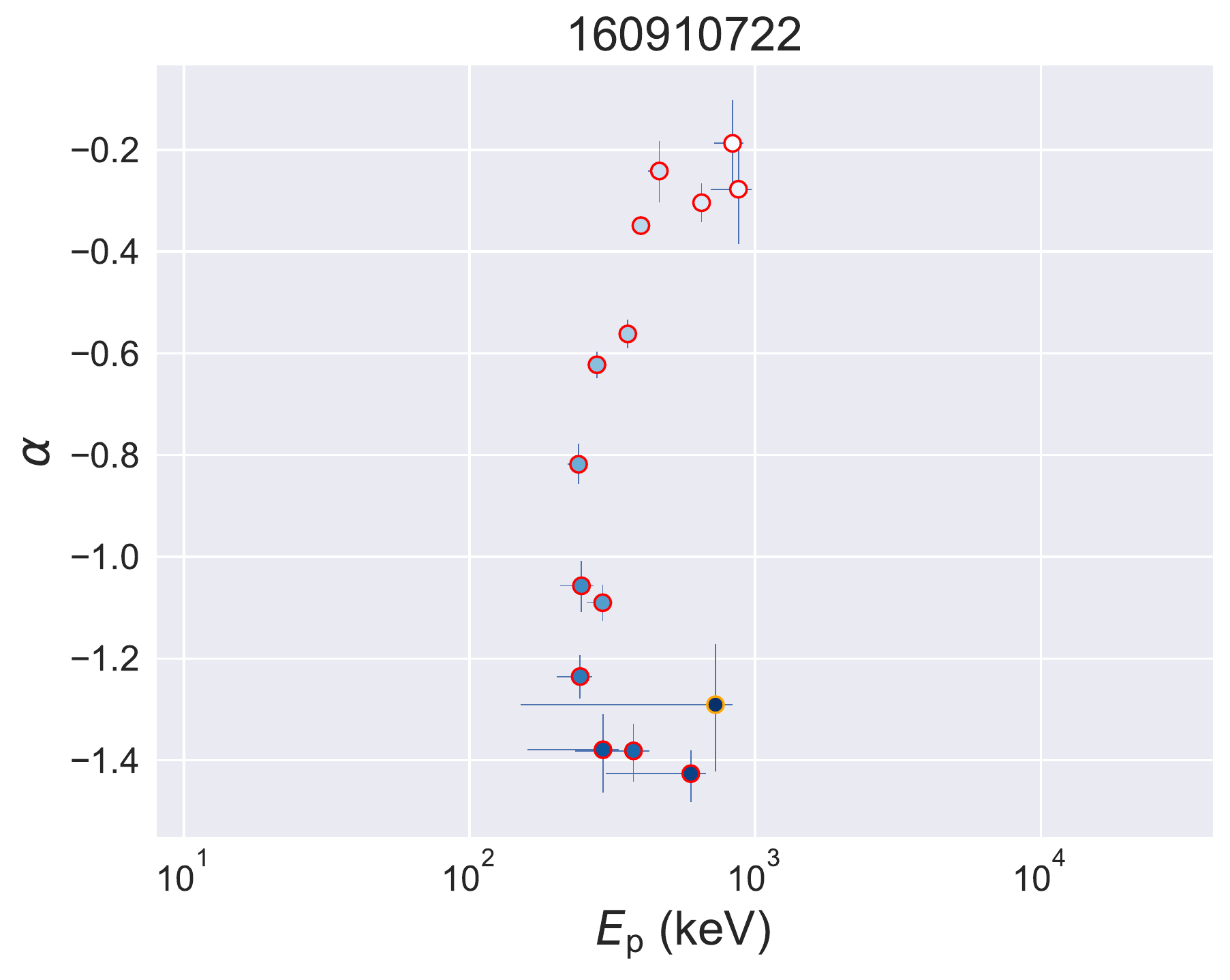}}
\subfigure{\includegraphics[width=0.3\linewidth]{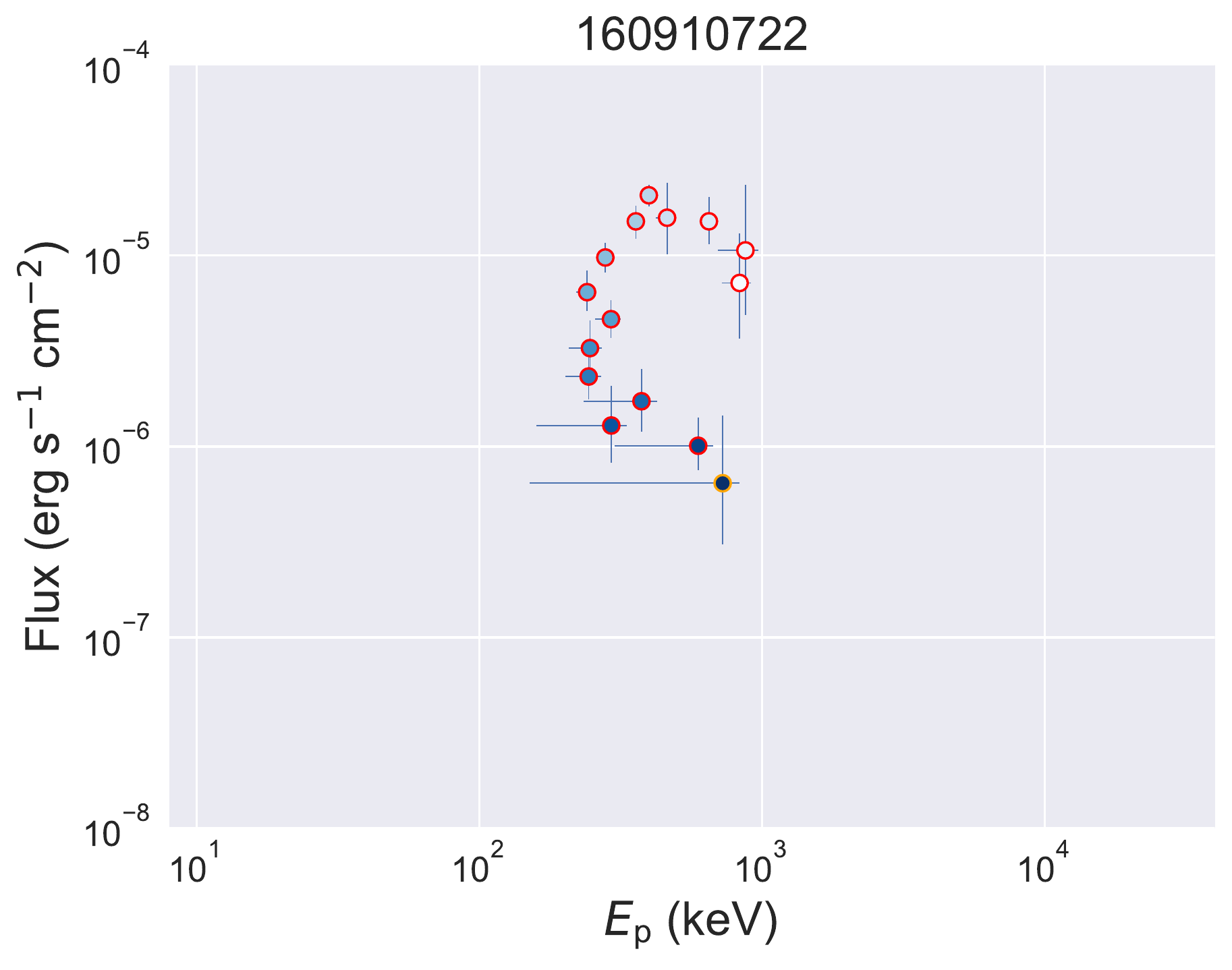}}
\subfigure{\includegraphics[width=0.3\linewidth]{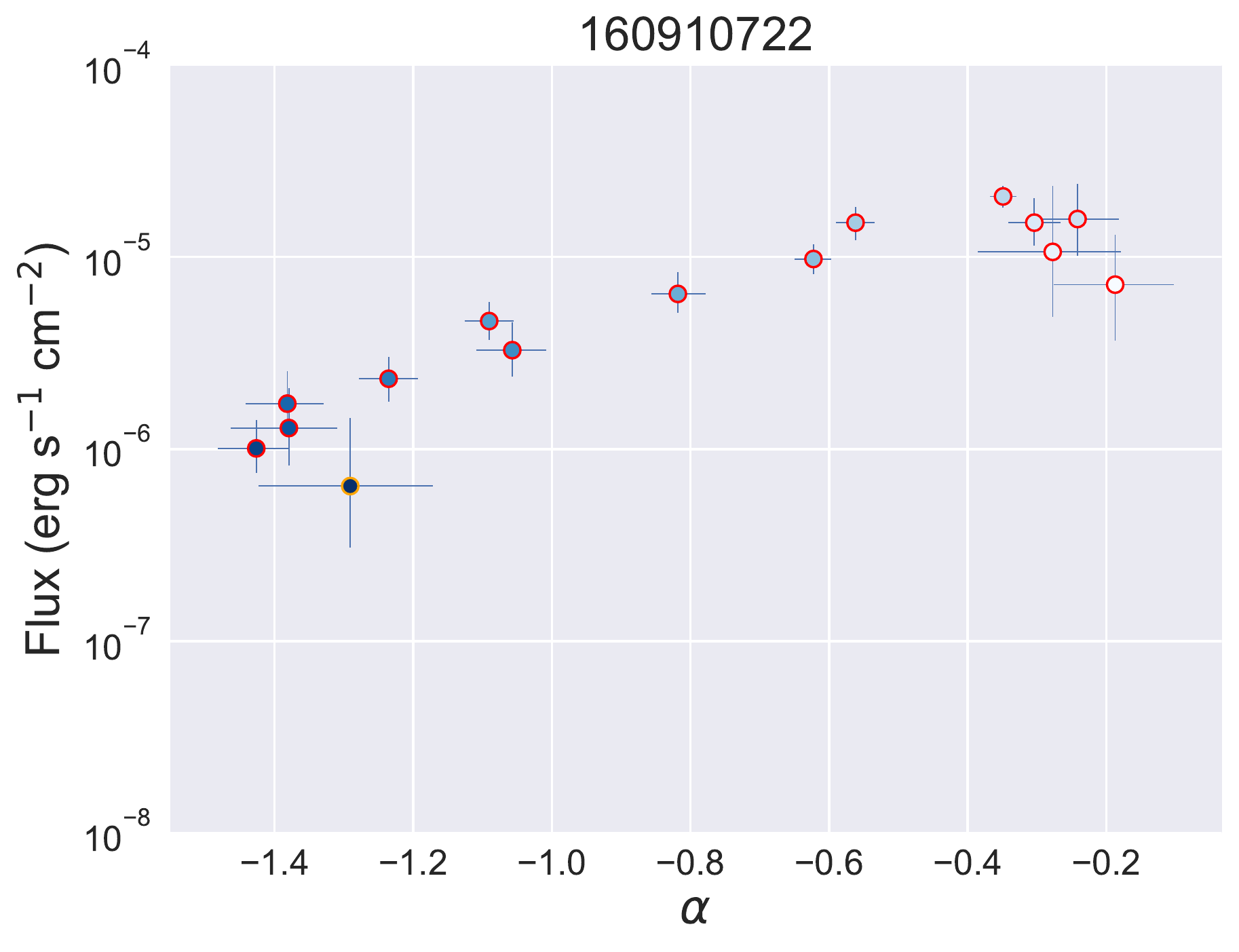}}

\caption{Same as Fig.~\ref{fig:correlation_group1}.
\label{fig:correlation_group9}}
\end{figure*}

\begin{figure*}

\subfigure{\includegraphics[width=0.3\linewidth]{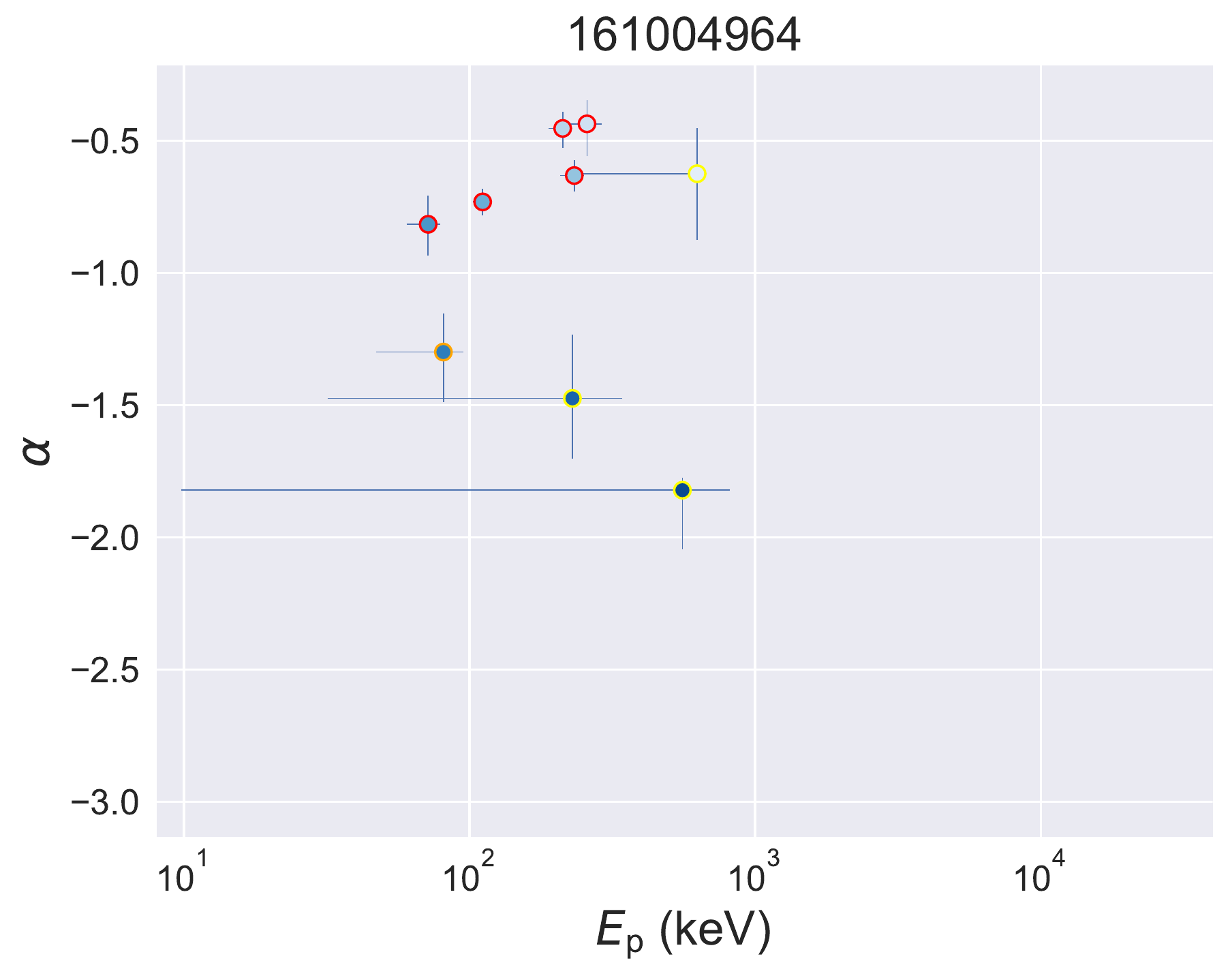}}
\subfigure{\includegraphics[width=0.3\linewidth]{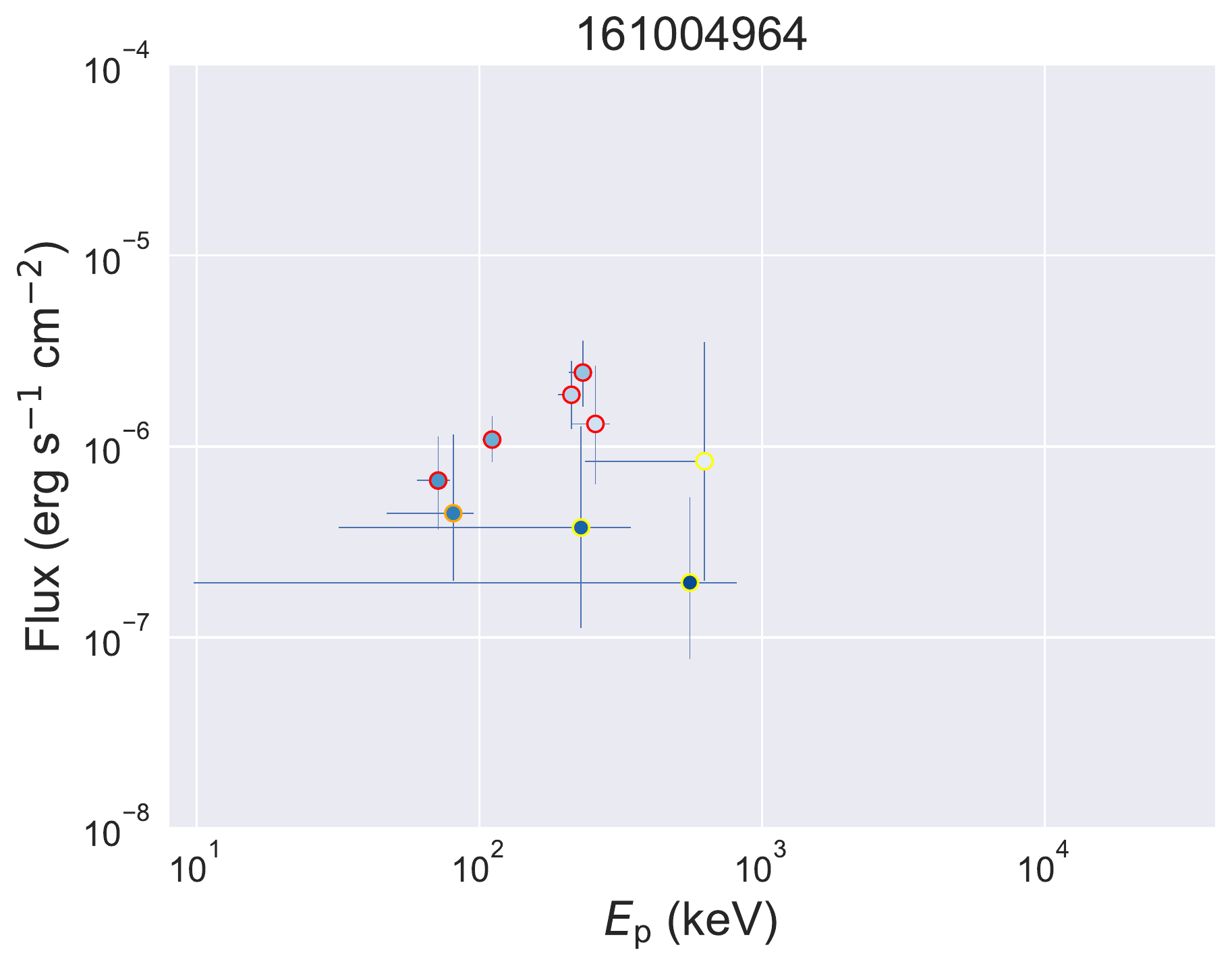}}
\subfigure{\includegraphics[width=0.3\linewidth]{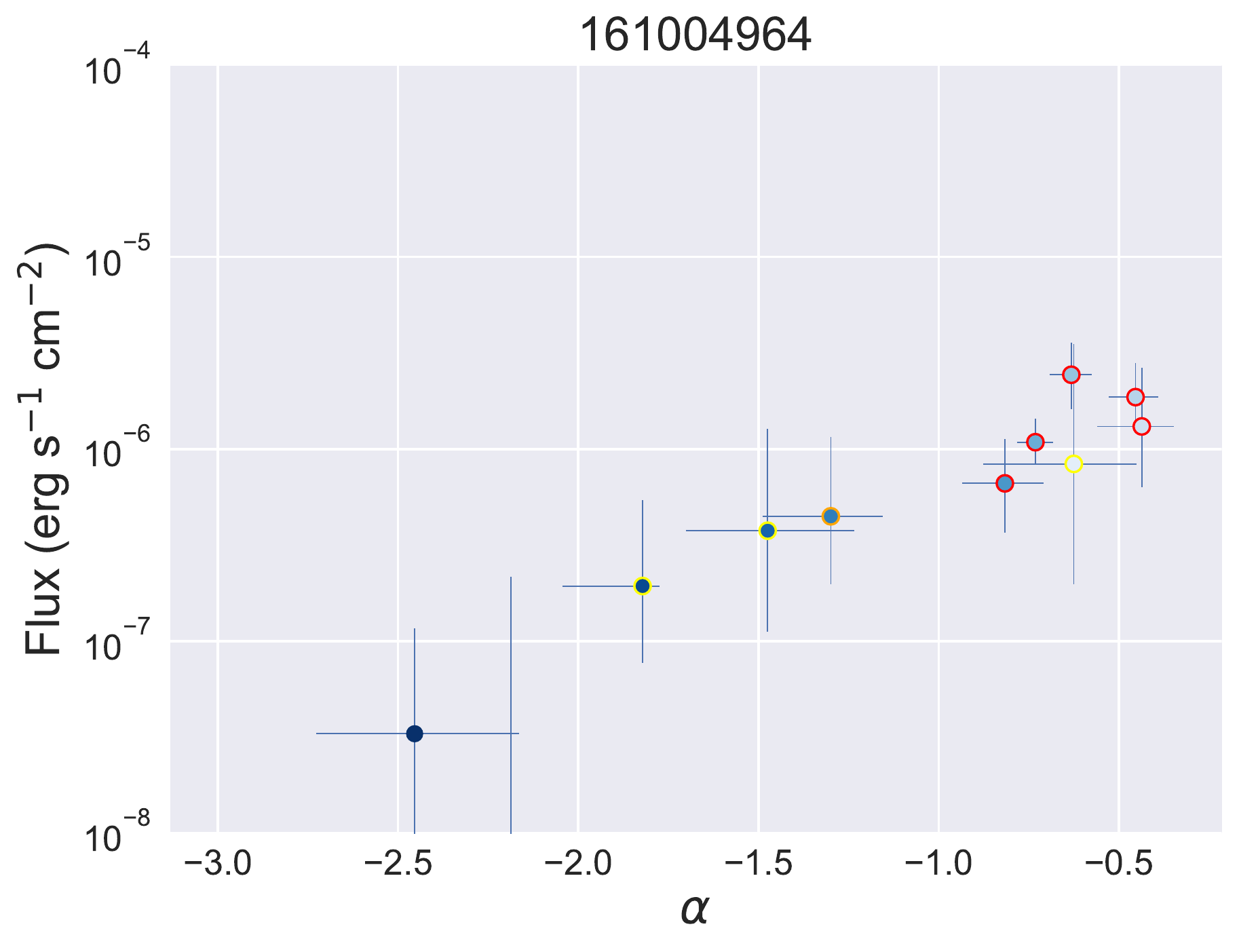}}

\subfigure{\includegraphics[width=0.3\linewidth]{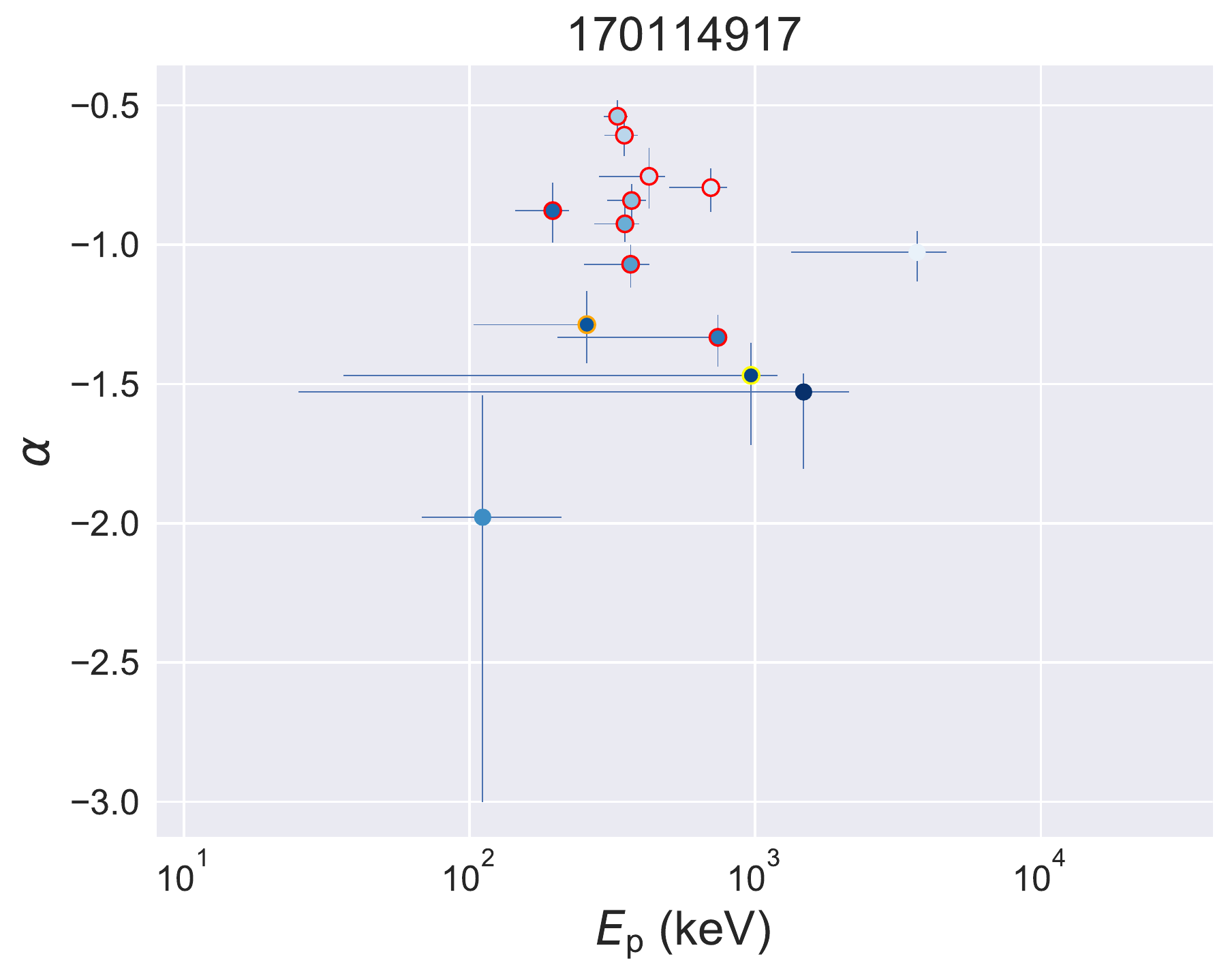}}
\subfigure{\includegraphics[width=0.3\linewidth]{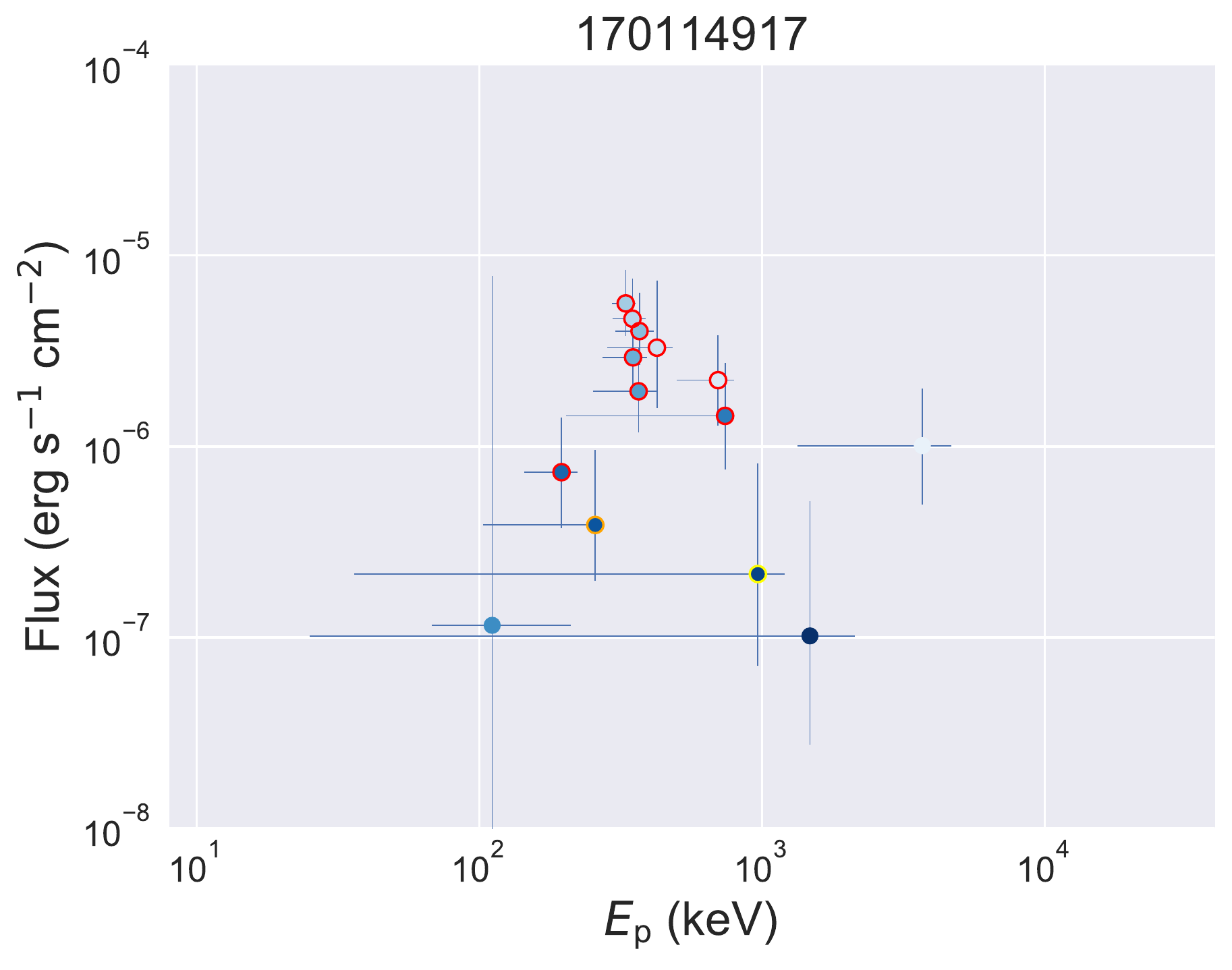}}
\subfigure{\includegraphics[width=0.3\linewidth]{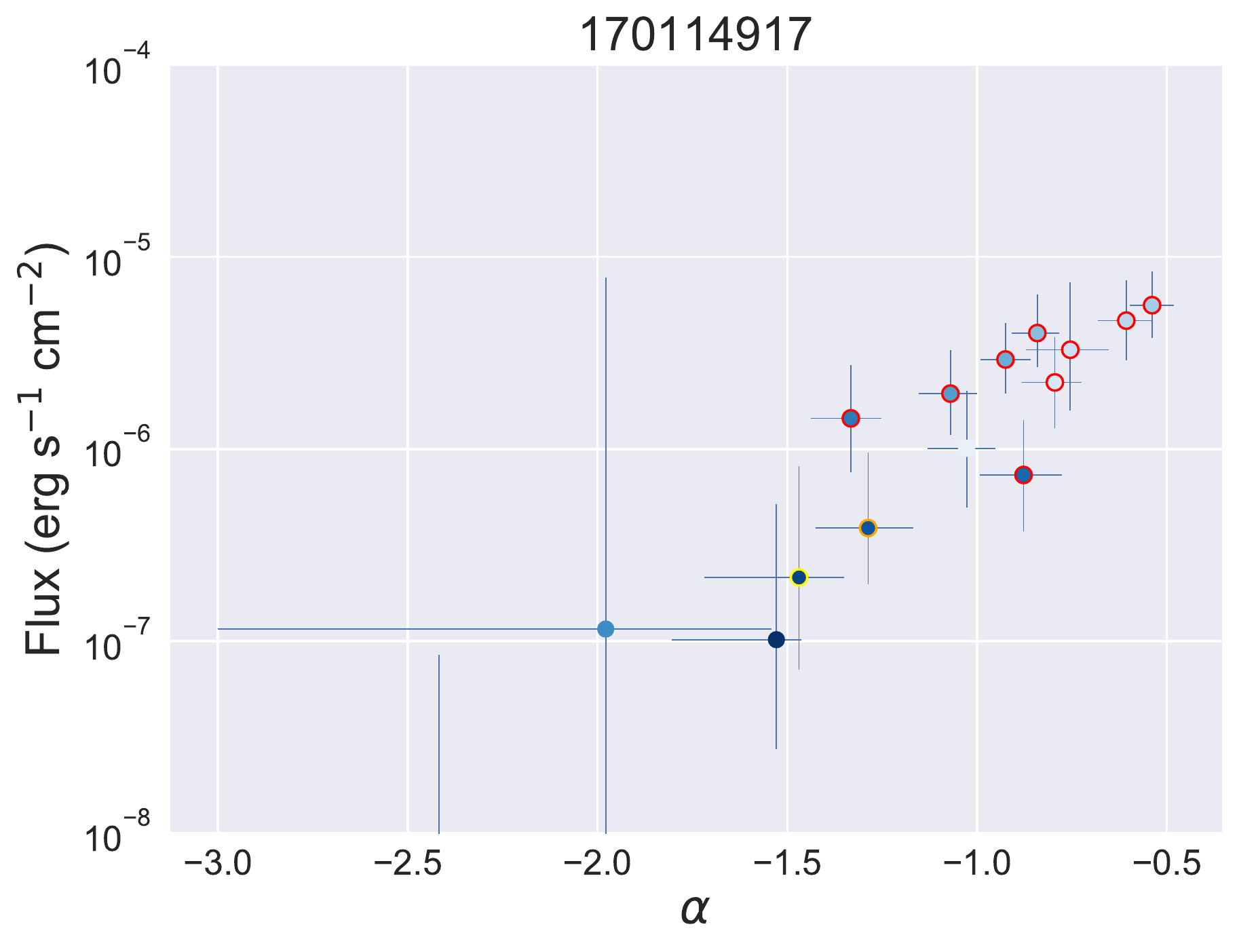}}

\caption{Same as Fig.~\ref{fig:correlation_group1}.
\label{fig:correlation_group10}}
\end{figure*}

\section{Tables for All Results}
\label{app:tables}

\movetabledown=3.6in
\begin{rotatetable}
\begin{deluxetable}{cccccccccccccccc}
\tabletypesize{\scriptsize}
\tablecaption{Time-resolved spectral analysis results of the first pulse of GRB081009140. \label{tab:fitresults}}
\setlength{\tabcolsep}{0pt}
\tablehead{ \colhead{$t_{\rm start}$} & \colhead{$t_{\rm stop}$} & \colhead{$S$} & \colhead{$K$}& \colhead{$\alpha$} & \colhead{$E_{\rm c}$} &\colhead{$E_{\rm p}$}& \colhead{$F$} & \colhead{$K_{\rm BAND}$} & \colhead{$\alpha_{\rm BAND}$} & \colhead{$\beta_{\rm BAND}$} & \colhead{$E_{\rm p,BAND}$} & \colhead{$F_{\rm BAND}$} & \colhead{$\Delta$DIC} & \colhead{$p_{\rm DIC}$} & \colhead{$p_{\rm DIC, BAND}$} \\
\colhead{(1)} & \colhead{(2)} & \colhead{(3)} & \colhead{(4)} & \colhead{(5)} & \colhead{(6)} & \colhead{(7)} & \colhead{(8)} & \colhead{(9)} & \colhead{(10)} & \colhead{(11)} & \colhead{(12)} & \colhead{(13)} & \colhead{(14)} & \colhead{(15)} & \colhead{(16)}
}
\startdata
0.00 & 0.37 & 21.26 & $ 1.35 ^{+ 0.45 }_{- 1.15 } \times 10^{ +2 }$ & $ -1.75 ^{+ 0.18 }_{- 0.38 }$ & $ 1485.52 ^{+ 786.01 }_{- 1455.41 }$ & $ 375.08 ^{+ 198.46 }_{- 367.47 }$ & $ 1.17 ^{+ 5.33 }_{- 0.87 } \times 10^{ -6 }$ & $ 4.78 ^{+ 1.48 }_{- 4.77 } \times 10^{ +1 }$ & $ 0.76 ^{+ 0.80 }_{- 0.68 }$ & $ -2.58 ^{+ 0.24 }_{- 0.14 }$ & $ 29.55 ^{+ 2.61 }_{- 4.25 }$ & $ 9.83 ^{+ 104.80 }_{- 8.85 } \times 10^{ -7 }$ & -921.87 & -125.85 & -1030.44 \\
0.37 & 1.11 & 47.27 & $ 2.14 ^{+ 0.54 }_{- 1.04 } \times 10^{ +2 }$ & $ -1.63 ^{+ 0.15 }_{- 0.20 }$ & $ 51.68 ^{+ 8.36 }_{- 15.47 }$ & $ 18.99 ^{+ 3.07 }_{- 5.68 }$ & $ 1.40 ^{+ 1.53 }_{- 0.69 } \times 10^{ -6 }$ & $ 5.55 ^{+ 0.59 }_{- 4.81 } \times 10^{ -1 }$ & $ -1.15 ^{+ 0.25 }_{- 0.45 }$ & $ -3.45 ^{+ 0.72 }_{- 0.17 }$ & $ 21.43 ^{+ 2.88 }_{- 1.85 }$ & $ 1.43 ^{+ 3.75 }_{- 1.08 } \times 10^{ -6 }$ & -236.43 & -12.66 & -242.95 \\
1.11 & 1.36 & 39.67 & $ 1.60 ^{+ 0.34 }_{- 0.94 } \times 10^{ +2 }$ & $ -1.46 ^{+ 0.18 }_{- 0.19 }$ & $ 73.67 ^{+ 9.67 }_{- 27.70 }$ & $ 40.13 ^{+ 5.27 }_{- 15.09 }$ & $ 2.49 ^{+ 3.50 }_{- 1.37 } \times 10^{ -6 }$ & $ 7.73 ^{+ 2.00 }_{- 6.66 } \times 10^{ -1 }$ & $ -1.04 ^{+ 0.19 }_{- 0.49 }$ & $ -3.51 ^{+ 1.00 }_{- 0.44 }$ & $ 35.90 ^{+ 4.34 }_{- 3.65 }$ & $ 2.79 ^{+ 6.19 }_{- 2.00 } \times 10^{ -6 }$ & -212.93 & -12.42 & -220.78 \\
1.36 & 1.66 & 57.40 & $ 4.82 ^{+ 1.12 }_{- 2.20 } \times 10^{ +1 }$ & $ -0.92 ^{+ 0.13 }_{- 0.16 }$ & $ 43.56 ^{+ 4.64 }_{- 7.33 }$ & $ 47.02 ^{+ 5.01 }_{- 7.91 }$ & $ 3.49 ^{+ 3.35 }_{- 1.80 } \times 10^{ -6 }$ & $ 7.49 ^{+ 1.21 }_{- 2.69 } \times 10^{ -1 }$ & $ -0.87 ^{+ 0.12 }_{- 0.16 }$ & $ -4.38 ^{+ 0.18 }_{- 0.62 }$ & $ 46.32 ^{+ 1.91 }_{- 2.19 }$ & $ 3.49 ^{+ 1.78 }_{- 1.17 } \times 10^{ -6 }$ & 4.66 & -6.30 & 1.16 \\
1.66 & 1.89 & 58.50 & $ 3.61 ^{+ 0.82 }_{- 1.81 } \times 10^{ +1 }$ & $ -0.73 ^{+ 0.16 }_{- 0.16 }$ & $ 36.90 ^{+ 3.91 }_{- 5.64 }$ & $ 46.98 ^{+ 4.98 }_{- 7.18 }$ & $ 4.30 ^{+ 4.30 }_{- 2.28 } \times 10^{ -6 }$ & $ 1.58 ^{+ 0.27 }_{- 0.72 } \times 10^{ +0 }$ & $ -0.61 ^{+ 0.16 }_{- 0.18 }$ & $ -4.34 ^{+ 0.20 }_{- 0.66 }$ & $ 45.78 ^{+ 1.72 }_{- 2.10 }$ & $ 4.20 ^{+ 2.67 }_{- 1.57 } \times 10^{ -6 }$ & 5.61 & -8.96 & -1.60 \\
1.89 & 2.04 & 58.07 & $ 1.18 ^{+ 0.26 }_{- 0.51 } \times 10^{ +2 }$ & $ -1.14 ^{+ 0.12 }_{- 0.14 }$ & $ 67.09 ^{+ 7.75 }_{- 14.36 }$ & $ 57.95 ^{+ 6.69 }_{- 12.40 }$ & $ 6.06 ^{+ 5.55 }_{- 2.76 } \times 10^{ -6 }$ & $ 7.44 ^{+ 1.06 }_{- 2.88 } \times 10^{ -1 }$ & $ -1.05 ^{+ 0.14 }_{- 0.16 }$ & $ -4.10 ^{+ 0.29 }_{- 0.89 }$ & $ 55.36 ^{+ 3.00 }_{- 3.72 }$ & $ 6.12 ^{+ 3.20 }_{- 2.23 } \times 10^{ -6 }$ & 0.58 & -4.09 & -0.11 \\
2.04 & 2.72 & 140.80 & $ 5.96 ^{+ 0.75 }_{- 0.96 } \times 10^{ +1 }$ & $ -0.91 ^{+ 0.05 }_{- 0.05 }$ & $ 67.28 ^{+ 3.93 }_{- 4.50 }$ & $ 73.36 ^{+ 4.28 }_{- 4.90 }$ & $ 8.06 ^{+ 2.44 }_{- 1.72 } \times 10^{ -6 }$ & $ 9.21 ^{+ 0.81 }_{- 0.98 } \times 10^{ -1 }$ & $ -0.90 ^{+ 0.05 }_{- 0.05 }$ & $ -4.43 ^{+ 0.19 }_{- 0.55 }$ & $ 72.69 ^{+ 1.77 }_{- 1.87 }$ & $ 8.19 ^{+ 1.28 }_{- 1.01 } \times 10^{ -6 }$ & -2.09 & 1.96 & 3.26 \\
2.72 & 3.00 & 102.35 & $ 5.87 ^{+ 0.93 }_{- 1.32 } \times 10^{ +1 }$ & $ -0.88 ^{+ 0.07 }_{- 0.07 }$ & $ 76.95 ^{+ 5.66 }_{- 7.33 }$ & $ 85.94 ^{+ 6.32 }_{- 8.19 }$ & $ 1.01 ^{+ 0.45 }_{- 0.31 } \times 10^{ -5 }$ & $ 1.04 ^{+ 0.11 }_{- 0.16 } \times 10^{ +0 }$ & $ -0.86 ^{+ 0.07 }_{- 0.08 }$ & $ -4.37 ^{+ 0.20 }_{- 0.63 }$ & $ 84.43 ^{+ 2.85 }_{- 3.59 }$ & $ 1.05 ^{+ 0.23 }_{- 0.17 } \times 10^{ -5 }$ & -0.25 & 1.28 & 3.00 \\
3.00 & 3.29 & 91.89 & $ 9.24 ^{+ 1.64 }_{- 2.72 } \times 10^{ +1 }$ & $ -0.98 ^{+ 0.09 }_{- 0.10 }$ & $ 52.22 ^{+ 4.26 }_{- 5.97 }$ & $ 53.52 ^{+ 4.36 }_{- 6.11 }$ & $ 6.73 ^{+ 4.00 }_{- 2.33 } \times 10^{ -6 }$ & $ 1.13 ^{+ 0.15 }_{- 0.27 } \times 10^{ +0 }$ & $ -0.93 ^{+ 0.09 }_{- 0.10 }$ & $ -4.33 ^{+ 0.21 }_{- 0.67 }$ & $ 52.48 ^{+ 1.75 }_{- 1.96 }$ & $ 7.06 ^{+ 2.10 }_{- 1.70 } \times 10^{ -6 }$ & -0.22 & -0.24 & 2.45 \\
3.29 & 4.84 & 181.02 & $ 8.87 ^{+ 0.94 }_{- 1.23 } \times 10^{ +1 }$ & $ -0.98 ^{+ 0.05 }_{- 0.05 }$ & $ 42.72 ^{+ 1.86 }_{- 2.14 }$ & $ 43.54 ^{+ 1.89 }_{- 2.18 }$ & $ 5.32 ^{+ 1.33 }_{- 1.01 } \times 10^{ -6 }$ & $ 1.02 ^{+ 0.09 }_{- 0.12 } \times 10^{ +0 }$ & $ -0.96 ^{+ 0.05 }_{- 0.05 }$ & $ -4.65 ^{+ 0.09 }_{- 0.35 }$ & $ 43.23 ^{+ 0.74 }_{- 0.67 }$ & $ 5.40 ^{+ 0.98 }_{- 0.76 } \times 10^{ -6 }$ & 0.44 & 2.06 & 2.95 \\
4.84 & 5.38 & 93.03 & $ 5.77 ^{+ 0.98 }_{- 1.55 } \times 10^{ +1 }$ & $ -0.96 ^{+ 0.08 }_{- 0.08 }$ & $ 54.84 ^{+ 4.14 }_{- 5.96 }$ & $ 56.95 ^{+ 4.30 }_{- 6.19 }$ & $ 4.85 ^{+ 2.37 }_{- 1.56 } \times 10^{ -6 }$ & $ 7.54 ^{+ 0.96 }_{- 1.59 } \times 10^{ -1 }$ & $ -0.91 ^{+ 0.08 }_{- 0.09 }$ & $ -4.11 ^{+ 0.55 }_{- 0.49 }$ & $ 55.71 ^{+ 1.79 }_{- 1.86 }$ & $ 4.92 ^{+ 1.46 }_{- 1.02 } \times 10^{ -6 }$ & -1.90 & -0.04 & 2.73 \\
5.38 & 5.65 & 58.32 & $ 3.50 ^{+ 0.72 }_{- 1.73 } \times 10^{ +1 }$ & $ -0.79 ^{+ 0.15 }_{- 0.15 }$ & $ 41.98 ^{+ 4.19 }_{- 6.92 }$ & $ 50.80 ^{+ 5.07 }_{- 8.38 }$ & $ 3.97 ^{+ 3.64 }_{- 2.05 } \times 10^{ -6 }$ & $ 1.20 ^{+ 0.19 }_{- 0.53 } \times 10^{ +0 }$ & $ -0.66 ^{+ 0.16 }_{- 0.18 }$ & $ -4.24 ^{+ 0.24 }_{- 0.76 }$ & $ 49.09 ^{+ 2.10 }_{- 2.14 }$ & $ 3.88 ^{+ 2.18 }_{- 1.47 } \times 10^{ -6 }$ & 3.14 & -7.04 & -1.53 \\
5.65 & 6.35 & 75.49 & $ 1.68 ^{+ 0.37 }_{- 0.70 } \times 10^{ +1 }$ & $ -0.51 ^{+ 0.13 }_{- 0.14 }$ & $ 27.01 ^{+ 2.07 }_{- 2.79 }$ & $ 40.16 ^{+ 3.08 }_{- 4.15 }$ & $ 2.56 ^{+ 2.03 }_{- 1.10 } \times 10^{ -6 }$ & $ 1.76 ^{+ 0.31 }_{- 0.64 } \times 10^{ +0 }$ & $ -0.46 ^{+ 0.14 }_{- 0.14 }$ & $ -4.65 ^{+ 0.09 }_{- 0.35 }$ & $ 39.63 ^{+ 1.08 }_{- 1.01 }$ & $ 2.66 ^{+ 1.46 }_{- 0.93 } \times 10^{ -6 }$ & 8.07 & -8.16 & -0.82 \\
6.35 & 6.89 & 57.49 & $ 3.42 ^{+ 0.77 }_{- 1.81 } \times 10^{ +1 }$ & $ -0.69 ^{+ 0.19 }_{- 0.18 }$ & $ 21.74 ^{+ 2.21 }_{- 2.78 }$ & $ 28.55 ^{+ 2.90 }_{- 3.65 }$ & $ 1.92 ^{+ 2.27 }_{- 1.02 } \times 10^{ -6 }$ & $ 2.03 ^{+ 0.35 }_{- 1.16 } \times 10^{ +0 }$ & $ -0.55 ^{+ 0.18 }_{- 0.23 }$ & $ -4.57 ^{+ 0.12 }_{- 0.43 }$ & $ 28.22 ^{+ 1.34 }_{- 1.06 }$ & $ 2.10 ^{+ 2.03 }_{- 1.05 } \times 10^{ -6 }$ & 1.07 & -13.85 & -13.43 \\
6.89 & 7.22 & 38.60 & $ 2.12 ^{+ 0.39 }_{- 1.45 } \times 10^{ +2 }$ & $ -1.45 ^{+ 0.24 }_{- 0.23 }$ & $ 33.53 ^{+ 4.42 }_{- 10.41 }$ & $ 18.58 ^{+ 2.45 }_{- 5.77 }$ & $ 1.70 ^{+ 2.94 }_{- 1.02 } \times 10^{ -6 }$ & $ 1.11 ^{+ 0.16 }_{- 0.96 } \times 10^{ +0 }$ & $ -0.99 ^{+ 0.22 }_{- 0.45 }$ & $ -4.07 ^{+ 0.44 }_{- 0.77 }$ & $ 20.50 ^{+ 2.78 }_{- 2.13 }$ & $ 1.63 ^{+ 3.84 }_{- 1.18 } \times 10^{ -6 }$ & -129.60 & -24.08 & -149.93 \\
7.22 & 7.71 & 31.50 & $ 1.84 ^{+ 0.31 }_{- 1.51 } \times 10^{ +2 }$ & $ -1.24 ^{+ 0.33 }_{- 0.38 }$ & $ 15.83 ^{+ 1.77 }_{- 5.22 }$ & $ 12.00 ^{+ 1.34 }_{- 3.95 }$ & $ 9.38 ^{+ 24.68 }_{- 6.49 } \times 10^{ -7 }$ & $ 6.45 ^{+ 0.27 }_{- 5.95 } \times 10^{ +0 }$ & $ -0.44 ^{+ 0.40 }_{- 0.36 }$ & $ -4.40 ^{+ 0.19 }_{- 0.60 }$ & $ 14.57 ^{+ 1.40 }_{- 1.42 }$ & $ 1.07 ^{+ 3.98 }_{- 0.85 } \times 10^{ -6 }$ & -83.98 & -65.80 & -146.92 \\
7.71 & 7.95 & 15.08 & $ 6.11 ^{+ 3.88 }_{- 1.32 } \times 10^{ +2 }$ & $ -2.02 ^{+ 0.13 }_{- 0.36 }$ & $ 36.51 ^{+ 3.66 }_{- 24.80 }$ & $ -0.90 ^{+ 0.09 }_{- 0.61 }$ & $ 6.17 ^{+ 12.35 }_{- 3.80 } \times 10^{ -7 }$ & $ 2.41 ^{+ 0.26 }_{- 2.39 } \times 10^{ +2 }$ & $ 0.65 ^{+ 0.70 }_{- 0.34 }$ & $ -4.01 ^{+ 0.66 }_{- 0.41 }$ & $ 12.68 ^{+ 1.19 }_{- 1.66 }$ & $ 6.96 ^{+ 65.80 }_{- 6.29 } \times 10^{ -7 }$ & -122.81 & -32.70 & -152.33 \\
7.95 & 8.95 & 15.27 & $ 4.70 ^{+ 1.57 }_{- 3.90 } \times 10^{ +2 }$ & $ -2.03 ^{+ 0.20 }_{- 0.49 }$ & $ 19.81 ^{+ 1.29 }_{- 9.81 }$ & $ -0.61 ^{+ 0.04 }_{- 0.30 }$ & $ 3.22 ^{+ 9.24 }_{- 2.25 } \times 10^{ -7 }$ & $ 5.17 ^{+ 0.94 }_{- 5.15 } \times 10^{ +1 }$ & $ 0.18 ^{+ 0.58 }_{- 0.58 }$ & $ -4.41 ^{+ 0.17 }_{- 0.59 }$ & $ 11.59 ^{+ 0.59 }_{- 1.38 }$ & $ 2.92 ^{+ 36.42 }_{- 2.68 } \times 10^{ -7 }$ & -672.65 & -34.82 & -707.46 \\
8.95 & 10.00 & 5.45 & $ 3.36 ^{+ 1.38 }_{- 1.93 } \times 10^{ +2 }$ & $ -2.76 ^{+ 0.06 }_{- 0.24 }$ & $ 4946.77 ^{+ 1892.03 }_{- 4908.87 }$ & $ -3771.55 ^{+ 1442.54 }_{- 3742.65 }$ & $ 1.21 ^{+ 1.48 }_{- 0.68 } \times 10^{ -7 }$ & $ 2.94 ^{+ 0.88 }_{- 2.94 } \times 10^{ +2 }$ & $ 1.61 ^{+ 0.81 }_{- 0.44 }$ & $ -3.79 ^{+ 0.79 }_{- 0.74 }$ & $ 16.30 ^{+ 2.83 }_{- 4.33 }$ & $ 8.36 ^{+ 99.21 }_{- 7.87 } \times 10^{ -8 }$ & -70.97 & 0.73 & -70.73
\enddata
\end{deluxetable}
\tablecomments{Time-resolved spectral analysis results of the first pulse of GRB081009140. Columns (1) and (2) list the start and stop times (in units of s) of the Bayesian block time bins. Column (3) lists the significance of the bin. Columns (4) - (6) list the best-fit parameters for the CPL model. Column (7) lists the derived values of $E_{\rm p}$ for the CPL model. Column (8) lists the derived CPL energy flux. Columns (9) - (12) list the best-fit parameters for the BAND model. Column (13) lists the derived BAND energy flux. Column (14) list the difference between the Deviance Information Criterion (DIC) for the CPL and BAND model, $\Delta{\rm DIC}={\rm DIC}_{\rm BAND}-{\rm DIC}_{\rm CPL}$. Columns (15) and (16) list the effective number of parameters for the CPL and BAND model, respectively. All time parameters have units of s, normalisations have units of ph~s$^{-1}$~cm$^{-2}$~keV$^{-1}$, energies have units of keV, and fluxes have units of erg~s$^{-1}$~cm$^{-2}$. N/A means that a reliable value of the flux could not be computed due to large errors in the fitted parameters.}
\end{rotatetable}

\movetabledown=3.6in
\begin{rotatetable}
\begin{deluxetable}{cccccccccccccccc}
\tabletypesize{\scriptsize}
\tablecaption{Time-resolved spectral analysis results of the second pulse of GRB081009140.}
\setlength{\tabcolsep}{0pt}
\tablehead{ \colhead{$t_{\rm start}$} & \colhead{$t_{\rm stop}$} & \colhead{$S$} & \colhead{$K$}& \colhead{$\alpha$} & \colhead{$E_{\rm c}$} &\colhead{$E_{\rm p}$}& \colhead{$F$} & \colhead{$K_{\rm BAND}$} & \colhead{$\alpha_{\rm BAND}$} & \colhead{$\beta_{\rm BAND}$} & \colhead{$E_{\rm p,BAND}$} & \colhead{$F_{\rm BAND}$} & \colhead{$\Delta$DIC} & \colhead{$p_{\rm DIC}$} & \colhead{$p_{\rm DIC, BAND}$} \\
\colhead{(1)} & \colhead{(2)} & \colhead{(3)} & \colhead{(4)} & \colhead{(5)} & \colhead{(6)} & \colhead{(7)} & \colhead{(8)} & \colhead{(9)} & \colhead{(10)} & \colhead{(11)} & \colhead{(12)} & \colhead{(13)} & \colhead{(14)} & \colhead{(15)} & \colhead{(16)}
}
\startdata
33.00 & 34.81 & 1.91 & $ 4.83 ^{+ 1.87 }_{- 4.83 } \times 10^{ +1 }$ & $ -2.60 ^{+ 0.08 }_{- 0.36 }$ & $ 4853.16 ^{+ 1902.21 }_{- 4842.68 }$ & $ -2896.23 ^{+ 1135.18 }_{- 2889.98 }$ & $ 2.55 ^{+ 5.50 }_{- 1.87 } \times 10^{ -8 }$ & $ 3.12 ^{+ 1.11 }_{- 3.12 } \times 10^{ +2 }$ & $ 1.91 ^{+ 0.71 }_{- 0.49 }$ & $ -3.98 ^{+ 0.27 }_{- 1.02 }$ & $ 13.99 ^{+ 0.53 }_{- 3.99 }$ & $ 2.09 ^{+ 35.25 }_{- 1.94 } \times 10^{ -8 }$ & -4.04 & 0.65 & -2.47 \\
34.81 & 37.40 & 8.37 & $ 4.41 ^{+ 1.68 }_{- 1.47 } \times 10^{ +2 }$ & $ -2.84 ^{+ 0.07 }_{- 0.16 }$ & $ 4776.67 ^{+ 1812.37 }_{- 4750.37 }$ & $ -4023.38 ^{+ 1526.55 }_{- 4001.22 }$ & $ 1.37 ^{+ 1.20 }_{- 0.59 } \times 10^{ -7 }$ & $ 2.54 ^{+ 0.78 }_{- 2.53 } \times 10^{ +1 }$ & $ 0.39 ^{+ 0.82 }_{- 0.67 }$ & $ -3.96 ^{+ 0.44 }_{- 0.93 }$ & $ 14.18 ^{+ 2.15 }_{- 2.57 }$ & $ 6.98 ^{+ 127.50 }_{- 6.52 } \times 10^{ -8 }$ & 766.57 & 0.00 & -805.18 \\
37.40 & 39.07 & 15.72 & $ 3.33 ^{+ 0.08 }_{- 3.15 } \times 10^{ +1 }$ & $ -1.11 ^{+ 0.61 }_{- 0.38 }$ & $ 19.88 ^{+ 2.22 }_{- 8.80 }$ & $ 17.76 ^{+ 1.98 }_{- 7.86 }$ & $ 3.37 ^{+ 14.91 }_{- 2.91 } \times 10^{ -7 }$ & $ 2.26 ^{+ 0.70 }_{- 2.24 } \times 10^{ +1 }$ & $ 0.48 ^{+ 0.58 }_{- 0.71 }$ & $ -4.28 ^{+ 0.24 }_{- 0.72 }$ & $ 19.50 ^{+ 1.76 }_{- 1.88 }$ & $ 2.48 ^{+ 23.19 }_{- 2.26 } \times 10^{ -7 }$ & -643.38 & -172.52 & -815.60 \\
39.07 & 39.96 & 26.34 & $ 5.63 ^{+ 0.80 }_{- 4.50 } \times 10^{ +1 }$ & $ -1.05 ^{+ 0.34 }_{- 0.31 }$ & $ 17.05 ^{+ 1.80 }_{- 4.85 }$ & $ 16.12 ^{+ 1.70 }_{- 4.59 }$ & $ 6.42 ^{+ 16.03 }_{- 4.46 } \times 10^{ -7 }$ & $ 2.73 ^{+ 0.09 }_{- 2.61 } \times 10^{ +0 }$ & $ -0.47 ^{+ 0.45 }_{- 0.46 }$ & $ -4.45 ^{+ 0.16 }_{- 0.55 }$ & $ 17.36 ^{+ 1.80 }_{- 1.49 }$ & $ 5.75 ^{+ 29.26 }_{- 4.62 } \times 10^{ -7 }$ & -166.21 & -48.08 & -214.19 \\
\enddata
\end{deluxetable}
\tablecomments{All columns are the same as Table~\ref{tab:fitresults}.}
\end{rotatetable}

\movetabledown=3.6in
\begin{rotatetable}
\begin{deluxetable}{cccccccccccccccc}
\tabletypesize{\scriptsize}
\tablecaption{Time-resolved spectral analysis results of GRB081125496. }
\setlength{\tabcolsep}{0pt}
\tablehead{ \colhead{$t_{\rm start}$} & \colhead{$t_{\rm stop}$} & \colhead{$S$} & \colhead{$K$}& \colhead{$\alpha$} & \colhead{$E_{\rm c}$} &\colhead{$E_{\rm p}$}& \colhead{$F$} & \colhead{$K_{\rm BAND}$} & \colhead{$\alpha_{\rm BAND}$} & \colhead{$\beta_{\rm BAND}$} & \colhead{$E_{\rm p,BAND}$} & \colhead{$F_{\rm BAND}$} & \colhead{$\Delta$DIC} & \colhead{$p_{\rm DIC}$} & \colhead{$p_{\rm DIC, BAND}$} \\
\colhead{(1)} & \colhead{(2)} & \colhead{(3)} & \colhead{(4)} & \colhead{(5)} & \colhead{(6)} & \colhead{(7)} & \colhead{(8)} & \colhead{(9)} & \colhead{(10)} & \colhead{(11)} & \colhead{(12)} & \colhead{(13)} & \colhead{(14)} & \colhead{(15)} & \colhead{(16)}
}
\startdata
-5.00 & -1.21 & 1.30 & $ 2.06 ^{+ 1.57 }_{- 2.06 } \times 10^{ +0 }$ & $ -1.32 ^{+ 0.67 }_{- 0.16 }$ & $ 5146.64 ^{+ 2695.51 }_{- 3451.13 }$ & $ 3510.65 ^{+ 1838.68 }_{- 2354.10 }$ & $ 8.67 ^{+ 148.26 }_{- 8.55 } \times 10^{ -8 }$ & $ 8.94 ^{+ 3.63 }_{- 5.95 } \times 10^{ -4 }$ & $ -0.20 ^{+ 0.23 }_{- 1.11 }$ & $ -3.22 ^{+ 1.61 }_{- 0.61 }$ & $ 3317.41 ^{+ 774.78 }_{- 3032.88 }$ & N/A & 63.87 & -78.34 & -6.75 \\
-1.21 & 0.42 & 10.40 & $ 1.96 ^{+ 0.40 }_{- 1.72 } \times 10^{ -1 }$ & $ -0.56 ^{+ 0.15 }_{- 0.21 }$ & $ 522.15 ^{+ 76.35 }_{- 250.93 }$ & $ 749.99 ^{+ 109.67 }_{- 360.42 }$ & $ 1.05 ^{+ 2.85 }_{- 0.78 } \times 10^{ -6 }$ & $ 1.17 ^{+ 0.11 }_{- 0.26 } \times 10^{ -2 }$ & $ -0.45 ^{+ 0.17 }_{- 0.24 }$ & $ -3.78 ^{+ 0.40 }_{- 1.22 }$ & $ 643.50 ^{+ 110.40 }_{- 207.01 }$ & $ 1.08 ^{+ 0.44 }_{- 0.31 } \times 10^{ -6 }$ & 27.71 & -26.41 & 2.66 \\
0.42 & 1.66 & 25.90 & $ 7.02 ^{+ 1.72 }_{- 3.51 } \times 10^{ -1 }$ & $ -0.55 ^{+ 0.10 }_{- 0.13 }$ & $ 227.03 ^{+ 28.13 }_{- 52.84 }$ & $ 329.41 ^{+ 40.82 }_{- 76.66 }$ & $ 2.24 ^{+ 2.67 }_{- 1.27 } \times 10^{ -6 }$ & $ 6.36 ^{+ 0.55 }_{- 1.75 } \times 10^{ -2 }$ & $ -0.41 ^{+ 0.12 }_{- 0.21 }$ & $ -2.69 ^{+ 0.81 }_{- 0.09 }$ & $ 276.15 ^{+ 42.81 }_{- 55.80 }$ & $ 2.21 ^{+ 0.92 }_{- 0.62 } \times 10^{ -6 }$ & 2.73 & -6.73 & 0.51 \\
1.66 & 2.11 & 26.30 & $ 6.68 ^{+ 1.60 }_{- 3.56 } \times 10^{ -1 }$ & $ -0.38 ^{+ 0.11 }_{- 0.12 }$ & $ 186.47 ^{+ 21.33 }_{- 33.35 }$ & $ 301.49 ^{+ 34.49 }_{- 53.92 }$ & $ 3.76 ^{+ 4.43 }_{- 1.99 } \times 10^{ -6 }$ & $ 1.11 ^{+ 0.12 }_{- 0.17 } \times 10^{ -1 }$ & $ -0.32 ^{+ 0.12 }_{- 0.14 }$ & $ -3.95 ^{+ 0.34 }_{- 1.04 }$ & $ 286.05 ^{+ 23.92 }_{- 30.76 }$ & $ 3.84 ^{+ 0.97 }_{- 0.69 } \times 10^{ -6 }$ & 9.52 & -6.75 & 3.14 \\
\enddata
\end{deluxetable}
\tablecomments{All columns are the same as Table~\ref{tab:fitresults}.}
\end{rotatetable}

\movetabledown=3.6in
\begin{rotatetable}
\begin{deluxetable}{cccccccccccccccc}
\tabletypesize{\scriptsize}
\tablecaption{Time-resolved spectral analysis results of GRB081224887.}
\setlength{\tabcolsep}{0pt}
\tablehead{ \colhead{$t_{\rm start}$} & \colhead{$t_{\rm stop}$} & \colhead{$S$} & \colhead{$K$}& \colhead{$\alpha$} & \colhead{$E_{\rm c}$} &\colhead{$E_{\rm p}$}& \colhead{$F$} & \colhead{$K_{\rm BAND}$} & \colhead{$\alpha_{\rm BAND}$} & \colhead{$\beta_{\rm BAND}$} & \colhead{$E_{\rm p,BAND}$} & \colhead{$F_{\rm BAND}$} & \colhead{$\Delta$DIC} & \colhead{$p_{\rm DIC}$} & \colhead{$p_{\rm DIC, BAND}$} \\
\colhead{(1)} & \colhead{(2)} & \colhead{(3)} & \colhead{(4)} & \colhead{(5)} & \colhead{(6)} & \colhead{(7)} & \colhead{(8)} & \colhead{(9)} & \colhead{(10)} & \colhead{(11)} & \colhead{(12)} & \colhead{(13)} & \colhead{(14)} & \colhead{(15)} & \colhead{(16)}
}
\startdata
0.00 & 0.35 & 14.01 & $ 2.32 ^{+ 0.59 }_{- 1.66 } \times 10^{ -1 }$ & $ -0.51 ^{+ 0.12 }_{- 0.14 }$ & $ 1042.06 ^{+ 196.60 }_{- 403.76 }$ & $ 1551.67 ^{+ 292.75 }_{- 601.22 }$ & $ 2.84 ^{+ 5.48 }_{- 1.74 } \times 10^{ -6 }$ & $ 1.88 ^{+ 0.15 }_{- 0.18 } \times 10^{ -2 }$ & $ -0.39 ^{+ 0.12 }_{- 0.21 }$ & $ -3.67 ^{+ 0.80 }_{- 0.91 }$ & $ 1281.64 ^{+ 183.09 }_{- 405.02 }$ & $ 3.22 ^{+ 0.99 }_{- 0.67 } \times 10^{ -6 }$ & 22.64 & -22.11 & 2.14 \\
0.35 & 1.76 & 48.64 & $ 1.00 ^{+ 0.19 }_{- 0.25 } \times 10^{ -1 }$ & $ -0.13 ^{+ 0.05 }_{- 0.05 }$ & $ 344.76 ^{+ 22.13 }_{- 24.22 }$ & $ 644.75 ^{+ 41.39 }_{- 45.30 }$ & $ 6.46 ^{+ 3.20 }_{- 1.97 } \times 10^{ -6 }$ & $ 5.41 ^{+ 0.16 }_{- 0.19 } \times 10^{ -2 }$ & $ -0.12 ^{+ 0.05 }_{- 0.05 }$ & $ -4.18 ^{+ 0.27 }_{- 0.81 }$ & $ 636.82 ^{+ 25.30 }_{- 29.47 }$ & $ 6.28 ^{+ 0.48 }_{- 0.41 } \times 10^{ -6 }$ & 5.46 & -1.54 & 3.29 \\
1.76 & 3.69 & 78.71 & $ 5.79 ^{+ 0.64 }_{- 0.79 } \times 10^{ -1 }$ & $ -0.37 ^{+ 0.03 }_{- 0.03 }$ & $ 257.03 ^{+ 11.76 }_{- 12.31 }$ & $ 417.76 ^{+ 19.11 }_{- 20.01 }$ & $ 6.30 ^{+ 1.44 }_{- 1.33 } \times 10^{ -6 }$ & $ 1.03 ^{+ 0.03 }_{- 0.03 } \times 10^{ -1 }$ & $ -0.37 ^{+ 0.03 }_{- 0.04 }$ & $ -4.26 ^{+ 0.23 }_{- 0.74 }$ & $ 412.26 ^{+ 12.56 }_{- 14.28 }$ & $ 6.16 ^{+ 0.34 }_{- 0.33 } \times 10^{ -6 }$ & 1.89 & 2.12 & 3.24 \\
3.69 & 4.65 & 46.14 & $ 1.31 ^{+ 0.23 }_{- 0.26 } \times 10^{ +0 }$ & $ -0.57 ^{+ 0.04 }_{- 0.07 }$ & $ 223.78 ^{+ 20.22 }_{- 20.26 }$ & $ 320.07 ^{+ 28.91 }_{- 28.97 }$ & $ 3.78 ^{+ 1.60 }_{- 1.23 } \times 10^{ -6 }$ & $ 9.68 ^{+ 0.49 }_{- 1.04 } \times 10^{ -2 }$ & $ -0.54 ^{+ 0.05 }_{- 0.07 }$ & $ -3.36 ^{+ 1.10 }_{- 0.43 }$ & $ 299.20 ^{+ 27.32 }_{- 21.41 }$ & $ 3.81 ^{+ 0.60 }_{- 0.50 } \times 10^{ -6 }$ & 2.13 & -2.90 & 2.81 \\
\enddata
\end{deluxetable}
\tablecomments{All columns are the same as Table~\ref{tab:fitresults}.}
\end{rotatetable}

\movetabledown=3.6in
\begin{rotatetable}
\begin{deluxetable}{cccccccccccccccc}
\tabletypesize{\scriptsize}
\tablecaption{Time-resolved spectral analysis results of GRB090530760.}
\setlength{\tabcolsep}{0pt}
\tablehead{ \colhead{$t_{\rm start}$} & \colhead{$t_{\rm stop}$} & \colhead{$S$} & \colhead{$K$}& \colhead{$\alpha$} & \colhead{$E_{\rm c}$} &\colhead{$E_{\rm p}$}& \colhead{$F$} & \colhead{$K_{\rm BAND}$} & \colhead{$\alpha_{\rm BAND}$} & \colhead{$\beta_{\rm BAND}$} & \colhead{$E_{\rm p,BAND}$} & \colhead{$F_{\rm BAND}$} & \colhead{$\Delta$DIC} & \colhead{$p_{\rm DIC}$} & \colhead{$p_{\rm DIC, BAND}$} \\
\colhead{(1)} & \colhead{(2)} & \colhead{(3)} & \colhead{(4)} & \colhead{(5)} & \colhead{(6)} & \colhead{(7)} & \colhead{(8)} & \colhead{(9)} & \colhead{(10)} & \colhead{(11)} & \colhead{(12)} & \colhead{(13)} & \colhead{(14)} & \colhead{(15)} & \colhead{(16)}
}
\startdata
-1.00 & 0.69 & 5.80 & $ 3.52 ^{+ 0.73 }_{- 3.09 } \times 10^{ -1 }$ & $ -0.79 ^{+ 0.18 }_{- 0.24 }$ & $ 1033.51 ^{+ 15.16 }_{- 806.18 }$ & $ 1249.56 ^{+ 18.33 }_{- 974.71 }$ & $ 6.48 ^{+ 26.20 }_{- 5.03 } \times 10^{ -7 }$ & $ 8.91 ^{+ 1.06 }_{- 3.14 } \times 10^{ -3 }$ & $ -0.46 ^{+ 0.27 }_{- 0.39 }$ & $ -3.29 ^{+ 1.41 }_{- 0.67 }$ & $ 628.27 ^{+ 3.46 }_{- 342.34 }$ & $ 6.08 ^{+ 5.35 }_{- 2.14 } \times 10^{ -7 }$ & 68.73 & -87.93 & -15.18 \\
0.69 & 2.89 & 15.90 & $ 1.58 ^{+ 0.35 }_{- 0.83 } \times 10^{ -1 }$ & $ -0.35 ^{+ 0.11 }_{- 0.12 }$ & $ 183.52 ^{+ 20.10 }_{- 32.07 }$ & $ 302.58 ^{+ 33.14 }_{- 52.88 }$ & $ 1.07 ^{+ 1.23 }_{- 0.55 } \times 10^{ -6 }$ & $ 3.31 ^{+ 0.26 }_{- 0.71 } \times 10^{ -2 }$ & $ -0.24 ^{+ 0.12 }_{- 0.18 }$ & $ -3.43 ^{+ 1.18 }_{- 0.61 }$ & $ 274.80 ^{+ 31.45 }_{- 31.46 }$ & $ 1.07 ^{+ 0.31 }_{- 0.23 } \times 10^{ -6 }$ & 12.37 & -11.32 & 1.65 \\
2.89 & 4.98 & 21.80 & $ 1.34 ^{+ 0.30 }_{- 0.52 } \times 10^{ -1 }$ & $ -0.07 ^{+ 0.10 }_{- 0.09 }$ & $ 90.56 ^{+ 6.80 }_{- 8.88 }$ & $ 174.60 ^{+ 13.11 }_{- 17.12 }$ & $ 1.14 ^{+ 0.95 }_{- 0.51 } \times 10^{ -6 }$ & $ 1.98 ^{+ 0.40 }_{- 0.79 } \times 10^{ -1 }$ & $ 0.39 ^{+ 0.22 }_{- 0.23 }$ & $ -2.58 ^{+ 0.35 }_{- 0.04 }$ & $ 136.40 ^{+ 9.41 }_{- 15.79 }$ & $ 1.26 ^{+ 0.70 }_{- 0.46 } \times 10^{ -6 }$ & -9.97 & -6.56 & -11.53 \\
4.98 & 12.66 & 50.40 & $ 2.44 ^{+ 0.33 }_{- 0.48 } \times 10^{ -1 }$ & $ -0.10 ^{+ 0.05 }_{- 0.05 }$ & $ 66.42 ^{+ 2.36 }_{- 3.54 }$ & $ 125.90 ^{+ 4.47 }_{- 6.72 }$ & $ 1.03 ^{+ 0.39 }_{- 0.26 } \times 10^{ -6 }$ & $ 2.40 ^{+ 0.24 }_{- 0.38 } \times 10^{ -1 }$ & $ 0.16 ^{+ 0.07 }_{- 0.09 }$ & $ -2.76 ^{+ 0.13 }_{- 0.10 }$ & $ 109.14 ^{+ 3.16 }_{- 3.91 }$ & $ 1.17 ^{+ 0.22 }_{- 0.19 } \times 10^{ -6 }$ & -31.86 & 0.70 & 2.93 \\
\enddata
\end{deluxetable}
\tablecomments{All columns are the same as Table~\ref{tab:fitresults}.}
\end{rotatetable}

\movetabledown=3.6in
\begin{rotatetable}
\begin{deluxetable}{cccccccccccccccc}
\tabletypesize{\scriptsize}
\tablecaption{Time-resolved spectral analysis results of GRB090620400.}
\setlength{\tabcolsep}{0pt}
\tablehead{ \colhead{$t_{\rm start}$} & \colhead{$t_{\rm stop}$} & \colhead{$S$} & \colhead{$K$}& \colhead{$\alpha$} & \colhead{$E_{\rm c}$} &\colhead{$E_{\rm p}$}& \colhead{$F$} & \colhead{$K_{\rm BAND}$} & \colhead{$\alpha_{\rm BAND}$} & \colhead{$\beta_{\rm BAND}$} & \colhead{$E_{\rm p,BAND}$} & \colhead{$F_{\rm BAND}$} & \colhead{$\Delta$DIC} & \colhead{$p_{\rm DIC}$} & \colhead{$p_{\rm DIC, BAND}$} \\
\colhead{(1)} & \colhead{(2)} & \colhead{(3)} & \colhead{(4)} & \colhead{(5)} & \colhead{(6)} & \colhead{(7)} & \colhead{(8)} & \colhead{(9)} & \colhead{(10)} & \colhead{(11)} & \colhead{(12)} & \colhead{(13)} & \colhead{(14)} & \colhead{(15)} & \colhead{(16)}
}
\startdata
-1.00 & 0.05 & 0.70 & $ 3.44 ^{+ 0.51 }_{- 3.41 } \times 10^{ -1 }$ & $ -0.92 ^{+ 0.17 }_{- 0.40 }$ & $ 2615.36 ^{+ 693.83 }_{- 2531.76 }$ & $ 2827.30 ^{+ 750.05 }_{- 2736.93 }$ & $ 1.79 ^{+ 16.10 }_{- 1.50 } \times 10^{ -7 }$ & $ 2.23 ^{+ 0.40 }_{- 2.04 } \times 10^{ -2 }$ & $ 1.00 ^{+ 0.94 }_{- 1.31 }$ & $ -3.52 ^{+ 0.50 }_{- 1.48 }$ & $ 294.32 ^{+ 9.06 }_{- 146.31 }$ & $ 2.43 ^{+ 8.00 }_{- 1.92 } \times 10^{ -7 }$ & -64.25 & -53.28 & -110.60 \\
0.05 & 0.96 & 11.50 & $ 1.38 ^{+ 0.19 }_{- 1.23 } \times 10^{ -2 }$ & $ 0.36 ^{+ 0.21 }_{- 0.26 }$ & $ 95.87 ^{+ 14.10 }_{- 19.49 }$ & $ 225.99 ^{+ 33.25 }_{- 45.93 }$ & $ 8.44 ^{+ 33.85 }_{- 6.65 } \times 10^{ -7 }$ & $ 1.11 ^{+ 0.06 }_{- 0.69 } \times 10^{ -1 }$ & $ 1.01 ^{+ 0.33 }_{- 0.65 }$ & $ -3.49 ^{+ 1.08 }_{- 0.57 }$ & $ 189.02 ^{+ 20.18 }_{- 24.69 }$ & $ 8.48 ^{+ 7.76 }_{- 3.99 } \times 10^{ -7 }$ & 72.07 & -85.92 & -16.81 \\
0.96 & 3.21 & 30.00 & $ 9.37 ^{+ 2.08 }_{- 3.27 } \times 10^{ -2 }$ & $ 0.06 ^{+ 0.08 }_{- 0.10 }$ & $ 74.75 ^{+ 5.19 }_{- 6.20 }$ & $ 154.11 ^{+ 10.69 }_{- 12.78 }$ & $ 1.04 ^{+ 0.75 }_{- 0.43 } \times 10^{ -6 }$ & $ 1.76 ^{+ 0.25 }_{- 0.43 } \times 10^{ -1 }$ & $ 0.30 ^{+ 0.12 }_{- 0.15 }$ & $ -2.83 ^{+ 0.33 }_{- 0.11 }$ & $ 135.85 ^{+ 6.24 }_{- 9.03 }$ & $ 1.19 ^{+ 0.38 }_{- 0.26 } \times 10^{ -6 }$ & -1.88 & -3.86 & 2.36 \\
3.21 & 4.21 & 28.90 & $ 2.42 ^{+ 0.46 }_{- 0.85 } \times 10^{ -1 }$ & $ -0.22 ^{+ 0.07 }_{- 0.08 }$ & $ 144.33 ^{+ 11.04 }_{- 15.84 }$ & $ 257.31 ^{+ 19.69 }_{- 28.23 }$ & $ 2.39 ^{+ 1.67 }_{- 0.93 } \times 10^{ -6 }$ & $ 1.01 ^{+ 0.10 }_{- 0.17 } \times 10^{ -1 }$ & $ -0.10 ^{+ 0.09 }_{- 0.13 }$ & $ -2.99 ^{+ 0.69 }_{- 0.15 }$ & $ 230.88 ^{+ 19.19 }_{- 19.83 }$ & $ 2.42 ^{+ 0.55 }_{- 0.44 } \times 10^{ -6 }$ & 0.91 & -1.99 & 2.84 \\
\enddata
\end{deluxetable}
\tablecomments{All columns are the same as Table~\ref{tab:fitresults}.}
\end{rotatetable}

\movetabledown=3.6in
\begin{rotatetable}
\begin{deluxetable}{cccccccccccccccc}
\tabletypesize{\scriptsize}
\tablecaption{Time-resolved spectral analysis results of GRB090626189.}
\setlength{\tabcolsep}{0pt}
\tablehead{ \colhead{$t_{\rm start}$} & \colhead{$t_{\rm stop}$} & \colhead{$S$} & \colhead{$K$}& \colhead{$\alpha$} & \colhead{$E_{\rm c}$} &\colhead{$E_{\rm p}$}& \colhead{$F$} & \colhead{$K_{\rm BAND}$} & \colhead{$\alpha_{\rm BAND}$} & \colhead{$\beta_{\rm BAND}$} & \colhead{$E_{\rm p,BAND}$} & \colhead{$F_{\rm BAND}$} & \colhead{$\Delta$DIC} & \colhead{$p_{\rm DIC}$} & \colhead{$p_{\rm DIC, BAND}$} \\
\colhead{(1)} & \colhead{(2)} & \colhead{(3)} & \colhead{(4)} & \colhead{(5)} & \colhead{(6)} & \colhead{(7)} & \colhead{(8)} & \colhead{(9)} & \colhead{(10)} & \colhead{(11)} & \colhead{(12)} & \colhead{(13)} & \colhead{(14)} & \colhead{(15)} & \colhead{(16)}
}
\startdata
30.00 & 31.74 & 7.90 & $ 1.36 ^{+ 0.27 }_{- 0.85 } \times 10^{ +1 }$ & $ -1.78 ^{+ 0.15 }_{- 0.12 }$ & $ 5059.38 ^{+ 2651.07 }_{- 3816.49 }$ & $ 1103.24 ^{+ 578.09 }_{- 832.22 }$ & $ 2.34 ^{+ 2.94 }_{- 1.33 } \times 10^{ -7 }$ & $ 7.98 ^{+ 5.46 }_{- 7.96 } \times 10^{ -1 }$ & $ -0.47 ^{+ 0.67 }_{- 1.45 }$ & $ -2.68 ^{+ 1.04 }_{- 0.28 }$ & $ 1175.52 ^{+ 1053.67 }_{- 1165.50 }$ & N/A & -251384.33 & 1.10 & -251380.21 \\
31.74 & 33.23 & 16.20 & $ 3.41 ^{+ 0.83 }_{- 2.14 } \times 10^{ +0 }$ & $ -1.08 ^{+ 0.15 }_{- 0.18 }$ & $ 272.08 ^{+ 20.10 }_{- 147.60 }$ & $ 250.55 ^{+ 18.51 }_{- 135.92 }$ & $ 6.74 ^{+ 12.97 }_{- 4.29 } \times 10^{ -7 }$ & $ 2.76 ^{+ 0.38 }_{- 1.14 } \times 10^{ -2 }$ & $ -0.93 ^{+ 0.19 }_{- 0.23 }$ & $ -3.50 ^{+ 1.06 }_{- 0.95 }$ & $ 235.61 ^{+ 19.37 }_{- 109.80 }$ & $ 6.77 ^{+ 4.73 }_{- 2.46 } \times 10^{ -7 }$ & -10.43 & -19.18 & -26.08 \\
33.23 & 34.24 & 29.20 & $ 8.23 ^{+ 1.75 }_{- 3.24 } \times 10^{ +0 }$ & $ -1.08 ^{+ 0.10 }_{- 0.11 }$ & $ 173.74 ^{+ 24.04 }_{- 45.63 }$ & $ 159.22 ^{+ 22.03 }_{- 41.82 }$ & $ 1.33 ^{+ 1.24 }_{- 0.61 } \times 10^{ -6 }$ & $ 5.97 ^{+ 0.70 }_{- 1.42 } \times 10^{ -2 }$ & $ -1.03 ^{+ 0.10 }_{- 0.13 }$ & $ -3.78 ^{+ 0.45 }_{- 1.16 }$ & $ 145.70 ^{+ 13.71 }_{- 21.60 }$ & $ 1.37 ^{+ 0.39 }_{- 0.29 } \times 10^{ -6 }$ & 1.04 & -1.67 & 2.71 \\
34.24 & 34.24 & 7.10 & $ 5.32 ^{+ 3.41 }_{- 2.53 } \times 10^{ +2 }$ & $ -1.85 ^{+ 0.06 }_{- 0.24 }$ & $ 4286.23 ^{+ 1868.80 }_{- 4248.75 }$ & $ 653.19 ^{+ 284.79 }_{- 647.48 }$ & $ 5.02 ^{+ 8.28 }_{- 3.38 } \times 10^{ -6 }$ & $ 3.25 ^{+ 2.12 }_{- 2.72 } \times 10^{ -1 }$ & $ -1.60 ^{+ 0.09 }_{- 0.40 }$ & $ -3.38 ^{+ 1.32 }_{- 0.75 }$ & $ 3584.20 ^{+ 1796.29 }_{- 3572.22 }$ & N/A & -63.58 & -0.53 & -58.58 \\
\enddata
\end{deluxetable}
\tablecomments{All columns are the same as Table~\ref{tab:fitresults}.}
\end{rotatetable}

\movetabledown=3.6in
\begin{rotatetable}
\begin{deluxetable}{cccccccccccccccc}
\tabletypesize{\scriptsize}
\tablecaption{Time-resolved spectral analysis results of GRB090719063.}
\setlength{\tabcolsep}{0pt}
\tablehead{ \colhead{$t_{\rm start}$} & \colhead{$t_{\rm stop}$} & \colhead{$S$} & \colhead{$K$}& \colhead{$\alpha$} & \colhead{$E_{\rm c}$} &\colhead{$E_{\rm p}$}& \colhead{$F$} & \colhead{$K_{\rm BAND}$} & \colhead{$\alpha_{\rm BAND}$} & \colhead{$\beta_{\rm BAND}$} & \colhead{$E_{\rm p,BAND}$} & \colhead{$F_{\rm BAND}$} & \colhead{$\Delta$DIC} & \colhead{$p_{\rm DIC}$} & \colhead{$p_{\rm DIC, BAND}$} \\
\colhead{(1)} & \colhead{(2)} & \colhead{(3)} & \colhead{(4)} & \colhead{(5)} & \colhead{(6)} & \colhead{(7)} & \colhead{(8)} & \colhead{(9)} & \colhead{(10)} & \colhead{(11)} & \colhead{(12)} & \colhead{(13)} & \colhead{(14)} & \colhead{(15)} & \colhead{(16)}
}
\startdata
-1.00 & -0.08 & 1.60 & $ 2.18 ^{+ 0.17 }_{- 2.18 } \times 10^{ +1 }$ & $ -2.13 ^{+ 0.33 }_{- 0.87 }$ & $ 5257.07 ^{+ 4595.63 }_{- 1738.30 }$ & $ -667.32 ^{+ 583.36 }_{- 220.65 }$ & $ 2.50 ^{+ 167.31 }_{- 2.42 } \times 10^{ -8 }$ & $ 5.92 ^{+ 5.62 }_{- 5.92 } \times 10^{ +1 }$ & $ -0.27 ^{+ 3.26 }_{- 0.98 }$ & $ -3.37 ^{+ 0.63 }_{- 1.61 }$ & $ 2805.21 ^{+ 1135.99 }_{- 2795.21 }$ & N/A & -25615329.02 & -7.40 & -25615332.41 \\
-0.08 & 0.72 & 26.50 & $ 7.25 ^{+ 1.77 }_{- 4.49 } \times 10^{ -2 }$ & $ -0.10 ^{+ 0.11 }_{- 0.12 }$ & $ 306.98 ^{+ 36.09 }_{- 47.80 }$ & $ 582.21 ^{+ 68.44 }_{- 90.66 }$ & $ 4.14 ^{+ 6.90 }_{- 2.36 } \times 10^{ -6 }$ & $ 4.11 ^{+ 0.26 }_{- 0.34 } \times 10^{ -2 }$ & $ -0.03 ^{+ 0.13 }_{- 0.14 }$ & $ -3.67 ^{+ 0.98 }_{- 0.63 }$ & $ 548.66 ^{+ 40.88 }_{- 54.99 }$ & $ 4.05 ^{+ 0.76 }_{- 0.60 } \times 10^{ -6 }$ & 25.42 & -22.68 & 3.10 \\
0.72 & 1.38 & 37.90 & $ 8.96 ^{+ 2.07 }_{- 4.22 } \times 10^{ -2 }$ & $ 0.09 ^{+ 0.10 }_{- 0.10 }$ & $ 145.41 ^{+ 11.79 }_{- 13.97 }$ & $ 303.86 ^{+ 24.65 }_{- 29.19 }$ & $ 4.75 ^{+ 4.14 }_{- 2.42 } \times 10^{ -6 }$ & $ 1.38 ^{+ 0.11 }_{- 0.18 } \times 10^{ -1 }$ & $ 0.18 ^{+ 0.10 }_{- 0.13 }$ & $ -3.73 ^{+ 0.93 }_{- 0.55 }$ & $ 287.27 ^{+ 17.53 }_{- 17.09 }$ & $ 4.41 ^{+ 0.83 }_{- 0.66 } \times 10^{ -6 }$ & 13.80 & -9.93 & 3.06 \\
1.38 & 3.80 & 80.70 & $ 3.10 ^{+ 0.41 }_{- 0.61 } \times 10^{ -1 }$ & $ -0.12 ^{+ 0.05 }_{- 0.05 }$ & $ 122.69 ^{+ 5.09 }_{- 6.39 }$ & $ 230.98 ^{+ 9.58 }_{- 12.04 }$ & $ 3.97 ^{+ 1.54 }_{- 1.04 } \times 10^{ -6 }$ & $ 1.98 ^{+ 0.12 }_{- 0.18 } \times 10^{ -1 }$ & $ -0.04 ^{+ 0.05 }_{- 0.07 }$ & $ -3.34 ^{+ 0.52 }_{- 0.21 }$ & $ 217.03 ^{+ 9.38 }_{- 8.06 }$ & $ 3.99 ^{+ 0.48 }_{- 0.41 } \times 10^{ -6 }$ & -2.96 & 0.98 & 3.62 \\
\enddata
\end{deluxetable}
\tablecomments{All columns are the same as Table~\ref{tab:fitresults}.}
\end{rotatetable}

\movetabledown=3.6in
\begin{rotatetable}
\begin{deluxetable}{cccccccccccccccc}
\tabletypesize{\scriptsize}
\tablecaption{Time-resolved spectral analysis results of GRB090804940.}
\setlength{\tabcolsep}{0pt}
\tablehead{ \colhead{$t_{\rm start}$} & \colhead{$t_{\rm stop}$} & \colhead{$S$} & \colhead{$K$}& \colhead{$\alpha$} & \colhead{$E_{\rm c}$} &\colhead{$E_{\rm p}$}& \colhead{$F$} & \colhead{$K_{\rm BAND}$} & \colhead{$\alpha_{\rm BAND}$} & \colhead{$\beta_{\rm BAND}$} & \colhead{$E_{\rm p,BAND}$} & \colhead{$F_{\rm BAND}$} & \colhead{$\Delta$DIC} & \colhead{$p_{\rm DIC}$} & \colhead{$p_{\rm DIC, BAND}$} \\
\colhead{(1)} & \colhead{(2)} & \colhead{(3)} & \colhead{(4)} & \colhead{(5)} & \colhead{(6)} & \colhead{(7)} & \colhead{(8)} & \colhead{(9)} & \colhead{(10)} & \colhead{(11)} & \colhead{(12)} & \colhead{(13)} & \colhead{(14)} & \colhead{(15)} & \colhead{(16)}
}
\startdata
-1.00 & -0.30 & -0.40 & $ 7.70 ^{+ 1.26 }_{- 7.70 } \times 10^{ +0 }$ & $ -2.46 ^{+ 0.09 }_{- 0.54 }$ & $ 4942.49 ^{+ 1849.83 }_{- 4930.70 }$ & $ -2273.54 ^{+ 850.92 }_{- 2268.12 }$ & $ 2.07 ^{+ 27.42 }_{- 1.98 } \times 10^{ -9 }$ & $ 4.34 ^{+ 3.56 }_{- 4.34 } \times 10^{ +1 }$ & $ 1.41 ^{+ 1.59 }_{- 0.14 }$ & $ -3.33 ^{+ 0.70 }_{- 1.56 }$ & $ 357.09 ^{+ 222.58 }_{- 347.09 }$ & N/A & -3194287.41 & -2.31 & -3194288.73 \\
-0.30 & -0.10 & 5.60 & $ 5.24 ^{+ 1.16 }_{- 5.11 } \times 10^{ +0 }$ & $ -1.24 ^{+ 0.13 }_{- 0.37 }$ & $ 2270.36 ^{+ 142.62 }_{- 2192.91 }$ & $ 1730.50 ^{+ 108.71 }_{- 1671.46 }$ & $ 6.42 ^{+ 39.03 }_{- 5.10 } \times 10^{ -7 }$ & $ 2.77 ^{+ 0.57 }_{- 2.76 } \times 10^{ +0 }$ & $ 1.27 ^{+ 1.72 }_{- 0.63 }$ & $ -2.76 ^{+ 1.05 }_{- 0.25 }$ & $ 101.98 ^{+ 10.20 }_{- 44.01 }$ & $ 8.57 ^{+ 105.57 }_{- 7.99 } \times 10^{ -7 }$ & -983.42 & -46.91 & -1025.45 \\
-0.10 & 0.62 & 22.40 & $ 1.00 ^{+ 0.22 }_{- 0.65 } \times 10^{ +0 }$ & $ -0.41 ^{+ 0.17 }_{- 0.19 }$ & $ 62.46 ^{+ 7.24 }_{- 13.07 }$ & $ 99.57 ^{+ 11.54 }_{- 20.83 }$ & $ 8.35 ^{+ 14.29 }_{- 4.96 } \times 10^{ -7 }$ & $ 2.00 ^{+ 0.38 }_{- 0.85 } \times 10^{ -1 }$ & $ -0.21 ^{+ 0.19 }_{- 0.21 }$ & $ -3.64 ^{+ 0.94 }_{- 0.45 }$ & $ 90.95 ^{+ 5.39 }_{- 7.58 }$ & $ 8.84 ^{+ 5.81 }_{- 3.10 } \times 10^{ -7 }$ & 21.90 & -24.31 & -0.70 \\
0.62 & 0.97 & 23.80 & $ 1.02 ^{+ 0.24 }_{- 0.63 } \times 10^{ +0 }$ & $ -0.22 ^{+ 0.15 }_{- 0.19 }$ & $ 49.11 ^{+ 5.16 }_{- 7.86 }$ & $ 87.60 ^{+ 9.21 }_{- 14.02 }$ & $ 1.39 ^{+ 1.97 }_{- 0.83 } \times 10^{ -6 }$ & $ 4.57 ^{+ 0.78 }_{- 1.92 } \times 10^{ -1 }$ & $ -0.06 ^{+ 0.18 }_{- 0.21 }$ & $ -4.09 ^{+ 0.31 }_{- 0.89 }$ & $ 83.44 ^{+ 4.01 }_{- 5.30 }$ & $ 1.38 ^{+ 0.80 }_{- 0.49 } \times 10^{ -6 }$ & 15.98 & -15.92 & -0.08 \\
\enddata
\end{deluxetable}
\tablecomments{All columns are the same as Table~\ref{tab:fitresults}.}
\end{rotatetable}

\movetabledown=3.6in
\begin{rotatetable}
\begin{deluxetable}{cccccccccccccccc}
\tabletypesize{\scriptsize}
\tablecaption{Time-resolved spectral analysis results of GRB090820027.}
\setlength{\tabcolsep}{0pt}
\tablehead{ \colhead{$t_{\rm start}$} & \colhead{$t_{\rm stop}$} & \colhead{$S$} & \colhead{$K$}& \colhead{$\alpha$} & \colhead{$E_{\rm c}$} &\colhead{$E_{\rm p}$}& \colhead{$F$} & \colhead{$K_{\rm BAND}$} & \colhead{$\alpha_{\rm BAND}$} & \colhead{$\beta_{\rm BAND}$} & \colhead{$E_{\rm p,BAND}$} & \colhead{$F_{\rm BAND}$} & \colhead{$\Delta$DIC} & \colhead{$p_{\rm DIC}$} & \colhead{$p_{\rm DIC, BAND}$} \\
\colhead{(1)} & \colhead{(2)} & \colhead{(3)} & \colhead{(4)} & \colhead{(5)} & \colhead{(6)} & \colhead{(7)} & \colhead{(8)} & \colhead{(9)} & \colhead{(10)} & \colhead{(11)} & \colhead{(12)} & \colhead{(13)} & \colhead{(14)} & \colhead{(15)} & \colhead{(16)}
}
\startdata
25.00 & 28.68 & -2.40 & $ 6.12 ^{+ 0.85 }_{- 6.12 } \times 10^{ +0 }$ & $ -2.56 ^{+ 0.08 }_{- 0.44 }$ & $ 4955.59 ^{+ 1860.13 }_{- 4945.22 }$ & $ -2781.45 ^{+ 1044.04 }_{- 2775.62 }$ & $ 1.42 ^{+ 11.83 }_{- 1.34 } \times 10^{ -9 }$ & $ 8.16 ^{+ 4.61 }_{- 8.16 } \times 10^{ +1 }$ & $ 2.29 ^{+ 0.71 }_{- 0.10 }$ & $ -3.60 ^{+ 0.44 }_{- 1.40 }$ & $ 24.91 ^{+ 3.10 }_{- 14.91 }$ & $ 9.53 ^{+ 869.02 }_{- 9.49 } \times 10^{ -10 }$ & -79.09 & -1.90 & -81.83 \\
28.68 & 29.77 & 6.90 & $ 1.63 ^{+ 0.37 }_{- 1.01 } \times 10^{ +1 }$ & $ -1.78 ^{+ 0.13 }_{- 0.16 }$ & $ 4664.39 ^{+ 1841.06 }_{- 4570.12 }$ & $ 1048.80 ^{+ 413.97 }_{- 1027.60 }$ & $ 2.56 ^{+ 3.86 }_{- 1.39 } \times 10^{ -7 }$ & $ 6.91 ^{+ 2.14 }_{- 6.90 } \times 10^{ +1 }$ & $ 1.68 ^{+ 1.32 }_{- 0.38 }$ & $ -2.40 ^{+ 0.58 }_{- 0.12 }$ & $ 34.21 ^{+ 5.48 }_{- 8.44 }$ & $ 2.38 ^{+ 69.33 }_{- 2.31 } \times 10^{ -7 }$ & -1353.45 & 0.67 & -1348.86 \\
29.77 & 30.35 & 16.90 & $ 2.56 ^{+ 0.52 }_{- 1.74 } \times 10^{ +0 }$ & $ -0.88 ^{+ 0.18 }_{- 0.17 }$ & $ 236.08 ^{+ 18.74 }_{- 115.27 }$ & $ 264.87 ^{+ 21.02 }_{- 129.33 }$ & $ 1.24 ^{+ 2.62 }_{- 0.83 } \times 10^{ -6 }$ & $ 5.06 ^{+ 0.63 }_{- 1.97 } \times 10^{ -2 }$ & $ -0.71 ^{+ 0.17 }_{- 0.23 }$ & $ -3.36 ^{+ 1.30 }_{- 0.71 }$ & $ 200.22 ^{+ 30.47 }_{- 51.03 }$ & $ 1.17 ^{+ 0.66 }_{- 0.37 } \times 10^{ -6 }$ & 18.11 & -20.85 & 0.86 \\
30.35 & 30.63 & 20.20 & $ 2.67 ^{+ 0.64 }_{- 1.48 } \times 10^{ +0 }$ & $ -0.75 ^{+ 0.11 }_{- 0.15 }$ & $ 198.39 ^{+ 29.73 }_{- 56.96 }$ & $ 248.92 ^{+ 37.31 }_{- 71.48 }$ & $ 2.55 ^{+ 3.63 }_{- 1.42 } \times 10^{ -6 }$ & $ 8.76 ^{+ 1.13 }_{- 2.02 } \times 10^{ -2 }$ & $ -0.67 ^{+ 0.14 }_{- 0.15 }$ & $ -3.87 ^{+ 0.36 }_{- 1.13 }$ & $ 227.05 ^{+ 23.48 }_{- 39.50 }$ & $ 2.44 ^{+ 0.79 }_{- 0.55 } \times 10^{ -6 }$ & 5.30 & -4.32 & 2.93 \\
\enddata
\end{deluxetable}
\tablecomments{All columns are the same as Table~\ref{tab:fitresults}.}
\end{rotatetable}

\movetabledown=3.6in
\begin{rotatetable}
\begin{deluxetable}{cccccccccccccccc}
\tabletypesize{\scriptsize}
\tablecaption{Time-resolved spectral analysis results of GRB100122616.}
\setlength{\tabcolsep}{0pt}
\tablehead{ \colhead{$t_{\rm start}$} & \colhead{$t_{\rm stop}$} & \colhead{$S$} & \colhead{$K$}& \colhead{$\alpha$} & \colhead{$E_{\rm c}$} &\colhead{$E_{\rm p}$}& \colhead{$F$} & \colhead{$K_{\rm BAND}$} & \colhead{$\alpha_{\rm BAND}$} & \colhead{$\beta_{\rm BAND}$} & \colhead{$E_{\rm p,BAND}$} & \colhead{$F_{\rm BAND}$} & \colhead{$\Delta$DIC} & \colhead{$p_{\rm DIC}$} & \colhead{$p_{\rm DIC, BAND}$} \\
\colhead{(1)} & \colhead{(2)} & \colhead{(3)} & \colhead{(4)} & \colhead{(5)} & \colhead{(6)} & \colhead{(7)} & \colhead{(8)} & \colhead{(9)} & \colhead{(10)} & \colhead{(11)} & \colhead{(12)} & \colhead{(13)} & \colhead{(14)} & \colhead{(15)} & \colhead{(16)}
}
\startdata
-5.00 & -3.77 & -1.73 & $ 5.42 ^{+ 0.91 }_{- 5.42 } \times 10^{ +0 }$ & $ -2.52 ^{+ 0.10 }_{- 0.48 }$ & $ 4996.18 ^{+ 1810.29 }_{- 4986.08 }$ & $ -2583.68 ^{+ 936.15 }_{- 2578.45 }$ & $ 1.63 ^{+ 13.66 }_{- 1.53 } \times 10^{ -9 }$ & $ 7.81 ^{+ 5.02 }_{- 7.81 } \times 10^{ +1 }$ & $ 2.16 ^{+ 0.84 }_{- 0.03 }$ & $ -3.52 ^{+ 0.53 }_{- 1.47 }$ & $ 31.31 ^{+ 4.49 }_{- 21.30 }$ & N/A & -280.71 & -1.98 & -283.90 \\
-3.77 & 5.86 & 8.44 & $ 7.61 ^{+ 3.35 }_{- 3.87 } \times 10^{ +0 }$ & $ -1.73 ^{+ 0.04 }_{- 0.23 }$ & $ 4534.50 ^{+ 1866.17 }_{- 4496.42 }$ & $ 1244.62 ^{+ 512.22 }_{- 1234.17 }$ & $ 1.10 ^{+ 2.34 }_{- 0.73 } \times 10^{ -7 }$ & $ 2.01 ^{+ 0.61 }_{- 2.01 } \times 10^{ +0 }$ & $ 0.39 ^{+ 1.09 }_{- 0.88 }$ & $ -2.37 ^{+ 0.63 }_{- 0.01 }$ & $ 37.66 ^{+ 11.17 }_{- 17.88 }$ & N/A & -4627.56 & -16.72 & -4642.11 \\
5.86 & 18.54 & 1.36 & $ 2.11 ^{+ 0.34 }_{- 2.11 } \times 10^{ +1 }$ & $ -2.40 ^{+ 0.16 }_{- 0.60 }$ & $ 4860.23 ^{+ 1759.48 }_{- 4814.88 }$ & $ -1934.25 ^{+ 700.23 }_{- 1916.20 }$ & $ 1.38 ^{+ 9.24 }_{- 1.15 } \times 10^{ -8 }$ & $ 1.75 ^{+ 0.02 }_{- 1.75 } \times 10^{ +2 }$ & $ 2.29 ^{+ 0.71 }_{- 0.15 }$ & $ -3.11 ^{+ 1.47 }_{- 0.53 }$ & $ 18.84 ^{+ 1.41 }_{- 8.05 }$ & $ 1.26 ^{+ 25.40 }_{- 1.20 } \times 10^{ -8 }$ & -48.13 & -2.20 & -52.09 \\
18.54 & 19.48 & 12.45 & $ 1.79 ^{+ 1.00 }_{- 0.73 } \times 10^{ +2 }$ & $ -2.12 ^{+ 0.03 }_{- 0.32 }$ & $ 4014.90 ^{+ 2044.29 }_{- 4000.95 }$ & $ -464.76 ^{+ 236.65 }_{- 463.15 }$ & $ 3.71 ^{+ 6.60 }_{- 2.50 } \times 10^{ -7 }$ & $ 6.05 ^{+ 0.75 }_{- 6.01 } \times 10^{ +0 }$ & $ -0.06 ^{+ 0.70 }_{- 0.58 }$ & $ -2.47 ^{+ 0.24 }_{- 0.10 }$ & $ 17.71 ^{+ 4.21 }_{- 4.07 }$ & $ 3.77 ^{+ 41.07 }_{- 3.51 } \times 10^{ -7 }$ & -731.23 & -81.64 & -809.36 \\
\enddata
\end{deluxetable}
\tablecomments{All columns are the same as Table~\ref{tab:fitresults}.}
\end{rotatetable}

\movetabledown=3.6in
\begin{rotatetable}
\begin{deluxetable}{cccccccccccccccc}
\tabletypesize{\scriptsize}
\tablecaption{Time-resolved spectral analysis results of GRB100528075.}
\setlength{\tabcolsep}{0pt}
\tablehead{ \colhead{$t_{\rm start}$} & \colhead{$t_{\rm stop}$} & \colhead{$S$} & \colhead{$K$}& \colhead{$\alpha$} & \colhead{$E_{\rm c}$} &\colhead{$E_{\rm p}$}& \colhead{$F$} & \colhead{$K_{\rm BAND}$} & \colhead{$\alpha_{\rm BAND}$} & \colhead{$\beta_{\rm BAND}$} & \colhead{$E_{\rm p,BAND}$} & \colhead{$F_{\rm BAND}$} & \colhead{$\Delta$DIC} & \colhead{$p_{\rm DIC}$} & \colhead{$p_{\rm DIC, BAND}$} \\
\colhead{(1)} & \colhead{(2)} & \colhead{(3)} & \colhead{(4)} & \colhead{(5)} & \colhead{(6)} & \colhead{(7)} & \colhead{(8)} & \colhead{(9)} & \colhead{(10)} & \colhead{(11)} & \colhead{(12)} & \colhead{(13)} & \colhead{(14)} & \colhead{(15)} & \colhead{(16)}
}
\startdata
-5.00 & -0.19 & 7.81 & $ 2.82 ^{+ 0.96 }_{- 2.07 } \times 10^{ +0 }$ & $ -1.47 ^{+ 0.08 }_{- 0.26 }$ & $ 2652.71 ^{+ 636.59 }_{- 2586.43 }$ & $ 1414.36 ^{+ 339.41 }_{- 1379.02 }$ & $ 1.40 ^{+ 4.62 }_{- 0.90 } \times 10^{ -7 }$ & $ 3.61 ^{+ 2.63 }_{- 3.44 } \times 10^{ -2 }$ & $ -0.83 ^{+ 0.17 }_{- 0.83 }$ & $ -3.11 ^{+ 1.45 }_{- 0.63 }$ & $ 317.71 ^{+ 135.55 }_{- 285.07 }$ & $ 1.57 ^{+ 10.73 }_{- 1.18 } \times 10^{ -7 }$ & -5894.07 & -26.52 & -5914.73 \\
-0.19 & 1.78 & 15.86 & $ 1.79 ^{+ 0.42 }_{- 0.85 } \times 10^{ +0 }$ & $ -1.01 ^{+ 0.11 }_{- 0.13 }$ & $ 255.08 ^{+ 31.76 }_{- 106.18 }$ & $ 252.47 ^{+ 31.44 }_{- 105.09 }$ & $ 5.47 ^{+ 6.36 }_{- 2.82 } \times 10^{ -7 }$ & $ 3.69 ^{+ 0.44 }_{- 2.42 } \times 10^{ -2 }$ & $ -0.68 ^{+ 0.13 }_{- 0.39 }$ & $ -2.47 ^{+ 0.77 }_{- 0.00 }$ & $ 158.10 ^{+ 43.19 }_{- 68.39 }$ & $ 5.79 ^{+ 6.97 }_{- 2.80 } \times 10^{ -7 }$ & -108.48 & -9.26 & -111.77 \\
1.78 & 3.89 & 29.87 & $ 4.01 ^{+ 0.62 }_{- 0.84 } \times 10^{ +0 }$ & $ -1.12 ^{+ 0.05 }_{- 0.06 }$ & $ 519.85 ^{+ 71.83 }_{- 134.86 }$ & $ 456.44 ^{+ 63.07 }_{- 118.41 }$ & $ 1.34 ^{+ 0.63 }_{- 0.38 } \times 10^{ -6 }$ & $ 2.40 ^{+ 0.14 }_{- 0.30 } \times 10^{ -2 }$ & $ -1.08 ^{+ 0.05 }_{- 0.08 }$ & $ -3.06 ^{+ 1.23 }_{- 0.54 }$ & $ 399.11 ^{+ 85.74 }_{- 90.30 }$ & $ 1.33 ^{+ 0.25 }_{- 0.23 } \times 10^{ -6 }$ & -4.24 & 1.17 & 1.64 \\
3.89 & 5.77 & 35.16 & $ 3.15 ^{+ 0.49 }_{- 0.66 } \times 10^{ +0 }$ & $ -1.01 ^{+ 0.05 }_{- 0.05 }$ & $ 404.22 ^{+ 42.54 }_{- 82.16 }$ & $ 400.80 ^{+ 42.18 }_{- 81.46 }$ & $ 1.67 ^{+ 0.70 }_{- 0.53 } \times 10^{ -6 }$ & $ 3.80 ^{+ 0.52 }_{- 1.03 } \times 10^{ -2 }$ & $ -0.92 ^{+ 0.01 }_{- 0.14 }$ & $ -3.32 ^{+ 1.02 }_{- 1.08 }$ & $ 353.88 ^{+ 73.14 }_{- 32.27 }$ & $ 1.66 ^{+ 0.33 }_{- 0.26 } \times 10^{ -6 }$ & -62.89 & 1.23 & -64.42 \\
\enddata
\end{deluxetable}
\tablecomments{All columns are the same as Table~\ref{tab:fitresults}.}
\end{rotatetable}

\movetabledown=3.6in
\begin{rotatetable}
\begin{deluxetable}{cccccccccccccccc}
\tabletypesize{\scriptsize}
\tablecaption{Time-resolved spectral analysis results of GRB100612726.}
\setlength{\tabcolsep}{0pt}
\tablehead{ \colhead{$t_{\rm start}$} & \colhead{$t_{\rm stop}$} & \colhead{$S$} & \colhead{$K$}& \colhead{$\alpha$} & \colhead{$E_{\rm c}$} &\colhead{$E_{\rm p}$}& \colhead{$F$} & \colhead{$K_{\rm BAND}$} & \colhead{$\alpha_{\rm BAND}$} & \colhead{$\beta_{\rm BAND}$} & \colhead{$E_{\rm p,BAND}$} & \colhead{$F_{\rm BAND}$} & \colhead{$\Delta$DIC} & \colhead{$p_{\rm DIC}$} & \colhead{$p_{\rm DIC, BAND}$} \\
\colhead{(1)} & \colhead{(2)} & \colhead{(3)} & \colhead{(4)} & \colhead{(5)} & \colhead{(6)} & \colhead{(7)} & \colhead{(8)} & \colhead{(9)} & \colhead{(10)} & \colhead{(11)} & \colhead{(12)} & \colhead{(13)} & \colhead{(14)} & \colhead{(15)} & \colhead{(16)}
}
\startdata
-2.00 & 0.02 & 1.31 & $ 1.10 ^{+ 0.02 }_{- 1.05 } \times 10^{ +0 }$ & $ -1.36 ^{+ 0.24 }_{- 0.19 }$ & $ 4861.88 ^{+ 2056.35 }_{- 4150.41 }$ & $ 3100.70 ^{+ 1311.45 }_{- 2646.95 }$ & $ 1.15 ^{+ 3.87 }_{- 0.91 } \times 10^{ -7 }$ & $ 2.01 ^{+ 0.15 }_{- 1.07 } \times 10^{ -3 }$ & $ -0.97 ^{+ 0.08 }_{- 0.55 }$ & $ -3.29 ^{+ 1.48 }_{- 0.80 }$ & $ 3501.71 ^{+ 1337.66 }_{- 3372.09 }$ & N/A & -8.11 & -2.12 & -5.48 \\
0.02 & 0.99 & 12.51 & $ 1.22 ^{+ 0.25 }_{- 0.77 } \times 10^{ +0 }$ & $ -0.81 ^{+ 0.14 }_{- 0.16 }$ & $ 221.75 ^{+ 23.50 }_{- 97.26 }$ & $ 263.62 ^{+ 27.94 }_{- 115.62 }$ & $ 8.53 ^{+ 15.64 }_{- 5.17 } \times 10^{ -7 }$ & $ 3.66 ^{+ 0.24 }_{- 1.54 } \times 10^{ -2 }$ & $ -0.63 ^{+ 0.14 }_{- 0.25 }$ & $ -3.06 ^{+ 1.29 }_{- 0.53 }$ & $ 199.23 ^{+ 41.84 }_{- 53.90 }$ & $ 8.55 ^{+ 5.17 }_{- 3.35 } \times 10^{ -7 }$ & 7.59 & -15.42 & -4.30 \\
0.99 & 1.82 & 23.12 & $ 9.44 ^{+ 1.84 }_{- 3.26 } \times 10^{ -1 }$ & $ -0.59 ^{+ 0.08 }_{- 0.08 }$ & $ 189.57 ^{+ 19.82 }_{- 29.58 }$ & $ 267.63 ^{+ 27.98 }_{- 41.76 }$ & $ 1.96 ^{+ 1.37 }_{- 0.80 } \times 10^{ -6 }$ & $ 6.38 ^{+ 0.57 }_{- 0.88 } \times 10^{ -2 }$ & $ -0.55 ^{+ 0.08 }_{- 0.10 }$ & $ -3.82 ^{+ 0.39 }_{- 1.18 }$ & $ 254.79 ^{+ 21.77 }_{- 26.59 }$ & $ 1.99 ^{+ 0.39 }_{- 0.32 } \times 10^{ -6 }$ & 2.95 & -0.88 & 3.05 \\
1.82 & 2.86 & 33.02 & $ 7.64 ^{+ 1.58 }_{- 2.15 } \times 10^{ -1 }$ & $ -0.38 ^{+ 0.07 }_{- 0.08 }$ & $ 103.61 ^{+ 8.38 }_{- 11.25 }$ & $ 167.85 ^{+ 13.58 }_{- 18.23 }$ & $ 1.92 ^{+ 1.20 }_{- 0.77 } \times 10^{ -6 }$ & $ 2.36 ^{+ 0.32 }_{- 0.66 } \times 10^{ -1 }$ & $ -0.05 ^{+ 0.12 }_{- 0.15 }$ & $ -2.33 ^{+ 0.16 }_{- 0.10 }$ & $ 128.28 ^{+ 8.87 }_{- 11.36 }$ & $ 2.16 ^{+ 0.77 }_{- 0.57 } \times 10^{ -6 }$ & -20.52 & -0.76 & 0.81 \\
\enddata
\end{deluxetable}
\tablecomments{All columns are the same as Table~\ref{tab:fitresults}.}
\end{rotatetable}

\movetabledown=3.6in
\begin{rotatetable}
\begin{deluxetable}{cccccccccccccccc}
\tabletypesize{\scriptsize}
\tablecaption{Time-resolved spectral analysis results of GRB100707032.}
\setlength{\tabcolsep}{0pt}
\tablehead{ \colhead{$t_{\rm start}$} & \colhead{$t_{\rm stop}$} & \colhead{$S$} & \colhead{$K$}& \colhead{$\alpha$} & \colhead{$E_{\rm c}$} &\colhead{$E_{\rm p}$}& \colhead{$F$} & \colhead{$K_{\rm BAND}$} & \colhead{$\alpha_{\rm BAND}$} & \colhead{$\beta_{\rm BAND}$} & \colhead{$E_{\rm p,BAND}$} & \colhead{$F_{\rm BAND}$} & \colhead{$\Delta$DIC} & \colhead{$p_{\rm DIC}$} & \colhead{$p_{\rm DIC, BAND}$} \\
\colhead{(1)} & \colhead{(2)} & \colhead{(3)} & \colhead{(4)} & \colhead{(5)} & \colhead{(6)} & \colhead{(7)} & \colhead{(8)} & \colhead{(9)} & \colhead{(10)} & \colhead{(11)} & \colhead{(12)} & \colhead{(13)} & \colhead{(14)} & \colhead{(15)} & \colhead{(16)}
}
\startdata
0.00 & 0.18 & 6.39 & $ 1.14 ^{+ 0.21 }_{- 1.09 } \times 10^{ -1 }$ & $ -0.31 ^{+ 0.16 }_{- 0.29 }$ & $ 1309.69 ^{+ 346.51 }_{- 872.57 }$ & $ 2219.81 ^{+ 587.31 }_{- 1478.93 }$ & $ 3.36 ^{+ 18.08 }_{- 2.62 } \times 10^{ -6 }$ & $ 2.18 ^{+ 0.33 }_{- 0.65 } \times 10^{ -2 }$ & $ 0.92 ^{+ 0.39 }_{- 0.81 }$ & $ -1.90 ^{+ 0.22 }_{- 0.08 }$ & $ 651.55 ^{+ 88.35 }_{- 206.81 }$ & $ 4.42 ^{+ 4.94 }_{- 2.01 } \times 10^{ -6 }$ & 98.41 & -123.55 & -6.62 \\
0.18 & 0.36 & 16.26 & $ 6.15 ^{+ 1.03 }_{- 5.53 } \times 10^{ -2 }$ & $ -0.07 ^{+ 0.14 }_{- 0.20 }$ & $ 636.40 ^{+ 105.64 }_{- 215.69 }$ & $ 1229.94 ^{+ 204.16 }_{- 416.85 }$ & $ 7.57 ^{+ 24.20 }_{- 5.50 } \times 10^{ -6 }$ & $ 3.69 ^{+ 0.36 }_{- 0.50 } \times 10^{ -2 }$ & $ 0.34 ^{+ 0.23 }_{- 0.32 }$ & $ -2.88 ^{+ 0.73 }_{- 0.16 }$ & $ 842.05 ^{+ 106.33 }_{- 206.74 }$ & $ 8.31 ^{+ 4.19 }_{- 2.75 } \times 10^{ -6 }$ & 59.75 & -63.40 & 1.22 \\
0.36 & 0.70 & 33.40 & $ 9.85 ^{+ 1.73 }_{- 7.01 } \times 10^{ -3 }$ & $ 0.46 ^{+ 0.13 }_{- 0.13 }$ & $ 282.43 ^{+ 22.67 }_{- 38.47 }$ & $ 693.81 ^{+ 55.70 }_{- 94.50 }$ & $ 1.23 ^{+ 2.12 }_{- 0.77 } \times 10^{ -5 }$ & $ 7.03 ^{+ 0.46 }_{- 0.54 } \times 10^{ -2 }$ & $ 0.67 ^{+ 0.15 }_{- 0.20 }$ & $ -2.86 ^{+ 0.46 }_{- 0.13 }$ & $ 612.93 ^{+ 41.20 }_{- 51.60 }$ & $ 1.24 ^{+ 0.31 }_{- 0.21 } \times 10^{ -5 }$ & 69.27 & -70.88 & 3.63 \\
0.70 & 1.19 & 54.18 & $ 2.02 ^{+ 0.39 }_{- 0.90 } \times 10^{ -2 }$ & $ 0.42 ^{+ 0.08 }_{- 0.08 }$ & $ 235.82 ^{+ 14.13 }_{- 17.02 }$ & $ 569.99 ^{+ 34.15 }_{- 41.14 }$ & $ 1.77 ^{+ 1.46 }_{- 0.85 } \times 10^{ -5 }$ & $ 1.67 ^{+ 0.07 }_{- 0.20 } \times 10^{ -1 }$ & $ 0.86 ^{+ 0.11 }_{- 0.20 }$ & $ -2.37 ^{+ 0.12 }_{- 0.08 }$ & $ 441.67 ^{+ 29.27 }_{- 27.40 }$ & $ 1.59 ^{+ 0.31 }_{- 0.28 } \times 10^{ -5 }$ & -13.51 & -14.18 & 0.84 \\
\enddata
\end{deluxetable}
\tablecomments{All columns are the same as Table~\ref{tab:fitresults}.}
\end{rotatetable}

\movetabledown=3.6in
\begin{rotatetable}
\begin{deluxetable}{cccccccccccccccc}
\tabletypesize{\scriptsize}
\tablecaption{Time-resolved spectral analysis results of GRB101126198.}
\setlength{\tabcolsep}{0pt}
\tablehead{ \colhead{$t_{\rm start}$} & \colhead{$t_{\rm stop}$} & \colhead{$S$} & \colhead{$K$}& \colhead{$\alpha$} & \colhead{$E_{\rm c}$} &\colhead{$E_{\rm p}$}& \colhead{$F$} & \colhead{$K_{\rm BAND}$} & \colhead{$\alpha_{\rm BAND}$} & \colhead{$\beta_{\rm BAND}$} & \colhead{$E_{\rm p,BAND}$} & \colhead{$F_{\rm BAND}$} & \colhead{$\Delta$DIC} & \colhead{$p_{\rm DIC}$} & \colhead{$p_{\rm DIC, BAND}$} \\
\colhead{(1)} & \colhead{(2)} & \colhead{(3)} & \colhead{(4)} & \colhead{(5)} & \colhead{(6)} & \colhead{(7)} & \colhead{(8)} & \colhead{(9)} & \colhead{(10)} & \colhead{(11)} & \colhead{(12)} & \colhead{(13)} & \colhead{(14)} & \colhead{(15)} & \colhead{(16)}
}
\startdata
-5.00 & -2.99 & -0.27 & $ 3.42 ^{+ 1.00 }_{- 3.42 } \times 10^{ +1 }$ & $ -2.55 ^{+ 0.06 }_{- 0.45 }$ & $ 4734.42 ^{+ 1802.13 }_{- 4722.94 }$ & $ -2622.13 ^{+ 998.10 }_{- 2615.78 }$ & $ 1.73 ^{+ 6.51 }_{- 1.31 } \times 10^{ -8 }$ & $ 2.12 ^{+ 0.19 }_{- 2.12 } \times 10^{ +2 }$ & $ 2.23 ^{+ 0.77 }_{- 0.14 }$ & $ -3.62 ^{+ 0.46 }_{- 1.34 }$ & $ 18.21 ^{+ 1.47 }_{- 8.21 }$ & $ 1.67 ^{+ 29.81 }_{- 1.58 } \times 10^{ -8 }$ & -33.54 & 0.03 & -37.10 \\
-2.99 & 0.04 & 7.32 & $ 1.36 ^{+ 0.24 }_{- 0.79 } \times 10^{ +1 }$ & $ -1.91 ^{+ 0.16 }_{- 0.12 }$ & $ 5051.06 ^{+ 2838.22 }_{- 3678.39 }$ & $ 431.50 ^{+ 242.46 }_{- 314.24 }$ & $ 1.26 ^{+ 1.60 }_{- 0.71 } \times 10^{ -7 }$ & $ 8.27 ^{+ 4.04 }_{- 8.27 } \times 10^{ +0 }$ & $ 0.44 ^{+ 0.66 }_{- 0.70 }$ & $ -1.97 ^{+ 0.17 }_{- 0.11 }$ & $ 16.89 ^{+ 1.51 }_{- 6.63 }$ & $ 1.20 ^{+ 16.64 }_{- 1.14 } \times 10^{ -7 }$ & -1732.25 & 0.80 & -1733.70 \\
0.04 & 6.64 & 17.63 & $ 4.83 ^{+ 1.22 }_{- 2.10 } \times 10^{ +0 }$ & $ -1.26 ^{+ 0.12 }_{- 0.15 }$ & $ 128.43 ^{+ 19.82 }_{- 47.10 }$ & $ 94.76 ^{+ 14.62 }_{- 34.75 }$ & $ 2.54 ^{+ 3.00 }_{- 1.21 } \times 10^{ -7 }$ & $ 2.03 ^{+ 0.14 }_{- 0.99 } \times 10^{ -2 }$ & $ -1.15 ^{+ 0.10 }_{- 0.20 }$ & $ -3.67 ^{+ 0.45 }_{- 1.32 }$ & $ 83.34 ^{+ 11.42 }_{- 12.34 }$ & $ 2.66 ^{+ 1.66 }_{- 0.95 } \times 10^{ -7 }$ & -39.53 & -6.45 & -41.50 \\
6.64 & 8.78 & 32.96 & $ 7.21 ^{+ 1.12 }_{- 1.51 } \times 10^{ +0 }$ & $ -1.15 ^{+ 0.06 }_{- 0.06 }$ & $ 215.36 ^{+ 23.76 }_{- 45.51 }$ & $ 183.70 ^{+ 20.27 }_{- 38.82 }$ & $ 1.09 ^{+ 0.58 }_{- 0.33 } \times 10^{ -6 }$ & $ 4.17 ^{+ 0.41 }_{- 0.69 } \times 10^{ -2 }$ & $ -1.07 ^{+ 0.07 }_{- 0.08 }$ & $ -2.77 ^{+ 0.71 }_{- 0.03 }$ & $ 157.04 ^{+ 16.18 }_{- 24.29 }$ & $ 1.18 ^{+ 0.22 }_{- 0.22 } \times 10^{ -6 }$ & -3.30 & 1.30 & 2.26 \\
\enddata
\end{deluxetable}
\tablecomments{All columns are the same as Table~\ref{tab:fitresults}.}
\end{rotatetable}

\movetabledown=3.6in
\begin{rotatetable}
\begin{deluxetable}{cccccccccccccccc}
\tabletypesize{\scriptsize}
\tablecaption{Time-resolved spectral analysis results of GRB110721200.}
\setlength{\tabcolsep}{0pt}
\tablehead{ \colhead{$t_{\rm start}$} & \colhead{$t_{\rm stop}$} & \colhead{$S$} & \colhead{$K$}& \colhead{$\alpha$} & \colhead{$E_{\rm c}$} &\colhead{$E_{\rm p}$}& \colhead{$F$} & \colhead{$K_{\rm BAND}$} & \colhead{$\alpha_{\rm BAND}$} & \colhead{$\beta_{\rm BAND}$} & \colhead{$E_{\rm p,BAND}$} & \colhead{$F_{\rm BAND}$} & \colhead{$\Delta$DIC} & \colhead{$p_{\rm DIC}$} & \colhead{$p_{\rm DIC, BAND}$} \\
\colhead{(1)} & \colhead{(2)} & \colhead{(3)} & \colhead{(4)} & \colhead{(5)} & \colhead{(6)} & \colhead{(7)} & \colhead{(8)} & \colhead{(9)} & \colhead{(10)} & \colhead{(11)} & \colhead{(12)} & \colhead{(13)} & \colhead{(14)} & \colhead{(15)} & \colhead{(16)}
}
\startdata
-1.00 & -0.04 & 1.50 & $ 3.59 ^{+ 0.02 }_{- 3.59 } \times 10^{ -2 }$ & $ -0.71 ^{+ 0.14 }_{- 0.26 }$ & $ 44330.70 ^{+ 13665.87 }_{- 35093.13 }$ & $ 57236.29 ^{+ 17644.29 }_{- 45309.48 }$ & $ 1.29 ^{+ 7.20 }_{- 1.15 } \times 10^{ -7 }$ & $ 4.02 ^{+ 2.89 }_{- 4.02 } \times 10^{ +1 }$ & $ 1.89 ^{+ 1.11 }_{- 0.17 }$ & $ -2.62 ^{+ 1.02 }_{- 0.32 }$ & $ 57.39 ^{+ 4.09 }_{- 47.39 }$ & N/A & -9299.93 & -87.81 & -9419.53 \\
-0.04 & 0.08 & 11.60 & $ 3.50 ^{+ 0.76 }_{- 1.08 } \times 10^{ +0 }$ & $ -1.09 ^{+ 0.05 }_{- 0.06 }$ & $ 39270.76 ^{+ 7371.12 }_{- 27853.91 }$ & $ 35792.92 ^{+ 6718.32 }_{- 25387.15 }$ & $ 2.82 ^{+ 1.47 }_{- 0.95 } \times 10^{ -6 }$ & $ 2.27 ^{+ 0.18 }_{- 0.18 } \times 10^{ -2 }$ & $ -0.97 ^{+ 0.05 }_{- 0.05 }$ & $ -2.43 ^{+ 0.83 }_{- 0.07 }$ & $ 8498.14 ^{+ 1501.81 }_{- 331.29 }$ & $ 3.13 ^{+ 0.31 }_{- 0.27 } \times 10^{ -6 }$ & -3.25 & 0.68 & 1.04 \\
0.08 & 0.47 & 32.40 & $ 3.04 ^{+ 0.33 }_{- 0.40 } \times 10^{ +0 }$ & $ -0.93 ^{+ 0.02 }_{- 0.02 }$ & $ 5842.96 ^{+ 573.67 }_{- 792.64 }$ & $ 6266.72 ^{+ 615.27 }_{- 850.13 }$ & $ 6.17 ^{+ 1.19 }_{- 1.03 } \times 10^{ -6 }$ & $ 4.23 ^{+ 0.12 }_{- 0.12 } \times 10^{ -2 }$ & $ -0.91 ^{+ 0.03 }_{- 0.03 }$ & $ -2.91 ^{+ 0.88 }_{- 0.25 }$ & $ 5479.07 ^{+ 830.16 }_{- 753.46 }$ & $ 6.16 ^{+ 0.30 }_{- 0.29 } \times 10^{ -6 }$ & -5.34 & 2.77 & 3.25 \\
0.47 & 1.96 & 77.80 & $ 4.91 ^{+ 0.30 }_{- 0.32 } \times 10^{ +0 }$ & $ -0.94 ^{+ 0.01 }_{- 0.02 }$ & $ 1836.34 ^{+ 118.05 }_{- 136.09 }$ & $ 1941.63 ^{+ 124.82 }_{- 143.89 }$ & $ 7.61 ^{+ 0.86 }_{- 0.75 } \times 10^{ -6 }$ & $ 7.01 ^{+ 0.16 }_{- 0.20 } \times 10^{ -2 }$ & $ -0.85 ^{+ 0.02 }_{- 0.03 }$ & $ -2.18 ^{+ 0.11 }_{- 0.07 }$ & $ 1156.62 ^{+ 100.78 }_{- 132.28 }$ & $ 7.79 ^{+ 0.40 }_{- 0.43 } \times 10^{ -6 }$ & -44.32 & 2.90 & 3.87 \\
\enddata
\end{deluxetable}
\tablecomments{All columns are the same as Table~\ref{tab:fitresults}.}
\end{rotatetable}

\movetabledown=3.6in
\begin{rotatetable}
\begin{deluxetable}{cccccccccccccccc}
\tabletypesize{\scriptsize}
\tablecaption{Time-resolved spectral analysis results of GRB110817191.}
\setlength{\tabcolsep}{0pt}
\tablehead{ \colhead{$t_{\rm start}$} & \colhead{$t_{\rm stop}$} & \colhead{$S$} & \colhead{$K$}& \colhead{$\alpha$} & \colhead{$E_{\rm c}$} &\colhead{$E_{\rm p}$}& \colhead{$F$} & \colhead{$K_{\rm BAND}$} & \colhead{$\alpha_{\rm BAND}$} & \colhead{$\beta_{\rm BAND}$} & \colhead{$E_{\rm p,BAND}$} & \colhead{$F_{\rm BAND}$} & \colhead{$\Delta$DIC} & \colhead{$p_{\rm DIC}$} & \colhead{$p_{\rm DIC, BAND}$} \\
\colhead{(1)} & \colhead{(2)} & \colhead{(3)} & \colhead{(4)} & \colhead{(5)} & \colhead{(6)} & \colhead{(7)} & \colhead{(8)} & \colhead{(9)} & \colhead{(10)} & \colhead{(11)} & \colhead{(12)} & \colhead{(13)} & \colhead{(14)} & \colhead{(15)} & \colhead{(16)}
}
\startdata
-1.00 & -0.04 & 2.70 & $ 5.23 ^{+ 0.24 }_{- 4.90 } \times 10^{ +0 }$ & $ -1.61 ^{+ 0.24 }_{- 0.21 }$ & $ 5155.76 ^{+ 4754.14 }_{- 1842.40 }$ & $ 1987.10 ^{+ 1832.31 }_{- 710.09 }$ & $ 1.44 ^{+ 5.79 }_{- 1.09 } \times 10^{ -7 }$ & $ 1.52 ^{+ 1.02 }_{- 1.52 } \times 10^{ +1 }$ & $ 1.52 ^{+ 1.48 }_{- 0.42 }$ & $ -2.43 ^{+ 0.83 }_{- 0.04 }$ & $ 58.24 ^{+ 13.12 }_{- 40.51 }$ & $ 1.37 ^{+ 73.35 }_{- 1.35 } \times 10^{ -7 }$ & -4815.91 & -2.00 & -4818.33 \\
-0.04 & 0.43 & 12.90 & $ 7.50 ^{+ 1.50 }_{- 5.71 } \times 10^{ -1 }$ & $ -0.68 ^{+ 0.14 }_{- 0.17 }$ & $ 416.64 ^{+ 49.08 }_{- 190.89 }$ & $ 550.55 ^{+ 64.86 }_{- 252.24 }$ & $ 1.99 ^{+ 3.99 }_{- 1.33 } \times 10^{ -6 }$ & $ 3.08 ^{+ 0.36 }_{- 0.60 } \times 10^{ -2 }$ & $ -0.56 ^{+ 0.15 }_{- 0.20 }$ & $ -3.47 ^{+ 1.23 }_{- 0.78 }$ & $ 447.50 ^{+ 55.56 }_{- 125.04 }$ & $ 1.88 ^{+ 0.75 }_{- 0.45 } \times 10^{ -6 }$ & 29.23 & -29.77 & 1.77 \\
0.43 & 0.81 & 24.90 & $ 6.19 ^{+ 1.38 }_{- 2.69 } \times 10^{ -1 }$ & $ -0.42 ^{+ 0.08 }_{- 0.10 }$ & $ 242.06 ^{+ 27.02 }_{- 38.64 }$ & $ 381.58 ^{+ 42.59 }_{- 60.91 }$ & $ 4.46 ^{+ 3.99 }_{- 2.14 } \times 10^{ -6 }$ & $ 8.64 ^{+ 0.65 }_{- 0.98 } \times 10^{ -2 }$ & $ -0.38 ^{+ 0.09 }_{- 0.11 }$ & $ -3.56 ^{+ 1.19 }_{- 0.72 }$ & $ 360.18 ^{+ 31.42 }_{- 36.24 }$ & $ 4.27 ^{+ 0.79 }_{- 0.66 } \times 10^{ -6 }$ & 7.42 & -5.31 & 2.97 \\
0.81 & 1.94 & 56.70 & $ 1.28 ^{+ 0.19 }_{- 0.23 } \times 10^{ +0 }$ & $ -0.45 ^{+ 0.04 }_{- 0.05 }$ & $ 150.52 ^{+ 8.13 }_{- 9.71 }$ & $ 233.72 ^{+ 12.62 }_{- 15.08 }$ & $ 4.26 ^{+ 1.67 }_{- 1.07 } \times 10^{ -6 }$ & $ 1.67 ^{+ 0.09 }_{- 0.11 } \times 10^{ -1 }$ & $ -0.43 ^{+ 0.04 }_{- 0.05 }$ & $ -3.93 ^{+ 0.67 }_{- 0.72 }$ & $ 228.04 ^{+ 8.89 }_{- 8.95 }$ & $ 4.27 ^{+ 0.42 }_{- 0.38 } \times 10^{ -6 }$ & 0.73 & 1.54 & 3.24 \\
\enddata
\end{deluxetable}
\tablecomments{All columns are the same as Table~\ref{tab:fitresults}.}
\end{rotatetable}

\movetabledown=3.6in
\begin{rotatetable}
\begin{deluxetable}{cccccccccccccccc}
\tabletypesize{\scriptsize}
\tablecaption{Time-resolved spectral analysis results of GRB110920546.}
\setlength{\tabcolsep}{0pt}
\tablehead{ \colhead{$t_{\rm start}$} & \colhead{$t_{\rm stop}$} & \colhead{$S$} & \colhead{$K$}& \colhead{$\alpha$} & \colhead{$E_{\rm c}$} &\colhead{$E_{\rm p}$}& \colhead{$F$} & \colhead{$K_{\rm BAND}$} & \colhead{$\alpha_{\rm BAND}$} & \colhead{$\beta_{\rm BAND}$} & \colhead{$E_{\rm p,BAND}$} & \colhead{$F_{\rm BAND}$} & \colhead{$\Delta$DIC} & \colhead{$p_{\rm DIC}$} & \colhead{$p_{\rm DIC, BAND}$} \\
\colhead{(1)} & \colhead{(2)} & \colhead{(3)} & \colhead{(4)} & \colhead{(5)} & \colhead{(6)} & \colhead{(7)} & \colhead{(8)} & \colhead{(9)} & \colhead{(10)} & \colhead{(11)} & \colhead{(12)} & \colhead{(13)} & \colhead{(14)} & \colhead{(15)} & \colhead{(16)}
}
\startdata
-1.00 & 1.05 & 7.80 & $ 3.05 ^{+ 0.52 }_{- 2.90 } \times 10^{ -2 }$ & $ -0.32 ^{+ 0.12 }_{- 0.26 }$ & $ 1034.29 ^{+ 315.54 }_{- 527.38 }$ & $ 1737.51 ^{+ 530.08 }_{- 885.95 }$ & $ 1.02 ^{+ 3.67 }_{- 0.78 } \times 10^{ -6 }$ & $ 5.17 ^{+ 0.66 }_{- 0.90 } \times 10^{ -3 }$ & $ 0.18 ^{+ 0.31 }_{- 0.48 }$ & $ -3.53 ^{+ 1.08 }_{- 0.73 }$ & $ 985.96 ^{+ 107.98 }_{- 431.86 }$ & $ 1.06 ^{+ 0.98 }_{- 0.40 } \times 10^{ -6 }$ & 97.07 & -107.51 & -8.44 \\
1.05 & 2.02 & 10.90 & $ 3.24 ^{+ 0.53 }_{- 2.77 } \times 10^{ -2 }$ & $ -0.15 ^{+ 0.18 }_{- 0.22 }$ & $ 554.43 ^{+ 67.27 }_{- 244.40 }$ & $ 1023.66 ^{+ 124.20 }_{- 451.25 }$ & $ 2.01 ^{+ 6.46 }_{- 1.50 } \times 10^{ -6 }$ & $ 1.27 ^{+ 0.09 }_{- 0.19 } \times 10^{ -2 }$ & $ 0.09 ^{+ 0.18 }_{- 0.31 }$ & $ -3.44 ^{+ 1.15 }_{- 0.78 }$ & $ 795.72 ^{+ 123.56 }_{- 196.33 }$ & $ 2.04 ^{+ 0.99 }_{- 0.60 } \times 10^{ -6 }$ & 102.55 & -104.42 & -1.25 \\
2.02 & 4.78 & 29.50 & $ 4.62 ^{+ 1.25 }_{- 2.03 } \times 10^{ -2 }$ & $ -0.12 ^{+ 0.08 }_{- 0.09 }$ & $ 325.39 ^{+ 34.28 }_{- 39.43 }$ & $ 612.87 ^{+ 64.56 }_{- 74.26 }$ & $ 2.69 ^{+ 2.46 }_{- 1.38 } \times 10^{ -6 }$ & $ 2.51 ^{+ 0.12 }_{- 0.14 } \times 10^{ -2 }$ & $ -0.09 ^{+ 0.09 }_{- 0.10 }$ & $ -3.98 ^{+ 0.34 }_{- 1.00 }$ & $ 593.99 ^{+ 38.39 }_{- 47.25 }$ & $ 2.69 ^{+ 0.39 }_{- 0.32 } \times 10^{ -6 }$ & 28.04 & -24.30 & 3.20 \\
4.78 & 6.16 & 26.70 & $ 3.03 ^{+ 0.60 }_{- 1.71 } \times 10^{ -2 }$ & $ 0.09 ^{+ 0.10 }_{- 0.11 }$ & $ 229.97 ^{+ 19.37 }_{- 28.29 }$ & $ 480.63 ^{+ 40.49 }_{- 59.12 }$ & $ 3.66 ^{+ 4.19 }_{- 1.96 } \times 10^{ -6 }$ & $ 4.33 ^{+ 0.26 }_{- 0.37 } \times 10^{ -2 }$ & $ 0.16 ^{+ 0.12 }_{- 0.13 }$ & $ -3.62 ^{+ 1.03 }_{- 0.57 }$ & $ 457.32 ^{+ 31.26 }_{- 34.59 }$ & $ 3.43 ^{+ 0.61 }_{- 0.44 } \times 10^{ -6 }$ & 29.26 & -25.32 & 3.07 \\
\enddata
\end{deluxetable}
\tablecomments{All columns are the same as Table~\ref{tab:fitresults}.}
\end{rotatetable}

\movetabledown=3.6in
\begin{rotatetable}
\begin{deluxetable}{cccccccccccccccc}
\tabletypesize{\scriptsize}
\tablecaption{Time-resolved spectral analysis results of GRB111017657.}
\setlength{\tabcolsep}{0pt}
\tablehead{ \colhead{$t_{\rm start}$} & \colhead{$t_{\rm stop}$} & \colhead{$S$} & \colhead{$K$}& \colhead{$\alpha$} & \colhead{$E_{\rm c}$} &\colhead{$E_{\rm p}$}& \colhead{$F$} & \colhead{$K_{\rm BAND}$} & \colhead{$\alpha_{\rm BAND}$} & \colhead{$\beta_{\rm BAND}$} & \colhead{$E_{\rm p,BAND}$} & \colhead{$F_{\rm BAND}$} & \colhead{$\Delta$DIC} & \colhead{$p_{\rm DIC}$} & \colhead{$p_{\rm DIC, BAND}$} \\
\colhead{(1)} & \colhead{(2)} & \colhead{(3)} & \colhead{(4)} & \colhead{(5)} & \colhead{(6)} & \colhead{(7)} & \colhead{(8)} & \colhead{(9)} & \colhead{(10)} & \colhead{(11)} & \colhead{(12)} & \colhead{(13)} & \colhead{(14)} & \colhead{(15)} & \colhead{(16)}
}
\startdata
-5.00 & -1.80 & 2.60 & $ 2.67 ^{+ 0.17 }_{- 2.56 } \times 10^{ +0 }$ & $ -1.66 ^{+ 0.28 }_{- 0.20 }$ & $ 4980.60 ^{+ 1936.42 }_{- 4740.17 }$ & $ 1691.75 ^{+ 657.74 }_{- 1610.09 }$ & $ 6.92 ^{+ 25.80 }_{- 5.60 } \times 10^{ -8 }$ & $ 7.79 ^{+ 3.29 }_{- 7.79 } \times 10^{ +1 }$ & $ 2.17 ^{+ 0.83 }_{- 0.22 }$ & $ -2.08 ^{+ 0.48 }_{- 0.02 }$ & $ 27.99 ^{+ 3.26 }_{- 17.65 }$ & $ 4.61 ^{+ 136.77 }_{- 4.44 } \times 10^{ -8 }$ & -675.72 & -2.34 & -680.65 \\
-1.80 & 0.77 & 9.00 & $ 1.02 ^{+ 0.26 }_{- 0.54 } \times 10^{ +0 }$ & $ -1.20 ^{+ 0.09 }_{- 0.12 }$ & $ 2971.81 ^{+ 492.87 }_{- 2326.54 }$ & $ 2376.02 ^{+ 394.06 }_{- 1860.11 }$ & $ 3.30 ^{+ 4.23 }_{- 1.70 } \times 10^{ -7 }$ & $ 4.02 ^{+ 0.10 }_{- 0.78 } \times 10^{ -3 }$ & $ -1.13 ^{+ 0.08 }_{- 0.17 }$ & $ -3.55 ^{+ 0.51 }_{- 1.45 }$ & $ 2019.01 ^{+ 331.90 }_{- 1578.53 }$ & $ 3.57 ^{+ 0.93 }_{- 0.73 } \times 10^{ -7 }$ & -2.15 & -1.05 & 0.19 \\
0.77 & 2.68 & 18.10 & $ 9.45 ^{+ 2.27 }_{- 3.86 } \times 10^{ -1 }$ & $ -0.93 ^{+ 0.07 }_{- 0.10 }$ & $ 1255.60 ^{+ 209.60 }_{- 728.21 }$ & $ 1337.33 ^{+ 223.24 }_{- 775.62 }$ & $ 1.22 ^{+ 1.16 }_{- 0.55 } \times 10^{ -6 }$ & $ 1.29 ^{+ 0.09 }_{- 0.14 } \times 10^{ -2 }$ & $ -0.87 ^{+ 0.08 }_{- 0.10 }$ & $ -3.23 ^{+ 1.17 }_{- 0.55 }$ & $ 975.25 ^{+ 121.16 }_{- 382.96 }$ & $ 1.23 ^{+ 0.26 }_{- 0.20 } \times 10^{ -6 }$ & 4.63 & -7.98 & 1.30 \\
2.68 & 3.60 & 23.20 & $ 8.29 ^{+ 2.01 }_{- 2.89 } \times 10^{ -1 }$ & $ -0.76 ^{+ 0.07 }_{- 0.09 }$ & $ 614.21 ^{+ 89.70 }_{- 180.22 }$ & $ 759.33 ^{+ 110.90 }_{- 222.80 }$ & $ 2.06 ^{+ 1.42 }_{- 0.88 } \times 10^{ -6 }$ & $ 2.41 ^{+ 0.16 }_{- 0.21 } \times 10^{ -2 }$ & $ -0.73 ^{+ 0.07 }_{- 0.09 }$ & $ -3.63 ^{+ 0.52 }_{- 1.33 }$ & $ 698.81 ^{+ 85.79 }_{- 144.97 }$ & $ 2.14 ^{+ 0.33 }_{- 0.33 } \times 10^{ -6 }$ & 2.91 & -2.42 & 2.75 \\
\enddata
\end{deluxetable}%
\tablecomments{All columns are the same as Table~\ref{tab:fitresults}.}
\end{rotatetable}

\end{document}